\documentclass[A4,10pt]{article}
\usepackage[margin=1in,dvips]{geometry}
\usepackage{graphicx,cancel}
\usepackage{graphics}
\usepackage{color, soul}
\usepackage{amsmath, amsthm, slashed}
\usepackage{amssymb}
\usepackage{marvosym}
\usepackage{rotating}
\usepackage{mathrsfs}
\usepackage{epsfig}
\usepackage{comment}
\usepackage[affil-it]{authblk}
\usepackage{rotating}
\usepackage{graphicx}
\usepackage{accents}
\usepackage{harmony, slashed}
\usepackage{tikz}
\usetikzlibrary{patterns}
\usepackage{scalerel}

\usepackage{enumerate}
\usepackage{stackengine,wasysym}

\newtheorem{definition}{Definition}[section]
\newtheorem{theorem}{Theorem}[section]
\newtheorem{conjecture}{Conjecture}
\newtheorem*{conjecture*}{Conjecture}

\newtheorem{corollary}{Corollary}[section]
\newtheorem*{theorem*}{Theorem}
\newtheorem*{corollary*}{Corollary}
\newtheorem{proposition}{Proposition}[section]
\newtheorem{lemma}{Lemma}[section]
\newtheorem{remark}{Remark}[section]

\DeclareMathAlphabet\mathbfcal{OMS}{cmsy}{b}{n}

\title{Naked Singularities for the Einstein Vacuum Equations: \\ The Interior Solution}

\date\today
\author[1,2]{Yakov Shlapentokh-Rothman}

\affil[1]{\small University of Toronto, Department of Mathematics, 40 St.~George~Street, Toronto, ON, Canada\vskip.2pc \ }
\begin{document}
\affil[2]{\small University of Toronto Mississauga, Department of Mathematical and Computational Sciences, 3359 Mississauga Road, Mississauga, ON, Canada\vskip.2pc \ }

\maketitle
\begin{abstract}In a previous work~\cite{nakedone} we constructed solutions to the Einstein vacuum equations in $3+1$ dimensions which corresponded to the exterior region of a naked singularity. In this work we construct solutions which correspond to the interior region and show that the two solutions may be glued together to produce a naked singularity. Fundamental to our construction is the novel type of self-similarity for the Einstein vacuum equations that we introduced in our previous work~\cite{nakedone} and also the study of a new class of quasilinear PDE's of mixed degenerate elliptic-hyperbolic type. 
\end{abstract}
\tableofcontents
\section{Introduction}
Central to the modern understanding of classical General Relativity in $3+1$ dimensions is Penrose's famous Weak Cosmic Censorship Conjecture~\cite{Penroseergo}:
\begin{conjecture}[Weak Cosmic Censorship: Informal Version]\label{cosmic}Spacetimes $\left(\mathcal{M},g\right)$ which solve the Einstein vacuum equations
\[{\rm Ric}\left(g\right) = 0\]
and arise as maximal developments of  suitably regular, asympotically flat Cauchy (or characteristic) data, \underline{generically} have their singularities hidden inside of a black hole region which is causally disconnected from a regular exterior region.
\end{conjecture}
The importance of this conjecture stems from the fact that if it holds, then despite the presence of singular solutions in General Relativity (such as the Schwarzschild black hole~\cite{schwarzschild1916}), as long as one does not venture into the ``black hole region,''  the physics of isolated self-gravitating systems can be understood without needing to access any possible modifications that General Relativity must undergo near a singularity. The Weak Cosmic Censorship Conjecture may be loosely thought of as the fundamental ``global existence conjecture'' for General Relativity.

We have emphasized the word ``generic'' in the statement of Conjecture~\ref{cosmic}. However, this word was \emph{not} present in the original formulations of Conjecture~\ref{cosmic}. Indeed, it is primarily due to a sequence of works of Christodoulou~\cite{ChristBV,ChristNaked,Christodoulou4}, which established the analogue of Conjecture~\ref{cosmic} for the \emph{spherically-symmetric} Einstein--scalar field system, that we understand both a heuristic mechanism by which singularities and black hole formation may be so inextricable linked and also that one \underline{does} expect the existence of singular spacetimes which do not contain a black hole region, even though these solutions should be non-generic. These singular spacetimes which do not contain a black hole region are called \emph{naked singularities}. 

In an earlier work, Christodoulou also constructed naked singularity solutions for the spherically symmetric Einstein--null dust model~\cite{dustnaked}, and later there were heuristic and numerical studies of spherically symmetric naked singularities for fluid models which allow for pressure (see~\cite{nakedfluidoripiran,nakedfluidjoshidwivedi}). Very recently Guo--Hadzic--Jang~\cite{nakedeulereinstin} have provided a rigorous construction of the solutions studied in~\cite{nakedfluidoripiran}.  There is also a large heuristic and numerical literature concerning naked singularities associated to \emph{critical phenomenon} in spherical symmetry (see~\cite{chop,critrotate,critsurv}) and naked singularities which solve the  Einstein vacuum equations in higher dimensions (see~\cite{blackstringnakedsing,nakedsinganzhang,ultraspinviolate}).\footnote{In higher dimensions, naked singularity formation is often associated with the occurrence of Gregory--Laflamme instabilities~\cite{gregflam}.}

Given Christodoulou's result for the spherically symmetric Einstein-scalar field system, it is natural to ask whether we can construct such naked singularities for the Einstein vacuum equations in $3+1$ dimensions. As is well-known, due to Birkhoff's Theorem~\cite{birkhofftheo} the Einstein vacuum equations do not possess any dynamical degrees of freedom in spherical symmetry. In particular, we cannot expect to directly use the naked singularities of Christodoulou. Furthermore, the strategy behind Christodoulou's construction fundamentally relies on a reduction of the spherically symmetric Einstein-scalar field system by way of a certain ``$k$-self-similarity'' to a two dimensional autonomous system. There does not exist any such dramatic reduction for the Einstein vacuum equations which is consistent with the requirement of asymptotic flatness; thus a completely different strategy is needed. 

Before we turn directly to our main results, we give here a Penrose diagram corresponding to a naked singularity:
\begin{center}
\begin{tikzpicture}[scale = 1]
\fill[lightgray] (0,0)--(0,-4)--(4,0)  -- (2,2) -- (0,0);
\draw (0,0) -- (0,-4);
\draw (0,0) -- (2,-2) node[sloped,above,midway]{\footnotesize $\mathcal{H}$};
\draw (0,-4) -- (4,0) node[sloped,below,midway]{\footnotesize asymptotically flat cone};
\draw[dashed] (4,0) -- (2,2) node[sloped,above,midway]{\footnotesize $\mathcal{I}^+$};
\draw[dashed] (2,2) -- (0,0) node[sloped,above,midway]{\footnotesize $\mathcal{W}$};
\path [draw=black,fill=white] (0,0) circle (1/16); 
\draw (-.8,0) node[above]{\footnotesize singularity};
\draw (.75,-2) node[above]{\footnotesize $\mathcal{M}_{\rm int}$};
\draw (2,0) node[above]{\footnotesize $\mathcal{M}_{\rm ext}$};
\end{tikzpicture}
\\

\text{Figure 1}
\end{center}
Here $\mathcal{H}$ denotes the past line cone of the singularity and $\mathcal{W}$ represents the future light cone of the singularity (which is not actually part of the spacetime). The singularity is naked if there are a sequence of ingoing null vectors which are parallel transported to the asymptotically flat end of the initial cone and so that the corresponding ingoing null geodesics $\gamma$ all intersect $\mathcal{W}$ in affine time less than $A$ for some fixed constant $A$. In this case we also say that \emph{future null infinity is incomplete}.\footnote{This definition is in the spirit of that given in~\cite{Chrmil}. In particular, we do not rely on an explicit conformal compactification.} We call the region $\mathcal{M}_{\rm ext}$ to the future of the cone $\mathcal{H}$, the ``exterior region'' of the naked singularity, and the region $\mathcal{M}_{\rm int}$ to the past of the cone $\mathcal{H}$, the ``interior region.'' As we shall see, it is natural to consider separately the problem of constructing the solution in the exterior and interior region.

In~\cite{nakedone} we initiated the study of naked singularities for the Einstein vacuum equations in $3+1$ dimensions by constructing solutions which correspond to the exterior region of a naked singularity. A key idea in our construction was the introduction  of a new type of self-similarity for the Einstein vacuum equations which generalizes the type of self-similar solutions associated to the \emph{Ambient Metric} of Fefferman--Graham~\cite{FG1,FG2,scaleinvariant}. By necessity this construction required us to consider large data; on the past cone $\mathcal{H}$ of the singularity we had that the outgoing shear satisfies $\Omega^{-1}\hat{\chi}\sim \epsilon^{-1}$ where $0 < \epsilon \ll 1$. However, we identified a mechanism, consistent with the underlying self-similarity, by which one may arrange for $\Omega^{-1}\hat{\chi}$ to become small quickly as moves away from this cone. This gain of smallness allowed us to close our arguments.

In this paper we turn to the construction of the corresponding interior solution. Before we state the main results of the paper, it is convenient to introduce the convention that $g$ is said to be ``in the double-null form'' if 
\begin{equation}\label{32iojjiooi4}
	g = -2\Omega^2\left(du\otimes dv + dv\otimes du\right) + \slashed{g}_{AB}\left(d\theta^A - b^Adu\right)\otimes \left(d\theta^B - b^Bdu\right)
	\end{equation}
for some choice of coordinates $\left(u,v,\theta^A\right) \in \mathcal{U} \times \mathbb{S}^2$ where $\mathcal{U} \subset \mathbb{R}^2$. Our main theorem concerning the interior solution is the following.
\begin{theorem}\label{maintheohere}Let $N \gg 1$ be any sufficiently large integer and let $0 < \epsilon \ll \gamma \ll 1$ be sufficiently small (possibly depending on $N$). Then there exists a spacetime $\left(\mathcal{M}_{\rm int},g\right)$ which solves the Einstein vacuum equations and so that the following hold:
\begin{enumerate}
	\item The manifold $\mathcal{M}_{\rm int}$ is diffeomorphic to $\mathcal{C} \doteq \{(t,x,y,z) \in \mathbb{R}^4 : t < 0 \text{ and }\sqrt{x^2+y^2+z^2} \leq -t\}$. We then write $\mathcal{M}_{\rm int}$  as a disjoint union of $\mathscr{A}$, $\mathring{\mathcal{M}}$, and $\mathcal{H}$, where $\mathscr{A}$ corresponds to $\{x = y = z = 0\}$, $\mathcal{H}$ corresponds to $\left\{\sqrt{x^2+y^2+z^2} = -t\right\}$, and $\mathring{\mathcal{M}}$ corresponds to $0 < \sqrt{x^2+y^2 + z^2} < -t$. The Penrose diagram of the spacetime is as follows:
	
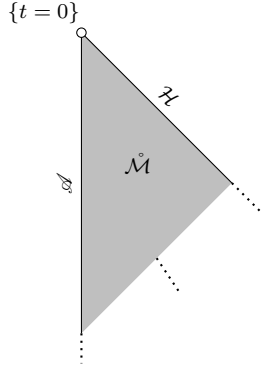
\begin{center}
\begin{tikzpicture}[scale = 1]
\fill[lightgray] (0,0)--(0,-4)--(2,-2);
\draw (0,-4) -- (0,0) node[sloped,above,midway]{\footnotesize $\mathscr{A}$};
\draw [dotted,thick] (0,-4)--(0,-4.4);
\draw (0,0) -- (2,-2) node[sloped,above,midway]{\footnotesize $\mathcal{H}$};
\draw [dotted,thick] (2,-2) -- (2.4,-2.4);
\draw (.75,-2) node[above]{\footnotesize $\mathring{\mathcal{M}}$};
\path [draw=black,fill=white] (0,0) circle (1/16); 
\draw (-.5,0) node[above]{\footnotesize $\{t = 0\}$};
\draw [dotted,thick] (1,-3) -- (1.3,-3.45);
\end{tikzpicture}
\end{center}

	\item The manifold $\mathring{\mathcal{M}}$ is diffeomorphic to $\mathcal{U} \times \mathbb{S}^2$ where  $\mathcal{U} \doteq \left\{\left(u,v\right) : u \in (-\infty,0),\ v \in (u,0)\right\}$. The metric $g$ takes the double-null form within $\mathcal{U} \times \mathbb{S}^2$ and is $C^N$. The vector field 
	\[K \doteq u\partial_u + v\partial_v\]
	is conformally Killing and satisfies
	\[\mathcal{L}_Kg = 2g.\]
	\item Define a function $\hat{v} : \mathring{\mathcal{M}} \to \mathbb{R}$ by
	\[\hat{v}\left(v,u,\theta^A\right) \doteq -\int_v^0 \Omega^2\left(\tilde{v},u,\theta^A\right)\, d\tilde{v},\]
	where $\Omega^2$ is the lapse function from the double-null expression~\eqref{32iojjiooi4}. Then $\left(\hat{v},u,\theta^A\right)$ may be used to define a coordinate system on $\mathring{\mathcal{M}}$. In $\left(\hat{v},u,\theta^A\right)$ coordinates, the metric $g$ extends to $\mathring{\mathcal{M}} \cup \{\hat{v} = 0\}$ which we identify with $\mathring{\mathcal{M}}\cup \mathcal{H}$. The boundary hypersurface $\{\hat{v} = 0\}$ is null, and there exists an $s > 0$ so that this extended metric is $C^{1,s}_{\hat{v}}C^N_{u,\theta^A}$. In the $\left(\hat{v},u,\theta^A\right)$ coordinates, the vector field $K$ takes the form $u\partial_u + \hat{v}\partial_{\hat{v}}$. 
		\item Let $\underline{\mathring{\mathcal{M}}} \doteq \mathring{\mathcal{M}} \cap \{x^2+y^2 + z^2 \leq t^2/2\}$. Then $\underline{\mathring{\mathcal{M}}} \cup \mathscr{A}$ may be covered with self-similar wave coordinates $\{\xi^{\alpha}\}$ satisfying
	\[\Box_g\xi^{\alpha} = 0,\qquad \mathcal{L}_K\xi^{\alpha} = \xi^{\alpha}.\]
	The metric $g$ is smooth when expressed in $\{\xi^{\alpha}\}$ coordinates.
	\item The cone $\mathcal{H}$ satisfies the following properties:
	\begin{enumerate}
		\item\label{coneprop1} $\lim_{v\to 0}\left|\Omega\hat{\underline{\chi}}\right|_{\slashed{g}} \sim \epsilon \frac{\left|a\right|}{-u} $, where $a: \mathbb{S}^2 \to [0,\infty)$ satisfies ${\rm Area}\left(\theta^A \in \mathbb{S}^2 : a \leq 1/2\right) \lesssim \gamma$ (where the area is computed with respect to the round metric).
		\item\label{coneprop2} Letting $m\left(u\right)$ denote the Hawking mass of $\mathbb{S}^2_{u,0}$, we have $\frac{m(u)}{\sqrt{\rm Area\left(\mathbb{S}^2_{u,0}\right)}} \sim \epsilon^2$.
	\end{enumerate}
	\end{enumerate}
\end{theorem}

We have furthermore arranged our interior construction so that we may glue the solution from Theorem~\ref{maintheohere} to one of the exterior solutions produced in~\cite{nakedone}. Before stating this explicitly as a theorem, we quickly recall the parts of the main result from~\cite{nakedone} which will be relevant to us here.
\begin{theorem}\label{exteriornaked}\cite{nakedone} Let $N \gg 1$ be any sufficiently large integer and let $0 < \epsilon \ll \gamma \ll 1$ be sufficiently small (possibly depending on $N$).  Then there exists a spacetime $\left(\mathcal{M}_{\rm ext},g\right)$ which solves the Einstein vacuum equations such that the following holds:
\begin{enumerate}
	\item $\left(\mathcal{M}_{\rm ext},g\right)$ is covered by double-null coordinates $\left\{\left(u,v,\theta^A\right) \in [-1,0) \times [0,\infty) \times \mathbb{S}^2\right\}$ where the metric takes the double-null form~\eqref{32iojjiooi4}. The null hypersurface $\{u = -1\}$ is asymptotically flat and the Penrose diagram of $\left(\mathcal{M}_{\rm ext},g\right)$ is as follows:
	
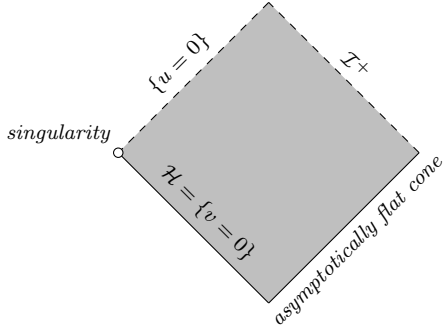
\begin{center}
\begin{tikzpicture}[scale = 1]
\fill[lightgray] (0,0)--(2,-2)--(4,0)  -- (2,2) -- (0,0);
\draw (0,0) -- (2,-2) node[sloped,above,midway]{\footnotesize $\mathcal{H} = \{v = 0\}$};
\draw (2,-2) -- (4,0) node[sloped,below,midway]{\footnotesize asymptotically flat cone};
\draw[dashed] (4,0) -- (2,2) node[sloped,above,midway]{\footnotesize $\mathcal{I}^+$};
\draw[dashed] (2,2) -- (0,0) node[sloped,above,midway]{\footnotesize $\{u = 0\}$};
\path [draw=black,fill=white] (0,0) circle (1/16); 
\draw (-.8,0) node[above]{\footnotesize singularity};
\end{tikzpicture}
\end{center}
\item In the region $\{v > 0\}$, the metric $g$ is $C^N$. If we include $\{v = 0\}$ then there exists an $s > 0$ so that the metric  is $C^{1,s}_{\hat{v}}C^N_{u,\theta^A}$.
\item The cone $\{ v= 0\}$ satisfies the properties~\ref{coneprop1} and~\ref{coneprop2} from Theorem~\ref{maintheohere}.
\item Future null infinity is incomplete. 
\end{enumerate}
\end{theorem}

This next theorem is the statement that these two spacetimes may be glued together along the hypersurface $\mathcal{H}$.
\begin{theorem}\label{glueityay}Let $N \gg 1$ be any sufficiently large integer and let $0 < \epsilon \ll \gamma \ll 1$ be sufficiently small (possibly depending on $N$).  Then we can pick spacetimes $\left(\mathcal{M}_{\rm int},g\right)$ and $\left(\mathcal{M}_{\rm ext},g\right)$ produced by Theorems~\ref{maintheohere} and~\ref{exteriornaked} so that $\left(\mathcal{M}_{\rm int}\cap \{u \geq -1\},g\right)$ may be glued to $\left(\mathcal{M}_{\rm ext},g\right)$ by a suitable identification of the hypersurfaces $\mathcal{H}$ with each other. 
\begin{enumerate}
\item Away from the hypersurface $\mathcal{H}$, the resulting metric $g$ is $C^N$. If we include $\mathcal{H}$ then there exists an $s > 0$ so that the metric  is $C^{1,s}_{\hat{v}}C^N_{u,\theta^A}$ for a suitable coordinate system $\left(\hat{v},u,\theta^A\right)$ where $\mathcal{H} = \{\hat{v} = 0\}$. 
\item The spacetime has the Penrose diagram depicted in Figure 1 above. 
\item The cone $\mathcal{H}$ satisfies the properties~\ref{coneprop1} and~\ref{coneprop2} from Theorem~\ref{maintheohere}.
\item The resulting spacetime is the maximal development associated to the complete asymptotically flat cone $\{u = -1\}$. Future null infinity is incomplete, and thus the spacetime represents a naked singularity. 
\end{enumerate}

\end{theorem}

\begin{remark}\label{k2iji5}(Comparison with the Naked Singularities of Christodoulou) The solution produced by Theorem~\ref{glueityay} shares many essential qualitative features with Christodoulou's naked singularities~\cite{ChristNaked}. We direct the reader to the discussion in Section 1.2 of~\cite{nakedone}.
\end{remark}

\begin{remark}(Regularity of the Initial Data) The spacetime produced by Theorem~\ref{glueityay} does not arise from smooth characterstic data and the reader may wonder why these spacetimes are regular enough to justify being called naked singularities (cf.~the word ``regular'' in Conjecture~\ref{cosmic}). We refer the reader to the discussion at the end of Section 1.1.3 and in Section 1.2 of~\cite{nakedone}. We quickly summarize the relevant points here: 
\begin{enumerate}
	\item Christodoulou's naked singularities are also not smooth: the derivative of the scalar field is only H\"{o}lder continuous. However, the solutions are more regular than the class of so-called ``bounded variation solutions'' for which Christodoulou~\cite{ChristBV} established a well-posedness result,\footnote{Also in~\cite{ChristBV} Christodoulou showed that Minkowski space is stable to small perturbations in the bounded variation solution class.} and hence we may consider the naked singularities to be sufficiently regular.
	\item The solutions produced by Theorem~\ref{glueityay} may be constructed so that the initial characteristic data has as much regularity in the angular directions as one wishes. The only singular behavior is in the null direction where the metric is $C^{1,s}$ for some $s > 0$. This is thus formally analogous to Christodoulou's naked singularity solutions. Moreover, one expects a well-posedness result may be established for the Einstein vacuum equations which encompasses characterstic data with high angular regularity and which are $C^{1,s}$ regular in the null direction (cf.~\cite{impulsivefirst} and~\cite{impulsive}).
\end{enumerate}
\end{remark}

\begin{remark}(Nature of the Singularity) Christodoulou's naked singularity solutions are inextendible as solutions of bounded variation. As discussed in Section 1.2 of~\cite{nakedone}, these may seen by either noting that $\frac{m}{r}$ does not converge to $0$ as we approach the singularity (where $m$ denotes the Hawking mass) or the fact that $\partial_u\phi$ (where $\phi$ is the scalar field) is not locally integrable along the past light cone of the singularity. The properties~\ref{coneprop1} and~\ref{coneprop2} from Theorem~\ref{maintheohere} formally correspond to these two properties (see the discussion in Section 1.2 of~\cite{nakedone}). 

Though we will not give the proofs here, it is also possible to show that for certain of the solutions produced by Theorem~\ref{glueityay}, there exists future inextendible null geodesic $\varphi(s) : [a,b) \to \mathcal{M}$ so that 
	\begin{enumerate}
	\item There exists a Jacobi field $J$ along $\varphi$ with $\left|J\right| \to \infty$ as $s\to b$. This has the physical interpretation that there are infinite tidal forces at the singularity.
	\item We have  a ``strong singularity'' along $\varphi$ as $s\to b$ in the sense of~\cite{TiperStrongSing} (see also Chapter 8.2 of the textbook~\cite{clarkbook}).\footnote{We recall the definition for the convenience of the reader: If $\varphi(s) : [a,b)$ is a future inextendible null geodesic, then we say that the geodesic terminates in a ``strong singularity'' if for every $c \in (a,b)$ and every linearly independent pair of Jacobi fields $J_1$ and $J_2$ which are perpendicular to $\varphi'$, satisfy $J_1(c) = J_2(c) = 0$, and satisfy $\left|J_1'(c)\right| = \left|J_2'(c)\right| = 1$, we have that the norm of the corresponding volume form $\left(J_1\right)_{\flat} \wedge \left(J_2\right)_{\flat}$ converges to $0$ as $s\to b$.} This has the physical interpretation that any object traveling along the geodesic $\varphi$ will be crushed to zero volume as $s\to b$.
	\item We have that 
	\[\int_a^b \left|{\rm tf}\left(\mathcal{L}_{\varphi'}\slashed{g}\right)\right|_{\slashed{g}}\, ds = \infty. \]
	Here ${\rm tf}$ denotes the trace-free part. This is a reparametrization invariant way to state that the shear along $\varphi$ is not integrable as $s\to b$. 
	\end{enumerate}
	
	These facts are all formally consistent with the occurrence of a $C^0$-type singularity of the metric at $(u,v) = (0,0)$. We plan to carry out a more systematic study of the singularity in a future work.
	\end{remark}

In the rest of this introductory section we will discuss at a high level various aspects of the proofs of these theorems.

\subsection{Relation to the Ambient Metric of Fefferman--Graham}
It is instructive to compare Theorem~\ref{maintheohere} with analogous results in the context of Fefferman and Graham's  Ambient Metric~\cite{FG1,FG2}. We start with a quick review of the relevant aspects of the Ambient Metric construction: We let $n \geq 2$ and take as our seed data for the construction a Reimannian metric $\left(\mathbb{S}^n,\slashed{g}_0\right)$ and a $\slashed{g}_0$-trace free symmetric $(0,2)$-tensor $k$ on $\mathbb{S}^n$.\footnote{One may in fact replace $\mathbb{S}^n$ with any compact manifold of dimension $n$, and let $\slashed{g}_0$ be Pseudo-Riemannian.} The Ambient Metric is then equivalent\footnote{Fefferman and Graham did not work in double-null coordinates, but for the purpose of comparison with the current paper, it is useful to present their work in the double-null gauge. (See the introduction of~\cite{scaleinvariant}.)} to a formal power series expansion in $\frac{v}{u}$ corresponding to a Ricci flat $n+2$-dimensional Lorentzian metric $\left(\mathcal{M},g\right)$ in the double-null form~\eqref{32iojjiooi4} with $\mathcal{U} = \{u \in (-\infty,0)\text{ and }-\epsilon \leq \frac{v}{-u} \leq \epsilon\}$ for $0 < \epsilon \ll 1$ and so that
\begin{enumerate}
	\item The metric $g$ is self-similar in that $K = u\partial_u+v\partial_v$ is conformally Killing and we have that 
	\begin{equation}\label{32ioij92}
	\Omega\left(v,u,\theta^A\right) = \tilde{\Omega}\left(\frac{v}{u},\theta^A\right),\ b_A\left(v,u,\theta^B\right) = u\tilde{b}_A\left(\frac{v}{u},\theta^B\right),\ \slashed{g}_{AB}\left(u,v,\theta^C\right) = u^2\tilde{\slashed{g}}_{AB}\left(\frac{v}{u},\theta^C\right),
	\end{equation}
	for suitable $\tilde{\Omega}$, $\tilde{b}$, and $\tilde{g}_{AB}$. 
	\item We have that 
	\[g|_{v = 0} = -2\left(du\otimes dv + dv\otimes du\right) + u^2\left(\slashed{g}_0\right)_{AB}d\theta^A\otimes d\theta^B,\qquad {\rm tf}\left(\mathcal{L}_{\partial_v}^{\frac{n}{2}}\slashed{g}\right)_{AB}|_{v=0} = u k_{AB},\]
	where ${\rm tf}$ refers to the trace-free part and, if $n$ is odd, $\mathcal{L}_{\partial_v}^{\frac{n}{2}}$ is defined in terms of suitable coefficient in the power series expansion for $\slashed{g}$. 
\end{enumerate}
The following diagram depicts the formal region covered by the power series expansions:
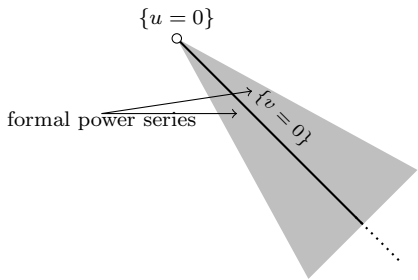
\begin{center}
\begin{tikzpicture}
\fill[lightgray] (0,0)--(3.2,-1.75)--(1.75,-3.2)  -- (0,0);
\draw[thick] (0,0) -- (2.475,-2.475) node[sloped,above,midway]{\scriptsize $\{v = 0\}$}; 
\path [draw=black,fill=white] (0,0) circle (1/16); 
\draw (0,0) node[above]{\footnotesize $\{u = 0\}$};
\draw[dotted,thick] (2.375,-2.375)--(3,-3) ;

\draw[->] (-1,-1) -- (.8,-1);
\draw[->] (-1,-1) -- (1,-.7);
\draw  (-1,-1.1) node{\footnotesize formal power series};

\end{tikzpicture}
\end{center}

These power series expansions are the starting point for many interesting questions in mathematical physics (see the survey paper~\cite{andersonsurvey}) but we will here focus on the particular question of finding a past complete ``fill-in'' for the null cone $\{v = 0\}$:
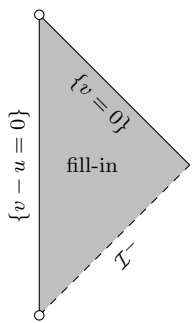
\begin{center}
\begin{tikzpicture}
\fill[lightgray] (0,0)--(0,-4)--(2,-2)  -- (0,0);

\draw (0,-4)--(0,0) node[sloped,above,midway]{\footnotesize $\{v-u = 0\}$};
\draw (0,0) -- (2,-2) node[sloped,below,midway]{\footnotesize $\{v = 0\}$};
\draw [dashed] (0,-4) -- (2,-2) node[sloped,below,midway]{\footnotesize $\mathcal{I}^-$};
\path [draw=black,fill=white] (0,0) circle (1/16); 
\path [draw = black,fill = white] (0,-4) circle (1/16);
\draw (.7,-2) node{\footnotesize $\text{fill-in}$};

\end{tikzpicture}
\end{center}
It is natural to attempt to quotient out by the self-similarity and produce an equivalent problem on an $n+1$ dimensional manifold. As is well-known, Fefferman and Graham~\cite{FG1,FG2} discovered a remarkable way to carry out such a \emph{Kaluza--Klein} reduction. Namely, they showed that if one assumes a so-called ``straightness condition''\footnote{In terms of the double-null gauge, the straightness condition is the assumption that the lapse $\Omega$ is identically $1$ everywhere and that the shift vector $b$ vanishes identically.} then by considering the induced metric along a hyperbola $\{uv = 1\}$, the existence of a past-complete fill-in metric is equivalent to finding a complete $n+1$ dimensional (Riemannian) Poincar\'{e}--Einstein metric which induces the metric $\slashed{g}_0$ on the copy of $\mathbb{S}^n$ which lies on the boundary of a suitable conformal compactification of the hyperbola. A key fact behind the success of this procedure is that the straightness condition forces the self-similar vector field $K = u\partial_u+v\partial_v$ to be strictly \emph{timelike} in the region $\{v < 0\}$. Hence it is reasonable to expect the quotient metric to be Riemannian, and the resulting PDE problems may be expected to be (degenerate) elliptic. 

The Poincar\'{e}--Einstein reduction was exploited by Graham--Lee in the work~\cite{grahamlee} where they showed that when $n \geq 3$ there exists a past complete fill-in metric whenever $\slashed{g}_0$ is sufficiently close to the round metric on $\mathbb{S}^n$. (When $n = 2$, in view of the fact that the Ricci tensor determines the curvature tensor for a three dimensional manifold, the Poincar\'{e}--Einstein reduction can only result in obtaining fill-in metrics which are flat.) After the introduction of suitable (generalized) harmonic coordinates, Graham--Lee were able to treat the existence problem for the Poincar\'{e}--Einstein metric as a quasilinear degenerate elliptic (compactified) boundary value problem. In this framework, the tensor $\slashed{g}_0$ corresponds to Dirichlet data along the boundary, and the tensor $k$ corresponds to the Neumann data along the boundary. 

We now compare and contrast the above with the setting of the current paper. We first note that we work with a generalization of the self-similarity of Fefferman--Graham where~\eqref{32ioij92} only holds for $v \neq 0$ and the metric extends to $\{v = 0\}$ instead in an alternative coordinate system (see Section 3 of~\cite{nakedone} and Section~\ref{iojoij284982} below for a more detailed discussion). Our Theorem~\ref{maintheohere} plays a similar role for this new type of self-similarity as the result of Graham--Lee~\cite{grahamlee} plays for the self-similarity of Fefferman--Graham. Most importantly, our Theorem~\ref{maintheohere} is able to produce a non-flat solution in the case of $n = 2$, and thus may be used to construct naked singularities in $3+1$ dimensions. There are, however, at least four fundamental differences which significantly complicate the problem in our setting:
\begin{enumerate}
	\item\label{3joijo2} Loss of ellipticity:  A key geometric difference in the two types of self-similarity is that the self-similar vector field $K$ is null along $\{v = 0\}$ for the Fefferman--Graham spacetimes while for our new ``twisted'' self-similarity the self-similar vector field $K$ must be \emph{spacelike} on some portions of $\{v = 0\}$. In particular, when one considers the analogous problem of finding a fill-in metric, by continuity, the vector field $K$ will be spacelike is some of regions of the fill-in. Thus one cannot expect to reduce the problem to a degenerate elliptic PDE. In fact, as we will see, the vector field $K$ is also timelike in other regions of the fill-in metric, and thus we will need to solve a \underline{mixed degenerate elliptic-hyperbolic problem}.
	\item\label{1ij2oijoi41} Loss of smallness near past light-cone of the singularity: For Fefferman--Graham spacetimes, if one assumes that boundary metric is sufficiently small and regular, then the formal expansions for the corresponding Poincar\'{e}--Einstein metric $g$ are consistent with metric $g$ being small in the H\"{o}lder space $C^{n-1,\alpha}$ defined with respect to a suitable coordinate system which includes the boundary. In particular, $g$ may be expected to lie in and be small in $C^{1,\alpha}$ and this is consistent with estimating the nonlinear terms in a purely perturbative fashion. In contrast, in the twisted self-similar setting of the current paper, the analogous formal expansions show that even if the tangential boundary data are size $\epsilon$, then the first normal derivative of the metric to the boundary  may be size $\epsilon^{-1}$ near the past light cone of the singularity. As in our work~\cite{nakedone}, ultimately we find that this largeness dissipates quickly away from the cone, and derivatives of $g$ turn out to be small in $L^p$ for $p \lesssim 1$. However, with only this $L^p$ smallness to work with, the specific structure of the nonlinear terms becomes crucial for the validity of the argument. Related to this largeness as $\epsilon\to 0$ is the fact is that while the Dirichlet to Neumann map may be linearized around the trivial solution in the context of Poincar\'{e}--Einstein metrics (see~\cite{GrahamDtoN} for some explicit formulas), in the setting of this paper, the corresponding linearization of the Dirichlet to Neumann map at $\epsilon = 0$ cannot even be defined.
	\item No clean Kaluza--Klein reduction: Our new twisted self-similarity is specifically generated by allowing a non-trivial shift vector along $\{v = 0\}$. In particular, there is no analogue of the straightness condition of Fefferman--Graham. It is still of course possible to quotient out by the self-similarity to produce an equivalent system of equations along an $n+1$ dimensional hyperbola. However, due to the non-triviality of the lapse and shift vectors, one will not get a clean curvature equation for the induced metric, and the reduced equations for the lapse and shift do not appear to exhibit any particularly useful simplifications. (As we will see later, we instead use self-similarity to reduce the the system to a set of equations along a suitable null hypersurface.)
	\item Constrained boundary data: One of the appealing geometric features of the Fefferman--Graham theory is the clean parametrization of the free data in the construction in terms of the tensors $\left(\slashed{g}_0,k\right)$. In contrast, seed data for our new twisted self-similarity consists of a suitable solution to the null constraint equations (see Section 3 of~\cite{nakedone}). These constraint equations depend themselves in a nonlinear fashion on the solution, so when we prove Theorem~\ref{maintheohere} we will have to simultaneously solve equations in the ``bulk'' $\{v < 0\}$ and also equations along the ``boundary'' $\{v = 0\}$. A naive treatment of these boundary equations results in a loss of derivatives,\footnote{This is essentially the familiar loss of derivatives inherent in studying the characteristic initial value problem for a hyperbolic equation.} and we will thus need to carefully design the scheme for combining the bulk and boundary equations. 
\end{enumerate}

\subsection{A Mixed Elliptic-Hyperbolic Model Second Order Equation}
As we have mentioned in the previous section, a key aspect in the proof of the main result Theorem~\ref{maintheohere} is that we will need to consider PDE's of mixed elliptic-hyperbolic type. We give here a model linear equation of a type which will often show up in this paper and explain some of the relevant features. (We will also have to solve other types of equations such as elliptic equations, parabolic equations, and various types of degenerate transport equations.)

We let $\psi \left(v,\theta^A\right) : (-1,0) \times \mathbb{S}^2 \to \mathbb{R}$ be our unknown, $H\left(v,\theta^A\right) :(-1,0) \times\mathbb{S}^2 \to \mathbb{R}$ be given, and consider the following equation:
\begin{equation}\label{i3oijio4}
(-v)\mathcal{L}_{\partial_v}^2\psi + \left(\frac{A_1}{v+1} + A_2\right)\mathcal{L}_{\partial_v}\psi + \mathcal{P}_{\ell \in \mathscr{A}}\mathcal{L}_b\mathcal{L}_{\partial_v}\psi + \mathcal{P}_{\ell \in \mathscr{A}}\left(\Omega^2\left(\slashed{\Delta} + A_3\left(v+1\right)^{-2}\right)\psi\right) = H.
\end{equation}
Here $A_1 \geq 0$, $A_2 \geq 0$, and $A_3$ are suitable real numbers, $b^A\left(v,\theta^C\right)$ is a $v$-dependent vector field on $\mathbb{S}^2$ with $\left|b\right| \ll 1$, $\slashed{\Delta}$ is the Laplacian associated to a family of metrics $\slashed{g}_{AB}\left(v,\theta^C\right)$ with $\left|\left(v+1\right)^{-2}\slashed{g}_{AB} - \mathring{\slashed{g}}_{AB}\right|_{\mathring{\slashed{g}}} \ll 1$ where $\mathring{\slashed{g}}_{AB}$ is the round metric on $\mathbb{S}^2$, $\Omega^2$ is a function which satisfies $\left|\Omega^2-1\right| \ll_c 1$ on any interval $v \in [-1,c]$ for $c < 0$ but may have certain singularities as $v\to 0$, and $\mathcal{P}_{\ell \in \mathscr{A}}$ denotes a projection to spherical harmonics associated to eigenvalues $\ell\left(\ell+1\right)$ for $\ell$ lying in a set $\mathscr{A}$ which will be the complement in $\mathbb{Z}_{\geq 0}$ of some finite set. We will desire to solve a boundary value problem where $\psi|_{v = 0} = h$ is given. Near $v = -1$ we will require that $\psi \to 0$ as $v\to -1$.  

The reader should think of $\left(v,\theta^A\right) \in (-1,0)\times \mathbb{S}^2$ as parametrizing the null hypersurface $\{u = -1\}$ and the equation~\eqref{i3oijio4} arising from a suitable reduction of the Einstein equations via self-similarity to an equation along $\{u =-1\}$. In particular, since we expect double-null coordinates to break down near the axis (analogously to the breakdown of  spherical coordinates at $r = 0$) it is natural to see singular terms in the equation as $v\to -1$. These double-null coordinates turn out to also be irregular at $v = 0$, and this is reflected in the singular behavior of some of the coefficients of~\eqref{i3oijio4} as $v\to 0$. 
\subsubsection{Global Aspects of~\eqref{i3oijio4}}\label{oi3ji2oi4}
If the vector field $b$ were to vanish, the equation~\eqref{i3oijio4} would be degenerate elliptic. If $b$ is not identically $0$ however, then the equation will transition from elliptic to hyperbolic when $(-v) \sim |b|$.  A proto-typical example of such a type changing equation is 
\begin{equation}\label{tricomieqn}
y\partial_x^2u + \partial_y^2u = 0.
\end{equation}
The study of such mixed elliptic-hyperbolic equations \underline{in $1+1$ dimensions} has a long history. Tricomi~\cite{tricomi} initiated the study of the equation~\eqref{tricomieqn} (which bears his name) and later various generalizations of~\eqref{tricomieqn} received a great deal of attention due to their connection with transonic potential flows (see~\cite{notransonic1,notransonic2,notransonic3,prottertricomiexist,protagmonnirenberg,transonicsurvey}).\footnote{Such mixed elliptic-hyperbolic problems also show up in many other settings such as the prescribed curvature problem for surfaces with sign changing Gauss curvature, see the references in the texts~\cite{mixedellhyp1,mixedellhyp2}.} One insight that may be drawn from these aforementioned works is that one generally does not expect a well-posed Dirichlet problem for equations like~\eqref{tricomieqn} when data is prescribed \emph{everywhere} on the boundary of a domain which includes regions where the equation is both elliptic and hyperbolic. 

It is traditional to refer to the curve where a $1+1$ dimensional equation changes from elliptic type to hyperbolic type as the ``sonic line.'' (This terminology is related to the aforementioned applications to fluid mechanics.) Though the equation~\eqref{tricomieqn} may appear to be of a very special form, in fact, any second order $1+1$ dimensional equation whose type changes from hyperbolic to elliptic and whose characteristics satisfy a suitable non-degeneracy condition along the sonic line will admit a change of variables which locally, near the sonic line, puts the equation in the form~\eqref{tricomieqn} up to lower order terms (see the discussions in Sections 3.1 and 3.2 of~\cite{mixedellhyp2}). However, despite the fact that~\eqref{tricomieqn} thus captures the local behavior of (many) mixed elliptic-hyperbolic problems, it is not a good model for the global geometry underlying the equation~\eqref{i3oijio4}. For this purpose, a better $1+1$ dimensional model problem is 
\begin{equation}\label{i3o9912}
(-v)\partial_v^2u + 2\epsilon \partial_{\phi}\partial_vu + \partial_{\phi}^2u = 0,\qquad u|_{v=0} = h,\qquad u|_{v=-1} = 0,
\end{equation}
in the domain $(v,\phi) \in (-1,0) \times \mathbb{S}^1$ and where $0 < \epsilon \ll 1$ and $h:\mathbb{S}^1 \to \mathbb{R}$ is given. This equation is hyperbolic for $v \in (-\epsilon^2,0)$ and elliptic for $v \in (-1,-\epsilon^2)$. Moreover, the boundary curve $\{v = 0\}$ is characteristic. We may look for a solution in the form 
\[u \doteq \sum_{k \in \mathbb{Z}}e^{ik\phi}\hat{u}_k(v),\]
which leads to
\begin{equation}\label{32ijoijoi24}
(-v)\frac{d^2\hat{u}_k}{dv^2} +2ik\epsilon\frac{d\hat{u}_k}{dv} -k^2 \hat{u}_k = 0.
\end{equation}
It is straightforward to see from an indicial analysis near $v = 0$ that any solution of~\eqref{32ijoijoi24} will satisfy that $\frac{d\hat{u}_k}{dv}$ blows-up at most at a logarithmic rate as $v\to 0$ and that $\hat{u}_k$ extends continuously to $v = 0$. Furthermore, we can find a solution to~\eqref{32ijoijoi24} which satisfies the boundary conditions $\hat{u}_k|_{v=-1} = 0$ and $\hat{u}_k|_{v=0} = \hat{h}_k$ for any constant $\hat{h}_k$ if and only if the only solution to~\eqref{32ijoijoi24} which vanishes at $v = -1$ and $v = 0$ is the trivial solution. This latter fact may be established as follows. Assume that $\hat{u}_k$ solves~\eqref{32ijoijoi24} and satisfies $\hat{u}_k|_{v=-1} = 0$. We rewrite~\eqref{32ijoijoi24} as
\[\frac{d}{dv}\left((-v)\frac{d\hat{u}_k}{dv} + 2ik\epsilon \hat{u}_k +\hat{u}_k\right) - k^2\hat{u}_k = 0,\]
contract with $-\overline{\left(2ik\epsilon \hat{u}_k +\hat{u}_k\right)}$, take the real part, and then integrate by parts over $v \in (-1,0)$. We obtain that 
\begin{equation}\label{3oijo2ij3}
\int_{-1}^0\left[(-v)\left|\frac{d\hat{u}_k}{dv}\right|^2 + k^2\left|\hat{u}_k\right|^2\right]\, dv = \frac{1}{2}\left|2ik\epsilon \hat{u}_k + \hat{u}_k\right|^2|_{v=0}.
\end{equation}
This a priori estimate thus yields the desired existence result for~\eqref{32ijoijoi24} and after summing in $k$ yields an existence result and a priori estimate for~\eqref{i3o9912}.  

The equation~\eqref{i3oijio4} is not separable (and also not $1+1$ dimensional), so we cannot expect to directly reduce to an ordinary differential equation. Nevertheless we will be able to use an elliptic regularization and a variant of the above a priori estimate to establish our desired existence statements.
\subsubsection{Local Analysis Near $v = -1$ and $v = 0$}\label{i3iojjoij91}
The discussion in Section~\ref{oi3ji2oi4} neglects certain aspects of the local analysis of~\eqref{i3oijio4} near $v = -1$ and $v = 0$. Near $v = -1$ the equation~\eqref{i3oijio4} is well approximated by the operator 
\begin{equation}\label{23ioj4oij2o}
\mathcal{L}_{\partial_v}^2\psi + \frac{A_1}{v+1}\mathcal{L}_{\partial_v}\psi  + \left(v+1\right)^{-2}\left(\mathring{\Delta} + A_3\right)\psi = H,
\end{equation}
where $\mathring{\Delta}$ denotes the round Laplacian on $\mathbb{S}^2$. Our basic estimate for the operator~\eqref{23ioj4oij2o} near $v  = -1$ is obtained by contracting with $\left(v+1\right)^{-2p}\psi$, integrating by parts, and using (sharp) Hardy inequalities. For \emph{suitable ranges} of $p$ (depending on $A_1$, $A_3$, and the spherical harmonics for which $\psi$ is supported on) and $c > 0$ sufficiently small, this eventually leads to an estimate of the following form:\footnote{This can be awkward to actually do directly for the equation~\eqref{23ioj4oij2o} as one needs to argue that the boundary terms at $v = -1$ vanish. Instead, we actually work first with a regularized version of~\eqref{23ioj4oij2o} where the boundary terms automatically vanish, and after establishing the corresponding estimates, take a limit which recovers the desired equation.}
\begin{align*}
&\int_{-1}^{-1+c}\int_{\mathbb{S}^2}\left[\left(v+1\right)^{-2p+2}\left(\mathcal{L}_{\partial_v}^2\psi\right)^2 + \left(v+1\right)^{-2p}\left(\mathcal{L}_{\partial_v}\psi\right)^2 + \left(v+1\right)^{-2p}\left|\slashed{\nabla}\psi\right|^2\right]\, dv\, \mathring{\rm dVol} \lesssim 
\\ \nonumber &\qquad \qquad \int_{-1}^{-1+c}\int_{\mathbb{S}^2}\left(v+1\right)^{-2p+2}\left|H\right|^2\, dv\, \mathring{\rm dVol} + \int_{-1+c}^{-1+2c}\int_{\mathbb{S}^2}\left[\left(\partial_v\psi\right)^2 + \left|\slashed{\nabla}\psi\right|^2\right]\, dv\, \mathring{\rm dVol}.
\end{align*}
Given this estimate (and suitable higher order versions), we may apply the fundamental theorem of calculus in $v$ and also obtain $L^{\infty}_vL^2\left(\mathbb{S}^2\right)$ estimates for (angular derivatives of) $\mathcal{L}_{\partial_v}\psi$ and $\psi$.

For most of the arguments in the paper near $v = -1$, the above estimates will suffice; however, due to a potential logarithmic divergence there is a slight loss in the $L^{\infty}$ estimates relative to the true optimal weights as $v\to -1$. When we turn to constructing a regular coordinate system near the axis, we will need to improve this estimate. This is done as follows. Expanding $\psi = \sum_{\ell \in \mathbb{Z}}\psi_{\ell}$ in terms of spherical harmonics leads to
\begin{equation}\label{3ijo2}
\frac{d^2\psi_{\ell}}{dv^2}+ \frac{A_1}{v+1}\frac{d\psi_{\ell}}{dv} + \left(v+1\right)^{-2}\left(A_3-\ell\left(\ell+1\right)\right)\psi_{\ell} = H_{\ell}.
\end{equation}
We have a regular singularity at $v = -1$ and, assuming $\left|H_{\ell}\right| \sim \left(v+1\right)^{\beta}$ as $v\to -1$, an indicial analysis will yield two numbers $\alpha_1\left(\ell\right)$ and $\alpha_2\left(\ell\right)$ (with $\alpha_1 \to -\infty$ and $\alpha_2 \to \infty$ as $\ell \to \infty$) so that $\psi$ is asymptotic to a sum of $\left(v+1\right)^{{\rm min}\left(\alpha_1\left(\ell\right),\beta+2\right)}$ and $\left(v+1\right)^{{\rm min}\left(\alpha_2\left(\ell\right),\beta+2\right)}$ as $v\to -1$.\footnote{In certain situations one must also consider logarithmic corrections to these decay rates.} In favorable situations Our a priori estimates will eliminate the $\alpha_1$ branch and this eventually allows us to establish sharp estimates for $\psi$ near $v = -1$. There will be a few situations where for some of the low-$\ell$ spherical harmonics, the decay rate as $v\to -1$ determined by $\alpha_2\left(\ell\right)$ would not be fast enough for our applications. In these situations we will give up the freedom to pose a boundary condition at $v = 0$ and use the extra degree of freedom to also turn off the $\alpha_2$ branch and thus obtain a solution with faster decay as $v\to -1$. 

Now we turn to the local analysis near $v = 0$. In order to establish sharp asymptotics as $v\to 0$, the equation~\eqref{i3oijio4} turns out to be well-approximated by the equation 
\begin{equation}\label{23jhihuiuih4}
\left((-v)\mathcal{L}_{\partial_v}+\mathcal{L}_b+A_2\right)\mathcal{L}_{\partial_v}\psi  = H.
\end{equation}
We see that $\mathcal{L}_{\partial_v}\psi$ will satisfy an ordinary differential equation along the integral curves of $(-v)\mathcal{L}_{\partial_v} + \mathcal{L}_b$. In particular, the value of $A_2$ will strongly affect the behavior of $\mathcal{L}_{\partial_v}\psi$ with larger values of $A_2$ corresponding to better control of $\mathcal{L}_{\partial_v}\psi$ as $v\to 0$. Finally, we note that estimates for $\psi$ (and angular derivatives thereof) may be obtained from those of $\mathcal{L}_{\partial_v}\psi$ (and angular derivatives thereof) by the fundamental theorem of calculus. 
\subsubsection{Nonlinear Analysis and Hierarchies of Estimates}
So far we have discussed purely linear aspects of the equation~\eqref{i3oijio4}. However, the function $\Omega^2$, vector field $b$, and metric $\slashed{g}$ will themselves all be dynamic quantities and the study of~\eqref{i3oijio4} therefore involves a quasilinear analysis. The most serious difficulty here will involve the lapse function $\Omega$. When we study the equations modeled on~\eqref{i3oijio4} the best estimates we will have for the lapse $\Omega$ are the pointwise estimate $\Omega \sim (-v)^{-\kappa}$ as $v\to 0$ for $|\kappa| \lesssim \epsilon$ and also $\left|\mathring{\nabla}^j\log\Omega\right| \lesssim \epsilon (-v)^{-\epsilon}$.\footnote{The reason for these weak estimates may be explained as follows. The Ricci coefficient $\Omega\underline{\omega}$ will be obtained by solving an equation of the form~\eqref{i3o9912}. Then we will obtain $\log\Omega$ by integrating the transport equation $\left((-v)\mathcal{L}_{\partial_v}-\mathcal{L}_b\right)\log\Omega = 2\left(\Omega\underline{\omega}\right)$. In order for our gague to be regular at the axis, it is necessary to impose the boundary condition that $\log\Omega|_{v=-1} = 0$, but this then results in relatively weak control of the lapse $\Omega$ as $v\to 0$ even if we have very strong estimates for $\Omega\underline{\omega}$.} All of our multiplier estimates thus have to be carefully calibrated to as to be consistent with this singular behavior of $\Omega$. (It will turn out to be crucial that we have that $\mathcal{L}_b\log\Omega$ remains bounded as $v\to 0$.)

A consequence of the analysis of Sections~\ref{oi3ji2oi4} and~\ref{i3iojjoij91} is that the estimates we may establish for solutions $\psi$ to~\eqref{i3oijio4} will strongly depend on the values of the various constants $A_1$, $A_2$, and $A_3$ (and also how much regularity we have for the boundary value $\psi|_{v=0}$). A result of this is that we will have various hierarchies of estimates for different parts of the metric. 

There are also semilinear nonlinearities which we have hidden in the inhomogeneous term $H$ on the right hand side of~\eqref{i3oijio4}. It will be crucial for the analysis that these nonlinearities are consistent with the aforementioned hierarchies imposed by the linear analysis. As is familiar from the many other works which have studied the Einstein vacuum equations via null frame decompositions, so-called ``signature'' considerations (see~\cite{ck,klainrodtrap}) greatly constrain the types of quadratic terms one may observe. Even more important to our analysis will be the implicit consistency of the nonlinear terms with the existence of certain formal power series expansions (see the discussion below in Section~\ref{iojoij284982}).

\subsection{The Self-Similar Reduction and Constraint Propagation}\label{reveieww3}
The spacetime of Theorem~\ref{maintheohere} will be constructed in the double-null form~\eqref{32iojjiooi4} for $-1 < \frac{v}{-u} < 0$. As is well-known, if a spacetime $\left(\mathcal{M},g\right)$ is in the form~\eqref{32iojjiooi4}, then ${\rm Ric}\left(g\right) = 0$ if and only if the null-structure equations (see Section~\ref{doubleitup}) hold with all Ricci curvature terms set to $0$. (Note that we do not include the Bianchi equations for curvature in the null-structure equations.) Additionally, we will look for a self-similar solution, that is, we will require that~\eqref{32ioij92} holds for $-1 < \frac{v}{-u} < 0$. In view of~\eqref{32ioij92} the metric $g$ is completely determined by it's restriction to the null hypersurface $\{u = -1\}$. Conversely, given $\Omega$, $b^A$, and $\slashed{g}_{AB}$ defined along a $3$-dimensional hypersurface $\left(v,\theta^A\right) \in (-1,0)\times \mathbb{S}^2$, we may identify this hypersurface with $\{u = -1\}$ and then use the formula~\eqref{32ioij92} to uniquely define a corresponding self-similar metric $g$. In view of this equivalence, we may work entirely along the hypersurface $\{u = -1\}$. We can then express all of the null-structure equations as equations just involving $\Omega$, $b^A$, and $\slashed{g}_{AB}$ along $\left(v,\theta^A\right) \in (-1,0)\times \mathbb{S}^2$. We call these resulting equations the \emph{self-similar null-structure equations}.

An important complication in this approach to the proof of Theorem~\ref{maintheohere} is that, considered as a system of PDE's for $\left(\Omega,b^A,\slashed{g}_{AB}\right)$, the self-similar null-structure equations are overdetermined. Thus we will need to select a suitable subset of the null-structure equations to actually solve and determine a spacetime $\left(\mathcal{M},g\right)$. There are two fundamental difficulties associated to this procedure:
\begin{enumerate}
	\item In the standard approaches to estimates in the double-null gauge, one often takes advantage of all of the available equations. For example, when estimating $\hat{\chi}$ it is convenient sometimes to carry out elliptic estimates along $\mathbb{S}^2$ via the Codazzi equation~\eqref{tcod1} and other times to use the transport equation~\eqref{3hatchi}. We will lose this flexibility, and we thus have to choose our subset of equations carefully in order that we can obtain all of the necessary estimates. This will be difficult both from the point of view of obtaining estimates which do not lose derivatives at the top order and from the point of view of obtaining pointwise estimates at a lower level of regularity which yield sufficient control of the metric as $v\to 0$. 
	\item In view of the fact that we only directly solve a subset of the null-structure equations, the spacetime we produce at the end of this procedure is not manifestly Ricci flat. In order to show that the Ricci tensor vanishes, we must run a ``preservation of constraints'' argument. The inputs to this argument will be various relations amongst the components of the Ricci tensor implied by the equations we solve, boundary conditions which hold for the Ricci tensor as $v\to -1$ and $v\to 0$ which hold as a consequence of boundary conditions imposed when solving for $\left(\Omega,b^A,\slashed{g}_{AB}\right)$, and additional equations which hold for the Ricci tensor as a result of the Einstein tensor being divergence free.
\end{enumerate}
In order to overcome the two above difficulties it will also be convenient to introduce what we call ``artificial variables.'' These are extra unknowns (solving corresponding extra equations) which at the end of the preservation of constraints argument will eventually be shown to equal one of the standard Ricci coefficients or metric components. The benefit of introducing them is that they allow us to decouple different appearances $\Omega$, $b$, or $\slashed{g}$ and use different subsets of the null-structure equations in different situations. Of course, the price of doing this is an increase in the complexity of the constraint propagation argument.

\subsection{Formal Power Series Expansions and Regular Coordinate Systems}\label{iojoij284982}
The double-null coordinate system that we will originally construct our spacetime in will not be valid at $\{v = 0\}$ or along the axis $\{u = v\}$. In this section we discuss how we understand the behavior of our spacetime near $\{v = 0\}$ and $\{u = v\}$.
\subsubsection{Formal Expansions and Nonlinear Compatibility}\label{sososoformalwow}
Similarly to the power series expansions for the Ambient Metric developed by Fefferman and Graham~\cite{FG1,FG2} one may develop suitable expansions associated for the more general ``twisted'' self-similarity considered in this paper. We will not need a systematic development of these expansions for this paper (a formal treatment of them will appear in~\cite{generaltwisted}) but it will be conceptually clarifying to note some of the consequences of the formal theory. 

Our starting ansatz is that we have a metric $\left(\mathcal{M},g\right)$ in the double-null form~\eqref{32iojjiooi4} which satisfies~\eqref{32ioij92} for $-c < \frac{v}{-u}  < 0$ with $0 < c \ll 1$. We further assume that $g$ is $C^N$ for some $N \gg 1$ and that there exists $0 < \epsilon \ll 1$ such that every Ricci coefficient $\psi$ (except for $\omega$) and metric component $\phi$ satisfies a bound
\begin{equation}\label{joijoi32}
\left|\tilde{\psi}|_{u=-1}\right|_{\slashed{g}} \lesssim \epsilon (-v)^{-\epsilon},\qquad \left|\tilde{\phi}|_{u=-1}\right|_{\slashed{g}} \lesssim \epsilon (-v)^{-\epsilon},
\end{equation}
where $\tilde{\psi}$ and $\tilde{\phi}$ denote the difference of $\psi$ and $\phi$ with their Minkowski space values. One may then show that all such spacetimes admit suitable formal expansions.

The most relevant ``lesson'' from the formal theory is the following classification of Ricci coefficients and metric coefficients.
\begin{enumerate}
	\item The following are ``regular quantities'': $\left\{\slashed{g},b,\Omega\underline{\omega},\eta,\Omega\hat{\underline{\chi}},\Omega{\rm tr}\underline{\chi},\Omega^{-1}{\rm tr}\chi\right\}$. We expect all of these quantities (and their angular derivatives) to have limits as $v\to 0$ and moreover to be (at least) $C^{0,1-O\left(\epsilon\right)}$ in the double-null coordinates at $\{v = 0\}$. This expectation follows immediately from the fundamental theorem of calculus, the bounds~\eqref{joijoi32}, and that fact that in view of the null-structure equations all of the regular quantities  have an equation for their $\Omega\nabla_4$ derivative directly in terms of other Ricci coefficients. (Note that we have multiplied the Ricci coefficients with a suitable power of the lapse so that in the corresponding $\nabla_4$ equation, the term containing $\omega$ cancels. The appropriate power of the lapse is, in fact, determined by signature considerations.)
	\item The following are ``irregular quantities'': $\left\{\Omega,\Omega\omega,\Omega^{-1}\hat{\chi},\underline{\eta}\right\}$. These are all quantities which we expect to be (generically) less regular at $\{v = 0\}$ than the regular quantities. We now quickly discuss each of these.
	\begin{enumerate}
		\item The only equation which relates the lapse $\Omega$ along $\{u = -1\}$ with a regular quantity is 
		\begin{equation}\label{2ij3oijio452}
		\left((-v)\mathcal{L}_{\partial_v}-\mathcal{L}_b\right)\log\Omega = 2\Omega\underline{\omega}.
		\end{equation}
		(This may be derived from the definition of $\Omega\underline{\omega}$ and self-similarity.) This equation in general produces solutions which blow-up logarithmically. Correspondingly, $\Omega\omega$ is the derivative of a function which may blow-up logarithmically.
		\item From~\eqref{3hatchi} and self-similarity, one may derive the following equation for $\Omega^{-1}\hat{\chi}$ along $\{u =-1\}$:
		\begin{align}\label{3kj2iojoi32}
& v\Omega\nabla_4\left(\Omega^{-1}\hat{\chi}\right)_{AB} +\mathscr{L}\left(\Omega^{-1}\hat{\chi}\right)_{AB} +v\Omega{\rm tr}\chi\left(\Omega^{-1}\hat{\chi}\right)_{AB} - 2\left(\Omega\underline{\omega}\right)\left(\Omega^{-1}\hat{\chi}\right)_{AB}= {\rm regular},
 \end{align}
 where
 \[\mathscr{L}f_{AB} \doteq \mathcal{L}_bf_{AB}- \left(\slashed{\nabla}\hat{\otimes}b\right)^C_{\ \ (A}f_{B)C} -\frac{1}{2}\slashed{\rm div}bf_{AB}.\]
 An analysis of the transport equation on the left hand side of~\eqref{3kj2iojoi32} indicates that even in the event that $\Omega^{-1}\hat{\chi}$ extends continuously to $\{v = 0\}$, generically it will be at most $O\left(\epsilon\right)$-H\"{o}lder continuous.
	\item Finally, the irregularity of $\underline{\eta}$ follows from the above discussion of the lapse, the fact that $\eta$ is regular, and the relation $\underline{\eta} = -\eta +2\slashed{\nabla}_A\log\Omega$.   
	\end{enumerate}
	\end{enumerate}
	
	The ``nonlinear compatibility'' we have put in the title of this section refers to the fact that due to the validity of the formal expansions, we may expect the nonlinear terms in the null-structure equations to respect the dichotomy between regular and irregular quantities. This is most easily explained by way of an example. The equation~\eqref{3ueta} and~\eqref{tcod2} may be combined to establish an equation which involves $\eta$, $\underline{\eta}$, $\underline{\chi}$ and ${\rm tr}\underline{\chi}$. This is a mixture of regular and irregular quantities; however, our expectation is that it is possible to simply the equation in such a way that the irregular quantities will cancel each other out. This does indeed turn out to be possible and one may obtain  
	\begin{align}\label{2k3j2khk}
&\frac{v}{u}\mathcal{L}_{\partial_v}\eta_A - \mathcal{L}_b\eta_A -\eta_A\left(\Omega{\rm tr}\underline{\chi}\right)- 4\slashed{\nabla}_A\left(\Omega\underline{\omega}\right) =  \slashed{\nabla}^B\left(\Omega\hat{\underline{\chi}}\right)_{AB} - \frac{1}{2}\slashed{\nabla}_A\left(\Omega{\rm tr}\underline{\chi}\right) -\Omega{\rm Ric}_{3A}.
\end{align}
This type of nonlinear analysis will be used repeatedly in the paper.

\subsubsection{Coordinates Near $\{v = 0\}$ and Gluing the Interior to the Exterior}\label{oneone}
The double-null coordinates where most of the analysis in the paper takes place are not valid at $\{v = 0\}$. Instead we must introduce there what we call ``lapse-renormalized coordinates'' $\left(\hat{v},u,\theta^A\right)$ defined by 
\[\hat{v} \doteq -\int_v^0\Omega^2\, dv.\]
In $\left(\hat{v},u,\theta^A\right)$ coordinates, the metric becomes 
\begin{align}\label{2klj3oij1o}
&\qquad g = -2\left(du\otimes d\hat{v} + d\hat{v}\otimes du\right) + \slashed{g}_{AB}d\theta^A\otimes d\theta^B 
\\ \nonumber & + \left(-4\int_v^0\left(\Omega^2\partial_{\theta^A}\log\Omega\right)dv-b_A\right)\left(du\otimes d\theta^A + d\theta^A\otimes du\right)
 +\left(-8\int_v^0\left(\Omega^2\partial_u\log\Omega\right)dv + \left|b\right|^2\right)du\otimes du.
\end{align}

A calculation indicates the $g$ will extend to $\{\hat{v} = 0\}$ as a $C^{1,s}_{\hat{v}}C^N_{u,\theta^A}$ metric if in the original $\left(u,v,\theta^A\right)$ coordinates we have that $\eta$, $\Omega^{-1}{\rm tr}\chi$, $\Omega^{-1}\hat{\chi}$, and $\Omega\underline{\omega}$ extend to $\{v = 0\}$ in a $C^{0,s}C^N_{u,\theta^A}$ fashion and if the lapse blows-up in a sufficiently mild fashion as $v\to 0$. We will show that these conditions are satisfied if suitable boundary/seed data are prescribed for our construction along $\{v = 0\}$: The difficult terms to control in this part of the argument will be $\Omega^{-1}\hat{\chi}$ and $\log\Omega$, and our argument will fundamentally rely on insights from~\cite{nakedone} concerning the ``$\kappa$-singular transport equation'' which will allow us to establish favorable estimates for solutions to the degenerate transport equation~\eqref{3kj2iojoi32} when the vector field $b$ takes a suitable form as $v\to 0$. Analogues of these ideas will also be used to control the lapse $\Omega$ via an analysis of the transport equations~\eqref{2ij3oijio452} assuming again that $b$ takes a suitable form as $v\to 0$. The special boundary data we select for our solution are exactly chosen so that the necessary conditions on the vector field $b$ hold along $\{v = 0\}$.

Finally, the lapse-renormalized variables are also useful for gluing our interior solution with exterior solution constructed in~\cite{nakedone}. More concretely, if we identify the two solutions in lapse-renormalized coordinates along $\{\hat{v} = 0\}$, then the resulting spacetime will be $C^{1,s}_{\hat{v}}C^N_{u,\theta^A}$ if in the original $\left(u,v,\theta^A\right)$ coordinates,  the limits as $v\to 0^-$ and $v\to 0^+$ agree for all of the quantities $\{b,\slashed{g},\Omega\underline{\omega},\eta,\Omega^{-1}\hat{\chi},\Omega^{-1}{\rm tr}\chi\}$ and if the lapse $\Omega$ blows-up in a sufficiently mild fashion as $v\to 0$.

\subsubsection{Coordinates Near the Axis $\{v=u\}$}\label{twotwo}
The double-null coordinates $\left(u,v,\theta^A\right)$ break down along the axis $\{u = v\}$, and thus we need to construct an alternative coordinate system in this region. We will introduce the new coordinate system in two steps.

We first construct a $\left(t,x,y,z\right)$ coordinate system ``by hand.'' That is, we set $t = v+u$ and $r =v-u$, and then define the coordinates $\left(x,y,z\right)$ by using the standard formulas which link Cartesian coordinates $\left(x,y,z\right)$ in $\mathbb{R}^3$ with spherical coordinates $\left(r,\theta,\phi\right)$. After translating our estimates for the metric in the $\left(u,v,\theta^A\right)$ coordinates into the $\left(t,x,y,z\right)$ we will see that our metric is at least $C^{0,1}$.

In the second step we define set of self-similar wave coordinates $\left\{\xi^i\right\}_{i=0}^3$ by solving
\begin{equation}\label{3oij2oij}
K\xi^i = \xi^i,\qquad \Box_g\xi^i = 0,
\end{equation}
along with suitable boundary conditions. We will have that the self-similar vector field $K$ will be timelike near the axis, and after restricting to an appropriate neighborhood of the axis, we will be able to reduce~\eqref{3oij2oij} to a suitable elliptic equation along $\{t = 0\}$. In particular these new coordinates will be seen to lie in $W^{2,p}_{\rm loc}$ for all $p \geq 1$ along $\{t = 0\}$ and in these new coordinates we will have that $g$ lies in $W^{1,p}_{\rm loc}$ for all $p \geq 1$ along suitable spacelike hypersurfaces. Furthermore, using again that $K$ is timelike in this region, $g$ itself will satisfy an elliptic equation when expressed in the self-similar wave coordinates. Using $L^p$-elliptic theory, we will then be able to run a standard bootstrap argument to upgrade the $W^{1,p}$ regularity of $g$ to the statement that $g$ is smooth.
\subsection{Acknowledgements}
The author would like to acknowledge that this work would not exist without fundamental contributions from Igor Rodnianski. The author also acknowledges support from NSF grant DMS-1900288, from an Alfred P. Sloan Fellowship in Mathematics, and from NSERC discovery grants RGPIN-2021-02562 and DGECR-2021-00093.

\section{Outline of the Paper}
In this section we will give a guide to the remaining parts of the paper.
\subsection{Preliminaries}
In Section~\ref{3iojoijio4} we discuss various preparatory material.

 We start in Section~\ref{constantconstant} by discussing various useful conventions which appear throughout the paper. Importantly, we fix a hierarchy of smallness conditions for various constants and introduce some notation for a generic cut-off function.

Then in Section~\ref{doubleitup} we review the formalism of the double-null gauge. In particular, we state the null-structure equations for a spacetime $\left(\mathcal{M},g\right)$ which is not assumed to satisfy ${\rm Ric}\left(g\right) = 0$. We also state various commutation formulas and review the various operators $\slashed{D}_1$, ${}^*\slashed{D}_1$, $\slashed{D}_2$, ${}^*\slashed{D}_2$ defined along $\left(\mathbb{S}^2,\slashed{g}\right)$.

In Section~\ref{roundround} we review the Hodge decompositions of functions, $1$-forms, and $(0,2)$-tensors on the round sphere. We also discuss the corresponding (tensorial) spherical harmonic expansions.

Finally, in Section~\ref{smoothingopeatorsection} we state a lemma concerning the existence of a family of smoothing operators $\{\Pi\}_{\delta > 0}$ which act on $L^2\left(\mathbb{S}^2\right)$. 
\subsection{Bootstrap Norms and Basic Inequalities}
In Section~\ref{sectionbootstrapnorms} we introduce the various norms which will play a central role in the paper and discuss some of their basic properties.

We start in Section~\ref{notationfornorms} by introducing some useful notations. In particular, we define the norms $\left\vert\left\vert \cdot\right\vert\right\vert_{\mathscr{Q}_{a_1}^{a_2}\left(N,p_1,p_2\right)}$, $\left\vert\left\vert \cdot\right\vert\right\vert_{\mathscr{S}_{a_1}^{a_2}\left(N,p_1,p_2\right)}$, and $\left\vert\left\vert \cdot\right\vert\right\vert_{\check{\mathscr{S}}_{a_1}^{a_2}\left(N,p_1,p_2,p_3\right)}$. These norms are all defined for $\mathbb{S}^2_{-1,v}$ tensors where $v \in (-1,0)$ and are weighted versions of either $L^2_vL^2\left(\mathbb{S}^2\right)$ or $L^{\infty}_vL^2\left(\mathbb{S}^2\right)$ (or a combination thereof) which also incorporate commutation with the standard Killing fields along $\mathbb{S}^2$. The norm $\check{\mathscr{S}}$ moreover allows for the weight to change depending on the Killing field commutation. 

Then in Section~\ref{bootthenormbootthenorm} we will define the main bootstrap norms which will be used throughout the paper.  These are norms which will be applied to various manifestations of the lapse $\Omega$, shift $b$, and $\mathbb{S}^2_{-1,v}$ metric $\slashed{g}$. The norms are split into two main families: norms which involves the maximal number (or close to the maximal) number of derivatives and are (mostly) weighted $L^2_vL^2_{\theta^A}$ norms and norms which involve a much lower number of derivatives and are (mostly) weighted $L^{\infty}_vL^2_{\theta^A}$ norms. Furthermore, some of the norms involved improved estimates for various special derivatives. 

Lastly, in Section~\ref{inequalitysectionsection} we present various general inequalities for the previously introduced norms. Most of these are straightforward consequence of Sobolev inequalities along $\mathbb{S}^2$. Of particular importance to us will be the interpolation inequality Lemma~\ref{3m2omo4} which will allow to slightly improve the weights in certain $L^2$-estimates at the expense of losing one angular derivative. 
\subsection{Degenerate Transport and Elliptic Equations}
In Section~\ref{ij3oin4in234} we study various degenerate transport (and degenerate transport/elliptic) equations. 

In Section~\ref{2omomo2} we focus on degenerate transport equations of the form
\begin{equation}\label{oipip234}
(-v)\nabla_v\phi + A_3\mathcal{P}_{\ell \in \mathscr{A}}\mathcal{L}_b\phi + \left(\frac{(-v)A_1}{v+1} + A_2\right)\phi = H,\qquad \forall v \in (-1,0),
\end{equation}
where $\phi$ and $H$ are suitable $\mathbb{S}^2_{-1,v}$ tensors for $v\in (-1,0)$, $A_1$, $A_2$, and $A_3$ are suitable constants with $A_1,A_2 \geq 0$, $\mathcal{P}_{\ell \in \mathscr{A}}$ denotes a projection away from finitely many spherical harmonics, and the $\mathbb{S}^2_{-1,v}$ vector field $b$ is assumed to satisfy various ``bootstrap'' assumptions. Assuming $H$ to be given, we will establish basic existence results for $\phi$ solving~\eqref{oipip234} as well as corresponding a priori estimates. We will establish both weighted $L^2_vL^2_{\theta^A}$ and $L^{\infty}_vL^2_{\theta^A}$ estimates which will be proved via the use of suitable multipliers. Using also some of the theory developed in~\cite{nakedone} we will establish results which concern the $v\to 0$ limit of solutions $\phi$ to~\eqref{oipip234}.

In Section~\ref{iojoijiojo21} we combine the transport equation analysis from Section~\ref{2omomo2} with an analysis of certain degenerate elliptic equations. A prototype of the degenerate elliptic equation we study is the following:
\begin{equation}\label{2l3kjl2jkl3}
	 \mathcal{P}_{\ell \geq 1}\slashed{\rm div}\left(\Omega^{-2}P\right) =  \mathcal{P}_{\ell \geq 1}H,\qquad  \mathring{\Pi}_{\rm curl}P^A = - \mathring{\slashed{\epsilon}}^{AB}\mathring{\nabla}_Bf,
	\end{equation}
	where $P$ is an unknown $\mathbb{S}^2_{-1,v}$ vector field for $v \in (-1,0)$, $H$ and $f$ are given functions, $\mathcal{P}_{\ell \geq 1}$ denotes the projection to spherical harmonics satisfying $\ell \geq 1$, $\mathring{\Pi}_{\rm curl}$ denotes the projection of $P$ onto it's ``curl part,'' and the lapse function $\Omega$ is assumed to satisfy various bootstrap assumptions. The key difficulty in the study of~\eqref{2l3kjl2jkl3} is the presence of the lapse $\Omega$ which has potentially a quite singular behavior as $v\to 0$; in particular, angular derivatives of $\Omega$ may blow-up faster as $v\to 0$ than $\Omega$ itself. Despite this difficulty, we will establish various existence results and a priori estimates. Finally, we treat equations for an unknown $\mathbb{S}^2_{-1,v}$ vector field $P$ where operators of the form of the left hand side of~\eqref{2l3kjl2jkl3} applied to $P$, themselves satisfy degenerate transport transport equations of the type~\eqref{oipip234}.  Again we establish suitable existence results and a priori estimates.

\subsection{General Theory for the Model Second Order Equation}
In Section~\ref{secondordertheorysection} we study equations of the form~\eqref{i3oijio4} discussed in the introduction above where we assume that $\Omega$, $b$, and $\slashed{g}$ satisfy suitable bootstrap assumptions. We refer to these equations as model second order equations and further introduce the notion of a model second order equation of type $I$, $II$, or $III$ which depends on the various values of the constants $A_1$, $A_2$, and $A_3$ and also the set $\mathscr{A}$. These different types will correspond to different manifestations of the model second order equation which will appear in the paper.

In Section~\ref{makeitregularyay} we introduced a regularized version of the model second order equation corresponding to every triple $\left(\tilde{\delta}_1,\tilde{\delta}_2,\tilde{\delta}_3\right)$ with $0 \leq \tilde{\delta}_i \ll 1$. Namely, we restrict the range of $v$ to lie in the interval $\left(-1+\tilde{\delta}_2,\tilde{\delta}_1\right)$ and apply a smoothing operator $\Pi_{\tilde{\delta}_3}$, which converges to the identity as $\tilde{\delta}_3\to 0$, to the term involving $\mathcal{L}_b\mathcal{L}_{\partial_v}$ in~\eqref{i3oijio4}. When each of the $\tilde{\delta}_i$ is positive, the corresponding equation is uniformly elliptic and we will then be able to exploit elliptic theory to establish an existence theory for the regularized equation. We also then establish a prior estimates which are independent of the parameters $\tilde{\delta}_i$, and this allows us to eventually run a limiting argument which takes each $\tilde{\delta}_i \to 0$  and produces a solution $\psi$ to the original equation~\eqref{i3oijio4}. The estimates which involve the top number of derivatives are proven via the use of suitable multipliers while weighted $L^{\infty}_vL^2_{\theta^A}$ estimates which are sharper as $v\to 0$ are proven by treating~\eqref{i3oijio4} as a degenerate transport equation for $\mathcal{L}_{\partial_v}$ and treating the angular derivatives as an inhomogeneity.

In Section~\ref{soclosetovminusone} we focus on establishing local estimates near $\{v = -1\}$ for solutions $\psi$ to~\eqref{i3oijio4}\footnote{In reality, for certain technical reasons, in part of Section~\ref{soclosetovminusone} we also work with the regularized version of the equation~\eqref{i3oijio4}.} which have improved weights. For our $L^2_vL^2_{\theta^A}$ estimates we use multiplier estimates in conjunction with Hardy inequalities (with explicit constant). Our $L^{\infty}_vL^2_{\theta^A}$ estimates are then obtained by the fundamental theorem of calculus in $v$.

In Section~\ref{finalestestestsce} we group all of the analysis of the model second order equation together and introduce concise notation so that we may easily refer to the various established estimates. 

Finally, in Section~\ref{bitofanextensionsection} we discuss a certain modification of the estimates established for equations of type $III$. These modified estimates are useful later when we run the propagation of constraints argument. 
\subsection{A Parabolic Equation}
In Section~\ref{paraparasec} we establish existence results and a priori estimates for the parabolic equation
\begin{equation}\label{ijo1ijoi2}
\mathcal{L}_{\partial_v}\Theta +\frac{2}{v+1}\Theta- 4\mathcal{P}_{\ell \geq 1}\left(\Omega^2\left(v+1\right)^{-2}\mathring{\Delta}\Theta\right) = H,
\end{equation}
where $H: (-1,0) \times \mathbb{S}^2 \to \mathbb{R}$ is a given function satisfying $\mathcal{P}_{\ell = 0}H = 0$, $\Theta: (-1,0) \times \mathbb{S}^2 \to \mathbb{R}$ is an unknown function satisfying $\mathcal{P}_{\ell = 0}\Theta = 0$, and the lapse function $\Omega$ satisfies suitable bootstrap assumptions. This equation is more straightforward to treat than the model second order equation; however, a special analysis is needed as $v\to -1$, due to the singular coefficients, and as $v\to 0$, due to the singular behavior of the lapse $\Omega$ as $v\to 0$. 
\subsection{Algebraic Consequences of Self-Similarity}\label{4jk8uu88u92}
In Section~\ref{alloftheequationsinselfsimilar} we establish various algebraic consequences of the null-structure equations when the underlying spacetime is also assumed to be self-similar. In particular, the analysis of this section establishes the fact that if $\left(\mathcal{M},g\right)$ is both self-similar and Ricci flat, then the Ricci coefficient $\Omega\underline{\omega}$ and the metric component $\slashed{\rm curl}b$ satisfy model second order equations with  inhomogeneities which consist of purely \emph{nonlinear} combinations of Ricci coefficients. This linear decoupling is one of the central observations which allow us to unlock the structure of the system of self-similar null-structure equations.
\subsection{The Main Iteration Argument}\label{oijjio23ijoioj}
In Sections~\ref{2omoo09882uhh3hh}-\ref{foim2io34o5u92hj5991} we run a large family of iteration arguments. As explained above in Section~\ref{reveieww3} the outcome of these iteration arguments will be a spacetime $\left(\mathcal{M},g\right)$ in the double-null form~\eqref{32iojjiooi4} from $-1 < \frac{v}{-u} < 0$ and which solves a certain subset of the Ricci flat null-structure equations. These equations that we solve will all have been derived using the algebraic consequences of self-similarity as discussed above in Section~\ref{4jk8uu88u92}. We will leave the detailed overview of these sections to Section~\ref{2omoo09882uhh3hh} which outlines these iteration arguments in full detail.
\subsection{Propagation of Constraints}
In Section~\ref{propagatetheconstraintsforever} we take the spacetime $\left(\mathcal{M},g\right)$ produced in the fashion described in Section~\ref{oijjio23ijoioj} above and show that $g$ is a Ricci flat metric. This is accomplished in the following fashion: As a consequence of the equations used to solve for $g$, we will have various relations among the components of the Ricci curvature. Furthermore, the fact that the Einstein tensor is divergence free yields more relations for the Ricci components.\footnote{This is, of course, reminiscent of the ``constraint propagation'' part of the usual local existence construction for the Einstein equations.} Together these all yield a system of equations for the different components of the Ricci curvature. It will also be useful to note that a consequence of the construction of $\left(\mathcal{M},g\right)$ will be that certain combinations of Ricci coefficients must vanish as $v\to 0$. 

We then show that suitable combinations of Ricci coefficients satisfy model second order equations with a right hand side schematically of the form $\epsilon \cdot {\rm Ricci}$. (We also derive some parabolic equations.)  This then leads us to define a suitable norm $\left\vert\left\vert {\rm Ric}\right\vert\right\vert_{\mathscr{Z}}$ and use our theory for the model second order equations to eventually establish an estimate of the form
\[\left\vert\left\vert {\rm Ric}\right\vert\right\vert_{\mathscr{Z}} \lesssim \epsilon \left\vert\left\vert {\rm Ric}\right\vert\right\vert_{\mathscr{Z}}.\]
This, of course, implies that ${\rm Ric} = 0$. In order for this strategy to be successful we also have to establish various estimates along the boundary $\{v = 0\}$, and it will be important that we have certain a priori estimate for ${\rm Ric}$ as $v\to -1$. These needed estimates will already be a consequence of the argument which produces $\left(\mathcal{M},g\right)$. 
\subsection{Building the Regular Coordinate Systems}
In Sections~\ref{axiscoordinates} and~\ref{toexterior} we extend our solution $\left(\mathcal{M},g\right)$ both to the axis, which corresponds to the limit $\frac{v}{-u} \to -1$, and to the past light cone of the singularity which corresponds to $\frac{v}{-u} \to 0$. Finally, we also establish that we can glue this interior solution to one of the exterior solutions produced in the work~\cite{nakedone}. We have already outlined how these arguments will go in Sections~\ref{oneone} and~\ref{twotwo} above.
\section{Preliminaries}\label{3iojoijio4}
\subsection{Constants and Cut-offs}\label{constantconstant}
In this section we introduce various useful conventions for constants and cut-offs.
\subsubsection{Fixed Constants}\label{fixedfixedfixedfixedconst}
We will have various constants which play a recurring role throughout the paper. 

First we have a positive integer $N_1$ which is assumed sufficiently large (but is otherwise arbitrary) and is then fixed. This represents the maximal number of angular derivatives which we shall use for commutation. We will not provide an explicit estimate for how large we must take $N_1$; nevertheless, it would be straightforward, if tedious, to run through the proof and produce a quantitative bound. Having fixed $N_1$, we then define another positive integer $N_2 \doteq \lfloor \frac{3}{4}N_1 \rfloor$. The integer $N_2$ will represent the maximal number of angular derivatives we use in certain (weighted) $L^{\infty}_vL^2_{\theta^A}$ estimates. 

Next we have a positive integer $\ell_0$ which is assumed sufficiently large and then fixed. This represents a certain spherical harmonic number which we use when decomposing tensors and functions into a ``bounded frequency part'' by applying $\mathcal{P}_{\ell \leq \ell_0}$ and a ``high frequency part'' by applying $\mathcal{P}_{\ell > \ell_0}$. (For an explanation of this spherical harmonic projection notation, see Section~\ref{roundround}.)

Next we have a positive constant $\check{\delta} > 0$ which is assumed to be sufficiently small and then fixed. The role of the constant $\check{\delta}$ is to modify various weights in order to avoid logarithmic divergences. In particular, the implied constants on the right hand sides of any given estimate may blow-up as $\check{\delta} \to 0$. 

We also have a constant $\check{p} > 0$ which is assumed to sufficiently small, depending on $N_1$, and then fixed. The small constant $\check{p}$ governs the allowed blow-up rates in our estimates of various quantities as $v\to 0$.\footnote{For the final solution produced in our main theorems, these estimates will turn out not to be sharp.} 

Finally, we have our fundamental smallness constant $\epsilon$. The constant $\epsilon$ is assumed to be positive and sufficiently small, depending on the values of the constants $N_1$, $\check{\delta}$, and $\check{p}$. Unless noted otherwise, all constants in the paper are considered independent of $\epsilon$. 

\subsubsection{Implied Constants and Cut-offs}\label{cutcutcut}
Many of our inequalities will be written with a $\lesssim$ which indicates the presence of an implied constants. Unless said otherwise these constants are always assumed independent of $\epsilon$ and also independent of any quantities involved in the inequality. Similarly, unless said otherwise all constants which appear explicitly are assumed to independent of $\epsilon$ and the quantities in the corresponding inequality unless said otherwise. We will employ the usual practice that the constant $C$ may change value from line to line. 

We will often desire to use a cut-off function of $v$. We introduce the definition that  $\xi(v)$ denotes a smooth function $v$ which is identically $0$ for $v \leq -1/2$ and is identically $1$ for $v \geq -1/4$.

\subsection{The Null Structure Equations for a General $3+1$ Dimensional Lorentzian Manifold}\label{doubleitup}
In this section we let $\left(\mathcal{M},g\right)$ denote a $3+1$ dimensional Lorentzian manifold where $\mathcal{M}$ is covered by coordinates $(u,v,\theta^A) \in \mathcal{U} \times \mathbb{S}^2$,
for some open set $\mathcal{U} \subset \mathbb{R}^2$, and $g$ takes the form
\begin{equation}\label{doublenullisg}
g = -2\Omega^2\left(du\otimes dv + dv\otimes du\right) + \slashed{g}_{AB}\left(d\theta^A - b^Adu\right)\otimes \left(d\theta^B - b^Bdu\right).
\end{equation}
\emph{Let us emphasize that throughout this section we do \underline{not} assume that the Einstein equations hold!}

We introduce the convention that Greek indices correspond to all spacetime directions while Latin indices only correspond to directions tangent to an $\mathbb{S}^2_{u,v}$. (See~\cite{KN} for a detailed introduction to the $\mathbb{S}^2_{u,v}$-tensor formalism in general relativity.) Since we will not be assuming that the Einstein vacuum equations hold, we avoid the use of the standard curvature symbols $\alpha$, $\beta$, etc., since there could be an ambiguity about which curvature component they would be referring to. It is also convenient to introduce the notation that for any symmetric $\mathbb{S}^2_{u,v}$ tensor $\Phi_{AB}$, $\hat{\Phi}_{AB}$ refers to the trace-free part. Thus, we have the following formula
\[\Phi_{AB} = \frac{1}{2}\left(\slashed{g}^{AB}\Phi_{AB}\right)\slashed{g}_{AB} + \hat{\Phi}_{AB}.\]
We also use the following standard convention for symmetrization and anti-symmetrization:
\[\Phi_{(AB)} \doteq \frac{1}{2}\left(\Phi_{AB} + \Phi_{BA}\right),\qquad \Phi_{[AB]} \doteq \frac{1}{2}\left(\Phi_{AB} - \Phi_{BA}\right).\]

We will denote the spacetime covariant derivative by $D$, the Lie-derivative by $\mathcal{L}$, and the induced covariant derivative on each $\mathbb{S}^2_{u,v}$ by $\slashed{\nabla}$. Finally, we adopt the curvature convention that
\[D_{\alpha}D_{\beta}\vartheta_{\gamma}  - D_{\beta}D_{\alpha}\vartheta_{\gamma} \doteq R_{\alpha\beta\gamma\delta}\vartheta^{\delta}. \]
Then we have that
\[{\rm Ric}_{\alpha\beta} \doteq g^{\gamma\delta}R_{\alpha\gamma\beta\delta},\qquad {\rm R} \doteq g^{\alpha\beta}{\rm Ric}_{\alpha\beta}.\]
We denote the corresponding induced curvature tensors on each $\mathbb{S}^2_{u,v}$ by $\slashed{R}_{ABCD}$, $\slashed{\rm Ric}_{AB}$, $\slashed{R}$, and the Gauss curvature of each $\mathbb{S}^2_{u,v}$ by $K$. (We have $K = \frac{1}{2}\slashed{R}$.)

We now introduce the Ricci coefficients:
\[\chi_{AB} \doteq g\left(D_Ae_4,e_B\right),\qquad \underline{\chi}_{AB} = g\left(D_Ae_3,e_B\right),\]
\[\eta_A \doteq -\frac{1}{2}g\left(D_3e_A,e_4\right),\qquad \underline{\eta}_A \doteq -\frac{1}{2}g\left(D_4e_A,e_3\right),\]
\[\omega \doteq -\frac{1}{4}g\left(D_4e_3,e_4\right),\qquad \underline{\omega} \doteq -\frac{1}{4}g\left(D_3e_4,e_3\right),\]
\[\zeta_A \doteq \frac{1}{2}g\left(D_Ae_4,e_3\right).\]
We define the null vectors $e_3$ and $e_4$ by
\[e_4 \doteq \Omega^{-1}\partial_v,\qquad e_3 \doteq \Omega^{-1}\left(\partial_u + b\cdot\slashed{\nabla}\right).\]
These will satisfy $g\left(e_3,e_4\right) = -2$. We denote the $\mathbb{S}^2_{u,v}$-projections of $D_{e_3}$ and $D_{e_4}$ by $\nabla_3$ and $\nabla_4$. It will also be convenient to set $\nabla_v \doteq \Omega\nabla_4$. 

It will also be useful to split $\chi$ and $\underline{\chi}$ into their trace-free parts $\hat{\chi}$ and $\hat{\underline{\chi}}$ and pure trace parts ${\rm tr}\chi$ and ${\rm tr}\underline{\chi}$ with respect to $\slashed{g}$:
\[\chi_{AB} = \hat{\chi}_{AB} + \frac{1}{2}{\rm tr}\chi\slashed{g}_{AB},\qquad \underline{\chi}_{AB} = \hat{\underline{\chi}}_{AB} + \frac{1}{2}{\rm tr}\underline{\chi}\slashed{g}_{AB}.\]

The next lemma relates the Ricci coefficients to derivatives of the metric components $\Omega$, $b$ and $\slashed{g}$.
\begin{lemma}\label{dermetrcomp}
\[\omega = -\frac{1}{2}\nabla_4\log\Omega,\qquad \underline{\omega} = -\frac{1}{2}\nabla_3\log\Omega,\qquad \mathcal{L}_{\partial_v}b^A = -4\Omega^2\zeta^A,\]
\[\eta_A = \zeta_A + \slashed{\nabla}_A\log\Omega,\qquad \underline{\eta}_A = -\zeta_A + \slashed{\nabla}_A\log\Omega,\]
\[\mathcal{L}_{e_4}\slashed{g}_{AB} = 2\chi_{AB},\qquad \mathcal{L}_{e_3}\slashed{g}_{AB} = 2\underline{\chi}_{AB}.\]
\end{lemma}
It is sometimes useful to note that we can re-write the relation between $b$ and $\zeta$ as
\begin{equation}\label{3m33zeta}
\nabla_4b_A - \chi_A^{\ \ B}b_B = -4\Omega\zeta_A.
\end{equation}

Now we recall the well-known formulas which relate $D$ to $\slashed{\nabla}$, $\nabla_3$, $\nabla_4$, and the Ricci coefficients.
\begin{lemma}\label{Dtonabla}
\[D_4e_4 = - 2\omega e_4,\qquad D_4e_3 = 2\omega e_3 + 2\underline{\eta}^Ae_A,\qquad D_4e_A = \underline{\eta}_Ae_4 + \nabla_4e_A, \]
\[D_3e_4 = 2\underline{\omega}e_4 + 2\eta^Ae_A,\qquad D_3e_3 = -2\underline{\omega}e_3,\qquad D_3e_A = \eta_Ae_3 + \nabla_3e_A,\]
\[D_Ae_4 = -\zeta_A e_4 + \chi_A^{\ B}e_B,\qquad D_Ae_3 = \zeta_A e_3 + \underline{\chi}_A^{\ B}e_B,\qquad D_Ae_B = \frac{1}{2}\underline{\chi}_{AB}e_4 + \frac{1}{2}\chi_{AB}e_3 + \slashed{\nabla}_Ae_B.\]
\end{lemma}

We next recall a few standard definitions. For $1$-forms $\nu_A$ and symmetric trace-free $(0,2)$-tensors $\vartheta_{AB}$, we have
\begin{align*}
\left(\nu^{(1)}\hat{\otimes}\nu^{(2)}\right)_{AB} &\doteq \nu^{(1)}_A\nu^{(2)}_B + \nu^{(1)}_B\nu^{(2)}_A - \slashed{g}^{CD}\nu^{(1)}_C\nu^{(2)}_D\slashed{g}_{AB},
\\ \nonumber \left(\slashed{\nabla}\hat{\otimes}\nu\right)_{AB} &\doteq \slashed{\nabla}_A\nu_B + \slashed{\nabla}_B\nu_A - \slashed{g}^{CD}\slashed{\nabla}_C\nu_D \slashed{g}_{AB},
\\ \nonumber \vartheta^{(1)}\wedge \vartheta^{(2)} &\doteq \slashed{\epsilon}^{AC}\slashed{g}^{BD}\vartheta^{(1)}_{AB}\vartheta^{(2)}_{CD},
\\ \nonumber  \nu^{(1)}\wedge \nu^{(2)} &\doteq \slashed{\epsilon}^{AB}\nu^{(1)}_A\nu^{(2)}_B,
\\ \nonumber \slashed{\rm div}\nu &\doteq \slashed{g}^{AB}\slashed{\nabla}_A\nu_B,
\\ \nonumber \slashed{\rm curl}\nu &\doteq \slashed{\epsilon}^{AB}\slashed{\nabla}_A\nu_B,
\\ \nonumber \slashed{\rm div}\vartheta_A &\doteq \slashed{g}^{BC}\slashed{\nabla}_B\vartheta_{CA}.
\end{align*}
Here $\slashed{\epsilon}$ denotes the volume form corresponding to $\slashed{g}$.

Now we present the null structure equations which relate the derivatives of Ricci coefficients to curvature components.

\begin{proposition}\label{thenullstructeqns}
\begin{align}
\label{4trchi}\nabla_4{\rm tr}\chi + \frac{1}{2}\left({\rm tr}\chi\right)^2 &=-{\rm Ric}_{44} -\left|\hat{\chi}\right|^2 - 2\omega{\rm tr}\chi,
\\ \label{4hatchi} \nabla_4\hat{\chi}_{AB}+{\rm tr}\chi \hat{\chi}_{AB} &= -\widehat{R_{A4B4}} -2\omega\hat{\chi}_{AB},
\\ \label{3truchi} \nabla_3{\rm tr}\underline{\chi} + \frac{1}{2}\left({\rm tr}\underline{\chi}\right)^2 &=-{\rm Ric}_{33} -\left|\hat{\underline{\chi}}\right|^2 - 2\underline{\omega}{\rm tr}\underline{\chi},
\\ \label{3hatuchi} \nabla_3\underline{\hat{\chi}}_{AB}+{\rm tr}\underline{\chi} \underline{\hat{\chi}}_{AB} &= -\widehat{R_{A3B3}} -2\underline{\omega}\underline{\hat{\chi}}_{AB},
\\ \label{3hatchi} \nabla_3\hat{\chi}_{AB} +\frac{1}{2}{\rm tr}\underline{\chi}\hat{\chi}_{AB}&= \widehat{{\rm Ric}}_{AB} + 2\underline{\omega}\hat{\chi}_{AB} + \left(\slashed{\nabla}\hat\otimes \eta\right)_{AB} + \left(\eta\hat\otimes \eta\right)_{AB} - \frac{1}{2}{\rm tr}\chi \hat{\underline{\chi}}_{AB},
\\ \label{3trchi}\nabla_3{\rm tr}\chi + \frac{1}{2}{\rm tr}\chi{\rm tr}\underline{\chi} &= \slashed{g}^{AB}R_{3AB4}+ 2\underline{\omega}{\rm tr}\chi + 2\slashed{\rm div}\eta + 2\left|\eta\right|^2 - \hat{\chi}\cdot\hat{\underline{\chi}},
\\ \label{4hatuchi} \nabla_4\hat{\underline{\chi}}_{AB} + \frac{1}{2}{\rm tr}\chi \hat{\underline{\chi}}_{AB}&= \widehat{{\rm Ric}}_{AB} + 2\omega\hat{\underline{\chi}}_{AB} + \left(\slashed{\nabla}\hat\otimes \underline\eta\right)_{AB} + \left(\underline\eta\hat\otimes \underline\eta\right)_{AB} - \frac{1}{2}{\rm tr}\underline{\chi} \hat{\chi}_{AB},
\\ \label{4truchi} \nabla_4{\rm tr}\underline{\chi} + \frac{1}{2}{\rm tr}\chi{\rm tr}\underline{\chi} &= \slashed{g}^{AB}R_{3AB4}+ 2\omega{\rm tr}\underline{\chi} + 2\slashed{\rm div}\underline{\eta} + 2\left|\underline\eta\right|^2 - \hat{\chi}\cdot\hat{\underline{\chi}},
\\ \label{4eta} \nabla_4\eta &= -\chi\cdot\left(\eta-\underline{\eta}\right) -\frac{1}{2}R_{A434},
\\ \label{3ueta} \nabla_3\underline{\eta} &= -\underline{\chi}\cdot\left(\underline\eta-\eta\right) -\frac{1}{2}R_{A343},
\\ \label{curleta} \slashed{\nabla}_A\eta_B-\slashed{\nabla}_B\eta_A &= R_{4[AB]3} + \hat{\underline{\chi}}^C_{\ \ [A}\hat{\chi}_{B]C},
\\ \label{curlueta} \slashed{\nabla}_A\underline{\eta}_B-\slashed{\nabla}_B\underline{\eta}_A &= -R_{4[AB]3} - \hat{\underline{\chi}}^C_{\ \ [A}\hat{\chi}_{B]C},
\\ \label{4uomega} \nabla_4\underline{\omega} &= \frac{1}{4}\left({\rm Ric}_{34} + \slashed{g}^{AB}R_{3AB4}\right) + 2\underline{\omega}\omega + \frac{1}{2}\left|\eta\right|^2 - \eta\cdot\underline{\eta},
\\ \label{3omega} \nabla_3\omega &= \frac{1}{4}\left({\rm Ric}_{34} + \slashed{g}^{AB}R_{3AB4}\right) + 2\underline{\omega}\omega + \frac{1}{2}\left|\underline\eta\right|^2 - \eta\cdot\underline{\eta},
\end{align}
and

\begin{align}\label{genGauss}
\slashed{R}_{ABCD} &= R_{ABCD} + \frac{1}{2}\left(\underline{\chi}_{BC}\chi_{AD} + \chi_{BC}\underline{\chi}_{AD} - \underline{\chi}_{AC}\chi_{BD} - \chi_{AC}\underline{\chi}_{BD}\right),
\\ \label{useful} \widehat{\slashed{g}^{CD}R_{ACBD}} &= 0,
\\ \label{ricequality} \widehat{{\rm Ric}}_{AB} &= \widehat{R_{3(AB)4}},
\\ \label{slashr} K &= \frac{1}{2}\slashed{g}^{AB}R_{3A4B} + \frac{1}{2}R +\frac{1}{2}{\rm Ric}_{34}+\frac{1}{2}\hat{\chi}\cdot\hat{\underline{\chi}} -\frac{1}{4}{\rm tr}\chi{\rm tr}\underline{\chi},
\\ \label{cod1} \slashed{\nabla}_A\chi_{BC} - \slashed{\nabla}_B\chi_{AC} &= R_{ABC4} +\chi_{AC}\zeta_B - \chi_{BC}\zeta_A,
\\ \label{cod2} \slashed{\nabla}_A\underline{\chi}_{BC} - \slashed{\nabla}_B\underline{\chi}_{AC} &= R_{ABC3} -\underline{\chi}_{AC}\zeta_B + \underline{\chi}_{BC}\zeta_A,
\\ \label{tcod1} \slashed{\nabla}^B\hat{\chi}_{AB}-\frac{1}{2}\slashed{\nabla}_A{\rm tr}\chi &=  -\frac{1}{2}R_{A434}+ {\rm Ric}_{4A}+ \frac{1}{2}{\rm tr}\chi \zeta_A - \zeta^B\hat{\chi}_{AB},
\\ \label{tcod2} \slashed{\nabla}^B\hat{\underline\chi}_{AB}-\frac{1}{2}\slashed{\nabla}_A{\rm tr}\underline\chi  &= -\frac{1}{2}R_{A343} + {\rm Ric}_{3A}- \frac{1}{2}{\rm tr}\underline\chi \zeta_A + \zeta^B\underline{\hat{\chi}}_{AB}.
\end{align}
\end{proposition}
\begin{proof}Equations~\eqref{4trchi},~\eqref{4hatchi},~\eqref{3truchi},~\eqref{3hatuchi},~\eqref{3trchi},~\eqref{4truchi},~\eqref{4eta},~\eqref{3ueta},~\eqref{curleta},~\eqref{curlueta},~\eqref{4uomega}, and~\eqref{3omega} follow in a straightforward fashion from the usual derivative of the null structure equations and the definition of the Riemann curvature and Ricci tensor.

Next we consider the equations~\eqref{3hatchi} and~\eqref{4hatuchi}. The usual derivation would give
\[\nabla_3\hat{\chi}_{AB} +\frac{1}{2}{\rm tr}\underline{\chi}\hat{\chi}_{AB}= \widehat{R_{3(AB)4}} + 2\underline{\omega}\hat{\chi}_{AB} + \left(\slashed{\nabla}\hat\otimes \eta\right)_{AB} + \left(\eta\hat\otimes \eta\right)_{AB} - \frac{1}{2}{\rm tr}\chi \hat{\underline{\chi}}_{AB},\]
\[\nabla_4\hat{\underline{\chi}}_{AB} + \frac{1}{2}{\rm tr}\chi \hat{\underline{\chi}}_{AB}= \widehat{R_{3(AB)4}} + 2\omega\hat{\underline{\chi}}_{AB} + \left(\slashed{\nabla}\hat\otimes \underline{\eta}\right)_{AB} + \left(\underline{\eta}\hat\otimes \underline{\eta}\right)_{AB} - \frac{1}{2}{\rm tr}\underline{\chi} \hat{\chi}_{AB}.\]
So,~\eqref{3hatchi} and~\eqref{4hatuchi} will follow if we can show~\eqref{ricequality}.

In order to derive~\eqref{ricequality} we first note that~\eqref{genGauss} is simply  the Gauss equation which relates the induced curvature tensor $\slashed{R}_{ABCD}$ on the $\mathbb{S}^2$'s with the spacetime curvature.

If we trace~\eqref{genGauss} then we obtain
\begin{align}
\nonumber \slashed{Ric}_{AB} &= \slashed{g}^{CD}R_{ACBD} + \frac{1}{2}\left(\underline{\chi}_B^{\ \ C}\chi_{AC} + \underline{\chi}_A^{\ \ C}\chi_{BC} - \underline{\chi}_{AB}{\rm tr}\chi - \chi_{AB}{\rm tr}\underline{\chi}\right)
\\ \label{tracegenGauss} &=\slashed{g}^{CD}R_{ACBD} + \hat{\underline{\chi}}^C_{\ \ (A}\hat{\chi}_{B)C} - \frac{1}{4}\slashed{g}_{AB}{\rm tr}\chi{\rm tr}\underline{\chi}.
\end{align}
By dimensional considerations, the trace-free part of $\slashed{Ric}_{AB}$ must vanish. We also recall that in two dimensions, the trace-free part of a symmetrized product of symmetric trace-free matrices must vanish. These two facts imply that if we take the trace-free part of~\eqref{tracegenGauss} we obtain
\[\widehat{\slashed{g}^{CD}R_{ACBD}} = 0\Rightarrow \widehat{{\rm Ric}}_{AB} = \widehat{R_{3(AB)4}},\]
where in the last implication we simply used the definition of Ricci curvature. This establishes~\eqref{useful} and~\eqref{ricequality} and hence we also obtain~\eqref{3hatchi} and~\eqref{4hatuchi}.

Tracing~\eqref{tracegenGauss} then leads to~\eqref{slashr}. Finally,~\eqref{cod1} and~\eqref{cod2} are the Codazzi equations relative to $e_4$ and $e_3$ respectively. Then~\eqref{tcod1} and~\eqref{tcod2} are obtained by suitable traces.

\end{proof}

We next take the opportunity to recall the familiar formulas for the commutator of $\nabla_4$ and $\nabla_3$ and $\mathcal{L}$ with $\slashed{\nabla}$ :
\begin{align}\label{acommut}
\left[\Omega\nabla_4,\slashed{\nabla}_A\right]\phi_{B_1\cdots B_k} &= \Omega\sum_{i=1}^k\left(R_{4AB_i}^{\ \ \ \ \ C}-\chi_A^{\ \ C}\underline{\eta}_{B_i}+\chi_{B_iA}\underline{\eta}^C\right)\phi_{B_1\cdots \hat{B_i}C\cdots B_k} 
\\ \nonumber &\qquad  - \Omega\chi_A^{\ \ C}\slashed{\nabla}_C\phi_{B_1\cdots B_k}
\\ \nonumber &=  \sum_{i=1}^k\left(-\slashed{\nabla}_{B_i}\left(\Omega\chi\right)_A^{\ \ C} + \slashed{\nabla}^C\left(\Omega\chi\right)_{AB_i}\right)\phi_{B_1\cdots \hat{B_i}C\cdots B_k} 
\\ \nonumber &\qquad  - \Omega\chi_A^{\ \ C}\slashed{\nabla}_C\phi_{B_1\cdots B_k},
\end{align}
\begin{align}\label{acommut2}
\left[\Omega\nabla_3,\slashed{\nabla}_A\right]\phi_{B_1\cdots B_k} &= \Omega\sum_{i=1}^k\left(R_{3AB_i}^{\ \ \ \ \ C}-\underline{\chi}_A^{\ \ C}\eta_{B_i}+\underline{\chi}_{B_iA}\eta^C\right)\phi_{B_1\cdots \hat{B_i}C\cdots B_k} 
\\ \nonumber &\qquad  - \Omega\underline{\chi}_A^{\ \ C}\slashed{\nabla}_C\phi_{B_1\cdots B_k}
\\ \nonumber &= \sum_{i=1}^k\left(-\slashed{\nabla}_{B_i}\left(\Omega\underline{\chi}\right)_A^{\ \ C} + \slashed{\nabla}^C\left(\Omega\underline{\chi}\right)_{AB_i}\right)\phi_{B_1\cdots \hat{B_i}C\cdots B_k} 
\\ \nonumber &\qquad  - \Omega\underline{\chi}_A^{\ \ C}\slashed{\nabla}_C\phi_{B_1\cdots B_k},
\end{align}
\begin{align}\label{acommut3}
\left[\mathcal{L}_X,\slashed{\nabla}_A\right]\phi_{B_1\cdots B_k} &= -\sum_{i=1}^k{}^{(X)}\Gamma_{B_iAC}\phi_{B_1\cdots\ \cdots B_K}^{\ \ \ \ \ C},
\end{align}
\begin{align}\label{32oj2oj4}
{}^{(X)}\Gamma_{ABC} = \frac{1}{2}\left(\slashed{\nabla}_A{}^{(X)}\pi_{BC} + \slashed{\nabla}_B{}^{(X)}\pi_{AC} -\slashed{\nabla}_C{}^{(X)}\pi_{AB} \right),
\end{align}
where ${}^{(X)}\pi$ denotes the deformation tensor of a vector field $X$ which is tangent to $\mathbb{S}^2$. (A proof of the formula~\eqref{acommut3} may be found in Lemma 7.1.3 of~\cite{ck}.)

It will later be useful to us to have the following formula which relates the $\partial_v$ derivative of the Gaussian curvature $K$ to suitable derivatives of $\hat{\chi}$ and ${\rm tr}\chi$.
\begin{lemma}We have
\begin{align}\label{ko2o3k4}
\mathcal{L}_{\partial_v}K + \Omega{\rm tr}\chi K &= \slashed{\rm div}\slashed{\rm div}\left(\Omega\hat{\chi}\right) - \frac{1}{2}\slashed{\Delta}\left(\Omega{\rm tr}\chi\right)
\\ \nonumber &= \slashed{\rm div}\slashed{\rm div}\left(\Omega\chi\right) -\slashed{\Delta}\left(\Omega{\rm tr}\chi\right).
\end{align}
We also have the following ``linearized'' version of this formula
\begin{align}\label{3l3omo29}
&\mathcal{L}_{\partial_v}\left(K - \frac{1}{(v+1)^2}\right) + \frac{2\Omega^2}{v+1}\left(K-\frac{1}{(v+1)^2}\right) + \Omega^2\left(\Omega^{-1}{\rm tr}\chi - \frac{2}{v+1}\right)\left(K - \frac{1}{(v+1)^2}\right) =
\\ \nonumber &\qquad \frac{2\left(1-\Omega^2\right)}{(v+1)^3} - \frac{\Omega^2}{(v+1)^2}\left(\Omega^{-1}{\rm tr}\chi - \frac{2}{v+1}\right) + \slashed{\rm div}\slashed{\rm div}\left(\Omega\hat{\chi}\right) - \left(\slashed{\Delta}\log\Omega\right)\Omega^2\left(\Omega^{-1}{\rm tr}\chi\right)
\\ \nonumber&\qquad - \frac{1}{2}\Omega^2\slashed{\Delta}\left(\Omega^{-1}{\rm tr}\chi - \frac{2}{v+1}\right) -4\Omega^2\left(\slashed{\nabla}\log\Omega\right)\slashed{\nabla}\left(\Omega^{-1}{\rm tr}\chi - \frac{2}{v+1}\right).
\end{align}
\end{lemma}
\begin{proof}The equation \eqref{ko2o3k4} is formula 5.28 from~\cite{Chr}. Then~\eqref{3l3omo29} follows from a straightforward calculation.
\end{proof}

The following lemma will be useful.
\begin{lemma}\label{othercommutelemma}Let $Y^A$ be a vector-field, and set $\Theta_A \doteq \slashed{g}_{AB}\mathcal{L}_{\partial_v}Y^B$. Then
\begin{align}\label{2ijoij1oi2}
\Omega\nabla_4\slashed{\rm div}Y &= \slashed{\rm div}\Theta + \mathcal{L}_Y\left(\Omega{\rm tr}\chi\right),
\end{align}
\begin{align}\label{3k2ki2i3i}
\Omega\nabla_4\left(\slashed{\nabla}\hat{\otimes}Y\right)_{AB} &= \left(\slashed{\nabla}\hat{\otimes}\Theta\right)_{AB} + 2\mathcal{L}_Y\left(\Omega\hat{\chi}\right)_{AB} - 2\slashed{\rm div}Y\left(\Omega\hat{\chi}\right)_{AB} - 2\left(\Omega\hat{\chi}\right)^C_{\ \ (A}\left(\slashed{\nabla}\hat{\otimes}Y\right)_{B)C},
\end{align}
\begin{align*}
\Omega\nabla_4\slashed{\rm curl}Y &= \slashed{\rm curl}\Theta + 2\slashed{\epsilon}^{BC}\slashed{\nabla}_B\left(\Omega\chi\right)_{CA}Y^A + \slashed{\epsilon}^{AB}\left[\left(\Omega\hat{\chi}\right)_{BC}\slashed{\nabla}_AY^C - \left(\Omega\hat{\chi}\right)_{AC}\slashed{\nabla}^CY_B\right]
\end{align*}
\end{lemma}
\begin{proof}
We have
\begin{align}\nonumber
\Omega\nabla_4\left(\slashed{\nabla}_AY_B + \slashed{\nabla}_BY_A\right)&= 2\slashed{\nabla}_{(A}\left(\Omega\nabla_4\right)Y_{B)}  + \left[-2\slashed{\nabla}_{(A}\left(\Omega\chi\right)_{B)C} + 2\slashed{\nabla}_C\left(\Omega\chi\right)_{AB}\right]Y^C -2\left(\Omega\chi\right)_{C(A}\slashed{\nabla}^CY_{B)}
\\ \nonumber &= 2\slashed{\nabla}_{(A}\left[\Theta_{B)} + \left(\Omega\chi\right)_{B)C}Y^C\right]
\\ \nonumber &\qquad + \left[-2\slashed{\nabla}_{(A}\left(\Omega\chi\right)_{B)C} + 2\slashed{\nabla}_C\left(\Omega\chi\right)_{AB}\right]Y^C -2\left(\Omega\chi\right)_{C(A}\slashed{\nabla}^CY_{B)}
\\ \nonumber &= 2\slashed{\nabla}_{(A}\Theta_{B)} + 2\left(\Omega\chi\right)_B^{\ \ C}\slashed{\nabla}_{[A}Y_{C]} + 2\left(\Omega\chi\right)_A^{\ \ C}\slashed{\nabla}_{[B}Y_{C]} +2\slashed{\nabla}_C\left(\Omega\chi\right)_{AB} Y^C
\\  \label{dogmo4k3sfw33} &= 2\slashed{\nabla}_{(A}\Theta_{B)} + 2\mathcal{L}_Y\left(\Omega\chi\right)_{AB} -2\left(\Omega\chi\right)_B^{\ \ C}\slashed{\nabla}_{(A}Y_{C)}   -2\left(\Omega\chi\right)_A^{\ \ C}\slashed{\nabla}_{(B}Y_{C)}.
\end{align}

Tracing~\eqref{dogmo4k3sfw33} yields
\begin{align*}
\Omega\nabla_4\slashed{\rm div}Y &= \slashed{\rm div}\Theta + \mathcal{L}_Y\left(\Omega{\rm tr}\chi\right)\end{align*}

Then taking the trace-free part of~\eqref{dogmo4k3sfw33} yields
\begin{align*}
\Omega\nabla_4\left(\slashed{\nabla}\hat{\otimes}Y\right)_{AB} &= \left(\slashed{\nabla}\hat{\otimes}\Theta\right)_{AB} + 2\mathcal{L}_Y\left(\Omega\hat{\chi}\right)_{AB} - 2\slashed{\rm div}Y\left(\Omega\hat{\chi}\right)_{AB} - 2\left(\Omega\hat{\chi}\right)^C_{\ \ (A}\left(\slashed{\nabla}\hat{\otimes}Y\right)_{B)C}.
\end{align*}

The final formula follows from a similar analysis.
\end{proof}

It will be useful to have the following commutation formulas.
\begin{lemma}Let $X$ be a vector field and $\Theta$ be a trace-free symmetric $(0,2)$-tensor. Then we have
\begin{align}\label{io32ijo32ij}
\left[\mathcal{L}_X,\slashed{g}^{BC}\slashed{\nabla}_C\right]\Theta_{AB} &=  -\left(\slashed{\nabla}\hat{\otimes}X\right)^{BC}\slashed{\nabla}_B\Theta_{CA} - \left(\slashed{\rm div}X\right) \slashed{\nabla}^B\Theta_{BA}
\\ \nonumber &\qquad  -\frac{1}{2}\slashed{\nabla}_A\left(\slashed{\nabla}\hat{\otimes}X\right)_{BC}\Theta^{BC} - \frac{1}{2}\left(\slashed{\nabla}_A\slashed{\rm div}X\right){\rm tr}\Theta -\slashed{\nabla}^A\left(\slashed{\nabla}\hat{\otimes}X\right)_{BC}\Theta_A^{\ \ C}.
\end{align}
\end{lemma}
\begin{proof}Let 
\[\pi_{AB} = \slashed{\nabla}_AX_B + \slashed{\nabla}_BX_A = \left(\slashed{\nabla}\hat{\otimes}X\right)_{AB} + \slashed{\rm div}X \slashed{g}_{AB},\]
be the deformation tensor of $X$. We then use~\eqref{32oj2oj4} and find that 
\begin{align*}
&\left[\mathcal{L}_X,\slashed{g}^{BC}\slashed{\nabla}_C\right]\Theta_{AB} =
 -\pi^{BC}\slashed{\nabla}_C\Theta_{AB} - \frac{1}{2}\slashed{g}^{BC}\left(\slashed{\nabla}_C\pi_{AD} + \slashed{\nabla}_A\pi_{CD} - \slashed{\nabla}_D\pi_{AC}\right)\Theta^D_{\ \ B}
 \\ \nonumber &\qquad \qquad \qquad  - \frac{1}{2}\slashed{g}^{BC}\left(\slashed{\nabla}_C\pi_{BD} + \slashed{\nabla}_B\pi_{CD} - \slashed{\nabla}_D\pi_{BC}\right)\Theta_A^{\ \ D}.
\end{align*}
Then~\eqref{io32ijo32ij} follows after simplifying. 
\end{proof}

Lastly, we now recall some standard notation for differential operators defined with respect to a Riemannian metric $\left(\mathbb{S}^2,\slashed{g}\right)$ with Guass curvature $K$: For any symmetric $2$-tensor $\Phi_{AB}$ we set
\[\left(\slashed{\mathcal{D}}_2\Phi\right)_A \doteq \slashed{\nabla}^B\Phi_{AB},\]
and for any $1$-form $\Theta_A$ we set
\[\left({}^*\slashed{\mathcal{D}}_2\Theta\right)_{AB} \doteq -\frac{1}{2}\left(\slashed{\nabla}_A\Theta_B + \slashed{\nabla}_B\Theta_A - \slashed{g}_{AB}\slashed{\nabla}^C\Theta_C\right),\qquad \slashed{\mathcal{D}}_1\Theta \doteq \left(\slashed{\nabla}^A\Theta_A,\slashed{\epsilon}^{AB}\slashed{\nabla}_A\Theta_B\right).\]
The operator ${}^*\slashed{\mathcal{D}}_2$ is the $L^2$-adjoint of $\slashed{\mathcal{D}}_2$ acting on trace-free tensors. The adjoint of $\slashed{\mathcal{D}}_1$ is given by
\[{}^*\slashed{\mathcal{D}}_1\left(\rho,\sigma\right)_A = -\slashed{\nabla}_A\rho + \slashed{\epsilon}_A^{\ \ B}\slashed{\nabla}_B\sigma.\]
We have the following well-known formulas (see Section 2.2 of~\cite{ck})
\begin{equation}\label{d2ids}
{}^*\slashed{\mathcal{D}}_2\slashed{\mathcal{D}}_2 = -\frac{1}{2}\slashed{\Delta} + K,\qquad \slashed{\mathcal{D}}_2{}^*\slashed{\mathcal{D}}_2 = -\frac{1}{2}\left(\slashed{\Delta} + K\right),\qquad {}^*\slashed{\mathcal{D}}_1\slashed{\mathcal{D}}_1 = -\slashed{\Delta} + K,\qquad \slashed{\mathcal{D}}_1{}^*\slashed{\mathcal{D}}_1 = -\slashed{\Delta}.
\end{equation}
Two consequences of~\eqref{d2ids} that will be particularly important for us are 
\begin{equation}\label{2m3momo2}
\slashed{\rm curl}\slashed{\rm div}\slashed{\nabla}\hat{\otimes}\Theta =\left(\slashed{\Delta} + 2K\right)\slashed{\rm curl}\Theta + 2 \left(\slashed{\nabla}K\right)\wedge \Theta, 
\end{equation}
\begin{equation}\label{3o3oioi4}
\slashed{\rm div}\slashed{\rm div}\slashed{\nabla}\hat{\otimes}\Theta = \left(\slashed{\Delta} + 2K\right)\slashed{\rm div}\Theta + 2\left( \slashed{\nabla}K\right)\cdot\Theta,
\end{equation}
for any $1$-form $\Theta_A$. The following formula will also be convenient to cite later:
\begin{lemma}\label{jioio2o2lkj3lijoi392}Let $\Phi$ be a symmetric trace-free $(0,2)$-tensor. Then
\begin{align*}
\slashed{\Delta}^2\Phi = 2{}^*\slashed{\mathcal{D}}_2{}^*\slashed{\mathcal{D}}_1\slashed{\mathcal{D}}_1\slashed{\mathcal{D}}_2\Phi -4{}^*\slashed{\mathcal{D}}_2 \left(K \slashed{\mathcal{D}}_2\Phi\right)
-4K{}^*\slashed{\mathcal{D}}_2\slashed{\mathcal{D}}_2 \Phi + 2\slashed{\Delta}\left(K \Phi\right).
\end{align*}
\end{lemma}
\begin{proof}This follows by systematically using~\eqref{d2ids} to express ${}^*\slashed{\mathcal{D}}_1\slashed{\mathcal{D}}_1$ in terms of $\slashed{\mathcal{D}}_2{}^*\slashed{\mathcal{D}}_2$ and lower order terms, and then using~\eqref{d2ids} again to express the operator ${}^*\slashed{\mathcal{D}}_2  \slashed{\mathcal{D}}_2{}^*\slashed{\mathcal{D}}_2  \slashed{\mathcal{D}}_2$ in terms of $\frac{1}{4}\slashed{\Delta}^2$ and lower terms.
\end{proof}

We now recall some standard consequences of elliptic theory, the uniformization theorem, and the formulas~\eqref{d2ids}. The following lemma will be stated for smooth metrics $\slashed{g}$ with positive Gauss curvature $K$, but it is straightforward to weaken the regularity to cover the case when $\slashed{g}$ lies in a suitable Sobolev space.
\begin{lemma}\label{skfijo3}Let $\slashed{g}$ be a smooth metric on $\mathbb{S}^2$ with positive Gauss curvature $K$. Then
\begin{enumerate}
	\item ${\rm ker}\left({}^*\slashed{\mathcal{D}}_2\right)$ is 6-dimensional.
	\item ${\rm ker}\left(\slashed{\mathcal{D}}_2\right)$ is trivial.
	\item ${\rm ker}\left(\slashed{\mathcal{D}}_1\right)$ is trivial.
	\item ${\rm ker}\left({}^*\slashed{\mathcal{D}}_1\right)$ is spanned by pairs of constant functions.
\end{enumerate}
\end{lemma}

We also now introduce the convention that for a multi-index $\alpha = \left(\alpha_1,\alpha_2,\alpha_3\right)$ with $\alpha_i \in \mathbb{Z}_{\geq 0}$,  the notation $\mathcal{L}_{Z^{(\alpha)}}$ refers to the corresponding product of Lie-derivatives relative to the angular momentum operators:
\[Z^1 = \partial_{\phi},\qquad Z^2 = \cos\phi\partial_{\theta} - \cot\theta\sin\phi\partial_{\phi},\qquad Z^3 = -\sin\phi\partial_{\theta} - \cot\theta\cos\phi \partial_{\phi}.\]
Furthermore, for any tensor $h$ we define
\[h^{(\alpha)} \doteq \mathcal{L}_{Z^{(\alpha)}}h.\]
Recall that each $Z^i$ is Killing with respect to $\mathring{\slashed{g}}$ and that the tangent bundle of $\mathbb{S}^2$ is spanned by $\{Z^i\}_{i=1}^3$. If we do not specify otherwise, all multi-indices $\alpha$ will be of the form $\alpha = \left(\alpha_1,\alpha_2,\alpha_3\right)$ for $\alpha_i \in \mathbb{Z}_{\geq 0}$.

\subsection{Hodge Decompositions and Spherical Harmonic Expansions on the Round Sphere}\label{roundround}
Throughout this paper we will use $\mathring{\slashed{g}}$ to denote the round metric on $\mathbb{S}^2$. In standard spherical coordinates $\mathbb{S}^2 = \left\{(\theta,\phi) \in [0,\pi] \times [0,2\pi)\right\}$, this metric takes the familiar form
\[\mathring{\slashed{g}} = d\theta^2 +  \sin^2\theta d\phi^2.\]
We will generally use the symbol $\mathring{\ }$ to denote that a given geometric quantity is defined with respect to the round metric. For example, $\mathring{\nabla}$ will denote the covariant derivative associated to $\mathring{\slashed{g}}$, $\mathring{\rm dVol}$ will denote the volume form associated to $\mathring{\slashed{g}}$,  $\mathring{\slashed{\epsilon}}$ will denote the volume form on $\mathring{\slashed{g}}$, $\mathring{H}^j$ will denote Sobolev spaces defined with respect to $\mathring{\slashed{g}}$, etc.

\subsubsection{Spherical harmonics}\label{sphericalhrmow2}
Associated  to the round metric  $\mathring{\slashed{g}}$ are the spherical harmonics which are denoted by $Y_m^{\ell}$, $\ell = 0,1,\cdots$ and $m \in \{-\ell,-\ell +1,\cdots,0,\cdots,\ell-1,\ell\}$. We have
\[\mathring{\Delta} Y_m^{\ell} = -\ell\left(\ell+1\right)Y_m^{\ell},\]
where, as per our conventions, $\mathring{\Delta} \doteq \mathring{\slashed{g}}^{AB}\mathring{\nabla}_A\mathring{\nabla}_B$ is the Laplace--Beltrami operator associated to $\mathring{\slashed{g}}$.  

For a sufficiently regular function $f$, the spherical harmonic expansion takes the form
\begin{equation}\label{sphharmoexpan}
f\left(\theta^A\right) = \sum_{\ell \in \mathbb{Z}_{\geq 0},\ |m| \leq \ell}\hat{f}(m,\ell)Y_m^{\ell}\left(\theta^A\right),
\end{equation}
\[\hat{f}(m,\ell) \doteq \int_{\mathbb{S}^2}f Y_m^{\ell}\mathring{dVol}.\]
This is, of course, an orthonormal expansion relative to $\mathring{dVol}$. Associated to this expansion are various projections. We fix some notation for certain of these projections in the following definition. 
\begin{definition}\label{thescalarprojections}Let $f$ be a square integrable function on $\left(\mathbb{S}^2,\mathring{\slashed{g}}\right)$. For any  $d \in \mathbb{Z}_{\geq 0}$ we then define 
\[\mathcal{P}_{\{\ell = d\}}f \doteq \sum_{|m| \leq d}\left(\int_{\mathbb{S}^2}fY_m^{d}\mathring{dVol}\right)Y^d_m.\]

Similarly, for any subset $\mathscr{A} \subset \mathbb{Z}_{\geq 0}$ we may define 
\[\mathcal{P}_{\{\ell\in \mathscr{A}\}}f \doteq \sum_{d \in \mathscr{A}}\mathcal{P}_{\ell = d}f.\]

\end{definition}
\begin{remark}\label{23oij2391}When $\mathscr{A} = \{\ell : \ell \geq d\}$, $\{\ell : \ell > d\}$, $\{\ell : \ell \leq d\}$, $\{\ell : \ell < d\}$, etc., then we may replace $\{\ell \in \mathscr{A}\}$ with $\{\ell \geq d\}$, $\{\ell > d\}$, $\{\ell \leq d\}$, $\{\ell < d\}$, etc. We also may sometimes write $\mathcal{P}_{\mathscr{A}}$ instead of $\mathcal{P}_{\ell \in \mathscr{A}}$ when there is unlikely to be any confusion.
\end{remark}

The above facts allow us to make the following definition:
\begin{definition}For any  $F \in L^2\left(\mathbb{S}^2\right)$ with $\mathcal{P}_{\ell = 0}F = 0$, we define $\mathring{\Delta}^{-1}F$ to be the unique solution to $\mathring{\Delta}\left(\mathring{\Delta}^{-1}F\right) = F$ so that $\mathcal{P}_{\ell = 0}\left(\mathring{\Delta}^{-1}F\right) = 0$.

\end{definition}

\subsubsection{Hodge decomposition for a $1$-form}\label{sechodge1form}
In this section we review how the spherical harmonics may be used to build an orthonormal basis for $1$-forms on $\left(\mathbb{S}^2,\mathring{\slashed{g}}\right)$. We will use $\mathring{\slashed{\epsilon}}_{AB}$ to denote the volume form on $\left(\mathbb{S}^2,\mathring{\slashed{g}}\right)$. For any $1$-form $\Theta_A$ we then define
\[\mathring{\rm div}\Theta \doteq \mathring{\nabla}^A\Theta_A,\qquad \mathring{\rm curl}\Theta \doteq \mathring{\slashed{\epsilon}}^{AB}\mathring{\nabla}_A\Theta_B.\]
Any sufficiently regular $1$-form $\Theta_A$ may be written as
\begin{equation}\label{hodge1form}
\Theta_A = \mathring{\nabla}_Af + \mathring{\slashed{\epsilon}}_A^{\ \ B}\mathring{\nabla}_Bh,
\end{equation}
where the functions $f$ and $h$ are uniquely specified by requiring that they have a vanishing average and that they satisfy
\[\mathring{\Delta}f = \mathring{\rm div}\Theta,\qquad \mathring{\Delta}h = -\mathring{\rm curl}\Theta.\]
We note that~\eqref{hodge1form} is an orthogonal decomposition.

We may now make  a definition.
\begin{definition}Let $\Theta_A$ be a $1$-form with the decomposition~\eqref{hodge1form}. Then we define 
\[\mathring{\Pi}_{{\rm div}}\Theta_A \doteq \mathring{\nabla}_Af,\qquad \mathring{\Pi}_{\rm curl}\Theta_A\doteq \mathring{\slashed{\epsilon}}_A^{\ \ B}\nabla_Bh.\]
\end{definition}
It is easily verified that 
\[\mathring{\rm curl}\left(\mathring{\Pi}_{{\rm div}}\Theta\right) = 0,\qquad \mathring{\rm div}\left(\mathring{\Pi}_{\rm curl}\Theta\right) = 0.\]

Using the spherical harmonic expansions~\eqref{sphharmoexpan} for the functions $f$ and $h$ in~\eqref{hodge1form} we obtain a expansion for any square integrable $1$-form $\Theta_A$ in terms of $\{\mathring{\nabla}_AY_m^{\ell}\}_{\ell \geq 1}$ and $\{\mathring{\slashed{\epsilon}}_A^{\ \ B}\mathring{\nabla}_BY_m^{\ell}\}_{\ell \geq 1}$:
\begin{align}
\nonumber \Theta_A &= \sum_{\ell \in \mathbb{Z}_{\geq 1},\ |m| \leq \ell}\left[\left(\int_{\mathbb{S}^2}fY_m^{\ell}\mathring{dVol}\right)\mathring{\nabla}_AY_m^{\ell} + \left(\int_{\mathbb{S}^2}hY_m^{\ell}\mathring{dVol}\right)\mathring{\slashed{\epsilon}}_A^{\ \ B}\mathring{\nabla}_BY_m^{\ell}\right]
\\ \label{sphharmoexpan1form} &= \sum_{\ell \in \mathbb{Z}_{\geq 1},\ |m| \leq \ell}\frac{1}{\ell(\ell+1)}\left[\left(\int_{\mathbb{S}^2}\Theta^A\mathring{\nabla}_AY_m^{\ell}\mathring{dVol}\right)\mathring{\nabla}_AY_m^{\ell} + \left(\int_{\mathbb{S}^2}\Theta^A\mathring{\slashed{\epsilon}}_A^{\ \ B}\mathring{\nabla}_BY_m^{\ell}\mathring{dVol}\right)\mathring{\slashed{\epsilon}}_A^{\ \ B}\mathring{\nabla}_BY_m^{\ell}\right].
\end{align}
This expansion is orthonormal. One may show (see~\eqref{d2ids}) that 
\[\mathring{\Delta}\mathring{\nabla}_AY_m^{\ell} = \left(1-\ell\left(\ell+1\right)\right)\mathring{\nabla}_AY_m^{\ell},\qquad \mathring{\Delta}\left( \mathring{\slashed{\epsilon}}_{AB}\mathring{\nabla}^BY_m^{\ell}\right) = \left(1-\ell\left(\ell+1\right)\right)\mathring{\slashed{\epsilon}}_{AB}\mathring{\nabla}^BY_m^{\ell}.\]

Next we have the analogue of Definition~\ref{thescalarprojections}.
\begin{definition}\label{theformprojections}Let $\Theta_A$ be a square integrable $1$-form on $\left(\mathbb{S}^2,\mathring{\slashed{g}}\right)$. For any  $d \in \mathbb{Z}_{\geq 1}$ we then define 
\[\mathcal{P}_{\{\ell = d\}}\Theta_A \doteq\frac{1}{d(d+1)}\sum_{ |m| \leq d}\left[\left(\int_{\mathbb{S}^2}\Theta^A\mathring{\nabla}_AY_m^d\mathring{dVol}\right)\mathring{\nabla}_AY_m^d + \left(\int_{\mathbb{S}^2}\Theta^A\mathring{\slashed{\epsilon}}_A^{\ \ B}\mathring{\nabla}_BY_m^d\mathring{dVol}\right)\mathring{\slashed{\epsilon}}_A^{\ \ B}\mathring{\nabla}_BY_m^d\right].\]

Similarly, for any subset $\mathscr{A} \subset \mathbb{Z}_{\geq 1}$ we may define 
\[\mathcal{P}_{\{\ell\in \mathscr{A}\}}\Theta \doteq \sum_{d \in \mathscr{A}}\mathcal{P}_{\ell = d}\Theta.\]

\end{definition}
The analogue of Remark~\ref{23oij2391} holds also in this setting.

We observe the following commutation relations:
\[\left[\mathring{\Pi}_{\rm div},\mathcal{P}_{\ell = d}\right]\Theta = \left[\mathring{\Pi}_{\rm curl},\mathcal{P}_{\ell = d}\right]\Theta = 0.\]

Lastly, we observe that by raising and lowering indices with respect to $\mathring{\slashed{g}}$, we can extend the considerations of this section to vector fields $X^A$. 
\subsubsection{Hodge decomposition for a  $(0,2)$-tensor}\label{sechodge2form}
In this section we review how the spherical harmonics may be used to build an orthonormal basis for $2$-tensors on $\left(\mathbb{S}^2,\mathring{\slashed{g}}\right)$.  First of all, any symmetric $2$-tensor $\Phi_{AB}$ can be decomposed into a pure-trace part, symmetric trace-free part, and an anti-symmetric part:
\begin{equation}\label{tracetracefreefred}
\Phi_{AB} = \frac{1}{2}\left(\mathring{\slashed{g}}^{CD}\Phi_{CD}\right)\mathring{\slashed{g}}_{AB} + \hat{\Phi}_{AB} + \frac{1}{2}\left(\mathring{\slashed{\epsilon}}^{CD}\Phi_{CD}\right)\mathring{\slashed{\epsilon}}_{AB}.
\end{equation}
The functions $\mathring{\slashed{g}}^{CD}\Phi_{CD}$ and $\mathring{\slashed{\epsilon}}^{CD}\Phi_{CD}$ may be expanded into spherical harmonics via~\eqref{sphharmoexpan}. 

For the trace-free part symmetric part $\hat{\Phi}$, we start by recalling that such a tensor  can be represented by
\begin{equation}\label{hodge2ten}
\hat{\Phi}_{AB} = \mathring{\nabla}_A\Theta_B + \mathring{\nabla}_B\Theta_A - \mathring{\slashed{g}}_{AB}\mathring{\nabla}^C\Theta_C,
\end{equation}
where the $1$-form $\Theta$ may by uniquely specified by requiring that $\mathcal{P}_{\ell = 1}\Theta = 0$ and that
\[\mathring{\Delta}\Theta_A + \Theta_A = \mathring{\nabla}^B\hat{\Phi}_{AB}.\]
Expanding $\Theta$ into spherical harmonics then yields a corresponding expansion for $\hat{\Phi}$ in terms of $\left\{\left(\mathring{\nabla}\hat{\otimes} \mathring{\nabla} Y_m^{\ell}\right)_{AB}\right\}_{\ell \geq 2}$ and $\left\{\left(\mathring{\nabla}\hat{\otimes} {}^*\mathring{\nabla}Y_m^{\ell}\right)_{AB}\right\}_{\ell \geq 2}$. One may show (see~\eqref{d2ids}) that
\[\mathring{\Delta}\left(\mathring{\nabla}\hat{\otimes} \mathring{\nabla} Y_m^{\ell},\mathring{\nabla}\hat{\otimes} {}^*\mathring{\nabla} Y_m^{\ell}\right) =\left(4-\ell(\ell+1)\right)\left(\mathring{\nabla}\hat{\otimes} \mathring{\nabla} Y_m^{\ell},\mathring{\nabla}\hat{\otimes}{}^*\mathring{\nabla} Y_m^{\ell}\right). \]
The analogue of~\eqref{sphharmoexpan1form} is then the following orthonormal expansion
\begin{align}
\nonumber \Phi &= \frac{1}{2}\sum_{\ell \in\mathbb{Z}_{\geq 0}} \left(\int_{\mathbb{S}^2}\mathring{\slashed{g}}^{AB}\Phi_{AB} Y_m^{\ell}\mathring{dVol}\right)Y_m^{\ell} \mathring{\slashed{g}}+ \frac{1}{2}\sum_{\ell \in\mathbb{Z}_{\geq 0}} \left(\int_{\mathbb{S}^2}\mathring{\slashed{\epsilon}}^{AB}\Phi_{AB} Y_m^{\ell}\mathring{dVol}\right)Y_m^{\ell} \mathring{\slashed{\epsilon}}
\\ \label{sphharmoexpan1form} &\qquad \sum_{\ell \in \mathbb{Z}_{\geq 2},\ |m| \leq \ell}\frac{1}{c(\ell)}\left[\left(\int_{\mathbb{S}^2}\Phi\cdot\mathring{\nabla}\hat{\otimes} \mathring{\nabla}Y_m^{\ell}\mathring{dVol}\right)\mathring{\nabla}\hat{\otimes}\mathring{\nabla}Y_m^{\ell} + \left(\int_{\mathbb{S}^2}\Phi\cdot \mathring{\nabla}\hat{\otimes}{}^*\mathring{\nabla} Y_m^{\ell}\mathring{dVol}\right)\mathring{\nabla}\hat{\otimes}{}^* \mathring{\nabla}Y_m^{\ell}\right],
\end{align}
where $c(\ell) \doteq 2\ell(\ell+1)\left(\ell(\ell+1)-2\right)$.  Again, the analogue of Remark~\ref{23oij2391} holds also in this setting.

\subsection{Smoothing Operators on $\mathbb{S}^2$}\label{smoothingopeatorsection}
It will be convenient to have a family of regularizing operators which interact well with geometry of the round sphere.
\begin{lemma}\label{thesmoothlemma}There exists a a family of operators $\{\Pi_{\delta}\}_{0 \leq \delta \ll 1} : L^2(\mathbb{S}^2) \to L^2(\mathbb{S}^2)$, where $L^2(\mathbb{S}^2)$ is defined with respect to the round metric, which satisfy the following properties for any $(0,k)$ tensors $f_{A_1\cdots A_k},h_{A_1\cdots A_k} \in L^2(\mathbb{S}^2)$ with $k \in \{0,1,2\}$:
\begin{enumerate}
	\item $\Pi_0$ is the identity operator. 
	\item $\Pi_{\delta}f \in C^{\infty}(\mathbb{S}^2)$ if $\delta > 0$, and for ever positive integer $s$, we have
	\begin{equation}\label{smoothitprop1}
	\left\vert\left\vert \Pi_{\delta}f\right\vert\right\vert_{\mathring{H}^s} \lesssim_s \delta^{-\frac{s}{2}} \left\vert\left\vert f\right\vert\right\vert_{L^2}. 
	\end{equation}
	\item We have
	\begin{equation}\label{smoothitprop2}
	\int_{\mathbb{S}^2}\left(\Pi_{\delta}f\right)h\, \mathring{\rm dVol} = \int_{\mathbb{S}^2}f\left(\Pi_{\delta}h\right)\, \mathring{\rm dVol}.
	\end{equation}
	\item For every Killing field $Z$ of the round metric, spherical harmonic projector $\mathcal{P}_{\ell \in \mathscr{A}}$, and spherical laplacian $\mathring{\Delta}$, we have
	\begin{equation}\label{smoothitprop3}
	\left[Z,\Pi_{\delta}\right] = \left[\mathcal{P}_{\ell \in \mathscr{A}},\Pi_{\delta}\right] = \left[\mathring{\Delta},\Pi_{\delta}\right] = 0.
	\end{equation}
	\item We have $\lim_{\delta \to 0}\Pi_{\delta}f = f$ in $L^2$. Moreover, if $f \in \mathring{H}^s$ for some positive integer $s$, then we also have
	\begin{equation}\label{smoothitprop4}
	\left\vert\left\vert f - \Pi_{\delta}f\right\vert\right\vert_{\mathring{H}^s} = 0 \text{ as }\delta\to 0,\qquad \left\vert\left\vert \Pi_{\delta}f - f\right\vert\right\vert_{\mathring{H}^{s-1}(\mathbb{S}^2)} \lesssim \delta^{1/2} \left\vert\left\vert f\right\vert\right\vert_{\mathring{H}^s}. 
	\end{equation}
	\item For every vector field $X$ on $\mathbb{S}^2$ and non-negative integer $s$, we have
	\begin{align}\label{smoothitprop5}
	&\left\vert\left\vert \left[\mathcal{L}_X,\Pi_{\delta}\right]f\right\vert\right\vert_{\mathring{H}^s} \lesssim \\ \nonumber &\qquad {\rm min}\left(\left\vert\left\vert X\right\vert\right\vert_{\mathring{H}^{1+s}}\left\vert\left\vert f\right\vert\right\vert_{\mathring{W}^{\lceil \frac{s+3}{2}\rceil,\infty}} + \left\vert\left\vert X\right\vert\right\vert_{\mathring{W}^{\lceil \frac{s+3}{2}\rceil,\infty}}\left\vert\left\vert f\right\vert\right\vert_{\mathring{H}^s}, \left\vert\left\vert X\right\vert\right\vert_{\mathring{W}^{s+1,\infty}}\left\vert\left\vert f\right\vert\right\vert_{\mathring{H}^s}\right). 
	\end{align}
	Similarly, for every function $y : \mathbb{S}^2 \to \mathbb{R}$ and non-negative $s \in \mathbb{Z}_{\geq 0}$, we have 
	\begin{align}\label{smoothitprop6}
	&\left\vert\left\vert \left[y,\Pi_{\delta}\right]f\right\vert\right\vert_{\mathring{H}^s} \lesssim
	\\ \nonumber &\qquad {\rm min}\left( \left\vert\left\vert y\right\vert\right\vert_{\mathring{H}^{1+s}}\left\vert\left\vert f\right\vert\right\vert_{\mathring{W}^{\lceil\frac{s}{2} \rceil+ 1,\infty}} + \left\vert\left\vert y\right\vert\right\vert_{\mathring{W}^{\lceil\frac{s}{2}\rceil+1,\infty}}\left\vert\left\vert f\right\vert\right\vert_{\mathring{H}^{{\rm max}(s-1,0)}},\left\vert\left\vert y\right\vert\right\vert_{\mathring{W}^{1+s,\infty}}\left\vert\left\vert f\right\vert\right\vert_{\mathring{H}^{{\rm max}(s-1,0)}}\right). 
	\end{align}
\end{enumerate}
\end{lemma}
\begin{proof}For any tensor $f$ along  $\mathbb{S}^2$ we have a corresponding solution $w$ to the heat equation with initial data determined by $f$:
\begin{equation}\label{m3omo4mo3o}
\partial_tw - \mathring{\Delta}w = 0,\qquad w|_{t=0} = f.
\end{equation}
We now define 
\[\Pi_{\delta}f \doteq w|_{t=\delta}.\]
The properties of $\Pi_{\delta}$ are then straightforward consequences of parabolic regularity and energy estimates.
\end{proof}

 \section{The Main Bootstrap Norms and Basic Inequalities}\label{sectionbootstrapnorms}

In this section we will define some norms which will serve as ``bootstrap'' norms in various contexts throughout the paper. We will then also record some basic estimates which will be used repeatedly.

Throughout this section we let $\tilde{\kappa}$ be a constant satisfying $\left|\tilde{\kappa}\right| \lesssim \epsilon$ and $\tilde{\Omega} : (-1,0) \times \mathbb{S}^2 \to (0,\infty)$ denote a function which is required to satisfy
\begin{equation}\label{k2moo39}
\tilde{\Omega}^2 \sim (-v)^{-2\tilde{\kappa}}.
\end{equation}

\subsection{Norm Notation}\label{notationfornorms}
 In order to keep our notation compact, it will be convenient to introduce various notations for weighted norms which commonly appear. Unless said otherwise, we will assume that the tensors we consider are $\mathbb{S}^2_{-1,v}$ tensors defined for $v \in (-1,0)$. We may equip each $\mathbb{S}^2_{-1,v}$ with the metric $\left(v+1\right)^2\mathring{\slashed{g}}$ (where we recall that $\mathring{\slashed{g}}$ denotes a choice of a round metric). Unless said otherwise, throughout the definitions given in this section, norms of $\mathbb{S}^2_{-1,v}$ tensors are computed with respect to $\left(v+1\right)^2\mathring{\slashed{g}}$ and also indices are raised and lowered with respect to $\left(v+1\right)^2\mathring{\slashed{g}}$.

 Our first norm we define is a weighted-$L^2$ norm in $v$ and along $\mathbb{S}^2$. Large numbers of derivatives of our solution will need to be controlled in such a norm. 
 \begin{definition}For any $\mathbb{S}^2_{-1,v}$ tensor $\phi$, non-negative integer $N \geq 0$, $(p_1,p_2) \in \mathbb{R}^2$, and $-1 \leq a_1 < a_2 \leq 0$ we set
 \[\left\vert\left\vert \phi\right\vert\right\vert^2_{\mathscr{Q}_{a_1}^{a_2}\left(N,p_1,p_2\right)} \doteq \sum_{\left|\alpha\right| \leq N}\int_{a_1}^{a_2}\int_{\mathbb{S}^2}\left(v+1\right)^{2p_1}(-v)^{2p_2}\left|\phi^{(\alpha)}\right|^2\, dv\, \mathring{\rm dVol}.\]
 If we write $\mathscr{Q}\left(N,p_1,p_2\right)$ instead of $\mathscr{Q}_{a_1}^{a_2}\left(N,p_1,p_2\right)$, then we take $a_1 = -1$ and $a_2 = 0$. We then define the space $\mathscr{Q}_{a_1}^{a_2}\left(N,p_1,p_2\right)[k]$ to be the corresponding completion of smooth $\mathbb{S}^2_{-1,v}$ $(0,k)$-tensors.
 \end{definition} 
 
 The next norm we define is a weighted $L^{\infty}_v$ and $L^2$ along $\mathbb{S}^2$ norm. Norms of this type will be used to  control the solution if there are not too many angular derivatives.
 \begin{definition}For any $\mathbb{S}^2_{-1,v}$ tensor $\phi$, non-negative integer $N \geq 0$, $(p_1,p_2) \in \mathbb{R}^2$, and $-1 \leq a_1 < a_2 \leq 0$ we set
 \[\left\vert\left\vert \phi\right\vert\right\vert^2_{\mathscr{S}_{a_1}^{a_2}\left(N,p_1,p_2\right)} \doteq \sum_{\left|\alpha\right| \leq N}\sup_{v\in (a_1,a_2)}\int_{\mathbb{S}^2}\left(v+1\right)^{2p_1}(-v)^{2p_2}\left|\phi^{(\alpha)}\right|^2\, \mathring{\rm dVol}.\]
 If we write $\mathscr{S}\left(N,p_1,p_2\right)$ instead of $\mathscr{S}_{a_1}^{a_2}\left(N,p_1,p_2\right)$, then we take $a_1 = -1$ and $a_2 = 0$. We then define the space $\mathscr{S}_{a_1}^{a_2}\left(N,p_1,p_2\right)[k]$ to be the corresponding completion of smooth $\mathbb{S}^2_{-1,v}$ $(0,k)$-tensors.
 \end{definition}
 
 The following norm is a variant of the $\mathscr{S}$-norm. The first key feature is that the control as $v\to 0$ is allowed to change for every angular derivative one takes (generally the estimate will become weaker). The second key feature is the addition of a term integrated in $v$ with the weight $(-v)^{-1+2p_2}$. This represents a slight strengthening of the estimate near $v = 0$ which is often useful in applications.
 \begin{definition}For any $\mathbb{S}^2_{-1,v}$ tensor $\phi$, non-negative integer $N \geq 0$, $(p_1,p_2,p_3) \in \mathbb{R} \times \mathbb{R}\times \mathbb{R}$ and $-1 \leq a_1 < a_2 \leq 0$, we set
 \begin{align*}
& \left\vert\left\vert \phi\right\vert\right\vert^2_{\check{\mathscr{S}}_{a_1}^{a_2}\left(N,p_1,p_2,p_3\right)} \doteq \sum_{\left|\alpha\right| \leq N}\sup_{v\in (a_1,a_2)}\int_{\mathbb{S}^2}\left(v+1\right)^{2p_1}(-v)^{2p_2+2\left|\alpha\right|p_3}\left|\phi^{(\alpha)}\right|^2\,\mathring{\rm dVol} 
\\ \nonumber &\qquad \qquad \qquad \qquad \qquad +\sum_{\left|\alpha\right| \leq N} \int_{-1/2}^{{\rm max}\left(a_2,-1/2\right)}\int_{\mathbb{S}^2}(-v)^{-1+2p_2+2\left|\alpha\right|p_3}\left|\phi^{(\alpha)}\right|^2\, dv\, \mathring{\rm dVol}.
 \end{align*}
 If we write $\check{\mathscr{S}}\left(N,p_1,p_2,p_3\right)$ instead of $\check{\mathscr{S}}_{a_1}^{a_2}\left(N,p_1,p_2,p_3\right)$, then we take $a_1 = -1$ and $a_2 = 0$. We then define the space $\check{\mathscr{S}}_{a_1}^{a_2}\left(N,p_1,p_2,p_3\right)[k]$ to be the corresponding completion of smooth $\mathbb{S}^2_{-1,v}$ $(0,k)$-tensors.
 \end{definition}

\subsection{Bootstrap Norms}\label{bootthenormbootthenorm}
We now present the norms which we shall employ as bootstrap norms in many of our arguments. We start with the norm which involves the maximal number of derivatives.
\begin{definition}\label{om3om039cji2}Let $\tilde{b}^A$ be an $\mathbb{S}^2_{-1,v}$ vector field and $\tilde{\slashed{g}}_{AB}$ be a Riemannian metric on each $\mathbb{S}^2_{-1,v}$ for $v \in (-1,0)$. Then, for any function $\phi\left(v,\theta^A\right): (-1,0) \times \mathbb{S}^2 \to \mathbb{R}$, $\mathbb{S}^2_{-1,v}$ vector field $h^A$ for $v \in (-1,0)$, and a positive definite symmetric $(0,2)$-$\mathbb{S}^2_{-1,v}$ tensor $p_{AB}$ for $v \in (-1,0)$, we define
\begin{align}\label{2k3om2o3949}
&\left\vert\left\vert \phi \right\vert\right\vert_{\mathscr{A}_0\left(\tilde{\kappa},\tilde{b}\right)} \doteq \left\vert\left\vert \phi\right\vert\right\vert_{\mathscr{Q}\left(N_1,-3/2+\check{\delta},-\tilde{\kappa}\right)}+\left\vert\left\vert \phi \right\vert\right\vert_{\mathscr{Q}\left(N_1-1,-3/2+\check{\delta},-1/2+\check{\delta}\right)}
\\ \nonumber &+\left\vert\left\vert \left((-v)\mathcal{L}_{\partial_v}-\mathcal{L}_{\tilde{b}}\right)\phi \right\vert\right\vert_{\mathscr{Q}\left(N_1-1,-1/2+\check{\delta},-1/2+\check{\delta}\right)} 
\\ \nonumber &+\sum_{j=0}^1 \left\vert\left\vert \mathcal{L}^{1+j}_{\partial_v}\left((-v)\mathcal{L}_{\partial_v}-\mathcal{L}_{\tilde{b}}\right)\phi\right\vert\right\vert_{\mathscr{Q}\left(N_1-2-j,1/2+\check{\delta}+j,\tilde{\kappa}+j\right)}
\\ \nonumber &+\sum_{j=0}^1 \left\vert\left\vert \mathcal{L}^{1+j}_{\partial_v}\left((-v)\mathcal{L}_{\partial_v}-\mathcal{L}_{\tilde{b}}\right)\phi\right\vert\right\vert_{\mathscr{Q}_{-1/2}^0\left(N_1-3-j,0,-\sqrt{\check{p}}+j\right)},
\end{align}
\begin{align}\label{2pk300jj0}
&\left\vert\left\vert h\right\vert\right\vert_{\mathscr{A}^-_1\left(\tilde{\kappa}\right)} \doteq  \left\vert\left\vert h \right\vert\right\vert_{\mathscr{Q}\left(N_1,-3/2+\check{\delta}, -\tilde{\kappa} \right)} +  \left\vert\left\vert h\right\vert\right\vert_{\mathscr{Q}\left(N_1-1,-3/2+\check{\delta}, -1/2+\check{\delta} \right)} 
\\ \nonumber &\qquad +\sum_{j=0}^1\left\vert\left\vert \mathcal{L}^{1+j}_{\partial_v}h\right\vert\right\vert_{\mathscr{Q}\left(N_1-1-j,-1/2+\check{\delta}+j,1/2+j\right)} 
 + \left\vert\left\vert \left(1,v\mathcal{L}_{\partial_v}\right)\mathcal{L}_{\partial_v}h\right\vert\right\vert_{\mathscr{Q}_{-1/2}^0\left(N_1-2,0,\tilde{\kappa}\right)}
 \\ \nonumber &\qquad + \left\vert\left\vert \left(1,v\mathcal{L}_{\partial_v}\right)\mathcal{L}_{\partial_v}h\right\vert\right\vert_{\mathscr{Q}_{-1/2}^0\left(N_1-3,0,-\sqrt{\check{p}}\right)},
\end{align}
\begin{align}\label{2pk300jj0123}
&\left\vert\left\vert h\right\vert\right\vert_{\mathscr{A}_1\left(\tilde{\kappa},\tilde{\slashed{g}}\right)} \doteq &
\\ \nonumber &\qquad  \left\vert\left\vert h\right\vert\right\vert_{\mathscr{Q}\left(N_1,-3/2+\check{\delta},-\tilde{\kappa}\right)} + \sum_{j=0}^1\left\vert\left\vert \mathcal{L}_{\partial_v}^{1+j}h\right\vert\right\vert_{\mathscr{Q}\left(N_1-1-j,-1/2+\check{\delta}+j,\tilde{\kappa} +j \right)}+\left\vert\left\vert \left(1,\widetilde{\slashed{\nabla}\hat{\otimes}},\widetilde{\slashed{\rm div}}\right)h\right\vert\right\vert_{\mathscr{Q}\left(N_1-1,-1/2+\check{\delta},-1/2+\check{\delta}\right)}
\\ \nonumber &\qquad +\sum_{j=0}^1\left\vert\left\vert \mathcal{L}_{\partial_v}^{1+j}h\right\vert\right\vert_{\mathscr{Q}_{-1/2}^0\left(N_1-2-j,0,-\sqrt{\check{p}}+j \right)},
\end{align}
where the $\widetilde{\slashed{\nabla}\hat{\otimes}}$ and $\widetilde{\slashed{\rm div}}$ are defined with respect to $\tilde{\slashed{g}}$.

We next have 

\begin{align}\label{2i2j2i1i2i33}
&\left\vert\left\vert p\right\vert\right\vert_{\mathscr{A}^-_2\left(\tilde{\kappa},\tilde{b}\right)} \doteq  \left\vert\left\vert \left(p - \left(v+1\right)^2\mathring{\slashed{g}}\right)\right\vert\right\vert_{\mathscr{Q}\left(N_1,-3/2+\check{\delta}, -\tilde{\kappa} \right)} 
\\ \nonumber&\qquad +  \left\vert\left\vert \left(p - \left(v+1\right)^2\mathring{\slashed{g}}\right)\right\vert\right\vert_{\mathscr{Q}\left(N_1-1,-3/2+\check{\delta}, -1/2+\check{\delta} \right)}
\\ \nonumber &\qquad  +\sum_{j=0}^1\left\vert\left\vert \mathcal{L}^{1+j}_{\partial_v}\left(p-(v+1)^2\mathring{\slashed{g}}\right)\right\vert\right\vert_{\mathscr{Q}\left(N_1-1-j,-1/2+\check{\delta}+j,1/2+j\right)}
\\ \nonumber &\qquad + \left\vert\left\vert \left(1,v\mathcal{L}_{\partial_v},\mathcal{L}_{\tilde{b}}\right)\mathcal{L}_{\partial_v}\left(p-(v+1)^2\mathring{\slashed{g}}\right)\right\vert\right\vert_{\mathscr{Q}_{-1/2}^0\left(N_1-2,0,\tilde{\kappa}\right)}
\\ \nonumber &\qquad + \left\vert\left\vert \left(1,v\mathcal{L}_{\partial_v},\mathcal{L}_{\tilde{b}}\right)\mathcal{L}_{\partial_v}\left(p-(v+1)^2\mathring{\slashed{g}}\right)\right\vert\right\vert_{\mathscr{Q}_{-1/2}^0\left(N_1-3,0,-\sqrt{\check{p}}\right)},
\end{align}
\begin{align}\label{2i2j2i1i2i3312333}
&\left\vert\left\vert p\right\vert\right\vert_{\mathscr{A}_2\left(\tilde{\kappa},\tilde{b},\tilde{\Omega}\right)} \doteq 
\left\vert\left\vert p\right\vert\right\vert_{\mathscr{A}_2^-\left(\tilde{\kappa},\tilde{b}\right)} + \left\vert\left\vert \mathfrak{A}\right\vert\right\vert_{\mathscr{Q}\left(N_1,-1/2+\check{\delta},50\check{p}\right)}
\\ \nonumber &\qquad + \left\vert\left\vert \mathfrak{A}\right\vert\right\vert_{\mathscr{Q}_{-1/2}^0\left(N_1-1,0,-1/2+\check{\delta}\right)} +\left\vert\left\vert \mathcal{L}_{\tilde{b}}K\left[p\right] \right\vert\right\vert_{\mathscr{Q}_{-1/2}^0\left(N_1-3,0,-1/2+\check{\delta}\right)}
\\ \nonumber &\qquad +\sum_{j=0}^1 \left\vert\left\vert \mathcal{L}^{1+j}_{\partial_v}\mathfrak{A}\right\vert\right\vert_{\mathscr{Q}\left(N_1-1-j,1/2+\check{\delta}+j,50\check{p}+1/2 +j \right)}+\sum_{j=0}^1 \left\vert\left\vert \left(1,\mathcal{L}_{\tilde{b}}\right)\mathcal{L}^{1+j}_{\partial_v}\mathfrak{A}\right\vert\right\vert_{\mathscr{Q}_{-1/2}^0\left(N_1-2-j,0,50\check{p}+j \right)},
\end{align}
where $K\left[p\right]$ denotes the Gaussian curvature of $p$, and we define
\begin{equation}\label{23ok3om}
\mathfrak{A} \doteq \frac{1}{2}\tilde{\Omega}^{-2}\left(p^{-1}\right)^{AB}\mathcal{L}_{\partial_v}p_{AB}-2(v+1)^{-1}.
\end{equation}
\end{definition}

Next we give norms which involve less derivatives but provide stronger control.
\begin{definition}\label{2im2om10o3}Let  $\tilde{b}$ be an $\mathbb{S}^2_{-1,v}$ vector field for $v \in (-1,0)$. Then, for any spherically symmetric function $t:(-1,0) \times \mathbb{S}^2 \to \mathbb{R}$, function $\phi : (-1,0)\times \mathbb{S}^2 \to \mathbb{R}$, $\mathbb{S}^2_{-1,v}$ vector field $h^A$ for $v \in (-1,0)$, and positive definite symmetric $\mathbb{S}^2_{-1,v}$ tensor $p_{AB}$  for $v \in (-1,0)$, we define
\begin{align}\label{2om2om2om3}
&\left\vert\left\vert t\right\vert\right\vert_{\mathscr{B}_{00}\left(\tilde{\kappa}\right)} \doteq \sum_{j=0}^3\left\vert\left\vert \mathcal{L}^j_{\partial_v}t \right\vert\right\vert_{\mathscr{S}_{-1}^{-1/2}\left(0,-1+\check{\delta}+j,0\right)} 
+ \sum_{j=0}^1\left\vert\left\vert \left(v\mathcal{L}_{\partial_v}\right)^j\left(t+\tilde{\kappa}\log\left(-v\right)\right)\right\vert\right\vert_{\mathscr{S}_{-1/2}^0\left(0,0,0\right)}
\\ \nonumber &+\sum_{j=0}^1\left\vert\left\vert \left(v\mathcal{L}_{\partial_v}\right)^j\mathcal{L}_{\partial_v}\left((-v)\mathcal{L}_{\partial_v}t\right) \right\vert\right\vert_{\check{\mathscr{S}}_{-1/2}^0\left(0,0,500\check{p}\left(1+j\right)+2\tilde{\kappa}\right)} + \sum_{j=0}^3\left\vert\left\vert  \mathcal{L}_{\partial_v}t\right\vert\right\vert_{\mathscr{Q}_{-1}^{-1/2}\left(0, -3/2+\check{\delta}+j,0\right)},
\end{align}

\begin{align}\label{2ok3oj29je4}
&\left\vert\left\vert\phi\right\vert\right\vert_{\mathscr{B}_{01}\left(\tilde{\kappa},\tilde{b}\right)}\doteq 
\\ \nonumber &\sup_{\left(v,\theta^A\right) \in (-1/2,0)\times \mathbb{S}^2}\left|\left((-v)\mathcal{L}_{\partial_v},\mathcal{L}_{\tilde{b}},1\right)\phi\right|
\\ \nonumber &+\sum_{j=0}^1\left\vert\left\vert \left(v\mathcal{L}_{\partial_v}\right)^j\phi\right\vert\right\vert_{\mathscr{S}_{-1/2}^0\left(N_1-2-j,0,\epsilon^{\frac{9}{10}}\right)} +\sum_{j=0}^1\left\vert\left\vert \left(v\mathcal{L}_{\partial_v}\right)^j\left((-v)\mathcal{L}_{\partial_v}-\mathcal{L}_{\tilde{b}}\right)\phi\right\vert\right\vert_{\mathscr{S}_{-1/2}^0\left(N_1-2-j,0,-1/4\right)}
\\ \nonumber &+\sum_{j=0}^1\left\vert\left\vert \left(v\mathcal{L}_{\partial_v}\right)^j\mathcal{L}_{\partial_v}\left((-v)\mathcal{L}_{\partial_v}-\mathcal{L}_{\tilde{b}}\right)\phi\right\vert\right\vert_{\check{\mathscr{S}}_{-1/2}^0\left(N_2-2-j,0,500\check{p}\left(1+j\right)+2\tilde{\kappa},500\check{p}\right)}
\\ \nonumber &+\sum_{j=0}^3\left\vert\left\vert \mathcal{L}_{\partial_v}^j\phi\right\vert\right\vert_{\mathscr{S}_{-1}^{-1/2}\left(N_1-1-j,-1+\check{\delta}+j,0\right)},
\end{align}

\begin{align}\label{iuwihu3iuh2hnjiojiojio}
&\left\vert\left\vert h\right\vert\right\vert_{\mathscr{B}^-_1\left(\tilde{\kappa}\right)} \doteq \sum_{j=0}^2\left\vert\left\vert \mathcal{L}_{\partial_v}^jh\right\vert\right\vert_{\mathscr{S}_{-1}^{-1/2}\left(N_1-1-j,-1+\check{\delta}+j,0\right)}+\sum_{j=0}^1\left\vert\left\vert \left(v\mathcal{L}_{\partial_v}\right)^jh\right\vert\right\vert_{\mathscr{S}_{-1/2}^0\left(N_1-1-j,0,\frac{\check{p}}{100}\right)} 
\\ \nonumber & + \sum_{j=0}^1\left\vert\left\vert \left(v\mathcal{L}_{\partial_v}\right)^jh\right\vert\right\vert_{\mathscr{S}_{-1/2}^0\left(N_1-2-j,0,0\right)} + \sum_{j=0}^1\left\vert\left\vert \left(v\mathcal{L}_{\partial_v}\right)^j\mathcal{L}_{\partial_v}h\right\vert\right\vert_{\check{\mathscr{S}}_{-1/2}^0\left(N_2-1-j,0,100\check{p}+50\check{p}\left(1+j\right)+2\tilde{\kappa},50\check{p}\right)},
\end{align}
\begin{align}\label{iuwihu3iuh2hn}
&\left\vert\left\vert h\right\vert\right\vert_{\mathscr{B}_1\left(\tilde{\kappa}\right)} \doteq \sum_{j=0}^2\left\vert\left\vert \mathcal{L}_{\partial_v}^jh\right\vert\right\vert_{\mathscr{S}_{-1}^{-1/2}\left(N_1-1-j,-1+\check{\delta}+j,0\right)}
\\ \nonumber & + \sum_{j=0}^1\left\vert\left\vert \left(v\mathcal{L}_{\partial_v}\right)^jh\right\vert\right\vert_{\mathscr{S}_{-1/2}^0\left(N_1-1-j,0,0\right)} + \sum_{j=0}^1\left\vert\left\vert \left(v\mathcal{L}_{\partial_v}\right)^j\mathcal{L}_{\partial_v}h\right\vert\right\vert_{\check{\mathscr{S}}_{-1/2}^0\left(N_2-1-j,0,50\check{p}\left(1+j\right)+2\tilde{\kappa},50\check{p}\right)},
\end{align}

\begin{align}\label{12345tgfwwijnbghj}
&\left\vert\left\vert p\right\vert\right\vert_{\mathscr{B}^-_2\left(\tilde{\kappa}\right)} \doteq \sum_{j=0}^2\left\vert\left\vert \mathcal{L}_{\partial_v}^j\left(p - (v+1)^2\mathring{\slashed{g}}\right)\right\vert\right\vert_{\mathscr{S}_{-1}^{-1/2}\left(N_1-1-j,-1+\check{\delta}+j,0\right)}
 \\ \nonumber &+ \sum_{j=0}^1\left\vert\left\vert \left(v\mathcal{L}_{\partial_v}\right)^j\left(p  - (v+1)^2\mathring{\slashed{g}}\right)\right\vert\right\vert_{\mathscr{S}_{-1/2}^0\left(N_1-1-j,0,\frac{\check{p}}{100}\right)}
 \\ \nonumber &+ \sum_{j=0}^1\left\vert\left\vert \left(v\mathcal{L}_{\partial_v}\right)^j\left(p  - (v+1)^2\mathring{\slashed{g}}\right)\right\vert\right\vert_{\mathscr{S}_{-1/2}^0\left(N_1-2-j,0,0\right)}
\\ \nonumber &+ \sum_{j=0}^1\left\vert\left\vert \left(v\mathcal{L}_{\partial_v}\right)^j\mathcal{L}_{\partial_v}\left(p - (v+1)^2\mathring{\slashed{g}}\right)\right\vert\right\vert_{\check{\mathscr{S}}_{-1/2}^0\left(N_2-j,0,\check{p}+2\tilde{\kappa},0\right)}, 
\\ \nonumber &\left\vert\left\vert p\right\vert\right\vert_{\mathscr{B}_2\left(\tilde{\kappa},\tilde{b},\tilde{\Omega}\right)} \doteq  \left\vert\left\vert p\right\vert\right\vert_{\mathscr{B}^-_2\left(\tilde{\kappa}\right)}+\left\vert\left\vert \left(1,\mathcal{L}_{\tilde{b}}\right)\mathfrak{A}\right\vert\right\vert_{\mathscr{S}_{-1/2}^0\left(N_1-2,0,0\right)}
\\ \nonumber &+\sum_{j=0}^1\left\vert\left\vert \left(v\mathcal{L}_{\partial_v}\right)^j\mathcal{L}_{\partial_v}\mathfrak{A}\right\vert\right\vert_{\check{\mathscr{S}}_{-1/2}^0\left(N_2-1-j,0,10\check{p}+2\tilde{\kappa},0\right)}+\sum_{j=0}^2\left\vert\left\vert \mathcal{L}_{\partial_v}^j\mathfrak{A}\right\vert\right\vert_{\mathscr{S}_{-1}^{-1/2}\left(N_1-1-j,\check{\delta}+j,0\right)},
\end{align}
where we define $\mathfrak{A}$ as in~\eqref{23ok3om}.

\end{definition}

\subsection{Heuristic Discussion of Norms}
In order for the reader to orient themself with respect to these norms, they should consider the following correspondences, which indicate how these norms will be used in the later bootstrap argument:
\[t\leftrightarrow \log\Omega_{\rm sing}, \qquad \phi \leftrightarrow \log\Omega_{\rm bound},\qquad h \leftrightarrow b,\qquad p \leftrightarrow \slashed{g} - (v+1)^2\mathring{\slashed{g}},\]
where $\log\Omega$ has a decomposition
\begin{equation}\label{2om2o4o294h3949}
\log\Omega = \log\Omega_{\rm sing} + \log\Omega_{\rm boun},
\end{equation}
defined by
\[(-v)\mathcal{L}_{\partial_v}\log\Omega_{\rm sing} = 2\mathcal{P}_{\ell = 0}\left(\Omega\underline{\omega}\right),\qquad \left((-v)\mathcal{L}_{\partial_v}-\mathcal{L}_b\right)\log\Omega_{\rm boun} = 2\mathcal{P}_{\ell \geq 1}\left(\Omega\underline{\omega}\right).\]

We have already discussed in Section~\ref{sososoformalwow} the basic heuristics for the pointwise behavior of these quantities as $v\to 0$. We note, however, that the norms here reflect an expected improvement over the generic behavior discussed in Section~\ref{sososoformalwow} in that since we will enforce a boundary condition $\mathcal{P}_{\ell \geq 1}\left(\Omega\underline{\omega}\right) = 0$, we thus expect (slightly) improved estimates for $\Omega_{\rm boun}$ as $v\to 0$ versus $\Omega_{\rm sing}$. The reason our $L^2$ estimate for the highest number of derivatives of $\slashed{g}$ has a weaker weight than for the other quantities is related to the fact that we will have less regularity for $\slashed{g}$ along $\{v = 0\}$ than for the other quantities. Near the axis, when $v\to -1$, the basic expectation is that $\Omega =1+ O\left(v+1\right)$, $b = O\left(v+1\right)$, and $\slashed{g}_{AB} = \left(v+1\right)^2\mathring{\slashed{g}}_{AB} + O\left(v+1\right)$.\footnote{We briefly explain the source of these expectations: The quantity $\left(v+1\right)^2\Omega\underline{\omega}$ will satisfy a model second order equation which linearly decouples from all other quantities. In view of the boundary conditions we pose and an indicial analysis at $v = 1$, we will find that the best decay we can expect for $\Omega\underline{\omega}$ is $\left|\Omega\underline{\omega}\right| \lesssim 1$. (The slow decay is associated to an $\ell = 1$ spherical harmonic.) Integrating from $v= -1$ leads to the expectation that $\left|\Omega- 1\right| \lesssim \left(v+1\right)$. An analogous analysis of $\slashed{\rm curl}b$ leads to the expectation that $\left|\slashed{\rm curl}b\right| \lesssim 1$, which is then consistent with $\left|b\right| \lesssim \left(v+1\right)$. (Again the obstruction to further decay is associated to an $\ell = 1$ spherical harmonic.) The quantity $\slashed{\rm div}b$ will satisfy a transport equation which couples linearly only to $\Omega\underline{\omega}$, and in view of its boundary conditions at the axis we thus expect $\left|\slashed{\rm div}b\right| \lesssim \left(v+1\right)^2$. Finally, the equations for $\slashed{g} - \left(v+1\right)^2\mathring{\slashed{g}}$ will couple linearly to the lapse $\Omega$, and this coupling leads to the expectation that $\left|\slashed{g}-\left(v+1\right)^2\mathring{\slashed{g}}\right| \lesssim \left(v+1\right)$.} (Our bootstrap norms have a $\check{\delta}$-loss with respect to these expectations because our $L^{\infty}_v$ estimates will be obtained from weighted $L^2_v$ estimates and the fundamental theorem of calculus.)

 Of course, the reasons for the specific definitions of the norms involve more than the expected asymptotics of the various quantities as $v\to -1$ and $v\to 0$; for example, the definitions also take into account certain linear and nonlinear hierarchies of our later estimates and regularity considerations for the data along the boundary $\{v = 0\}$. We will discuss these aspects later as they become relevant.

We quickly note one of the relevant hierarchical aspects. We have denoted some of the norms above with a ``$-$'' superscript. This denotes a more minimal norm which will suffice to control certain subsets of the nonlinear terms which will arise, while the stronger estimate without the ``$-$''  will be necessary for certain other terms. We have these two separate norms because for some equations we will have hierarchies of iteration schemes which allow us to deal with these various nonlinear terms sequentially. While this has the cost of complicating the argument, it allows us to partially exploit certain useful nonlinear structures of the equations even as we work with iterates.

\subsection{Some Inequalities}\label{inequalitysectionsection}
In this section we gather a few common inequalities in a single place for easier reference. 
\subsubsection{Nonlinear Estimates}\label{iio98987923}
We start by consider estimates for $fg$ in various of our norms where $fg$ stands for any type of tensor product or contraction of $\mathbb{S}^2_{-1,v}$ tensors $f$ and $g$ with respect to $\left(v+1\right)^2\mathring{\slashed{g}}$ . By repeatedly applying these inequalities, we immediately obtain statements for nonlinear expressions involving a tensor product or contraction of any finite number of tensors.

The first estimate concerns our high-regularity $L^2$-norm.
\begin{lemma}Let $-1 \leq a < b \leq 0$, $(p,q) \in \mathbb{R}^2$, $N \in \mathbb{Z}_{\geq 0}$, and $f$ and $h$ be $\mathbb{S}^2_{-1,v}$ tensors. Then, for any choice of $\{(p_i,q_i)\}_{i=1}^4$ with $(p_i,q_i) \in \mathbb{R}^2$ and so that $p_1+p_2 = p_3+p_4 = p$ and $q_1+q_2 = q_3+q_4 = q$, we have 
\[\left\vert\left\vert f h\right\vert\right\vert_{\mathscr{Q}_a^b\left(N,p,q\right)} \lesssim_N \left\vert\left\vert f\right\vert\right\vert_{\mathscr{Q}_a^b\left(N,p_1,q_1\right)}\left\vert\left\vert h\right\vert\right\vert_{\mathscr{S}\left(\lceil \frac{N}{2}\rceil +2,p_2,q_2\right)} +\left\vert\left\vert h\right\vert\right\vert_{\mathscr{Q}_a^b\left(N,p_3,q_3\right)}\left\vert\left\vert f\right\vert\right\vert_{\mathscr{S}\left(\lceil \frac{N}{2}\rceil +2,p_4,q_4\right)}.    \]
\end{lemma}
\begin{proof}This is an immediate consequence of Sobolev inequalities on $\mathbb{S}^2$.
\end{proof}

This second inequality concerns the $\mathscr{S}$-norm
\begin{lemma}Let $-1 \leq a < b \leq 0$, $(p,q) \in \mathbb{R}^2$, $N \in \mathbb{Z}_{\geq 0}$, and $f$ and $h$ be $\mathbb{S}^2_{-1,v}$ tensors. Then, for any choice of $\{(p_i,q_i)\}_{i=1}^4$ with $(p_i,q_i) \in \mathbb{R}^2$ and so that $p_1+p_2 = p_3+p_4 = p$ and $q_1+q_2 = q_3+q_4 = q$, we have 
\[\left\vert\left\vert f h\right\vert\right\vert_{\mathscr{S}_a^b\left(N,p,q\right)} \lesssim_N \left\vert\left\vert f \right\vert\right\vert_{\mathscr{S}_a^b\left(N,p_1,q_1\right)}\left\vert\left\vert h \right\vert\right\vert_{\mathscr{S}_a^b\left(\lceil \frac{N}{2}\rceil + 2,p_2,q_2\right)}   + \left\vert\left\vert h\right\vert\right\vert_{\mathscr{S}_a^b\left(N,p_3,q_3\right)}\left\vert\left\vert f \right\vert\right\vert_{\mathscr{S}_a^b\left(\lceil \frac{N}{2}\rceil ,p_4,q_4\right)}.\]
\end{lemma}
\begin{proof}This is an immediate consequence of Sobolev inequalities on $\mathbb{S}^2$.
\end{proof}

The third inequality concerns the $\check{\mathscr{S}}$-norm.
\begin{lemma}Let $-1 \leq a < b \leq 0$, $(p,q,r) \in \mathbb{R}^3$, $N \in \mathbb{Z}_{\geq 0}$, and $f$ and $h$ be $\mathbb{S}^2_{-1,v}$ tensors. Then, for any choice of $\{p_i\}_{i=1}^2$ with $p_i \in \mathbb{R}$ and so that $p_1+p_2 = p$, we have 
\[\left\vert\left\vert f h\right\vert\right\vert_{\check{\mathscr{S}}_a^b\left(N,p,q,r\right)} \lesssim_N \left\vert\left\vert f \right\vert\right\vert_{\check{\mathscr{S}}_a^b\left(N,p_1,q,r\right)}\left\vert\left\vert h \right\vert\right\vert_{\mathscr{S}_a^b\left(\lceil \frac{N}{2}\rceil + 2,p_2,0\right)}   + \left\vert\left\vert h\right\vert\right\vert_{\mathscr{S}_a^b\left(N,p_2,0\right)}\left\vert\left\vert f \right\vert\right\vert_{\check{\mathscr{S}}_a^b\left(\lceil \frac{N}{2}\rceil ,p_1,q,r\right)}.\]
\end{lemma}
\begin{proof}This is an immediate consequence of Sobolev inequalities on $\mathbb{S}^2$.
\end{proof}

\subsubsection{Linear Estimates}
In this section we provide two inequalities which we apply frequently during the paper.

The first inequality is the straightforward observation that we can control $\check{\mathscr{S}}$-norms in terms of suitable $\mathscr{S}$-norms. 
\begin{lemma}Let $-1 \leq a < b \leq 0$, $(p,q,r) \in \mathbb{R}^3$ with $r \geq 0$, $N \in \mathbb{Z}_{\geq 0}$, and $f$ be an $\mathbb{S}^2_{-1,v}$ tensor. Then, for any $\tilde{q}$ with $q-\tilde{q}  \gtrsim 1$, we have that
\[\left\vert\left\vert f\right\vert\right\vert_{\check{\mathscr{S}}_a^b\left(N,p,q,r\right)} \lesssim \left\vert\left\vert f\right\vert\right\vert_{\mathscr{S}_a^b\left(N,p,\tilde{q}\right)}.\]
\end{lemma}
\begin{proof}This is immediate from the definitions of the norms.
\end{proof}

This next lemma allows to control a $\mathscr{Q}$-norm with an improved weight if we control a suitable $\mathscr{Q}$-norm with one extra derivative and a suitable $\mathscr{S}$-norm with no derivatives.
\begin{lemma}\label{3m2omo4}Let $N \in \mathbb{Z}_{\geq 0}$ with $N\lesssim 1$ and $q \in \mathbb{R}$ such that $1/2-q \gtrsim 1$ and $c$ be a constant which satisfies $\left|c\right| \ll 1$.

 Then for every $\mathbb{S}^2_{-1,v}$ tensor $f$ we have
\[\left\vert\left\vert f\right\vert\right\vert_{\mathscr{Q}_{-1/2}^0\left(N,0,-\sqrt{\check{p}} \right)} \lesssim \left\vert\left\vert f\right\vert\right\vert_{\mathscr{Q}_{-1/2}^0\left(N+1,0,c\right)}+\left\vert\left\vert f\right\vert\right\vert_{\mathscr{S}_{-1/2}^0\left(0,0,q\right)}\]
\end{lemma}
\begin{proof}Our starting point is the following standard interpolation inequality along $\mathbb{S}^2$ which holds for any  $f \in \mathring{H}^{N+1}\left(\mathbb{S}^2\right)$:
\[\left\vert\left\vert f\right\vert\right\vert_{\mathring{H}^N}\lesssim \sum_{j=0}^N\left\vert\left\vert f\right\vert\right\vert^{\frac{(N+1)-j}{N+1}}_{L^2}\left\vert\left\vert f\right\vert\right\vert^{\frac{j}{N+1}}_{\mathring{H}^{N+1}}.\]
Combining this with H\"{o}lder's inequality, we thus have
\begin{align*}
\left\vert\left\vert f\right\vert\right\vert^2_{\mathscr{Q}_{-1/2}^0\left(N,0,-\sqrt{\check{p}}\right)} &\lesssim \sum_{j=0}^N\int_{-1/2}^0\left\vert\left\vert f\right\vert\right\vert_{L^2(\mathbb{S}^2)}^{\frac{2((N+1)-j)}{N+1}}\left\vert\left\vert f\right\vert\right\vert_{\mathring{H}^{N+1}\left(\mathbb{S}^2\right)}^{\frac{2j}{N+1}}(-v)^{-2\sqrt{\check{p}}}\, dv
\\ \nonumber &\lesssim \sum_{j=0}^N \left\vert\left\vert f\right\vert\right\vert^{\frac{2j}{N+1}}_{\mathscr{Q}_{-1/2}^0\left(N+1,0,c\right)}\left\vert\left\vert f\right\vert\right\vert^{\frac{2(N+1-j)}{N+1}}_{\mathscr{Q}_{-1/2}^0\left(0,0,-\sqrt{\check{p}}\frac{N+1}{N+1-j}-c\frac{j}{N+1-j}\right)}
\\ \nonumber &\lesssim \sum_{j=0}^N\left\vert\left\vert f\right\vert\right\vert^{\frac{2j}{N+1}}_{\mathscr{Q}_{-1/2}^0\left(N+1,0,c\right)}\left\vert\left\vert f\right\vert\right\vert^{\frac{2(N+1-j)}{N+1}}_{\mathscr{S}_{-1/2}^0\left(0,0,q\right)}\left(\int_{-1/2}^0(-v)^{2\left(-\sqrt{\check{p}}\frac{N+1}{N+1-j}-c\frac{j}{N+1-j}\right) -2q}\, dv\right)^{\frac{N+1-j}{N+1}}
\\ \nonumber &\lesssim \sum_{j=0}^N\left\vert\left\vert f\right\vert\right\vert^{\frac{2j}{N+1}}_{\mathscr{Q}_{-1/2}^0\left(N+1,0,c\right)}\left\vert\left\vert f\right\vert\right\vert^{\frac{2(N+1-j)}{N+1}}_{\mathscr{S}_{-1/2}^0\left(0,0,q\right)}
\\ \nonumber &\lesssim \left\vert\left\vert f\right\vert\right\vert^2_{\mathscr{Q}_{-1/2}^0\left(N+1,0,c\right)}+\left\vert\left\vert f\right\vert\right\vert^2_{\mathscr{S}_{-1/2}^0\left(0,0,q\right)}.
\end{align*}
\end{proof}

\section{Degenerate Transport and Elliptic Equations}\label{ij3oin4in234}
In this section we will study some transport equations with various types of degenerations, some elliptic equations, and some combined transport/elliptic equations. We will assume that we have an $\mathbb{S}^2_{-1,v}$ vector field $b^A$ for $v \in (-1,0)$, a function $\Omega = \Omega_{\rm sing}\Omega_{\rm boun}$ defined for $(v,\theta^A) \in (-1,0) \times \mathbb{S}^2$ with $\Omega_{\rm sing}$ spherically symmetric, and an $\mathbb{S}^2_{-1,v}$ symmetric $(0,2)$-tensors $\slashed{g}_{AB}$ for $v \in (-1,0)$ so that
\begin{align}\label{boottransport}
&\left\vert\left\vert \log\Omega_{\rm boun}\right\vert\right\vert_{\mathscr{A}_0\left(\kappa,\tilde{b}\right)}+\left\vert\left\vert b\right\vert\right\vert_{\mathscr{A}^{-}_1\left(\tilde{b},\kappa\right)} + \left\vert\left\vert \slashed{g}\right\vert\right\vert_{\mathscr{A}_2\left(\kappa,\tilde{b},\tilde{\Omega}\right)}+ \left\vert\left\vert b\right\vert\right\vert_{\mathscr{B}^-_1\left(\kappa\right)} 
\\ \nonumber &\qquad \qquad +\left\vert\left\vert \log\Omega_{\rm boun} \right\vert\right\vert_{\mathscr{B}_{01}\left(\kappa,\tilde{b}\right)}+\left\vert\left\vert \slashed{g}\right\vert\right\vert_{\mathscr{B}_2\left(\kappa,\tilde{b},\tilde{\Omega}\right)} +\left\vert\left\vert \log\Omega_{\rm sing} \right\vert\right\vert_{\mathscr{B}_{00}\left(\kappa\right)} \lesssim \epsilon,
\end{align}
for a suitable  vector field $\tilde{b}$ and constant $\kappa$ satisfying
\begin{equation}\label{3oijo2i94}
\sup_{(v,\theta^A)}\left|\tilde{b}\right| \lesssim \epsilon,\qquad \left|\kappa\right| \lesssim \epsilon,
\end{equation}
and function $\tilde{\Omega}$ satisfying~\eqref{k2moo39} with $\tilde{\kappa} = \kappa$. We emphasize that none of the results in this section depend on the implied constants in~\eqref{boottransport} or~\eqref{3oijo2i94} (though by our conventions for $\epsilon$, we may assume that $\epsilon$ is sufficiently small depending on the implied constants).

\subsection{Degenerate Transport Equations}\label{2omomo2}
We start with the main definition of the section.

\begin{definition}\label{okm3mo4o2222m34o3}We say that an $\mathbb{S}^2_{-1,v}$ tensor $\phi$ satisfies a degenerate-$(A_1,A_2,A_3)$ transport equation if
\begin{equation}\label{anicetransposrtequationse}
(-v)\nabla_v\phi + A_3\mathcal{P}_{\ell \in \mathscr{A}}\mathcal{L}_b\phi + \left(\frac{(-v)A_1}{v+1} + A_2\right)\phi = H,\qquad \forall v \in (-1,0).
\end{equation}
where $A_1 \in \mathbb{R}$ with $|A_1| \lesssim 1$, $A_2 \in [0,2]$, $A_3 \in \mathbb{R}$ with $|A_3| \lesssim 1$, $\mathscr{A} = \{\ell \in \mathbb{Z}_{\geq \ell_1}\}$ for some $\ell_1 \in \mathbb{Z}$ satisfying $|\ell_1| \lesssim 1$, $H$ is a $C^1$ $\mathbb{S}^2_{-1,v}$ tensor satisfying $\left(1-\mathcal{P}_{\ell \in \mathscr{A}}\right)H = 0$ and $\left(v+1\right)^{A_1}H \in L^1_vC^0\left(\mathbb{S}^2\right)$, and $\phi$ is $C^1$ for $v \in (-1,0)$, and we have $\left(1-\mathcal{P}_{\ell \in \mathscr{A}}\right)\phi = 0$. 
\end{definition}

We start by stating a standard existence result.
\begin{lemma}\label{existitdoes}There exists a unique solution $\phi \in C^1$ to the degenerate-$(A_1,A_2,A_3)$ transport equation~\eqref{anicetransposrtequationse} for $v \in (-1,0)$ such that 
\[\lim_{v\to -1}\left|\phi\right|\left(v+1\right)^{A_1} = 0.\]

Moreover, for every $\delta > 0$, we may define $\phi(\delta)$ to solve~\eqref{anicetransposrtequationse} for $v \in (-1+\delta,0)$ with the boundary condition $\phi(\delta)|_{v=-1+\delta} = 0$. Then, for every $-1 < v_0 < v_1 < 0$, we will have that,
\[\lim_{\delta \to 0}\sup_{v \in (v_0,v_1)}\left|\phi-\phi(\delta)\right| = 0.\]
\end{lemma}
\begin{proof} If we re-write~\eqref{anicetransposrtequationse} in terms of $h \doteq \left(v+1\right)^{A_1}\phi$, then we obtain 
\begin{equation}\label{anicetransposrtequationse123}
(-v)\nabla_vh+ A_3\mathcal{P}_{\ell \in \mathscr{A}}\mathcal{L}_bh+ A_2 h = \left(v+1\right)^{A_1}H,\qquad \forall v \in (-1,0).
\end{equation}
For any $\delta > 0$, considered as an equation for $h$, the coefficients of~\eqref{anicetransposrtequationse123} are  $C^1$ for $v \in [-1+\delta,0)$. We may then solve~\eqref{anicetransposrtequationse123} in the region $v \in [-1+\delta,0)$ via an iteration process where we set $h^{(0)} = 0$ and then define $h^{(j)}$ for $j \geq 1$ by solving for $v \in [-1+\delta,0)$:
\begin{equation}\label{anicetransposrtequationse1234}
(-v)\nabla_vh^{(j)}+ A_3\mathcal{L}_bh^{(j)}+ A_2 h^{(j)} = \left(v+1\right)^{A_1}H + A_3\mathcal{P}_{\ell \leq d}\mathcal{L}_bh^{(j-1)},\qquad h^{(j)}|_{v=-1+\delta}  =0.
\end{equation}
The operator on the left hand side is now a standard transport operator and it is straightforward to extract a suitable convergent subsequence of $\left\{h^{(j)}\right\}$. Denoting this limit $h(\delta)$, we then set $\phi(\delta) = \left(v+1\right)^{-A_1}h(\delta)$.

As a consequences of our bootstrap assumptions~\eqref{boottransport}, we have that
\[\sqrt{\mathring{\slashed{g}}_{AB}b^Ab^B} \lesssim \left(v+1\right)^{-\check{\delta}}.\]
Since $\left(v+1\right)^{-\check{\delta}}$ is integrable near $v = -1$ the integral curves of $(-v)\mathcal{L}_{\partial_v} + A_3\mathcal{L}_b$ have uniformly bounded lengths as $v\to -1$.  Thus the statement about the convergence of the $\phi(\delta)$ to the function $\phi$ follows in a standard fashion by integrating along the integral curves of $(-v)\mathcal{L}_{\partial_v} + A_3\mathcal{L}_b$ and applying Gr\"{o}nwall's inequality. The asserted regularity of $\phi$ then follows form the corresponding regularity statements for $\phi(\delta)$.

\end{proof}

The next lemma establishes weighted $L_v^{\infty}L^2\left(\mathbb{S}^2\right)$  estimates solutions to degenerate-$(A_1,A_2,A_3)$ transport equations.
\begin{lemma}\label{linftofkwp3}Let $\phi$ be a solution  to a degenerate-$(A_1,A_2, A_3)$ transport equation such that
\[\lim_{v\to -1}\left(v+1\right)^{A_1}\left|\phi\right| = 0.\]
 Let $q_1 \in \mathbb{R}$ satisfy $A_2+q_1 \gtrsim 1$, $q_2 \geq 0$, $p \in \mathbb{R}$ satisfy $A_1 - p \gtrsim 1$,  $N \in \mathbb{Z}_{\geq 0}$ satisfy $N \leq N_1-3$, and $\tilde{v} \in (-1,0)$. Then, assuming $H$ lies in the closure of smooth functions under the norm on the right hand of the corresponding estimate, we have
	\begin{equation}\label{mlmclmsl}
	\sum_{j=0}^1\left\vert\left\vert \left(v\nabla_v\right)^j\phi\right\vert\right\vert_{\check{\mathscr{S}}_{-1}^{\tilde{v}}\left(N-j,p+j,q_1+jq_2,q_2\right)} \lesssim \left\vert\left\vert H\right\vert\right\vert_{\check{\mathscr{S}}_{-1}^{\tilde{v}}\left(N,p+1,q_1,q_2\right)}.
	\end{equation}

If we weaken the assumption that $A_2+q_1 \gtrsim 1$ to just that $A_2+q_1 \gtrsim \epsilon^{1-\check{\delta}}$,  then, for any $c \gtrsim 1$, we have the estimate
\begin{equation}\label{23om3omo2}
\sum_{j=0}^1\left\vert\left\vert \left(v\nabla_v\right)^j\phi\right\vert\right\vert_{\mathscr{S}_{-1}^{\tilde{v}}\left(N-j,p+j,q_1\right)} \lesssim \left\vert\left\vert H\right\vert\right\vert_{\mathscr{S}_{-1}^{\tilde{v}}\left(N,p+1,q_1-c\right)}.
\end{equation}

Finally, under the assumption that $A_2 \gtrsim 1$, we also have the following estimate
\begin{equation}\label{21jnkjnkjn}
\sum_{j=0}^1\left\vert\left\vert \left(v\nabla_v\right)^j\phi\right\vert\right\vert_{\mathscr{S}_{-1}^{\tilde{v}}\left(N-j,p+j,0\right)} \lesssim \left\vert\left\vert H\right\vert\right\vert_{\mathscr{S}_{-1}^{\tilde{v}}\left(N,p+1,0\right)}.
\end{equation}

When $N = 0$ we drop the sums in $j$ in all of these estimates.

\end{lemma}
\begin{proof}By Lemma~\ref{existitdoes} and an approximation argument, it suffices to prove these estimates with $\phi$ replaced by $\phi(\delta)$ and $-1$ replaced by $-1+\delta$ where we require $0 < \delta \ll 1$.

Set $W = \left(v+1\right)^{A_1}(-v)^q$. Then we have 
\begin{align}\label{momgm3o30cw222}
&(-v)\nabla_v\left(W\phi(\delta)\right) + A_3\mathcal{P}_{\ell \in \mathscr{A}}\mathcal{L}_b\left(W\phi(\delta)\right)  + \left(A_2+q\right)\left(W\phi(\delta)\right) = WH.
\end{align}
We may then contract~\eqref{momgm3o30cw222} with $2W\phi (-v)^{-1}$ and integrate over $\mathbb{S}^2$ to obtain that
\begin{align}\label{ldm3omi9}
&\frac{d}{dv}\left(\int_{\mathbb{S}^2}\left|W\phi(\delta)\right|_{\slashed{g}}^2\mathring{\rm dVol}\right) + 2\left(A_2+q-(1+v)^{-\delta}\epsilon^{1-\check{\delta}^2}\right)(-v)^{-1}\int_{\mathbb{S}^2}\left|W\phi(\delta)\right|_{\slashed{g}}^2\mathring{\rm dVol}
 \leq 
 \\ \nonumber &\qquad \qquad \qquad 2(-v)^{-1}\int_{\mathbb{S}^2}\left|WH\right|_{\slashed{g}}\cdot\left|W\phi(\delta)\right|_{\slashed{g}}\mathring{\rm dVol}.
\end{align}
Note that in the case that $A_2+q \gtrsim \epsilon^{1-\check{\delta}}$, we will have $A_2+q -\epsilon^{1-\check{\delta}^2} \gtrsim \epsilon^{1-\check{\delta}}$. Both~\eqref{mlmclmsl} and~\eqref{23om3omo2} (with the modifications for $\phi(\delta)$) in the case when $N = 0$ follow from integrating~\eqref{ldm3omi9} for $v \in (-1+\delta,\hat{v})$ for any $\hat{v} \in (-1+\delta,\tilde{v})$ (and applying Gr\"{o}nwall's inequality for $v$ near $-1$). 

In order to prove~\eqref{21jnkjnkjn}, we consider~\eqref{momgm3o30cw222} with $W$ replaced by $\tilde{W} = \left(v+1\right)^{A_1}(-v)^{-A_2/2}$. Then~\eqref{momgm3o30cw222} becomes   
\begin{align}\label{kmlm1l2km3}
&(-v)\nabla_v\left(\tilde{W}\phi(\delta)\right) + A_3\mathcal{P}_{\ell \in \mathscr{A}}\mathcal{L}_b\left(\tilde{W}\phi(\delta)\right)  + \frac{1}{2}A_2\left(\tilde{W}\phi(\delta)\right) = \tilde{W}H.
\end{align}
The analogue of~\eqref{ldm3omi9} is then the fact that for $v \in [-1/2,0)$ we have
\begin{align}\label{ioewiojwtijo}
&\frac{d}{dv}\left(\int_{\mathbb{S}^2}\left|\tilde{W}\phi(\delta)\right|_{\slashed{g}}^2\mathring{\rm dVol}\right) + A_2(-v)^{-1}\int_{\mathbb{S}^2}\left|\tilde{W}\phi(\delta)\right|_{\slashed{g}}^2\mathring{\rm dVol}
 \leq 
 \\ \nonumber &\qquad \qquad \qquad 2(-v)^{-1}\int_{\mathbb{S}^2}\left|\tilde{W}H\right|_{\slashed{g}}\cdot\left|\tilde{W}\phi(\delta)\right|_{\slashed{g}}\mathring{\rm dVol}.
\end{align}
We now drop the second term on the left hand side of~\eqref{ioewiojwtijo}, integrate both sides, use our previous estimate, and obtain, for any $\tilde{v} \in [-1/2,0)$:
\begin{align}\label{3oij3ijo4}
\sup_{v \in [-1/2,\tilde{v}]} (-v)^{-A_2}\int_{\mathbb{S}^2}\left|\phi(\delta)\right|_{\slashed{g}}^2\mathring{\rm dVol} \lesssim \left(\tilde{v}\right)^{-A_2}\sup_{v\in [-1,\tilde{v}]}\int_{\mathbb{S}^2}\left|H\right|_{\slashed{g}}^2\left(v+1\right)^{2A_1}\mathring{\rm dVol}.
\end{align}
This then yields~\eqref{21jnkjnkjn} when $N = 0$. 

When $N > 0$, we simply induct on $N$, commute the equation with $\mathcal{L}_{Z^{(\alpha)}}$, and repeat the above estimate (after adjusting the weight depending on $q_2$). The terms generated by the commutator of $\mathcal{L}_{Z^{(\alpha)}}$ are easily controlled with Sobolev inequalities on $\mathbb{S}^2$ and the bootstrap assumptions on $b$ and $\slashed{g}$. The estimate for $v\mathcal{L}_{\partial_v}\phi(\delta)$ is obtained directly from the equation. The qualitative regularity required to justify the commutation is immediately obtained by standard transport equation theory.

\end{proof}

The next lemma establishes weighted $L^2$ estimates for solutions to degenerate-$(A_1,A_2, A_3)$ transport equations.
\begin{lemma}\label{l2l2degtranstrans}Let $\phi$ be a solution  to a degenerate-$(A_1,A_2, A_3)$ transport equation such that
\[\lim_{v\to -1}\left(v+1\right)^{A_1}\left|\phi\right| = 0.\]
Let $p \in \mathbb{R}$ satisfy $A_1 - (p+1/2) \gtrsim 1$, $q \in \mathbb{R}$ satisfy $1+2q \gtrsim 1$, $\tilde{v} \in (-1,0)$, and $N \in \mathbb{Z}_{\geq 0}$ satisfy $N \leq N_1-2$. Then, if $H$ lies in the closure of smooth functions under the norm on the right hand side of~\eqref{asdfko4m}, we have that 
\begin{align}\label{asdfko4m}
\sum_{j=0}^1\left\vert\left\vert \left(v\nabla_v\right)^j\phi\right\vert\right\vert_{\mathscr{Q}_{-1}^{\tilde{v}}\left(N-j,p+j,q\right)} \lesssim \left\vert\left\vert H\right\vert\right\vert_{\mathscr{Q}_{-1}^{\tilde{v}}\left(N,p+1,q\right)}.
\end{align}
\end{lemma}
\begin{proof}As in the proof of Lemma~\ref{linftofkwp3} it suffices to prove these statements for $\phi(\delta)$ and with all ranges of $v$ starting at $-1+\delta$ instead of $-1$ for an arbitrary $0 < \delta \ll 1$. 

Let $W \doteq \left(v+1\right)^{p+1/2}$. Conjugating by $W$ yields
\begin{align}\label{momgm3o30cw222ke3k230fk}
&(-v)\nabla_v\left(W\phi(\delta)\right) + A_3\mathcal{P}_{\ell \geq \ell_1}\mathcal{L}_b\left(W\phi(\delta)\right)+ \frac{(-v)\left(A_1-(p+1/2)\right)}{v+1}W\phi(\delta) + A_2\left(W\phi\right) = WH.
\end{align}
We may contract~\eqref{momgm3o30cw222ke3k230fk} with $(-v)^{2q}W\phi$, integrate by parts over $dv\mathring{\rm dVol}$ for $v \in (-1+\delta,\tilde{v})$, and apply Cauchy-Schwarz to easily obtain (the $\delta$ modified version of)~\eqref{asdfko4m} for $N = 0$. 

The case of higher $N$ in the case when $N \leq N_1-3$ follows from a straightforward induction argument in $N$, commuting with $\mathcal{L}_{Z^{(\alpha)}}$, using the bootstrap assumptions and Sobolev inequalities on $\mathbb{S}^2$ to control the errors generated by the commutators, and using the equation~\eqref{momgm3o30cw222ke3k230fk} directly to estimate $v\mathcal{L}_{\partial_v}\phi(\delta)$.

\end{proof}

\begin{remark}\label{2km2om3o} It is a straightforward consequence of these estimates and the contraction mapping principle, that Lemmas~\ref{existitdoes},~\ref{linftofkwp3}, and~\ref{l2l2degtranstrans} continue to hold if, in the definition~\eqref{anicetransposrtequationse} of a degenerate-$(A_1,A_2,A_3)$ transport equation, we replace the term $\frac{(-v)A_1}{v+1}$ with $\frac{(-v)A_1\left(\slashed{g}^{AB}\mathcal{L}_{\partial_v}\slashed{g}_{AB}\right)}{v+1}$.

\end{remark}

It is now useful to state a proposition which follows in a straightforward fashion from Proposition 4.5 and its proof from our previous~\cite{nakedone}.
\begin{proposition}\label{somestuimdie}Let $k \in \mathbb{Z}_{\geq 0}$. Suppose that $P \gtrsim 1$ is a constant, $s \geq 2$, $d$ is an $\mathring{H}^{s+1}(\mathbb{S}^2)$ vector field with $\left\vert\left\vert d\right\vert\right\vert_{\mathring{H}^{s+1}} \lesssim 1$ and $\left\vert\left\vert d\right\vert\right\vert_{C^1} \ll 1$, $W$ is an $\mathring{H}^s\left(\mathbb{S}^2\right)$ tensor with $\left\vert\left\vert W\right\vert\right\vert_{\mathring{H}^s} \lesssim 1$ and $\left\vert\left\vert W\right\vert\right\vert_{L^{\infty}} \ll 1$,  and that $H$ is a $H^s(\mathbb{S}^2)$ $(0,m)$-tensor on $\mathbb{S}^2$. Then,  assuming the contraction of $W$ with a $(0,m)$-tensor produces another $(0,m)$-tensor, there exists a $\mathring{H}^s(\mathbb{S}^2)$  solution $\phi$ to
\begin{equation}\label{ii29jj93j2j2io3}
\mathcal{L}_d\phi + P\phi +W\phi = H,
\end{equation}
which satisfies
\begin{equation}\label{ij2938h482h4}
\left\vert\left\vert \left(1,\mathcal{L}_d\right)\phi\right\vert\right\vert_{\mathring{H}^s} \lesssim \left\vert\left\vert H\right\vert\right\vert_{\mathring{H}^s}.
\end{equation}
Moreover, this solution is unique among all solutions to~\eqref{ii29jj93j2j2io3} for which $\phi$ and $\mathcal{L}_d\phi$ are in $L^2$. \end{proposition}
\begin{remark}\label{m2om1o1qw}It is straightforward to extend the proof of Proposition 4.5 from~\cite{nakedone} (and hence Proposition~\ref{somestuimdie} above) to include the case when $\mathcal{L}_d$ is replaced by $\mathcal{P}_{\ell \in \mathscr{A}}\mathcal{L}_d$, $W\phi$ is replaced by $\mathcal{P}_{\ell \in \mathscr{A}}\left(W\phi\right)$, and $H$ satisfies $\left(1-\mathcal{P}_{\ell \in \mathscr{A}}\right)H = 0$. Then the resulting solution $\phi$ will also satisfy $\left(1-\mathcal{P}_{\ell \in \mathscr{A}}\right)\phi = 0$. Finally, we can also then replace with or $\mathcal{P}_{\ell \in \mathscr{A}}\mathcal{L}_d$ with $\mathcal{P}_{\ell \in \mathscr{A}}\Pi_{\delta}\mathcal{L}_d\Pi_{\delta}$, but in the estimate~\eqref{ij2938h482h4} we must then replace $\mathcal{L}_d$ with $\Pi_{\delta}\mathcal{L}_d\Pi_{\delta}$ (where $\Pi_{\delta}$ are as in Lemma~\ref{thesmoothlemma}).
\end{remark}
We will now use Proposition~\ref{somestuimdie} to understand the $v\to 0$ limits of solutions to degenerate-$(A_1,A_2,A_3)$ transport equations (in the case when $A_2 \gtrsim 1$).
\begin{lemma}\label{2om2omo4}Let $\phi$ satisfy a degenerate-$(A_1,A_2,A_3)$ transport equation, and suppose that $A_2 \gtrsim 1$, $H(0) = \lim_{v\to 0}H$ exists, and $\left\vert\left\vert H(0)\right\vert\right\vert_{H^s} < \infty$ for some $s$ such that $2 \leq s \leq N_1-3$. Keeping in mind that the bootstrap assumptions imply that $b$ extends to $v = 0$ as a $\mathring{H}^{N_1-2}$ tensor, we may use Proposition~\ref{somestuimdie} and Remark~\ref{m2om1o1qw} to define $\phi(0)$ by solving 
\[\mathcal{P}_{\ell \geq \ell_1}\mathcal{L}_{b(0)}\phi(0) + A_2\phi(0) = H(0).\]

Let $c \in \mathbb{R}$ satisfy $c \gtrsim 1$, $A_2 - c \gtrsim 1$, and $1 -c \gtrsim 1$, and $\tilde{s} \in \mathbb{Z}$ satisfy $2 \leq \tilde{s} \leq N_2-2$. Assuming that $H$ lies in the closure of smooth functions under the norms on the right hand side of~\eqref{i3momo2}, we then have
\begin{align}\label{i3momo2}
&\sum_{j=0}^1\left\vert\left\vert \left(v\Omega\nabla_4\right)^j\left(\phi-\phi(0)\right)\right\vert\right\vert_{\mathscr{S}_{-1/2}^0\left(\tilde{s}-j,0,-c\right)} \lesssim
\\ \nonumber &\qquad  \left\vert\left\vert H-H(0)\right\vert\right\vert_{\mathscr{S}_{-1/2}^0\left(\tilde{s},0,-c-\check{\delta}\right)} + \left\vert\left\vert \phi\right\vert\right\vert_{\mathscr{S}_{-3/4}^{-1/2}\left(\tilde{s},0,0\right)} + \left\vert\left\vert H(0)\right\vert\right\vert_{\mathring{H}^{\tilde{s}+1}},
\end{align}
where we extend $\phi(0)$ and $H(0)$ to be constant in $v$ in the coordinate frame.
\end{lemma}
\begin{proof}From Proposition~\ref{somestuimdie} and Remark~\ref{m2om1o1qw}, we will have the estimate
\begin{equation}\label{2oo4i292}
\left\vert\left\vert \left(\phi(0),\mathcal{L}_{b(0)}\phi(0)\right)\right\vert\right\vert_{\mathring{H}^s} \lesssim \left\vert\left\vert H(0)\right\vert\right\vert_{\mathring{H}^s}.
\end{equation}
Set $\tilde{\phi} \doteq \xi\left(\phi - \phi(0)\right)$ and $\tilde{H} \doteq \xi\left(H-H(0)\right)$ (recall that $\xi$ is defined in Section~\ref{cutcutcut}). Then we have
\begin{align}\label{k3mom}
&(-v)\nabla_v\tilde{\phi} + A_3\mathcal{P}_{\ell \geq \ell_1}\mathcal{L}_b\tilde{\phi} + \left(\frac{(-v)A_1}{v+1} + A_2\right)\tilde{\phi} = 
\\ \nonumber &\qquad (-v)\nabla_v\phi(0) + \tilde{H} + \frac{vA_1\xi \phi(0)}{v+1} + A_3\mathcal{P}_{\ell \geq \ell_1}\left(\mathcal{L}_b-\mathcal{L}_{b(0)}\right)\phi(0)+ (-v)\xi' \phi.
\end{align}
The result then follows by applying Lemma~\ref{linftofkwp3} to~\eqref{k3mom} and the fact that the bootstrap assumptions on $b$ imply that $\left\vert\left\vert b(v)-b(0)\right\vert\right\vert_{\mathring{H}^{N_2-1}} \lesssim (-v)^{1-q}$ for some $q$ which satisfies $0 < q \ll 1$.

\end{proof}
\subsection{An Elliptic  Equation with Degenerating Principle Symbol}\label{iojoijiojo21}
In this section we will study some elliptic equations. We will use these estimates later in Section~\ref{k2m3mo492} when we study the equations used to solve for the shift $b$. The reason these equations will need a special treatment is that the lapse $\Omega$ will be present in the principle symbol, and due to our weak control of the lapse as $v\to 0$ (see~\eqref{boottransport}, \eqref{2om2om2om3}, and~\eqref{2ok3oj29je4}) we will need to be careful to avoid angular derivatives falling on $\Omega$ unnecessarily. We will only use these estimates in the region $v \geq -1/2$. It will be useful to introduce the constant $\mathscr{U}$ to stand for the left hand side of~\eqref{boottransport}.

\begin{lemma}\label{om3om2}We consider either of the following two elliptic equations for an unknown $\mathbb{S}^2_{-1,v}$ vector field $P^A$ in the region $v \geq -1/2$, in terms of an already known $\mathbb{S}^2_{-1,v}$ $1$-form $ \left(H_1\right)_A$ and functions $H_2,f: [-1/2,0) \times \mathbb{S}^2 \to \mathbb{R}$ with $\mathcal{P}_{\ell = 0}f  = 0$ and a $(2,0)-\mathbb{S}^2_{-1,v}$ tensor $\slashed{h}^{AB}$ for $v \geq -1/2$ with 
\[\sum_{j=0}^1\sum_{\left|\alpha\right| +j \leq 2}\left\vert\left\vert \left(v\mathcal{L}_{\partial_v}\right)^j\slashed{h}^{(\alpha)}\right\vert\right\vert_{L^{\infty}_vL^{\infty}\left(\mathbb{S}^2\right)} \lesssim 1:\]
	\begin{equation}\label{m3omo3}
	 \mathcal{P}_{\ell \geq 1}\slashed{\rm div}\left(\Omega^{-2}P\right) =  \mathcal{P}_{\ell \geq 1}\left(\slashed{h}^{AB}\mathring{\nabla}_A\left(H_1\right)_B + \Omega^{-2}H_2\right) \doteq H,\qquad  \mathring{\Pi}_{\rm curl}P^A = - \mathring{\slashed{\epsilon}}^{AB}\mathring{\nabla}_Bf,
	 \end{equation}
	 \begin{equation}\label{m3omo31234}
	 \mathcal{P}_{\ell \geq 1}\slashed{\rm curl}\left(\Omega^{-2}P\right) =  \mathcal{P}_{\ell \geq 1}\left(\slashed{h}^{AB}\mathring{\nabla}_A\left(H_1\right)_B + \Omega^{-2}H_2\right)\doteq H,\qquad  \mathring{\Pi}_{\rm div}P^A = \mathring{\slashed{g}}^{AB}\mathring{\nabla}_Bf.
	 \end{equation}
	
We let $q \in \mathbb{R}$ satisfy $\epsilon^{\frac{9}{10}} \ll q \ll 1$. Then, in the case of~\eqref{m3omo3}, assuming that $H$ and $\mathring{\Pi}_{\rm curl}P$ lie in the closure of smooth functions under the norm on the right hand side of~\eqref{3m4omo}, there exists a unique solution $P$ to~\eqref{m3omo3} which satisfies:
	 \begin{equation}\label{3m4omo}
	 \left\vert\left\vert \mathring{\Pi}_{\rm div}P\right\vert\right\vert_{\check{\mathscr{S}}_{-1/2}^0\left(0,0,q,0\right)} \lesssim \underbrace{\left\vert\left\vert  H_1\right\vert\right\vert_{\check{\mathscr{S}}_{-1/2}^0\left(0,0,q-2\kappa,0\right)} + \left\vert\left\vert  H_2\right\vert\right\vert_{\check{\mathscr{S}}_{-1/2}^0\left(0,0,q,0\right)} + \mathscr{U} \left\vert\left\vert\mathring{\Pi}_{\rm curl}P \right\vert\right\vert_{\check{\mathscr{S}}_{-1/2}^0\left(0,0,q,0\right)}}_{\doteq \mathscr{E}_1\left[q\right]}
	 \end{equation}
	We may also commute with $v\mathcal{L}_{\partial_v}$ to obtain that, for suitable regular $H_1$, $H_2$, and $\mathring{\Pi}_{\rm curl}P$,
	  \begin{align}\label{3m4omoininiq}
	 &\left\vert\left\vert \left(v\mathcal{L}_{\partial_v}\right)\mathring{\Pi}_{\rm div}P\right\vert\right\vert_{\check{\mathscr{S}}_{-1/2}^0\left(0,0,2q,0\right)} \lesssim \mathscr{E}_1\left[q\right]+
	 \\ \nonumber & \underbrace{\left\vert\left\vert \left(v\mathcal{L}_{\partial_v}\right) H_1\right\vert\right\vert_{\check{\mathscr{S}}_{-1/2}^0\left(0,0,2q-2\kappa,0\right)} + \left\vert\left\vert  \left(v\mathcal{L}_{\partial_v}\right)H_2\right\vert\right\vert_{\check{\mathscr{S}}_{-1/2}^0\left(0,0,2q,0\right)} + \mathscr{U}\left\vert\left\vert\left(v\mathcal{L}_{\partial_v}\right)\mathring{\Pi}_{\rm curl}P \right\vert\right\vert_{\check{\mathscr{S}}_{-1/2}^0\left(0,0,2q,0\right)}}_{\mathscr{E}_2\left[q\right]} .
	 \end{align}
	 We then have the following estimate for $\slashed{\rm div}P$:
	 \begin{align}\label{2kn2kn3}
		&\left\vert\left\vert \slashed{\rm div}P\right\vert\right\vert_{\check{\mathscr{S}}_{-1/2}^0\left(0,0,2q,0\right)} \lesssim \left\vert\left\vert \left(\Omega^2H\right) \right\vert\right\vert_{\check{\mathscr{S}}_{-1/2}^0\left(0,0,2q,0\right)}+\mathscr{E}_1\left[q\right]
		 \end{align}
	 For~\eqref{m3omo31234}, the same estimates hold except with the switching of the role of ${\rm div}$ with ${\rm curl}$. 
	 
	 We then also have higher order estimates in the case of~\eqref{m3omo3} (with the obvious adjustments for~\eqref{m3omo31234}): For any $1 \leq N \leq N_2-2$, we also have 
	 	 \begin{align}\label{2lm3om2o}
		&\sum_{j=0}^1 \left\vert\left\vert \left(v\mathcal{L}_{\partial_v}\right)^j\slashed{\rm div}P\right\vert\right\vert_{\check{\mathscr{S}}_{-1/2}^0\left(N-j,0,q(2+j),q\right)} \lesssim \mathscr{E}_1\left[q\right] + \mathscr{E}_2\left[q\right]
						\\ \nonumber &\qquad +\sum_{j=0}^1 \left\vert\left\vert \left(v\mathcal{L}_{\partial_v}\right)^j\left(\Omega^2H\right) \right\vert\right\vert_{\check{\mathscr{S}}_{-1/2}^0\left(N-j,0,q(2+j),q\right)}+\mathscr{U}\sum_{j=0}^1\left\vert\left\vert \left(v\mathcal{L}_{\partial_v}\right)^j\mathring{\Pi}_{\rm curl}P \right\vert\right\vert_{\check{\mathscr{S}}_{-1/2}^0\left(N-j,0,q(1+j),q\right)}
						\\ \nonumber &\doteq \mathscr{E}_3\left[q,N\right].
		 \end{align}
		 Moreover, for any $N_2-1 \leq N \leq N_1-2$, we have, for any $\tilde{q}$ satisfying $\left|\tilde{q}\right| \ll 1$:
		 \begin{align}\label{3om3omo4}
		&\sum_{j=0}^1 \left\vert\left\vert \left(v\mathcal{L}_{\partial_v}\right)^j\slashed{\rm div}P\right\vert\right\vert_{\mathscr{Q}_{-1/2}^0\left(N-j,0,\tilde{q}\right)}\lesssim
		\\ \nonumber & \sum_{j=0}^1\left\vert\left\vert \left(v\mathcal{L}_{\partial_v}\right)^j\Omega^2 H\right\vert\right\vert_{\mathscr{Q}_{-1/2}^0\left(N-j,0,\tilde{q}\right)} 
		+ \mathscr{U} \sum_{j=0}^1\left\vert\left\vert \left(v\mathcal{L}_{\partial_v}\right)^j\mathring{\Pi}_{\rm curl}P\right\vert\right\vert_{\mathscr{Q}_{-1/2}^0\left(N-j,0,\tilde{q}\right)} + \mathscr{E}_3\left[q,N_2-1\right].
		 \end{align}
		 There is also the alternative estimate, where we weaken the weight at the top level of derivatives except for a privileged derivative $\mathcal{L}_X$. For any $\tilde{y} > \tilde{q}$, we have 
		  \begin{align}\label{knk1nknk2}
		& \left\vert\left\vert \slashed{\rm div}P\right\vert\right\vert_{\mathscr{Q}_{-1/2}^0\left(N,0,\tilde{y}\right)}+ \sum_{\left|\alpha\right| \leq N_1-1}\left\vert\left\vert \left(1,\mathcal{L}_X,v\mathcal{L}_{\partial_v}\right)\slashed{\rm div}P^{(\alpha)}\right\vert\right\vert_{\mathscr{Q}_{-1/2}^0\left(0,0,\tilde{q}\right)} \lesssim \left\vert\left\vert \Omega^2 H\right\vert\right\vert_{\mathscr{Q}_{-1/2}^0\left(N,0,\tilde{y}\right)} 
		\\ \nonumber &\qquad +\sum_{\left|\alpha\right| \leq N_1-1}\left\vert\left\vert \left(1,\mathcal{L}_X,v\mathcal{L}_{\partial_v}\right)\left(\Omega^2 H\right)^{(\alpha)}\right\vert\right\vert_{\mathscr{Q}_{-1/2}^0\left(0,0,\tilde{q}\right)} +\sum_{\left|\alpha\right| \leq N_1-1}\mathscr{U} \left\vert\left\vert \left(1,\mathcal{L}_X,v\mathcal{L}_{\partial_v}\right)\mathring{\Pi}_{\rm curl}P^{(\alpha)}\right\vert\right\vert_{\mathscr{Q}_{-1/2}^0\left(0,0,\tilde{q}\right)} 
		\\ \nonumber &\qquad+ \mathscr{E}_3\left[q,N_2-1\right],
		 \end{align}
		 here $X$ is any vector field so that $\left\vert\left\vert X\right\vert\right\vert_{C^1\left(\mathbb{S}^2\right)} \lesssim 1$. 
		 	 
	\end{lemma}
\begin{proof}We start with the existence statement for $P$ and the estimate~\eqref{3m4omo}. Without loss of generality, we consider the equation~\eqref{m3omo3}. If it exists, we may represent $P^A = \slashed{\mathring{g}}^{AB}\mathring{\nabla}_Bw - \mathring{\slashed{\epsilon}}^{AB}\mathring{\nabla}_Bf$ for functions $w$ and $f$ which satisfy $\mathcal{P}_{\ell = 0}w = \mathcal{P}_{\ell =0}f = 0$. This leads us to re-write~\eqref{m3omo3} as
\begin{align}\label{3momo2}
 &\mathring{\nabla}_A\left(\Omega^{-2}\slashed{\mathring{g}}^{AB}\mathring{\nabla}_Bw\right) = 
 \\ \nonumber &\qquad \mathcal{P}_{\ell \geq 1}\underbrace{\left[-\left(\slashed{\nabla}_A-\mathring{\nabla}_A\right)\left(\Omega^{-2}P^A\right)+\slashed{h}^{AB}\mathring{\nabla}_A\left(H_1\right)_B + \Omega^{-2}H_2\right]}_{\doteq \mathscr{F}}+\mathring{\nabla}_A\left(\left(\Omega^{-2}-(-v)^{2\kappa}\right)\mathring{\slashed{\epsilon}}^{AB}\mathring{\nabla}_Bf\right).
\end{align}

The operator on the left hand side of~\eqref{3momo2} is of divergence form and self-adjoint with respect to $\mathring{\rm dVol}$. Moreoever, it is clear that the kernel and co-kernel of the operator consists of constant functions and that the right hand side of~\eqref{3momo2} is thus orthogonal to the co-kernel. We next note that we may generate a priori estimates for $w$ in terms of $\mathscr{F}$ and $f$ by multiplying both sides with $-l(v)w $ for any positive function $l(v)$ and integrating by parts over $\mathring{\rm dVol}$.  With this all understood, it is straightforward to run an iteration argument and find the unique solution $w$ to~\eqref{3momo2} with $\mathcal{P}_{\ell = 0}w$ so that~\eqref{3m4omo} is satisfied.  To establish~\eqref{3m4omoininiq}, we simply commute with $v\mathcal{L}_{\partial_v}$ and repeat the above the estimate.

The higher order estimates~\eqref{2lm3om2o},~\eqref{3om3omo4}, and~\eqref{knk1nknk2} now follow easily from an induction argument and Lemma~\ref{3m2omo4} since we may re-write~\eqref{3momo2} as
\[\slashed{\rm div}P = 2\left(\slashed{\nabla}_A\log\Omega\right)P^A + \Omega^2H,\]
and then inductively commute and carry out higher order elliptic estimates (with~\eqref{3m4omo} as our base case), and use the bootstrap assumptions to control the growth of the angular derivatives of the lapse.
\end{proof}
\begin{remark}\label{23mo0904}When we apply Lemma~\ref{om3om2}, it is useful to keep in mind that any function $W$ with $\mathcal{P}_{\ell = 0}W = 0$ may be written as $W = \slashed{\mathring{g}}^{AB}\mathring{\nabla}_A\tilde{W}_B$ by setting $\tilde{W}_B \doteq \mathring{\nabla}_B\mathring{\Delta}^{-1}W$. In this way we can use Lemma~\ref{om3om2} to always ``gain a derivative'' even if the right hand side does not explicitly contain a derivative. 
\end{remark}

In this last lemma of the section, we will study a combined elliptic transport equation. We will use this lemma later in Section~\ref{22oiiojioj32}.
\begin{lemma}\label{32m2omo4}We consider the following equation for an unknown $\mathbb{S}^2_{-1,v}$ vector field $P^A$ for $v \in [-1/2,0)$, in terms of already known $\mathbb{S}^2_{-1,v}$ vector fields $ (H_1)^A$, $(H_2)^A$, $(H_3)^A$, and $(H_4)^A$ defined for $v \in (-1,0)$ and which vanish for $v  < -1/2$ and a function $f: (-1,0) \times \mathbb{S}^2 \to \mathbb{R}$ with $\mathcal{P}_{\ell = 0}f  = 0$ which vanishes for $v < -1/2$:
\begin{align}\label{3o3om2om3o}
\mathcal{P}_{\ell > \ell_0}\slashed{\rm curl}\slashed{\rm div}\left[\left((-v)\nabla_v + A_1\mathcal{L}_b + A_2\right)\slashed{\nabla}\hat{\otimes}\left(\Omega^{-2}P\right)\right] &= \mathcal{P}_{\ell > \ell_0}H_1 
\\ \nonumber &= \mathcal{P}_{\ell > \ell_0}H_2 + \mathcal{P}_{\ell > \ell_0}\left((-v)\nabla_4+A_1\mathcal{L}_b\right)H_3,
\\ \nonumber \mathcal{P}_{1 \leq \ell \leq \ell_0}\slashed{\rm curl}\left[\left((-v)\nabla_v + A_3\mathcal{L}_b + A_4\right)\left(\Omega^{-2}P\right)\right] &= \mathcal{P}_{1 \leq \ell \leq \ell_0}H_4,
\\ \nonumber \mathring{\Pi}_{\rm div}P^A = \slashed{\mathring{g}}^{AB}\mathring{\nabla}_Bf.
\end{align}
Here $\sum_{i=1}^4 |A_i| + |A_3| \lesssim 1$ and $A_2,A_4 \geq 0$. 

Then, assuming $H_1$, $H_4$, and $\mathring{\Pi}_{\rm div}P$ lie in the closure of smooth functions under the norm on the right hand side of~\eqref{jil2ij2ijij}, there exists a unique solution $P$ to~\eqref{3o3om2om3o} which vanishes for $v < -1/2$ and which satisfies, for any $q \in \mathbb{R}$ satisfying $\epsilon^{\frac{9}{10}} \ll q \ll 1$:
 \begin{equation}\label{jil2ij2ijij}
	 \left\vert\left\vert \mathring{\Pi}_{\rm curl}P\right\vert\right\vert_{\check{\mathscr{S}}_{-1/2}^0\left(0,0,q+2\kappa,0\right)} \lesssim \underbrace{\left\vert\left\vert  \left(\mathring{\Delta}^{-2}H_1,H_4\right)\right\vert\right\vert_{\check{\mathscr{S}}_{-1/2}^0\left(1,0,q,0\right)}+ \mathscr{U} \left\vert\left\vert\mathring{\Pi}_{\rm div}P \right\vert\right\vert_{\check{\mathscr{S}}_{-1/2}^0\left(0,0,q+2\kappa,0\right)}}_{\doteq \mathscr{E}_1\left[q\right]}.
	 \end{equation}
	We may also commute with $v\mathcal{L}_{\partial_v}$ to obtain that
	  \begin{align}\label{ij3ijoiooi20}
	 &\left\vert\left\vert \left(v\mathcal{L}_{\partial_v}\right)\mathring{\Pi}_{\rm curl}P\right\vert\right\vert_{\check{\mathscr{S}}_{-1/2}^0\left(0,0,2q+2\kappa,0\right)} \lesssim \mathscr{E}_1\left[q\right]+
	 \\ \nonumber &\qquad \underbrace{\left\vert\left\vert \left(v\mathcal{L}_{\partial_v}\right)\left(\mathring{\Delta}^{-2}H_1,H_4\right)\right\vert\right\vert_{\check{\mathscr{S}}_{-1/2}^0\left(1,0,2q,0\right)} +\mathscr{U} \left\vert\left\vert\left(v\mathcal{L}_{\partial_v}\right)\mathring{\Pi}_{\rm div}P \right\vert\right\vert_{\check{\mathscr{S}}_{-1/2}^0\left(0,0,2q+2\kappa,0\right)}}_{\doteq \mathscr{E}_2\left[q\right]}.
	 \end{align}
	 We will then have the following estimate for $\slashed{\rm curl}P$:
	 \begin{align}\label{jiooijoij129090}
		&\left\vert\left\vert \slashed{\rm curl}P\right\vert\right\vert_{\check{\mathscr{S}}_{-1/2}^0\left(0,0,q+2\kappa,0\right)} \lesssim \mathscr{E}_1\left[q\right] +\left\vert\left\vert \left(\mathring{\Delta}^{-1}H_1,H_4\right) \right\vert\right\vert_{\check{\mathscr{S}}_{-1/2}^0\left(0,0,q,0\right)}.
		 \end{align}	 
\end{lemma}

	 We then also have higher order estimates: For any $1 \leq N \leq N_2-2$ we also have 
	 	 \begin{align}\label{jk32ji23iu32oi}
		&\sum_{j=0}^1 \left\vert\left\vert \left(v\mathcal{L}_{\partial_v}\right)^j\slashed{\rm curl}P\right\vert\right\vert_{\check{\mathscr{S}}_{-1/2}^0\left(N-j,0,q(2+j)+2\kappa,q\right)} \lesssim \mathscr{E}_1\left[q\right] + \mathscr{E}_2\left[q\right]
						\\ \nonumber &\qquad +\sum_{j=0}^1 \left\vert\left\vert \left(v\mathcal{L}_{\partial_v}\right)^j\left(\mathring{\Delta}^{-1}H_1,H_4\right) \right\vert\right\vert_{\check{\mathscr{S}}_{-1/2}^0\left(N-j,0,q(2+j),q\right)}+\mathscr{U}\sum_{j=0}^1\left\vert\left\vert \left(v\mathcal{L}_{\partial_v}\right)^j\mathring{\Pi}_{\rm div}P \right\vert\right\vert_{\check{\mathscr{S}}_{-1/2}^0\left(N+1-j,0,q(1+j)+2\kappa,q\right)}
						\\ \nonumber &\doteq \mathscr{E}_3\left[q,N\right].
		 \end{align}
		 Moreover, for any $N_2-1 \leq N \leq N_1-2$, we have, for $|\tilde{q}| \ll 1$, 
		 \begin{align}\label{ijo23ijo2oij23oi}
		 &\sum_{j=0}^1\left\vert\left\vert \left(v\mathcal{L}_{\partial_v}\right)^j\slashed{\rm curl}P\right\vert\right\vert_{\mathscr{Q}_{-1/2}^0\left(N-j,0,\tilde{q}+2\kappa\right)} \lesssim \sum_{j=0}^1\left\vert\left\vert \left(v\mathcal{L}_{\partial_v}\right)^j \left(\mathring{\Delta}^{-1}H_2,\mathring{\Delta}^{-1}H_3,H_4\right)\right\vert\right\vert_{\mathscr{Q}\left(N-j,0,\tilde{q}\right)} 
		 \\ \nonumber &\qquad + \mathscr{U} \sum_{j=0}^1\left\vert\left\vert \left(v\mathcal{L}_{\partial_v}\right)^j\mathring{\Pi}_{\rm div}P\right\vert\right\vert_{\mathscr{Q}\left(N+1-j,0,\tilde{q}+2\kappa\right)} + \mathscr{E}_3\left[q,N_2-1\right].
		 \end{align}
As in Lemma~\ref{om3om2} we also have a version where we weaken the weight at the top level of derivatives, for $\tilde{y} > \tilde{q}$:
 \begin{align}\label{ijo2o1o1o1049}
		 &\left\vert\left\vert \slashed{\rm curl}P\right\vert\right\vert_{\mathscr{Q}_{-1/2}^0\left(N,0,\tilde{y}+2\kappa\right)}+\left\vert\left\vert \left(1,\mathcal{L}_b,v\mathcal{L}_{\partial_v}\right)\slashed{\rm curl}P\right\vert\right\vert_{\mathscr{Q}_{-1/2}^0\left(N-1,0,\tilde{q}+2\kappa\right)} \lesssim 
		 \\ \nonumber &\qquad \left\vert\left\vert  \left(\mathring{\Delta}^{-1}H_2,\mathring{\Delta}^{-1}H_3,H_4\right)\right\vert\right\vert_{\mathscr{Q}\left(N-j,0,\tilde{y}\right)} +\left\vert\left\vert \left(1,\mathcal{L}_b,v\mathcal{L}_{\partial_v}\right)\left(\mathring{\Delta}^{-1}H_2,\mathring{\Delta}^{-1}H_3,H_4\right)\right\vert\right\vert_{\mathscr{Q}\left(N-1,0,\tilde{q}\right)} 
		 \\ \nonumber &\qquad +\mathscr{U}\left\vert\left\vert \mathring{\Pi}_{\rm div}P\right\vert\right\vert_{\mathscr{Q}\left(N+1,0,\tilde{y}+2\kappa\right)}+ \mathscr{U}\left\vert\left\vert \left(1,\mathcal{L}_b,v\mathcal{L}_{\partial_v}\right)\mathring{\Pi}_{\rm div}P\right\vert\right\vert_{\mathscr{Q}\left(N,0,\tilde{q}+2\kappa\right)} + \mathscr{E}_3\left[q,N_2-1\right],
		 \end{align}
		 where $X$ is a vector field which satisfies $\left\vert\left\vert X\right\vert\right\vert_{C^1} \lesssim 1$.
\begin{proof}

We solve for $P$ via an iteration process.

We start by rewriting~\eqref{3o3om2om3o} as  
\begin{align}\label{2kn2knk4}
&\left((-v)\nabla_v + A_1\mathcal{P}_{\ell > \ell_0}\mathcal{L}_b + A_2\right)\left(\left(v+1\right)^2\mathcal{P}_{\ell > \ell_0}\slashed{\rm curl}\slashed{\rm div}\slashed{\nabla}\hat{\otimes}\left(\Omega^{-2}P\right)\right)
\\ \nonumber &\qquad  + \left[\left(v+1\right)^2\mathcal{P}_{\ell > \ell_0}\slashed{\rm curl}\slashed{\rm div},(-v)\nabla_v + A_1\mathcal{P}_{\ell > \ell_0}\mathcal{L}_b + A_2\right]\slashed{\nabla}\hat{\otimes}\left(\Omega^{-2}P\right) = \left(v+1\right)^2\mathcal{P}_{\ell > \ell_0}H_1,
\end{align}
\begin{align}\label{2o3omo2}
&\left((-v)\nabla_v + A_1\mathcal{P}_{1 \leq \ell \leq \ell_0}\mathcal{L}_b + A_2\right)\left(\left(v+1\right)\mathcal{P}_{1 \leq \ell \leq \ell_0}\slashed{\rm curl}\left(\Omega^{-2}P\right)\right)
\\ \nonumber &\qquad  + \left[\left(v+1\right)\mathcal{P}_{1 \leq \ell \leq \ell_0}\slashed{\rm curl},(-v)\nabla_v + A_1\mathcal{P}_{1 \leq \ell \leq \ell_0}\mathcal{L}_b + A_2\right]\left(\Omega^{-2}P\right) = \left(v+1\right)\mathcal{P}_{1 \leq \ell \leq \ell_0}H_4.
\end{align}

We will also introduce two scalar auxiliary unknowns $\mathfrak{A}$ and $\mathfrak{B}$ (which will eventually be seen to correspond to $\left(v+1\right)^2\mathcal{P}_{\ell > \ell_0}\slashed{\rm curl}\slashed{\rm div}\slashed{\nabla}\hat{\otimes}\left(\Omega^{-2}P\right)$ and $\left(v+1\right)\mathcal{P}_{1 \leq \ell \leq \ell_0}\slashed{\rm curl}\left(\Omega^{-2}P\right)$ respectively). We define the sequence $\{P^{(i)},\mathfrak{A}^{(i)},\mathfrak{B}^{(i)}\}_{i=0}^{\infty}$ as follows. We start by setting
\[P^{(0)} = 0,\qquad \mathfrak{A}^{(0)} = \mathfrak{B}^{(0)} = 0.\]
For $i \geq 1$, we set 
\begin{align}\label{2o3omomo2ji2939}
&\left((-v)\nabla_v + A_1\mathcal{P}_{\ell > \ell_0}\mathcal{L}_b + A_2\right)\mathring{\Delta}^{-2}\mathfrak{A}^{(i)} + \mathcal{P}_{\ell > \ell_0}\left[\mathring{\Delta}^{-2},A_1\mathcal{L}_b \right]\mathfrak{A}^{(i-1)}
\\ \nonumber &\qquad  + \mathring{\Delta}^{-2}\left(\left[\left(v+1\right)^2\mathcal{P}_{\ell > \ell_0}\slashed{\rm curl}\slashed{\rm div},(-v)\nabla_v + A_1\mathcal{P}_{\ell > \ell_0}\mathcal{L}_b + A_2\right]\slashed{\nabla}\hat{\otimes}\left(\Omega^{-2}P^{(i-1)}\right) \right)= \mathring{\Delta}^{-2}\left(\left(v+1\right)^2\mathcal{P}_{\ell > \ell_0}H_1\right),
\end{align}
\[\mathcal{P}_{\ell \leq \ell_0}\mathfrak{A}^{(i)} = 0,\]
\begin{align}\label{oijjioijo32oi}
&\left((-v)\nabla_v + A_1\mathcal{P}_{1 \leq \ell \leq \ell_0}\mathcal{L}_b + A_2\right)\left(\left(v+1\right)\mathring{\Delta}^{-1}\mathfrak{B}^{(i)}\right) + \mathcal{P}_{1 \leq \ell \leq \ell_0}\left[\mathring{\Delta}^{-1},A_1\mathcal{L}_b\right]\mathfrak{B}^{(i-1)}
\\ \nonumber &\qquad  + \mathring{\Delta}^{-1}\left(\left[\left(v+1\right)\mathcal{P}_{1 \leq \ell \leq \ell_0}\slashed{\rm curl},(-v)\nabla_v + A_1\mathcal{P}_{1 \leq \ell \leq \ell_0}\mathcal{L}_b + A_2\right]\left(\Omega^{-2}P^{(i-1)}\right)\right) = \mathring{\Delta}^{-1}\left(\left(v+1\right)\mathcal{P}_{1 \leq \ell \leq \ell_0}H_4\right),
\end{align}
\[\left(1-\mathcal{P}_{1 \leq \ell \leq \ell_0}\right)\mathfrak{B}^{(i)} = 0,\]
\begin{align}\label{2oko3mo4}
(v+1)^2\mathcal{P}_{\ell > \ell_0}\left(\slashed{\Delta}+2K\right)\slashed{\rm curl}\left(\Omega^{-2}P^{(i)}\right) = \mathfrak{A}^{(i)} - 2(v+1)^2\mathcal{P}_{\ell > \ell_0}\slashed{\nabla}K\wedge \left(\Omega^{-2}P^{(i-1)}\right),
\end{align}
\begin{align}\label{3oij4ijo4io}
\left(v+1\right)\mathcal{P}_{1\leq \ell \leq \ell_0}\slashed{\rm curl}\left(\Omega^{-2}P^{(i)}\right) = \mathfrak{B}^{(i)},
\end{align}
\[\mathring{\Pi}_{\rm div}\left(P^{(i)}\right)^A = \slashed{\mathring{g}}^{AB}\mathring{\nabla}_Bf.\]

It is then straightforward, if tedious, to analyze this system and the corresponding iterates with an amalgamation of the transport estimates from Section~\ref{2omomo2}, elliptic estimates on $\mathbb{S}^2$, and Lemma~\ref{om3om2} (keeping Remark~\ref{23mo0904} in mind). Finally, after establishing that the iterates converge, it may show that one obtains a solution to the original system. We omit the details.
\end{proof}
\section{General Theory for the Model Second Order Equation}\label{secondordertheorysection}
Throughout this section we will assume that we have an $\mathbb{S}^2_{-1,v}$ vector field $b^A$ for $v\in (-1,0)$, function $\Omega = \Omega_{\rm sing}\Omega_{\rm boun}$ defined for $(v,\theta^A) \in (-1,0) \times \mathbb{S}^2$ with $\Omega_{\rm sing}$ spherically symmetric, and an $\mathbb{S}^2_{-1,v}$ positive definite symmetric $(0,2)$-tensor $\slashed{g}_{AB}$ for $v \in (-1,0)$ so that
\begin{align}\label{thisisisismom3o}
&\left\vert\left\vert \log\Omega_{\rm boun}\right\vert\right\vert_{\mathscr{A}_0\left(\kappa,b\right)}+\left\vert\left\vert b\right\vert\right\vert_{\mathscr{A}^{-}_1\left(\kappa\right)} + \left\vert\left\vert \slashed{g}\right\vert\right\vert_{\mathscr{A}_2\left(\kappa,\tilde{b},\tilde{\Omega}\right)} + \left\vert\left\vert b\right\vert\right\vert_{\mathscr{B}^-_1\left(\kappa\right)} 
\\ \nonumber &\qquad \qquad + \left\vert\left\vert \log\Omega_{\rm sing} \right\vert\right\vert_{\mathscr{B}_{00}\left(\kappa\right)} +\left\vert\left\vert \log\Omega_{\rm boun} \right\vert\right\vert_{\mathscr{B}_{01}\left(\kappa,b\right)}+\left\vert\left\vert \slashed{g}\right\vert\right\vert_{\mathscr{B}_2\left(\kappa,\tilde{b},\tilde{\Omega}\right)} \lesssim \epsilon,
\end{align}
for an $\mathbb{S}^2_{-1,v}$ vector field $\tilde{b}$ and a constant $\kappa$ satisfying
\begin{equation}\label{3oi909889}
\sup_{(v,\theta^A)}\left|\tilde{b}\right| \lesssim \epsilon,\qquad \left|\kappa\right| \lesssim \epsilon,
\end{equation}
and function $\tilde{\Omega}$ satisfying~\eqref{k2moo39} with $\tilde{\kappa} = \kappa$.  We note that because the $\mathscr{A}$ and $\mathscr{B}$ norm for $\log\Omega$ involves the vector field $b$ (as opposed to $\tilde{b}$), the assumption~\eqref{thisisisismom3o} includes a ``compatibility'' condition between $b$ and $\Omega$. We emphasize that none of the results in this section depend on the implied constants in~\eqref{thisisisismom3o} or~\eqref{3oi909889} (though by our conventions for $\epsilon$, we may assume that $\epsilon$ is sufficiently small depending on the implied constants).

We start by introducing the type of equations we will study in this section.
\begin{definition}\label{themodeleqnqnenq}We say that a function $\phi(v,\theta^A) : (-1,0)\times \mathbb{S}^2 \to \mathbb{R}$ satisfies the model second order equation of type $I$, $II$, or $III$ with right hand side $H$ if
\begin{align}\label{wiiogjr9bj3}
&L\phi \doteq (-v)\mathcal{L}^2_{\partial_v}\phi + \frac{(-v)\mathcal{P}_{\ell \in \mathscr{A}}\left(\Omega^2A_1\mathcal{L}_{\partial_v}\phi\right)}{v+1} + A_2\mathcal{L}_{\partial_v}\phi + A_3\mathcal{P}_{\ell \in \mathscr{A}}\left(\mathcal{L}_b\mathcal{L}_{\partial_v}\phi\right)
\\ \nonumber &\qquad + \mathcal{P}_{\ell \in \mathscr{A}}\left(\left[\Omega^2\left(\slashed{\Delta}+A_4\left(v+1\right)^{-2}\right)+A_5(-v)\Omega^{2s}(v+1)^{-2}\right]\phi\right) = H,
\end{align}
where $\left(\Omega,b,\slashed{g}\right)$ satisfy~\eqref{thisisisismom3o}, $H$ is a given $C^0$ function for $\left(v,\theta^A\right) \in (-1,0)\times \mathbb{S}^2$ satisfying $\mathcal{P}_{\ell \in \mathscr{A}} H = H$, $\phi$ is $C^2$  and satisfies $\mathcal{P}_{\ell \in \mathscr{A}} \phi = \phi$, $|A_3| \lesssim 1$, $|s| \lesssim 1$, and we have one of 
\begin{enumerate}[I.]
	\item   $A_1 = 0$, $A_2 = 1$, $A_4 = 0$, $A_5 = 0$, and $\mathscr{A} = \{1 \leq \ell\}$,
	\item   $A_1 = 4$, $A_2 = 1$, $A_4 = 2$, $A_5 = 0$, and $\mathscr{A} = \{2 \leq \ell \leq \ell_0\}$,
	\item  $ 0 \leq A_1 \lesssim 1$, $0 \leq A_2 \lesssim 1$, $|A_4| \ll \ell_0 $, $|A_5| \ll \ell_0$, and $\mathscr{A} = \{\ell_0 \leq \ell\}$.
\end{enumerate}
(We recall that the large positive integer $\ell_0$ has been set in Section~\ref{fixedfixedfixedfixedconst}.)
\end{definition}

The following Hardy inequality (with explicit constant) will be useful in what follows.
\begin{lemma}\label{HardyHardyHardy}Let $0 \leq a_0 \ll 1$, $a_1 > -1+a_0$, and $f(v) : (-1+a_0,a_1) \to \mathbb{R}$ be a $C^1$ function. Let $p \in \mathbb{R}$ and $p \neq 1/2$. Furthermore:
\begin{enumerate}
\item If $p < 1/2$ then assume that $\left(v+1\right)^{2p-1}f(v)$ extends continuously to $v = -1+a_0$ where it vanishes.
\item If $p > 1/2$ then assume that $f$ extends continuously to $v = a_1$ where it vanishes. 
\end{enumerate}
Then we have 
\[\left\vert\left\vert f\right\vert\right\vert_{\mathscr{Q}_{-1+a_0}^{a_1}\left(0,p-1,0\right)}^2 \leq \frac{4}{\left(2p-1\right)^2}\left\vert\left\vert \mathcal{L}_{\partial_v}f\right\vert\right\vert_{\mathscr{Q}_{-1+a_0}^{a_1}\left(0,p,0\right)}^2. \]

\end{lemma}
\begin{proof}We follow the standard integration by parts argument:
\begin{align*}
\int_{-1+a_0}^{a_1}\left(v+1\right)^{2p-2}f^2\, dv &= \int_{-1+a_0}^{a_1}\left(2p-1\right)^{-1}\frac{d}{dv}\left(\left(v+1\right)^{2p-1}\right)f^2\, dv
\\ \nonumber &\leq \frac{2}{|2p-1|}\int_{-1+a_0}^{a_1}\left(v+1\right)^{2p-1}|f||\partial_vf|\, dv
\\ \nonumber &\leq \frac{2}{|2p-1|}\left(\int_{-1+a_0}^{a_1}\left(v+1\right)^{2p-2}f^2\, dv\right)^{1/2}\left(\int_{-1+a_0}^{a_1}\left(v+1\right)^{2p}\left(\partial_vf\right)^2\, dv\right)^{1/2}.
\end{align*}
\end{proof}

\subsection{Regularization and a First Estimate and Existence Result}\label{makeitregularyay}
First we define some useful function spaces:
\begin{definition}Let $0 \leq \tilde{\delta}_1 \ll 1$ and $0 \leq \tilde{\delta}_2 \ll 1$ be two small positive constants. Then we define $L^{2,\mathscr{A}}$, $H_0^{1,\mathscr{A}}$, and $H_0^{2,\mathscr{A}}$ to be the closures of the set of smooth compactly supported functions $f(v,\theta^A): (-1+\tilde{\delta}_2,-\tilde{\delta}_1) \times \mathbb{S}^2 \to \mathbb{R}$ satisfying $\left(1-\mathcal{P}_{\mathscr{A}}\right)f = 0$ under the respective norms 
\[\left\vert\left\vert f\right\vert\right\vert_{\mathscr{Q}_{-1+\tilde{\delta}_2}^{-\tilde{\delta}_1}\left(0,0,0\right)},\qquad \left\vert\left\vert \left(\mathcal{L}_{\partial_v}f,\mathring{\nabla}f\right)\right\vert\right\vert_{\mathscr{Q}_{-1+\tilde{\delta}_2}^{-\tilde{\delta}_1}\left(0,0,0\right)},\qquad \sum_{j+k\leq 2}\left\vert\left\vert \mathcal{L}_{\partial_v}^j\mathring{\nabla}^kf\right\vert\right\vert_{\mathscr{Q}_{-1+\tilde{\delta}_2}^{-\tilde{\delta}_1}\left(0,0,0\right)}.\]

If we drop the assumption of compact support, then we will refer to the spaces by $H^{1,\mathscr{A}}$ and $H^{2,\mathscr{A}}$. Finally we may also define the corresponding spaces $L^{2,\mathscr{A}}_{\rm loc}$, $H^{1,\mathscr{A}}_{\rm loc}$, and $H^{2,\mathscr{A}}_{\rm loc}$ by a suitable restriction to compact subsets of $(-1+\tilde{\delta}_2,-\tilde{\delta}_1) \times \mathbb{S}^2$.
\end{definition}
\begin{remark}
For notational simplicity and because there is unlikely to be confusion in practice, we suppress the dependence of these function spaces on the parameters $\tilde{\delta}_1$ and $\tilde{\delta}_2$.
\end{remark}
We now define a regularization of the model second order equations.

\begin{definition}\label{regularizedmodel}
Consider a model second order equation as in Definition~\ref{themodeleqnqnenq}. Let $0 \leq \tilde{\delta}_1 \ll 1$, $0\leq \tilde{\delta}_2 \ll 1$, and $0 \leq \tilde{\delta}_3 \ll 1$ be given. Then we say that $\phi(v,\theta^A) \in H^{2,\mathscr{A}}$ satisfies the $(\tilde{\delta}_1,\tilde{\delta}_2,\tilde{\delta}_3)$-regularized model second order equation of type $I$, $II$, or $III$ with right hand side $H \in L^{2,\mathscr{A}}$ if for $\left(v,\theta^A\right) \in \left(-1+\tilde{\delta}_2,-\tilde{\delta}_1\right)\times\mathbb{S}^2$ 
\begin{align}\label{wiiogjr9bj3kmeo3}
&L^{(\tilde{\delta}_1,\tilde{\delta}_2,\tilde{\delta}_3)}\phi \doteq (-v)\mathcal{L}^2_{\partial_v}\phi + \frac{(-v)\mathcal{P}_{\ell \in \mathscr{A}}\left(\Omega^2A_1\mathcal{L}_{\partial_v}\phi\right)}{v+1} + A_2\mathcal{L}_{\partial_v}\phi \\ \nonumber &\qquad + A_3\mathcal{P}_{\ell \in \mathscr{A}}\left(\Pi_{\tilde{\delta}_3}\mathcal{L}_b\Pi_{\tilde{\delta}_3}\mathcal{L}_{\partial_v}\phi\right)
 + \mathcal{P}_{\ell \in \mathscr{A}}\left(\left[\Omega^2\left(\slashed{\Delta}+A_4\left(v+1\right)^{-2}\right)+A_5(-v)\Omega^{2s}(v+1)^{-2}\right]\phi\right) = H.
\end{align}
\end{definition}
\begin{remark}A fundamental property of the regularized equation~\eqref{wiiogjr9bj3kmeo3} is that if $\tilde{\delta}_1,\tilde{\delta}_2,\tilde{\delta}_3  > 0$, then, since the term $A_3{P}_{\ell \in \mathscr{A}}\Pi_{\tilde{\delta}_3}\mathcal{L}_b\Pi_{\tilde{\delta}_3}\mathcal{L}_{\partial_v}\phi$ may be controlled in $L^2(\mathbb{S}^2)$ by the $L^2(\mathbb{S}^2)$ norm of $\mathcal{L}_{\partial_v}\phi$, the operator~\eqref{wiiogjr9bj3kmeo3} is uniformly elliptic on the cylinder $\left(v,\theta^A\right) \in \left(-1+\tilde{\delta}_2,-\tilde{\delta}_1\right)\times\mathbb{S}^2$. 
\end{remark}

We start by establishing a global a priori estimate for the regularized equation. In particular, this will show that the operator $L^{(\tilde{\delta}_1,\tilde{\delta}_2,\tilde{\delta}_3)}$ has a trivial kernel.
\begin{proposition}\label{wqiqiu1u8nkk}Let $0 < \tilde{\delta}_1,\tilde{\delta}_2,\tilde{\delta}_3 \ll 1$, $H_1 \in L^{2,\mathscr{A}}$,  and $H_2\in H_0^{1,\mathscr{A}}$. Further assume that $\phi$ is a $H_0^{2,\mathscr{A}}$ solution of a $(\tilde{\delta}_1,\tilde{\delta}_2,\tilde{\delta}_3)$-regularized model second order equation with right hand side $H \doteq H_1 + \mathcal{L}_{\partial_v}H_2$. 

Then $\phi$ satisfies
\begin{align}\label{ijwiji293}
&\left\vert\left\vert \mathcal{L}_{\partial_v}\phi\right\vert\right\vert_{\mathscr{Q}_{-1+\tilde{\delta}_2}^{-\tilde{\delta}_1}\left(0,A_1,1/2\right)}^2 + \left\vert\left\vert \phi\right\vert\right\vert_{\mathscr{Q}_{-1+\tilde{\delta}_2}^{-\tilde{\delta}_1}\left(1,A_1-1,-\kappa\right)}^2 \lesssim
\\ \nonumber &\qquad  \left\vert\left\vert H_1\right\vert\right\vert_{\mathscr{Q}_{-1+\tilde{\delta}_2}^{-\tilde{\delta}_1}\left(0,A_1+1,1/2-\check{\delta} \right)}^2 +  \left\vert\left\vert \mathcal{L}_{\partial_v}H_2\right\vert\right\vert_{\mathscr{Q}_{-1+\tilde{\delta}_2}^{-\tilde{\delta}_1}\left(0,A_1,1/2 \right)}^2 + \left\vert\left\vert  H_2\right\vert\right\vert_{\mathscr{Q}_{-1+\tilde{\delta}_2}^{-\tilde{\delta}_1}\left(1,A_1-1,-\kappa \right)}^2. 
\end{align}
Furthermore, in the case of an equation of type $III$, we can replace $A_1$ in the estimate~\eqref{ijwiji293} with any constant $\tilde{A}$ which satisfies $\left|\tilde{A}\right| \ll \ell_0$. 
\end{proposition}
\begin{proof}It will be convenient to introduce the convention in this proof that $q$ stands a positive constant which may change from line to line and may be assumed smaller than any other given fixed constant in the proof.

The following facts are straightforward consequences of the fundamental theorem of calculus and the fact that $\phi \in H_0^{2,\mathscr{A}}$:
\[\phi \in C^1_{v \in (-1+\tilde{\delta}_2,-\tilde{\delta}_1)}L^2\left(\mathbb{S}^2\right),\]
\[\lim_{v\to -\tilde{\delta}_1}\int_{\mathbb{S}^2}\left|\phi\right|^2\, \mathring{\rm dVol} = \lim_{v\to -1+\tilde{\delta}_2}\int_{\mathbb{S}^2}\left|\phi\right|^2\, \mathring{\rm dVol} = 0.\]
We will use these properties without comment in the rest of the proof.

Let $W$ be defined  by solving
\begin{equation}\label{2om43o294}
\partial_vW - \frac{A_1\Omega^2}{v+1}W = 0,\qquad \lim_{v\to -1}W\left(v+1\right)^{-A_1} = 1.
\end{equation}
We note that $W(v,\theta^A)$ will be everywhere non-negative and satisfy $W \sim \left(v+1\right)^{A_1}$. (In the case when $A_1 = 0$, then $W$ will be identically $1$.) We then re-write~\eqref{wiiogjr9bj3kmeo3} as
\begin{align}\label{wiiogjr9bj3kmeo3123456543}
&\mathcal{L}_{\partial_v}\left((-v)W\mathcal{L}_{\partial_v}\phi + W(1+A_2)\phi + A_3\xi W\mathcal{P}_{\ell \in \mathscr{A}}\Pi_{\tilde{\delta}_3}\mathcal{L}_b\Pi_{\tilde{\delta}_3}\phi\right)
\\ \nonumber &\qquad  - A_1(1+A_2)W\Omega^2(v+1)^{-1}\phi- \xi A_1A_3W\Omega^2(v+1)^{-1}\mathcal{P}_{\mathscr{A}}\Pi_{\tilde{\delta}_3}\mathcal{L}_b\Pi_{\tilde{\delta}_3}\phi
 \\ \nonumber &\qquad + W\mathcal{P}_{\ell \in \mathscr{A}}\left[\Omega^2\left(\slashed{\Delta}+A_4\left(v+1\right)^{-2}\right)+A_5(-v)\Omega^{2s}(v+1)^{-2}\right]\phi = 
 \\ \nonumber &\qquad \qquad W\Bigg(H + \frac{(-v)\mathcal{P}_{\ell \not\in \mathscr{A}}(\left(1-\Omega^2\right)\mathcal{L}_{\partial_v}\phi)A_1}{v+1} - A_3\mathcal{P}_{\ell \in \mathscr{A}}\Pi_{\tilde{\delta}_3}\left[\xi \mathcal{L}_b,\partial_v\right]\Pi_{\tilde{\delta}_3}\phi
 \\ \nonumber &\qquad \qquad \qquad \qquad \qquad \qquad \qquad \qquad \qquad \qquad -\left(1-\xi\right)A_3\mathcal{P}_{\ell \in \mathscr{A}}\left(\Pi_{\tilde{\delta}_3}\mathcal{L}_b\Pi_{\tilde{\delta}_3}\mathcal{L}_{\partial_v}\phi\right)\Bigg).
\end{align}
Now we contract~\eqref{wiiogjr9bj3kmeo3123456543} with $-\left(W(1+A_2)\phi + \xi A_3W\mathcal{P}_{\ell \in \mathscr{A}}\Pi_{\tilde{\delta}_3}\mathcal{L}_b\Pi_{\tilde{\delta}_3}\phi\right)$ and integrate by parts with respect to $dv\mathring{\rm dVol}$. We eventually obtain (after a bit of calculation)
\begin{align}\label{ckoeom3p}
&\int_{-1+\tilde{\delta}_2}^{-\tilde{\delta}_1}\int_{\mathbb{S}^2}\Bigg[(-v)W^2(1+A_2+O\left(q\right))\left(\mathcal{L}_{\partial_v}\phi\right)^2 + \Omega^2W^2(1+A_2+O\left(q\right))\left(\left|\slashed{\nabla}\phi\right|^2
 -(v+1)^{-2}A_4\phi^2\right)  \\ \nonumber &\qquad + A_5\Omega^4(-v)(v+1)^{-2}\phi^2
\\ \nonumber &\qquad + A_1(1+A_2)\left((1+A_2)W^2\Omega^2(v+1)^{-1}
 -\frac{1}{2}\partial_v\left[\frac{W^2(-v)}{v+1}\Omega^2\right]\right)\phi^2\Bigg]\, dv\, \mathring{\rm dVol}
\\ \nonumber & \lesssim \int_{-1+\tilde{\delta}_2}^{-\tilde{\delta}_1}\int_{\mathbb{S}^2}\left[W^2\left|H\right|\left|(1+A_2)\phi + \xi A_3\mathcal{P}_{\ell \in \mathscr{A}}\Pi_{\tilde{\delta}_3}\mathcal{L}_b\Pi_{\tilde{\delta}_3}\phi\right|+O\left(q\right)\Omega^2W^2(-v)^{-\sqrt{\epsilon}-2\check{p}}\left(\phi^2+\left(\mathcal{L}_b\phi\right)^2\right)\right]\, dv\, \mathring{\rm dVol},
\end{align}
Let us  denote the right hand side of~\eqref{ckoeom3p} by $\mathscr{U}$. We note moreover that the implied constant here is independent of $\ell_0$. For an equation of type $I$ where we have $A_1 = A_4 = A_5 = 0$ or in the case of a model equation of type $III$ where we can exploit the largeness of $\ell_0$, we immediately see that the left hand side of~\eqref{ckoeom3p} controls the left hand side of~\eqref{ijwiji293}. (It is also immediate that instead of conjugating with $\left(v+1\right)^{A_1}$ we could use $\left(v+1\right)^{\tilde{A}}$ for any constant $\tilde{A}$ which satisfies $\left|\tilde{A}\right| \ll \ell_0$.) Thus for now, we focus on the case of a model equation of type $II$.

For case $II$ we have that $A_1 = 4$, $A_2 = 1$, $A_4 = 2$, $A_5 = 0$, and that $\phi$ is supported on spherical harmonics with $\ell \geq 2$.  We may then compute that the left hand side of~\eqref{ckoeom3p} controls
\begin{align}\label{2kmo2}
&\int_{-1+\tilde{\delta}_2}^{-\tilde{\delta}_1}\Big[2\left(1+O\left(q\right)\right)(-v)(v+1)^8\left(\mathcal{L}_{\partial_v}\phi\right)^2 
\\ \nonumber &\qquad + \left(\Omega^2(8+O\left(q\right))(v+1)^6 +20(1+O\left(q\right))\left(v+1\right)^7\Omega^2 -\left(28+O\left(q\right)\right)\left(v+1\right)^6(-v)\Omega^4 \right)\phi^2\Big]\, dv\, \mathring{\rm dVol}.
\end{align}
The final term which multiplies $\phi^2$ is not positive; nevertheless, we will now show that this is a coercive estimate for $\phi$. Focusing on the problematic term, we have
\begin{align*}
&\int_{-1+\tilde{\delta}_2}^{-\tilde{\delta}_1} \left(28+O\left(q\right)\right)\left(v+1\right)^6(-v)\Omega^2\phi^2\, dv = 
\\ \nonumber &\qquad \int_{-1+\tilde{\delta}_2}^{-\tilde{\delta}_1} \left(4+O\left(q\right)\right)\frac{d}{dv}\left(\left(v+1\right)^7\right)(-v)\Omega^2 \phi^2\, dv \leq
\\ \nonumber &\qquad \int_{-1+\tilde{\delta}_2}^{-\tilde{\delta}_1}\left(4+O\left(q\right)\right)\left(v+1\right)^7\Omega^2\phi^2\, dv + \int_{-1+\tilde{\delta}_2}^{-\tilde{\delta}_1}\left(8+O\left(q\right)\right)\left(v+1\right)^7(-v)\Omega^2\left|\phi\right|\left|\mathcal{L}_{\partial_v}\phi\right|\, dv \leq
\\ \nonumber &\qquad  \int_{-1+\tilde{\delta}_2}^{-\tilde{\delta}_1}\left(4+O\left(q\right)\right)\left(v+1\right)^7\Omega^2\phi^2\, dv +  \int_{-1+\tilde{\delta}_2}^{-\tilde{\delta}_1}\left(8+O\left(q\right)\right)\left(v+1\right)^6(-v)\Omega^4\phi^2\, dv 
\\ \nonumber &\qquad \qquad +\int_{-1+\tilde{\delta}_2}^{-\tilde{\delta}_1}\left(2+O\left(q\right)\right)\left(v+1\right)^8(-v)\left(\mathcal{L}_{\partial_v}\phi\right)^2\, dv.
\end{align*}
In particular, combining this with~\eqref{2kmo2} yields that    
\begin{align}\label{3km2lm23}
&\int_{-1+\tilde{\delta}_2}^{-\tilde{\delta}_1}\Big[O\left(q\right)(-v)(v+1)^8\left(\mathcal{L}_{\partial_v}\phi\right)^2 
\\ \nonumber &\qquad + \left(-\Omega^2O\left(q\right)(v+1)^6 +\left(v+1\right)^7\Omega^2 \right)\phi^2- 30\left(v+1\right)^6(-v)\left|\Omega^2-\Omega^4\right|\phi^2\Big]\, dv\, \mathring{\rm dVol} \lesssim \mathscr{U}.
\end{align}
Away from $v = -1+\tilde{\delta}_2$ and $v = -\tilde{\delta}_1$ this estimate is coercive (modulo the $O(q)$ term multiplying $\mathcal{L}_{\partial_v}\phi$).  We now establish localized estimates near $v = -1+\tilde{\delta}_2$ and $v = -\tilde{\delta}_1$. We return to the expression~\eqref{2kmo2}, and observe that if we choose $c$ sufficiently small, then we obtain
\begin{align}\label{123213}
&\int_{-1+\tilde{\delta}_2}^{-1+\tilde{\delta}_2+c}\tilde{\xi}(v)\Big[2\left(1+O\left(q\right)\right)(v+1)^8\left(\mathcal{L}_{\partial_v}\phi\right)^2 - (20+O\left(q\right))(v+1)^6\left|\phi\right|^2\Big]\, dv\, \mathring{\rm dVol}
\\ \nonumber &\qquad +\int_{-\tilde{\delta}_1-c}^{-\tilde{\delta}_1}\left[(-v)\left(\mathcal{L}_{\partial_v}\phi\right)^2 + \Omega^2\left|\phi\right|^2\right]\, dv
 \lesssim 
\int_{-1+\tilde{\delta}_2+c}^{-\tilde{\delta}_1-c}\left(v+1\right)^6(-v)\Omega^2\phi^2\, dv +  \mathscr{U}.
\end{align}
where $\tilde{\xi}(v)$ is identically $1$ for $v \in [-1+\tilde{\delta}_2,-1+\tilde{\delta}_2+c/2]$ and vanishes for $v > -1+\tilde{\delta}_2+c$. We may then apply Lemma~\ref{HardyHardyHardy} with $p = 4$ and use that $80/49 < 2$ to obtain from~\eqref{123213} that
\begin{align}\label{kmo1mo23}
&\int_{-1+\tilde{\delta}_2}^{-1+\tilde{\delta}_2+c/2}\Big[(v+1)^8\left(\mathcal{L}_{\partial_v}\phi\right)^2 +(v+1)^6\left|\phi\right|^2\Big]\, dv\, \mathring{\rm dVol}  +\int_{-\tilde{\delta}_1-c}^{-\tilde{\delta}_1}\left[(-v)\left(\mathcal{L}_{\partial_v}\phi\right)^2 + \Omega^2\left|\phi\right|^2\right]\, dv \lesssim 
\\ \nonumber &\qquad c^{-1}\int_{-1+\tilde{\delta}_2+c/2}^{-\tilde{\delta}_1-c}(-v)\Omega^2(v+1)^6\left|\phi\right|^2\, dv\, \mathring{\rm dVol}+ \mathscr{U}.
\end{align}
Now, finally, it is clear that we may multiply~\eqref{kmo1mo23} with a suitably small constant and combine with~\eqref{3km2lm23} and~\eqref{2kmo2} to obtain the desired estimate
\begin{align}\label{2om3o4}
\int_{-1+\tilde{\delta}_2}^{-\tilde{\delta}_1}\Big[(v+1)^8(-v)\left(\mathcal{L}_{\partial_v}\phi\right)^2 +(v+1)^6\Omega^2\left|\phi\right|^2\Big]\, dv\, \mathring{\rm dVol}  \lesssim \mathscr{U}
\end{align} 
In particular, we can return to the estimate~\eqref{ckoeom3p}, and also  combine with cases of type $I$ or $III$ to obtain that 
\begin{align}\label{2knmkm1l2}
\int_{-1+\tilde{\delta}_2}^{-\tilde{\delta}_1}\Big[W^2(-v)\left(\mathcal{L}_{\partial_v}\phi\right)^2 +W^2\Omega^2\left|\slashed{\nabla}\phi\right|^2\Big]\, dv\, \mathring{\rm dVol}  \lesssim \mathscr{U}.
\end{align}

We now turn to an estimate which is only relevant in the region where $|v| \ll 1$. We first re-write~\eqref{wiiogjr9bj3kmeo3} in the following form:
\begin{align}\label{jgji39cj30jfow}
&\mathcal{L}_{\partial_v}\left((-v)\mathcal{L}_{\partial_v}\phi+\left(A_2+1\right)\phi+A_3\mathcal{P}_{\ell \in \mathscr{A}}\Pi_{\tilde{\delta}_3}\mathcal{L}_b\Pi_{\tilde{\delta}_3}\phi\right) + \frac{(-v)A_1}{v+1}\mathcal{P}_{\ell \in \mathscr{A}}\left(\Omega^2\mathcal{L}_{\partial_v}\phi\right) \\ \nonumber &\qquad  + 
  \mathcal{P}_{\ell \in \mathscr{A}}\left[\Omega^2\left(\slashed{\Delta}+A_4\left(v+1\right)^{-2}\right)+A_5(-v)\Omega^4(v+1)^{-2}\right]\phi = H - A_3\mathcal{P}_{\ell \in \mathscr{A}}\Pi_{\tilde{\delta}_3}\left[\mathcal{L}_b,\partial_v\right]\Pi_{\tilde{\delta}_3}\phi.
 \end{align}
 Now we let $c$ satisfy $0 < \tilde{\delta}_1 < c \ll 1$ and $\xi(v)$ be a cut-off function which is identically $1$ for $v \in [-c,-\tilde{\delta}_1]$ and vanishes identically for $v \in [-1,-2c]$. Now we contract~\eqref{jgji39cj30jfow} with $\xi(v) (-v)^{2\check{\delta}}\left(\left(A_2+1\right)\phi+A_3\mathcal{P}_{\ell \in \mathscr{A}}\Pi_{\tilde{\delta}_3}\mathcal{L}_b\Pi_{\tilde{\delta}_3}\phi\right)$, integrate by parts, use the estimate~\eqref{2knmkm1l2}, and finally appeal to Proposition~\ref{somestuimdie} (with $k=0$) and Remark~\ref{m2om1o1qw}. We obtain
 \begin{align}\label{kljgerjki39j}
& \int_{-c}^{-\tilde{\delta}_1}\int_{\mathbb{S}^2}(-v)^{-1+2\check{\delta}}\left[\left|\phi\right|^2 + \left|\Pi_{\tilde{\delta}_3}\mathcal{L}_b\Pi_{\tilde{\delta}_3}\phi\right|^2\right]\, dv\, \mathring{\rm dVol} \lesssim_{c,\check{\delta}} \mathscr{U}
 \end{align}
 After a suitable application of Cauchy-Schwarz we then obtain the estimate~\eqref{ijwiji293} in the case when $H_2= 0$. 
 
 Now we consider the estimate~\eqref{ijwiji293} in the case  when $H_2$ does not necessarily vanish. The proof is similar to the case when $H_2 = 0$, except that instead of using~\eqref{wiiogjr9bj3kmeo3123456543}, we write 
 \begin{align}\label{wiiogjr9bj3kmeo3123456543asdfe2doo}
&\mathcal{L}_{\partial_v}\left((-v)W\mathcal{L}_{\partial_v}\phi + W(1+A_2)\phi + A_3\xi W\mathcal{P}_{\ell \in \mathscr{A}}\Pi_{\tilde{\delta}_3}\mathcal{L}_b\Pi_{\tilde{\delta}_3}\phi- WH_2\right)
\\ \nonumber &\qquad  - A_1(1+A_2)W\Omega^2(v+1)^{-1}\phi- \xi A_1A_3W\Omega^2(v+1)^{-1}\mathcal{P}_{\mathscr{A}}\Pi_{\tilde{\delta}_3}\mathcal{L}_b\Pi_{\tilde{\delta}_3}\phi
 \\ \nonumber &\qquad + W\mathcal{P}_{\ell \in \mathscr{A}}\left[\Omega^2\left(\slashed{\Delta}+A_4\left(v+1\right)^{-2}\right)+A_5(-v)\Omega^{2s}(v+1)^{-2}\right]\phi = 
 \\ \nonumber &\qquad \qquad W\Bigg(H + \frac{(-v)\mathcal{P}_{\ell \not\in \mathscr{A}}(\left(1-\Omega^2\right)\mathcal{L}_{\partial_v}\phi)A_1}{v+1} - A_3\mathcal{P}_{\ell \in \mathscr{A}}\Pi_{\tilde{\delta}_3}\left[\xi \mathcal{L}_b,\partial_v\right]\Pi_{\tilde{\delta}_3}\phi
 \\ \nonumber &\qquad \qquad \qquad \qquad \qquad \qquad \qquad \qquad -\left(1-\xi\right)A_3\mathcal{P}_{\ell \in \mathscr{A}}\left(\Pi_{\tilde{\delta}_3}\mathcal{L}_b\Pi_{\tilde{\delta}_3}\mathcal{L}_{\partial_v}\phi\right)\Bigg) - \frac{A_1W\Omega^2}{v+1}H_2,
\end{align}
and, instead of contracting with $-\left(W(1+A_2)\phi + \xi A_3W\mathcal{P}_{\ell \in \mathscr{A}}\Pi_{\tilde{\delta}_3}\mathcal{L}_b\Pi_{\tilde{\delta}_3}\phi\right)$, we contract with 
\[-\left(W(1+A_2)\phi + \xi A_3W\mathcal{P}_{\ell \in \mathscr{A}}\Pi_{\tilde{\delta}_3}\mathcal{L}_b\Pi_{\tilde{\delta}_3}\phi-WH_2\right).\]
All other aspects of the estimate proceed as before.
\end{proof}

Now we can use our a priori estimates to obtain an existence result for the $(\tilde{\delta}_1,\tilde{\delta}_2,\tilde{\delta}_3)$-regularized equation~\eqref{wiiogjr9bj3kmeo3}. 
\begin{corollary}\label{akcoem3o}For any $0 < \tilde{\delta}_1,\tilde{\delta}_2,\tilde{\delta}_3 \ll 1$ and $H \in L^{2,\mathscr{A}}$, there exists a solution $\phi \in H_0^{2,\mathscr{A}}$ to~\eqref{wiiogjr9bj3kmeo3}. 

If $H$ is further assumed to be smooth, then $\phi$ lies in the closure of smooth functions of compact support in $(v,\theta^A) \in \left(-1+\tilde{\delta}_2,-\tilde{\delta}_1\right)\times \mathbb{S}^2$ under the norm
\[\sum_{ 0 \leq k \leq 2}\left\vert\left\vert \mathcal{L}_{\partial_v}^k\phi\right\vert\right\vert_{\mathscr{Q}_{-1+\tilde{\delta}_2}^{-\tilde{\delta}_1}\left(N_1+1-k,0,0\right)}^2.\]
\end{corollary}

\begin{proof}

Let $\lambda \gg 1$ be a constant which is sufficiently large depending on $\tilde{\delta}_1$, $\tilde{\delta}_2$, and $\tilde{\delta}_3$. We then define an operator 
\[\mathscr{P} \doteq -\mathcal{L}_{\partial_v}^2 - \mathcal{P}_{\ell \in \mathscr{A}}\left((-v)^{-1}\Omega^2\slashed{\Delta}\right) + (-v)^{-1}\lambda,\]
and an associated bilinear form $\mathscr{K}\left(\cdot,\cdot\right): H_0^{1,\mathscr{A}} \times H_0^{1,\mathscr{A}} \to \mathbb{R}$:
\begin{align}\label{2oo948jh2}
&\mathscr{K}\left(f,h\right) = \int_{-1+\tilde{\delta}_2}^{-\tilde{\delta}_1}\int_{\mathbb{S}^2}\Bigg[\left(\mathcal{L}_{\partial_v}f\right)\left(\mathcal{L}_{\partial_v}h\right) +
  (-v)^{-1}\sum_{A,B}\Bigg[\Omega^2\slashed{g}^{AB}\slashed{\partial}_Af\slashed{\partial}_Bh +\\ \nonumber &\qquad \frac{\sqrt{\slashed{g}}}{\sqrt{\mathring{\slashed{g}}}}\slashed{\partial}_A\left(\Omega^2 \frac{\sqrt{\mathring{\slashed{g}}}}{\sqrt{\slashed{g}}}\right)\slashed{g}^{AB}f\slashed{\partial}_Bh\Bigg]+(-v)^{-1}\lambda f h\Bigg]\, dv\, \mathring{\rm dVol},
\end{align}
where $\theta^A$ and $\theta^B$ as usual denote local coordiantes on $\mathbb{S}^2$ with associated derivatives $\slashed{\partial}_A$ and $\slashed{\partial}_B$, and $\frac{\sqrt{\mathring{\slashed{g}}}}{\sqrt{\slashed{g}}}$ denotes the ratio of the volume of $\mathring{\slashed{g}}$ with the volume for of $\slashed{g}$. Using, in particular, the largeness of $\lambda$, it is immediate that $\mathscr{K}$ satisfies the assumptions of the Lax-Milgram Theorem, that is,
\begin{enumerate}
    \item $\left|\mathscr{K}\left(f,h\right)\right| \leq C\left(\lambda\right) \left\vert\left\vert f\right\vert\right\vert_{H_0^{1,\mathscr{A}}}\left\vert\left\vert h\right\vert\right\vert_{H_0^{1,\mathscr{A}}}$.
    \item $\mathscr{K}\left(f,f\right) \geq c\left(\lambda\right) \left\vert\left\vert f\right\vert\right\vert_{H_0^{1,\mathscr{A}}}^2$.
\end{enumerate}
We thus obtain that for every $\tilde{H} \in H_0^{1,\mathscr{A}}$, there exists a $\psi \in H_0^{1,\mathscr{A}}$ so that
\[\mathscr{K}\left(\psi,h\right) = \langle \tilde{H},h\rangle_{H_0^{1,\mathscr{A}}},\qquad \forall h \in H_0^{1,\mathscr{A}}.\]
Now consider $\check{H} \in L^{2,\mathscr{A}}$. Since $h \mapsto \langle \check{H},h\rangle_{L^{2,\mathscr{A}}}$ is a bounded linear functional on $H_0^{1,\mathscr{A}}$, by the Riesz Representation Theorem and the above, we can find $\tilde{H} \in H_0^{1,\mathscr{A}}$ and $\psi \in H_0^{1,\mathscr{A}}$ so that 
\[\langle \check{H},h\rangle_{L^{2,\mathscr{A}}} = \langle \tilde{H},h\rangle_{H_0^{1,\mathscr{A}}} = \mathscr{K}\left(\psi,h\right) ,\qquad \forall h \in H_0^{1,\mathscr{A}}.\]
Then, it is straightforward to use elliptic regularity to show that $\psi$ in fact lies in $H_0^{2,\mathscr{A}}$ and is the unique strong solution to $-\mathscr{P}\psi= \check{H}$. Thus we have successfully defined 
\[\mathscr{P}^{-1} : L^{2,\mathscr{A}} \to H^{2,\mathscr{A}}_0.\]

Now we are ready to turn to the actual operator of interest. We may write
\[L^{(\tilde{\delta}_1,\tilde{\delta}_2,\tilde{\delta}_3)} = (-v)\mathscr{P} + \mathcal{W},\]
where $\mathcal{W}: H_0^{1,\mathscr{A}} \to L^{2,\mathscr{A}}$. Using this we see that
\[L^{(\tilde{\delta}_1,\tilde{\delta}_2,\tilde{\delta}_3)}\phi = H \Leftrightarrow \phi + K\phi = \left((-v)\mathscr{P}\right)^{-1}H,\]
where 
\[K \doteq \left((-v)\mathscr{P}\right)^{-1}\mathcal{W}\phi.\]
From the Rellich--Kondrachov theorem we see that $K$ is a compact operator from $H_0^{1,\mathscr{A}} \to H_0^{1,\mathscr{A}}$. Thus, by the Fredholm alternative and elliptic regularity, it suffices to complete the proof that we check  that ${\rm ker}\left(L^{(\tilde{\delta}_1,\tilde{\delta}_2,\tilde{\delta}_3)}\right) =  \{0\}$ acting on $H^{2,\mathscr{A}}_0$ functions $\phi$. This fact follows immediately from Proposition~\ref{kwmogmomog003}.
\end{proof}

The estimate for $\mathcal{L}_{\partial_v}\phi$ that follows from Proposition~\ref{wqiqiu1u8nkk} is quite weak in that we do not even obtain an estimate on the $L^1$ norm of $\mathcal{L}_{\partial_v}\phi$ (which is uniform as $\tilde{\delta}_1,\tilde{\delta}_2,\tilde{\delta}_3 \to 0$). This is remedied in the following.
\begin{proposition}\label{2om2omo412kini2o2}Let $0 < \tilde{\delta}_1,\tilde{\delta}_2,\tilde{\delta}_3 \ll 1$ and $H \in L^{2,\mathscr{A}}$. Further assume that $\phi$ is a $H_0^{2,\mathscr{A}}$ solution of a $(\tilde{\delta}_1,\tilde{\delta}_2,\tilde{\delta}_3)$-regularized model second order equation with right hand side $H$. Then $\phi$ satisfies
\begin{align}\label{2l2omo3}
&\left\vert\left\vert \mathcal{L}_{\partial_v}\phi \right\vert\right\vert_{\mathscr{Q}_{-1/4}^{-\tilde{\delta}_1}\left(0,0,\kappa\right)}^2 + \left\vert\left\vert \slashed{\nabla}\phi\right\vert\right\vert_{\mathscr{Q}_{-1/4}^{-\tilde{\delta}_1}\left(0,0,-1/2+\check{\delta}\right)}^2 \lesssim 
\\ \nonumber &\qquad \left\vert\left\vert \mathcal{L}_{\partial_v}\phi \right\vert\right\vert_{\mathscr{Q}_{-3/4}^{-1/4}\left(0,0,0\right)}^2 + \left\vert\left\vert \slashed{\nabla}\phi\right\vert\right\vert_{\mathscr{Q}_{-1/2}^{-\tilde{\delta}_1}\left(0,0,0\right)}^2  +\left\vert\left\vert H\right\vert\right\vert^2_{\mathscr{Q}_{-1/2}^{-\tilde{\delta}_1}\left(0,0,\kappa\right)}.
\end{align}
\end{proposition}
\begin{proof}
We contract the equation~\eqref{wiiogjr9bj3kmeo3} with $\left(1-r\left(-v\right)^{2\check{\delta}}\right)\xi(v)\Omega^{-2}\mathcal{L}_{\partial_v}\phi$ for some $0 < r \ll 1$ and integrate by parts. Keeping in mind that the boundary term proportional to $\left(\mathcal{L}_{\partial_v}\phi\right)^2$ comes with a good sign at $v = -\tilde{\delta}_1$, the estimate~\eqref{2l2omo3} immediately follows.
\end{proof}

In the next corollary, we observe that we may treat the equation~\eqref{wiiogjr9bj3kmeo3} as a degenerate transport equation for $\mathcal{L}_{\partial_v}\phi$ and apply the estimates from Section~\ref{ij3oin4in234} in order to obtain (weighted) $L^{\infty}_v$ estimates for $\mathcal{L}_{\partial_v}\phi$.
\begin{corollary}\label{2im3om2o492}Let $0 < \tilde{\delta}_1, \tilde{\delta}_2,\tilde{\delta}_3 \ll 1$  and $H \in L^{2,\mathscr{A}}$. Further assume that $\phi$ is a $H_0^{2,\mathscr{A}}$ solution of a $(\tilde{\delta}_1,\tilde{\delta}_2,\tilde{\delta}_3)$-regularized model second order equation with right hand side $H$. Let $J$ be an integer such that  $0 \leq J \leq N_2-2$. Then, assuming that $H$ lies in the closure of smooth functions under the norm on the right hand side of~\eqref{ok204i2o4}, we have, for every $\tilde{v} \in [-1/2,0)$, $\check{s} \geq \check{p}$, and $\check{q} \geq 0$:
\begin{align}\label{ok204i2o4}
&\sum_{j=0}^1\left\vert\left\vert \left(v\mathcal{L}_{\partial_v}\right)^j\mathcal{L}_{\partial_v}\phi\right\vert\right\vert_{\check{\mathscr{S}}_{-1/2}^{\tilde{v}}\left(J-j,0,\check{s}+\check{q}j+2\kappa,\check{q}\right)}^2 \lesssim 
\\ \nonumber &\qquad \sum_{j=0}^1\left\vert\left\vert \left(v\mathcal{L}_{\partial_v}\right)^j\mathcal{L}_{\partial_v}\phi\right\vert\right\vert_{\mathscr{Q}_{-3/4}^{-1/2}\left(J-j,0,0\right)}^2+\left\vert\left\vert H\right\vert\right\vert_{\check{\mathscr{S}}_{-1/4}^{\tilde{v}}\left(J,0,\check{s}+2\kappa,\check{q}\right)}^2  + \sum_{\left|\alpha\right| \leq 2}\left\vert\left\vert \phi^{(\alpha)} \right\vert\right\vert_{\mathscr{S}_{-1/4}^{\tilde{v}}\left(J,0,0\right)}^2.
\end{align}
If $J = 0$ then we omit the sum in $j$. 

\end{corollary} 
\begin{proof}The estimate~\eqref{ok204i2o4} is a straightforward consequences of applying the proof of Lemma~\ref{linftofkwp3}  and using the bootstrap assumptions. 
\end{proof}

Now we are ready for higher order estimates.

\begin{proposition}\label{kwmogmomog003}Let $0 < \tilde{\delta}_1,\tilde{\delta}_2,\tilde{\delta}_3 \ll 1$, $H(v,\theta^A): (-1+\tilde{\delta}_2,-\tilde{\delta}_1) \times\mathbb{S}^2 \to \mathbb{R}$ be a smooth function satisfying $\left(1-\mathcal{P}_{\mathscr{A}}\right)H = 0$, and $\phi$ be a $H_0^{2,\mathscr{A}}$ solution of a $(\tilde{\delta}_1,\tilde{\delta}_2,\tilde{\delta}_3)$-regularized model second order equation with right hand side $H$. Let $J$ be a positive integer less than or equal to $N_1-2$. Then we have the following estimates:
\begin{enumerate}
\item
\begin{align}\label{ji2jo0193}
&\sum_{j=0}^1\left\vert\left\vert \left(v\mathcal{L}_{\partial_v}\right)^j\mathcal{L}_{\partial_v}\phi\right\vert\right\vert_{\mathscr{Q}_{-1+\tilde{\delta}_2}^{-\tilde{\delta}_1}\left(J-j,A_1+j,1/2\right)}^2 + \left\vert\left\vert \phi\right\vert\right\vert_{\mathscr{Q}_{-1+\tilde{\delta}_2}^{-\tilde{\delta}_1}\left(J+1,A_1-1,-\kappa\right)}^2
\\ \nonumber &\qquad +\left\vert\left\vert \left(1,v\mathcal{L}_{\partial_v},\mathcal{L}_b\right)\mathcal{L}_{\partial_v}\phi\right\vert\right\vert_{\mathscr{Q}_{-1/2}^{-\tilde{\delta}_1}\left(J-1,0,\kappa\right)}^2 + \left\vert\left\vert \left(1,\mathcal{L}_b\right)\phi\right\vert\right\vert_{\mathscr{Q}_{-1/2}^{-\tilde{\delta}_1}\left(J,0,-1/2+\check{\delta}\right)}^2 \lesssim
\\ \nonumber &\sum_{\left|\alpha\right| = J}\Bigg[\left\vert\left\vert H_1(\alpha) \right\vert\right\vert_{\mathscr{Q}_{-1+\tilde{\delta}_2}^{-\tilde{\delta}_1}\left(0,A_1+1,1/2-\check{\delta}\right)}^2 +\left\vert\left\vert \mathcal{L}_{\partial_v}H_2(\alpha) \right\vert\right\vert_{\mathscr{Q}_{-1+\tilde{\delta}_2}^{-\tilde{\delta}_1}\left(0,A_1,1/2\right)}^2
\\ \nonumber &\qquad  +\left\vert\left\vert H_2(\alpha) \right\vert\right\vert_{\mathscr{Q}_{-1+\tilde{\delta}_2}^{-\tilde{\delta}_1}\left(1,A_1-1,-\kappa\right)}^2\Bigg] +\left\vert\left\vert \left(1,\mathcal{L}_b\right)H\right\vert\right\vert_{\mathscr{Q}_{-1+\tilde{\delta}_2}^{-\tilde{\delta}_1}\left(J-1,A_1+1,\kappa\right)}^2, 
\end{align}

where $H_1(\alpha)$ and $H_2(\alpha)$ represent anyway of writing $H^{(\alpha)}  = H_1(\alpha) + \mathcal{L}_{\partial_v}H_2(\alpha)$ for smooth $H_1(\alpha)$ and $H_2(\alpha)$ and so that $\lim_{v\to 0}H_2(\alpha) = 0$ and $H_2(\alpha)$ is supported for $v \geq -1/2$. 

\item 
\begin{align}\label{k32om2oo}
&\sum_{j=0}^1\left\vert\left\vert \left(v\mathcal{L}_{\partial_v}\right)^j\mathcal{L}_{\partial_v}\phi\right\vert\right\vert_{\mathscr{Q}_{-1+\tilde{\delta}_2}^{-\tilde{\delta}_1}\left(J-j,A_1+j,\kappa\right)}^2 + \left\vert\left\vert \phi\right\vert\right\vert_{\mathscr{Q}_{-1+\tilde{\delta}_2}^{-\tilde{\delta}_1}\left(J+1,A_1-1,-1/2+\check{\delta}\right)}^2 \lesssim 
 \left\vert\left\vert H\right\vert\right\vert_{\mathscr{Q}_{-1+\tilde{\delta}_2}^{-\tilde{\delta}_1}\left(J,A_1+1,\kappa\right)}^2 
 \end{align}
\end{enumerate}

Finally, it is immediate from a density argument using Proposition~\ref{wqiqiu1u8nkk} and Corollary~\ref{akcoem3o}, that the same result holds if $H$, $H_1(\alpha)$, and $H_2(\alpha)$ lie in the closure of smooth compactly supported functions which are annihilated by $\left(1-\mathcal{P}_{\mathscr{A}}\right)$ under the norms on the right hand sides of~\eqref{ji2jo0193} and~\eqref{k32om2oo}.

\end{proposition}

\begin{proof}We will only discuss the proof of~\eqref{ji2jo0193} as the proof of~\eqref{k32om2oo} is completely analogous (and in fact simpler).  We will need to commute to the equation~\eqref{wiiogjr9bj3kmeo3} with suitable $\mathcal{L}_{Z^{(\alpha)}}$. We note that the qualitative regularity required for the calculations which follow are easily justified by elliptic regularity.

We now start with the proof of~\eqref{ji2jo0193}.  Let us denote the right hand side of~\eqref{ji2jo0193} by $\mathcal{E}\left(J\right)$ and the left hand side by $\mathcal{P}\left(J\right)$. We proceed by induction on $J$. Thus we assume that $J$ is a positive integer and that~\eqref{ji2jo0193} holds with $J$ replaced by $J-1$. (When $J = 0$ we remove the sum in $j$ and observe that this has been established in Proposition~\ref{wqiqiu1u8nkk}.)

As a consequence of this induction hypothesis we may apply the fundamental theorem of calculus to obtain various $L^{\infty}_vL^2\left(\mathbb{S}^2\right)$ estimates for $\phi^{(\alpha)}$. In particular, for any $\tilde{v} \in [-1/2,-\tilde{\delta}_1)$ we have
\begin{align}\label{2km4o242}
\left\vert\left\vert \phi\right\vert\right\vert_{\mathscr{S}_{-1/2}^{\tilde{v}}\left(J-1,0,0\right)}^2 \lesssim \left\vert\left\vert \phi\right\vert\right\vert_{\mathscr{Q}_{-3/4}^{-1/4}\left(J-1,0,0\right)}^2 + \left\vert\left\vert \mathcal{L}_{\partial_v}\phi\right\vert\right\vert_{\mathscr{Q}_{-1/2}^{\tilde{v}}\left(J-1,0,1/2\right)}^2\log\left(\tilde{v}\right) \lesssim \log\left(\tilde{v}\right)\mathcal{E}\left(J-1\right).
\end{align}

For $R = J-1$ or $R = J$ and for any multi-index $\alpha$ with $|\alpha| \leq R$, we may commute our equation~\eqref{wiiogjr9bj3kmeo3} with $\mathcal{L}_{Z^{(\alpha)}}$ to end up with the following equation:
\begin{align}\label{jklwfeknjekjn}
L^{(\tilde{\delta}_1,\tilde{\delta}_2,\tilde{\delta}_3)}\mathcal{L}_{Z^{(\alpha)}}\phi = H_1(\alpha) + \mathcal{L}_{\partial_v}H_2(\alpha) + \left[L^{(\tilde{\delta}_1,\tilde{\delta}_2,\tilde{\delta}_3)},\mathcal{L}_{Z^{(\alpha)}}\right]\phi,
\end{align}
where we just write $H_1(\alpha) + \mathcal{L}_{\partial_v}H_2(\alpha) = H^{(\alpha)}$ if $\left|\alpha\right| = J-1$.

There are two main types of terms generated by the commutator $\left[L^{(\tilde{\delta}_1,\tilde{\delta}_2,\tilde{\delta}_3)},\mathcal{L}_{Z^{(\alpha)}}\right]\phi$. The first are generated by  
\begin{equation}\label{3km999492jfi3}
\left[\mathcal{P}_{\mathscr{A}}\Pi_{\tilde{\delta}_3}\mathcal{L}_b\Pi_{\tilde{\delta}_3}\mathcal{L}_{\partial_v},\mathcal{L}_{Z^{(\alpha)}}\right]\phi,
\end{equation}
 and may be written as a sum of terms which may include any of the following:
\begin{align}\label{scheshcem2}
&\bigcup_{|\alpha_1| = R} \xi(v)\mathcal{P}_{\mathscr{A}}\left(\Pi_{\delta_3}(b^{(\alpha_1)})^A\mathring{\nabla}_A\Pi_{\delta_3}\mathcal{L}_{\partial_v}\phi\right)
\\ \nonumber &\qquad \bigcup_{|\alpha_1|+|\alpha_2| = R \atop 1 \leq  |\alpha_1| \leq R-1}\left( \xi(v)\mathcal{L}_{\partial_v}\left(\mathcal{P}_{\mathscr{A}}\left(\Pi_{\delta_3}(b^{(\alpha_1)})^A\mathring{\nabla}_A\Pi_{\delta_3}\phi^{(\alpha_2)}\right)\right) , \mathcal{P}_{\mathscr{A}}\left(\Pi_{\delta_3}(\mathcal{L}_{\partial_v}b^{(\alpha_1)})^A\mathring{\nabla}_A\Pi_{\delta_3}\phi^{(\alpha_2)}\right)\right)
\\ \nonumber &\qquad + \bigcup_{|\alpha_1|+|\alpha_2| = R \atop |\alpha_1| \geq 1} \left(1-\xi(v)\right)\left(\mathcal{P}_{\mathscr{A}}\left(\Pi_{\delta_3}(b^{(\alpha_1)})^A\mathring{\nabla}_A\Pi_{\delta_3}\mathcal{L}_{\partial_v}\phi^{(\alpha_2)}\right)\right) 
\\ \nonumber &\qquad \doteq \xi(v)\mathcal{L}_{\partial_v}H_3(\alpha) + H_4(\alpha).
\end{align}

The second are generated by the commutator \[H_5(\alpha) \doteq \left[\mathcal{P}_{\ell \in \mathscr{A}}\left(\Omega^2\left(\slashed{\Delta}+A_4\left(v+1\right)^{-2}\right)+A_5(-v)\Omega^4(v+1)^{-2}\right),\mathcal{L}_{Z^{(\alpha)}}\right]\phi.\]
This commutator can further be written as a product of terms which involve at least one of $\Omega^{(\alpha_1)}$ or $\left[\mathcal{L}_{\mathcal{Z}^{(\alpha_1)}},\slashed{\Delta}-\mathring{\Delta}\right]$ for some $|\alpha_1| \neq 0$. We note furthermore that $H_5(\alpha)$ will involve at most $|\alpha|+1$ derivatives of $\slashed{g}$ and $\phi$. 

Using  the induction hypothesis, the bootstrap assumptions~\eqref{thisisisismom3o}, the nonlinear estimates from Section~\ref{iio98987923}, and~\eqref{2km4o242}, it is then straightforward to establish the desired induction step by applying Proposition~\ref{wqiqiu1u8nkk} and summing over all $\left|\alpha\right| \leq J$, applying Proposition~\ref{2om2omo412kini2o2} and summing over all $\left|\alpha\right| \leq J-1$,  carrying out one additional commutation with $\mathcal{L}_b$ and applying Proposition~\ref{2om2omo412kini2o2} and summing over all $\left|\alpha\right| \leq J-1$, and finally using the commuted equation directly to estimate $v\mathcal{L}_{\partial_v}^2\phi^{(\alpha)}$.

\end{proof}
\begin{remark}The motivation for considering an improved estimate for $\mathcal{L}_b\phi$ in the context of~\eqref{ji2jo0193} will become more apparent after the next lemma. See Remark~\ref{2kn1inini}.
\end{remark}

In this next lemma, we extend the above existence results and estimates to allow more general boundary conditions at $v = -\tilde{\delta}_1$. It is then straightforward to take the limits as $\tilde{\delta}_1,\tilde{\delta}_3\to 0$. 
\begin{lemma}\label{vmoem3029222ece}Let $0 < \tilde{\delta}_2 \ll 1$, $H(v,\theta^A): (-1+\tilde{\delta}_2,0] \to \mathbb{R}$ be smooth and compactly supported, and $h(\theta^A): \mathbb{S}^2 \to \mathbb{R}$ be a smooth function satisfying $\left(1-\mathcal{P}_{\mathscr{A}}\right)H = \left(1-\mathcal{P}_{\mathscr{A}}\right)h = 0$. There there exists a $H^{2,\mathscr{A}}_{\rm loc}$ solution $\phi$ to~\eqref{wiiogjr9bj3kmeo3} with $\tilde{\delta}_1 = \tilde{\delta}_3 = 0$ such that $\phi|_{v = 0} = h$, $\phi|_{v = -1+\tilde{\delta}_2} = 0$, and which moreover satisfies~\eqref{ji2jo0193} and~\eqref{k32om2oo} (with $\tilde{\delta}_1 = \tilde{\delta}_3 = 0$) if we add to the right hand side of~\eqref{ji2jo0193}:
\[\left\vert\left\vert \left(1,\mathcal{L}_{b|_{v=0}}\right)h\right\vert\right\vert_{\mathring{H}^J\left(\mathbb{S}^2\right)}^2\]
and add to the right hand side of~\eqref{k32om2oo}:
\[\left\vert\left\vert h\right\vert\right\vert_{\mathring{H}^{J+1}\left(\mathbb{S}^2\right)}.\]

We next observe that this lemma continues to hold under the assumption that $H$, $H_1(\alpha)$ and $H_2(\alpha)$ lie in the closure of smooth functions under the norm on the right hand of these estimates if we also add the following term to the right hand of the estimate~\eqref{ji2jo0193}:
\[\sum_{\left|\alpha\right| = J}\left\vert\left\vert H_2(\alpha)|_{v=0}\right\vert\right\vert^2_{L^2\left(\mathbb{S}^2\right)}.\]

We have the following quantitative sense in which $\phi$ obtains its boundary data:
\begin{equation}\label{knein2ied2ie92}
\limsup_{v\to 0}\left[(-v)^{-1/4}\left\vert\left\vert \phi-h\right\vert\right\vert_{\mathring{H}^{J-2}\left(\mathbb{S}^2\right)}^2\right]  = 0.
\end{equation}
We have the following uniqueness statement: If $\tilde{\phi}$ is another $H^{2,\mathscr{A}}_{\rm loc}$ solution to~\eqref{wiiogjr9bj3kmeo3} with $\tilde{\delta}_1 = \tilde{\delta}_3 = 0$ such that
\begin{equation}\label{2m20k00302k04k02}
\left\vert\left\vert \phi\right\vert\right\vert_{\mathscr{Q}_{-1+\tilde{\delta}_2}^0\left(2,0,0\right)} < \infty,\qquad \lim_{v\to 0}\left\vert\left\vert \tilde{\phi} - h\right\vert\right\vert_{L^2\left(\mathbb{S}^2\right)}= 0
\end{equation}
holds and so that $\tilde{\phi}|_{v=-1+\tilde{\delta}_2} = 0$, then we have that $\tilde{\phi} = \phi$.

Finally, we have the following extension of the estimate~\eqref{k32om2oo} in the case when $\tilde{\delta}_1 = \tilde{\delta}_3 = 0$. Suppose that
\[H = \tilde{H}_1 + \left((-v)\mathcal{L}_{\partial_v} + A_3\mathcal{P}_{\ell \in \mathscr{A}}\mathcal{L}_b\right)\tilde{H}_2,\]
for $\tilde{H}_2$ which vanishes for $v \leq -1/2$. Then we may replace norms involving $H$ on the right hand side of~\eqref{k32om2oo} with 
\begin{align}\label{ik2om3om34}
& \left\vert\left\vert \tilde{H}_1\right\vert\right\vert_{\mathscr{Q}_{-1+\tilde{\delta}_2}^0\left(J,A_1+1,\kappa\right)}^2 +\left\vert\left\vert \tilde{H}_2\right\vert\right\vert_{\mathscr{Q}_{-1+\tilde{\delta}_2}^0\left(J,0,\kappa\right)}^2 +\left\vert\left\vert \tilde{H}_2\right\vert\right\vert_{\mathscr{Q}_{-1+\tilde{\delta}_2}^0\left(J+1,0,1/2-\check{\delta}\right)}^2
\end{align}
Similarly, in~\eqref{ji2jo0193} we may replace $\left\vert\left\vert \left(1,\mathcal{L}_b\right)H\right\vert\right\vert_{\mathscr{Q}_{-1/2}^{-\tilde{\delta}_1}\left(J-1,0,\kappa\right)}^2 $ on the right hand side by 
\begin{align}\label{ini2ni1}
& \left\vert\left\vert \left(1,\mathcal{L}_b\right)\tilde{H}_1\right\vert\right\vert_{\mathscr{Q}_{-1+\tilde{\delta}_2}^0\left(J-1,A_1+1,\kappa\right)}^2 +\left\vert\left\vert \left(1,\mathcal{L}_b\right)\tilde{H}_2\right\vert\right\vert_{\mathscr{Q}_{-1+\tilde{\delta}_2}^0\left(J-1,0,\kappa\right)}^2 +\left\vert\left\vert \left(1,\mathcal{L}_b\right)\tilde{H}_2\right\vert\right\vert_{\mathscr{Q}_{-1+\tilde{\delta}_2}^0\left(J,0,1/2-\check{\delta}\right)}^2
\end{align}

\end{lemma}
\begin{proof}We start by letting $0 < \tilde{\delta}_1, \tilde{\delta}_3 \ll 1$. For the existence statement with $H$ and $h$ smooth, we may use the standard cut-off trick: We define $\check{H} =  H - L^{(\tilde{\delta}_1,\tilde{\delta}_2,\tilde{\delta}_3)}\left[\xi(v)h\right]$. Then we apply Corollary~\ref{akcoem3o} to define $\check{\phi}$ to be a solution to~\eqref{wiiogjr9bj3kmeo3} corresponding to $\tilde{\delta}_1,\tilde{\delta}_2,\tilde{\delta}_3$ and right hand side $\check{H}$. We may then define a function $\phi$ by setting $\phi = \check{\phi} +\xi(v)\mathcal{P}_{\ell \in \mathscr{A}}h$. This will satisfy $\phi|_{v = -\tilde{\delta}_1} = h$. It is furthermore immediate from our uniform estimates and a compactness argument that we may take $\tilde{\delta}_1 \to 0$.  

Next we would like to show that $\phi$ satisfies the desired estimates by repeating the proof of Proposition~\ref{kwmogmomog003}  \emph{mutatis mutandis} while tracking the boundary terms at $\{v = 0\}$. In order to do this we must first establish $\phi$ attains its boundary data at $v = 0$ in a suitable sense. We will have, by the (limits as $\tilde{\delta}_1\to0$) of the estimates from Proposition~\ref{kwmogmomog003} that
\begin{equation}\label{omomom3}
\left\vert\left\vert \mathcal{L}_{\partial_v}\check{\phi}\right\vert\right\vert_{\mathscr{Q}_{-1/2}^0\left(J,0,\kappa\right)} +\left\vert\left\vert \check{\phi}\right\vert\right\vert_{\mathscr{Q}_{-1/2}^0\left(J+1,0,-1/2+\check{\delta}\right)} < \infty. 
\end{equation}
In particular, it follows easily from the fundamental theorem of calculus that $\check{\phi} \to 0$ as $v\to 0$ in $\mathring{H}^J$. Since $\delta_3 > 0$ we also have that $\Pi_{\tilde{\delta}_3}\mathcal{L}_b\Pi_{\tilde{\delta}_3}\check{\phi}$ converges to $0$ in $\mathring{H}^J$ as $v\to 0$. Furthermore, it is an immediate consequence of~\eqref{omomom3} that there exists a sequence $\{v_i\}_{i=1}^{\infty}$ such that $v_i \to 0$ and $\left[(-v)^{1/2}\mathcal{L}_{\partial_v}\check{\phi}\right]|_{v=v_i} \to 0$ in $\mathring{H}^J$. With these facts established we may now re-run the proof of Proposition~\ref{kwmogmomog003} while tracking the boundary terms that arise. In particular, we find that~\eqref{ji2jo0193} and~\eqref{k32om2oo} hold with $\tilde{\delta}_1 = 0$ if we add to the right hand side of~\eqref{ji2jo0193}:
\[\left\vert\left\vert \left(1,\Pi_{\tilde{\delta}_3}\mathcal{L}_{b|_{v=0}}\Pi_{\tilde{\delta}_3}\right)h\right\vert\right\vert_{\mathring{H}^J\left(\mathbb{S}^2\right)}^2\]
and add to the right hand side of~\eqref{k32om2oo}:
\[\left\vert\left\vert h\right\vert\right\vert_{\mathring{H}^{J+1}\left(\mathbb{S}^2\right)}.\] 
Finally, as we have observed before, a straightforward compactness argument allows us to take $\tilde{\delta}_3 \to 0$.

The case of non-smooth and non-compactly supported $H$ and $h$ follows by similar methods and with a straightforward limiting argument. Then~\eqref{knein2ied2ie92} is a consequence of the fundamental theorem of calculus. We omit the straightforward details.

We next need to establish the uniqueness statement. The idea is to consider the equation for the difference $\phi - \tilde{\phi}$ and run (the proof of) Proposition~\ref{wqiqiu1u8nkk} in the setting where $\tilde{\delta}_1 = \tilde{\delta}_3 = 0$ to conclude that $\phi - \tilde{\phi} = 0$. However, the proof of Proposition~\ref{wqiqiu1u8nkk} involves various integration by parts which could, in principle, produce boundary terms at $v = 0$. To see that these boundary terms vanish it would suffice to have that
\[\limsup_{v\to 0}(-v)\int_{\mathbb{S}^2}\left|\mathcal{L}_{\partial_v}\tilde{\phi}\right|^2\mathring{\rm dVol} = 0.\]
However, this is easily established by considering the equation for $\tilde{\phi}$ as a degenerate transport equation for $\mathcal{L}_{\partial_v}\tilde{\phi}$, applying Lemma~\ref{linftofkwp3}, and using the assumption~\eqref{2m20k00302k04k02}. 

Lastly, we come to~\eqref{ik2om3om34} (\eqref{ini2ni1} may be handled with an analogous argument). It will suffice to consider the case when $J = 0$. In this case after applying the multiplier from Proposition~\ref{2om2omo412kini2o2}, we have on the right hand side a term
\[\int_{-1/2}^0\int_{\mathbb{S}^2}\left(1-r(-v)^{\check{\delta}}\right)\Omega^{-2}\left((-v)\mathcal{L}_{\partial_v}+A_3\mathcal{P}_{\ell \in \mathscr{A}}\mathcal{L}_b\right)\tilde{H}_2 \left(\mathcal{L}_{\partial_v}\phi\right)\, dv\, \mathring{\rm dVol}.\]
We now simply integrate by parts with the operator $\left((-v)\mathcal{L}_{\partial_v}+A_3\mathcal{P}_{\ell \in \mathscr{A}}\mathcal{L}_b\right)$, use the bootstrap assumptions on $\Omega$, and use the equation to substitute $\left((-v)\mathcal{L}_{\partial_v}+A_3\mathcal{P}_{\ell \in \mathscr{A}}\mathcal{L}_b\right)\mathcal{L}_{\partial_v}\phi$. The result then follows from a suitable application of Cauchy-Schwarz.
\end{proof}
\begin{remark}\label{2kn1inini}We may now see one motivation for considering separately estimates for $\mathcal{L}_b\phi$ in Proposition~\ref{kwmogmomog003}. Namely, our improved estimate for $\mathcal{L}_b\phi$ requires us to put $\mathcal{L}_bh$ in $\mathring{H}^{N_1-1}$; however, our estimate for the highest number of derivatives of $\phi$ already requires this, so we do not lose anything with regards to the estimates on $h$ by adding in the stronger estimate for $\mathcal{L}_b\phi$. (Of course, there is a price to pay in that we must ask that the right hand side of our equation satisfies a suitable estimate after an application of $\mathcal{L}_b$.)
\end{remark}

\subsection{Analysis Near $\{v = -1\}$}\label{soclosetovminusone}
In this section we  obtain families of different weighted estimates near $v = -1+\tilde{\delta}_2$ and then we will discuss taking the limit as $\tilde{\delta}_2 \to 0$. We note that the Hardy inequality of Lemma~\ref{HardyHardyHardy} will play an important role in this section.

\begin{proposition}\label{2m3o}Let $0 < \tilde{\delta}_2 \ll 1$ and $\tilde{\delta}_1 = \tilde{\delta}_3 = 0$, and let $\phi$ be a $\mathcal{H}_0^{2,\mathscr{A}}$ solution to~\eqref{wiiogjr9bj3kmeo3} such that $\phi$ vanishes identically for $v \geq -1/4$ and so that the right hand $H$ is a smooth function with $\left(1-\mathcal{P}_{\mathscr{A}}\right)H = 0$. 

Then we have, for any $0 \leq J \leq N_1-2$:
\begin{align}\label{emrf0g4e00292123123}
\sum_{j=0}^2\left\vert\left\vert \mathcal{L}^j_{\partial_v}\phi\right\vert\right\vert^2_{\mathscr{Q}_{-1+\tilde{\delta}_2}^0\left(J+1-j ,-p-1+j,0\right)} \lesssim \sum_{j=0}^1\left\vert\left\vert \mathcal{L}^j_{\partial_v}\phi\right\vert\right\vert^2_{\mathscr{Q}_{-1+\tilde{\delta}_2}^0\left(J+1-j, -p+j,0\right)}  + \left\vert\left\vert H\right\vert\right\vert^2_{\mathscr{Q}_{-1+\tilde{\delta}_2}^0\left(J,-p+1,0\right)},
\end{align}
where if $J = 0$ the sum on the left hand side just goes up to $j = 1$. 
\begin{enumerate}
	\item For an equation of type $I$, we may take $p \in [-3/2+\check{\delta},3/2-\check{\delta}]$.
	\item For an equation of type $II$, we may take $p \in [-9/2+\check{\delta},1/2-\check{\delta}]$.
	\item For an equation of type $III$, we may take any $p$ satisfying $|p|\lesssim 1$.
\end{enumerate}
As usual, all statements of the proposition remain true if $H$ lies in the closure of smooth functions which are annihilated by $\left(1-\mathcal{P}_{\mathscr{A}}\right)$ under the norm determined by the right hand side of~\eqref{emrf0g4e00292123123}.
\end{proposition}
\begin{proof}Throughout the proof we let $q$ denote a positive constant which may be assumed suitably small.

We start with the case $J = 0$. Contracting~\eqref{wiiogjr9bj3kmeo3} with $-\left(v+1\right)^{-2p}\phi$ and integrating by parts leads to 
\begin{align}\label{eok3ommsmw3}
&\int_{-1+\tilde{\delta}_2}^0\int_{\mathbb{S}^2}\Bigg[\left(v+1\right)^{-2p}\left(\mathcal{L}_{\partial_v}\phi\right)^2 \\ \nonumber &\qquad \qquad \qquad - \left[\left(2p+1\right)\left(p+\frac{1}{2}A_1\right)+A_4+A_5+O\left(q\right)\right]\left(v+1\right)^{-2p-2}\phi^2 
 + \left(v+1\right)^{-2p}\left|\slashed{\nabla}\phi\right|^2\Bigg]\, dv\, \mathring{\rm dVol}\lesssim 
\\ \nonumber &\qquad \int_{-1+\tilde{\delta}_2}^0\int_{\mathbb{S}^2}\left[\left(v+1\right)^{-2p+1}\left(\mathcal{L}_{\partial_v}\phi\right)^2 +\left(v+1\right)^{-2p-1}\phi^2 + \left(v+1\right)^{-2p+1}\left|\slashed{\nabla}\phi\right|^2\right]\, dv\, \mathring{\rm dVol} - 
\\ \nonumber &\qquad  \int_{-1+\tilde{\delta}_2}^0\left(v+1\right)^{-2p}H\phi\, dv\, \mathring{\rm dVol}.
\end{align}
We note that the implied constants are independent of the choice of the constant $\ell_0$. 

Let $\ell_{\rm min}$ be the smallest integer in $\mathscr{A}$ (this will be $1$ for type $I$, $2$ for type $II$, and $\ell_0 \gg 1$ for type $III$). Then we may then use use Lemma~\ref{HardyHardyHardy}, and a short calculation, to obtain that the first line of~\eqref{eok3ommsmw3} controls, under the hypothesis of the proposition,
\begin{align}\label{32o4ok20}
&\int_{-1+\tilde{\delta}_2}^0\int_{\mathbb{S}^2}\Bigg(\frac{(2p+1)^2}{4} + \ell_{\rm min}\left(\ell_{\rm min}+1\right) 
\\ \nonumber &\qquad \qquad - \left[\left(2p+1\right)\left(p+\frac{1}{2}A_1\right)+A_4+A_5\right]+O\left(q\right)\Bigg)\phi^2\left(v+1\right)^{-2p-2}\, dv\, \mathring{\rm dVol}
\\ \nonumber &\qquad \qquad \gtrsim \int_{-1+\tilde{\delta}_2}^0\int_{\mathbb{S}^2}\phi^2\left(v+1\right)^{-2p-2}\, dv\, \mathring{\rm dVol}.
\end{align} 
After combining this with~\eqref{eok3ommsmw3}, an application of Cauchy-Schwarz then establishes the proposition when $J = 0$. The case of higher $J$ follows in a straightforward fashion by commuting with $\mathcal{L}_{\mathcal{Z}^{(\alpha)}}$,  repeating the above estimate, and then using the equation directly to estimate $\mathcal{L}_{\partial_v}^2\phi^{(\alpha)}$.

\end{proof}

\subsection{The Final Estimates}\label{finalestestestsce}
In this section we will take $\tilde{\delta}_2 \to 0$ and put all of our estimates together. We start by defining our main norm for the solutions $\phi$ to the model second order equations.
\begin{definition}\label{i32ij3o990oinkijijo}For any smooth function $\phi(v,\theta^A): (-1,0)\times\mathbb{S}^2 \to \mathbb{R}$, positive integer $J_1$ satisfying $N_2 \leq J_1 \leq N_1-1$, and choice of $I$, $II$, $III\left(\tilde{p},\tilde{s},\tilde{q}\right)$, or $III'(\tilde{p},\tilde{s},\tilde{q})$ (where $\tilde{s} \geq \check{p}$, $\tilde{q} \geq 0$, and $\tilde{p} \not\in \{1/2,3/2\}$), we define $ J_2 = J_2\left(J_1\right) = N_2-1 - (N_1-1-J_1)$ and then we set 
\begin{enumerate}\item
\begin{align*}
&\left\vert\left\vert \phi\right\vert\right\vert^2_{L,I,J_1} \doteq \sum_{j=0}^1 \left\vert\left\vert \mathcal{L}_{\partial_v}^{1+j}\phi\right\vert\right\vert_{\mathscr{Q}\left(J_1-1-j,-3/2+\check{\delta}+j,\kappa+j\right) }^2 + \sum_{j=0}^1 \left\vert\left\vert \mathcal{L}_{\partial_v}^{1+j}\phi\right\vert\right\vert_{\mathscr{Q}_{-1/2}^0\left(J_1-2-j,0,-\sqrt{\check{p}}+j\right) }^2 
\\ \nonumber &\qquad + \left\vert\left\vert \phi\right\vert\right\vert_{\mathscr{Q}\left(J_1,-5/2+\check{\delta},-1/2+\check{\delta}\right)}^2 + \sum_{j=0}^2\left\vert\left\vert \mathcal{L}_{\partial_v}^j\phi \right\vert\right\vert^2_{\mathscr{S}_{-1}^{-1/2}\left(J_1-1-j,-2+\check{\delta}+j,0\right)} \\ \nonumber &\qquad +\sum_{j=0}^1\left\vert\left\vert \left(v\mathcal{L}_{\partial_v}\right)^j\mathcal{L}_{\partial_v}\phi\right\vert\right\vert_{\check{\mathscr{S}}_{-1/2}^0\left(J_2-1-j,0,500\check{p}\left(1+j\right)+2\kappa,500\check{p}\right)}^2
  + \sum_{j=0}^1\left\vert\left\vert \left(v\mathcal{L}_{\partial_v}\right)^j\phi\right\vert\right\vert_{\mathscr{S}_{-1/2}^0\left(J_1-1-j,0,0\right)}^2.
\end{align*} 
\item
\begin{align*}
&\left\vert\left\vert \phi\right\vert\right\vert^2_{L,II,J_1} \doteq \sum_{j=0}^1 \left\vert\left\vert \mathcal{L}_{\partial_v}^{1+j}\phi\right\vert\right\vert_{\mathscr{Q}\left(J_1-1-j,1/2+\check{\delta}+j,\kappa+j\right) }^2 +\sum_{j=0}^1 \left\vert\left\vert \mathcal{L}_{\partial_v}^{1+j}\phi\right\vert\right\vert_{\mathscr{Q}_{-1/2}^0\left(J_1-2-j,0,-\sqrt{\check{p}}+j\right) }^2 
\\ \nonumber &\qquad + \left\vert\left\vert \phi\right\vert\right\vert_{\mathscr{Q}\left(J_1,-1/2+\check{\delta},-1/2+\check{\delta}\right)}^2+\sum_{j=0}^2\left\vert\left\vert \mathcal{L}_{\partial_v}^j\phi\right\vert\right\vert_{\mathscr{S}_{-1}^{-1/2}\left(J_1-1-j,\check{\delta}+j,0\right)}
\\ \nonumber &\qquad +\sum_{j=0}^1\left\vert\left\vert \left(v\mathcal{L}_{\partial_v}\right)^j\mathcal{L}_{\partial_v}\phi\right\vert\right\vert_{\check{\mathscr{S}}_{-1/2}^0\left(J_2-1-j,0,50\check{p}\left(1+j\right)+50\check{p}+2\kappa,50\check{p}\right)}^2
+ \sum_{j=0}^1\left\vert\left\vert \left(v\mathcal{L}_{\partial_v}\right)^j\phi\right\vert\right\vert_{\mathscr{S}_{-1/2}^0\left(J_1-1-j,0,0\right)}^2,
\end{align*}
\item 
\begin{align*}
&\left\vert\left\vert \phi\right\vert\right\vert^2_{L,III',J_1} \doteq \sum_{j=0}^1 \left\vert\left\vert \mathcal{L}_{\partial_v}^{1+j}\phi\right\vert\right\vert_{\mathscr{Q}\left(J_1-1-j,-\tilde{p}+j,\kappa+j\right) }^2 +\sum_{j=0}^1 \left\vert\left\vert \mathcal{L}_{\partial_v}^{1+j}\phi\right\vert\right\vert_{\mathscr{Q}_{-1/2}^0\left(J_1-2-j,0,-\sqrt{\check{p}}+j\right) }^2 
\\ \nonumber &\qquad + \left\vert\left\vert \phi\right\vert\right\vert_{\mathscr{Q}\left(J_1,-\tilde{p}-1,-1/2+\check{\delta}\right)}^2+\sum_{j=0}^2\left\vert\left\vert \mathcal{L}_{\partial_v}^j\phi\right\vert\right\vert^2_{\mathscr{S}_{-1}^{-1/2}\left(J_1-1-j,-\tilde{p}-1/2+j,0\right)}
\\ \nonumber &\qquad +\sum_{j=0}^1\left\vert\left\vert \left(v\mathcal{L}_{\partial_v}\right)^j\mathcal{L}_{\partial_v}\phi\right\vert\right\vert_{\check{\mathscr{S}}_{-1/2}^0\left(J_2-1-j,0,\tilde{s} + \tilde{q}j+2\kappa,\tilde{q}\right)}^2
  + \sum_{j=0}^1\left\vert\left\vert \left(v\mathcal{L}_{\partial_v}\right)^j\phi\right\vert\right\vert_{\mathscr{S}_{-1/2}^0\left(J_1-1-j,0,0\right)}^2.
\end{align*}

\item
\begin{align*}
&\left\vert\left\vert \phi\right\vert\right\vert^2_{L,III,J_1} \doteq \sum_{j=0}^1\left\vert\left\vert \mathcal{L}_{\partial_v}^{1+j}\phi\right\vert\right\vert_{\mathscr{Q}\left(J_1-1-j,-\tilde{p}+j,1/2+j\right)}^2 + \left\vert\left\vert \phi\right\vert\right\vert_{\mathscr{Q}\left(J_1,-\tilde{p}-1,-\kappa\right)}^2
\\ \nonumber &\qquad +\left\vert\left\vert\left(1,v\mathcal{L}_{\partial_v},\mathcal{L}_b\right)\mathcal{L}_{\partial_v}\phi\right\vert\right\vert_{\mathscr{Q}_{-1/2}^0\left(J_1-2,0,\kappa\right)}^2 +\left\vert\left\vert\left(1,v\mathcal{L}_{\partial_v},\mathcal{L}_b\right)\mathcal{L}_{\partial_v}\phi\right\vert\right\vert_{\mathscr{Q}_{-1/2}^0\left(J_1-3,0,-\sqrt{\check{p}}\right)}^2
\\ \nonumber &\qquad + \left\vert\left\vert \left(1,\mathcal{L}_b\right)\phi\right\vert\right\vert_{\mathscr{Q}_{-1/2}^0\left(J_1-1,0,-1/2+\check{\delta}\right)}^2 + \sum_{j=0}^2\left\vert\left\vert \mathcal{L}_{\partial_v}^j\phi\right\vert\right\vert_{\mathscr{S}_{-1}^{-1/2}\left(J_1-1-j,-\tilde{p}-1/2+j,0\right)}^2
\\ \nonumber &\qquad + \sum_{j=0}^1\left\vert\left\vert \left(v\mathcal{L}_{\partial_v}\right)^j \mathcal{L}_{\partial_v}\phi\right\vert\right\vert_{\check{\mathscr{S}}_{-1/2}^0\left(J_2-1-j,0,\tilde{s}+\tilde{q}j+2\kappa,\tilde{q}\right)}^2+\sum_{j=0}^1\left\vert\left\vert \left(v\mathcal{L}_{\partial_v}\right)^j\phi\right\vert\right\vert_{\mathscr{S}_{-1/2}^0\left(J_1-2-j,0,0\right)}^2.
\end{align*}
\end{enumerate}

We refer to the corresponding spaces of functions by $\mathscr{X}_{L,(1,II,III,III'),J_1}$.
\end{definition}
\begin{remark}The ``L'' in the norm reminds us that this is the norm for the left hand side of our estimates.
\end{remark}
\begin{remark}\label{ook2o3o49}When we want to emphasize the vector field $b$ which is involved in the definition of the $\left(L,III,J_1\right)$ norm, we will then write $\left\vert\left\vert \cdot\right\vert\right\vert_{L(b),III,J_1}$.
\end{remark}

Next, we define the norm for the right hand side of model second order equations.
\begin{definition}\label{2kmo2o3}For any smooth function $H(v,\theta^A): (-1,0)\times\mathbb{S}^2 \to \mathbb{R}$ such that $\left(1-\mathcal{P}_{\mathscr{A}}\right)H = 0$, smooth function $h: \mathbb{S}^2 \to \mathbb{R}$ such that $\left(1-\mathcal{P}_{\mathscr{A}}\right)h = 0$, positive integer $J_1$ satisfying $N_2 \leq J_1 \leq N_1$, and choice of $I$, $II$, $III\left(\tilde{p},\tilde{s},\tilde{q}\right)$, or $III'(\tilde{p},\tilde{s},\tilde{q})$ (where $\tilde{s} \geq \check{p}$, $\tilde{q} \geq 0$, and $\tilde{p} \not\in \{1/2,3/2\}$), we define $ J_2 = J_2\left(J_1\right) = N_2 - (N_1-J_1)$ and then we set 
\begin{align*}
&\left\vert\left\vert \left(H_1(\alpha),H_2(\alpha),\tilde{H}_1,\tilde{H}_2,h\right) \right\vert\right\vert^2_{R,III,J_1-1}  \doteq \left\vert\left\vert H\right\vert\right\vert_{\mathscr{Q}\left(J_1-2,{\rm min}\left(-\tilde{p},A_1\right)+1,\kappa\right)}^2+\left\vert\left\vert \left(1,\mathcal{L}_b|_{v=0}\right)h\right\vert\right\vert_{\mathring{H}^{J_1-1}\left(\mathbb{S}^2\right)}^2\\ \nonumber &\qquad  + \sum_{\left|\alpha\right| = J-1}\left[\left\vert\left\vert H_1(\alpha)\right\vert\right\vert_{\mathscr{Q}\left(0,-\tilde{p}+1,1/2-\check{\delta}\right)}^2 + \left\vert\left\vert \mathcal{L}_{\partial_v}H_2(\alpha)\right\vert\right\vert_{\mathscr{Q}\left(0,0,1/2\right)}^2  + \left\vert\left\vert H_2(\alpha)\right\vert\right\vert_{\mathscr{Q}\left(1,0,-\kappa\right)}^2\right]
\\ \nonumber &\qquad +\left\vert\left\vert H\right\vert\right\vert_{\check{\mathscr{S}}_{-1/2}^0\left(J_2-1,0,\check{s}+2\kappa,\check{q}\right)}^2 +\left\vert\left\vert H\right\vert\right\vert^2_{\mathscr{S}_{-1}^{-1/2}\left(J_1-3,-\tilde{p}+3/2,0\right)} + \sum_{\left|\alpha\right| = J-1}\left\vert\left\vert H_2(\alpha)|_{v=0}\right\vert\right\vert_{L^2\left(\mathbb{S}^2\right)}^2
\\ \nonumber &\qquad +\left\vert\left\vert \left(1,\mathcal{L}_b\right)\tilde{H}_1\right\vert\right\vert_{\mathscr{Q}_{-1/2}^0\left(J-1,0,\kappa\right)}^2 +\left\vert\left\vert \left(1,\mathcal{L}_b\right)\tilde{H}_2\right\vert\right\vert_{\mathscr{Q}_{-1/2}^0\left(J-1,0,\kappa\right)}^2 +\left\vert\left\vert \left(1,\mathcal{L}_b\right)\tilde{H}_2\right\vert\right\vert_{\mathscr{Q}\left(J,0,1/2-\check{\delta}\right)}^2
\end{align*}
where, for any $\left|\alpha\right| = J_1-1$, $H_1(\alpha)$ and $H_2(\alpha)$ represent anyway of writing $H(\alpha)= H_1(\alpha) + \mathcal{L}_{\partial_v}H_2(\alpha)$ for smooth $H_1(\alpha)$ and $H_2(\alpha)$ and so that $\lim_{v\to 0}H_2$ exists and $H_2$ is supported for $v \geq -1/2$,  $\tilde{H}_1$ and $\tilde{H}_2$ represent anyway of writing $H = \tilde{H}_1 + \left((-v)\mathcal{L}_{\partial_v}-\mathcal{P}_{\ell \in \mathscr{A}}\mathcal{L}_b\right)\tilde{H}_2$ such that $\tilde{H}_2$ vanishes for $v \leq -1/2$.

We next set 
\begin{enumerate}
	\item \begin{align*}
&\left\vert\left\vert \left(\tilde{H}_1,\tilde{H}_2,h\right) \right\vert\right\vert^2_{R,I,J_1-1}  \doteq \left\vert\left\vert H\right\vert\right\vert_{\mathscr{Q}_{-1}^{-1/2}\left(J_1-1,-1/2+\check{\delta},\kappa\right)}^2 +\left\vert\left\vert H\right\vert\right\vert_{\check{\mathscr{S}}_{-1/2}^0\left(J_2-1,-1,500\check{p}+2\kappa,500\check{p}\right)}^2
\\ \nonumber &\qquad +\left\vert\left\vert H\right\vert\right\vert^2_{\mathscr{S}_{-1}^{-1/2}\left(J_1-3,\check{\delta},0\right)} + \left\vert\left\vert \tilde{H}_1\right\vert\right\vert_{\mathscr{Q}\left(J,0,\kappa\right)}^2 +\left\vert\left\vert \tilde{H}_2\right\vert\right\vert_{\mathscr{Q}\left(J,0,\kappa\right)}^2 +\left\vert\left\vert \tilde{H}_2\right\vert\right\vert_{\mathscr{Q}\left(J+1,0,1/2-\check{\delta}\right)}^2+\left\vert\left\vert h\right\vert\right\vert_{\mathring{H}^{J_1}}^2.
\end{align*}
\item \begin{align*}
&\left\vert\left\vert \left(\tilde{H}_1,\tilde{H}_2,h\right) \right\vert\right\vert^2_{R,II,J_1-1}  \doteq \left\vert\left\vert H\right\vert\right\vert_{\mathscr{Q}_{-1}^{-1/2}\left(J_1-1,1,\kappa\right)}^2 +\left\vert\left\vert H\right\vert\right\vert_{\check{\mathscr{S}}\left(J_2-1,1-\check{\delta},100\check{p}+2\kappa,50\check{p}\right)}^2
\\ \nonumber &\qquad + \left\vert\left\vert \tilde{H}_1\right\vert\right\vert_{\mathscr{Q}_{-1/2}^0\left(J,0,\kappa\right)}^2 +\left\vert\left\vert \tilde{H}_2\right\vert\right\vert_{\mathscr{Q}_{-1/2}^0\left(J,0,\kappa\right)}^2 +\left\vert\left\vert \tilde{H}_2\right\vert\right\vert_{\mathscr{Q}_{-1/2}^0\left(J+1,0,1/2-\check{\delta}\right)}^2+\left\vert\left\vert h\right\vert\right\vert_{\mathring{H}^{J_1}}^2.
\end{align*}
\item \begin{align*}
&\left\vert\left\vert \left(\tilde{H}_1,\tilde{H}_2,h\right) \right\vert\right\vert^2_{R,III',J_1-1}  \doteq \left\vert\left\vert H\right\vert\right\vert_{\mathscr{Q}_{-1}^{-1/2}\left(J_1-1,-\tilde{p}+1,\kappa\right)}^2 +\left\vert\left\vert H\right\vert\right\vert_{\check{\mathscr{S}}\left(J_2-1,-\tilde{p}+1,\tilde{s}+2\kappa,\tilde{q}\right)}^2
\\ \nonumber &\qquad + \left\vert\left\vert \tilde{H}_1\right\vert\right\vert_{\mathscr{Q}\left(J,-p+1,\kappa\right)}^2 +\left\vert\left\vert \tilde{H}_2\right\vert\right\vert_{\mathscr{Q}\left(J,0,\kappa\right)}^2 +\left\vert\left\vert \tilde{H}_2\right\vert\right\vert_{\mathscr{Q}\left(J+1,0,1/2-\check{\delta}\right)}^2+\left\vert\left\vert h\right\vert\right\vert_{\mathring{H}^{J_1}}^2.
\end{align*}
\end{enumerate}

We refer to the corresponding spaces of functions by $\mathscr{X}_{R,(1,II,III,III'),J_1}$ (where we may also think of this as a norm defined on $H$ by taking an infimum over all possible ways of decomposing the right hand side $H$).
\end{definition}
\begin{remark}The ``R'' in the norm reminds us that this is the norm for the right hand side of our estimates.
\end{remark}
\begin{remark}\label{22omo1o3}When we want to emphasize the vector field $b$ which is involved in the definition of the $\left(R,III,J_1\right)$ norm, we will then write $\left\vert\left\vert \cdot\right\vert\right\vert_{R(b),III,J_1}$.
\end{remark}
\begin{remark}For model second order equations of type $II$, due to the fact that $\phi$ and $H$ are supported on a finite set of spherical harmonics, we could simplify the presentation of the norms by using the fact that estimates for angular derivatives of $\phi$ and $H$ may be deduced directly from estimates for $\phi$ and $H$ (with a dependence of $\ell_0$). Such a choice of norm would not affect our ability to prove the main theorems in the paper. However, we choose to work with the norms above because there is actually no need for the estimate to depend on $\ell_0$, and we believe it is clarifying to note that with these definitions of the norms, the fundamental estimate of Theorem~\ref{fo3p39iwu88u} below for equations of type $II$ is independent of $\ell_0$.
\end{remark}

Now we are ready to state our main result.
\begin{theorem}\label{fo3p39iwu88u}Let $J_1$ be a positive integer which satisfies $N_2 \leq J_1 \leq N_1$ and choose one of $I$, $II$, $III$, or $III'$. Then let $(H,h)$ lie in the closure of smooth functions which are annihilated by $\left(1-\mathcal{P}_{\mathscr{A}}\right)$ under the corresponding norm $\left\vert\left\vert \left(H,h\right) \right\vert\right\vert_{R,\left(I,II,III,III'\right)J_1-1}$. Then there exists a solution $\phi$ to the model second order equation of type $I$, $II$, or $III$ (depending on the original choice of $I$, $II$, $III$, or $III'$ with $III$ being associated to $III'$) such that 
\begin{align}\label{lm3l3m}
&\left\vert\left\vert \phi\right\vert\right\vert_{L,\left(I,II,III,III'\right),J_1} \lesssim 
\\ \nonumber &\qquad \left\vert\left\vert \left(H_1(\alpha),H_2(\alpha),\tilde{H}_1,\tilde{H}_2,h\right) \right\vert\right\vert_{R,III,J_1-1} \text{ or }\left\vert\left\vert \left(\tilde{H}_1,\tilde{H}_2,h\right) \right\vert\right\vert_{R,\left(I,II,III'\right),J_1-1},
\end{align}
(where we pick the suitable norm on the right and make a choice of $H_1$, $H_2$, $\tilde{H}_1$ and $\tilde{H}_2$ if appropriate) and such that 
\begin{equation}\label{oj2om3o2}
\sum_{\left|\alpha\right| \leq 10}\limsup_{v\to 0}(-v)^{-1/4}\int_{\mathbb{S}^2}\left(\phi^{(\alpha)}-h^{(\alpha)}\right)^2\mathring{\rm dVol} < \infty,
\end{equation}

Moreover, if $\tilde{\phi}$ is any other $H^{2,\mathscr{A}}_{\rm loc}$ solution to the model second order equation with right hand side $H$ which satisfies~\eqref{oj2om3o2} and so that
\begin{equation}\label{ijio2}
\sum_{\left|\alpha\right| \leq 1}\int_{-1}^0\int_{\mathbb{S}^2}\left[\Omega^{-2}\left(\mathcal{L}_{\partial_v}\tilde{\phi}^{(\alpha)}\right)^2 + \Omega^2\left|\slashed{\nabla}\tilde{\phi}^{(\alpha)}\right|^2\right]\, dv\, \mathring{\rm dVol} < \infty,
\end{equation}
then we must have $\phi = \tilde{\phi}$.  
\end{theorem}
\begin{proof} This is analogous to the proof of Lemma~\ref{vmoem3029222ece} with three modifications:
\begin{enumerate}
	\item Our estimates now follow from Proposition~\ref{2m3o}, Lemma~\ref{vmoem3029222ece}, and Lemma~\ref{3m2omo4}.
	\item We estimate $\phi$ and $\mathcal{L}_{\partial_v}\phi$ in the $\mathscr{S}_{-1}^{-1/2}$ norm by applying the fundamental theorem of calculus in the $v$-direction and applying Cauchy-Schwarz. The corresponding estimate for $\mathcal{L}_{\partial_v}^2\phi$ is then obtain directly from the equation. (In the case of an equation of type $III$ or $III'$ we have disallowed $\tilde{p} = 1/2$, $3/2$ which would lead to a logarithmic divergence here.)
	\item For the uniqueness statement, we only need to justify that a suitable boundary condition holds now along $\{v = -1\}$ so that we may apply the proof of Proposition~\ref{wqiqiu1u8nkk}. For this it suffices to note as a consequence of  the left hand side of~\eqref{ijio2} being finite, there exists a sequence $\{v_i\}_{i=1}^{\infty}$ such that $v_i \to -1$ and so that $\limsup_{i\to \infty}\int_{\mathbb{S}^2}\left[\left(v+1\right)\left(\mathcal{L}_{\partial_v}\phi\right)^2 + \left(v+1\right)\left|\slashed{\nabla}\phi\right|^2\right]\mathring{\rm dVol} = 0$.
	
\end{enumerate}

\end{proof}

Lastly, it will be useful to observe that the theory we have developed for the model second order equations is robust to lower-order (possibly nonlinear) perturbation of the coefficients.
\begin{lemma}\label{2omomo393}Let $J_1$ be a positive integer which satisfies $N_2 \leq J_1 \leq N_1$ and choose one of $I$, $II$, $III$, or $III'$. Then choose $0 < \tilde{\delta} \ll 1$ to be sufficiently small and assume that we have a mapping 
\[\mathscr{E} : B_{\tilde{\delta}}\left(\mathscr{X}_{L,\left(I,II,III,III'\right),J_1}\right) \to B_{\tilde{\delta}}\left(\mathscr{X}_{R,\left(I,II,III,III'\right),J_1-1}\right),\]
where $B_{\tilde{\delta}}$ denotes the ball of radius $\tilde{\delta}$. Furthermore, suppose that for $\phi_1,\phi_2 \in B_{\tilde{\delta}}\left(\mathscr{X}_{L,\left(I,II,III,III'\right),J_1}\right) $ we have that
\[\left\vert\left\vert\mathscr{E}\left(\phi_1\right) - \mathscr{E}\left(\phi_2\right)\right\vert\right\vert_{R,\left(I,II,III,III'\right),J_1-1} \leq F\left(\tilde{\delta}\right)\left\vert\left\vert \phi_1-\phi_2\right\vert\right\vert_{L,\left(I,II,III,III'\right),J_1},\]
for $\lim_{\tilde{\delta}\to 0}F\left(\tilde{\delta}\right) = 0$. Then for every $H$ which is sufficiently small in the appropriate $\mathscr{X}$ norm, there exists a solution to
\[\mathscr{L}\phi = H + \mathscr{E}\left(\phi\right),\]
where $\mathscr{L}$ is the corresponding model second order equation, so that~\eqref{lm3l3m} holds. 
\end{lemma}
\begin{proof}This is an immediate consequence of the contraction mapping principle.
\end{proof}
\begin{remark}One may, of course, easily formulate a generalization of Lemma~\ref{2omomo393} to suitable systems of model second order equations.
\end{remark}

\subsection{A Useful Extension}\label{bitofanextensionsection}
It will turn out to be useful to have a version of the estimate from Theorem~\ref{fo3p39iwu88u}  where, for an equation of type $III$, we allow the top order norm for $\mathcal{L}_{\partial_v}\phi$ to further degenerate as $v\to 0$ (and we also drop the $\sup_v$ estimates from the left and right hand sides). This extension will only be used in the propagation of constraints argument in Section~\ref{propagatetheconstraintsforever}.
\begin{lemma}\label{asdf2nini3i9iun12}Let $\phi$ be a solution to a model second order equation of type $III$ such that $\phi$, $\mathcal{L}_{\partial_v}\phi$, and the right hand side $H$ lie in $\mathring{H}^{10}\left(\mathbb{S}^2_{-1,v}\right)$ for each $v \in (-1,0)$, and which moreover satisfies the following boundary conditions:
\begin{align}\label{oj2ono23}
\lim_{v\to 0}(-v)\left\vert\left\vert \mathcal{L}_{\partial_v}\phi\right\vert\right\vert_{\mathring{H}^3\left(\mathbb{S}^2_{-1,v}\right)} = 0,\qquad \lim_{v\to 0}\left\vert\left\vert \left(1,\mathcal{L}_b\right)\left(\phi - h\right)\right\vert\right\vert_{\mathring{H}^3\left(\mathbb{S}^2_{-1,v}\right)} = 0,
\end{align}
\begin{align}\label{inoio2j4}
&\lim_{v\to -1} \left(v+1\right)^{A_1}\left\vert\left\vert \mathcal{L}_{\partial_v}\phi\right\vert\right\vert_{\mathring{H}^3\left(\mathbb{S}^2_{-1,v}\right)} = 0,\qquad \lim_{v\to -1}\left(v+1\right)^{A_1}\left\vert\left\vert \phi\right\vert\right\vert_{\mathring{H}^3\left(\mathbb{S}^2_{-1,v}\right)} = 0
\end{align}
\begin{align}
\lim_{v\to -1}\left(v+1\right)^{-2p}\left\vert\left\vert \mathcal{L}_{\partial_v}\phi\right\vert\right\vert_{\mathring{H}^3\left(\mathbb{S}^2_{-1,v}\right)}\left\vert\left\vert \phi\right\vert\right\vert_{\mathring{H}^3\left(\mathbb{S}^2_{-1,v}\right)}  = 0. 
\end{align}
where $h : \mathbb{S}^2 \to \mathbb{R}$ and $p \in \mathbb{R}$ satisfies $|p| \lesssim 1$. 

Then, for every $\check{q}$ satisfying $\check{\delta} \leq \check{q} \leq 1/2-\kappa$, we have the following estimate for $1 \leq N \leq 3$:
\begin{align}\label{1nk2nl1l3}
&\sum_{j=0}^1\left\vert\left\vert \mathcal{L}_{\partial_v}^{j+1}\phi\right\vert\right\vert_{\mathscr{Q}\left(N-j,-p+j,1/2 +\check{q}+j\right)}^2 + \left\vert\left\vert \phi\right\vert\right\vert_{\mathscr{Q}\left(N+1,-p-1,-\kappa\right)}^2 +\left\vert\left\vert \left(1,v\mathcal{L}_{\partial_v}\mathcal{L}_b\right)\mathcal{L}_{\partial_v}\phi\right\vert\right\vert^2_{\mathscr{Q}_{-1/2}^0\left(N-1,0,\kappa\right)} 
\\ \nonumber &\qquad + \left\vert\left\vert \left(1,\mathcal{L}_b\right)\phi\right\vert\right\vert_{\mathscr{Q}\left(N,-p-1,-1/2+\check{\delta}\right)}^2 \lesssim \left\vert\left\vert H\right\vert\right\vert^2_{\mathscr{Q}\left(N-1,-p+1,\kappa\right)} + \sum_{\left|\alpha\right| = N}\Bigg[\left\vert\left\vert H_1(\alpha) \right\vert\right\vert_{\mathscr{Q}\left(0,-p+1,1/2-\check{\delta}\right)}^2   
\\ \nonumber &\qquad +\left\vert\left\vert \mathcal{L}_{\partial_v}H_2(\alpha)\right\vert\right\vert_{\mathscr{Q}\left(0,0,1/2+\check{q}\right)}^2 + \left\vert\left\vert H_2(\alpha)\right\vert\right\vert_{\mathscr{Q}\left(1,0,-\kappa\right)}^2  + \left\vert\left\vert H_2(\alpha)|_{v=0}\right\vert\right\vert^2_{\mathring{H}^N\left(\mathbb{S}^2\right)}\Bigg]+ \left\vert\left\vert \left(\mathcal{L}_b|_{v=0},1\right)h\right\vert\right\vert_{\mathring{H}^N\left(\mathbb{S}^2\right)}^2
\end{align}
(The reader should note the presence of the $\check{q}$ in the first term of the first line of~\eqref{1nk2nl1l3} and in the first term of the third line of~\eqref{1nk2nl1l3}.) As in our previous estimates, $H_1(\alpha)$ and $H_2(\alpha)$ represent anyway of writing $H(\alpha)= H_1(\alpha) + \mathcal{L}_{\partial_v}H_2(\alpha)$ for  $H_1(\alpha)$ and $H_2(\alpha)$ so that the right hand side of~\eqref{1nk2nl1l3} is well-defined and so that $H_2(\alpha)$ vanishes for $v \leq -1/2$.
\end{lemma}
\begin{proof}We will only explicitly discuss the proof of 
\begin{align}\label{klnl2knlnlk12}
&\left\vert\left\vert \mathcal{L}_{\partial_v}\phi\right\vert\right\vert_{\mathscr{Q}\left(0,A_1,1/2+\check{q}\right)}^2 + \left\vert\left\vert \phi\right\vert\right\vert_{\mathscr{Q}\left(1,A_1-1,-\kappa\right)}^2 \lesssim
\\ \nonumber &\left\vert\left\vert H_1(0) \right\vert\right\vert_{\mathscr{Q}\left(0,A_1+1,1/2-\check{\delta}\right)}^2 +\left\vert\left\vert \mathcal{L}_{\partial_v}H_2(0) \right\vert\right\vert_{\mathscr{Q}\left(0,A_1,1/2+\check{q}\right)}^2
  +\left\vert\left\vert H_2(0) \right\vert\right\vert_{\mathscr{Q}\left(1,A_1-1,-\kappa\right)}^2
\\ \nonumber &\qquad +\left\vert\left\vert \left(1,\mathcal{L}_{b|_{v=0}}\right)h\right\vert\right\vert_{L^2\left(\mathbb{S}^2\right)}^2+\left\vert\left\vert H_2(0)|_{v=0}\right\vert\right\vert^2_{L^2\left(\mathbb{S}^2\right)}.
\end{align}
The commutation process to obtain the corresponding higher order versions of~\eqref{klnl2knlnlk12} is analogous to that described in the proof of Proposition~\ref{kwmogmomog003}. Once this ``global'' estimate is established, then one may combine with the proof Proposition~\ref{2m3o} and the proof of the estimate~\eqref{k32om2oo} to improve the weights at $\{v = -1\}$ and $\{v = 0\}$ and thus establish the desired result.

We will revisit the multiplier technique used in the proof of Proposition~\ref{wqiqiu1u8nkk}. As in the proof of Proposition~\ref{wqiqiu1u8nkk}, we write the equation in the form~\eqref{wiiogjr9bj3kmeo3123456543asdfe2doo} (except that now there are no projections $\Pi_{\tilde{\delta}_3}$).  Instead of contracting with $-\left(W(1+A_2)\phi + \xi A_3W\mathcal{P}_{\ell \in \mathscr{A}}\Pi_{\tilde{\delta}_3}\mathcal{L}_b\Pi_{\tilde{\delta}_3}\phi-WH_2(0)\right)$ we contract with 
\[-\left((-v)W\mathcal{L}_{\partial_v}\phi+W(1+A_2)\phi + \xi A_3W\mathcal{P}_{\ell \in \mathscr{A}}\mathcal{L}_b\phi-WH_2(0)\right) \doteq -W\mathfrak{P}\left[\phi,H_2(0)\right]\]
We then carry out the analogous integration by parts. Our assumptions on $\phi$ guarantee that there are no boundary contributions at $\{v = -1\}$ and $(-v)\mathcal{L}_{\partial_v}\phi$ does not produce a boundary term at $\{v = 0\}$. We end up with the following:
\begin{align}\label{2omoin2oijn4}
&\left\vert\left\vert \phi \right\vert\right\vert_{\mathscr{Q}\left(1,A_1-1,-\kappa\right)}^2 \lesssim \left\vert\left\vert H_2(0)\right\vert\right\vert_{\mathscr{Q}\left(1,A_1-1,-\kappa\right)}^2+\left\vert\left\vert \left(\mathcal{L}_b|_{v=0},1\right)h\right\vert\right\vert_{L^2\left(\mathbb{S}^2\right)}^2 + \left\vert\left\vert H_2(0)|_{v=0}\right\vert\right\vert^2_{L^2\left(\mathbb{S}^2\right)}
\\ \nonumber &\qquad  \int_{-1}^0\int_{\mathbb{S}^2}W^2\Big[\Omega^2(v+1)^{-1}\left|\phi\right| + \epsilon \Omega^2\left|\slashed{\nabla}\phi\right|
\\ \nonumber &\qquad \qquad \qquad +(-v)(v+1)^{-1}\left|\mathcal{P}_{\ell \not\in \mathscr{A}}\left((1-\Omega^2)\mathcal{L}_{\partial_v}\phi\right)\right|+H_1(0)\Big]\left|\mathfrak{P}\left[\phi,H_2(0)\right]\right|\, dv \, \mathring{\rm dVol},
\end{align}
where we note the constant does not depend on $\ell_0$.

We then (roughly) follow the analogous second step in the proof of Proposition~\ref{wqiqiu1u8nkk}; that is, we now contract with  $W(-v)^{2\check{\delta}}\mathfrak{P}\left[\phi,H_2(0)\right]$ and integrate by parts.  This leads to
\begin{align}\label{2m3om2omo3}
&\left\vert\left\vert \mathfrak{P}\left[\phi,H_2(0)\right]\right\vert\right\vert_{\mathscr{Q}\left(0,A_1,-1/2+\check{\delta}\right)}^2 \lesssim \left\vert\left\vert \phi \right\vert\right\vert_{\mathscr{Q}\left(1,A_1-1,-\kappa+\check{\delta}\right)}^2+ \left\vert\left\vert H_2(0)\right\vert\right\vert_{\mathscr{Q}\left(1,A_1-1,-\kappa+\check{\delta}\right)}^2
\\ \nonumber &\qquad+ \int_{-1}^0\int_{\mathbb{S}^2}W^2(-v)^{2\check{\delta}}\Big[\Omega^2(v+1)^{-1}\left|\phi\right| + \epsilon \Omega^2\left|\slashed{\nabla}\phi\right|
\\ \nonumber &\qquad \qquad \qquad +(-v)(v+1)^{-1}\left|\mathcal{P}_{\ell \not\in \mathscr{A}}\left((1-\Omega^2)\mathcal{L}_{\partial_v}\phi\right)\right|+H_1(0)\Big]\left|\mathfrak{P}\left[\phi,H_2(0)\right]\right|\, dv \, \mathring{\rm dVol}.
\end{align}
Since we have 
\[\left\vert\left\vert \mathcal{L}_{\partial_v}\phi\right\vert\right\vert_{\mathscr{Q}\left(0,A_1,1-\kappa\right)}^2 \lesssim \left\vert\left\vert \mathfrak{P}\left[\phi,H_2(0)\right]\right\vert\right\vert_{\mathscr{Q}\left(0,A_1,-1/2+\check{\delta}\right)}^2 +\left\vert\left\vert \phi \right\vert\right\vert_{\mathscr{Q}\left(1,A_1-1,-\kappa\right)}^2+ \left\vert\left\vert H_2(0)\right\vert\right\vert_{\mathscr{Q}\left(1,A_1-1,-\kappa\right)}^2,\]
we may exploit the smallness of $1-\Omega^2$, $\epsilon$, and $\ell_0^{-1}$ to combine~\eqref{2omoin2oijn4} and~\eqref{2m3om2omo3} and end up with 
\begin{align}\label{2oj3oijnoi}
&\left\vert\left\vert \mathcal{L}_{\partial_v}\phi\right\vert\right\vert_{\mathscr{Q}\left(0,A_1,1-\kappa\right)}^2 + \left\vert\left\vert \phi \right\vert\right\vert_{\mathscr{Q}\left(1,A_1-1,-\kappa\right)}^2 +\left\vert\left\vert \mathfrak{P}\left[\phi,H_2(0)\right]\right\vert\right\vert_{\mathscr{Q}\left(0,A_1,-1/2+\check{\delta}\right)}^2 \lesssim 
\\ \nonumber &\qquad \left\vert\left\vert H_2(0)\right\vert\right\vert_{\mathscr{Q}\left(1,A_1-1,-\kappa\right)}^2 + \left\vert\left\vert H_1(0)\right\vert\right\vert_{\mathscr{Q}\left(0,A_1,1/2-\check{\delta}\right)}^2
+\left\vert\left\vert \left(\mathcal{L}_b|_{v=0},1\right)h\right\vert\right\vert_{L^2\left(\mathbb{S}^2\right)}^2 + \left\vert\left\vert H_2(0)|_{v=0}\right\vert\right\vert^2_{L^2\left(\mathbb{S}^2\right)}.
\end{align} 
We note if $\check{q} = 1/2-\kappa$, then we are already done. However, if $\check{q} < 1/2-\kappa$, then we need to improve the weight on $\mathcal{L}_{\partial_v}\phi$ near $v = 0$. 

We now contract with 
\[- (-v)^{2\check{q}}\left(W(1+A_2)\phi + \xi A_3W\mathcal{P}_{\ell \in \mathscr{A}}\mathcal{L}_b\phi-WH_2(0)\right) \doteq  -(-v)^{2\check{q}}W\mathfrak{Q}\left[\phi,H_2(0)\right],\]
and integrate by parts. Note that we have $\mathfrak{P}\left[\phi,H_2(0)\right] = (-v)\mathcal{L}_{\partial_v}\phi + \mathfrak{Q}\left[\phi,H_2(0)\right]$ and also
\begin{align}
&-(-v)^{2\check{q}}\mathcal{L}_{\partial_v}\left(W\mathfrak{P}\right)\left(W\mathfrak{Q}\right) = \mathcal{L}_{\partial_v}\left(-(-v)^{2\check{q}}\left(W\mathfrak{P}\right)\left(W\mathfrak{Q}\right)\right)-2\check{q}(-v)^{-1+2\check{q}}\left(W\mathfrak{P}\right)\left(W\mathfrak{Q}\right)
\\ \nonumber &\qquad +(-v)^{2\check{q}+1}W\mathcal{L}_{\partial_v}\phi\mathcal{L}_{\partial_v}\left(W(1+A_2)\phi + \xi A_3W\mathcal{P}_{\ell \in \mathscr{A}}\mathcal{L}_b\phi-WH_2(0)\right)
\\ \nonumber &\qquad +\frac{1}{2}\mathcal{L}_{\partial_v}\left((-v)^{2\check{q}}\left(W\mathfrak{Q}\right)^2\right) + \check{q}(-v)^{-1+2\check{q}}\left(W\mathfrak{Q}\right)^2.
\end{align}
In particular, after combining with~\eqref{2oj3oijnoi}, we eventually obtain the estimate
\begin{align}\label{kljkjlkj12}
&\left\vert\left\vert \mathcal{L}_{\partial_v}\phi\right\vert\right\vert_{\mathscr{Q}\left(0,A_1,1-\kappa\right)}^2 + \left\vert\left\vert \phi \right\vert\right\vert_{\mathscr{Q}\left(1,A_1-1,-\kappa\right)}^2 +\left\vert\left\vert \mathfrak{P}\left[\phi,H_2(0)\right]\right\vert\right\vert_{\mathscr{Q}\left(0,A_1,-1/2+\check{\delta}\right)}^2 + \left\vert\left\vert \mathfrak{Q}\right\vert\right\vert_{\mathscr{Q}\left(0,A_1,-1/2+\check{q}\right)}^2 \lesssim 
\\ \nonumber &\qquad \left\vert\left\vert H_2(0)\right\vert\right\vert_{\mathscr{Q}\left(0,A_1,1/2+\check{q}\right)}^2+\left\vert\left\vert H_2(0)\right\vert\right\vert_{\mathscr{Q}\left(1,A_1-1,-\kappa\right)}^2 + \left\vert\left\vert H_1(0)\right\vert\right\vert_{\mathscr{Q}\left(0,A_1,1/2-\check{\delta}\right)}^2
\\ \nonumber &\qquad +\left\vert\left\vert \left(\mathcal{L}_b|_{v=0},1\right)h\right\vert\right\vert_{L^2\left(\mathbb{S}^2\right)}^2 + \left\vert\left\vert H_2(0)|_{v=0}\right\vert\right\vert^2_{L^2\left(\mathbb{S}^2\right)}.
\end{align} 
This then yields~\eqref{klnl2knlnlk12}.
\end{proof}

\section{A Parabolic Equation}\label{paraparasec}
Throughout this section we will assume that we have a function $\Omega = \Omega_{\rm sing}\Omega_{\rm boun}$ defined for $(v,\theta^A) \in (-1,0) \times \mathbb{S}^2$ with $\Omega_{\rm sing}$ spherically symmetric so that
\begin{align}\label{2kn2kn2k}
&\left\vert\left\vert \log\Omega_{\rm boun}\right\vert\right\vert_{\mathscr{A}_0\left(\kappa,\tilde{b}\right)} + \left\vert\left\vert \log\Omega_{\rm sing} \right\vert\right\vert_{\mathscr{B}_{00}\left(\kappa\right)} +\left\vert\left\vert \log\Omega_{\rm boun} \right\vert\right\vert_{\mathscr{B}_{01}\left(\kappa,\tilde{b}\right)} \lesssim \epsilon,
\end{align}
for a suitable $\mathbb{S}^2_{-1,v}$ vector field $\tilde{b}$ and constant $\kappa$ satisfying
\begin{equation}\label{3popojp1}
\sup_{(v,\theta^A)}\left|\tilde{b}\right| \lesssim \epsilon,\qquad \left|\kappa\right| \lesssim \epsilon.
\end{equation}
We emphasize that none of the results in this section depend on the implied constants in~\eqref{2kn2kn2k} or~\eqref{3popojp1} (though by our conventions for $\epsilon$, we may assume that $\epsilon$ is sufficiently small depending on the implied constants).

In this section we will study the parabolic equation:
\begin{equation}\label{skngi3ngoi3n}
\mathcal{L}_{\partial_v}\Theta +\frac{2}{v+1}\Theta- 4\mathcal{P}_{\ell \geq 1}\left(\Omega^2\left(v+1\right)^{-2}\mathring{\Delta}\Theta\right) = H,
\end{equation} 
for $H$ such that $\mathcal{P}_{\ell = 0}H = 0$. This equation will later be used to solve for $\slashed{\rm div}b$. If $v$ is bounded away from $v = -1$ or $v =0$, then~\eqref{skngi3ngoi3n} may treated with standard parabolic equation theory. However, as $v\to -1$ or $v\to 0$, then equation~\eqref{skngi3ngoi3n} degenerates, and thus we need to carry out a special analysis of~\eqref{skngi3ngoi3n}.

We start by stating an existence statement for solutions to~\eqref{skngi3ngoi3n}.
\begin{lemma}\label{3ljooi11193}Let $H : (-1,0)\times \mathbb{S}^2 \to \mathbb{R}$ be a smooth function which vanishes near $v= -1$. Then there exists a unique function $\Theta : (-1,0)\times \mathbb{S}^2 \to \mathbb{R}$ solving~\eqref{skngi3ngoi3n} so that $\Theta$ vanishes near $v = -1$ and satisfies $\Theta \in \cap_{j=0}^2 \left(C^j_{v \in (0,1)}\mathring{H}^{N_1+1-j}\left(\mathbb{S}^2\right) \cap H^j_{v,{\rm loc}}\mathring{H}^{N_1+2-j}\left(\mathbb{S}^2\right)\right)$.
\end{lemma}
\begin{proof}Given a suitable weak solution, the regularity statement for $\Theta$ will follow immediately from parabolic regularity. Thus we focus on the existence statement.

Let us define the operator $\mathscr{P} \doteq \mathcal{L}_{\partial_v} + \frac{2}{v+1} - 4\left(v+1\right)^{-2}\Omega^2\mathring{\Delta}$. Then, for an arbitrary $0 < c \ll 1$, define $\mathcal{U}_c$ to be the completion of smooth functions $\phi : [-1+c,-c] \times \mathbb{S}^2$ which vanish near $-1+c$ under the norm 
\[\left\vert\left\vert \phi\right\vert\right\vert_{C^0_v\mathring{H}^1\left(\mathbb{S}^2\right)} + \left\vert\left\vert \phi\right\vert\right\vert_{L^2_v\mathring{H}^2\left(\mathbb{S}^2\right)} + \left\vert\left\vert \phi\right\vert\right\vert_{H^1_vL^2\left(\mathbb{S}^2\right)}.\]
In view of the bootstrap assumptions on $\Omega$ and standard parabolic theory we may define 
\[\mathscr{P}^{-1} : L^2\left((-1+c,-c) \times \mathbb{S}^2\right) \to \mathcal{U}_c,\]
(where the norm of $\mathscr{P}^{-1}$ will depend, in principle, on $c$). We can then re-write the equation of interest~\eqref{skngi3ngoi3n} as
\begin{equation}\label{oijioijo32}
\Theta = \mathscr{P}^{-1}H + \mathscr{P}^{-1}\mathscr{K}\Theta,
\end{equation}
where $\mathscr{K}$ is the map
\[\Theta \mapsto 4\mathcal{P}_{\ell = 0}\left(\Omega^2\left(v+1\right)^{-2}\mathring{\Delta}\Theta\right).\]
Since the operator $\mathcal{P}^{-1}\mathscr{K}$ extends to a compact operator on $L^2\left((-1+c,-c) \times \mathbb{S}^2\right)$, by the Fredholm alternative the existence of a solution to~\eqref{oijioijo32} will follow if we show that the operator $\left(1 - \mathscr{P}^{-1}\mathscr{K}\right)$ has trivial kernel. However, this fact follows immediately from a standard parabolic energy estimate. 

We have thus shown that~\eqref{skngi3ngoi3n} has a solution whenever $H \in L^2\left((-1+c,-c) \times \mathbb{S}^2\right)$. Let us now additionally assume that $H$ is smooth, vanishes near $v = -1$, and satisfies $\mathcal{P}_{\ell = 0}H = 0$. Applying $\mathcal{P}_{\ell = 0}$ to the equation yields
\begin{equation}\label{3ojojo1}
 \mathcal{L}_{\partial_v}\mathcal{P}_{\ell = 0}\Theta +\frac{2}{v+1}\mathcal{P}_{\ell = 0}\Theta = 0.
 \end{equation}
As a consequence of our existence argument, we have that $\Theta|_{\{v \leq -1+c\}} = 0$ for some sufficiently small $c > 0$. Combining this with~\eqref{3ojojo1} also implies that $\mathcal{P}_{\ell = 0}\Theta$ vanishes everywhere. \end{proof}

We will present two sets of estimates for this equation. In the following lemma, our estimates will be optimized so as to minimize the number of $\mathcal{L}_{\partial_v}$ derivatives we apply to the right hand side $H$.
\begin{lemma}\label{2kn2k2nii3332kn4}Let $H : (-1,0) \times \mathbb{S}^2 \to \mathbb{R}$ be a function which lies in the closure of smooth functions under the norms determined by   the right hand side of~\eqref{k2knkn2kn} and~\eqref{2lm3mo3}. Then there exists a solution $\Theta$ to~\eqref{skngi3ngoi3n} so that for all $0 \leq J \leq N_1-2$
\begin{align}\label{k2knkn2kn}
&\sum_{j=0}^1\left\vert\left\vert \mathcal{L}_{\partial_v}^j\Theta\right\vert\right\vert_{\mathscr{S}\left(J+1-j,-3/2+5\check{\delta}+j,j\right)}^2+\sum_{j=0}^1\left\vert\left\vert \mathcal{L}^{1+j}_{\partial_v}\Theta\right\vert\right\vert^2_{\mathscr{Q}\left(J-j,-1/2+5\check{\delta}+j,\kappa+j\right)}
\\ \nonumber &\qquad 
+ \sum_{j=0}^1\left\vert\left\vert \mathcal{L}_{\partial_v}^j\Theta\right\vert\right\vert_{\mathscr{Q}\left(J+2-j,-5/2+5\check{\delta}+j,-\kappa+j\right)}^2  \lesssim \sum_{j=0}^1\left\vert\left\vert \mathcal{L}_{\partial_v}^jH\right\vert\right\vert_{\mathscr{Q}\left(J-j,-1/2+5\check{\delta}+j,\kappa+j\right)}^2,
\end{align}
where we may omit the sum in $j$ from each term if desired (and we must omit the sum in $j$ if $J = 0$). 

We also have, for any $\check{s}_1,\check{s}_2 > 0$ which satisfy $\epsilon^{\frac{9}{10}} \ll \check{s}_i \ll 1$, and $0 \leq J \leq N_2-2$:
\begin{align}\label{2lm3mo3}
&\sum_{j=0}^1\left\vert\left\vert \mathcal{L}^{1+j}_{\partial_v}\Theta\right\vert\right\vert_{\check{\mathscr{S}}_{-1/2}^0\left(J-j,j,\check{s}_1+\check{s}_2j+2\kappa+j,\check{s}_2\right)}^2 \lesssim 
\\ \nonumber &\qquad  \sum_{j=0}^1\left\vert\left\vert \mathcal{L}_{\partial_v}^jH\right\vert\right\vert_{\mathscr{Q}\left(J+1-j,-1/2+5\check{\delta}+j,\kappa+j\right)}^2+\sum_{j=0}^1\left\vert\left\vert \mathcal{L}^j_{\partial_v}H\right\vert\right\vert_{\check{\mathscr{S}}_{-1/2}^0\left(J-j,j,\check{s}_1+\check{s}_2j+2\kappa+j,\check{s}_2\right)}^2,
\end{align}
where, as usual, when $J = 0$ we omit the sum in $j$.

Finally, we have the following estimate near $v = -1$ for any $0 \leq J \leq N_1-3$:
\begin{align}\label{o01kj9uh332}
&\left\vert\left\vert \mathcal{L}_{\partial_v}^2\Theta\right\vert\right\vert^2_{\mathscr{S}_{-1}^{-1/2}\left(J-1,3/2+5\check{\delta},0\right)} \lesssim 
\\ \nonumber &\qquad \sum_{j=0}^1\left\vert\left\vert \mathcal{L}_{\partial_v}^jH\right\vert\right\vert^2_{\mathscr{Q}_{-1}^{-1/2}\left(J+1-j,-1/2+5\check{\delta}+j,0\right)}+\sum_{j=0}^1\left\vert\left\vert \mathcal{L}_{\partial_v}^jH\right\vert\right\vert^2_{\mathscr{S}_{-1}^{-1/2}\left(J-1,1/2+5\check{\delta}+j,0\right)}.
\end{align}
\end{lemma}
\begin{proof}

For the existence of $\Theta$ and proof of estimates~\eqref{k2knkn2kn} and~\eqref{2lm3mo3}, we may assume, by an approximation argument, without loss of generality that $H$ is smooth and vanishes near $v = -1$. We may thus apply Lemma~\ref{3ljooi11193} to obtain the existence of a solution $\Theta$. 

We start with~\eqref{k2knkn2kn}. For any multi-index $\alpha$ with $\left|\alpha\right| \leq N_1-2$, we may commute~\eqref{skngi3ngoi3n} with $\mathcal{L}_{Z^{(\alpha)}}$, divide the resulting equation by $\Omega^2$,  and then re-write the resulting equation as
\begin{align}\label{2kn1knk2}
&\Omega^{-2}\mathcal{L}_{\partial_v}\Theta^{(\alpha)} +\frac{2\Omega^{-2}}{v+1}\Theta^{(\alpha)}- 4\left(v+1\right)^{-2}\mathring{\Delta}\Theta^{(\alpha)} = 
\\ \nonumber &\qquad 4\Omega^{-2}\mathcal{P}_{\ell = 0}\left(\Omega^2\left(v+1\right)^{-2}\mathring{\Delta}\Theta^{(\alpha)}\right) +\Omega^{-2}H^{(\alpha)} + \Omega^{-2}\sum_{\left|\alpha_1\right| + \left|\alpha_2\right| = \left|\alpha\right| \atop \left|\alpha_1\right| \neq 0}c_{\alpha_1,\alpha_2}\mathcal{P}_{\ell \geq 1}\left(\left(\Omega^2\right)^{(\alpha_1)}\left(v+1\right)^{-2}\mathring{\Delta}\Theta^{(\alpha_2)}\right),
\end{align} 
for suitable constants $c_{\alpha_1,\alpha_2}$.

We may then establish a standard energy estimate by working inductively in $\left|\alpha\right|$ and multiplying~\eqref{2kn1knk2} by $\left(v+1\right)^{-1+2\check{\delta}}e^{-Cv}\mathcal{L}_{\partial_v}\Theta$ for a sufficiently large constant $C$ and integrating by parts with respect to $dv\, \mathring{\rm dVol}$. Combining also with elliptic estimates along $\mathbb{S}^2$ using directly the equation~\eqref{skngi3ngoi3n} we immediately obtain~\eqref{k2knkn2kn} without the sum in $j$. To obtain the estimate with the sum in $j$, we simply commute~\eqref{skngi3ngoi3n} with $\left(v+1\right)(-v)\mathcal{L}_{\partial_v}$ and then repeat the estimate. The terms generated by the commutator are easily seen to be controlled by the previous estimate. Then, for~\eqref{2lm3mo3}, we simply use the equation directly to estimate $\mathcal{L}_{\partial_v}\Theta$ and use~\eqref{skngi3ngoi3n} (and repeat after commutation with $v\mathcal{L}_{\partial_v}$).

We now turn to establishing~\eqref{o01kj9uh332} and continue to take $H$ to be smooth and to vanish near $v = -1$. Commuting twice with $\left(v+1\right)\mathcal{L}_{\partial_v}$ and carrying out a short computation shows that if we set
\begin{equation}\label{oiioio2}
P \doteq \left(v+1\right)\mathcal{L}_{\partial_v}\left(\left(v+1\right)\mathcal{L}_{\partial_v}\Theta\right) - \left(v+1\right)^2\mathcal{L}_{\partial_v}H,
\end{equation}
we may derive 
\begin{equation}\label{iojio23o}
\mathcal{L}_{\partial_v}P + \frac{2}{v+1}P - 4\Omega^2\left(v+1\right)^{-2}\mathring{\Delta} P = \tilde{H},
\end{equation}
where 
\begin{align}\label{ioioijo3}
&\left\vert\left\vert \tilde{H}\right\vert\right\vert_{\mathscr{Q}_{-1}^{-1/2}\left(J-2,1/2+5\check{\delta},0\right) }^2 \lesssim 
\\ \nonumber &\qquad \sum_{j=0}^1\left\vert\left\vert \mathcal{L}_{\partial_v}^jH\right\vert\right\vert_{\mathscr{Q}_{-1}^{-1/2}\left(J,-1/2+5\check{\delta}+j,0\right)} + \sum_{j=0}^1\left\vert\left\vert \mathcal{L}_{\partial_v}^j\Theta\right\vert\right\vert_{\mathscr{Q}_{-1}^{-1/2}\left(J,-3/2+5\check{\delta}+j,0\right)} + q\left\vert\left\vert \mathcal{L}_{\partial_v}^2\Theta\right\vert\right\vert_{\mathscr{Q}_{-1}^{-1/2}\left(0,3/2+5\check{\delta},0\right)},
\end{align}
where $|q| \ll 1$. In particular, commuting suitably with $\mathcal{L}_{Z^{(\alpha)}}$ for $\left|\alpha\right| \leq J-2$, contracting~\eqref{iojio23o} with $\left(v+1\right)^{1+10\check{\delta}}e^{-Cv}\mathcal{L}_{\partial_v}P^{(\alpha)}$ for a sufficiently large constant $C$, and integrating by parts leads to
\begin{equation}\label{3oj1po2040010100003}
\left\vert\left\vert P\right\vert\right\vert_{\mathscr{Q}_{-1}^{-1/2}\left(J,-3/2+5\check{\delta},0\right)}^2 + \left\vert\left\vert P\right\vert\right\vert_{\mathscr{S}_{-1}^{-1/2}\left(J-1,-1/2+5\check{\delta},0\right)}^2 \lesssim \left\vert\left\vert \tilde{H}\right\vert\right\vert_{\mathscr{Q}_{-1}^{-1/2}\left(J-2,1/2+5\check{\delta},0\right) }^2.
\end{equation}
By a density argument, we now may allow $H$ to lie in the closure of smooth functions under the norm determined by the first term on the right hand side of~\eqref{ioioijo3} and the right hand sides of the estimates~\eqref{k2knkn2kn} and~\eqref{2lm3mo3}. We will still have that~\eqref{3oj1po2040010100003} holds.  The estimate~\eqref{o01kj9uh332} now follows from~\eqref{oiioio2},~\eqref{ioioijo3},~\eqref{3oj1po2040010100003}, and the previous estimate~\eqref{k2knkn2kn}.

\end{proof}

In this next lemma we again consider the equation~\eqref{skngi3ngoi3n}. This time we establish estimates which are optimized to minimize the number of angular derivatives applied to $H$ and which require weaker weights near $v = -1$. The price we pay is that we will need more $\mathcal{L}_{\partial_v}$ derivatives of $H$ on the right hand side of our estimates. 
\begin{lemma}\label{3m3om2}Let $\check{H} : (-1,0) \times \mathbb{S}^2 \to \mathbb{R}$ be a function which satisfies $\mathcal{P}_{\ell = 0}\check{H} = 0$ and is in the closure of smooth functions under the norms determined by  the right hand side of~\eqref{3o3omoo21231i12i},~\eqref{3o3omoo21231i12i123},~\eqref{3o3oomo2}, and~\eqref{2om3omomo2}.

Then there exists a solution $\Theta$ to~\eqref{skngi3ngoi3n}, with $H \doteq \mathcal{P}_{\ell \geq 1}\left(\Omega^2\check{H}\right)$, so that for all $0 \leq J \leq N_1-2$:
\begin{equation}\label{3o3omoo21231i12i}
\left\vert\left\vert \Theta \right\vert\right\vert_{\mathscr{S}\left(J,1/2+\check{\delta},0\right)}^2 + \left\vert\left\vert \Theta \right\vert\right\vert_{\mathscr{Q}\left(J+1,-1/2+\check{\delta},-\kappa\right)}^2 \lesssim \left\vert\left\vert \mathring{\Delta}^{-1}H\right\vert\right\vert_{\mathscr{Q}\left(J+1,3/2+\check{\delta},\kappa\right)}^2.
\end{equation}
Moreover for all $0 \leq J \leq N_1-3$:
\begin{align}\label{3o3omoo21231i12i123}
&\sum_{j=0}^1\left\vert\left\vert \mathcal{L}_{\partial_v}^{1+j}\Theta \right\vert\right\vert_{\mathscr{Q}\left(J+1-j,1/2 + \check{\delta}+j,\kappa+j\right)}^2 + \left\vert\left\vert \Theta\right\vert\right\vert_{\mathscr{Q}_{-1/2}^0\left(J+2,0,-1/2+\check{\delta}\right)} \lesssim 
\\ \nonumber &\qquad  \sum_{j=0}^1\left\vert\left\vert \mathring{\Delta}^{-1}\mathcal{L}_{\partial_v}^{1+j}\check{H}\right\vert\right\vert_{\mathscr{Q}\left(J+1-j,5/2+\check{\delta}+j,\kappa+j\right)}^2+ \left\vert\left\vert \mathring{\Delta}^{-1}H\right\vert\right\vert_{\mathscr{Q}\left(J+2,3/2+\check{\delta},\kappa\right)}^2+ \left\vert\left\vert \check{H}\right\vert\right\vert_{\mathscr{Q}_{-1/2}^0\left(J,0,-1/2+\check{\delta}\right)}^2,
\end{align}
where we omit the sum in $j$ if $J = 0$.

It is convenient to split up our main $\sup_v$ estimates into the ones which concern the reigon $\{v \leq -1/2\}$ and the ones which concern $\{v \geq -1/2\}$. We have, for all $0 \leq J \leq N_1-3$:
\begin{align}\label{3o3oomo2}
&\sum_{j=0}^2\left\vert\left\vert \mathcal{L}_{\partial_v}^j\Theta\right\vert\right\vert_{\mathscr{S}_{-1}^{-1/2}\left(J+2-j,\check{\delta}+j,0\right)}^2 \lesssim \sum_{j=0}^2\left\vert\left\vert \mathcal{L}_{\partial_v}^jH\right\vert\right\vert_{\mathscr{S}_{-1}^{-1/2}\left(J-j,2+\check{\delta}+j,0\right)}^2 +\mathcal{E}
\\ \nonumber &\qquad +\sum_{j=0}^2\left\vert\left\vert \mathring{\Delta}^{-1}\mathcal{L}_{\partial_v}^{1+j}H\right\vert\right\vert_{\mathscr{Q}_{-1}^{-1/2}\left(J+1-j,5/2+\check{\delta}+j,0\right)}^2.
\end{align}
where $\mathcal{E}$ denotes the sum of the right hand sides of~\eqref{3o3omoo21231i12i} and~\eqref{3o3omoo21231i12i123} with $J = N_1-2$ and $J = N_1-3$ respectively.  For the region $v \geq -1/2$,  for any $\check{s}_1,\check{s}_2 > 0$ which satisfy $\epsilon^{\frac{9}{10}} \ll \check{s}_i \ll 1$ and $0 \leq J \leq N_2-1$, we have
\begin{align}\label{2om3omomo2}
&\sum_{j=0}^1\left\vert\left\vert \mathcal{L}^{1+j}_{\partial_v}\Theta\right\vert\right\vert_{\check{\mathscr{S}}_{-1/2}^0\left(J-j,j,\check{s}_1+\check{s}_2j+2\kappa+j,\check{s}_2\right)}^2 \lesssim \sum_{j=0}^1\left\vert\left\vert \mathcal{L}^j_{\partial_v}\check{H}\right\vert\right\vert_{\check{\mathscr{S}}_{-1/2}^0\left(J-j,j,\check{s}_1+\check{s}_2j+2\kappa+j,\check{s}_2\right)}^2 + \mathcal{E}.
\end{align}

\end{lemma}
\begin{proof}
As in the proof of Lemma~\ref{2kn2k2nii3332kn4}, we may assume without loss of generality that for the proof of the estimates~\eqref{3o3omoo21231i12i} and \eqref{3o3omoo21231i12i123}, that $H$ vanishes near $v = -1$ and thus may apply Lemma~\ref{3ljooi11193} to obtain the existence of $\Theta$. We may then carry out a standard energy estimate for $\Theta$ by contracting the equation with $\left(v+1\right)^{1+2\check{\delta}}\Theta$ and integrating by parts with respect to $dv\, \mathring{\rm dVol}$. We immediately obtain:
\begin{equation}\label{3o3omoo2}
\left\vert\left\vert \Theta \right\vert\right\vert_{\mathscr{S}\left(0,1/2+\check{\delta},0\right)}^2 + \left\vert\left\vert \Theta\right\vert\right\vert_{\mathscr{Q}\left(1,-1/2+\check{\delta},-\kappa\right)}^2 \lesssim \left\vert\left\vert \mathring{\Delta}^{-1}H\right\vert\right\vert_{\mathscr{Q}\left(1,3/2+\check{\delta},\kappa\right)}^2.
\end{equation}
It is then straightforward to commute with $\mathcal{L}_{Z^{(\alpha)}}$ and repeat \emph{mutatis mutandis} to also obtain~\eqref{3o3omoo21231i12i}. Next, we may commute with $\left(v+1\right)^j\mathcal{L}_{\partial_v}^j$ for $j \in \{1,2\}$ and repeat the above energy estimate except with $v$ restricted to $v \in [-1,-1/2]$. We obtain for all $0 \leq J \leq N_1-3$:
\begin{align}\label{eknkwnke}
&\sum_{j=0}^2\left\vert\left\vert \mathcal{L}_{\partial_v}^{1+j}\Theta\right\vert\right\vert_{\mathscr{S}_{-1}^{-1/2}\left(J-j,3/2+\check{\delta}+j,0\right)}^2+ \sum_{j=0}^2\left\vert\left\vert \mathcal{L}_{\partial_v}^{1+j}\Theta \right\vert\right\vert_{\mathscr{Q}_{-1}^{-1/2}\left(J+1-j,1/2+\check{\delta}+j,0\right)}^2
\\ \nonumber &\qquad  \lesssim \sum_{j=0}^2\left\vert\left\vert \mathring{\Delta}^{-1}\mathcal{L}_{\partial_v}^{1+j}H\right\vert\right\vert_{\mathscr{Q}_{-1}^{-1/2}\left(J+1-j,5/2+\check{\delta}+j,0\right)}^2,
\end{align}
where we omit the sum in $j$ if $J = 0$ and run the sum up to $j = 1$ if $J = 1$. It is also clear that we may also establish a version of~\eqref{eknkwnke} where each sum in $j$ only goes up to $1$.

For the analogous estimate in the region $v \in (-1/2,0)$ we cannot proceed in such a straightforward fashion because commutation with $\mathcal{L}_{\partial_v}$ will produce a term with $\mathcal{L}_{\partial_v}\log\Omega \sim (-v)^{-1}$. Instead, we re-write the equation~\eqref{skngi3ngoi3n} as 
\begin{align}\label{2o3ijij3}
&\Omega^{-2}\mathcal{L}_{\partial_v}\Theta +\frac{2\Omega^{-2}}{v+1}\Theta-4\Omega^{-2}\left[\mathcal{P}_{\ell \geq 1},\Omega^2\right]\left(\left(v+1\right)^{-2}\mathring{\Delta}\Theta\right)- 4\left(v+1\right)^{-2}\mathring{\Delta}\Theta  = 
\\ \nonumber &\qquad \mathcal{P}_{\ell \geq 1}\check{H} + \Omega^{-2}\left[\mathcal{P}_{\ell \geq 1},\Omega^2\right]\check{H}.
\end{align} 
Then we apply $\mathcal{L}_{\partial_v}$ and obtain
\begin{align}\label{2nj4jn2j}
&\mathcal{L}_{\partial_v}\left(\Omega^{-2}\mathcal{G}\right) - 4(v+1)^{-2}\mathring{\Delta}\mathcal{L}_{\partial_v}\Theta  =  -8(v+1)^{-3}\mathring{\Delta}\Theta +\mathcal{P}_{\ell \geq 1}\mathcal{L}_{\partial_v}\check{H},
\end{align}
where
\[\mathcal{G} \doteq \mathcal{L}_{\partial_v}\Theta +\frac{2}{v+1}\Theta-4\left[\mathcal{P}_{\ell \geq 1},\Omega^2\right]\left(\left(v+1\right)^{-2}\mathring{\Delta}\Theta\right)- \left[\mathcal{P}_{\ell \geq 1},\Omega^2\right]\check{H}.\]
We now multiply~\eqref{2nj4jn2j} with $\xi(v+1/4)\Omega^{-2}\mathcal{G}$, integrate by parts, and use Hardy inequalities.\footnote{More specifically, we use the Hardy inequality $\int_{-3/4}^0(-v)^{-p}f^2\, dv \lesssim  \int_{-3/4}^{-1/4}f^2\, dv + \int_{-3/4}^0(-v)^{2-p}\left(\mathcal{L}_{\partial_v}f\right)^2\, dv$, which holds whenever $p < 1$.} We obtain the estimate
\begin{align}\label{2moo20992j}
&\left\vert\left\vert \mathcal{L}_{\partial_v}\Theta\right\vert\right\vert^2_{\mathscr{Q}_{-1/2}^0\left(1,0,\kappa\right)} + \left\vert\left\vert \mathcal{G}\right\vert\right\vert_{\mathscr{S}_{-1/2}^0\left(0,0,2\kappa\right)}^2 \lesssim  \\ \nonumber &\qquad \left\vert\left\vert \Theta\right\vert\right\vert_{\mathscr{Q}_{-3/4}^0\left(1,0,-\kappa\right)}^2 + \left\vert\left\vert \mathcal{L}_{\partial_v}\Theta \right\vert\right\vert_{\mathscr{Q}_{-3/4}^{-1/2}\left(1,0,0\right)}^2+\left\vert\left\vert \Theta \right\vert\right\vert_{\mathscr{S}\left(0,0,0\right)}^2 +\left\vert\left\vert \mathring{\Delta}^{-1}\mathcal{L}_{\partial_v}\check{H}\right\vert\right\vert_{\mathscr{Q}_{-3/4}^0\left(1,0,\kappa\right)}^2 + \left\vert\left\vert \check{H}\right\vert\right\vert_{\mathscr{Q}_{-3/4}^{-1/4}\left(0,0,0\right)}^2.
\end{align}
Next, we can revisit the equation~\eqref{2o3ijij3}, use the estimate on $\mathcal{G}$ from~\eqref{2moo20992j} and elliptic estimates to obtain
\begin{align}\label{3jomomo2}
&\left\vert\left\vert \Theta\right\vert\right\vert^2_{\mathscr{Q}_{-1/2}^0\left(2,0,-1/2+\check{\delta}\right)} \lesssim \left\vert\left\vert \mathcal{G}\right\vert\right\vert^2_{\mathscr{S}_{-1/2}^0\left(0,0,2\kappa\right)} + \left\vert\left\vert \check{H}\right\vert\right\vert^2_{\mathscr{Q}_{-1/2}^0\left(0,0,-1/2+\check{\delta}\right)} + \mathscr{P},
\end{align}
where $\mathscr{P}$ denotes the right hand of~\eqref{2moo20992j}. In particular, it is clear that by commuting~\eqref{2nj4jn2j} with $\left(v\mathcal{L}_{\partial_v}\right)^j\mathcal{L}_{Z^{(\alpha)}}$ for $j \in \{0,1\}$ and suitable $\left|\alpha\right| \leq N_1-2-j$, re-running the estimate which lead to~\eqref{2moo20992j} and~\eqref{3jomomo2}, and then combining with~\eqref{eknkwnke} (the version where the sum only goes up to $j=1$) and~\eqref{3o3omoo2} leads to~\eqref{3o3omoo21231i12i123}.  By a density argument, we may now assume that $H$ lies in the closure of smooth functions under the norms on the right hand sides of~\eqref{3o3omoo21231i12i}-\eqref{2om3omomo2} and that the estimates we have established so far in the proof continue to hold.

We now turn to the $\sup_v$ estimates, first focusing in the region $v \in [-1,-1/2]$. It is here that we use the version of~\eqref{eknkwnke} where the sum goes up to $j = 2$. We re-write~\eqref{skngi3ngoi3n} as 
\begin{equation}\label{qkn2knk}
 4\mathcal{P}_{\ell \geq 1}\left(\Omega^2\left(v+1\right)^{-2}\mathring{\Delta}\Theta\right) = -H + \left(\mathcal{L}_{\partial_v}\Theta +\frac{2}{v+1}\Theta\right),
\end{equation} 
and use elliptic estimates along $\mathbb{S}^2$ to estimate $\Theta$ in terms of the right hand side. After suitable commutation and combination with~\eqref{eknkwnke}, this leads immediately to~\eqref{3o3oomo2}. Finally,~\eqref{2om3omomo2} is a straightforward consequence of using the equation~\eqref{2o3ijij3} directly to estimate $\mathcal{L}_{\partial_v}\Theta$ and derivatives thereof in a suitably $v$-weighted $L_v^{\infty}L^2\left(\mathbb{S}^2\right)$ and using the fundamental theorem of calculus to estimate angular derivatives of $\Theta$ and derivatives thereof with the norms on the left hand of~\eqref{3o3omoo21231i12i} and~\eqref{3o3omoo21231i12i123}.

\end{proof}

We close the section by defining some norms which will allow us to package the above two lemmas in a notationally concise fashion.
\begin{definition}\label{2kn3nmo2}Let $H_1$ and $H_2$ be smooth functions defined on $(-1,0) \times \mathbb{S}^2$ which satisfy $\mathcal{P}_{\ell = 0}H_i = 0$ and $s_1$, $s_2$ satisfy $\epsilon^{\frac{9}{10}} \ll \check{s}_i \ll 1$. We then define
\begin{align*}
&\left\vert\left\vert \left(H_1,H_2\right)\right\vert\right\vert^2_{\mathscr{PR}\left(\Omega,\kappa,\check{s}_1,\check{s}_2\right)} \doteq \sum_{j=0}^1\left\vert\left\vert \mathcal{L}_{\partial_v}^jH_1\right\vert\right\vert_{\mathscr{Q}\left(N_1-2-j,-1/2+5\check{\delta}+j,\kappa+j\right)}^2
\\ \nonumber &\qquad + \sum_{j=0}^1\left\vert\left\vert \mathcal{L}_{\partial_v}^jH_1\right\vert\right\vert^2_{\mathscr{S}_{-1}^{-1/2}\left(N_1-3,1/2+5\check{\delta}+j,0\right)}  +\sum_{j=0}^1\left\vert\left\vert \mathcal{L}^j_{\partial_v}H_1\right\vert\right\vert_{\check{\mathscr{S}}_{-1/2}^0\left(N_2-2-j,0,\check{s}_1+\check{s}_2j+2\kappa+j,\check{s}_2\right)}^2
\\ \nonumber &\qquad +\left\vert\left\vert H_2\right\vert\right\vert_{\mathscr{Q}\left(N_1-3,3/2+\check{\delta},-1/2+\check{\delta}\right)}^2 + \sum_{j=0}^1\left\vert\left\vert\mathcal{L}_{\partial_v}^{1+j}H_2\right\vert\right\vert_{\mathscr{Q}\left(N_1-4-j,5/2+\check{\delta}+j,\kappa+j\right)}^2
\\ \nonumber &\qquad + \sum_{j=0}^2\left\vert\left\vert\mathcal{L}_{\partial_v}^{1+j}H_2\right\vert\right\vert_{\mathscr{Q}_{-1}^{-1/2}\left(N_1-4-j,5/2+\check{\delta}+j,0\right)}^2 +\sum_{j=0}^2\left\vert\left\vert \mathcal{L}_{\partial_v}^jH_2\right\vert\right\vert_{\mathscr{S}_{-1}^{-1/2}\left(N_1-3-j,2+\check{\delta}+j,0\right)}^2
\\ \nonumber &\qquad +\sum_{j=0}^1\left\vert\left\vert \mathcal{L}^j_{\partial_v}H_2\right\vert\right\vert_{\check{\mathscr{S}}_{-1/2}^0\left(N_2-1-j,j,\check{s}_2+\check{s}_1j+2\kappa+j, \check{s}_1\right)}^2.
\end{align*}

Let $\Theta$ be a smooth function defined on $(-1,0)\times \mathbb{S}^2$ which satisfies $\mathcal{P}_{\ell = 0}\Theta = 0$. We then define
\begin{align*}
&\left\vert\left\vert \Theta\right\vert\right\vert^2_{\mathscr{P}\mathscr{L}\left(\kappa,\check{s}_1,\check{s}_2\right)} \doteq  \left\vert\left\vert \Theta\right\vert\right\vert^2_{\mathscr{Q}\left(N_1-1,-1/2+\check{\delta},-1/2+\check{\delta}\right)} + \sum_{j=0}^1\left\vert\left\vert \mathcal{L}^{1+j}_{\partial_v}\Theta\right\vert\right\vert^2_{\mathscr{Q}\left(N_1-2-j,1/2+\check{\delta}+j,\kappa+j\right)}
\\ \nonumber &\qquad+\sum_{j=0}^1\left\vert\left\vert \mathcal{L}^{1+j}_{\partial_v}\Theta\right\vert\right\vert^2_{\mathscr{Q}_{-1/2}^0\left(N_1-3-j,0,-\sqrt{\check{p}}+j\right)}+\sum_{j=0}^2\left\vert\left\vert \mathcal{L}_{\partial_v}^j\Theta\right\vert\right\vert^2_{\mathscr{S}_{-1}^{-1/2}\left(N_1-2,\check{\delta}+j,0\right)} 
\\ \nonumber &\qquad + \sum_{j=0}^1\left\vert\left\vert \mathcal{L}_{\partial_v}^{1+j}\Theta\right\vert\right\vert^2_{\check{\mathscr{S}}_{-1/2}^0\left(N_2-2-j,0,\check{s}_1+\check{s}_2j+2\kappa+j,\check{s}_2\right)}.
\end{align*}

\end{definition}
\begin{remark}\label{3o2ojio42}It will be useful later to note the following consequence of Lemma~\ref{3m2omo4}:
\[\sum_{j=0}^1\left\vert\left\vert \left(\mathcal{L}_{\partial_v}^jH_1,(-v)^{-2\kappa}\mathcal{L}_{\partial_v}^jH_2\right)\right\vert\right\vert^2_{\mathscr{Q}_{-1/2}^0\left(N_1-3-j,0,-\sqrt{\check{p}}+j\right)}\lesssim \left\vert\left\vert \left(H_1,H_2\right)\right\vert\right\vert^2_{\mathscr{PR}\left(\Omega,\kappa,\check{s}_1,\check{s}_2\right)}.\]
\end{remark}

We may now combine Lemmas~\ref{2kn2k2nii3332kn4} and~\ref{3m3om2} into the following:
\begin{proposition}\label{2km2om1o}Let $H_1$ and $H_2$ lie in the closure of smooth functions satisfying $\mathcal{P}_{\ell = 0}\left(H_1,H_2\right) = 0$ under the norm $\left\vert\left\vert \cdot\right\vert\right\vert_{\mathscr{P}\mathscr{R}}$. Then there exists a solution $\Theta$ to the equation
\begin{equation}\label{2o2jo4e89u49u8u89u89}
\mathcal{L}_{\partial_v}\Theta +\frac{2}{v+1}\Theta- 4\mathcal{P}_{\ell \geq 1}\left(\Omega^2\left(v+1\right)^{-2}\mathring{\Delta}\Theta\right) = H_1 + \mathcal{P}_{\ell \geq 1}\left(\Omega^2H_2\right),
\end{equation}
such that moreover $\Theta$ satisfies $\mathcal{P}_{\ell = 0}\Theta = 0$, $\Theta$ lies in the closure of smooth functions under the norm $\left\vert\left\vert \cdot\right\vert\right\vert_{\mathscr{P}\mathscr{L}}$, and $\Theta$ satisfies the estimate
\[\left\vert\left\vert \Theta\right\vert\right\vert_{\mathscr{P}\mathscr{L}\left(\kappa,\check{s}_1,\check{s}_2\right)} \lesssim \left\vert\left\vert \left(H_1,H_2\right)\right\vert\right\vert_{\mathscr{P}\mathscr{R}\left(\Omega,\kappa,\check{s}_1,\check{s}_2\right)}.\] 
Here we must have $\check{s}_i \gg \epsilon^{\frac{9}{10}}$. 
\end{proposition}  
\begin{proof}Since the equation~\eqref{2o2jo4e89u49u8u89u89} is linear, we may consider separately the cases when $H_1 = 0$ or $H_2 = 0$, and may thus directly apply either Lemma~\ref{2kn2k2nii3332kn4} or~\ref{3m3om2}. It is straightforward to check that the resulting estimate combined with the interpolation Lemma~\ref{3m2omo4} implies the desired estimate for $\Theta$. 
\end{proof}

In the context of the constraint propagation argument (see Section~\ref{om1o1mo2}) the following will be useful.
\begin{lemma}\label{2oj3omo1mo323}Suppose we have a function $X : (-1,0) \times \mathbb{S}^2 \to \mathbb{S}^2$ so that $\mathcal{L}_{\partial_v}^2X$, $\mathcal{L}_{\partial_v}X$, and $X$ all lie in $\mathring{H}^{10}\left(\mathbb{S}^2\right)$ for each $v$, we have $\left(1-\mathcal{P}_{\ell \geq 1}\right)X = 0$, we have
\begin{equation}\label{1om2om1o4}
\left((-v)\mathcal{L}_{\partial_v}-\mathcal{P}_{\ell \geq 1}\mathcal{L}_b\right)\left(\mathcal{L}_{\partial_v}X - \mathcal{P}_{\ell \geq 1}\left(\Omega^2\left(v+1\right)^{-2}\mathring{\Delta}X\right)\right) = H,
\end{equation}
and the following boundary conditions at $\{v = -1\}$ and $\{ v= 0\}$ are satisfied:
\begin{align}\label{omom2o4}
&\limsup_{v\to -1}\sup_{ \mathbb{S}_{-1,v}^2}\sum_{\left|\alpha\right| \leq 10}\left[\left(v+1\right)^{-2}\left|X^{(\alpha)}\right| + \left|\mathcal{L}_{\partial_v}X^{(\alpha)}\right|\right] = 0,
\\ \nonumber &\qquad \qquad \qquad \limsup_{v\to 0}\sup_{\theta^A \in \mathbb{S}^2}\sum_{\left|\alpha\right| \leq 10}\left|(-v)\mathcal{L}_{\partial_v}X^{(\alpha)}\right| = 0.
\end{align}
Then, assuming that
\begin{equation}\label{1mo2mo4o2}
\sup_{(v,\theta^A) \in (-1,-1/2) \times \mathbb{S}^2}\sum_{\left|\alpha\right| \leq 10}\left[\left(v+1\right)^{-1+100\check{\delta}}\left|H^{(\alpha)}\right| +\left(v+1\right)^{100\check{\delta}}\left|\mathcal{L}_{\partial_v}H^{(\alpha)}\right|\right]< \infty,
\end{equation}
and that $H$ lies in the closure of smooth functions under the norm on the right hand side of~\eqref{kmmo13}, we have
\begin{align}\label{kmmo13}
&\left\vert\left\vert X  \right\vert\right\vert_{\mathscr{Q}_{-1}^0\left(4,-4,-\kappa\right)} +\left\vert\left\vert X\right\vert\right\vert_{\mathscr{Q}_{-1/2}^0\left(3,0,-1/2+\check{\delta}\right)} 
 +\left\vert\left\vert \mathcal{L}_{\partial_v}X\right\vert\right\vert_{\mathscr{Q}_{-1}^0\left(3,-3,1/2+\sqrt{\check{\delta}}\right)}
 \\ \nonumber &\qquad + \left\vert\left\vert \left((-v)\mathcal{L}_{\partial_v}-\mathcal{L}_b\right)X\right\vert\right\vert_{\mathscr{Q}_{-1}^0\left(4,-3,-\kappa\right)} + \left\vert\left\vert \left((-v)\mathcal{L}_{\partial_v}-\mathcal{L}_b\right)X\right\vert\right\vert_{\mathscr{Q}_{-1/2}^0\left(3,0,-1/2+\check{\delta}\right)} +  \left\vert\left\vert \mathcal{L}_{\partial_v}X\right\vert\right\vert_{\mathscr{Q}_{-1}^0\left(2,-3,\kappa\right)}
 \\ \nonumber &\qquad + \left\vert\left\vert \mathcal{L}_{\partial_v} \left((-v)\mathcal{L}_{\partial_v}-\mathcal{L}_b\right)X\right\vert\right\vert_{\mathscr{Q}_{-1}^0\left(2,-2,\kappa\right)}+\left\vert\left\vert \left(1,\mathcal{L}_b\right)X|_{v=0}\right\vert\right\vert_{\mathring{H}^3\left(\mathbb{S}^2\right)} \lesssim 
 \\ \nonumber &\qquad \left\vert\left\vert H\right\vert\right\vert_{\mathscr{Q}\left(2,-1,-1/2+10\sqrt{\check{\delta}}\right)}+\left\vert\left\vert \left(1,\mathcal{L}_b\right)H\right\vert\right\vert_{\mathscr{Q}\left(1,-1,-1/2+10\check{\delta}\right)}.
 \end{align}
\end{lemma}
\begin{proof}
Let's set 
\begin{equation}\label{m2omo12}
 \mathcal{L}_{\partial_v}X - \mathcal{P}_{\ell \geq 1}\left(\Omega^2\left(v+1\right)^{-2}\mathring{\Delta}X\right) \doteq \mathcal{Z} .
\end{equation}
 In view of the boundary condition~\eqref{omom2o4}, we may apply Lemma~\ref{l2l2degtranstrans} (along with a commutation by $\mathcal{L}_b$) and obtain the following estimates
 \begin{align}\label{2mo2omo2}
\left\vert\left\vert \mathcal{Z}\right\vert\right\vert_{\mathscr{Q}\left(2,-2,-1/2+10\sqrt{\check{\delta}}\right)} &\lesssim \left\vert\left\vert H\right\vert\right\vert_{\mathscr{Q}\left(2,-1,-1/2+10\sqrt{\check{\delta}}\right)},
\\ \nonumber \left\vert\left\vert \left(1,\mathcal{L}_b\right)\mathcal{Z}\right\vert\right\vert_{\mathscr{Q}\left(1,-2,-1/2+10\check{\delta}\right)} &\lesssim \left\vert\left\vert \left(1,\mathcal{L}_b\right)H\right\vert\right\vert_{\mathscr{Q}\left(1,-1,-1/2+10\check{\delta}\right)}.
\end{align}
Combining also directly with the equation~\eqref{1om2om1o4} leads to the following two estimates
\begin{align}\label{1om2oj39r82hin}
&\left\vert\left\vert \mathcal{L}_{\partial_v}\mathcal{Z}\right\vert\right\vert_{\mathscr{Q}\left(1,-1,1/2+\sqrt{\check{\delta}}+2\kappa\right)}+\left\vert\left\vert \mathcal{Z}\right\vert\right\vert_{\mathscr{Q}\left(2,-2,-1/2+10\sqrt{\check{\delta}}\right)} +  \left\vert\left\vert \left(1,\mathcal{L}_b\right)\mathcal{Z}\right\vert\right\vert_{\mathscr{Q}\left(1,-2,-1/2+10\check{\delta}\right)} \lesssim 
\\ \nonumber &\qquad \left\vert\left\vert H\right\vert\right\vert_{\mathscr{Q}\left(2,-1,-1/2+10\sqrt{\check{\delta}}\right)}+ \left\vert\left\vert \left(1,\mathcal{L}_b\right)H\right\vert\right\vert_{\mathscr{Q}\left(1,-1,-1/2+10\check{\delta}\right)}.
\end{align}
With this estimate for $\mathcal{Z}$ established, we would like to carry out energy estimates using the equation~\eqref{m2omo12}. However, first it will be necessary to obtain an improvement of the boundary condition~\eqref{omom2o4}.

In order to establish the improvement of~\eqref{omom2o4} we start by noting that a consequence of Lemma~\ref{linftofkwp3}, the equation~\eqref{1om2om1o4}, and the assumption~\eqref{1mo2mo4o2}, we have that
\begin{equation}\label{betterforzzzz}
\left\vert\left\vert \mathcal{Z}\right\vert\right\vert_{\mathscr{S}_{-1}^{-1/2}\left(10, -2+100\check{\delta},0\right)}  < \infty.
\end{equation}  
Now we contract~\eqref{m2omo12} with $X$ and integrate by parts over $\mathbb{S}^2$. In view of the fact that $\mathcal{P}_{\ell = 0}X = 0$ and the bound~\eqref{betterforzzzz}, we obtain the following inequality for $\mathscr{A} \doteq \int_{\mathbb{S}^2}\left|X\right|^2\mathring{\rm dVol}$:
\begin{equation}\label{pokpokpo21}
\frac{d}{dv}\mathscr{A} + \left(v+1\right)^{-2}\mathscr{A} \lesssim \mathscr{A}^{1/2}\left(v+1\right)^{2-100\check{\delta}}
\end{equation}
In view of the fact that
\[\frac{d}{dv}\mathscr{A} + \left(v+1\right)^{-2}\mathscr{A} = \exp\left(\left(v+1\right)^{-1}\right)\frac{d}{dv}\left(\exp\left(-\left(v+1\right)^{-1}\right)\mathscr{A}\right),\]
and that for any $s \in \mathbb{R}$:
\[\left|\exp\left(\left(v+1\right)^{-1}\right)\int_{-1}^v\left(x+1\right)^s\exp\left(-\left(x+1\right)^{-1}\right)\, dx\right| \lesssim \left(v+1\right)^{s+2},\]
it is a straightforward consequence of~\eqref{pokpokpo21} and~\eqref{omom2o4} that 
\begin{equation*}
\sup_{v \in (-1,-1/2)}\left(\mathscr{A}^{1/2}\left(v+1\right)^{-4+200\check{\delta}}\right) < \infty. 
\end{equation*}
It is then clear that we may repeat this procedure after commutation with suitable $\mathcal{L}_{Z^{(\alpha)}}$ and obtain that 
\begin{equation}\label{oijijojiioj2}
\left\vert\left\vert X\right\vert\right\vert_{\mathscr{S}_{-1}^{-1/2}\left(10,-4+200\check{\delta},0\right)} < \infty.
\end{equation}
Finally, we may also commute with $\mathcal{L}_{\partial_v}$ and repeat the same procedure \emph{mutatis mutandis} to obtain 
\begin{equation}\label{2lkj3ljl1}
\left\vert\left\vert \mathcal{L}_{\partial_v}X\right\vert\right\vert_{\mathscr{S}_{-1}^{-1/2}\left(8,-3+300\check{\delta},0\right)} < \infty.
\end{equation}

Having established~\eqref{oijijojiioj2}, we may now carry out a standard energy estimates with~\eqref{m2omo12} by contracting with $e^{-Cv}\left(v+1\right)^{-6}X$ (for some $C \gg 1$) and integrating by parts (we use~\eqref{oijijojiioj2} to see that the boundary term at $v = -1$ vanishes). After commuting with $\mathcal{L}_{Z^{(\alpha)}}$ and repeating we obtain the following estimate (which is similar to the energy estimates carried out in Lemma~\ref{3m3om2})
\begin{align}\label{1momo180223}
\left\vert\left\vert X|_{v=0}\right\vert\right\vert_{\mathring{H}^3\left(\mathbb{S}^2\right)}+\left\vert\left\vert \mathfrak{X}\right\vert\right\vert_{\mathscr{S}\left(3,-3,0\right)} + \left\vert\left\vert \mathfrak{X}\right\vert\right\vert_{\mathscr{Q}\left(4,-4,-\kappa\right)} &\lesssim \left\vert\left\vert \mathcal{Z}\right\vert\right\vert_{\mathscr{Q}\left(2,-2,-\kappa\right)} 
\\ \nonumber &\lesssim \left\vert\left\vert H\right\vert\right\vert_{\mathscr{Q}\left(2,-1,-1/2+10\sqrt{\check{\delta}}\right)}+ \left\vert\left\vert \left(1,\mathcal{L}_b\right)H\right\vert\right\vert_{\mathscr{Q}\left(1,-1,-1/2+10\check{\delta}\right)}.
\end{align}
With~\eqref{1momo180223} established, we may use the equation~\eqref{m2omo12} directly to then obtain the following estimate for $\mathcal{L}_{\partial_v}X$:
\[\left\vert\left\vert \mathcal{L}_{\partial_v}X\right\vert\right\vert_{\mathscr{Q}\left(2,-2,\kappa\right)} \lesssim \left\vert\left\vert H\right\vert\right\vert_{\mathscr{Q}\left(2,-1,-1/2+10\sqrt{\check{\delta}}\right)}+ \left\vert\left\vert \left(1,\mathcal{L}_b\right)H\right\vert\right\vert_{\mathscr{Q}\left(1,-1,-1/2+10\check{\delta}\right)}.\]

Next, we commute~\eqref{m2omo12} with $\mathcal{L}_{\partial_v}$ to obtain  
\begin{equation}\label{2k3ml2}
 \mathcal{L}_{\partial_v}\mathcal{L}_{\partial_v}X - \mathcal{P}_{\ell \geq 1}\left(\Omega^2\left(v+1\right)^{-2}\mathring{\Delta}\mathcal{L}_{\partial_v}X\right) = \mathcal{L}_{\partial_v}\mathcal{Z} + \left[\mathcal{L}_{\partial_v},\mathcal{P}_{\ell \geq 1}\left(\Omega^2\left(v+1\right)^{-2}\mathring{\Delta}\right]X\right) .
\end{equation}
We contract this with $e^{-Cv}(-v)^{1+2\sqrt{\check{\delta}}+2\kappa}\left(v+1\right)^{-4}\mathcal{L}_{\partial_v}$, integrate by parts, use~\eqref{2lkj3ljl1} to argue that the boundary term at $v = -1$ vanishes, carry out additional commutations with $\mathcal{L}_{Z^{(\alpha)}}$, and combine with~\eqref{1om2oj39r82hin} and~\eqref{1momo180223} to obtain
\begin{align}\label{2o23oo2p}
\left\vert\left\vert \mathcal{L}_{\partial_v}X\right\vert\right\vert_{\mathscr{Q}\left(3,-3,1/2+\sqrt{\check{\delta}}\right)} &\lesssim \left\vert\left\vert\mathcal{L}_{\partial_v}\mathcal{Z}\right\vert\right\vert_{\mathscr{Q}\left(1,-1,1/2+\sqrt{\check{\delta}}+2\kappa\right)} + \left\vert\left\vert X\right\vert\right\vert_{\mathscr{Q}\left(3,-4,-1/2+\sqrt{\check{\delta}}+2\kappa\right)}
\\ \nonumber &\lesssim \left\vert\left\vert H\right\vert\right\vert_{\mathscr{Q}\left(2,-1,-1/2+10\sqrt{\check{\delta}}\right)}+ \left\vert\left\vert \left(1,\mathcal{L}_b\right)H\right\vert\right\vert_{\mathscr{Q}\left(1,-1,-1/2+10\check{\delta}\right)}.
\end{align}

The remaining estimates we need to establish concern $\left((-v)\mathcal{L}_{\partial_v}-\mathcal{L}_b\right)X$. From~\eqref{1om2om1o4} we have 
\begin{align}\label{o2om2oom4}
\left(\mathcal{L}_{\partial_v} - \mathcal{P}_{\ell \geq 1}\left(\Omega^2\left(v+1\right)^{-2}\mathring{\Delta}\cdot\right)\right)\left((-v)\mathcal{L}_{\partial_v}-\mathcal{P}_{\ell \geq 1}\mathcal{L}_b\right)X = \mathcal{J},
\end{align}
where, as a consequence of the previous estimates, 
\[\left\vert\left\vert \mathcal{J}\right\vert\right\vert_{\mathscr{Q}\left(2,-1,\kappa\right)} \lesssim\left\vert\left\vert H\right\vert\right\vert_{\mathscr{Q}\left(2,-1,-1/2+10\sqrt{\check{\delta}}\right)}+ \left\vert\left\vert \left(1,\mathcal{L}_b\right)H\right\vert\right\vert_{\mathscr{Q}\left(1,-1,-1/2+10\check{\delta}\right)}.\]
Given this, we can establish the remaining estimates by carrying out, yet again, energy estimates for the equation~\eqref{o2om2oom4}. We omit the straightforward details.

\end{proof}

The following lemma will be useful in Section~\ref{propagatetheconstraintsforever}. 
\begin{lemma}\label{k4oij2oijo3}Suppose that $f : (-1,0) \times \mathbb{S}^2 \to \mathbb{R}$ satisfies
\begin{equation}\label{o3joj3o294}
\mathcal{L}_{\partial_v}f - 4\left(v+1\right)^{-2}\mathring{\Delta}f = \mathcal{P}_{\ell \geq 1}H,\qquad \mathcal{P}_{\ell = 0}f = 0,
\end{equation}
for a function $H : (-1,0) \times \mathbb{S}^2 \to \mathbb{R}$ such that 
\[\sup_{v \in (-1,-1/2)}\sum_{\left|\alpha\right| \leq 4}\left(v+1\right)^p\left|H^{(\alpha)}\right| < \infty,\]
for some $p \in \mathbb{R}$. 

Then, if $f$ satisfies 
\begin{equation}\label{3iojoijo420492}
\sup_{v \in (-1,-1/2)}\left(v+1\right)^q\left|f\right| < \infty,
\end{equation}
for some $q \in \mathbb{R}$, we will have that 
\begin{equation}\label{3ojo2ijio5998871782}
\sup_{v \in (-1,-1/2)}\left[\left(v+1\right)^p\left|\mathcal{L}_{\partial_v}f\right|+\sum_{\left|\alpha\right| \leq 2}\left(v+1\right)^{p-2}\left|f^{(\alpha)}\right|\right] \lesssim \sup_{v \in (-1,-1/2)}\sum_{\left|\alpha\right| \leq 4}\left(v+1\right)^p\left|H^{(\alpha)}\right| .
\end{equation}
\end{lemma}
\begin{proof}Let $f_{m\ell}$ and $H_{m\ell}$ denote the coefficient of the projection of $f$ and $H$ onto the $S_{m\ell}$ spherical harmonic, for $\ell \in \mathbb{Z}_{\geq 1}$ and $m \in \mathbb{Z}_{[-\ell,\ell]}$. Then the equation~\eqref{o3joj3o294} becomes the following ordinary differential equation:
\begin{align}\label{3iojoij9489198}
&\frac{d f_{m\ell}}{dv} + \frac{4\ell\left(\ell+1\right)}{(v+1)^2}f_{m\ell} = H_{m\ell} \Leftrightarrow 
\\ \nonumber &\qquad \exp\left(4\ell\left(\ell+1\right)\left(v+1\right)^{-1}\right)\frac{d}{dv}\left(\exp\left(-4\ell\left(\ell+1\right)\left(v+1\right)^{-1}\right)f_{m\ell}\right) = H_{m\ell}.
\end{align}
In view of~\eqref{3iojoijo420492}, we may integrate~\eqref{3iojoij9489198} to obtain
\begin{equation}\label{3ioj2oij4ioj2oij4}
f_{m\ell}\left(v\right) = \exp\left(4\ell\left(\ell+1\right)\left(v+1\right)^{-1}\right)\int_{-1}^v\exp\left(-4\ell\left(\ell+1\right)\left(s+1\right)^{-1}\right)H_{m\ell}(s)\, ds.
\end{equation}
In view of the fact that
\[\exp\left(-4\ell\left(\ell+1\right)\left(s+1\right)^{-1}\right) = \left(4\ell\left(\ell+1\right)\right)^{-1}\left(s+1\right)^2\frac{d}{ds}\exp\left(-4\ell\left(\ell+1\right)\left(s+1\right)^{-1}\right),\]
we obtain from~\eqref{3ioj2oij4ioj2oij4} that for any $v \in (-1,-1/2)$
\begin{equation}\label{3ij2oijio42}
\ell\left(\ell+1\right)\left|f_{m\ell}(v)\right| \lesssim \left(v+1\right)^{-p+2}\sup_{s \in (-1,-1/2)}\left[\left(s+1\right)^p\left|H_{m\ell}(s)\right|\right].
\end{equation}
Returning back to~\eqref{3iojoij9489198}, we also obtain that
\begin{equation}\label{3oijoi2jioj2}
\left|\frac{df_{m\ell}}{dv}(v)\right| \lesssim \left(v+1\right)^{-2}\ell\left(\ell+1\right)\left|f_{m\ell}(v)\right| + \left|H_{m\ell}(v)\right| \lesssim \left(v+1\right)^{-p}\sup_{s \in (-1,-1/2)}\left[\left(s+1\right)^p\left|H_{m\ell}(s)\right|\right].
\end{equation}

We then have
\begin{align*}
&\sup_{\left(v,\theta^A\right)\in (-1,-1/2) \times \mathbb{S}^2}\left(\left(v+1\right)^{2p}\left[\left|\mathcal{L}_{\partial_v}f\right|^2 + \left(v+1\right)^{-2}\sum_{\left|\alpha\right| \leq 2}\left|f^{(\alpha)}\right|^2\right]\right) 
\\ \nonumber &\qquad \lesssim \sup_{v \in (-1,-1/2)}\left(\left(v+1\right)^{2p}\left[\left\vert\left\vert \mathcal{L}_{\partial_v}f\right\vert\right\vert^2_{\mathring{H}^2\left(\mathbb{S}^2_{-1,v}\right)} + \left(v+1\right)^{-2}\left\vert\left\vert f\right\vert\right\vert^2_{\mathring{H}^4\left(\mathbb{S}^2_{-1,v}\right)}\right]\right)
\\ \nonumber &\qquad \lesssim \sum_{m \ell}\sup_{v \in (-1,-1/2)}\left(\left(v+1\right)^{2p}\ell^4\left|\frac{df_{m\ell}}{dv}(v)\right|^2 + \left(v+1\right)^{2p-2}\ell^8\left|f_{m\ell}(v)\right|^2\right)
\\ \nonumber &\qquad \lesssim \sum_{m\ell}\ell^{-4}\left(\sup_v \left(v+1\right)^{2p}\ell^8\left|H_{m\ell}(v)\right|^2\right)
\\ \nonumber &\qquad \lesssim \sup_v \left(v+1\right)^{2p}\left\vert\left\vert H\right\vert\right\vert^2_{\mathring{H}^4\left(\mathbb{S}^2\right)}.
\end{align*}
 
\end{proof}
\begin{remark}\label{3ioj2oijoi42} An integration by parts computation using $\frac{d}{dv}$ applied to the formula~\eqref{3ioj2oij4ioj2oij4} shows that if one additionally assumes that 
\[\sup_{v \in (-1,-1/2)}\sum_{\left|\alpha\right| \leq 4}\left(v+1\right)^{p+1}\left|\mathcal{L}_{\partial_v}H^{(\alpha)}\right| < \infty,\]
then one may improve~\eqref{3ojo2ijio5998871782} to
\begin{equation*}
\sup_{v \in (-1,-1/2)}\left[\left(v+1\right)^{p-1}\left|\mathcal{L}_{\partial_v}f\right|+\sum_{\left|\alpha\right| \leq 2}\left(v+1\right)^{p-2}\left|f^{(\alpha)}\right|\right] \lesssim \sum_{j=0}^1\sup_{v \in (-1,-1/2)}\sum_{\left|\alpha\right| \leq 4}\left(v+1\right)^{p+j}\left|\mathcal{L}_{\partial_v}^jH^{(\alpha)}\right| .
\end{equation*}
\end{remark}
\section{Algebraic Consequences of Self-Similarity}\label{alloftheequationsinselfsimilar}

We start with the relevant definition of self-similarity (cf. Definition 3.1 from~\cite{nakedone}.)
\begin{definition}\label{folselfsim}
We say that a $3+1$ dimensional Lorentzian  manifold $\left(\mathcal{M},g\right)$ given in the double-null form~\eqref{doublenullisg} and defined in the region $\{u < 0 \} \cap \{0 < \frac{v}{u} < 1\}$ has a self-similar foliation if in the coordinate frame we have
\begin{equation}\label{scaleinvrelations2}
\Omega\left(u,v,\theta^A\right) = \check{\Omega}\left(\frac{v}{u},\theta\right),\qquad b_A\left(u,v,\theta^A\right) = u\check{b}_A\left(\frac{v}{u},\theta^A\right),\qquad \slashed{g}_{AB}\left(u,v,\theta^A\right) = u^2\check{\slashed{g}}_{AB}\left(\frac{v}{u},\theta\right),
\end{equation}
for some functions $\check\Omega$, $\check{b}$, and $\check{\slashed{g}}$.
\end{definition}

As in Section~\ref{doubleitup}, we emphasize that throughout this section we do \underline{not} assume that the Einstein vacuum equations hold! Our goal will be to derive a sequence of useful equations for the Ricci coefficients which hold for spacetimes with a self-similar foliation.

\begin{lemma}\label{thefirstrelations}Let $\left(\mathcal{M},g\right)$ be a Lorentzian manifold with a self-similar foliation. Then we have 
\begin{equation}\label{20asdww}
\Omega{\rm tr}\underline{\chi} + \Omega \frac{v}{u}{\rm tr}\chi = \frac{2}{u} + \slashed{\rm div}b,\qquad \Omega\hat{\underline{\chi}} + \Omega \frac{v}{u}\hat{\chi} = \frac{1}{2}\slashed{\nabla}\hat{\otimes}b,
\end{equation}
\begin{equation}\label{okodwok22}
\Omega\underline{\omega} + \frac{v}{u}\Omega\omega +\frac{1}{2}\mathcal{L}_b\log\Omega = 0.
\end{equation}
\end{lemma}
\begin{proof}This is an immediate consequence of~\eqref{scaleinvrelations2} and the definitions of the Ricci coefficients. (See Lemma B.1 of~\cite{scaleinvariant}.)
\end{proof}

Next, we use self-similarity to re-write the $\nabla_3\hat{\chi}$ equation as a propagation equation for $\hat{\chi}$.
\begin{lemma}\label{03kdo3k5}Let $\left(\mathcal{M},g\right)$ be a Lorentzian manifold with a self-similar foliation. Then the following equation holds:
\begin{align}\label{kwdkodwok23dg}
& -\frac{v}{u}\Omega\nabla_4\left(\Omega\hat{\chi}\right)_{AB} +\mathscr{L}\left(\Omega\hat{\chi}\right)_{AB} -\frac{v}{u}\Omega{\rm tr}\chi\left(\Omega\hat{\chi}\right)_{AB} = 
\\ \nonumber &\qquad \Omega^2\left(\left(\slashed{\nabla}\hat\otimes \eta\right)_{AB} + \left(\eta\hat\otimes \eta\right)_{AB}\right) - \frac{1}{4}\left(\Omega{\rm tr}\chi\right)\left(\slashed{\nabla}\hat{\otimes}b\right) + \Omega^2\widehat{{\rm Ric}}_{AB},
 \end{align}
 where
 \[\mathscr{L}f_{AB} \doteq \mathcal{L}_bf_{AB}- \left(\slashed{\nabla}\hat{\otimes}b\right)^C_{\ \ (A}f_{B)C} -\frac{1}{2}\slashed{\rm div}bf_{AB}.\]
\end{lemma}
\begin{proof}Using self-similarity we have
\begin{align*}
&\Omega\nabla_3\left(\Omega\hat{\chi}\right)_{AB} 
\\ \nonumber &\qquad =\mathcal{L}_{\partial_u}\left(\Omega\hat{\chi}_{AB}\right) + \mathcal{L}_b\left(\Omega\hat{\chi}\right)_{AB} - 2\left(\Omega\underline{\chi}\right)^C_{\ \ (A}\left(\Omega\hat{\chi}\right)_{B)C} 
\\ \nonumber &\qquad = -\frac{v}{u}\Omega\nabla_4\left(\Omega\hat{\chi}\right)_{AB}  +u^{-1}\Omega\hat{\chi}_{AB}+ \mathcal{L}_b\left(\Omega\hat{\chi}\right)_{AB}- 2\left(\frac{v}{u}\Omega\chi + \Omega\underline{\chi}\right)^C_{\ \ (A}\left(\Omega\hat{\chi}\right)_{B)C}
\\ \nonumber &\qquad = -\frac{v}{u}\Omega\nabla_4\left(\Omega\hat{\chi}\right)_{AB}  -u^{-1}\Omega\hat{\chi}_{AB}+ \mathcal{L}_b\left(\Omega\hat{\chi}\right)_{AB}- \left(\slashed{\nabla}\hat{\otimes}b\right)^C_{\ \ (A}\left(\Omega\hat{\chi}\right)_{B)C} - \slashed{\rm div}b\left(\Omega\hat{\chi}\right)_{AB}.
\end{align*}
The proof is concluded by substituting this formula into~\eqref{3hatchi}.
\end{proof}

\begin{definition}\label{thetaisdefinedhere}We define a symmetric trace-free tensor $\Theta_{AB}$ by
\begin{align*}
& \Theta_{AB} \doteq -\frac{v}{u}\Omega\nabla_4\left(\Omega\hat{\chi}\right)_{AB}+\mathcal{L}_b\left(\Omega\hat{\chi}\right)_{AB} +v(\Omega{\rm tr}\chi)\left(\Omega\hat{\chi}\right)_{AB} -\left(\slashed{\nabla}\hat{\otimes}b\right)^C_{\ \ (A}(\Omega\hat{\chi})_{B)C}-\frac{1}{2}\slashed{\rm div}b (\Omega\hat{\chi})_{AB}
 \end{align*}
\end{definition}
\begin{remark}The significance of $\Theta_{AB}$ is that, due to Lemma~\ref{03kdo3k5}, we may expect better estimates for $\Theta_{AB}$ as $v\to 0$ then we would for a generic term involving $\hat{\chi}_{AB}$. 
\end{remark}

Next, we find a useful equation for $\slashed{\rm div}b$.
\begin{lemma}\label{divbequnaa}Let $\left(\mathcal{M},g\right)$ be a Lorentzian manifold with a self-similar foliation. Then the following equation holds:
\begin{align}\label{eqnyaydivb}
&\left(-\frac{v}{u}\mathcal{L}_{\partial_v} + b\cdot\slashed{\nabla}\right)\slashed{\rm div}b + \left(\frac{1}{u}-\frac{v}{u}\Omega{\rm tr}\chi\right)\slashed{\rm div}b  + \frac{1}{2}\left(\slashed{\rm div}b\right)^2 +8\left(\Omega\underline{\omega}\right)\left(u^{-1}+\frac{\slashed{\rm div}b}{2}\right) \\ \nonumber &\qquad  = \Omega\frac{v}{u}\hat{\chi}\cdot\slashed{\nabla}\hat{\otimes}b - \frac{1}{4}\left|\slashed{\nabla}\hat{\otimes}b\right|^2+ \Omega^2\frac{v}{u} b\cdot\slashed{\nabla}\left(\Omega^{-1}{\rm tr}\chi\right) - \Omega^2{\rm Ric}_{33} + \frac{v^2}{u^2}\Omega^2{\rm Ric}_{44}.
\end{align}
\end{lemma}
\begin{proof}We start with the $e_3$-Raychaudhuri equation which can be re-written as 
\begin{equation}\label{eqn1}
\left(\mathcal{L}_{\partial_u} + b\cdot\slashed{\nabla}\right)\Omega{\rm tr}\underline{\chi} + \frac{1}{2}\left(\Omega {\rm tr}\underline{\chi}\right)^2 + 4\left(\Omega\underline{\omega}\right)\left(\Omega{\rm tr}\underline{\chi}\right) = -\left|\Omega\hat{\underline{\chi}}\right|^2 - \Omega^2{\rm Ric}_{33}.
\end{equation}
Then, using Lemma~\ref{thefirstrelations}, we plug in the self-similar relation
\begin{equation}\label{eqn2}
\Omega{\rm tr}\underline{\chi} = \frac{2}{u} + \slashed{\rm div}b - \frac{v}{u}\Omega{\rm tr}\chi
\end{equation}
into~\eqref{eqn1}. We obtain
\begin{align}\label{eqn3}
&\left(\mathcal{L}_{\partial_u} + b\cdot\slashed{\nabla}\right)\left(\frac{2}{u} + \slashed{\rm div}b -\frac{v}{u}\Omega{\rm tr}\chi\right) + \frac{1}{2}\left(\frac{2}{u} + \slashed{\rm div}b -\frac{v}{u}\Omega{\rm tr}\chi\right)^2 + 4\left(\Omega\underline{\omega}\right)\left(\frac{2}{u} + \slashed{\rm div}b -\frac{v}{u}\Omega{\rm tr}\chi\right) = 
\\ \nonumber &\qquad -\left|\Omega\hat{\underline{\chi}}\right|^2 -\Omega^2{\rm Ric}_{33}.
\end{align}
Simplifying yields 
\begin{align}\label{eqn4}
&\left(\mathcal{L}_{\partial_u} + b\cdot\slashed{\nabla}\right)\left(\slashed{\rm div}b - \frac{v}{u}\Omega{\rm tr}\chi\right) + \frac{2}{u}\left(\slashed{\rm div}b - \frac{v}{u}\Omega{\rm tr}\chi\right) + \frac{1}{2}\left(\slashed{\rm div}b - \frac{v}{u}\Omega{\rm tr}\chi\right)^2 
\\ \nonumber &\qquad + 8\left(\Omega\underline{\omega}\right)u^{-1} + 4\left(\Omega\underline{\omega}\right)\left(\slashed{\rm div}b - \frac{v}{u}\Omega{\rm tr}\chi\right) = -\left|\Omega\hat{\underline{\chi}}\right|^2 - \Omega^2{\rm Ric}_{33}.
\end{align}

Next, self-similarity implies that 
\begin{align}\label{eqn5}
-\mathcal{L}_{\partial_u}\left(\frac{v}{u}\Omega{\rm tr}\chi\right) = \frac{v}{u^2}\Omega{\rm tr}\chi + \frac{v}{u}\mathcal{L}_{\partial_v}\left(\frac{v}{u}\Omega{\rm tr}\chi\right).
\end{align}
The $e_4$-Raychaudhuri equation may be re-written as
\begin{align}\label{eqn6}
\mathcal{L}_{\partial_v}\left(\Omega {\rm tr}\chi\right) + \frac{1}{2}\left(\Omega {\rm tr}\chi\right)^2 + 4\left(\Omega\omega\right)\left(\Omega{\rm tr}\chi\right) = -\left|\Omega\hat{\chi}\right|^2 - \Omega^2{\rm Ric}_{44}.
\end{align}
Combining~\eqref{eqn5} and~\eqref{eqn6} yields
\begin{equation}\label{eqn7}
-\mathcal{L}_{\partial_u}\left(\frac{v}{u}\Omega{\rm tr}\chi\right) = \frac{2v}{u^2}\left(\Omega{\rm tr}\chi\right) - \frac{v^2}{u^2}\left(\frac{1}{2}\left(\Omega{\rm tr}\chi\right)^2 + 4\left(\Omega\omega\right)\left(\Omega{\rm tr}\chi\right) + \left|\Omega\hat{\chi}\right|^2 + \Omega^2{\rm Ric}_{44}\right).
\end{equation}
Now we plug~\eqref{eqn7} into~\eqref{eqn4} and obtain
\begin{align}\label{eqn8}
&\left(\mathcal{L}_{\partial_u} + b\cdot\slashed{\nabla}\right)\slashed{\rm div}b + \left(\frac{2}{u}-\frac{v}{u}\Omega{\rm tr}\chi\right)\slashed{\rm div}b  + \frac{1}{2}\left(\slashed{\rm div}b\right)^2 +8\left(\Omega\underline{\omega}\right)u^{-1} - \frac{4v^2}{u^2}\left(\Omega\omega\right)\left(\Omega{\rm tr}\chi\right)
\\ \nonumber &\qquad + 4\left(\Omega\underline{\omega}\right)\left(\slashed{\rm div}b - \frac{v}{u}\Omega{\rm tr}\chi\right) = \frac{v^2}{u^2}\left|\Omega\hat{\chi}\right|^2 - \left|\Omega\hat{\underline{\chi}}\right|^2 + b\cdot\slashed{\nabla}\left(\frac{v}{u}\Omega{\rm tr}\chi\right) - \Omega^2{\rm Ric}_{33} + \frac{v^2}{u^2}\Omega^2{\rm Ric}_{44}.
\end{align}
Next, from Lemma~\ref{thefirstrelations}, we have the self-similar relation:
\begin{equation}\label{eqn9}
\Omega\underline{\omega} + \frac{v}{u}\Omega\omega + \frac{1}{2}b\cdot\slashed{\nabla}\log\Omega = 0.
\end{equation}
Plugging~\eqref{eqn9} into~\eqref{eqn8} yields
\begin{align}\label{eqn10}
&\left(\mathcal{L}_{\partial_u} + b\cdot\slashed{\nabla}\right)\slashed{\rm div}b + \left(\frac{2}{u}-\frac{v}{u}\Omega{\rm tr}\chi\right)\slashed{\rm div}b  + \frac{1}{2}\left(\slashed{\rm div}b\right)^2 +8\left(\Omega\underline{\omega}\right)\left(u^{-1}+\frac{\slashed{\rm div}b}{2}\right) \\ \nonumber &\qquad  = \frac{v^2}{u^2}\left|\Omega\hat{\chi}\right|^2 - \left|\Omega\hat{\underline{\chi}}\right|^2 + b\cdot\slashed{\nabla}\left(\frac{v}{u}\Omega{\rm tr}\chi\right) - 2\left(b\cdot\slashed{\nabla}\log\Omega\right)\frac{v}{u}\Omega{\rm tr}\chi- \Omega^2{\rm Ric}_{33} + \frac{v^2}{u^2}\Omega^2{\rm Ric}_{44}.
\end{align}
Self-similarity yields
\begin{equation}\label{eqn11}
\mathcal{L}_{\partial_u}\slashed{\rm div}b = -u^{-1}\slashed{\rm div}b -\frac{v}{u}\mathcal{L}_{\partial_v}\slashed{\rm div}b.
\end{equation}
Plugging~\eqref{eqn11} into~\eqref{eqn10} yields
\begin{align}\label{eqn12}
&\left(-\frac{v}{u}\mathcal{L}_{\partial_v} + b\cdot\slashed{\nabla}\right)\slashed{\rm div}b + \left(\frac{1}{u}-\frac{v}{u}\Omega{\rm tr}\chi\right)\slashed{\rm div}b  + \frac{1}{2}\left(\slashed{\rm div}b\right)^2 +8\left(\Omega\underline{\omega}\right)\left(u^{-1}+\frac{\slashed{\rm div}b}{2}\right) \\ \nonumber &\qquad  = \frac{v^2}{u^2}\left|\Omega\hat{\chi}\right|^2 - \left|\Omega\hat{\underline{\chi}}\right|^2 + b\cdot\slashed{\nabla}\left(\frac{v}{u}\Omega{\rm tr}\chi\right) - 2\left(b\cdot\slashed{\nabla}\log\Omega\right)\frac{v}{u}\Omega{\rm tr}\chi- \Omega^2{\rm Ric}_{33} + \frac{v^2}{u^2}\Omega^2{\rm Ric}_{44}.
\end{align}
Finally,~\eqref{eqnyaydivb} is obtained after using Lemma~\ref{thefirstrelations} for the relation between $\hat{\underline{\chi}}$ and $\hat{\chi}$.

\end{proof}
\begin{remark}
The equation~\eqref{eqnyaydivb} may be considered as a propagation equation for $\slashed{\rm div}b$. Note the important fact that it couples \underline{linearly} to the lapse via $\underline{\omega}$.
\end{remark}

Next, we derive an equation for the lapse $\Omega$.
\begin{lemma}\label{2m2om3o3923332}Let $\left(\mathcal{M},g\right)$ be a Lorentzian manifold with a self-similar foliation. Then the following equation holds:
\begin{align}\label{2mo3330ck2}
&\frac{1}{2}\left(\Omega^{-1}{\rm tr}\chi\right)\slashed{\rm div}b + \mathcal{L}_b\left(\Omega^{-1}{\rm tr}\chi\right) +\frac{1}{2}\left(\Omega^{-1}\hat{\chi}\right)\cdot\left(\slashed{\nabla}\hat{\otimes}b\right) =
\\ \nonumber &\qquad  4\Omega^{-2}\left(\Omega\nabla_4\right)\left(\Omega\underline{\omega}\right)+ 4\underline{\omega}{\rm tr}\chi + 2\slashed{\rm div}\eta + 4\eta\cdot\underline{\eta} -{\rm Ric}_{34}- \frac{v}{u}{\rm Ric}_{44}.
\end{align}
\end{lemma}
\begin{proof}We start with the equation~\eqref{3trchi} which may be re-written as
\begin{align}\label{2lm2omo4}
\Omega\nabla_3\left(\Omega^{-1}{\rm tr}\chi\right) + \frac{1}{2}{\rm tr}\chi{\rm tr}\underline{\chi} &= \slashed{g}^{AB}R_{3AB4}+ 4\underline{\omega}{\rm tr}\chi + 2\slashed{\rm div}\eta + 2\left|\eta\right|^2 - \hat{\chi}\cdot\hat{\underline{\chi}}.
\end{align}
Now we use self-similarity and the equation~\eqref{4trchi} to derive that
\begin{align*}
\Omega\nabla_3\left(\Omega^{-1}{\rm tr}\chi\right) &= -u^{-1}\left(\Omega^{-1}{\rm tr}\chi\right) + \mathcal{L}_b\left(\Omega^{-1}{\rm tr}\chi\right) - \frac{v}{u}\Omega\nabla_4\left(\Omega^{-1}{\rm tr}\chi\right)
\\ \nonumber &=  -u^{-1}\left(\Omega^{-1}{\rm tr}\chi\right) + \mathcal{L}_b\left(\Omega^{-1}{\rm tr}\chi\right) + \frac{v}{u}\left(\frac{1}{2}\left({\rm tr}\chi\right)^2 + \left|\hat{\chi}\right|^2 + {\rm Ric}_{44}\right).
\end{align*}
Now we plug this into~\eqref{2lm2omo4} and use Lemma~\ref{thefirstrelations} to derive 
\begin{align}\label{k2o2o01039}
&\frac{1}{2}\left(\Omega^{-1}{\rm tr}\chi\right)\slashed{\rm div}b + \mathcal{L}_b\left(\Omega^{-1}{\rm tr}\chi\right) +\frac{1}{2}\left(\Omega^{-1}\hat{\chi}\right)\cdot\left(\slashed{\nabla}\hat{\otimes}b\right) =
\\ \nonumber &\qquad  \slashed{g}^{AB}R_{3AB4}+ 4\underline{\omega}{\rm tr}\chi + 2\slashed{\rm div}\eta + 2\left|\eta\right|^2 - \frac{v}{u}{\rm Ric}_{44}
\end{align}

The equation~\eqref{4uomega} may be written as
\begin{align*}
\Omega^{-2} \left(\Omega\nabla_4\right)\left(\Omega\underline{\omega}\right) &= \frac{1}{4}\left({\rm Ric}_{34} + \slashed{g}^{AB}R_{3AB4}\right) + \frac{1}{2}\left|\eta\right|^2 - \eta\cdot\underline{\eta},
\end{align*}
We then substitute this into~\eqref{k2o2o01039} to finally derive~\eqref{2mo3330ck2}.
\end{proof}
\begin{remark}\label{2lm3omo2}There are two main reasons why the equation of Lemma~\ref{2m2om3o3923332} will be significant for us. First of all, after writing $\slashed{\rm div}\eta = \slashed{\Delta}\log\Omega +\slashed{\rm div}\zeta$, we can consider the equation~\eqref{2mo3330ck2} as a second order equation for $\log\Omega$ which couples \underline{linearly} to the divergence of the shift $b$. This suggests that we should consider~\eqref{2mo3330ck2} and~\eqref{eqnyaydivb} together as a system which determines (at the linear level) $\slashed{\rm div}b$ and $\log\Omega$. Second of all, this equation suggests that we may obtain uniform boundedness of $\slashed{\rm div}\eta$ as $v\to 0$ if we have that 
\[ 4\Omega^{-2}\left(\Omega\nabla_4\right)\left(\Omega\underline{\omega}\right) + 4\eta\cdot\underline{\eta} -\frac{1}{2}\left(\Omega^{-1}\hat{\chi}\right)\cdot\left(\slashed{\nabla}\hat{\otimes}b\right)  \]
is uniformly bounded as $v\to 0$. 
\end{remark}

The following lemma will be useful when we prove Lemma~\ref{2mo2o3o2} below.
\begin{lemma}\label{j32ioj3io21}We have 
\begin{align*}
&\mathcal{L}_{\partial_v}\left(\Omega\frac{v}{u}\hat{\chi}\cdot\slashed{\nabla}\hat{\otimes}b - \frac{1}{4}\left|\slashed{\nabla}\hat{\otimes}b\right|^2\right)  
\\ \nonumber &=\qquad -\left(\Omega\hat{\chi}\right)\cdot\slashed{\nabla}\hat{\otimes}b + \frac{v}{u}\Omega\nabla_4\left(\Omega\hat{\chi}\right)\cdot \slashed{\nabla}\hat{\otimes}b+\frac{v}{u}\left(\Omega\hat{\chi}\right)\cdot \Omega\nabla_4\left(\slashed{\nabla}\hat{\otimes}b\right)-\frac{1}{2}\left(\slashed{\nabla}\hat{\otimes}\left(\mathcal{L}_{\partial_v}b\right)\right)\cdot\slashed{\nabla}\hat{\otimes}b
\\ \nonumber &\qquad +\left(-\mathcal{L}_b\left(\Omega\hat{\chi}\right) +\slashed{\rm div}b\left(\Omega\hat{\chi}\right)+\left(\Omega\hat{\chi}\right)^C_{\ \ (A}\left(\slashed{\nabla}\hat{\otimes}b\right)_{B)C}\right)\cdot \slashed{\nabla}\hat{\otimes}b
\\ \nonumber &= \left(-\Theta + \left(\Omega\hat{\chi}\right)\left(u^{-1}+\frac{1}{2}\slashed{\rm div}b\right)\right)\cdot \left(\slashed{\nabla}\hat{\otimes}b\right) +\frac{v}{u}\left(\Omega\hat{\chi}\right)\cdot \Omega\nabla_4\left(\slashed{\nabla}\hat{\otimes}b\right)-\frac{1}{2}\left(\slashed{\nabla}\hat{\otimes}\left(\mathcal{L}_{\partial_v}b\right)\right)\cdot\slashed{\nabla}\hat{\otimes}b,
\end{align*}
where $\Theta_{AB}$ is defined in Definition~\ref{thetaisdefinedhere}.

\end{lemma}
\begin{proof}This is a corollary of the commutation formulas of Lemma~\ref{othercommutelemma}.
\end{proof}

The following will be important later. For $v < 0$ it may be considered as a propagation equation for $\eta$ (which, however, couples linearly to the other components of the metric). The equation is also useful along $\{ v = 0\}$ where, assuming $v\mathcal{L}_{\partial_v}\eta|_{v=0} = 0$, it may be used to determine $\eta$ from $\Omega\underline{\omega}$, $b$, and $\slashed{g}$. 
\begin{lemma}\label{2mo2o3o2}Let $\left(\mathcal{M},g\right)$ be a Lorentzian manifold with a self-similar foliation. Then the following equation holds:
\begin{align}\label{3pk2o294}
&\frac{v}{u}\mathcal{L}_{\partial_v}\eta_A - \mathcal{L}_b\eta_A -\eta_A\left(\Omega{\rm tr}\underline{\chi}\right)- 4\slashed{\nabla}_A\left(\Omega\underline{\omega}\right) =  \slashed{\nabla}^B\left(\Omega\hat{\underline{\chi}}\right)_{AB} - \frac{1}{2}\slashed{\nabla}_A\left(\Omega{\rm tr}\underline{\chi}\right) -\Omega{\rm Ric}_{3A}.
\end{align}
\end{lemma}
\begin{proof}From~\eqref{3ueta} and~\eqref{tcod2} we have
\begin{align}\label{ijo20k424}
\nabla_3\underline{\eta}_A &= -\underline{\chi}_A^{\ \ B}\cdot\left(\underline\eta-\eta\right)_B +\slashed{\nabla}^B\hat{\underline\chi}_{AB}-\frac{1}{2}\slashed{\nabla}_A{\rm tr}\underline\chi+ \frac{1}{2}{\rm tr}\underline\chi \zeta_A - \zeta^B\underline{\hat{\chi}}_{AB} - {\rm Ric}_{3A}.
\end{align}
Next, using~\eqref{acommut2}, we observe that
\begin{align}\label{3om4o204}
\Omega\nabla_3\underline{\eta}_A &= -\Omega\nabla_3\eta_A +2\Omega\nabla_3\slashed{\nabla}_A\log\Omega 
\\ \nonumber &= -\Omega\nabla_3\eta_A -4\slashed{\nabla}_A\left(\Omega\underline{\omega}\right) + 2\left[\Omega\nabla_3,\slashed{\nabla}_A\right]\log\Omega
\\ \nonumber &=   -\Omega\nabla_3\eta_A -4\slashed{\nabla}_A\left(\Omega\underline{\omega}\right)-2\left(\Omega\underline{\chi}\right)_A^{\ \ B}\slashed{\nabla}_B\log\Omega.
\end{align}
Using self-similarity, we may write in the coordinate frame $\eta_A\left(u,v,\theta^B\right) = H_A\left(\frac{v}{u},\theta^B\right)$ for a suitable $H_A$. We may then compute 
\begin{align}\label{2o4o22293}
\Omega\nabla_3\eta_A &= \mathcal{L}_{\partial_u}\left(H_A\left(\frac{v}{u},\theta^B\right)\right)+\mathcal{L}_b\eta_A -\left(\Omega\underline{\chi}\right)_A^{\ \ B}\eta_B
\\ \nonumber &= -\frac{v}{u}\mathcal{L}_{\partial_v}\eta_A + \mathcal{L}_b\eta_A - \left(\Omega\underline{\chi}\right)_A^{\ \ B}\eta_B
\end{align}
Multiplying~\eqref{ijo20k424} with $\Omega$ and using~\eqref{3om4o204} and~\eqref{2o4o22293} then leads to~\eqref{3pk2o294}.

\end{proof}

\begin{lemma}\label{2km2omo34}Let $\left(\mathcal{M},g\right)$ be a Lorentzian manifold with a self-similar foliation. Then the following equation holds:
\begin{align}\label{3i3ei2i4i24u}
&\left(-\frac{v}{u}\mathcal{L}_{\partial_v} + \mathcal{L}_b\right)\Big(4\Omega^{-1}\nabla_4\left(\Omega\underline{\omega}\right) + 4\left(\Omega^{-1}{\rm tr}\chi\right)\left(\Omega\underline{\omega}\right) + 4\eta\cdot\underline{\eta} - \left|\eta\right|^2
\\ \nonumber &\qquad \qquad  -\frac{1}{2}\left(\Omega^{-1}\hat{\chi}\right)\cdot\left(\slashed{\nabla}\hat{\otimes}b\right) - \frac{1}{2}\mathcal{L}_b\left(\Omega^{-1}{\rm tr}\chi\right)\Big)
\\ \nonumber &\qquad +\left(-\frac{v}{u}\Omega{\rm tr}\chi + \slashed{\rm div}b\right)\Big(4\Omega^{-1}\nabla_4\left(\Omega\underline{\omega}\right) + 4\left(\Omega^{-1}{\rm tr}\chi\right)\left(\Omega\underline{\omega}\right) + 4\eta\cdot\underline{\eta} - \left|\eta\right|^2
\\ \nonumber &\qquad \qquad  -\frac{1}{2}\left(\Omega^{-1}\hat{\chi}\right)\cdot\left(\slashed{\nabla}\hat{\otimes}b\right) - \frac{1}{2}\mathcal{L}_b\left(\Omega^{-1}{\rm tr}\chi\right) \Big) -4 \slashed{\Delta}\left(\Omega\underline{\omega}\right)
\\ \nonumber &\qquad + \left(u^{-1} + \frac{1}{2}\slashed{\rm div}b\right)\Big(4\Omega^{-1}\nabla_4\left(\Omega\underline{\omega}\right) + 4\left(\Omega^{-1}{\rm tr}\chi\right)\left(\Omega\underline{\omega}\right)  -\frac{1}{2}\left(\Omega^{-1}\hat{\chi}\right)\cdot\left(\slashed{\nabla}\hat{\otimes}b\right)+ 4\eta\cdot\underline{\eta}\Big)
\\ \nonumber &\qquad -\frac{v}{u}\left(\mathcal{L}_{\partial_v}b\right)^A\slashed{\nabla}_A\log\Omega\left(\Omega^{-1}{\rm tr}\chi\right) -2\mathcal{L}_b\left(\left(\Omega\underline{\omega}\right)\left(\Omega^{-1}{\rm tr}\chi\right)\right) + \mathcal{L}_b\log\Omega\left(\frac{v}{u}\mathcal{L}_{\partial_v}\left(\Omega^{-1}{\rm tr}\chi\right) -\frac{v}{u}\left(\Omega^{-1}{\rm tr}\chi\right)^2\right)
\\ \nonumber &\qquad \qquad  -2\left(\Omega\underline{\omega}\right)\left(\Omega^{-1}{\rm tr}\chi\right)\slashed{\rm div}b -\eta^A\slashed{\nabla}_A\slashed{\rm div}b + \underline{\eta}^A\slashed{\nabla}_A
\left(\frac{v}{u}\Omega{\rm tr}\chi\right) - 2\slashed{\nabla}^A\left(\Omega\hat{\underline{\chi}}\right)_{AB}\eta^B-2\frac{v}{u}\Omega{\rm tr}\chi\left( \eta\cdot\underline{\eta}\right)
\\ \nonumber &\qquad \qquad -\left(\slashed{\nabla}\hat{\otimes}\eta\right)\cdot\left(\slashed{\nabla}\hat{\otimes}b\right) - 2\frac{v}{u}\left(\Omega\hat{\chi}\right)\cdot\left(\slashed{\nabla}\hat{\otimes}\slashed{\nabla}\log\Omega\right) -\frac{1}{2}\left(\eta\hat{\otimes}\eta\right)\cdot\left(\slashed{\nabla}\hat{\otimes}b\right)-\frac{1}{2}\left(\underline{\eta}\hat{\otimes}\underline{\eta}\right)\cdot\left(\frac{v}{u}\Omega\hat{\chi}\right)
\\ \nonumber &\qquad  + \frac{1}{2}\Omega^{-2}\Theta \cdot\left(\slashed{\nabla}\hat{\otimes}b\right)
  -\frac{1}{2}\mathcal{L}_b\left(\Omega^{-1}{\rm tr}\chi\right)\slashed{\rm div}b+\frac{1}{4}\left(\Omega^{-1}{\rm tr}\chi\right)\left(\slashed{\rm div}b\right)^2 
 \\ \nonumber &\qquad + \Omega^2\frac{v}{u}\left(\Omega^{-1}\hat{\chi}\right)\cdot\left(\slashed{\nabla}\hat{\otimes}b\right)\left(-\frac{1}{2}\Omega^{-1}{\rm tr}\chi\right) + \frac{1}{8}\left(\Omega^{-1}{\rm tr}\chi\right)\left|\slashed{\nabla}\hat{\otimes}b\right|^2 - \frac{1}{2}\Omega^2\left(\Omega^{-1}{\rm tr}\chi\right)\frac{v}{u}\mathcal{L}_b\left(\Omega^{-1}{\rm tr}\chi\right)
 \\ \nonumber &\qquad = -\frac{1}{2}\left(\Omega^{-2}\mathcal{L}_{\partial_v}\left(\Omega^2{\rm Ric}_{33}-\frac{v^2}{u^2}\Omega^2{\rm Ric}_{44}\right) + \Omega^{-1}{\rm tr}\chi \left(\Omega^2{\rm Ric}_{33}-\frac{v^2}{u^2}\Omega^2{\rm Ric}_{44}\right)\right) 
 \\ \nonumber &\qquad +\Omega\nabla_3\left({\rm Ric}_{34} + \frac{v}{u}{\rm Ric}_{44}\right) + \Omega{\rm tr}\underline{\chi}\left({\rm Ric}_{34} + \frac{v}{u}{\rm Ric}_{44}\right) + 2\left(\Omega{\rm Ric}_{3A}\right)\underline{\eta}^A.
\end{align}
\end{lemma}
\begin{proof}We start by writing 
\begin{align}\label{km2m4}
\slashed{\rm div}\eta &= \frac{1}{2}\Omega^{-2}\left(\slashed{\nabla}_A\log\Omega\right)\mathcal{L}_{\partial_v}b^A - \frac{1}{4}\Omega^{-2}\slashed{\nabla}_A\left(\mathcal{L}_{\partial_v}b^A\right) + \slashed{\Delta}\log\Omega
\\ \nonumber &= -\frac{1}{2}\left(\eta+\underline{\eta}\right)\cdot\left(\eta-\underline{\eta}\right) -\frac{1}{4}\Omega^{-2}\mathcal{L}_{\partial_v}\slashed{\rm div}b +\frac{1}{4}\Omega^{-2}\mathcal{L}_b\left(\Omega{\rm tr}\chi\right) + \slashed{\Delta}\log\Omega
\\ \nonumber &= -\frac{1}{2}\left|\eta\right|^2 +\frac{1}{2}\left|\underline{\eta}\right|^2  -\frac{1}{4}\Omega^{-2}\mathcal{L}_{\partial_v}\slashed{\rm div}b +\frac{1}{4}\Omega^{-2}\mathcal{L}_b\left(\Omega{\rm tr}\chi\right) + \slashed{\Delta}\log\Omega.
\end{align}
Here we used Lemma~\ref{othercommutelemma} when we went from the first line to the second line.

We now use~\eqref{km2m4} to substitute $\slashed{\rm div}\eta$ in~\eqref{2mo3330ck2} and then apply $\Omega\nabla_3$. We obtain 
\begin{align}\label{3om4omo}
&\Omega\nabla_3\Big(4\Omega^{-1}\nabla_4\left(\Omega\underline{\omega}\right) + 4\left(\Omega^{-1}{\rm tr}\chi\right)\left(\Omega\underline{\omega}\right) + 4\eta\cdot\underline{\eta} - \left|\eta\right|^2
\\ \nonumber &\qquad \qquad  -\frac{1}{2}\left(\Omega^{-1}\hat{\chi}\right)\cdot\left(\slashed{\nabla}\hat{\otimes}b\right) - \mathcal{L}_b\left(\Omega^{-1}{\rm tr}\chi\right) + \frac{1}{2}\Omega^{-2}\mathcal{L}_b\left(\Omega{\rm tr}\chi\right)\Big)
\\ \nonumber &\qquad +2\left(\Omega\nabla_3\underline{\eta}\right)\cdot\underline{\eta} + \Omega\nabla_3\left(-\frac{1}{2}\Omega^{-2}\mathcal{L}_{\partial_v}\slashed{\rm div}b -\frac{1}{2}\Omega^{-1}{\rm tr}\chi\slashed{\rm div}b\right)
\\ \nonumber &\qquad -4\slashed{\Delta}\left(\Omega\underline{\omega}\right) -4\slashed{\rm div}\left(\Omega\hat{\underline{\chi}}\right)\cdot\slashed{\nabla}\log\Omega -2\left(\Omega{\rm tr}\underline{\chi}\right)\slashed{\Delta}\log\Omega - 4\left(\Omega\hat{\underline{\chi}}\right)^{AB}\slashed{\nabla}^2_{AB}\log\Omega = \Omega\nabla_3\left({\rm Ric}_{34} + \frac{v}{u}{\rm Ric}_{44}\right).
\end{align}
Note that we have twice used the commutator formula~\eqref{acommut2} in the final line. Next, we use~\eqref{2mo3330ck2} and~\eqref{km2m4} to replace the term $-2\left(\Omega{\rm tr}\underline{\chi}\right)\slashed{\Delta}\log\Omega$ in~\eqref{3om4omo}. We obtain
\begin{align}\label{3m3omo203}
&\Omega\nabla_3\Big(4\Omega^{-1}\nabla_4\left(\Omega\underline{\omega}\right) + 4\left(\Omega^{-1}{\rm tr}\chi\right)\left(\Omega\underline{\omega}\right) + 4\eta\cdot\underline{\eta} - \left|\eta\right|^2
\\ \nonumber &\qquad \qquad  -\frac{1}{2}\left(\Omega^{-1}\hat{\chi}\right)\cdot\left(\slashed{\nabla}\hat{\otimes}b\right) - \mathcal{L}_b\left(\Omega^{-1}{\rm tr}\chi\right) + \frac{1}{2}\Omega^{-2}\mathcal{L}_b\left(\Omega{\rm tr}\chi\right)\Big)
\\ \nonumber &\qquad +\Omega{\rm tr}\underline{\chi}\Big(4\Omega^{-1}\nabla_4\left(\Omega\underline{\omega}\right) + 4\left(\Omega^{-1}{\rm tr}\chi\right)\left(\Omega\underline{\omega}\right) + 4\eta\cdot\underline{\eta} - \left|\eta\right|^2
\\ \nonumber &\qquad \qquad  -\frac{1}{2}\left(\Omega^{-1}\hat{\chi}\right)\cdot\left(\slashed{\nabla}\hat{\otimes}b\right) - \mathcal{L}_b\left(\Omega^{-1}{\rm tr}\chi\right) + \frac{1}{2}\Omega^{-2}\mathcal{L}_b\left(\Omega{\rm tr}\chi\right)\Big) +\left(2\left(\Omega\nabla_3\underline{\eta}\right)+\left(\Omega{\rm tr}\underline{\chi} \right)\underline{\eta}\right)\cdot\underline{\eta}
\\ \nonumber &\qquad + \Omega\nabla_3\left(-\frac{1}{2}\Omega^{-2}\mathcal{L}_{\partial_v}\slashed{\rm div}b -\frac{1}{2}\Omega^{-1}{\rm tr}\chi\slashed{\rm div}b\right)+\Omega{\rm tr}\underline{\chi}\left(-\frac{1}{2}\Omega^{-2}\mathcal{L}_{\partial_v}\slashed{\rm div}b -\frac{1}{2}\Omega^{-1}{\rm tr}\chi\slashed{\rm div}b\right)
\\ \nonumber &\qquad -4\slashed{\Delta}\left(\Omega\underline{\omega}\right) -4\slashed{\rm div}\left(\Omega\hat{\underline{\chi}}\right)\cdot\slashed{\nabla}\log\Omega  - 2\left(\Omega\hat{\underline{\chi}}\right)^{AB}\left(\slashed{\nabla}\hat{\otimes}\slashed{\nabla}\right)_{AB}\log\Omega =
\\ \nonumber &\qquad \qquad \qquad \qquad \qquad \qquad \qquad \qquad \qquad \qquad \qquad  \Omega\nabla_3\left({\rm Ric}_{34} + \frac{v}{u}{\rm Ric}_{44}\right) + \Omega{\rm tr}\underline{\chi}\left({\rm Ric}_{34} + \frac{v}{u}{\rm Ric}_{44}\right).
\end{align}

Now our plan will be to use Lemma~\ref{divbequnaa} (after an application of $\mathcal{L}_{\partial_v}$) to replace 
\[\Omega\nabla_3\left(-\frac{1}{2}\Omega^{-2}\mathcal{L}_{\partial_v}\slashed{\rm div}b -\frac{1}{2}\Omega^{-1}{\rm tr}\chi\slashed{\rm div}b\right).\]
We start by applying $\Omega^{-2}\mathcal{L}_{\partial_v}$ to~\eqref{eqnyaydivb} and using Lemma~\ref{j32ioj3io21} in order to obtain  
\begin{align}\label{2kmelml1l}
&\left(-\frac{v}{u}\mathcal{L}_{\partial_v} + b\cdot\slashed{\nabla}\right)\left(\Omega^{-2}\mathcal{L}_{\partial_v}\slashed{\rm div}b\right)+ \Omega^{-2}\left(\mathcal{L}_{\partial_v}b\right)^A\slashed{\nabla}_A\slashed{\rm div}b -\Omega^{-2}\mathcal{L}_{\partial_v}\left(\frac{v}{u}\Omega{\rm tr}\chi\slashed{\rm div}b \right) 
\\ \nonumber &\qquad + \Omega^{-2}\left(\mathcal{L}_{\partial_v}\slashed{\rm div}b\right)\slashed{\rm div}b+8\Omega^{-2}\mathcal{L}_{\partial_v}\left(\Omega\underline{\omega}\right)\left(u^{-1}+\frac{\slashed{\rm div}b}{2}\right) \\ \nonumber &\qquad  = \Omega^{-2}\left(-\Theta + \left(\Omega\hat{\chi}\right)\left(u^{-1}+\frac{1}{2}\slashed{\rm div}b\right)\right)\cdot \left(\slashed{\nabla}\hat{\otimes}b\right) +\frac{v}{u}\left(\Omega^{-1}\hat{\chi}\right)\cdot \Omega\nabla_4\left(\slashed{\nabla}\hat{\otimes}b\right)-\frac{1}{2}\Omega^{-2}\left(\slashed{\nabla}\hat{\otimes}\left(\mathcal{L}_{\partial_v}b\right)\right)\cdot\slashed{\nabla}\hat{\otimes}b
\\ \nonumber &\qquad +\Omega^{-2}\mathcal{L}_{\partial_v}\left( \Omega^2\frac{v}{u} b\cdot\slashed{\nabla}\left(\Omega^{-1}{\rm tr}\chi\right)\right) +\Omega^{-2}\mathcal{L}_{\partial_v}\left(- \Omega^2{\rm Ric}_{33} + \frac{v^2}{u^2}\Omega^2{\rm Ric}_{44}\right).
\end{align}
Using~\eqref{2kmelml1l}, ~\eqref{eqnyaydivb}, and self-similarity to simplify, we obtain
\begin{align}\label{km2km2o392}&\Omega\nabla_3\left(-\frac{1}{2}\Omega^{-2}\mathcal{L}_{\partial_v}\slashed{\rm div}b -\frac{1}{2}\Omega^{-1}{\rm tr}\chi\slashed{\rm div}b\right)+\Omega{\rm tr}\underline{\chi}\left(-\frac{1}{2}\Omega^{-2}\mathcal{L}_{\partial_v}\slashed{\rm div}b -\frac{1}{2}\Omega^{-1}{\rm tr}\chi\slashed{\rm div}b\right) =
\\ \nonumber &\left(-\frac{v}{u}\mathcal{L}_{\partial_v}+\mathcal{L}_b\right)\left(-\frac{1}{2}\Omega^{-2}\mathcal{L}_{\partial_v}\slashed{\rm div}b -\frac{1}{2}\Omega^{-1}{\rm tr}\chi\slashed{\rm div}b\right)
\\ \nonumber &\qquad +\left(\Omega{\rm tr}\underline{\chi}-\frac{2}{u}\right)\left(-\frac{1}{2}\Omega^{-2}\mathcal{L}_{\partial_v}\slashed{\rm div}b -\frac{1}{2}\Omega^{-1}{\rm tr}\chi\slashed{\rm div}b\right) =
\\ \nonumber &\qquad \frac{1}{2}\Omega^{-2}\left(\mathcal{L}_{\partial_v}b\right)^A\slashed{\nabla}_A\slashed{\rm div}b +\frac{1}{4}\Omega^{-2}\left(\slashed{\nabla}\hat{\otimes}\mathcal{L}_{\partial_v}b\right)\cdot\left(\slashed{\nabla}\hat{\otimes}b\right)+
\\ \nonumber &\qquad -\frac{v}{u}\left(\mathcal{L}_{\partial_v}\log\Omega\right)\mathcal{L}_b\left(\Omega^{-1}{\rm tr}\chi\right) -\frac{v}{u}\left(\mathcal{L}_{\partial_v}\log\Omega\right)\left(\Omega^{-1}{\rm tr}\chi\right)\slashed{\rm div}b
\\ \nonumber &\qquad 4\Omega^{-2}\mathcal{L}_{\partial_v}\left(\Omega\underline{\omega}\right)\left(u^{-1}+\frac{1}{2}\slashed{\rm div}b\right) 
 + \frac{1}{2}\Omega^{-2}\Theta \cdot\left(\slashed{\nabla}\hat{\otimes}b\right) -\frac{1}{2}\left(\Omega^{-1}\hat{\chi}\right)\cdot\left(\slashed{\nabla}\hat{\otimes}b\right)\left(u^{-1}+\frac{1}{2}\slashed{\rm div}b\right)
 \\ \nonumber &\qquad +4\left(\Omega\underline{\omega}\right)\left(\Omega^{-1}{\rm tr}\chi\right)\left(u^{-1}+\frac{1}{2}\slashed{\rm div}b\right) -\frac{1}{2}\mathcal{L}_b\left(\Omega^{-1}{\rm tr}\chi\right)\slashed{\rm div}b+\frac{1}{4}\left(\Omega^{-1}{\rm tr}\chi\right)\left(\slashed{\rm div}b\right)^2 
 \\ \nonumber &\qquad + \Omega^2\frac{v}{u}\left(\Omega^{-1}\hat{\chi}\right)\cdot\left(\slashed{\nabla}\hat{\otimes}b\right)\left(-\frac{1}{2}\Omega^{-1}{\rm tr}\chi\right) + \frac{1}{8}\left(\Omega^{-1}{\rm tr}\chi\right)\left|\slashed{\nabla}\hat{\otimes}b\right|^2 - \frac{1}{2}\Omega^2\left(\Omega^{-1}{\rm tr}\chi\right)\frac{v}{u}\mathcal{L}_b\left(\Omega^{-1}{\rm tr}\chi\right)
 \\ \nonumber &\qquad -\frac{1}{2}\mathcal{L}_{\partial_v}\left(\frac{v}{u}\mathcal{L}_b\left(\Omega^{-1}{\rm tr}\chi\right)\right) + \frac{1}{2}\left(\Omega^{-2}\mathcal{L}_{\partial_v}\left(\Omega^{-2}{\rm Ric}_{33}-\frac{v^2}{u^2}\Omega^2{\rm Ric}_{44}\right) + \Omega^{-1}{\rm tr}\chi \left(\Omega^{-2}{\rm Ric}_{33}-\frac{v^2}{u^2}\Omega^2{\rm Ric}_{44}\right)\right).
\end{align} 

Now we plug~\eqref{km2km2o392} into~\eqref{3m3omo203} and also use  Lemma~\ref{2mo2o3o2} (along with self-similarity) to replace the $\Omega\nabla_3\underline{\eta}$. After some calculations, we eventually obtain~\eqref{3i3ei2i4i24u}.

\end{proof}
\begin{remark}\label{2i4oij2oij452}The equation derived by Lemma~\ref{2km2omo34} will be significant for us for two key reasons. First of all, after setting the Ricci curvature components to vanish, it may be considered as a linear equation for $\Omega\underline{\omega}$ (and hence the lapse) which only couples nonlinearly to the other quantities. This should be contrasted with the equation for the lapse found in Lemma~\ref{2m2om3o3923332} (see Remark~\ref{2lm3omo2}).

Secondly, other than the terms contained in
\begin{equation}\label{2lm3omo2o3}
4\Omega^{-1}\nabla_4\left(\Omega\underline{\omega}\right)  -\frac{1}{2}\left(\Omega^{-1}\hat{\chi}\right)\cdot\left(\slashed{\nabla}\hat{\otimes}b\right)+ 4\eta\cdot\underline{\eta},
\end{equation}
the nonlinear terms are all ``regular'' as $v\to 0$ in the sense that, unless the term is multiplied by a $\frac{v}{u}$, we only see terms proportional to $\eta$, $b$, $\Omega\underline{\omega}$, ${\rm tr}\chi$, or $\Theta$ (see Definition~\ref{thetaisdefinedhere}). This is consistent with eventually showing that~\eqref{2lm3omo2o3} is uniformly bounded as $v\to 0$. In view of Remark~\ref{2lm3omo2}, this is consistent with showing that $\slashed{\rm div}\eta$ is bounded as $v\to 0$. 
\end{remark}

Next we will derive an equation for the torsion $\zeta$.
\begin{lemma}\label{3kdo2}Let $\left(\mathcal{M},g\right)$ be a Lorentzian manifold with a self-similar foliation. Then the following equation holds:
\begin{align}\label{thewavezetastartsojqj}
&2\frac{v}{u}\nabla_{\partial_v}\zeta_A -\mathcal{L}_b\zeta +\mathcal{L}_b\slashed{\nabla}_A\log\Omega -\left(u^{-1}+\frac{1}{2}\slashed{\rm div}b\right)\underline{\eta}_A- \frac{1}{2}\left(\slashed{\nabla}\hat{\otimes}b\right)_A^{\ \ B}\underline{\eta}_B  +2\Omega\left(\frac{v}{u}\chi_A^{\ \ B}-\underline{\chi}_A^{\ \ B}\right)\zeta_B = \\ \nonumber &\qquad \frac{1}{2}\slashed{\rm div}\left(\slashed{\nabla}\hat{\otimes}b\right)_A - \frac{1}{2}\slashed{\nabla}_A\slashed{\rm div}b +\zeta\left(\frac{v}{u}\Omega\hat{\chi} - \Omega\hat{\underline{\chi}}\right) -\frac{1}{2}\slashed{\nabla}\log\Omega\cdot\left(\slashed{\nabla}\hat{\otimes}b\right) 
\\ \nonumber &\qquad \qquad + \frac{1}{2}\zeta\left(\Omega{\rm tr}\underline{\chi} - \frac{v}{u}\Omega{\rm tr}\chi\right) + \slashed{\nabla}\log\Omega\left(u^{-1}+\frac{1}{2}\slashed{\rm div}b\right) - \frac{v}{u}\Omega{\rm Ric}_{4A} - \Omega{\rm Ric}_{3A}.
\end{align}
\end{lemma}
\begin{proof}Using self-similarity and computing in a coordinate frame, we have
\begin{align}\label{eqnforundeself}
\nabla_3\underline{\eta}_A &= \mathcal{L}_{e_3}\underline{\eta}_A - \underline{\chi}_A^{\ \ B}\underline{\eta}_B
\\ \nonumber &= -\frac{v}{u}\mathcal{L}_{e_4}\underline{\eta}_A + \Omega^{-1}\mathcal{L}_b\underline{\eta}_A - \underline{\chi}_A^{\ \ B}\underline{\eta}_B
\\ \nonumber &= -\frac{v}{u}\nabla_4\underline{\eta}_A + \Omega^{-1}\mathcal{L}_b\underline{\eta}_A -\Omega^{-1}\left(u^{-1}+\frac{1}{2}\slashed{\rm div}b\right)\underline{\eta}_A- \frac{1}{2}\Omega^{-1}\left(\slashed{\nabla}\hat{\otimes}b\right)_A^{\ \ B}\underline{\eta}_B.
\end{align}

The $\nabla_4$ equation for $\eta$ may be re-written as 
\begin{equation}\label{4etarewrute}
2\nabla_4\zeta_A + \nabla_4\underline{\eta}_A = -2\chi_A^{\ \ B}\zeta_B - \frac{1}{2}R_{A434}.
\end{equation}

Using~\eqref{eqnforundeself} we may re-write the $\nabla_3$ equation for $\underline{\eta}$ as
\begin{equation}\label{3etakkpw}
-\frac{v}{u}\nabla_4\underline{\eta}_A + \Omega^{-1}\mathcal{L}_b\underline{\eta}_A -\Omega^{-1}\left(u^{-1}+\frac{1}{2}\slashed{\rm div}b\right)\underline{\eta}_A- \frac{1}{2}\Omega^{-1}\left(\slashed{\nabla}\hat{\otimes}b\right)_A^{\ \ B}\underline{\eta}_B = 2\underline{\chi}_A^{\ \ B}\zeta_B - \frac{1}{2}R_{A343}.
\end{equation}

Now we multiply~\eqref{4etarewrute} by $\frac{v}{u}$, add the result to~\eqref{3etakkpw}, and multiply through by $\Omega$. We obtain
\begin{align}\label{thewavezetastarts}
&2\frac{v}{u}\nabla_{\partial_v}\zeta_A + \mathcal{L}_b\underline{\eta}_A -\left(u^{-1}+\frac{1}{2}\slashed{\rm div}b\right)\underline{\eta}_A- \frac{1}{2}\left(\slashed{\nabla}\hat{\otimes}b\right)_A^{\ \ B}\underline{\eta}_B  = \\ \nonumber &\qquad -2\Omega\left(\frac{v}{u}\chi_A^{\ \ B}-\underline{\chi}_A^{\ \ B}\right)\zeta_B -\frac{1}{2}\frac{v}{u}\Omega R_{A434} -\frac{1}{2} \Omega R_{A343}.
\end{align}

Next, we use the Codazzi equations~\eqref{tcod1} and~\eqref{tcod2} and Lemma~\ref{thefirstrelations} to simply $-\frac{1}{2}\frac{v}{u}\Omega R_{A434} -\frac{1}{2} \Omega R_{A343}$:
\begin{align}\label{simpbetabegag}
&-\frac{1}{2}\frac{v}{u}\Omega R_{A434} - \frac{1}{2}\Omega R_{A343} = 
\\ \nonumber &\qquad \frac{1}{2}\slashed{\rm div}\left(\slashed{\nabla}\hat{\otimes}b\right)_A - \frac{1}{2}\slashed{\nabla}_A\slashed{\rm div}b - \eta\left(\Omega\hat{\underline{\chi}}\right) -\frac{v}{u}\underline{\eta}\left(\Omega\hat{\chi}\right) + \frac{1}{2}\eta\left(\Omega{\rm tr}\underline{\chi}\right) + \frac{1}{2}\frac{v}{u}\underline{\eta}\left(\Omega{\rm tr}\chi\right) - \frac{v}{u}\Omega{\rm Ric}_{4A} - \Omega{\rm Ric}_{3A}=
\\ \nonumber & \qquad  \frac{1}{2}\slashed{\rm div}\left(\slashed{\nabla}\hat{\otimes}b\right)_A - \frac{1}{2}\slashed{\nabla}_A\slashed{\rm div}b +\zeta\left(\frac{v}{u}\Omega\hat{\chi} - \Omega\hat{\underline{\chi}}\right) -\frac{1}{2}\slashed{\nabla}\log\Omega\cdot\left(\slashed{\nabla}\hat{\otimes}b\right) 
\\ \nonumber &\qquad \qquad + \frac{1}{2}\zeta\left(\Omega{\rm tr}\underline{\chi} - \frac{v}{u}\Omega{\rm tr}\chi\right) + \slashed{\nabla}\log\Omega\left(u^{-1}+\frac{1}{2}\slashed{\rm div}b\right) - \frac{v}{u}\Omega{\rm Ric}_{4A} - \Omega{\rm Ric}_{3A}.
\end{align}

\end{proof}
\begin{remark}The most important aspect of this equation is that if we apply $\left(v+1\right)\slashed{\rm curl}$ to both sides of the equation, then we obtain a model second order equation for $\slashed{\rm curl}b$ which only couples nonlinearly to the other quantities. 
\end{remark}

The next identity will be useful when considered in conjunction with Lemma~\ref{3kdo2}.
\begin{lemma}\label{emfkeo3}Let $\left(\mathcal{M},g\right)$ be a Lorentzian manifold with a self-similar foliation. Then the following equations hold:
\begin{align}\label{mfm3m3o}
&\Omega^2\left(\slashed{\nabla}\hat{\otimes}\zeta\right) =-\frac{1}{4}\Omega\nabla_4\left(\slashed{\nabla}\hat{\otimes}b\right)+\frac{1}{2}\mathcal{L}_b\left(\Omega\hat{\chi}\right)-2\Omega^2\left(\slashed{\nabla}\log\Omega\hat{\otimes}\zeta\right)+ H_1,
\end{align}
\begin{align}\label{mkek3}
&\left(1-\frac{v}{u}\right)\slashed{\nabla}\hat{\otimes}\left[2\Omega \frac{v}{u}\nabla_4\zeta - \mathcal{L}_b\zeta\right] = 
\\ \nonumber &\qquad \left(-\frac{1}{2}\frac{v}{u}\Omega\nabla_4 + \frac{1}{2}\mathcal{L}_b\right)\left(\left(1-\frac{v}{u}\right)\left(\Omega^{-1}\nabla_4\left(\slashed{\nabla}\hat{\otimes}b\right) +8\left(\slashed{\nabla}\log\Omega\hat{\otimes}\zeta\right)-4\Omega^{-2}H_1\right) \right)
\\ \nonumber &\qquad + \left[\frac{v}{u}\Omega\nabla_4-\frac{1}{2}\mathcal{L}_b,\Omega^{-2}\left(1-\frac{v}{u}\right)\mathcal{L}_b\right]\left(\Omega\hat{\chi}\right) + 2\frac{v}{u}\left[\left(1-\frac{v}{u}\right)\slashed{\nabla}\hat{\otimes},\Omega\nabla_4\right]\zeta + \left[\mathcal{L}_b,\left(1-\frac{v}{u}\right)\slashed{\nabla}\hat{\otimes}\right]\zeta
\\ \nonumber &\qquad +2\Omega^{-2}\left(1-\frac{v}{u}\right)\mathcal{L}_b\log\Omega \left(\frac{v}{u}\Omega\nabla_4 - \frac{1}{2}\mathcal{L}_b\right)\left(\Omega\hat{\chi}\right)-\left(1-\frac{v}{u}\right)\mathcal{L}_b\left( \Omega^{-2}H_2+\widehat{\rm Ric}\right),
\end{align}
where we define
\[(H_1)_{AB} \doteq - \frac{1}{2}\slashed{\rm div}b\left(\Omega\hat{\chi}\right)_{AB} - \frac{1}{2}\left(\Omega\hat{\chi}\right)^C_{\ \ (A}\left(\slashed{\nabla}\hat{\otimes}b\right)_{B)C},\]
\begin{align*}
(H_2)_{AB} &\doteq  \left(\slashed{\nabla}\hat{\otimes}b\right)^C_{\ \ (A}\left(\Omega\hat{\chi}\right)_{B)C} 
 +\frac{1}{2}{\rm div}b\left(\Omega\hat{\chi}\right)_{AB}+\frac{v}{u}\Omega{\rm tr}\chi\left(\Omega \hat{\chi}\right)_{AB} 
 \\ \nonumber &\qquad +\Omega^2\left[\left(\left(\slashed{\nabla}\hat\otimes \slashed{\nabla}\log\Omega\right)_{AB} + \left(\eta\hat\otimes \eta\right)_{AB} - \frac{1}{2}\Omega^{-1}{\rm tr}\chi \left(\slashed{\nabla}\hat{\otimes}b\right)\right)\right].
 \end{align*}
\end{lemma}
\begin{proof}Equation~\eqref{mfm3m3o} is an immediate consequence of Lemma~\ref{othercommutelemma} and the fact from Lemma~\ref{dermetrcomp} that $\zeta^A = -\frac{1}{4}\Omega^{-2}\mathcal{L}_{\partial_v}b^A$. Using this, we can re-write~\eqref{kwdkodwok23dg} as 
\begin{align}\label{emmfo3mooes}
& -\frac{v}{u}\Omega\nabla_4\left(\Omega\hat{\chi}\right)_{AB} +\frac{1}{2}\mathcal{L}_b\left(\Omega\hat{\chi}\right)_{AB} =
\\ \nonumber &\qquad  -\frac{1}{4}\Omega\nabla_4\left(\slashed{\nabla}\hat{\otimes}b\right)_{AB} -2\Omega^2\left(\slashed{\nabla}\log\Omega\hat{\otimes}\zeta\right)+(H_1)_{AB} + (H_2)_{AB} + \Omega^2\widehat{{\rm Ric}}_{AB}. 
 \end{align}

 \begin{align*}
&\left(1-\frac{v}{u}\right)\slashed{\nabla}\hat{\otimes}\left[2\Omega \frac{v}{u}\nabla_4\zeta - \mathcal{L}_b\zeta\right] = 
\\ \nonumber &\qquad 2\frac{v}{u}\Omega\nabla_4\left(\left(1-\frac{v}{u}\right)\slashed{\nabla}\hat{\otimes}\zeta\right) - \mathcal{L}_b\left(\left(1-\frac{v}{u}\right)\slashed{\nabla}\hat{\otimes}\zeta\right)  + 2\frac{v}{u}\left[\left(1-\frac{v}{u}\right)\slashed{\nabla}\hat{\otimes},\Omega\nabla_4\right]\zeta + \left[\mathcal{L}_b,\left(1-\frac{v}{u}\right)\slashed{\nabla}\hat{\otimes}\right]\zeta.
\end{align*}
Combining with~\eqref{mfm3m3o} and~\eqref{kwdkodwok23dg} then leads to
\begin{align*}
&2\frac{v}{u}\Omega\nabla_4\left(\left(1-\frac{v}{u}\right)\slashed{\nabla}\hat{\otimes}\zeta\right) - \mathcal{L}_b\left(\left(1-\frac{v}{u}\right)\slashed{\nabla}\hat{\otimes}\zeta\right)   =
\\ \nonumber &\qquad  \frac{v}{u}\Omega\nabla_4\left(\left(1-\frac{v}{u}\right)\left(\Omega^{-2}\left(-\frac{1}{2}\Omega\nabla_4\left(\slashed{\nabla}\hat{\otimes}b\right)  +\mathcal{L}_b\left(\Omega\hat{\chi}\right) -4\Omega^2\left(\slashed{\nabla}\log\Omega\hat{\otimes}\zeta\right)+ 2H_1\right)\right)\right)
\\ \nonumber &\qquad - \frac{1}{2}\mathcal{L}_b\left(\left(1-\frac{v}{u}\right)\left(\Omega^{-2}\left(-\frac{1}{2}\Omega\nabla_4\left(\slashed{\nabla}\hat{\otimes}b\right) + \mathcal{L}_b\left(\Omega\hat{\chi}\right) -4\Omega^2\left(\slashed{\nabla}\log\Omega\hat{\otimes}\zeta\right)+ 2H_1\right)\right)\right)
\\ \nonumber &= \left(-\frac{1}{2}\frac{v}{u}\Omega\nabla_4 + \frac{1}{4}\mathcal{L}_b\right)\left(\left(1-\frac{v}{u}\right)\left(\Omega^{-1}\nabla_4\left(\slashed{\nabla}\hat{\otimes}b\right) +8\left(\slashed{\nabla}\log\Omega\hat{\otimes}\zeta\right)-4\Omega^{-2}H_1\right) \right)
\\ \nonumber &\qquad +\left[ \frac{v}{u}\Omega\nabla_4-\frac{1}{2}\mathcal{L}_b,\Omega^{-2}\left(1-\frac{v}{u}\right)\mathcal{L}_b\right]\left(\Omega\hat{\chi}\right) 
\\ \nonumber &\qquad +2\Omega^{-2}\left(1-\frac{v}{u}\right)\mathcal{L}_b\log\Omega \left(\frac{v}{u}\Omega\nabla_4 - \frac{1}{2}\mathcal{L}_b\right)\left(\Omega\hat{\chi}\right)
\\ \nonumber &\qquad +\left(1-\frac{v}{u}\right)\mathcal{L}_b\left(\frac{1}{4}\Omega^{-1}\nabla_4\left(\slashed{\nabla}\hat{\otimes}b\right) + 2\left(\slashed{\nabla}\log\Omega\hat{\otimes}\zeta\right)- \Omega^{-2}H_1 - \Omega^{-2}H_2 - \widehat{\rm Ric}\right). 
\end{align*}

\end{proof}

In this next equation we linearize the $e_4$-Raychaudhuri equation around the solution ${\rm tr}\chi = 2(v+1)^{-1}$. (This lemma does not require self-similarity.)
\begin{lemma}\label{2ijini2}Let $\left(\mathcal{M},g\right)$ be a Lorentzian manifold. Then the following equation holds along $u=-1$:
\begin{align}\label{1opj3opjoii9459o}
&\mathcal{L}_{\partial_v}\left(\Omega^{-1}{\rm tr}\chi - \frac{2}{v+1}\right) + \frac{2\Omega^2}{v+1}\left(\Omega^{-1}{\rm tr}\chi - \frac{2}{v+1}\right) =
\\ \nonumber &\qquad \frac{2\left(1-\Omega^2\right)}{(v+1)^2} -\frac{1}{2}\Omega^2\left(\Omega^{-1}{\rm tr}\chi-2(v+1)^{-1}\right)^2-\Omega^2\left|\Omega^{-1}\hat{\chi}\right|^2 -{\rm Ric}_{44}
\end{align}
\end{lemma}
\begin{proof}This is a straightforward consequence of~\eqref{4trchi}. 
\end{proof}

This next equation will be used to relate $\slashed{\rm div}\eta$ with the Gauss curvature.
\begin{lemma}\label{3kn2knk2k}Let $\left(\mathcal{M},g\right)$ be a Lorentzian manifold with a self-similar foliation. Then the following equations hold along $u=-1$:
\begin{align}\label{2momoo3}
&\left(-1+\slashed{\rm div}b + \frac{2v\Omega^2}{v+1} -4\Omega\underline{\omega}\right) \left(\Omega^{-1}{\rm tr}\chi-\frac{2}{v+1}\right) + \mathcal{L}_b\left(\Omega^{-1}{\rm tr}\chi-\frac{2}{v+1}\right)
\\ \nonumber &\qquad +\frac{v}{2}\Omega^2\left(\Omega^{-1}{\rm tr}\chi-2(v+1)^{-1}\right)^2
 -v\Omega^2\left|\Omega^{-1}\hat{\chi}\right|^2 
 - 2\left|\eta\right|^2
 +\frac{2}{v+1}\left(\slashed{\rm div}b - 4\Omega\underline{\omega}\right) - \frac{2v\left(1-\Omega^2\right)}{(v+1)^2}=  \\ \nonumber &\qquad
  2\slashed{\rm div}\eta -2\left(K-\frac{1}{(v+1)^2}\right)+ R +{\rm Ric}_{34} +v{\rm Ric}_{44}
  \end{align}
\end{lemma}
\begin{proof}From~\eqref{3trchi},~\eqref{slashr} and self-similarity, we may derive along $\{u=-1\}$:
\begin{align*}
&v\mathcal{L}_{\partial_v}\left(\Omega^{-1}{\rm tr}\chi\right) + \Omega^{-1}{\rm tr}\chi + \mathcal{L}_b\left(\Omega^{-1}{\rm tr}\chi\right) +{\rm tr}\chi{\rm tr}\underline{\chi} = -2K+ R +{\rm Ric}_{34}+ 4\underline{\omega}{\rm tr}\chi + 2\slashed{\rm div}\eta + 2\left|\eta\right|^2.
\end{align*}
We may further simplify to
\begin{align}\label{2omimrti92}
&v\mathcal{L}_{\partial_v}\left(\Omega^{-1}{\rm tr}\chi-\frac{2}{v+1}\right) + \left(\Omega^{-1}{\rm tr}\chi-\frac{2}{v+1}\right) + \mathcal{L}_b\left(\Omega^{-1}{\rm tr}\chi-\frac{2}{v+1}\right) 
\\ \nonumber &\qquad +\left(\Omega{\rm tr}\underline{\chi} -4\Omega\underline{\omega}\right)\left(\Omega^{-1}{\rm tr}\chi-\frac{2}{v+1}\right)=
  -2\left(K-\frac{1}{(v+1)^2}\right)+ R +{\rm Ric}_{34} + 2\slashed{\rm div}\eta + 2\left|\eta\right|^2
  \\ \nonumber &\qquad -\frac{2}{v+1}\left(\slashed{\rm div}b - 4\Omega\underline{\omega} +v\Omega^2\left(\Omega^{-1}{\rm tr}\chi - \frac{2}{v+1}\right)\right) + \frac{4v\left(1-\Omega^2\right)}{(v+1)^2}.
\end{align}
Finally, we may use Lemma~\ref{2ijini2} to obtain~\eqref{2momoo3}.

\end{proof}

\section{Overview of the Existence Argument}\label{2omoo09882uhh3hh}
In this section, we discuss the specific scheme we will use to construct our spacetime. 
\subsection{The Self-Similar Reduction}\label{ooiiouou832}
Let us consider a metric $g$ in the double-null form~\eqref{doublenullisg} in the open set 
\begin{equation}\label{2oj3omio2}
\mathscr{U} \doteq \left\{\left(u,v,\theta^A\right) \in (-\infty,0) \times (-\infty,0)\times \mathbb{S}^2 : v > u\right\}.
\end{equation}
It is then a consequence of Proposition~\ref{thenullstructeqns} that ${\rm Ric}\left(g\right) = 0$ if and only if the equations listed in Proposition~\ref{thenullstructeqns} all hold with the Ricci terms set to $0$. Let us now further suppose that $g$ is self-similar in the sense of Definition~\ref{folselfsim}. Then, because of the invariance of $\mathscr{U}$ under the map $\left(u,v,\theta^A\right) \mapsto \left(\lambda u, \lambda v,\theta^A\right)$ for any $\lambda > 0$, it is clear that we will have ${\rm Ric}\left(g\right) = 0$ if and only if the equations listed in Proposition~\ref{thenullstructeqns} hold, with the Ricci components set to $0$, in the region
\[\underline{\mathscr{U}} \doteq \mathscr{U} \cap \left\{u=-1\right\}.\]

In order to take advantage of this restriction to $\underline{\mathscr{U}}$, we would like to phrase these equations purely in terms of the restrictions of $\Omega$, $b$, and $\slashed{g}$ to $\underline{\mathscr{U}}$. We do this as follows: Suppose that $\Omega$, $b$, and $\slashed{g}$ are given as an $\mathbb{S}^2_{-1,v}$ function, $1$-form, and symmetric $(0,2)$-tensor respectively for $v \in (-1,0)$; that is, we suppose we are given $\Omega$, $b$, and $\slashed{g}$ along $\underline{\mathscr{U}}$. Then, via the formulas of Definition~\ref{folselfsim}, these induce unique self-similar extensions of $\Omega$, $b$, and $\slashed{g}$ to the entire region $\mathscr{U}$. In turn this allows us to define a metric $g$ on $\mathscr{U}$ from the formula~\eqref{doublenullisg}. We may then compute any double-null gauge quantity associated to $g$ in the region $\mathscr{U}$ by way of the usual formulas, that is, using Lemma~\ref{dermetrcomp}. For example, $\Omega\nabla_3\log\Omega$ is now well-defined, even though we originally only specified $\Omega$ as a function of $v$ and $\theta^A$. Because we have extended our metric components off of $\underline{\mathscr{U}}$ in a self-similar fashion, one may use the formulas from Definition~\ref{folselfsim} to relate the $\mathcal{L}_{\partial_u}$ derivative of any double-null quantity with the derivatives $\mathcal{L}_{\partial_v}$ and $\mathcal{L}_b$, which are tangential to $\underline{\mathscr{U}}$, and the quantity itself. Since all double-null quantities are defined in terms of $\Omega$, $b$, and $\slashed{g}$, by systematically replacing derivatives in this fashion, we may use Proposition~\ref{thenullstructeqns} to derive a set of equations along $(v,\theta^A) \in (-1,0)\times \mathbb{S}^2$ which hold for $\Omega$, $b$, and $\slashed{g}$ if and only if the induced self-similar metric $g$ will satisfy ${\rm Ric}(g) = 0$. We call these equations the \emph{self-similar double-null Einstein equations}.

As is well-known, when considered as a system of differential equations for $\Omega$, $b$, and $\slashed{g}$, the equations of the double-null gauge, that is, the equations of Proposition~\ref{thenullstructeqns} with all Ricci components set to $0$, are \emph{overdetermined}. Naturally, the overdetermined nature of the system is inherited by the self-similar double-null Einstein equations. Thus, our strategy is as follows. We first select a determined subset of the self-similar double-null Einstein equations. We then show that these may be solved by running suitable iteration arguments. Finally, we run a ``preservation of constraints'' argument (see Section~\ref{propagatetheconstraintsforever}) to show that in fact all of the components of the Ricci tensor must vanish. 

In Section~\ref{2om3om42o43984u4j2} below we list the specific set of equations which we use to solve for $\Omega$, $b$, and $\slashed{g}$. Then, in Section~\ref{2m2o2mo1o00050033} we list the specific sequence of iteration arguments which are used to solve these equations. Finally, in Section~\ref{moo3ok2o49} we state a theorem which includes all of the results of the iteration process. All together this serves as an outline for Sections~\ref{lapthesection} to \ref{foim2io34o5u92hj5991} of the paper. 

\subsection{The Specific Equations}\label{2om3om42o43984u4j2}
In this section  we will give explicitly the subset of the self-similar double null equations that we will solve. 
\subsubsection{Artificial Variables}
In addition to the quantities $\Omega$, $b$, $\slashed{g}$ we will also introduce two additional ``artificial variables'' in the equations we describe below. There are extra unknowns which during the propagation of constraints argument will eventually be shown to agree with a suitable quantity defined in terms of $\Omega$, $b$, $\slashed{g}$. The first artificial variable is an $\mathbb{S}^2_{-1,v}$ $1$-form $\mathfrak{n}$. After the propagation of constraints argument has concluded, we will see that $\mathfrak{n} = \eta$. The second artificial unkown is a $1$-form $\mathfrak{j}$ defined along $\mathbb{S}^2$. After the propagation of constraints argument has concluded, we will see that $\mathfrak{j} = \eta|_{v=0}$. 
\subsubsection{Seed Data}\label{seeddatasection}
The basic seed data for our construction are two scalar functions  $T_{\rm low}: \mathbb{S}^2 \to \mathbb{R}$ and $T_{\rm high}: \mathbb{S}^2 \to \mathbb{R}$ which are allowed to be arbitrary functions subject to the requirements that
\begin{equation}\label{oiiooi909198}
\left(1-\mathcal{P}_{1 \leq \ell \leq \ell_0}\right)T_{\rm low} = \left(1-\mathcal{P}_{\ell > \ell_0}\right)T_{\rm high} = 0, \qquad \left\vert\left\vert \left(T_{\rm low},T_{\rm high}\right)\right\vert\right\vert_{\mathring{H}^{N_1-3}\left(\mathbb{S}^2\right)} \lesssim 1.
\end{equation}
The functions $T_{\rm low}$ and $T_{\rm high}$ will enter as boundary conditions for our system via the following: 
\begin{equation}\label{2k24099j1nini4}
\mathcal{P}_{1 \leq \ell \leq \ell_0}\slashed{\rm curl}b|_{v=0} = \epsilon T_{\rm low},\qquad \mathcal{P}_{\ell > \ell_0}\slashed{\rm curl}\slashed{\rm div}\slashed{\nabla}\hat{\otimes}b|_{v=0} = \epsilon T_{\rm high}.
\end{equation}
\begin{remark}In view of the formula~\eqref{2m3momo2}, at the linear level, one sees that $T_{\rm high}$ will determine $\mathcal{P}_{\ell > \ell_0}\slashed{\rm curl}b|_{v=0}$. The advantage, however, of posing the value of $\mathcal{P}_{\ell > \ell_0}\slashed{\rm curl}\slashed{\rm div}\slashed{\nabla}\hat{\otimes}b$ as opposed to, say, $\mathcal{P}_{\ell > \ell_0}\slashed{\rm curl}b$, is that we will be able to avoid the presence in certain boundary equations (see Section~\ref{2o3om4o2mo4599j54}) of nonlinear terms proportional to a product of $\slashed{\nabla}K$ and $b$ (cf.~the formula~\eqref{2m3momo2}) which would pose some difficulties for our estimates from a regularity perspective (even those these are nonlinear terms).
\end{remark}
\subsubsection{Boundary Equations}\label{2o3om4o2mo4599j54}
In this section we will discuss various equations which will hold along the boundary $\{v = 0\}$ of our spacetime:
\begin{enumerate}
	\item We will require that the Gauss curvature $K$ satisfies the following equation along $\{v = 0\}$:
	 \begin{align}\label{2lkj34lj2lj4}
\mathcal{P}_{\ell > \ell_0}\slashed{\rm div}\mathfrak{j} -\mathcal{P}_{\ell > \ell_0}K &= \mathcal{P}_{\ell > \ell_0}\Bigg(\frac{1}{2}\left(-1+\slashed{\rm div}b  -4\Omega\underline{\omega}\right) \left(\Omega^{-1}{\rm tr}\chi-\frac{2}{v+1}\right) + \frac{1}{2}\mathcal{L}_b\left(\Omega^{-1}{\rm tr}\chi\right)
\\ \nonumber &\qquad \qquad \qquad
 - \left|\mathfrak{j}\right|^2
 +\frac{1}{v+1}\left(\slashed{\rm div}b - 4\Omega\underline{\omega}\right)\Bigg).
   \end{align}
  	In view of Lemma~\ref{3kn2knk2k} this equation is equivalent to 
	\begin{equation}\label{2om2omo3mo2o4}
	 \mathcal{P}_{\ell > \ell_0}\left(2\slashed{\rm div}\left(\mathfrak{j}-\eta\right) + 2\left|\mathfrak{j}\right|^2 - 2\left|\eta\right|^2\right)|_{v=0} = \mathcal{P}_{\ell > \ell_0}\left({\rm R}+{\rm Ric}_{34}\right)|_{v=0}.
	\end{equation}
	\item We will require that $\mathfrak{j}$ satisfies the following two equations along $\{v=0\}$:
	  \begin{align}\label{1i1o3ioiu43njnk}
&\mathcal{P}_{\ell \geq 1}\left( -\mathcal{L}_b\mathcal{P}_{\ell \geq 1}\slashed{\rm div}\mathfrak{j}+\left(2-\slashed{\rm div}b\right)\mathcal{P}_{\ell \geq 1}\slashed{\rm div}\mathfrak{j} +\left[\mathcal{P}_{\ell \geq 1}\slashed{\rm div}, -\mathcal{L}_b+\left(2-\slashed{\rm div}b\right)\right]\mathfrak{j}\right) = 
\\ \nonumber &\qquad \mathcal{P}_{\ell \geq 1}\left(K\slashed{\rm div}b  + \mathcal{L}_bK\right),
 \end{align}
 \begin{align}\label{kl23klj2lkj3j5}
 &\mathcal{P}_{\ell \geq 1}\left( -\mathcal{L}_b\mathcal{P}_{\ell \geq 1}\slashed{\rm curl}\mathfrak{j}+\left(2-\slashed{\rm div}b\right)\mathcal{P}_{\ell \geq 1}\slashed{\rm curl}\mathfrak{j} +\left[\mathcal{P}_{\ell \geq 1}\slashed{\rm curl}, -\mathcal{L}_b+\left(2-\slashed{\rm div}b\right)\right]\mathfrak{j}\right) = 
 \\ \nonumber &\qquad \frac{1}{2}\mathcal{P}_{\ell = 1}\left(\left(\slashed{\Delta}+2K\right)\slashed{\rm curl}b\right) + \mathcal{P}_{1 \leq \ell \leq \ell_0}\left(\slashed{\nabla}K\wedge b\right)+\frac{1}{2}\epsilon\left(\slashed{\Delta}+2K\right)T_{\rm low}+ \frac{1}{2}\epsilon T_{\rm high}
 \\ \nonumber &\qquad +\frac{1}{2}\left[\mathcal{P}_{2 \leq \ell \leq \ell_0},\slashed{\Delta}+2K\right]\slashed{\rm curl}b
 \end{align}
 In view of~\eqref{2k24099j1nini4}, Lemma~\ref{2mo2o3o2}, Lemma~\ref{thefirstrelations},~\eqref{2m3momo2}, and~\eqref{3o3oioi4}, and also the fact that we will have $\mathcal{P}_{\ell \geq 1}\left(\Omega\underline{\omega}\right)|_{v=0} = 0$ (see item~\ref{2k3om2o123} in Section~\ref{bulkbulkbulk} below), these equations are equivalent to
 \begin{align}\label{io32jiojio42}
&\mathcal{P}_{\ell \geq 1}\Big( -\mathcal{L}_b\mathcal{P}_{\ell \geq 1}\slashed{\rm div}\left(\mathfrak{j}-\eta\right)+\left(2-\slashed{\rm div}b\right)\mathcal{P}_{\ell \geq 1}\slashed{\rm div}\left(\mathfrak{j}-\eta\right)
\\ \nonumber &\qquad \qquad  \qquad +\left[\mathcal{P}_{\ell \geq 1}\slashed{\rm div}, -\mathcal{L}_b+\left(2-\slashed{\rm div}b\right)\right]\left(\mathfrak{j}-\eta\right)\Big)|_{v=0} =  \mathcal{P}_{\ell \geq 1}\slashed{\rm div}\left(\Omega{\rm Ric}_{3\cdot}\right)|_{v=0},
 \end{align}
 \begin{align}\label{afijijofiojoij}
 &\mathcal{P}_{\ell \geq 1}\Big( -\mathcal{L}_b\mathcal{P}_{\ell \geq 1}\slashed{\rm curl}\left(\mathfrak{j}-\eta\right)+\left(2-\slashed{\rm div}b\right)\mathcal{P}_{\ell \geq 1}\slashed{\rm curl}\left(\mathfrak{j}-\eta\right) 
 \\ \nonumber &\qquad \qquad \qquad +\left[\mathcal{P}_{\ell \geq 1}\slashed{\rm curl}, -\mathcal{L}_b+\left(2-\slashed{\rm div}b\right)\right]\left(\mathfrak{j}-\eta\right)\Big)|_{v=0} =  \mathcal{P}_{\ell \geq 1}\slashed{\rm curl}\left(\Omega{\rm Ric}_{3\cdot}\right)|_{v=0}.
 \end{align}

\end{enumerate}
\subsubsection{Bulk Equations}\label{bulkbulkbulk}
In this section we will give the equations which will be required to hold in the ``bulk'' $v \in (-1,0)$ as well as corresponding boundary conditions:
\begin{enumerate}
\item\label{in2ni13} For $\slashed{g}$, we shall require that the equation~\eqref{1opj3opjoii9459o} holds with the Ricci curvature terms dropped and that~\eqref{kwdkodwok23dg} holds with the Ricci curvature terms dropped and $\eta$ replaced with $\mathfrak{n}$. In view of Lemmas~\ref{2ijini2} and~\ref{03kdo3k5} these equations are equivalent to
\begin{equation}\label{2o2mom4o2}
{\rm Ric}_{44} = 0,
\end{equation}
\[\slashed{\nabla}\hat{\otimes}\left(\mathfrak{n}-\eta\right) + \mathfrak{n}\hat{\otimes}\mathfrak{n} - \eta\hat{\otimes}\eta = \widehat{\rm Ric}.\]
	\item\label{2k3om2o123} For the lapse $\Omega$, we will require that~\eqref{3i3ei2i4i24u} holds for $v \in (-1,0)$ with all of the Ricci curvature terms set to $0$. We will further impose the boundary condition that $\mathcal{P}_{\ell \geq 1}\left(\Omega\underline{\omega}\right)|_{v=0} = 0$. In view of Lemma~\ref{2km2omo34}, if we assume that~\eqref{2o2mom4o2} holds, this will be equivalent to the following equation holding for $v \in (-1,0)$:
	\begin{align}\label{kl2nlknlkn23}
&-\frac{1}{2}\Omega^{-2}\mathcal{L}_{\partial_v}\left(\Omega^2{\rm Ric}_{33}\right) -\frac{1}{2} \Omega^{-1}{\rm tr}\chi\left(\Omega^2{\rm Ric}_{33}\right) 
\\ \nonumber &\qquad - \left((-v)\mathcal{L}_{\partial_v}-\mathcal{L}_b\right){\rm Ric}_{34} + \left(\slashed{\rm div}b + v\Omega{\rm tr}\chi\right) {\rm Ric}_{34} + 2\left(\Omega{\rm Ric}_{3A}\right)\underline{\eta}^A = 0.
\end{align}
\item\label{k2n2oo3} For $\mathfrak{n}$, we will require that along $v \in (-1,0)$ we have $\mathcal{P}_{\ell \leq \ell_0}\mathfrak{n} = \mathcal{P}_{\ell \leq \ell_0}\eta$, we have that~\eqref{2momoo3} holds after applying $\mathcal{P}_{\ell > \ell_0}$ and replacing each instance of $\eta$ with $\mathfrak{n}$, and that 
\begin{align}\label{2kl3kljlk2}
&\Omega\nabla_4\left((v+1)^2\mathcal{P}_{\ell > \ell_0}\slashed{\rm curl}\mathfrak{n}\right)-\mathcal{P}_{\ell >\ell_0 }\slashed{\rm curl}\slashed{\rm div}\left(\Omega\hat{\chi}\right) =
\\ \nonumber &\qquad \qquad \mathcal{P}_{\ell >\ell_0}\Big( \left(v+1\right)^2\left[\Omega\nabla_4,(v+1)\slashed{\epsilon}^{CA}\slashed{\nabla}_C\right]\mathfrak{n}_A  -\left(v+1\right)^2\slashed{\epsilon}^{CA}\slashed{\nabla}_C\left(\Omega\hat{\chi}_{AB}\mathfrak{n}^B\right)
\\ \nonumber &\qquad +\frac{1}{2}\left(v+1\right)^2{}^*\slashed{\nabla}\left(\Omega{\rm tr}\chi\right)\left( -3\mathfrak{n}_A + 2\slashed{\nabla}_A\log\Omega\right) 
\\ \nonumber &\qquad + \frac{1}{4}\Omega{\rm tr}\chi\left(v+1\right)^2\slashed{\rm curl}\left(\Omega^{-2}\mathcal{L}_{\partial_v}b\right)-\frac{1}{2}\left(\Omega{\rm tr}\chi - 2(v+1)^{-1}\right)\left(v+1\right)^2\slashed{\rm curl}\mathfrak{n}\Big).
\end{align}
In view of Lemma~\ref{3kn2knk2k} and the equations~\eqref{4eta} and~\eqref{tcod1} these last two requirements for $\mathfrak{n}$ are equivalent to 
\begin{equation}\label{kl2j3kjlkjl2kj34}
\mathcal{P}_{\ell > \ell_0}\left(2\slashed{\rm div}\left(\mathfrak{n}-\eta\right) + 2\left|\mathfrak{n}\right|^2-2\left|\eta\right|^2 - \left(R + {\rm Ric}_{34}\right)\right) = 0,
\end{equation}
\begin{align}\label{kl2j3kljlk2j5lk2}
&\Omega\nabla_4\left((v+1)^2\mathcal{P}_{\ell > \ell_0}\slashed{\rm curl}\left(\mathfrak{n}-\eta\right)\right)
\\ \nonumber&\qquad +\mathcal{P}_{\ell > \ell_0}\frac{2\Omega^2}{v+1}\left(v+1\right)^2\mathcal{P}_{\ell > \ell_0}\slashed{\rm curl}\left(\mathfrak{n}-\eta\right)-(v+1)^2\mathcal{P}_{\ell \geq 1}\slashed{\rm curl}\left(\Omega{\rm Ric}_{\cdot 4}\right) = 
\\ \nonumber &\qquad  \mathcal{P}_{\ell \geq 1}\Big( \left(v+1\right)^2\left[\Omega\nabla_4,(v+1)\slashed{\epsilon}^{CA}\slashed{\nabla}_C\right]\left(\mathfrak{n}_A-\eta_A\right)  -\left(v+1\right)^2\slashed{\epsilon}^{CA}\slashed{\nabla}_C\left(\Omega\hat{\chi}_{AB}\left(\mathfrak{n}^B-\eta^B\right)\right)
\\ \nonumber &\qquad -\frac{3}{2}\left(v+1\right)^2\slashed{\rm curl}\left(\Omega{\rm tr}\chi\right)\left(\mathfrak{n}_A -\eta_A\right)-\frac{1}{2}\left(\Omega{\rm tr}\chi - 2(v+1)^{-1}\right)\left(v+1\right)^2\slashed{\rm curl}\left(\mathfrak{n}-\eta\right)\\ \nonumber &\qquad - \Omega{\rm tr}\chi\left(v+1\right)^2\mathcal{P}_{\ell \leq \ell_0}\slashed{\rm curl}\left(\mathfrak{n}-\eta\right) -\left(\Omega{\rm tr}\chi-2(v+1)^{-1}\right)\left(v+1\right)^2\mathcal{P}_{\ell > \ell_0}\slashed{\rm curl}\left(\mathfrak{n}-\eta\right)\Big),
\end{align}
We will also require that the boundary condition $\mathcal{P}_{\ell > \ell_0}\slashed{\rm curl}\mathfrak{n}|_{v=0} = \mathcal{P}_{\ell > \ell_0}\slashed{\rm curl}\mathfrak{j}|_{v=0}$ holds.
\item  For $\slashed{\rm div}b$, we will require that 
\begin{align*}
\mathcal{P}_{\ell \geq 1}\slashed{\nabla}_A\left(-\frac{1}{4}\Omega^{-2}\mathcal{L}_{\partial_v}b^A\right) -\frac{1}{4} \mathcal{P}_{\ell \geq 1}\left(\Omega^{-1}{\rm tr}\chi \slashed{\rm div}b\right) &+ \mathcal{P}_{\ell \geq 1}\left( \left(v+1\right)^{-2}\mathring{\Delta}\slashed{\rm div}b\right) = \mathcal{P}_{\ell \geq 1}\mathbb{H},
\end{align*}
where $\mathbb{H} \doteq \mathbb{H}^{(1)} + \mathbb{H}^{(2)}$ and 
\begin{align*}
&\mathbb{H}^{(1)} \doteq \frac{1}{4}\left(\Omega^{-1}{\rm tr}\chi\right)\slashed{\rm div}b + \frac{1}{2}\mathcal{L}_b\left(\Omega^{-1}{\rm tr}\chi\right) +\frac{1}{4}\left(\Omega^{-1}\hat{\chi}\right)\cdot\left(\slashed{\nabla}\hat{\otimes}b\right)
\\ \nonumber &\qquad \qquad -2\Omega^{-2}\mathcal{L}_{\partial_v}\left(\Omega\underline{\omega}\right)- 2\underline{\omega}{\rm tr}\chi - \slashed{\Delta}\log\Omega - 2\eta\cdot\underline{\eta}, 
\end{align*}
and $\mathbb{H}^{(2)}$ solves the following transport equation:
\begin{align}\label{2o3oiro2qo3}
&\left((-v)\mathcal{L}_{\partial_v}-\mathcal{P}_{\ell \geq 1}\mathcal{L}_b\right)\mathbb{H}^{(2)} + \left(1 + \frac{2(-v)}{v+1}+ (-v)\mathcal{P}_{\ell \geq 1}\Omega{\rm tr}\chi\right)\mathbb{H}^{(2)} =
\\ \nonumber &\qquad  \mathcal{P}_{\ell \geq 1}\left[\left(v+1\right)^{-2}\mathring{\Delta},\mathcal{L}_b\right]\slashed{\rm div}b  +\left(v+1\right)^{-2}\mathring{\Delta}\mathcal{P}_{\ell \geq 1}\Bigg(v\Omega\hat{\chi}\cdot\slashed{\nabla}\hat{\otimes}b - \frac{1}{4}\left|\slashed{\nabla}\hat{\otimes}b\right|^2
\\ \nonumber &\qquad \qquad \qquad \qquad \qquad \qquad+ \Omega^2v b\cdot\slashed{\nabla}\left(\Omega^{-1}{\rm tr}\chi\right)+\frac{1}{2}\left(\slashed{\rm div}b\right)^2 +8\left(\Omega\underline{\omega}\right)\left(u^{-1}+\frac{\slashed{\rm div}b}{2}\right)\Bigg),
\end{align}
with the boundary condition $\mathbb{H}^{(2)}\left(v+1\right)^2 \to 0$ as $v\to -1$. Assuming that~\eqref{2o2mom4o2} holds, then,  in view of Lemmas~\ref{existitdoes},~\ref{2m2om3o3923332}, and \ref{divbequnaa}, these equations will imply that
\begin{align*}
&\left(\left((-v)\mathcal{L}_{\partial_v}-\mathcal{P}_{\ell \geq 1}\mathcal{L}_b\right)+\left(1+(-v)\mathcal{P}_{\ell \geq 1} \Omega{\rm tr}\chi + \frac{2(-v)}{v+1}\right)\right)\mathcal{P}_{\ell \geq 1}{\rm Ric}_{34} 
\\ \nonumber &\qquad \qquad \qquad + \left(v+1\right)^{-2}\mathring{\Delta}\mathcal{P}_{\ell \geq 1}\left(\Omega^2{\rm Ric}_{33}\right) = 0,
\end{align*}
\begin{remark}This equation is a bit more unusual then the other equations that we use to solve for the components of $g$. The motivation to consider this equation is as follows. On the one hand the equation~\eqref{2mo3330ck2} is natural to use if one wants to make sure that it is possible to get an estimate which implies that $\slashed{\rm div}\eta$ is uniformly bounded as $v\to 0$. Unfortunately, the equation~\eqref{2mo3330ck2} cannot be used on its own to solve for $\slashed{\rm div}b$ because after splitting $\slashed{\rm div}\eta$ into $-\slashed{\rm div}\left(\frac{1}{4}\Omega^{-2}\mathcal{L}_{\partial_v}b\right)$ and $\slashed{\Delta}\log\Omega$, we would obtain a fatal loss of derivative from the $\slashed{\Delta}\log\Omega$ term. On the other hand, the equation~\eqref{eqnyaydivb} is natural to use if one wants to make sure it is possible to obtain estimates for $\slashed{\rm div}b$ which do not lose derivatives. In order to obtain good estimates for $\slashed{\rm div}\eta$ as $v\to 0$ from this equation one must differentiate with $\mathcal{L}_{\partial_v}$.  Unfortunately, it is difficult to use the resulting equation to conclude that $\slashed{\rm div}\eta$ is bounded as $v\to 0$. This leads us to combine the two equations into a single parabolic equation where we will be able to exploit the good features of both equations.
\end{remark}
\item\label{2oj4poj2} We will require that~\eqref{thewavezetastartsojqj} holds with the Ricci curvature terms set to $0$ after an application of $\mathcal{P}_{1 \leq \ell \leq \ell_0}\slashed{\rm curl}$. In view of Lemma~\ref{3kdo2} this is equivalent to
\[\mathcal{P}_{1 \leq \ell \leq \ell_0}\slashed{\rm curl}\left(v\Omega{\rm Ric}_{\cdot 4} - \Omega{\rm Ric}_{\cdot 3}\right) = 0.\] 

\item\label{2i4riojoij4oj22} We will require that~\eqref{thewavezetastartsojqj} holds (with the Ricci curvature terms set to $0$) after an application of $\mathcal{P}_{\ell > \ell_0}\slashed{\rm curl}\slashed{\rm div}\slashed{\nabla}\hat{\otimes}$ followed by the use of the equation~\eqref{mkek3} (with the Ricci curvature terms set to $0$) to the result of $\mathcal{P}_{\ell > \ell_0}\slashed{\rm curl}\slashed{\rm div}\slashed{\nabla}\hat{\otimes}$ applied to $2\Omega\nabla_4\zeta - \mathcal{L}_b\zeta$. In view of Lemmas~\ref{3kdo2} and~\ref{emfkeo3} this will be equivalent to 
\[\mathcal{P}_{\ell > \ell_0}\slashed{\rm curl}\slashed{\rm div}\left(\slashed{\nabla}\hat{\otimes}\left(v\Omega{\rm Ric}_{\cdot 4} - \Omega{\rm Ric}_{\cdot 3}\right) +\mathcal{L}_b\widehat{\rm Ric}_{\cdot \cdot}\right)= 0.\] 
\begin{remark}It may appear to be more natural to only use the equation~\eqref{thewavezetastartsojqj} instead of also invoking~\eqref{mkek3}. However, if we tried to avoid the use of the equation~\eqref{mkek3} and, for example, just applied $\mathcal{P}_{\ell > \ell_0}\slashed{\rm curl}$ to~\eqref{thewavezetastartsojqj}, we would run into the problem that 
\[\slashed{\rm curl}\slashed{\rm div}\slashed{\nabla}\hat{\otimes}b = \left(\slashed{\Delta}+2K\right)\slashed{\rm curl}b + 2\left(\slashed{\nabla}K\right)\wedge b.\]
The fact that we see three derivatives of the metric in the term $\slashed{\nabla}K$ would result in a fatal derivative loss in our existence scheme. By applying instead $\slashed{\rm curl}\slashed{\rm div}\slashed{\nabla}\hat{\otimes}$ and invoking~\eqref{mkek3} we avoid the appearance of nonlinear terms of this type.
\end{remark}
\end{enumerate}
\subsection{How We Solve the Equations: An Outline of Sections~\ref{lapthesection}-\ref{foim2io34o5u92hj5991}}\label{2m2o2mo1o00050033}
We will obtain the existence of solutions to the equations described in Section~\ref{2om3om42o43984u4j2} in a series of steps carried out in Sections~\ref{lapthesection}-\ref{foim2io34o5u92hj5991}. In this section we will give a high-level outline of how this procedure will work. (We will leave a detailed discussion of the various difficulties and their resolutions to the introductions of the actual sections.)

Before we begin, we will make a few general remarks about the existence proofs. We will generally solve the various nonlinear equations which arise via standard iteration arguments. This requires us to split the equation into a primary part (which will generally be a transport equation, linear parabolic equation, linear elliptic equation, or a model second order equation) and a perturbative part. This process of breaking up the equation has a risk of interfering with certain nonlinear structures that we wish to exploit. It turns out to be convenient, in order to preserve certain of these aforementioned nonlinear structures, to actually carry out multiple separate iteration arguments. When relevant, the specific advantages of this will be noted in the outlines below.

\subsubsection{Outline of Section~\ref{lapthesection}: Estimates for the lapse $\Omega$}\label{3ijoi901}
In Section~\ref{lapthesection} we start the analysis of the lapse equation. Throughout the section, we assume that $b$ and $\slashed{g}$ are arbitrary except for the requirement that they satisfy certain bootstrap assumptions. It will be useful to refer to the equation that $\Omega$ is supposed to satisfy (see item~\ref{2k3om2o123} in Section~\ref{bulkbulkbulk}) as ``the lapse equation.''

It turns out to be useful to decompose the lapse $\Omega$ into
\[\Omega \doteq \Omega_{\rm sing}\Omega_{\rm boun},\]
where $\Omega_{\rm sing}$ and $\Omega_{\rm boun}$ are defined by solving the transport equations
\begin{equation}\label{2oij4oij2o4}
(-v)\mathcal{L}_{\partial_v}\log\Omega_{\rm sing} = 2\mathcal{P}_{\ell = 0}\left(\Omega\underline{\omega}\right),\qquad \left((-v)\mathcal{L}_{\partial_v}-\mathcal{L}_b\right)\log\Omega_{\rm boun} = 2\mathcal{P}_{\ell \geq 1}\left(\Omega\underline{\omega}\right),
\end{equation}
with the boundary conditions
\[\log\Omega_{\rm sing}|_{v = -1} = 0,\qquad \log\Omega_{\rm boun}|_{v=-1} = 0.\]
We observe that $\Omega_{\rm sing}$ will be spherically symmetric; however, $\Omega_{\rm boun}$ is \underline{not} necessarily supported only on the $\ell \geq 1$ spherical harmonics. It is furthermore a consequence of~\eqref{2oij4oij2o4} that $\Omega$ will satisfy
\[\left((-v)\mathcal{L}_{\partial_v}-\mathcal{L}_b\right)\log\Omega = 2\left(\Omega\underline{\omega}\right),\]
as required by the definition of $\underline{\omega}$ and self-similarity. The names of $\Omega_{\rm sing}$ and $\Omega_{\rm boun}$ are motivated by the fact that $\Omega_{\rm sing}$ will generally blow-up as $v\to 0$ while $\Omega_{\rm boun}$ will remain pointwise bounded as $v\to 0$ (even though angular derivatives of $\Omega_{\rm boun}$ may blow-up as $v\to 0$).

If we set $\mathfrak{X} \doteq \left(v+1\right)^2\left(\Omega\underline{\omega}\right)$ then we can write the lapse equation as
\begin{equation}\label{klmkmlkml23}
\left((-v)\mathcal{L}_{\partial_v}-\mathcal{L}_b\right)\left(\mathcal{L}_{\partial_v}\mathfrak{X} + \mathscr{H}_2\right) + \left(1-\frac{3}{2}\slashed{\rm div}b+4\left(\Omega\underline{\omega}\right)\right)\left(\mathcal{L}_{\partial_v}\mathfrak{X} + \mathscr{H}_2\right) + \Omega^2\slashed{\Delta}\mathfrak{X} = \mathscr{H}_1,
\end{equation}
where $\mathscr{H}_1$ and $\mathscr{H}_2$ are suitable nonlinear expressions which involve at most two (respectively one) derivatives applied to any component of the metric. In view of the above decomposition of $\Omega$ in to $\Omega_{\rm sing}$ and $\Omega_{\rm boun}$, it is further natural to apply $\mathcal{P}_{\ell \geq 1}$ to~\eqref{klmkmlkml23} and end up with
\begin{align}\label{23kljl2}
&\left((-v)\mathcal{L}_{\partial_v}-\mathcal{P}_{\ell \geq 1}\mathcal{L}_b\right)\left(\mathcal{L}_{\partial_v}\mathfrak{X}_{\geq 1} + H_2\right) 
\\ \nonumber &\qquad + \mathcal{P}_{\ell \geq 1}\left(\left(1-\frac{3}{2}\slashed{\rm div}b+4\left(\Omega\underline{\omega}\right)\right)\left(\mathcal{L}_{\partial_v}\mathfrak{X}_{\geq 1} + H_2\right)\right) + \mathcal{P}_{\ell \geq 1}\Omega^2\slashed{\Delta}\mathfrak{X}_{\geq 1} = \mathcal{P}_{\ell \geq 1}H_1,
\end{align}
where $H_1$ and $H_2$ again denote various nonlinear terms and $X_{\geq 1} \doteq \mathcal{P}_{\ell \geq 1}\left(\Omega\underline{\omega}\right)$. In Section~\ref{lapthesection} we will consider such an equation except that we let $H_1$ and $H_2$ stand for arbitrary functions (which are finite under suitable norms), and then we use~\eqref{23kljl2} to solve for $\mathfrak{X}_{\geq 1}$, with the boundary condition $\mathfrak{X}_{\geq 1}|_{v=0} = 0$, and hence, via~\eqref{2oij4oij2o4}, also solve for $\Omega_{\rm boun}$. More concretely, we will obtain a map $\left(\slashed{g},b,H_1,H_2\right) \mapsto \left(\Omega_{\rm boun},\mathcal{P}_{\ell \geq 1}\left(\Omega\underline{\omega}\right)\right)$ and also obtain estimates for $\mathfrak{X}_{\geq 1}$ and $\Omega_{\rm boun}$ in terms of the free functions $H_1$ and $H_2$.

Also in Section~\ref{lapthesection} we will define $\Omega_{\rm sing}$ by solving a transport equation
\begin{equation}\label{3oj2om4o3}
\mathcal{L}_{\partial_v}\left((-v)\mathcal{L}_{\partial_v}\log\Omega_{\rm sing}\right) + \mathcal{P}_{\ell = 0}\left(\Omega{\rm tr}\chi\left((-v)\mathcal{L}_{\partial_v}\log\Omega_{\rm sing}\right)\right) = H_3,
\end{equation}
where $H_3$ is a free function which is only required to be finite under a suitable norm. More concretely, we will obtain a map $\left(\slashed{g},H_3\right) \mapsto \Omega_{\rm sing}$ and obtain estimates for $\Omega_{\rm sing}$ in terms of $H_3$. 

This procedure effectively allows us to decouple the process of solving for $\Omega_{\rm boun}$ and $\Omega_{\rm sing}$. We do this because we shall want to use $\Omega_{\rm sing}$ as weight in most of our bootstrap norms. Hence it is convenient to run an iteration argument where first one fixes some choice of $\Omega_{\rm sing}$, then one solves for all of the other unknowns in terms of $\Omega_{\rm sing}$, then one updates the choice of $\Omega_{\rm sing}$ by solving~\eqref{3oj2om4o3} for a suitable $H_3$, and then one repeats the process and show that the corresponding sequence has a suitably convergent subsequence.

One disadvantage of this decoupling between $\Omega_{\rm boun}$ and $\Omega_{\rm sing}$ is that we lose some potentially useful nonlinear structure in the equation for $\Omega$. Namely, if the lapse equation holds, then we would have expected the quantity
\begin{equation}\label{2o3om2k3joi2jor5h2o}
\Omega^{-1}\nabla_4\left(\Omega\underline{\omega}\right)+\Omega^{-1}{\rm tr}\chi \left(\Omega\underline{\omega}\right)+4\eta\cdot\underline{\eta} - \frac{1}{2}\left(\Omega^{-1}\hat{\chi}\right)\cdot\slashed{\nabla}\hat{\otimes}b
\end{equation}
to satisfy stronger pointwise estimates as $v\to 0$ than what one would obtain by estimating each term separately (see Remark~\ref{2i4oij2oij452}). However, because we plan to ``hold $\Omega_{\rm sing}$ fixed'' while solving for the other quantities, the transport equation structure which underlies this improved estimate will no longer hold. In order to circumvent this difficulty, we introduce a new artificial variable $Y$, which is defined in terms of $\Omega_{\rm boun}$, $b$, $\slashed{g}$, and another free function $H_4$ by solving the transport equation
\begin{equation}\label{2kn4knk2kn}
\left((-v)\mathcal{L}_{\partial_v}-\mathcal{L}_b\right)Y + \left((-v)\Omega{\rm tr}\chi +1 -\frac{3}{2}\slashed{\rm div}b\right)Y = -4\slashed{\Delta}\Omega\underline{\omega} + H_4,\qquad \left(v+1\right)^2Y \to 0\text{ as }v\to -1.
\end{equation}
(If $\Omega$ satisfies the lapse equation and $H_4$ denotes a suitable nonlinear expression then~\eqref{2kn4knk2kn} is satisfied with $Y$ given by~\eqref{2o3om2k3joi2jor5h2o}.) We thus obtain a map $\left(\slashed{g},b,\Omega_{\rm boun},H_4\right) \mapsto Y$, and also obtain estimates for $Y$ in terms of $H_4$ and $\Omega_{\rm boun}$. (Note that $\mathcal{P}_{\ell = 0}\Omega\underline{\omega}$ is not present on the right hand side of~\eqref{2kn4knk2kn}, and we thus do not include $\Omega_{\rm sing}$ as one of the inputs.) Since we have that $\mathcal{P}_{\ell = 0}\mathring{\Delta}\Omega\underline{\omega}$ vanishes, for $\mathcal{P}_{\ell = 0}Y$ we will have an estimate which depends quadratically on $\Omega_{\rm boun}$ if we also include a term which is quadratic in $\slashed{g} - \left(v+1\right)^2\mathring{\slashed{g}}$. Now, in the course of our iteration argument, when we desire to, we may replace the term~\eqref{2o3om2k3joi2jor5h2o} with $Y$ for which we will have improved pointwise estimates as $v\to 0$. At the end of the iteration argument, we will find that $Y$ is equal to the expression~\eqref{2o3om2k3joi2jor5h2o}.

Lastly, it is also convenient to define a function $\tilde{H}_6: (-1,0)\times \mathbb{S}^2 \to \mathbb{R}$ which satisfies $\mathcal{P}_{\ell =0}\tilde{H}_6 = 0$ by solving the following equation (which is very similar to~\eqref{2kn4knk2kn}): 
\begin{equation}\label{i2j3ij2o4ijo4}
\left((-v)\mathcal{L}_{\partial_v}-\mathcal{P}_{\ell \geq 1}\mathcal{L}_b\right)\tilde{H}_6 + \left(1+\frac{4(-v)}{v+1}\right)\tilde{H}_6 = 8(v+1)^{-2}\mathring{\Delta}\left(\Omega\underline{\omega}\right),\qquad \left(v+1\right)^2\tilde{H}_6 \to 0 \text{ as }v\to -1.
\end{equation}
This yields a map $\left(b,\mathcal{P}_{\ell \geq 1}\left(\Omega\underline{\omega}\right)\right) \mapsto \tilde{H}_6$ and a corresponding estimate for $\tilde{H}_6$ in terms of $\mathcal{P}_{\ell \geq 1}\left(\Omega\underline{\omega}\right)$. See the discussion Section~\ref{joijoij2oi34} for an explanation of the later use of $\tilde{H}_6$.

\subsubsection{Outline of Section~\ref{ij3jr9j3}: The $\left(\slashed{g},\mathfrak{n}\right)$ System}\label{1o2m3om2}
In Section~\ref{ij3jr9j3} we analyze the equations for $\slashed{g}$ and $\mathfrak{n}$. Throughout this section we allow $\Omega$ and $b$ to be arbitrary as long they satisfy suitable bootstrap assumptions. We also introduce five artificial variables $\pi$, $\mathfrak{o}$, $\mathfrak{w}$, $\mathfrak{q}$, and $\mathfrak{r}$ where where $\pi_A$ is an $\mathbb{S}^2_{-1,v}$ $1$-form for $v \in (-1,0)$ which satisfies $\left(1-\mathcal{P}_{\ell \leq \ell_0}\right)\pi = 0$, $\mathfrak{q} : (-1,0) \times \mathbb{S}^2\to \mathbb{R}$ is a function, $\mathfrak{r}$ is an $\mathbb{S}^2_{-1,v}$ symmetric $(0,2)$-tensor, and $\mathfrak{o}$, $\mathfrak{w}:\mathbb{S}^2\to \mathbb{R}$ satisfy $\left(1-\mathcal{P}_{\ell > \ell_0}\right)\mathfrak{o} = \left(1-\mathcal{P}_{\ell > \ell_0}\right)\mathfrak{w} = 0$.

We will define a certain system of equations in Section~\ref{ij3jr9j3} which we call the ``modified $\left(\slashed{g},\mathfrak{n}\right)$-system.'' See Definition~\ref{2n3inii1o2} for the specifics. For now we will just give a high level description of the modified system: Our starting point is the bulk equations that $\slashed{g}$ and $\mathfrak{n}$ are to satisfy as described in items~\ref{in2ni13} and~\ref{k2n2oo3} of Section~\ref{bulkbulkbulk}. We apply $\mathcal{P}_{\ell > \ell_0}\slashed{\rm div}^2$ and $\mathcal{P}_{\ell > \ell_0}\slashed{\rm div}\slashed{\rm curl}$ to the equation for $\nabla_3\hat{\chi}$ to derive two extra equations for $\mathcal{P}_{\ell > \ell_0}\slashed{\rm div}^2\left(\Omega\hat{\chi}\right)$ and $\mathcal{P}_{\ell > \ell_0}\slashed{\rm div}\slashed{\rm curl}\left(\Omega\hat{\chi}\right)$. Then we make three ``modifications'' to entire system. First of all, we add in certain projections $\Pi_{{\rm ker}\left({}^*\slashed{\rm D}_2\right)^{\perp}}$ and $\Pi_{{\rm ker}\left({}^*\slashed{\rm D}_1\right)^{\perp}}$ in various places. This projections allow us to more easily invert certain elliptic operators. Secondly, we replace in certain nonlinear terms, $\slashed{\rm div}b$ and $\slashed{\nabla}\hat{\otimes}b$ with the artificial unknowns $\mathfrak{q}$ and $\mathfrak{r}$. Finally, we replace the equation $\mathcal{P}_{\ell \leq \ell_0}\mathfrak{n} = \mathcal{P}_{\ell \leq \ell_0}\eta$ with the equation
\[\mathcal{P}_{\ell \leq \ell_0}\mathfrak{n} = \mathcal{P}_{\ell \leq \ell_0}\pi.\]
The role of the functions $\mathfrak{o}$ and $\mathfrak{w}$ is fix certain boundary conditions at $\{v = 0\}$. Again, we direct the reader to Section~\ref{ij3jr9j3} for the specifics. 

We will then show in Section~\ref{ij3jr9j3} that we may find $\slashed{g}$ and $\mathfrak{n}$ so that these equations hold. More concretely, the result will be a map $\left(\Omega,b,\pi,\mathfrak{o},\mathfrak{w},\mathfrak{q},\mathfrak{r}\right) \mapsto \left(\slashed{g},\mathfrak{n}\right)$ along with corresponding estimates for $\left(\slashed{g},\mathfrak{n}\right)$ in terms of suitable norms applied to $\left(\Omega,b,\pi,\mathfrak{o},\mathfrak{w}\right)$. (Our estimates will not depend on $\mathfrak{q}$ and $\mathfrak{r}$ because they only appear in nonlinear expressions contracted with a term involving the difference of $\slashed{g}$ and its Minkowski value.) We note that due to the linear coupling of all of the quantities $\left(\Omega,b,\pi,\mathfrak{o},\mathfrak{w}\right)$ to $\slashed{g}$ and $\mathfrak{n}$, the corresponding estimates for $\slashed{g}$ and $\mathfrak{n}$ will have a linear dependence on suitable norms of $\left(\Omega,b,\pi,\mathfrak{o},\mathfrak{w}\right)$. However, because $\pi$ is only supported on $\ell \leq \ell_0$ spherical harmonics, we will be able to show that when we estimate $\mathcal{P}_{\ell > \ell_0}\slashed{g}$, then the dependence on $\pi$ is quadratic. 

We will also show that given a solution to this modified system which happens to satisfy $\mathfrak{q} = \slashed{\rm div}b$ and $\mathfrak{r} = \slashed{\nabla}\hat{\otimes}b$, then the $\slashed{g}$ and $\mathfrak{n}$ we have found will actually solve the system of equations described in items~\ref{in2ni13} and~\ref{k2n2oo3} of Section~\ref{bulkbulkbulk} except that we still replace the equation $\mathcal{P}_{\ell \leq \ell_0}\mathfrak{n} = \mathcal{P}_{\ell \leq \ell_0}\eta$ with the equation
\[\mathcal{P}_{\ell \leq \ell_0}\mathfrak{n} = \mathcal{P}_{\ell \leq \ell_0}\pi.\] 
Moreover, we will then have the following boundary conditions:
\[\mathcal{P}_{\ell > \ell_0}K|_{v=0} = \mathfrak{o},\qquad \mathcal{P}_{\ell > \ell_0}\slashed{\rm curl}\mathfrak{n}|_{v=0} = \mathfrak{w}.\]

\subsubsection{Outline of Section~\ref{k2m3mo492}: Estimates for $\mathring{\Pi}_{\rm div}b$}\label{joijoij2oi34}
In Section~\ref{k2m3mo492} we analyze the equation which will hold for $\slashed{\rm div}b$. However, as in Section~\ref{lapthesection}, we postpone the analysis of most of the nonlinear terms. Throughout this section we let $\Omega$ and $\slashed{g}$ be arbitrary as long as they satisfy suitable bootstrap assumptions. We further assume that we are given a scalar function $\mathfrak{Y} : (-1,0) \times \mathbb{S}^2 \to \mathbb{R}$ which satisfies $\mathcal{P}_{\ell = 0}\mathfrak{Y} = 0$. We also assume that we are a given a  function $H_4$, and then, using the analysis from Section~\ref{lapthesection} we find the corresponding $Y$ satisfying~\eqref{2kn4knk2kn} and the corresponding $\tilde{H}_6$ satisfying~\eqref{i2j3ij2o4ijo4} 

The main result of Section~\ref{k2m3mo492} will concern the existence of solutions $b_A$ to the system of equations:
\begin{align}\label{klj3lk2jkl3j2}
&\mathcal{P}_{\ell \geq 1}\slashed{\nabla}_A\left(-\frac{1}{4}\Omega^{-2}\mathcal{L}_{\partial_v}b^A\right) -\frac{1}{4} \mathcal{P}_{\ell \geq 1}\left(\Omega^{-1}{\rm tr}\chi \slashed{\rm div}b\right) + \mathcal{P}_{\ell \geq 1}\left( \left(v+1\right)^{-2}\mathring{\Delta}\slashed{\rm div}b\right) 
\\ \nonumber &= \mathcal{P}_{\ell \geq 1}\left(\left(H_5 -\slashed{\Delta}\log\Omega_{\rm boun}-2Y\right) + H_6 + \tilde{H}_6\right),
\\ \nonumber  \qquad \mathring{\rm curl}b &= \mathfrak{Y},
\end{align}
where $H_5$ and $H_6$ are free functions which are only required to be finite under a suitable norm. More concretely, we obtain a map $\left(\Omega,\slashed{g},H_4,H_5,H_6,\mathfrak{Y}\right) \mapsto b$. We also obtain estimates for $b$ in terms of these quantities. In particular we obtain estimates for $\mathring{\Pi}_{\rm div}b$. These estimates for $\mathring{\Pi}_{\rm div}b$ will have a linear dependence on $H_4$, $H_5$, $H_6$, and $\Omega_{\rm boun}$ and a quadratic dependence on $\Omega_{\rm sing}$, $\slashed{g}$, and $\mathfrak{Y}$. (See~\eqref{3om4om2o4o32} for the specifics.)

\subsubsection{Outline of Section~\ref{2oj3o}: Estimates for $\mathring{\Pi}_{\rm curl}b$ and $\left(\mathfrak{o},\mathfrak{w}\right)$}
In this section we let $\Omega_{\rm sing}$ and $\pi$ be a free spherically symmetric function and an $\mathbb{S}^2_{-1,v}$ $1$-form satisfying $\left(1-\mathcal{P}_{\ell \leq \ell_0}\right) \pi =0$ such that both $\Omega_{\rm sing}$ and $\pi$ satisfy suitable bootstrap assumptions. 

Then, in Section~\ref{2oj3o} we show that for any choice of free functions $H_1$, $H_2$, $H_4$, $H_5$, $H_6$,  free $\mathbb{S}^2_{-1,v}$ $1$-forms $H_7$ and $H_8$   and a free $\mathbb{S}^2_{-1,v}$ $(0,2)$-tensor $H_9$ so that  $\left(1-\mathcal{P}_{\ell \geq 1}\right)\left(H_1,H_2\right) = 0$,  $H_2$ is supported for $v \geq -1/2$, and all of the free functions satisfy a suitable smallness condition, we may find $\slashed{g}$, $b$, and $\Omega = \Omega_{\rm sing}\Omega_{\rm boun}$ so that the following two equations hold:
\begin{align}\label{kl2j1o3jr5oij2o}
&\mathcal{P}_{1 \leq \ell \leq \ell_0}\slashed{\rm curl}\left[(-v)\nabla_{\partial_v}\zeta_A - \frac{1}{2}\mathcal{L}_b\zeta_A + \left(1+3(v+1)^{-1}(-v)\Omega^2\right)\zeta_A  -\frac{1}{4}\slashed{\rm div}\left(\slashed{\nabla}\hat{\otimes}b\right)_A \right] = \\ \nonumber &\qquad \mathcal{P}_{1 \leq \ell \leq \ell_0}\slashed{\rm curl}\left[H_7-\frac{1}{2}\mathcal{L}_b\slashed{\nabla}\log\Omega\right],
\end{align}
\begin{align}\label{1oitoi34joiqtoiq3oi}
&-\frac{1}{4}\mathcal{P}_{\ell > \ell_0}(v+1)^2\slashed{\rm curl}\slashed{\rm div}\Bigg[\left((-v)\nabla_v - \mathcal{L}_b\right)\left(\Omega^{-2}\nabla_v\left(\slashed{\nabla}\hat{\otimes}b\right)\right) + \left(1 + \frac{4(-v)\Omega^2}{v+1}\right)\Omega^{-2}\nabla_v\left(\slashed{\nabla}\hat{\otimes}b\right) 
\\ \nonumber &\qquad +\left(1 + 4\left(v+1\right)^{-1}(-v)\Omega^2\right)\left(-2\Omega^{-2}\mathcal{L}_b\left(\Omega\hat{\chi}\right) +8\slashed{\nabla}\log\Omega\hat{\otimes}\zeta -4 \Omega^{-2}\mathcal{E}\left[\slashed{g},b\right]\right)
\\ \nonumber &\qquad + \slashed{\nabla}\hat{\otimes}\left[\slashed{\rm div}\left(\slashed{\nabla}\hat{\otimes}b\right) - \slashed{\nabla}\slashed{\rm div}b\right]+\left((-v)\nabla_v-\mathcal{L}_b\right)\left(8\slashed{\nabla}\log\Omega\hat{\otimes}\zeta - 4\Omega^{-2}\mathcal{E}\left[\slashed{g},b\right]\right)\Bigg] 
\\ \nonumber &\qquad =\mathcal{P}_{\ell > \ell_0}(v+1)^2\slashed{\rm curl}\slashed{\rm div}\left(\slashed{\nabla}\hat{\otimes} \left[H_8-\frac{1}{2}\mathcal{L}_b\slashed{\nabla}\log\Omega\right] -2\Omega^{-2}\mathcal{L}_b\left(\Omega^2 \slashed{\nabla}\hat{\otimes}\slashed{\nabla}\log\Omega\right)+H_9 \right),
\end{align}
where
\begin{align}\label{l2k3jlk2jl23}
\mathcal{E}\left[\slashed{g},b\right]_{AB} \doteq -\frac{1}{2}\slashed{\rm div}b \left(\Omega\hat{\chi}\right)_{AB} - \frac{1}{2}\left(\Omega\hat{\chi}\right)^C_{\ \ (A}\left(\slashed{\nabla}\hat{\otimes}b\right)_{B)C},
\end{align}
and we moreover have that $\Omega_{\rm boun}$ satisfies~\eqref{23kljl2}, $b$ also satisfies the first equation of~\eqref{klj3lk2jkl3j2}, $\left(\slashed{g},\mathfrak{n}\right)$ solve the modified system described in Section~\ref{1o2m3om2} with $\mathfrak{q} = \slashed{\rm div}b$ and $\mathfrak{r} = \slashed{\nabla}\hat{\otimes}b$, and $\left(\mathfrak{o},\mathfrak{w}\right)$ satisfy the following equations:
  \begin{align}\label{2lk3jio4j2}
&\mathcal{P}_{\ell \geq 1}\left( -\mathcal{L}_b\mathcal{P}_{\ell \geq 1}\slashed{\rm div}\mathfrak{j}+\left(2-\slashed{\rm div}b\right)\mathcal{P}_{\ell \geq 1}\slashed{\rm div}\mathfrak{j} +\left[\mathcal{P}_{\ell \geq 1}\slashed{\rm div}, -\mathcal{L}_b+\left(2-\slashed{\rm div}b\right)\right]\mathfrak{j}\right) = 
\\ \nonumber &\qquad \mathcal{P}_{\ell \geq 1}\left(\left(\mathcal{P}_{\ell \leq \ell_0}K + \mathfrak{o}\right)\slashed{\rm div}b  + \mathcal{L}_b\left(\mathcal{P}_{\ell \leq \ell_0}K + \mathfrak{o}\right)\right),
 \end{align}
 \begin{align}\label{2lk3jl4j2jouo2}
 &\mathcal{P}_{\ell \geq 1}\left( -\mathcal{L}_b\mathcal{P}_{\ell \geq 1}\slashed{\rm curl}\mathfrak{j}+\left(2-\slashed{\rm div}b\right)\mathcal{P}_{\ell \geq 1}\slashed{\rm curl}\mathfrak{j} +\left[\mathcal{P}_{\ell \geq 1}\slashed{\rm curl}, -\mathcal{L}_b+\left(2-\slashed{\rm div}b\right)\right]\mathfrak{j}\right) = 
 \\ \nonumber &\qquad \frac{1}{2}\mathcal{P}_{\ell = 1}\left(\left(\slashed{\Delta}+2K\right)\slashed{\rm curl}b\right) + \mathcal{P}_{1 \leq \ell \leq \ell_0}\left(\slashed{\nabla}\left(\mathcal{P}_{\ell < \ell_0}K + \mathfrak{o}\right)\wedge b\right)+\frac{1}{2}\epsilon\left(\slashed{\Delta}+2K\right)T_{\rm low}+ \frac{1}{2}\epsilon T_{\rm high}
 \\ \nonumber &\qquad +\frac{1}{2}\left[\mathcal{P}_{2 \leq \ell \leq \ell_0},\slashed{\Delta}+2K\right]\slashed{\rm curl}b,
 \end{align}
   \begin{align}\label{3k2jlk2j3l2}
& \left(2-2\mathcal{P}_{\ell > \ell_0}\left(\mathcal{L}_b+\slashed{\rm div}b\right)\right)\mathfrak{o}  =
\left(2-\mathcal{P}_{\ell > \ell_0}\left(\mathcal{L}_b+\slashed{\rm div}b\right)\right)\mathcal{P}_{\ell > \ell_0}\mathscr{N} 
\\ \nonumber &\qquad + \mathcal{P}_{\ell > \ell_0}\left(\mathcal{L}_b\mathcal{P}_{\ell \leq \ell_0}\slashed{\rm div}\mathfrak{j} + \left(\slashed{\rm div}b -2\right)\mathcal{P}_{\ell \leq \ell_0}\slashed{\rm div}\mathfrak{j}\right)
		\\ \nonumber &\qquad + \mathcal{P}_{\ell > \ell_0}\left(\left[\mathcal{P}_{\ell \geq 1}\slashed{\rm div}, -\mathcal{L}_b+\left(2-\slashed{\rm div}b\right)\right]\mathfrak{j}\right)		
		+\mathcal{P}_{\ell > \ell_0}\left(\mathcal{P}_{\ell \leq \ell_0}K\slashed{\rm div}b  + \mathcal{L}_b\mathcal{P}_{\ell \leq \ell_0}K \right),
		\end{align}
		\begin{equation}\label{23pojmo2j4io2}
		\mathfrak{w} =\mathcal{P}_{\ell > \ell_0}\slashed{\rm curl}\mathfrak{j},
		\end{equation}
where
\begin{equation*}
\mathscr{N} = \frac{1}{2}(\Omega^{-1}{\rm tr}\chi-2(v+1)^{-1})\left(1+\left(4\Omega\underline{\omega}-\slashed{\rm div}b\right)\right)-\frac{1}{2}\mathcal{L}_b\left[\Omega^{-1}{\rm tr}{\rm tr}\chi\right]
	 - \left|\mathfrak{j}\right|^2.
\end{equation*}

We note that if $H_7$, $H_8$, and $H_9$ take the form of suitable nonlinear expressions, then~\eqref{kl2j1o3jr5oij2o} and~\eqref{1oitoi34joiqtoiq3oi} will exactly correspond to equations described in items~\ref{2oj4poj2} and~\ref{2i4riojoij4oj22} of Section~\ref{bulkbulkbulk}. Moreover, the equations~\eqref{2lk3jio4j2}-\eqref{23pojmo2j4io2} will imply that the boundary equations described in Section~\ref{2o3om4o2mo4599j54} hold. 

The end result of this section's analysis will be a map 
\begin{equation}\label{2i3in2i4}
\left(H_1,H_2,H_4,H_5,H_6,H_7,H_8,H_9,\Omega_{\rm sing},\pi\right) \mapsto \left(\Omega_{\rm boun},b,\slashed{g},\mathfrak{n},\mathfrak{o},\mathfrak{w},Y\right).
\end{equation}
Moreover, we will have estimates for the $\Omega_{\rm boun}$, $b$, $\slashed{g}$, $\mathfrak{o}$, and $\mathfrak{w}$ in terms of the inputs. These estimates will generally have a linear dependence on $\left(H_1,H_2,H_4,H_5,H_6,H_7,H_8,H_9,\Omega_{\rm sing},\pi\right)$ with a few exceptions: The dependence of all the solved for quantities on $\mathring{\Pi}_{\rm div}\left(H_7,H_8,H_9\right)$ will be quadratic and for all quantities except for $\slashed{g}$, the dependence on $\pi$ will be quadratic. 

We note that an important role is played in this section by the fact that $\mathcal{L}_b\hat{\chi}$ and $\mathcal{L}_bK$ satisfy improved higher order estimates compared to a generic angular derivative of $\hat{\chi}$ and $K$. However, we can only see this improvement if the ``$b$'' in the $\mathcal{L}_b$ derivative is the same $b$ used in the analysis discussed in Section~\ref{1o2m3om2}. At the same time however, in this section we are also solving for $b$. This requires us to set-up the corresponding iteration argument with care; in fact it turns out to be convenient to carry out two separate iteration arguments and an important role will be played by the fact that the estimates for $\left(\slashed{g},\mathfrak{n}\right)$ which are obtained in Section~\ref{1o2m3om2} do not require the full strength of the natural bootstrap assumptions for $b$. See Section~\ref{2oj3o} for the details.
\subsubsection{Outline of Section~\ref{2i3oij2o3}:  Adding in the Nonlinear Terms for $\Omega_{\rm boun}$ and $b$} 
In Section~\ref{2i3oij2o3}, we revisit the map defined in~\eqref{2i3in2i4} and iterate so that we eventually input the actual nonlinear expressions for $H_1$, $H_2$, $H_4$, $H_5$, $H_6$, $H_7,$ $H_8$, and $H_9$ which result in the full equations for $\Omega_{\rm boun}$, $Y$, and $b$ holding. 

We in fact do this in two steps. In the first step, we just consider the terms $H_7$, $H_8$, and $H_9$. Once we have closed this iteration argument, we will have, in particular, that the equations described in items~\ref{2oj4poj2} and~\ref{2i4riojoij4oj22} of Section~\ref{bulkbulkbulk} now hold. With the full nonlinear structure that these equations provide, it turns out to now be possible to combine with the equation~\eqref{klj3lk2jkl3j2} that $\slashed{\rm div}b$ satisfies, and establish that $\eta$ has an improved  estimate as $v\to 0$ beyond what one obtains by adding together the corresponding estimates for $\Omega^{-2}\mathcal{L}_{\partial_v}b$ and $\slashed{\nabla}\log\Omega$. 

Once we have this improved estimate for $\eta$ at our disposal, we then run another iteration argument to input the corresponding nonlinear expressions for $H_1$, $H_2$, $H_4$, $H_5$, and $H_6$. At the end of this section we will have a map
\begin{equation}\label{23ion2ono4}
\left(\Omega_{\rm sing},\pi\right) \mapsto \left(\Omega_{\rm boun},\slashed{g},\mathfrak{n},b,Y\right)
\end{equation}
\subsubsection{Outline of Section~\ref{foim2io34o5u92hj5991}: Solving for $\Omega_{\rm sing}$ and $\pi$ and Finishing the Proof of Theorem~\ref{thisiswhatishholdingatthenehdne}}
Finally, in Section~\ref{foim2io34o5u92hj5991} we revisit the map~\eqref{23ion2ono4} and run one more final iteration argument where we require that $\Omega_{\rm sing}$ now satisfies~\eqref{3oj2om4o3} where
\[H_3 = \mathcal{P}_{\ell = 0}\left(\Omega^2\left(\frac{1}{2}Y-2\Omega{\rm tr}\chi \mathcal{P}_{\ell \geq 1}\left(\Omega\underline{\omega}\right)-2\eta\cdot\underline{\eta} + \frac{1}{4}\left(\Omega^{-1}\hat{\chi}\right)\cdot\left(\slashed{\nabla}\hat{\otimes}b\right)\right)\right),\]
and we set $\pi \doteq \mathcal{P}_{\ell \leq \ell_0}\eta$. The iteration argument closes due to the fact that the estimates for $\mathcal{P}_{\ell = 0}Y$ have a quadratic dependence on $\Omega_{\rm sing}$ and that the estimates for $b$ and $\Omega_{\rm boun}$ have a quadratic dependence on $\pi$. The final result of the iteration is the solution we have been seeking.

In this section we will also complete the proof of Theorem~\ref{thisiswhatishholdingatthenehdne} (which is stated below in Section~\ref{moo3ok2o49}).

\subsection{Main Theorem}\label{moo3ok2o49}
In this section we state the key theorem which will be the result of all of our iteration arguments and whose proof is the taken up in Sections~\ref{lapthesection}-\ref{foim2io34o5u92hj5991} of the paper.

\begin{theorem}\label{thisiswhatishholdingatthenehdne}Let  $\left(T_{\rm low},T_{\rm high}\right) : \mathbb{S}^2 \to \mathbb{R}$ satisfy~\eqref{oiiooi909198}, and let $\epsilon > 0$ be any sufficiently small real number. Then there exists a function $\Omega  : (-1,0) \times \mathbb{S}^2 \to (0,\infty)$, $\mathbb{S}^2_{-1,v}$ vector fields $b$ and $\mathfrak{n}$ for $v \in (-1,0)$, an $\mathbb{S}^2_{-1,v}$ symmetric $(0,2)$-tensor $\slashed{g}_{AB}$, and a vector field $\mathfrak{j}$ along $\mathbb{S}^2$ so that the following hold, where $g$ denotes the  induced self-similar Lorentzian metric associated to $\left(\Omega,b,\slashed{g}\right)$ (see the discussion in Section~\ref{ooiiouou832}). 

\begin{enumerate}
\item\label{anitemitem1} There exists a constant $\kappa \in \mathbb{R}$  and functions $\Omega_{\rm sing}$ and $\Omega_{\rm boun}$ such that $\Omega_{\rm sing}$ is spherically symmetric, $\Omega = \Omega_{\rm sing}\Omega_{\rm boun}$, and so that we have
\begin{align}\label{ijoi9919njnkko2o201nbhjnbghj}
&\left|\kappa\right| + \left\vert\left\vert \log\Omega_{\rm boun}\right\vert\right\vert_{\mathscr{A}\left(\kappa,b\right)} + \left\vert\left\vert \log\Omega_{\rm boun}\right\vert\right\vert_{\mathscr{B}_{01}\left(\kappa,b\right)} +\left\vert\left\vert \log\Omega_{\rm sing}\right\vert\right\vert_{\mathscr{B}_{00}\left(\kappa\right)} +\left\vert\left\vert b\right\vert\right\vert_{\mathscr{A}_1\left(\kappa,\slashed{g}\right)}+\left\vert\left\vert b\right\vert\right\vert_{\mathscr{B}_1\left(\kappa\right)}
\\ \nonumber &\qquad +\left\vert\left\vert \slashed{g} \right\vert\right\vert_{\mathscr{A}_2\left(\kappa,b,\Omega\right)} + \left\vert\left\vert \slashed{g}\right\vert\right\vert_{\mathscr{B}_2\left(\kappa,b,\Omega\right)}
+\left\vert\left\vert \left(v+1\right)\mathfrak{n}\right\vert\right\vert_{L,III\left(0,100\check{p}N_1,0\right),N_1-1}+\left\vert\left\vert \left(\mathfrak{j},\mathcal{L}_{b|_{v=0}}\mathfrak{j}\right)\right\vert\right\vert_{\mathring{H}^{N_1-2}\left(\mathbb{S}^2\right)}
\\ \nonumber &\qquad +\sum_{j=0}^1\left\vert\left\vert \left(v\mathcal{L}_{\partial_v}\right)^j\eta\right\vert\right\vert_{\mathscr{S}_{-1/2}^0\left(N_2-1-j-k,0,0\right)}  +\sum_{j=0}^1\left\vert\left\vert \left(v\mathcal{L}_{\partial_v}\right)^j\mathcal{L}_{\partial_v}\eta\right\vert\right\vert_{\check{\mathscr{S}}_{-1/2}^0\left(N_2-1-j-k,0,500\check{p}\left(1+j\right)+2\tilde{\kappa},500\check{p}\right)}
\\ \nonumber &\qquad \lesssim \epsilon \left\vert\left\vert \left(T_{\rm low},T_{\rm high}\right)\right\vert\right\vert_{\mathring{H}^{N_1-3}\left(\mathbb{S}^2\right)},
\end{align}
\begin{align}\label{2om3om1oijtionhoin1}
&\left\vert\left\vert \left(1,\mathcal{L}_b\right)\left(\Omega^{-1}{\rm tr}\chi - 2(v+1)^{-1}\right)|_{v=0}\right\vert\right\vert_{\mathring{H}^{N_1-2}\left(\mathbb{S}^2\right)}   \lesssim \epsilon^2\left\vert\left\vert \left(T_{\rm low},T_{\rm high}\right)\right\vert\right\vert_{\mathring{H}^{N_1-3}}^2 .
\end{align}

\item\label{anitemitem2}  The following equations for the Ricci curvature of $g$ and $\mathfrak{n}$ hold for $(u,v) \in \{-1\} \times  (-1,0)$:
\begin{equation}\label{kjgrejierij392oi1}
{\rm Ric}_{44} = 0,
\end{equation}
\begin{align}\label{32oij32o3}
&-\frac{1}{2}\Omega^{-2}\mathcal{L}_{\partial_v}\left(\Omega^2{\rm Ric}_{33}\right) -\frac{1}{2} \Omega^{-1}{\rm tr}\chi\left(\Omega^2{\rm Ric}_{33}\right) 
\\ \nonumber &\qquad - \left((-v)\mathcal{L}_{\partial_v}-\mathcal{L}_b\right){\rm Ric}_{34} + \left(\slashed{\rm div}b + v\Omega{\rm tr}\chi\right) {\rm Ric}_{34} + 2\left(\Omega{\rm Ric}_{3A}\right)\underline{\eta}^A = 0,
\end{align}
\begin{align}\label{32oi32jio32123fuhn24}
&\left(\left((-v)\mathcal{L}_{\partial_v}-\mathcal{P}_{\ell \geq 1}\mathcal{L}_b\right)+\left(1+(-v)\mathcal{P}_{\ell \geq 1} \Omega{\rm tr}\chi + \frac{2(-v)}{v+1}\right)\right)\mathcal{P}_{\ell \geq 1}{\rm Ric}_{34} 
\\ \nonumber &\qquad \qquad \qquad + \left(v+1\right)^{-2}\mathring{\Delta}\mathcal{P}_{\ell \geq 1}\left(\Omega^2{\rm Ric}_{33}\right) = 0,
\end{align}
\begin{equation}\label{m3mini43jn2o}
\mathcal{P}_{1 \leq \ell \leq \ell_0}\slashed{\rm curl}\left[v\Omega{\rm Ric}_{4\cdot } - \Omega{\rm Ric}_{3\cdot }\right] = 0,
\end{equation}
\begin{equation}\label{plqowune33342}
 \mathcal{P}_{\ell_0 < \ell}\slashed{\rm curl}\slashed{\rm div}\left(\slashed{\nabla}\hat{\otimes}\left(v\Omega{\rm Ric}_{4\cdot}-\Omega{\rm Ric}_{3\cdot}\right)+\mathcal{L}_b\widehat{\rm Ric}\right)= 0,
\end{equation}
\begin{equation}\label{ini3moo}
(\slashed{\nabla}\hat{\otimes}\left(\mathfrak{n}-\eta\right))_{AB}+\left(\mathfrak{n}\hat{\otimes}\mathfrak{n}\right)_{AB} - \left(\eta\hat{\otimes}\eta\right)_{AB} - \widehat{\rm Ric}_{AB}  = 0,
\end{equation}
\begin{equation}\label{2om3pom2p}
\mathcal{P}_{\ell \leq \ell_0}\left(\mathfrak{n}-\eta\right) = 0,
\end{equation}
\begin{equation}\label{32ojojo1}
\mathcal{P}_{\ell > \ell_0}\left(2\slashed{\rm div}\left(\mathfrak{n}-\eta\right) + 2\left|\mathfrak{n}\right|^2-2\left|\eta\right|^2 - \left(R + {\rm Ric}_{34}\right)\right) = 0,
\end{equation}
\begin{align}\label{32ojpojp231}
&\Omega\nabla_4\left((v+1)^2\mathcal{P}_{\ell > \ell_0}\slashed{\rm curl}\left(\mathfrak{n}-\eta\right)\right)
\\ \nonumber&\qquad +\mathcal{P}_{\ell > \ell_0}\frac{2\Omega^2}{v+1}\left(v+1\right)^2\mathcal{P}_{\ell > \ell_0}\slashed{\rm curl}\left(\mathfrak{n}-\eta\right)-(v+1)^2\mathcal{P}_{\ell \geq 1}\slashed{\rm curl}\left(\Omega{\rm Ric}_{\cdot 4}\right) = 
\\ \nonumber &\qquad  \mathcal{P}_{\ell \geq 1}\Big( \left(v+1\right)^2\left[\Omega\nabla_4,(v+1)\slashed{\epsilon}^{CA}\slashed{\nabla}_C\right]\left(\mathfrak{n}_A-\eta_A\right)  -\left(v+1\right)^2\slashed{\epsilon}^{CA}\slashed{\nabla}_C\left(\Omega\hat{\chi}_{AB}\left(\mathfrak{n}^B-\eta^B\right)\right)
\\ \nonumber &\qquad -\frac{3}{2}\left(v+1\right)^2\slashed{\rm curl}\left(\Omega{\rm tr}\chi\right)\left(\mathfrak{n}_A -\eta_A\right)-\frac{1}{2}\left(\Omega{\rm tr}\chi - 2(v+1)^{-1}\right)\left(v+1\right)^2\slashed{\rm curl}\left(\mathfrak{n}-\eta\right)\\ \nonumber &\qquad - \mathfrak{a}\left(v+1\right)^2\mathcal{P}_{\ell \leq \ell_0}\slashed{\rm curl}\left(\mathfrak{n}-\eta\right) -\left(\Omega^{-2}\mathfrak{a}-2(v+1)^{-1}\right)\left(v+1\right)^2\mathcal{P}_{\ell > \ell_0}\slashed{\rm curl}\left(\mathfrak{n}-\eta\right)\Big) \doteq \mathfrak{H}
\end{align}
We note that the estimate~\eqref{ijoi9919njnkko2o201nbhjnbghj} implies that each of the individual terms in these expressions are well-defined.

\item\label{anitemitem3} Along $\{v = 0\}$ the following equations hold for the Ricci curvature of $g$, $b$, $\mathfrak{n}$, and $\mathfrak{j}$:
\begin{align}\label{2nk3knk2}
&\mathcal{P}_{\ell \geq 1}\Big( -\mathcal{L}_b\mathcal{P}_{\ell \geq 1}\slashed{\rm div}\left(\mathfrak{j}-\eta\right)+\left(2-\slashed{\rm div}b\right)\mathcal{P}_{\ell \geq 1}\slashed{\rm div}\left(\mathfrak{j}-\eta\right)
\\ \nonumber &\qquad \qquad  \qquad +\left[\mathcal{P}_{\ell \geq 1}\slashed{\rm div}, -\mathcal{L}_b+\left(2-\slashed{\rm div}b\right)\right]\left(\mathfrak{j}-\eta\right)\Big)|_{v=0} =  \mathcal{P}_{\ell \geq 1}\slashed{\rm div}\left(\Omega{\rm Ric}_{3\cdot}\right)|_{v=0},
 \end{align}
 \begin{align}\label{k1jk2ljl1kjkljl1}
 &\mathcal{P}_{\ell \geq 1}\Big( -\mathcal{L}_b\mathcal{P}_{\ell \geq 1}\slashed{\rm curl}\left(\mathfrak{j}-\eta\right)+\left(2-\slashed{\rm div}b\right)\mathcal{P}_{\ell \geq 1}\slashed{\rm curl}\left(\mathfrak{j}-\eta\right) 
 \\ \nonumber &\qquad \qquad \qquad +\left[\mathcal{P}_{\ell \geq 1}\slashed{\rm curl}, -\mathcal{L}_b+\left(2-\slashed{\rm div}b\right)\right]\left(\mathfrak{j}-\eta\right)\Big)|_{v=0} =  \mathcal{P}_{\ell \geq 1}\slashed{\rm curl}\left(\Omega{\rm Ric}_{3\cdot}\right)|_{v=0},
 \end{align}
 \begin{equation}\label{2ojoijo2}
 \mathcal{P}_{\ell > \ell_0}\left(2\slashed{\rm div}\left(\mathfrak{j}-\eta\right) + 2\left|\mathfrak{j}\right|^2 - 2\left|\eta\right|^2\right)|_{v=0} = \mathcal{P}_{\ell > \ell_0}\left({\rm R}+{\rm Ric}_{34}\right)|_{v=0},
 \end{equation}
 \begin{equation}\label{32ijo32ijo32ioj}
 \mathcal{P}_{\ell > \ell_0}\left(\slashed{\rm curl}\mathfrak{n}-\slashed{\rm curl}\mathfrak{j}\right)|_{v=0} = 0,
 \end{equation}
 \begin{equation}\label{3iojoijoij42090999012322sdff}
\mathcal{P}_{2 \leq \ell \leq \ell_0}\slashed{\rm curl}b|_{v=0} = \epsilon T_{\rm low},\qquad \mathcal{P}_{\ell > \ell_0}\slashed{\rm curl}\slashed{\rm div}\slashed{\nabla}\hat{\otimes}b|_{v=0} = \epsilon T_{\rm high}.
 \end{equation}
 We note that it is a consequence of~\eqref{ijoi9919njnkko2o201nbhjnbghj} that each of the individual terms in these expressions have well-defined limits at $\{v = 0\}$. 
\item\label{anitemitem4} Let $X$ be any smooth vector field along $\mathbb{S}^2_{-1,0}$, and let $\hat{\epsilon}$ be any constant satisfying $\hat{\epsilon} \in (0,\epsilon)$. Suppose that 
\begin{equation}\label{3kj2oi4iojir5i2io2}
\left\vert\left\vert \mathcal{L}_X\mathring{g}\right\vert\right\vert_{\mathring{H}^{N_1}} + \left\vert\left\vert \epsilon \left( \mathcal{L}_X T_{\rm low},\mathcal{L}_XT_{\rm high}\right)\right\vert\right\vert_{\mathring{H}^{N_1-4}} \lesssim \hat{\epsilon}.
\end{equation}
Then we can apply $\mathcal{L}_X$ to every term on the left hand sides of~\eqref{ijoi9919njnkko2o201nbhjnbghj} and replace the right hand side by $\hat{\epsilon}$ if we lower by $1$ the total number of angular derivatives in the definitions of all of the norms on the left hand side of~\eqref{ijoi9919njnkko2o201nbhjnbghj}.

\end{enumerate}
\end{theorem}

\section{Solving for the Lapse}\label{lapthesection}
We will decompose the lapse $\Omega$ into two pieces
\[\log\Omega = \log\Omega_{\rm sing} + \log\Omega_{\rm boun},\]
where $\Omega_{\rm sing}$ will be spherically symmetric and satisfy $\Omega_{\rm sing} \sim (-v)^{-\kappa}$ as $v\to -1$ for some $\left|\kappa\right| \lesssim \epsilon$. The other piece $\Omega_{\rm boun}$ will instead be bounded as $v\to -1$, though general angular derivatives of $\Omega_{\rm bound}$ may also potentially blow-up as $v\to -1$. In this section we will discuss the type of equations which will eventually be used to solve for $\Omega$. 

In the final iteration argument which solves for $\Omega$ our starting point will be the equation from Lemma~\ref{2km2omo34} with all of the Ricci curvature terms set to $0$. We will then define an unknown $\mathfrak{X} \doteq \left(v+1\right)^2\left(\Omega\underline{\omega}\right)$. After setting $u = -1$, this will lead to an equation schematically  of the form (see the discussion in Section~\ref{3ijoi901} above)
\begin{equation}\label{2pok3om23o}
\left((-v)\mathcal{L}_{\partial_v}-\mathcal{L}_b\right)\left(\mathcal{L}_{\partial_v}\mathfrak{X} + H_2\right) + \left(1-\frac{3}{2}\slashed{\rm div}b+4\left(\Omega\underline{\omega}\right)\right)\left(\mathcal{L}_{\partial_v}\mathfrak{X} + H_2\right) + \Omega^2\slashed{\Delta}\mathfrak{X} = H_1,
\end{equation}
where $H_1$ and $H_2$ are suitable nonlinear expressions. Thus in this section we will focus on studying (suitable projections of) equations of the form~\eqref{2pok3om23o}.

\subsection{Solving for $\Omega_{\rm boun}$}
Throughout this subsection we let $\Omega_{\rm sing}(v) : (-1,0) \to (0,\infty)$ be a given spherically symmetric function, $b^A$ be an  $\mathbb{S}^2_{-1,v}$ vector field for $v \in (-1,0)$, and $\slashed{g}_{AB}$ be a positive definite symmetric $(0,2)$-$\mathbb{S}^2_{-1,v}$ tensor for $v\in (-1,0)$. We will assume that for a suitable function $\tilde{\Omega}$ satisfying~\eqref{k2moo39} with $\tilde{\kappa} = \kappa$ satisfying $\left|\kappa\right| \lesssim \epsilon$, and some vector field $\tilde{b}$, we have 
\begin{equation}\label{jiojroijoijoi3r2}
 \left\vert\left\vert b\right\vert\right\vert_{\mathscr{A}^{-}_1(\kappa)} + \left\vert\left\vert \slashed{g}\right\vert\right\vert_{\mathscr{A}^-_2\left(\kappa,\tilde{b}\right)} + \left\vert\left\vert \log\Omega_{\rm sing}\right\vert\right\vert_{\mathscr{B}_{00}\left(\kappa\right)}+\left\vert\left\vert b\right\vert\right\vert_{\mathscr{B}^-_1(\kappa)} + \left\vert\left\vert \slashed{g}\right\vert\right\vert_{\mathscr{B}^-_2\left(\kappa\right)}  \lesssim \epsilon.
\end{equation}
We emphasize that none of the results in this section depend on the implied constants in~\eqref{jiojroijoijoi3r2} or in the inequality for $\kappa$ (though by our conventions for $\epsilon$, we may assume that $\epsilon$ is sufficiently small depending on the implied constants).

We can now define the equation of interest.
\begin{definition}\label{2km2o20392}We say that a function $\Omega_{\rm boun}: (-1,0) \times \mathbb{S}^2 \to (0,\infty)$ satisfies the $\Omega_{\rm boun}$-equation with right hand sides $H_1,H_2: (-1,0)\times \mathbb{S}^2 \to \mathbb{R}$ if  $\left(1-\mathcal{P}_{\ell \geq 1}\right)\left(H_1,H_2\right) = 0$, $H_2$ is supported for $v \geq -1/2$, and 
\begin{equation}\label{2o2o4ji2}
\left((-v)\mathcal{L}_{\partial_v}-\mathcal{L}_b\right)\log\Omega_{\rm boun} = \left(v+1\right)^{-2}\mathfrak{X}_{\geq 1},\qquad \log\Omega_{\rm boun}|_{v=-1} = 0,
\end{equation}
\begin{align}\label{2o2o4io2i4}
&\left((-v)\mathcal{L}_{\partial_v}-\mathcal{P}_{\ell \geq 1}\mathcal{L}_b\right)\mathcal{L}_{\partial_v}\mathfrak{X}_{\geq 1} + \mathcal{P}_{\ell \geq 1}\left[\left(1-\frac{3}{2}\slashed{\rm div}b + 4\left(\Omega\underline{\omega}\right)\right)\mathcal{L}_{\partial_v}\mathfrak{X}_{\geq 1}\right]+\mathcal{P}_{\ell \geq 1}\left(\Omega^2\slashed{\Delta}\mathfrak{X}_{\geq 1}\right)  =  
\\ \nonumber &\qquad \left((-v)\mathcal{L}_{\partial_v}-\mathcal{P}_{\ell \geq 1}\mathcal{L}_b\right)H_2 + \mathcal{P}_{\ell \geq 1}\left[\left(1-\frac{3}{2}\slashed{\rm div}b + 4\left(\Omega\underline{\omega}\right)\right)H_2\right]+H_1,
\\ \nonumber &\qquad \qquad \qquad \qquad \left(1-\mathcal{P}_{\ell \geq 1}\right)\mathfrak{X}_{\geq 1} = 0,\qquad \mathfrak{X}_{\geq 1}|_{v=0} = 0,
\end{align}
where we have $\log\Omega \doteq \log\Omega_{\rm sing} + \log\Omega_{\rm boun}$ and 
\[\left(\Omega\underline{\omega}\right) \doteq \frac{1}{2}\left((-v)\mathcal{L}_{\partial_v}-\mathcal{L}_b\right)\log\Omega_{\rm sing} + \frac{1}{2}\left(v+1\right)^{-2}\mathfrak{X}_{\geq 1}.\]
\end{definition}
\begin{remark}We note that this equation is quasilinear due to the  presence of the nonlinear term $\Omega_{\rm sing}^2\Omega_{\rm boun}^2\slashed{\Delta}\mathfrak{X}$. 
\end{remark}

We are now ready for our key result.
\begin{proposition}\label{23ini2999jn2j23i3j}Suppose that 
\[\left\vert\left\vert \left(H_1,H_2,0\right)\right\vert\right\vert_{R,I,N_1-2} \lesssim \epsilon.\]

Then there exists a solution $\Omega_{\rm boun}$ to the $\Omega_{\rm boun}$-equation which furthermore satisfies the estimate
\[\left\vert\left\vert \log\Omega_{\rm boun}\right\vert\right\vert_{\mathscr{A}\left(\kappa,b\right)} + \left\vert\left\vert \log\Omega_{\rm boun}\right\vert\right\vert_{\mathscr{B}_{01}\left(\kappa,b\right)} \lesssim \left\vert\left\vert \left(H_1,H_2,0\right)\right\vert\right\vert_{R,I,N_1-2}.\]
\end{proposition}
\begin{proof}We will solve the equation by carrying out a suitable iteration procedure. We define a sequence of functions $\{\mathfrak{X}^{(i)}_{\geq 1}\}_{i=-1}^{\infty}$ and $\{\Omega_{\rm boun}^{(i)}\}_{i=-1}^{\infty}$ by solving
\[\Omega_{\rm boun}^{(-1)} = 1,\qquad \mathfrak{X}^{(-1)}_{\geq 1} = 0,\]
\begin{align}\label{2k3o4r02j94}
&\mathscr{L}^{(i-1)}\mathfrak{X}^{(i)} \doteq \left((-v)\mathcal{L}_{\partial_v}-\mathcal{P}_{\ell \geq 1}\mathcal{L}_b\right)\mathcal{L}_{\partial_v}\mathfrak{X}^{(i)}_{\geq 1} + \mathcal{P}_{\ell \geq 1}\left[\left(1-\frac{3}{2}\slashed{\rm div}b + 4\left(\Omega\underline{\omega}\right)^{(i-1)}\right)\mathcal{L}_{\partial_v}\mathfrak{X}^{(i)}_{\geq 1}\right]
\\ \nonumber &\qquad \qquad +\mathcal{P}_{\ell \geq 1}\left(\left(\Omega^{(i-1)}\right)^2\slashed{\Delta}\mathfrak{X}^{(i)}_{\geq 1}\right) = 
\\ \nonumber &\qquad  \left((-v)\mathcal{L}_{\partial_v}-\mathcal{P}_{\ell \geq 1}\mathcal{L}_b\right)H_1 + \mathcal{P}_{\ell \geq 1}\left[\left(1-\frac{3}{2}\slashed{\rm div}b + 4\left(\Omega\underline{\omega}\right)^{(i-1)}\right)H_1\right]+H_2,\qquad \forall i\geq 0,
\end{align}
\[\Omega^{(i)} = \Omega_{\rm sing}\Omega_{\rm boun}^{(i)},\qquad \left(\Omega\underline{\omega}\right)^{(i)} \doteq \frac{1}{2}\left((-v)\mathcal{L}_{\partial_v}-\mathcal{L}_b\right)\log\Omega^{(i)},\]
\begin{equation}\label{2ol2om4om2o3}
\left((-v)\mathcal{L}_{\partial_v}-\mathcal{L}_b\right)\log\Omega_{\rm boun}^{(i)} = \left(v+1\right)^{-2}\mathfrak{X}^{(i)}_{\geq 1},\qquad \forall i \geq 0,
\end{equation}
with the boundary conditions
\[\log\Omega_{\rm boun}^{(i)}|_{v=-1} = 0,\qquad \mathfrak{X}_{\geq 1}^{(i)}|_{v=0} = 0.\]
We will now prove by induction on $i$ that 
\begin{align}\label{2o2mo4o23}
&\left\vert\left\vert \log\Omega^{(i)}_{\rm boun}\right\vert\right\vert_{\mathscr{A}\left(\kappa,b\right)} + \left\vert\left\vert \log\Omega^{(i)}_{\rm boun}\right\vert\right\vert_{\mathscr{B}_{01}\left(\kappa,b\right)} + \left\vert\left\vert \mathfrak{X}_{\geq 1}^{(i)}\right\vert\right\vert_{L,I,N_1-1}\leq C_{\rm boot}\left\vert\left\vert \left(H_1,H_2,0\right)\right\vert\right\vert_{R,I,N_1-2},
\end{align}
for a suitable bootstrap constant $C_{\rm boot}$, to be fixed later. In the rest of the proof, all constants are assumed independent of the to be chosen constant $C_{\rm boot}$. 

The base case $i = -1$ is immediate, so we assume that $j$ is a non-negative integer and that~\eqref{2o2mo4o23} holds for $i = j-1$. We may use Lemma~\ref{2omomo393} with vanishing boundary conditions at $\{v = 0\}$ (and straightforward nonlinear estimates) to solve~\eqref{2k3o4r02j94} and obtain a solution $\mathfrak{X}^{(j)}$ which solves~\eqref{2k3o4r02j94} with $i =j$ and satisfies
\[\left\vert\left\vert \mathfrak{X}_{\geq 1}^{(j)}\right\vert\right\vert_{L,I,N_1-1}\lesssim \left\vert\left\vert \left(H_1,H_2,0\right)\right\vert\right\vert_{R,I,N_1-2}.\]

We turn now to the estimates for $\log\Omega^{(j)}_{\rm boun}$. Keeping in mind that $\mathfrak{X}_{\geq 1}^{(j)}|_{v=0} = 0$, we may integrate along the integral curves of $(-v)\mathcal{L}_{\partial_v}-\mathcal{L}_b$ to obtain that $\log\Omega^{(j)}_{\rm boun}$ is well-defined and satisfies 
\begin{equation}\label{2lm4omo2o49}
\sup_{\left(v,\theta^A\right)\in (-1,0)\times\mathbb{S}^2}\left|\log\Omega^{(j)}_{\rm boun}\right|\left(v+1\right)^{-1+\check{\delta}} \lesssim \left\vert\left\vert \mathfrak{X}_{\geq 1}^{(j)}\right\vert\right\vert_{L,I,N_1-1}.
\end{equation}
For any $\left|\alpha\right| \leq 1$, we may commute~\eqref{2ol2om4om2o3} with $\mathcal{L}_{\mathcal{Z}^{(\alpha)}}$ to obtain  
\begin{equation}\label{2o3mo4o2}
\left((-v)\mathcal{L}_{\partial_v}-\mathcal{L}_b\right)\left(\log\Omega_{\rm boun}^{(j)}\right)^{(\alpha)} - \mathcal{L}_{[Z^{(\alpha)},b]}\log\Omega_{\rm boun}^{(j)} = \left(v+1\right)^{-2}\mathfrak{X}_{\geq 1}^{(j)}.
\end{equation}
Again we may integrate along the integral curves of $(-v)\mathcal{L}_{\partial_v} - \mathcal{L}_b$, sum the resulting estimate over $\left|\alpha\right| \leq 1$, and apply Gr\"{o}nwall's inequality to obtain 
\begin{equation}\label{ijo23joi2ijooij2r3i}
\sum_{\left|\alpha\right| \leq 1}\sup_{\left(v,\theta^A\right)\in (-1,0)\times\mathbb{S}^2}\left|\left(\log\Omega^{(j)}_{\rm boun}\right)^{(\alpha)}\right|\left(v+1\right)^{-1+\check{\delta}}(-v)^{\epsilon^{3/4}} \lesssim \left\vert\left\vert \mathfrak{X}_{\geq 1}^{(j)}\right\vert\right\vert_{L,I,N_1-1}.
\end{equation}
Next, we observe that we can commute with $\mathcal{L}_b$ to obtain that
\[\left((-v)\mathcal{L}_{\partial_v}-\mathcal{L}_b\right)\mathcal{L}_b\log\Omega^{(j)}_{\rm boun} = \left(v+1\right)^{-2}\mathcal{L}_b\mathfrak{X}_{\geq 1}^{(j)} + v\left(\mathcal{L}_{\partial_v}b\right)^A\slashed{\nabla}_A\log\Omega^{(j)}_{\rm boun}.\]
We may again integrate along the integral curves of $(-v)\mathcal{L}_{\partial_v}-\mathcal{L}_b$ and use~\eqref{ijo23joi2ijooij2r3i} to obtain the estimate~\eqref{2lm4omo2o49} with $\mathcal{L}_b\log\Omega^{(j)}_{\rm boun}$ replacing $\log\Omega^{(i)}_{\rm boun}$ and $(v,\theta^A) \in (-1/2,0) \times \mathbb{S}^2$ replacing $(v,\theta^A) \in (-1,0) \times \mathbb{S}^2$. Revisiting the transport equation defining $\log\Omega^{(j)}_{\rm boun}$ then also provides a bound for $(-v)\mathcal{L}_{\partial_v}\log\Omega^{(j)}_{\rm boun}$. All together, we have 
\begin{equation}\label{2lm4omo2o49123}
\sup_{\left(v,\theta^A\right)\in (-1/2,0)\times\mathbb{S}^2}\left|\left((-v)\mathcal{L}_{\partial_v},\mathcal{L}_b,1\right)\log\Omega^{(j)}_{\rm boun}\right| \lesssim  \left\vert\left\vert \mathfrak{X}_{\geq 1}^{(j)}\right\vert\right\vert_{L,I,N_1-1}.
\end{equation}

Next we consider higher derivative $\sup_v$ estimates for $\log\Omega^{(j)}_{\rm boun}$. It is an immediate consequence of Lemma~\ref{linftofkwp3} that
\begin{equation}\label{2j3o2o3}
\sum_{\left|\alpha\right|+k \leq N_1-2\atop k\in \{0,1,2\}}\sup_{v \in (-1,-1/2)} \left(v+1\right)^{-2+2\check{\delta}+2k}\int_{\mathbb{S}^2}\left|\mathcal{L}^k_{\partial_v}\left(\log\Omega^{(j)}_{\rm boun}\right)^{(\alpha)}\right|^2\mathring{\rm dVol} \lesssim \left\vert\left\vert \mathfrak{X}^{(j)}\right\vert\right\vert^2_{L,I,N_1-1}.
\end{equation}
In order to control $\left(\log\Omega^{(j)}_{\rm boun}\right)^{(\alpha)}$ with $\left|\alpha\right| = N_1-1$ however, we must argue a bit differently (note that the bound on $\left\vert\left\vert \mathfrak{X}^{(i)}\right\vert\right\vert_{I,N_1-1}$ only establishes that $\left(v+1\right)^{-2+\check{\delta}}\left(\mathfrak{X}^{(i)}\right)^{(\alpha)}$ is bounded in $L^{\infty}_{v \leq -1/2}L^2\left(\mathbb{S}^2\right)$ for $\left|\alpha\right| \leq N_1-2$). Commute~\eqref{2ol2om4om2o3} with $\left(v+1\right)^2\mathcal{P}_{\ell \geq 1}\left(\Omega^{(j-1)}\right)^2\slashed{\Delta}$ and use the equation~\eqref{2k3o4r02j94}. We may obtain
\begin{align}\label{2m2om4oo23iot4ioiot4}
&\left((-v)\mathcal{L}_{\partial_v}-\mathcal{L}_b\right)\left(\left(v+1\right)^2\mathcal{P}_{\ell \geq 1}\left(\left(\Omega^{(j-1)}\right)^2\slashed{\Delta}\left(\log\Omega^{(j)}_{\rm boun}\right)\right)\right) = 
\\ \nonumber &\qquad -\left((-v)\mathcal{L}_{\partial_v}-\mathcal{P}_{\ell \geq 1}\mathcal{L}_b\right)\left(\mathcal{L}_{\partial_v}\mathfrak{X}^{(j)}_{\geq 1}-H_1 \right)
 - \mathcal{P}_{\ell \geq 1}\left[\left(1-\frac{3}{2}\slashed{\rm div}b + 4\left(\Omega\underline{\omega}\right)^{(j-1)}\right)\left(\mathcal{L}_{\partial_v}\mathfrak{X}^{(j)}_{\geq 1}-H\right)\right]
\\ \nonumber &\qquad+ H_2 - \left[\left((-v)\mathcal{L}_{\partial_v}-\mathcal{L}_b\right),\left(v+1\right)^2\mathcal{P}_{\ell \geq 1}\left(\left(\Omega^{(j-1)}\right)^2\slashed{\Delta}\cdot\right)\right]\left(\log\Omega^{(j)}_{\rm bound}\right).
\end{align}
The equation~\eqref{2m2om4oo23iot4ioiot4} may be re-written as a transport equation for 
\[\left(v+1\right)^2\mathcal{P}_{\ell \geq 1}\left(\left(\Omega^{(j-1)}\right)^2\slashed{\Delta}\left(\log\Omega^{(j)}_{\rm boun}\right)\right)+ \left(\mathcal{L}_{\partial_v}\mathfrak{X}^{(j)}_{\geq 1}-H_1 \right).\]
where the terms on the right hand side involving only $\mathcal{L}_{\partial_v}\mathfrak{X}^{(j)}_{\geq 1}$ and, in the final nonlinear term, up to $2$ angular derivatives of $\log\Omega^{(j)}_{\rm boun}$. Thus, we can use Lemma~\ref{linftofkwp3},~\eqref{2j3o2o3}, the equation relating $\Omega^{(j)}_{\rm boun}$ with $\mathfrak{X}^{(j)}_{\geq 1}$, and elliptic estimates along $\mathbb{S}^2$ to obtain 
\begin{align}\label{kn2i3oim402}
& \sum_{j =0}^2\left\vert\left\vert \mathcal{L}_{\partial_v}^j\log\Omega^{(j)}_{\rm boun}\right\vert\right\vert_{\mathscr{S}_{-1}^{-1/2}\left(N_1-1-j,-1+\check{\delta}+j,0\right)}^2\lesssim \left\vert\left\vert \mathfrak{X}^{(j)}_{\geq 1}\right\vert\right\vert^2_{L,I,N_1-1}
\end{align}
This concludes the necessary $L^{\infty}_{v \leq -1/2}L^2\left(\mathbb{S}^2\right)$ estimates.

Now we consider $v \geq -1/2$. The fundamental theorem of calculus in $v$ and the fact that $\mathfrak{X}^{(j)}_{\geq 1}|_{v=0} = 0$ imply that 
\begin{equation}\label{mflwmfl3}
\left\vert\left\vert \mathfrak{X}^{(j)}_{\geq 1}\right\vert\right\vert_{\mathscr{S}_{-1/2}^0\left(N_1-2,0,-1/2\right)}^2 \lesssim \left\vert\left\vert \mathfrak{X}^{(j)}_{\geq 1}\right\vert\right\vert^2_{L,I,N_1-1} .
\end{equation}
It particular, it is immediate from Lemma~\ref{linftofkwp3} that
\begin{equation}\label{2om3om2o34923}
\sum_{j=0}^1\left\vert\left\vert \log\Omega^{(j)}_{\rm boun}\right\vert\right\vert_{\mathscr{S}_{-1/2}^0\left(N_1-2-j,0,\epsilon^{\frac{9}{20}}+j\right)}^2  \lesssim   \left\vert\left\vert \mathfrak{X}^{(j)}_{\geq 1}\right\vert\right\vert^2_{L,I,N_1-1}.
\end{equation}
Combining the available estimates, we have thus shown that
\[\left\vert\left\vert \log\Omega^{(j)}_{\rm boun}\right\vert\right\vert^2_{\mathscr{B}\left(b\right)} \lesssim \left\vert\left\vert \left(H_1,H_2,0\right)\right\vert\right\vert_{R,I,N_1-2}^2.\]
The argument to control $\left\vert\left\vert \log\Omega^{(j)}_{\rm boun}\right\vert\right\vert_{\mathscr{A}\left(\kappa,b\right)}$ given our estimates on $\mathfrak{X}^{(j)}_{\geq 1}$ involves analogous ideas to our previous use of~\eqref{2m2om4oo23iot4ioiot4} and the Hardy inequality
\begin{align*}
&\int_{-1/2}^0\int_{\mathbb{S}^2}(-v)^{-1+2\check{\delta}}f^2\, dv\mathring{\rm dVol} \lesssim 
\\ \nonumber &\qquad \int_{-1/2}^{-1/4}\int_{\mathbb{S}^2}f^2\, dv \mathring{\rm dVol} + \int_{-1/2}^0\int_{\mathbb{S}^2}(-v)^{-1+2\check{\delta}}\left((-v)\mathcal{L}_{\partial_v}f-\mathcal{L}_bf\right)^2\, dv \mathring{\rm dVol},
\end{align*}
which holds for any suitably regular function $f$. We thus omit the details. Thus, for a suitable bootstrap constant $C_{\rm boot}$, we find that we will have verified the induction assumption.

Finally, it is straightforward to use a compactness argument to extract a suitably convergent subsequence as $j \to \infty$ and observe that the limiting $\Omega_{\rm boun}$ and $\mathfrak{X}_{\geq 1}$ solve the desired equations  and satisfy the required bounds.

\end{proof}

We close with two lemmas which concerns solutions to transport equations sourced by the lapse.
\begin{lemma}\label{2ij23ji3ijo32}Let $H_4 : (-1,0) \times \mathbb{S}^2 \to \mathbb{R}$ be a function which lies in the closure of smooth functions under then norm determined by the right hand side of~\eqref{3ij32ji324i}. Then there exists a unique function $Y$ which solves
\begin{equation}\label{23ij32io32}
\left((-v)\mathcal{L}_{\partial_v}-\mathcal{L}_b\right)Y + \left((-v)\Omega{\rm tr}\chi +1 -\frac{3}{2}\slashed{\rm div}b\right)Y = -\slashed{\Delta}\Omega\underline{\omega} + H_4,\qquad \left(v+1\right)^2Y \to 0\text{ as }v\to -1.
\end{equation}
We also let $\tilde{H}_6 : (-1,0)\times \mathbb{S}^2 \to \mathbb{R}$ be the unique solution to 
\begin{equation}\label{3i4moo2}
\left((-v)\mathcal{L}_{\partial_v}-\mathcal{P}_{\ell \geq 1}\mathcal{L}_b\right)\tilde{H}_6 + \left(1+\frac{4(-v)}{v+1}\right)\tilde{H}_6 = 8(v+1)^{-2}\mathring{\Delta}\left(\Omega\underline{\omega}\right),\qquad \left(v+1\right)^4\tilde{H}_6 \to 0 \text{ as }v\to -1.
\end{equation}

Then we have the following estimates
\begin{align}\label{io4ijorjiw4etj}
&\sum_{j=0}^1\left[\left\vert\left\vert \mathcal{L}_{\partial_v}^j\mathcal{P}_{\ell = 0}Y\right\vert\right\vert_{\mathscr{S}\left(0,5\check{\delta}+j,j\right)} +\left\vert\left\vert \mathcal{L}_{\partial_v}^j\mathcal{P}_{\ell = 0}Y\right\vert\right\vert_{\mathscr{Q}_{-1}^{-1/2}\left(0,-1/2+5\check{\delta}+k,0\right)}\right]\lesssim 
\\ \nonumber &\qquad  \left\vert\left\vert \slashed{g}\right\vert\right\vert_{\mathscr{B}^-_2\left(\kappa\right)}^2  + \left\vert\left\vert \log\Omega_{\rm boun}\right\vert\right\vert^2_{\mathscr{B}_{01}\left(\kappa,b\right)}  + \left\vert\left\vert H_4\right\vert\right\vert_{\mathscr{Q}\left(N_1-2,1/2+5\check{\delta},-\kappa\right)}
\\ \nonumber &\qquad + \left\vert\left\vert H_4\right\vert\right\vert_{\mathscr{S}_{-1}^{-1/2}\left(N_1-3,1+5\check{\delta},0\right)}+ \left\vert\left\vert H_4\right\vert\right\vert_{\mathscr{S}_{-1/2}^0\left(N_2-2,0,0\right)} 
\end{align}
\begin{align}\label{3ij32ji324i}
&\sum_{j=0}^1\left\vert\left\vert \mathcal{L}_{\partial_v}^j\left(2Y+\slashed{\Delta}\log\Omega\right)\right\vert\right\vert_{\mathscr{Q}\left(N_1-2-j,-1/2+5\check{\delta}+j,-\kappa +j\right)} \\ \nonumber&\qquad + \sum_{j=0}^1\left\vert\left\vert \mathcal{L}_{\partial_v}^j\left(2Y+\slashed{\Delta}\log\Omega\right)\right\vert\right\vert_{\mathscr{S}_{-1}^{-1/2}\left(N_1-3-j,5\check{\delta}+j,0\right)} +\sum_{j=0}^1\left\vert\left\vert \mathcal{L}_{\partial_v}^jY\right\vert\right\vert_{\mathscr{S}_{-1/2}^0\left(N_2-2-j,0,j\right)}\lesssim 
\\ \nonumber &\qquad \left\vert\left\vert \log\Omega_{\rm boun}\right\vert\right\vert_{\mathscr{A}\left(\kappa,b\right)} + \left\vert\left\vert \log\Omega_{\rm boun}\right\vert\right\vert_{\mathscr{B}_{01}\left(\kappa,b\right)} + \left\vert\left\vert H_4\right\vert\right\vert_{\mathscr{Q}\left(N_1-2,1/2+5\check{\delta},-\kappa\right)}
\\ \nonumber &\qquad + \left\vert\left\vert H_4\right\vert\right\vert_{\mathscr{S}_{-1}^{-1/2}\left(N_1-3,1+5\check{\delta},0\right)}+ \left\vert\left\vert H_4\right\vert\right\vert_{\mathscr{S}_{-1/2}^0\left(N_2-2,0,0\right)} \Rightarrow 
\end{align}
\begin{align*}
&\left\vert\left\vert (-v)^{-2\kappa} \left(2Y+\slashed{\Delta}\log\Omega\right),0\right\vert\right\vert_{\mathscr{P}\mathscr{R}\left(\Omega,\kappa,50\check{p},50\check{p}\right)} \lesssim 
\\ \nonumber &\qquad \left\vert\left\vert \log\Omega_{\rm boun}\right\vert\right\vert_{\mathscr{A}\left(\kappa,b\right)} + \left\vert\left\vert \log\Omega_{\rm boun}\right\vert\right\vert_{\mathscr{B}_{01}\left(\kappa,b\right)} + \left\vert\left\vert H_4\right\vert\right\vert_{\mathscr{Q}\left(N_1-2,1/2+5\check{\delta},-\kappa\right)}
\\ \nonumber &\qquad + \left\vert\left\vert H_4\right\vert\right\vert_{\mathscr{S}_{-1}^{-1/2}\left(N_1-3,1+5\check{\delta},0\right)}+ \left\vert\left\vert H_4\right\vert\right\vert_{\mathscr{S}_{-1/2}^0\left(N_2-2,0,0\right)}.
\end{align*}
Similarly,
\begin{align}\label{3ij32ji324i123}
&\left\vert\left\vert \tilde{H}_6\right\vert\right\vert_{\mathscr{Q}\left(N_1-3,3/2+\check{\delta},-1/2+\check{\delta}\right)} +\sum_{j=0}^1\left\vert\left\vert \mathcal{L}_{\partial_v}^{1+j}\tilde{H}_6\right\vert\right\vert_{\mathscr{Q}\left(N_1-4-j,5/2+\check{\delta}+j,\kappa+j\right)}
\\ \nonumber &\qquad + \sum_{j=0}^1\left\vert\left\vert \mathcal{L}_{\partial_v}^j\tilde{H}_6\right\vert\right\vert_{\mathscr{S}\left(N_2-2-j,1+j,j\right)}
\\ \nonumber &\qquad +\sum_{j=0}^2\left\vert\left\vert \mathcal{L}_{\partial_v}^{1+j}\tilde{H}_6\right\vert\right\vert_{\mathscr{Q}_{-1}^{-1/2}\left(N_1-4-j,5/2+\check{\delta}+j,0\right)} + \sum_{j=0}^2\left\vert\left\vert \mathcal{L}_{\partial_v}^j\tilde{H}_6\right\vert\right\vert_{\mathscr{S}_{-1}^{-1/2}\left(N_1-3-j,2+\check{\delta}+j,0\right)} \lesssim 
\\ \nonumber &\qquad \left\vert\left\vert \log\Omega_{\rm boun}\right\vert\right\vert_{\mathscr{A}\left(\kappa,b\right)} + \left\vert\left\vert \log\Omega_{\rm boun}\right\vert\right\vert_{\mathscr{B}_{01}\left(\kappa,b\right)} \Rightarrow
\end{align}
\begin{align}\label{3ij2oj4}
&\left\vert\left\vert \left(0,\tilde{H}_6\right)\right\vert\right\vert_{\mathscr{P}\mathscr{R}\left(\Omega,\kappa,50\check{p},50\check{p}\right)} \lesssim 
\\ \nonumber &\qquad\left\vert\left\vert \log\Omega_{\rm boun}\right\vert\right\vert_{\mathscr{A}\left(\kappa,b\right)} + \left\vert\left\vert \log\Omega_{\rm boun}\right\vert\right\vert_{\mathscr{B}_{01}\left(\kappa,b\right)} + \left\vert\left\vert \slashed{g} \right\vert\right\vert^2_{\mathscr{A}_2\left(\kappa,b,\Omega\right)} + \left\vert\left\vert \slashed{g}\right\vert\right\vert^2_{\mathscr{B}_2\left(\kappa,b,\Omega\right)} +  \left\vert\left\vert \log\Omega_{\rm sing}\right\vert\right\vert_{\mathscr{B}_{00}\left(\kappa\right)}^2.
\end{align}
\end{lemma}
\begin{proof}The estimates~\eqref{3ij32ji324i} are~\eqref{3ij32ji324i123} are consequence of the transport equation estimates from Lemmas~\ref{linftofkwp3} and~\ref{l2l2degtranstrans} as well as the fact that $4\slashed{\Delta}\Omega\underline{\omega}$ may be written, to leading order as $v\to -1$ (see~\eqref{acommut2}), as 
\[\left((-v)\mathcal{L}_{\partial_v}-\mathcal{L}_b\right)\left(2\slashed{\Delta}\log\Omega\right) + (-v)\Omega{\rm tr}\chi\left(2\slashed{\Delta}\log\Omega\right).\]
For the $\tilde{H}_6$ estimate, we must also consider an additional commutation with $\mathcal{L}_{\partial_v}$. We omit the details.

\end{proof}

\subsection{Solving for $\Omega_{\rm sing}$}
Throughout this subsection we let $\Omega_{\rm boun} : (-1,0) \times \mathbb{S}^2 \to (0,\infty)$ be a given  function, $b^A$ be an  $\mathbb{S}^2_{-1,v}$ vector field for $v \in (-1,0)$, and $\slashed{g}_{AB}$ be a positive definite symmetric $(0,2)$-$\mathbb{S}^2_{-1,v}$ tensor for $v\in (-1,0)$. We will assume that for a suitable bootstrap constant function $\tilde{\Omega}$ satisfying~\eqref{k2moo39} with $\tilde{\kappa}$ satisfying $\left|\tilde{\kappa}\right| \lesssim \epsilon$ and some vector field $\tilde{b}$, we have 
\begin{align}\label{i2nini1i2i}
& \left\vert\left\vert b\right\vert\right\vert_{\mathscr{A}^{-}_1(\tilde{\kappa})} + \left\vert\left\vert \slashed{g}\right\vert\right\vert_{\mathscr{A}_2\left(\tilde{\kappa},\tilde{b},\tilde{\Omega}\right)} +\left\vert\left\vert \log\Omega_{\rm boun}\right\vert\right\vert_{\mathscr{A}\left(\tilde{\kappa},\tilde{b}\right)}
 \\ \nonumber &\qquad + \left\vert\left\vert \log\Omega_{\rm boun}\right\vert\right\vert_{\mathscr{B}_{01}\left(\tilde{\kappa},\tilde{b}\right)}+\left\vert\left\vert b\right\vert\right\vert_{\mathscr{B}^-_1(\tilde{\kappa})} + \left\vert\left\vert \slashed{g}\right\vert\right\vert_{\mathscr{B}_2\left(\tilde{\kappa},\tilde{b},\tilde{\Omega}\right)}  \lesssim \epsilon.
\end{align}
We emphasize that none of the results in this section depend on the implied constants in~\eqref{i2nini1i2i} or in the inequality for $\tilde{\kappa}$ (though by our conventions for $\epsilon$, we may assume that $\epsilon$ is sufficiently small depending on the implied constants).

We can now define the equation of interest.
\begin{definition}\label{3j2jiij23}We say that a spherically symmetric function $\Omega_{\rm sing}: (-1,0) \times \mathbb{S}^2 \to (0,\infty)$ satisfies the $\Omega_{\rm sing}$-equation with right hand side $H_3 : (-1,0)\times \mathbb{S}^2 \to \mathbb{R}$ satisfying $\left(1-\mathcal{P}_{\ell \geq 0}\right)H_3 = 0$  if 
\begin{equation}\label{2o2o4ji2}
\mathcal{L}_{\partial_v}\left((-v)\mathcal{L}_{\partial_v}\log\Omega_{\rm sing}\right) + \mathcal{P}_{\ell = 0}\left(\Omega{\rm tr}\chi\left((-v)\mathcal{L}_{\partial_v}\log\Omega_{\rm sing}\right)\right) = H_3.
\end{equation}
\end{definition}

Our main result of the subsection is the following.
\begin{proposition}\label{iojoij2}Suppose that $H_3$ lies in the closure of smooth functions under the norm determined by the right hand side of~\eqref{2om3om}.  

Then there exists a constant $\kappa$  and a solution $\Omega_{\rm sing}$ to the $\Omega_{\rm sing}$-equation which furthermore satisfies the estimate
\begin{align}\label{2om3om}
&\left\vert\left\vert \log\Omega_{\rm sing}\right\vert\right\vert_{\mathscr{B}_{00}\left(\kappa\right)} + \left|\kappa\right| 
\\ \nonumber &\qquad \lesssim \sum_{j=0}^1\left\vert\left\vert \mathcal{L}_{\partial_v}^jH_3\right\vert\right\vert_{\mathscr{S}_{-1}^{-1/2}\left(0,1+\check{\delta}+j,0\right)}+\sum_{j=0}^1\left\vert\left\vert \mathcal{L}_{\partial_v}^jH_3\right\vert\right\vert_{\mathscr{Q}_{-1}^{-1/2}\left(0,1/2+\check{\delta}+j,0\right)}+ \sum_{j=0}^1\left\vert\left\vert \left(v\mathcal{L}_{\partial_v}\right)^jH_3\right\vert\right\vert_{\check{\mathscr{S}}_{-1/2}^0\left(0,0,200\check{p}\left(1+j\right)\right)}.
\end{align}
\end{proposition}
\begin{proof}
This is an immediate consequence of standard o.d.e.~estimates.
\end{proof}

\section{The $\left(\slashed{g},\mathfrak{n}\right)$-system}\label{ij3jr9j3}
Throughout this section we will assume that $b^A$ is an  $\mathbb{S}^2_{-1,v}$ vector field for $v \in (-1,0)$, $\Omega_{\rm sing}$ and $\Omega_{\rm boun}$ are  given functions mapping $(-1,0)\times \mathbb{R}$ to $(0,\infty)$ with $\Omega_{\rm sing}$ spherically symmetric and $\Omega \doteq \Omega_{\rm sing}\Omega_{\rm boun}$, $\mathfrak{\pi}$ is a function on $(-1,0)\times\mathbb{S}^2$ satisfying $\left(1-\mathcal{P}_{\ell \leq \ell_0}\right)\pi = 0$, $\mathfrak{q}$ is a function on $(-1,0)\times \mathbb{S}^2$, $\mathfrak{r}$ is an $\mathbb{S}^2_{-1,v}$ symmetric $(0,2)$-tensor, and $\mathfrak{o}$ and $\mathfrak{w}$ are functions defined on $\mathbb{S}^2$ which satisfy $\left(1-\mathcal{P}_{\ell > \ell_0}\right)\mathfrak{w} =\left(1- \mathcal{P}_{\ell > \ell_0}\right)\mathfrak{o} = 0$. We will assume that for a suitable  function $\tilde{\Omega}$ satisfying~\eqref{k2moo39}, and constant $\kappa$ satisfying $\left|\kappa\right| \lesssim \epsilon$, we have 
\begin{align}\label{knefiomomo2}
&\left\vert\left\vert \log\Omega_{\rm boun}\right\vert\right\vert_{\mathscr{A}_0\left(\kappa,b\right)}+  \left\vert\left\vert b\right\vert\right\vert_{\mathscr{A}^-_1\left(\kappa\right)} + \left\vert\left\vert \log\Omega_{\rm sing}\right\vert\right\vert_{\mathscr{B}_{00}\left(\kappa\right)}+\left\vert\left\vert \log\Omega_{\rm boun}\right\vert\right\vert_{\mathscr{B}_{01}\left(\kappa,b\right)} +\left\vert\left\vert b\right\vert\right\vert_{\mathscr{B}^-_1\left(\kappa\right)} 
\\ \nonumber &+\left\vert\left\vert \pi\right\vert\right\vert_{\mathscr{S}\left(0,\check{\delta},0\right)} +\left\vert\left\vert \pi\right\vert\right\vert_{\mathscr{Q}_{-1}^{-1/2}\left(0,-1/2+\check{\delta},0\right)}+ \left\vert\left\vert \left(\mathfrak{o},\mathcal{L}_{b|_{v=0}}\mathfrak{o}\right)\right\vert\right\vert_{\mathring{H}^{N_1-3}\left(\mathbb{S}^2\right)} + \left\vert\left\vert \left(\mathfrak{w},\mathcal{L}_{b|_{v=0}}\mathfrak{w}\right)\right\vert\right\vert_{\mathring{H}^{N_1-3}\left(\mathbb{S}^2\right)}  \lesssim \epsilon,
\end{align}
\begin{align}\label{o3momo2o4}
\left\vert\left\vert \left(\mathfrak{q},\mathfrak{r}\right)\right\vert\right\vert_{\mathscr{Q}\left(N_1-1,-1/2+\check{\delta},-1/2+\check{\delta}\right)} + \left\vert\left\vert \left(\mathcal{L}_{\partial_v}\mathfrak{q},\mathcal{L}_{\partial_v}\mathfrak{r}\right)\right\vert\right\vert_{\mathscr{Q}\left(N_1-2,1/2+\check{\delta},\kappa\right)} + 
\left\vert\left\vert \left(\mathfrak{q},\mathfrak{r}\right)\right\vert\right\vert_{\mathscr{S}\left(N_2-1,\check{\delta},0\right)} \lesssim \epsilon,
\end{align}
It will be convenient to then set
\begin{align}\label{3o2omoo2010injnn}
&\mathcal{D} \doteq \left\vert\left\vert \log\Omega_{\rm boun}\right\vert\right\vert_{\mathscr{A}_0\left(\kappa,b\right)}+  \left\vert\left\vert b\right\vert\right\vert_{\mathscr{A}^-_1\left(\kappa\right)} + \left\vert\left\vert \log\Omega_{\rm sing}\right\vert\right\vert_{\mathscr{B}_{00}\left(\kappa\right)}+\left\vert\left\vert \log\Omega_{\rm boun}\right\vert\right\vert_{\mathscr{B}_{01}\left(\kappa,b\right)} +\left\vert\left\vert b\right\vert\right\vert_{\mathscr{B}^-_1\left(\kappa\right)}
\\ \nonumber &\qquad+\left\vert\left\vert \pi\right\vert\right\vert_{\mathscr{S}\left(0,\check{\delta},0\right)} +\left\vert\left\vert \pi\right\vert\right\vert_{\mathscr{Q}_{-1}^{-1/2}\left(0,-1/2+\check{\delta},0\right)}  + \left\vert\left\vert \left(\mathfrak{o},\mathcal{L}_{b|_{v=0}}\mathfrak{o}\right)\right\vert\right\vert_{\mathring{H}^{N_1-3}\left(\mathbb{S}^2\right)} + \left\vert\left\vert \left(\mathfrak{w},\mathcal{L}_{b|_{v=0}}\mathfrak{w}\right)\right\vert\right\vert_{\mathring{H}^{N_1-3}\left(\mathbb{S}^2\right)}.
\end{align}
We emphasize that none of the results in this section depend on the implied constants in~\eqref{knefiomomo2},~\eqref{o3momo2o4}, or in the inequality for $\kappa$ (though by our conventions for $\epsilon$, we may assume that $\epsilon$ is sufficiently small depending on the implied constants).

In this section we will introduce and solve a class of equations which will eventually be used to solve for $\slashed{g}$ and the artificial variable $\mathfrak{n}$.\footnote{See Section~\ref{2m2o2mo1o00050033} for a discussion of the role of artificial variables in our argument.}  We note further that we will actually define two separate systems of equations, the ``$\left(\slashed{g},\mathfrak{n}\right)$-system'' and the ``modified $\left(\slashed{g},\mathfrak{n}\right)$-system''. The $\left(\slashed{g},\mathfrak{n}\right)$ is the system we are primarily interested in solving, but the modified $\left(\slashed{g},\mathfrak{n}\right)$-system will be useful in various intermediate stages of our argument.

Before explicitly giving the relevant equations, we provide some high level motivation for where the equations of the $\left(\slashed{g},\mathfrak{n}\right)$-system come from. We will  solve for what will end up being $\Omega\chi$ and then recover $\slashed{g}$ by integrating in $v$. For $\Omega\chi$ we will solve separately for, what will eventually turn out to be, $\Omega{\rm tr}\chi$, $\slashed{\rm div}\slashed{\rm div}\left(\Omega\hat{\chi}\right)$, $\slashed{\rm curl}\slashed{\rm div}\left(\Omega\hat{\chi}\right)$. 
\begin{enumerate}
	\item There are three scalar artificial variables directly associated to the metric $\slashed{g}$: $\mathfrak{a}$, $\mathfrak{d}$, and $\mathfrak{e}$. They are related to $\slashed{g}$ as follows:
	\[\left(\slashed{g}\right)^{AB}\mathcal{L}_{\partial_v}\slashed{g}_{AB} = 2\mathfrak{a},\]
\[\slashed{\rm div}\slashed{\rm div}\left( \mathcal{L}_{\partial_v}\slashed{g} - \mathfrak{a}\slashed{g}\right)= 2\mathfrak{d},\qquad \slashed{\rm curl}\slashed{\rm div}\left( \mathcal{L}_{\partial_v}\slashed{g} - \mathfrak{a}\slashed{g}\right)= 2\mathfrak{e}.\]
In particular, we see that in the eventual solution we will have that $\mathfrak{a} = \Omega{\rm tr}\chi$, $\mathfrak{d} = \slashed{\rm div}\slashed{\rm div}\left(\Omega\hat{\chi}\right)$, and $\mathfrak{e} = \slashed{\rm curl}\slashed{\rm div}\left(\Omega\hat{\chi}\right)$.
\item Our equation for $\mathfrak{a}$ is provided by taking~\eqref{4trchi} and dropping the Ricci curvature terms. This will imply in the eventual solution that 
\[{\rm Ric}_{44} = 0.\]
\item Before we explain the equations for $\mathfrak{d}$ and $\mathfrak{e}$ we need to introduce one more artificial unknown $\mathfrak{k}$. In the eventual solution we will find that  $\mathfrak{k}$ will equal the Gauss curvature $K$. The equations for $\mathfrak{k}$ and $\mathfrak{n}$ are as follows:
\begin{enumerate}
	\item We use $\pi$ for the low-frequency part of $\mathfrak{n}$. That is, we set $\mathcal{P}_{\ell \leq \ell_0}\mathfrak{n} = \mathcal{P}_{\ell \leq \ell_0} \pi$.
	\item Using elliptic theory along $\mathbb{S}^2$, the vector $\mathcal{P}_{\ell > \ell_0}\mathfrak{n}$ will be then determined by $\mathcal{P}_{\ell >\ell_0}\slashed{\rm div}\mathfrak{n}$ and $\mathcal{P}_{\ell > \ell_0}\slashed{\rm curl}\mathfrak{n}$. 
	 \item We then relate $\mathcal{P}_{\ell > \ell_0}\slashed{\rm div}\mathfrak{n}$ and $\mathcal{P}_{\ell > \ell_0}\mathfrak{k}$ by starting with the equation~\eqref{2momoo3}, dropping all of the Ricci curvature terms, applying $\mathcal{P}_{\ell > \ell_0}$, and then replacing $\mathcal{P}_{\ell > \ell_0}K$ with $\mathcal{P}_{\ell > \ell_0}\mathfrak{k}$ and $\mathcal{P}_{\ell > \ell_0}\slashed{\rm div}\eta$ with $\mathcal{P}_{\ell >\ell_0}\slashed{\rm div}\mathfrak{n}$.
	 \item For $\mathcal{P}_{\ell > \ell_0}\slashed{\rm curl}\mathfrak{n}$ we start with the equation~\eqref{4eta}, substituting $-\frac{1}{2}R_{A434}$ with~\eqref{tcod1}, dropping the Ricci curvature terms, applying $\mathcal{P}_{\ell > \ell_0}\slashed{\rm curl}$, and replacing $\mathcal{P}_{\ell  > \ell_0}\slashed{\rm curl}\eta$ with $\mathcal{P}_{\ell > \ell_0}\slashed{\rm curl}\mathfrak{n}$ in a suitable fashion.
	 \item We relate $\mathfrak{k}$ to $\mathfrak{d}$ via the equation~\eqref{ko2o3k4} with  $\slashed{\rm div}\slashed{\rm div}\left(\Omega\hat{\chi}\right)$ replaced by $\mathfrak{d}$, and $\Omega{\rm tr}\chi$ replaced by $\mathfrak{a}$.  
\end{enumerate}
\item The equation for $\mathfrak{d}$ is then obtained by applying $\slashed{\rm div}\slashed{\rm div}$ to~\eqref{3hatchi}, replacing $\slashed{\rm div}\slashed{\rm div}\Omega\hat{\chi}$ with $\mathfrak{d}$, replacing $\eta$ with $\mathfrak{n}$, and then dropping the Ricci curvature terms. If we apply $\mathcal{P}_{\ell \leq \ell_0}$, then we may treat this equation as a transport equation for $\mathcal{P}_{\ell \leq \ell_0}\mathfrak{d}$. If we apply $\mathcal{P}_{\ell \geq \ell_0}$, then, in view of the relation of $\mathfrak{d}$ with $\mathfrak{k}$, this may be considered a model second order equation of type $III$ for $\mathcal{P}_{\ell > \ell_0}\mathfrak{k}$. In the eventual solution we will construct (and after we have that $\mathfrak{k} = K$) this will be seen to imply that 
\begin{equation}\label{2om3om4o29392}
\slashed{\nabla}^A\slashed{\nabla}^B\Omega^2\left[(\slashed{\nabla}\hat{\otimes}\left(\mathfrak{n}-\eta\right))_{AB}+\left(\mathfrak{n}\hat{\otimes}\mathfrak{n}\right)_{AB} - \left(\eta\hat{\otimes}\eta\right)_{AB} - \widehat{\rm Ric}_{AB}\right] = 0,
\end{equation}
\[\mathcal{P}_{\ell> \ell_0}\Big(2\slashed{\rm div}\left(\mathfrak{n}-\eta\right) + \left|\mathfrak{n}\right|^2-\left|\eta\right|^2 - \left(R + {\rm Ric}_{34}\right)\Big) = 0.\]
\item The equation for $\mathfrak{e}$ is then obtained by applying $\slashed{\rm curl}\slashed{\rm div}$ to~\eqref{3hatchi}, replacing $\slashed{\rm curl}\slashed{\rm div}\Omega\hat{\chi}$ with $\mathfrak{e}$, replacing $\slashed{\nabla}\hat{\otimes}\eta$ with $\slashed{\nabla}\hat{\otimes}\mathfrak{n}$, and then dropping the Ricci curvature terms. In the eventual solution we will construct (and after we have that $\mathfrak{k} = K$) this will be seen to imply that 
\begin{equation}\label{3om3o2m4}
\slashed{\epsilon}^{AC}\slashed{\nabla}_C\slashed{\nabla}^B\Omega^2\left[(\slashed{\nabla}\hat{\otimes}\left(\mathfrak{n}-\eta\right))_{AB}+\left(\mathfrak{n}\hat{\otimes}\mathfrak{n}\right)_{AB} - \left(\eta\hat{\otimes}\eta\right)_{AB} - \widehat{\rm Ric}_{AB}\right]  = 0,
\end{equation}
\begin{align*}
&\Omega\nabla_4\left((v+1)^2\mathcal{P}_{\ell \geq 1}\slashed{\rm curl}\left(\mathfrak{n}-\eta\right)\right)-(v+1)^2\mathcal{P}_{\ell \geq 1}\slashed{\rm curl}\left(\Omega{\rm Ric}_{\cdot 4}\right) = 
\\ \nonumber &\qquad  \mathcal{P}_{\ell \geq 1}\Big( \left(v+1\right)^2\left[\Omega\nabla_4,(v+1)\slashed{\epsilon}^{CA}\slashed{\nabla}_C\right]\left(\mathfrak{n}_A-\eta_A\right)  -\left(v+1\right)^2\slashed{\epsilon}^{CA}\slashed{\nabla}_C\left(\Omega\hat{\chi}_{AB}\left(\mathfrak{n}^B-\eta^B\right)\right)
\\ \nonumber &\qquad -\frac{3}{2}\left(v+1\right)^2\slashed{\rm curl}\left(\Omega{\rm tr}\chi\right)\left(\mathfrak{n}_A -\eta_A\right)-\frac{1}{2}\left(\Omega{\rm tr}\chi - 2(v+1)^{-1}\right)\left(v+1\right)^2\slashed{\rm curl}\left(\mathfrak{n}-\eta\right)\Big).
\end{align*}
\item We note, that as a consequence of~\eqref{2om3om4o29392}  and~\eqref{3om3o2m4} and elliptic estimates, we will have that 
\[(\slashed{\nabla}\hat{\otimes}\left(\mathfrak{n}-\eta\right))_{AB}+\left(\mathfrak{n}\hat{\otimes}\mathfrak{n}\right)_{AB} - \left(\eta\hat{\otimes}\eta\right)_{AB} - \widehat{\rm Ric}_{AB}  = 0.\]
	\end{enumerate}

We now explicitly define the equations of interest in this section.
\begin{definition}Let $\left(\Omega,b,\pi,\mathfrak{o},\mathfrak{w}\right)$ be given satisfying~\eqref{knefiomomo2}. Let $\mathfrak{a}$, $\mathfrak{d}$, $\mathfrak{e}$, and $\mathfrak{k}$ be suitably regular scalar functions on $(-1,0)\times \mathbb{S}^2$, let $\mathfrak{n}_A$ be a suitably regular $\mathbb{S}^2_{-1,v}$ $1$-forms for  $v \in (-1,0)$, and $\slashed{g}_{AB}$ be a suitably regular positive definite $\mathbb{S}^2_{-1,v}$ $(0,2)$-tensor for $v \in (-1,0)$. Then  we say that $\left(\slashed{g},\mathfrak{a},\mathfrak{d},\mathfrak{e},\mathfrak{k},\mathfrak{n}\right)$ satisfy the $\left(\slashed{g},\mathfrak{n}\right)$-system if
\begin{equation}\label{mdom23o3oi4}
\slashed{g}^{AB}\mathcal{L}_{\partial_v}\slashed{g}_{AB} = 2\mathfrak{a},
\end{equation}
\begin{equation}\label{ijfjiowfeiojewoi}
\left(v+1\right)^2\slashed{\rm div}\slashed{\rm div}\left( \mathcal{L}_{\partial_v}\slashed{g} - \mathfrak{a}\slashed{g}\right)= 2\mathfrak{d},\qquad \left(v+1\right)^2\slashed{\rm curl}\slashed{\rm div}\left( \mathcal{L}_{\partial_v}\slashed{g} - \mathfrak{a}\slashed{g}\right)= 2\mathfrak{e},
\end{equation}
\begin{equation}\label{k2mo1mo3mo2p3p}
\mathcal{L}_{\partial_v}\left(\Omega^{-2}\mathfrak{a}\right) + \frac{1}{2}\Omega^2\left(\Omega^{-2}\mathfrak{a}\right)^2  = -\Omega^{-2}\left|\left(\Omega\hat{\chi}\right)\right|_{\slashed{g}}^2,
\end{equation}
\begin{equation}\label{2om3omo2}
\mathcal{P}_{\ell \leq \ell_0}\mathfrak{n} = \pi,
\end{equation}
 \begin{align}\label{jfijoijoo2}
& v\Omega\nabla_4\mathfrak{e} +\mathcal{L}_b\mathfrak{e} +v\mathfrak{a}\mathfrak{e}= 
\\ \nonumber &\qquad \left[v\Omega\nabla_4+\mathcal{L}_b+v\mathfrak{a},(v+1)\slashed{\epsilon}^{BA}\slashed{\nabla}_B\right](v+1)\slashed{\nabla}^C\left(\Omega\hat{\chi}\right)_{AC}
\\ \nonumber &\qquad +(v+1)\slashed{\epsilon}^{DA}\slashed{\nabla}_D\Bigg(\left[v\Omega\nabla_4 + \mathcal{L}_b + v\mathfrak{a},(v+1)\slashed{\nabla}^B\right](\Omega\hat{\chi})_{AB} 
\\ \nonumber &\qquad + (v+1)\slashed{\nabla}^B\left[\left(\slashed{\nabla}\hat{\otimes}b\right)^C_{\ \ (A}(\Omega\hat{\chi})_{B)C} + \frac{1}{2}\slashed{\rm div}b(\Omega\hat{\chi})_{AB}\right]
\\ \nonumber  &\qquad (v+1)\slashed{\nabla}^B\left[\Omega^2\left(\left(\slashed{\nabla}\hat\otimes \mathfrak{n}\right)_{AB} + \left(\mathfrak{n}\hat\otimes \mathfrak{n}\right)_{AB}\right) - \frac{1}{4}\mathfrak{a}\left(\slashed{\nabla}\hat{\otimes}b\right)\right]\Bigg).
 \end{align}
\begin{align}\label{kjsdjfoijoi3}
& v\Omega\nabla_4\mathfrak{d}+\mathcal{L}_b\mathfrak{d} +v\mathfrak{a}\mathfrak{d} = 
\\ \nonumber &\qquad \left[v\Omega\nabla_4+\mathcal{L}_b+v\mathfrak{a},(v+1)\slashed{\nabla}^A\right](v+1)\slashed{\nabla}^B\left(\Omega\hat{\chi}\right)_{AB}
\\ \nonumber &\qquad +(v+1)\slashed{\nabla}^A\Bigg(\left[v\Omega\nabla_4 + \mathcal{L}_b + v\mathfrak{a},(v+1)\slashed{\nabla}^B\right](\Omega\hat{\chi})_{AB}
\\ \nonumber &\qquad  + (v+1)\slashed{\nabla}^B\left[\left(\slashed{\nabla}\hat{\otimes}b\right)^C_{\ \ (A}(\Omega\hat{\chi})_{B)C} + \frac{1}{2}\slashed{\rm div}b (\Omega\hat{\chi})_{AB}\right]
\\ \nonumber  &\qquad (v+1)\slashed{\nabla}^B\left[\Omega^2\left(\left(\slashed{\nabla}\hat\otimes \mathfrak{n}\right)_{AB} + \left(\mathfrak{n}\hat\otimes \mathfrak{n}\right)_{AB}\right) - \frac{1}{4}\mathfrak{a}\left(\slashed{\nabla}\hat{\otimes}b\right)\right]\Bigg),
 \end{align}
  \begin{align}\label{jivjoiewoipo3}
  &\left(v+1\right)^{-2}\left[\mathcal{L}_{\partial_v}\left(\left(v+1\right)^2\mathfrak{k} -1\right) -\mathfrak{d}\right]  =  \frac{2\left(1-\Omega^2\right)}{v+1}\left(\mathfrak{k}-\frac{1}{(v+1)^2}\right)
\\ \nonumber &\qquad \frac{2\left(1-\Omega^2\right)}{(v+1)^3} - \frac{\Omega^2}{(v+1)^2}\left(\Omega^{-2}\mathfrak{a} - \frac{2}{v+1}\right)  - \left(\slashed{\Delta}\log\Omega\right)\Omega^2\left(\Omega^{-2}\mathfrak{a}\right)- \Omega^2\left(\Omega^{-2}\mathfrak{a} - \frac{2}{v+1}\right)\left(\mathfrak{k} - \frac{1}{(v+1)^2}\right)
\\ \nonumber&\qquad - \frac{1}{2}\Omega^2\slashed{\Delta}\left(\Omega^{-2}\mathfrak{a} - \frac{2}{v+1}\right) -4\Omega^2\left(\slashed{\nabla}\log\Omega\right)\slashed{\nabla}\left(\Omega^{-2}\mathfrak{a} - \frac{2}{v+1}\right),
\end{align}

 \[\lim_{v\to 0}\mathcal{P}_{\ell > \ell_0}\mathfrak{k} = \mathfrak{o},\]
 \begin{align}\label{knvnieo2o23}
&\mathcal{P}_{\ell > \ell_0}\Bigg(\left(-1+\slashed{\rm div}b + \frac{2v\Omega^2}{v+1} -4\Omega\underline{\omega}\right) \left(\Omega^{-2}\mathfrak{a}-\frac{2}{v+1}\right) + \mathcal{L}_b\left(\Omega^{-2}\mathfrak{a}-\frac{2}{v+1}\right)
\\ \nonumber &\qquad +\frac{v}{2}\Omega^2\left(\Omega^{-2}\mathfrak{a}-2(v+1)^{-1}\right)^2
 -v\Omega^{-2}\left|\Omega\hat{\chi}\right|^2 
 - 2\left|\mathfrak{n}\right|^2
 +\frac{2}{v+1}\left(\slashed{\rm div}b - 4\Omega\underline{\omega}\right) - \frac{2v\left(1-\Omega^2\right)}{(v+1)^2}\Bigg)=  \\ \nonumber &\qquad
\mathcal{P}_{\ell > \ell_0}\Bigg(  2\slashed{\rm div}\mathfrak{n} -2\left(\mathfrak{k}-\frac{1}{(v+1)^2}\right)\Bigg),
  \end{align}
\begin{align}\label{akmso082hjoj2om3}
&\Omega\nabla_4\left((v+1)^2\mathcal{P}_{\ell > \ell_0}\slashed{\rm curl}\mathfrak{n}\right)+\mathcal{P}_{\ell > \ell_0}\frac{2\Omega^2}{v+1}\left(v+1\right)^2\mathcal{P}_{\ell > \ell_0}\slashed{\rm curl}\mathfrak{n} - \mathcal{P}_{\ell >\ell_0 }\mathfrak{e} =
\\ \nonumber &\qquad \qquad \mathcal{P}_{\ell >\ell_0}\Big( \left(v+1\right)^2\left[\Omega\nabla_4,(v+1)\slashed{\epsilon}^{CA}\slashed{\nabla}_C\right]\mathfrak{n}_A  -\left(v+1\right)^2\slashed{\epsilon}^{CA}\slashed{\nabla}_C\left(\Omega\hat{\chi}_{AB}\mathfrak{n}^B\right)
\\ \nonumber &\qquad +\frac{1}{2}\left(v+1\right)^2{}^*\slashed{\nabla}\mathfrak{a}\left( -3\mathfrak{n}_A + 2\slashed{\nabla}_A\log\Omega\right) 
\\ \nonumber &\qquad- \mathfrak{a}\left(v+1\right)^2\mathcal{P}_{\ell \leq \ell_0}\slashed{\rm curl}\mathfrak{n} -\left(\Omega^{-2}\mathfrak{a}-2(v+1)^{-1}\right)\left(v+1\right)^2\mathcal{P}_{\ell >\ell_0}\slashed{\rm curl}\mathfrak{n}-\frac{1}{2}\left(\mathfrak{a} - 2(v+1)^{-1}\right)\left(v+1\right)^2\slashed{\rm curl}\mathfrak{n}\Big),
\end{align}
\begin{equation}\label{2m3om2}
\lim_{v\to 0}\mathcal{P}_{\ell > \ell_0}\slashed{\rm curl}\mathfrak{n} = \mathfrak{w},
\end{equation}
where the quantity $\Omega\hat{\chi}$ is computed from the metric $\slashed{g}$.
\end{definition}

We are interested in eventually finding a solution to the $\left(\slashed{g},\mathfrak{n}\right)$-system satisfying various estimates and additional properties. Our main strategy will be to exploit the fact that, upon linearization, the equations~\eqref{jfijoijoo2}-\eqref{akmso082hjoj2om3} lead to model second order equations, degenerate transport equations, or elliptic equations  for certain of the unknowns. However, there are two important technical difficulties which will make it convenient to consider, as a preliminary step, a modification of the $\left(\slashed{g},\mathfrak{n}\right)$ system (which we will describe below):
\begin{enumerate}
	\item The first problem is that in order to solve for all of the unknowns in the $\left(\slashed{g},\mathfrak{n}\right)$-system we must invert certain elliptic equations on $\mathbb{S}^2$ with non-trivial co-kernel. In particular, one cannot expect to find a solution to~\eqref{ijfjiowfeiojewoi} for arbitrary $\mathfrak{d}$ and $\mathfrak{e}$. This makes it difficult to run directly an iteration argument to solve the $\left(\slashed{g},\mathfrak{n}\right)$-system.
	\item  The second problem concerns certain nonlinear terms on the right hand sides of~\eqref{jfijoijoo2} and~\eqref{kjsdjfoijoi3}. Specifically, there are terms proportional to a contraction of two angular derivative of $\slashed{\rm div}b$ or $\slashed{\nabla}\hat{\otimes}b$ with $\Omega{\hat{\chi}}$. Our scheme for estimating $\mathfrak{d}$ and $\mathfrak{e}$ will require that after applying $\mathcal{L}_b$, the resulting term may be controlled in the $\mathscr{Q}\left(N_1-4,0,-\kappa\right)$ norm. However, if all of the derivatives in this norm fall on the shift $b$, then because $\Omega{\hat{\chi}}$ is not pointwise bounded as $v\to 0$, this nonlinear term will not be controllable using the specific assumptions on $b$ we make in this section.\footnote{\label{2oopjo0199919198}Later when we construct the shift, we will eventually show that $\slashed{\nabla}\hat{\otimes}b$ and $\slashed{\rm div}b$ have improved estimates compared to a generic derivative of $b$ (see Proposition~\ref{3ij2oio23}). With this improved estimate, these nonlinear terms becomes easily controllable. However, because this improvement can only be understood once $\slashed{g}$ is fixed, it will be useful to close some preliminary estimates on $\slashed{g}$ which do not require this improved behavior.}
\end{enumerate}

Our modified $\left(\slashed{g},\mathfrak{n}\right)$-system thus consists of two main changes. First of all, we will add in certain projections which directly enforce the invertibility of the desired elliptic operators. Secondly, we will replace $\slashed{\rm div}b$ and $\slashed{\nabla}\hat{\otimes}b$ in the aforementioned nonlinear terms with the artificial variables $\mathfrak{q}$ and $\mathfrak{r}$.\footnote{Later in our construction (see Proposition~\ref{3ij2oio23}) we will show that one may in fact take $\mathfrak{q} = \slashed{\rm div}b$ and $\mathfrak{r} = \slashed{\nabla}\hat{\otimes}b$.}

Before we present this modified $\left(\slashed{g},\mathfrak{n}\right)$-system, we present a useful lemma which is proved in the Appendix~\ref{ioj2ojo2}.
\begin{lemma}\label{ijmom2om392}Let $\slashed{h}(\tau)$ be a $1$-parameter family of metrics on $\mathbb{S}^2$ for $\tau \in [\tau_0,\tau_1]$ which satisfy 
\begin{equation}\label{km2omo}
\sup_{\tau \in [\tau_0,\tau_1]}\sum_{\left|\alpha\right| \leq 2}\left\vert\left\vert \left(\slashed{h} - \mathring{\slashed{g}}\right)^{(\alpha)}\right\vert\right\vert_{L^{\infty}(\mathbb{S}^2)}^2  \ll 1.
\end{equation}
Then, for any $1$-form $\vartheta$ and non-negative integer $j$, we have the following for $\tau \in (\tau_0,\tau_1)$:
\begin{equation}\label{i3omromo3}
\left\vert\left\vert \Pi_{{\rm ker}\left({}^*\slashed{\mathcal{D}}_2\right)}\vartheta\right\vert\right\vert_{\mathring{H}^j}^2 \lesssim \left\vert\left\vert \vartheta\right\vert\right\vert_{L^2}^2\left(1+\left\vert\left\vert \slashed{h} - \mathring{\slashed{g}}\right\vert\right\vert_{\mathring{H}^{{\rm max}(j,5)}}^2\right)
\end{equation}
\begin{equation}\label{kjfewijoewfiowefio}
\left\vert\left\vert \left[\mathcal{L}_{\partial_{\tau}},\Pi_{{\rm ker}\left({}^*\slashed{\mathcal{D}}_2\right)}\right]\vartheta\right\vert\right\vert_{\mathring{H}^j}^2 \lesssim \left\vert\left\vert \mathcal{L}_{\partial_{\tau}}\left(\slashed{h} - \mathring{\slashed{g}}\right)\right\vert\right\vert_{\mathring{H}^{\lceil \frac{j}{2}\rceil +2}}^2\left\vert\left\vert \vartheta\right\vert\right\vert_{\mathring{H}^j}^2+\left\vert\left\vert \mathcal{L}_{\partial_{\tau}}\left(\slashed{h} - \mathring{\slashed{g}}\right)\right\vert\right\vert_{\mathring{H}^j}^2\left\vert\left\vert \vartheta\right\vert\right\vert_{\mathring{H}^{\lceil \frac{j}{2}\rceil +2}}^2,
\end{equation}

 where $\Pi_{{\rm ker}\left({}^*\slashed{\mathcal{D}}_2\right)}$ denotes the projection onto the kernel of the operator ${}^*\slashed{\mathcal{D}}_2$ associated to $\slashed{h}$.
 
Let $\tilde{\slashed{h}}$ also be a metric on $\mathbb{S}^2$ which satisfies~\eqref{km2omo} with $\slashed{h}$ replaced by $\tilde{\slashed{h}}$.  Let $\tilde{\slashed{\mathcal{D}}}_2$ correspond to $\tilde{\slashed{h}}$. Then, for any $1$-form $\vartheta$ and sufficiently large non-negative integer $j$ we have the following
\begin{equation}\label{mfo3mo2p1}
\left\vert\left\vert \left(\Pi_{{\rm ker}\left({}^*\slashed{\mathcal{D}}_2\right)}-\Pi_{{\rm ker}\left({}^*\tilde{\slashed{\mathcal{D}}}_2\right)}\right)\vartheta\right\vert\right\vert_{\mathring{H}^j}^2 \lesssim \left\vert\left\vert \slashed{h}-\tilde{\slashed{h}} \right\vert\right\vert_{\mathring{H}^j}^2\left\vert\left\vert \vartheta\right\vert\right\vert_{\mathring{H}^j}^2.
\end{equation}

We also have the analogous facts for the operator ${}^*\slashed{D}_1$ and functions $(f,w)$:
\begin{equation}\label{khuhiu3iu2}
\left\vert\left\vert \Pi_{{\rm ker}\left({}^*\slashed{\mathcal{D}}_1\right)}(f,w)\right\vert\right\vert_{\mathring{H}^j}^2 \lesssim \left\vert\left\vert (f,w)\right\vert\right\vert_{L^2}^2\left(1+\left\vert\left\vert \slashed{h} - \mathring{\slashed{g}}\right\vert\right\vert_{L^{\infty}}^2\right)
\end{equation}
\begin{equation}\label{ijnfomo2p3p2}
\left\vert\left\vert \left[\mathcal{L}_{\partial_{\tau}},\Pi_{{\rm ker}\left({}^*\slashed{\mathcal{D}}_1\right)}\right](f,w)\right\vert\right\vert_{\mathring{H}^j}^2 \lesssim  \left\vert\left\vert (f,w)\right\vert\right\vert_{L^2}^2\left\vert\left\vert \mathcal{L}_{\partial_{\tau}}\slashed{h}\right\vert\right\vert_{L^{\infty}}^2\left(1+\left\vert\left\vert \slashed{h} - \mathring{\slashed{g}}\right\vert\right\vert_{L^{\infty}}^2\right)\end{equation}
\begin{equation}\label{ljfjii2p3o}
\left\vert\left\vert \left(\Pi_{{\rm ker}\left({}^*\slashed{\mathcal{D}}_1\right)}-\Pi_{{\rm ker}\left({}^*\tilde{\slashed{\mathcal{D}}}_1\right)}\right)(f,w)\right\vert\right\vert_{\mathring{H}^j}^2 \lesssim \left\vert\left\vert (f,w)\right\vert\right\vert_{L^2}^2\left\vert\left\vert \slashed{h}-\tilde{\slashed{h}}\right\vert\right\vert_{L^{\infty}}^2.
\end{equation}

\end{lemma}

The following lemma, which follows immediately from elliptic theory, will also be useful.
\begin{lemma}\label{2im3om2o}Suppose that $\slashed{g}$ is a positive definite $\mathbb{S}^2_{-1,v}$ symmetric $(0,2)$-tensor for $v \in (-1,0)$ which satisfies
\begin{equation}\label{iiojioj392}
\left\vert\left\vert \slashed{g}\right\vert\right\vert_{\mathscr{A}^{-}_2\left(b,\kappa\right)} + \left\vert\left\vert \slashed{g}\right\vert\right\vert_{\mathscr{B}^{-}_2\left(\kappa\right)} \lesssim  \epsilon.
\end{equation}

Then, for any $0 \leq J \leq N_1-2$ and $|q| \lesssim 1$, the operator $\left((v+1)^2\slashed{\Delta}\right)^{-1}$ extends from smooth $\mathbb{S}^2_{-1,v}$ $1$-forms to a bounded map:
\begin{equation}\label{2om3omo}
\left((v+1)^2\slashed{\Delta}\right)^{-1}: \mathscr{Q}_{-1/2}^0\left(J,0,q\right)[1] \to \mathscr{Q}_{-1/2}^0\left(J+2,0,q\right)[1].
\end{equation}
Furthermore, the commutator $\left[\left((v+1)^2\slashed{\Delta}\right)^{-1},\mathcal{L}_{\partial_v}\right]$ extends to a bounded map
\begin{align}\label{2km3om4o}
\left[\left((v+1)^2\slashed{\Delta}\right)^{-1},\mathcal{L}_{\partial_v}\right]: &\mathscr{Q}_{-1/2}^0\left(J,0,q-N_2\check{p}\right)[1] \cap \mathscr{S}_{-1/2}^0\left(\lfloor \frac{2}{3}J\rfloor,0,0\right)[1] 
\\ \nonumber &\qquad \to \mathscr{Q}_{-1/2}^0\left(J+2,0,q\right)[1].
\end{align}
The constants in these estimates are independent of the implied constant in~\eqref{iiojioj392}. (Recall that these function spaces are defined in Section~\ref{notationfornorms}.)
\end{lemma}

Now we are ready to define our modified system.
\begin{definition}\label{2n3inii1o2}We say that $\left(\slashed{g},\mathfrak{a},\mathfrak{d},\mathfrak{e},\mathfrak{k},\mathfrak{n},\Phi\right)$ satisfy the modified $\left(\slashed{g},\mathfrak{n}\right)$-system if $\mathfrak{a}$, $\mathfrak{d}$, $\mathfrak{e}$, and $\mathfrak{k}$ are scalar functions on $(-1,0)\times \mathbb{S}^2$, $\mathfrak{n}_A$ is an $\mathbb{S}^2_{-1,v}$ $1$-form for $v \in (-1,0)$, $\slashed{g}$ is an $\mathbb{S}^2_{-1,v}$ positive definite symmetric $(0,2)$-tensor for $v \in (-1,0)$, $\Phi_{AB}$ is an $\mathbb{S}^2_{-1,v}$ $\slashed{g}$-trace free symmetric $(0,2)$-tensor for $v \in (-1,0)$, equations~\eqref{k2mo1mo3mo2p3p}-\eqref{2m3om2} hold with the exception that we replace every $\Omega\hat{\chi}$ on the right hand sides of these equations with $\Phi$, we furthermore define $\Phi$ in terms of $\mathfrak{d}$ and $\mathfrak{e}$ by  
\begin{equation}\label{2k3mr4i2i4}
\Phi \doteq \slashed{\mathcal{D}}_2^{-1}\Pi_{{\rm ker}\left({}^*\slashed{\mathcal{D}}_2\right)^{\perp}}\slashed{\mathcal{D}}_1^{-1}\Pi_{{\rm ker}\left({}^*\slashed{\mathcal{D}}_1\right)^{\perp}}\left(-\left(v+1\right)^{-2}\mathfrak{d},\left(v+1\right)^{-2}\mathfrak{e}\right),
\end{equation}
we require that (as a replacement for~\eqref{mdom23o3oi4} and~\eqref{ijfjiowfeiojewoi})
\begin{align}\label{2k3jk2j}
\mathcal{L}_{\partial_v}\slashed{g}_{AB} = \mathfrak{a}\slashed{g}_{AB} + 2\Phi_{AB},
\end{align}
and finally, on the right hand sides of~\eqref{jfijoijoo2} and~\eqref{kjsdjfoijoi3} we replace every $\slashed{\rm div}b$ and $\slashed{\nabla}\hat{\otimes}b$ which is contracted with $\Omega\hat{\chi}$ with $\mathfrak{q}$ and $\mathfrak{r}$ respectively. We further include in this prescription the replacing of all occurrences of the $\slashed{\rm div}b$ and $\slashed{\nabla}\hat{\otimes}b$ with $\mathfrak{q}$ and $\mathfrak{r}$ that one sees when the commutator $\left[\mathcal{L}_b,\slashed{\rm div}\right]$ and $\left[\mathcal{L}_b,\slashed{\rm curl}\right]$ acts on $\Omega\hat{\chi}$. 

 We will say that $\left(\slashed{g},\mathfrak{a},\mathfrak{d},\mathfrak{e},\mathfrak{k},\mathfrak{n},\Phi\right)$ satisfies the modified $\left(\slashed{g},\mathfrak{n}\right)$ ``sub-system'' if we no longer require the equation~\eqref{2k3jk2j}  and instead just let $\slashed{g}_{AB}$ be any given family of Riemannian metrics along $\mathbb{S}^2_{-1,v}$ for $v \in (-1,0)$ such that 
 \begin{equation}\label{2om2omo492849294124}
\left\vert\left\vert \slashed{g}\right\vert\right\vert_{\mathscr{A}^{-}_2\left(b,\kappa\right)} + \left\vert\left\vert \slashed{g}\right\vert\right\vert_{\mathscr{B}^{-}_2\left(\kappa\right)} \lesssim  \epsilon.
\end{equation}
  
When we wish to refer to one of the equations~\eqref{jfijoijoo2}-\eqref{2m3om2} with $\Omega\hat{\chi}$ replaced by $\Phi$, and the various $\slashed{\rm div}b$'s and $\slashed{\nabla}\hat{\otimes}b$ replaced with $\mathfrak{q}$'s and $\mathfrak{r}$'s, we will refer to the equation as being ``modified.'' 
\end{definition}
\begin{remark}We observe that the $\mathfrak{q}$ and $\mathfrak{r}$ are only present in nonlinear terms where they are contracted against $\Phi$. In particular,  since we assume that $\mathfrak{q}$ and $\mathfrak{r}$ satisfy~\eqref{o3momo2o4} we may expect our estimates for a solution to the modified $\left(\slashed{g},\mathfrak{n}\right)$-system to not depend on $\mathfrak{q}$ and $\mathfrak{r}$. 
\end{remark}

We start by deriving some additional formulas which must hold for any solution to the modified $\left(\slashed{g},\mathfrak{n}\right)$ sub-system. This next lemma derives a useful equation for $\mathfrak{k}$.
\begin{lemma}\label{2kno1o2}Suppose that $\left(\slashed{g},\mathfrak{a},\mathfrak{d},\mathfrak{e},\mathfrak{k},\mathfrak{n},\Phi\right)$ solves the modified $\left(\slashed{g},\mathfrak{n}\right)$ sub-system. We may derive an equation for $\mathcal{P}_{\ell > \ell_0}\mathfrak{k}$ whose linearization is a model second order equation of type $III$ as follows:
\begin{enumerate}
	\item We apply $\mathcal{P}_{\ell > \ell_0}$ to the equation~\eqref{kjsdjfoijoi3}. 
	\item We use the following general formula\footnote{\label{genform}By general formula, we mean that this formula holds for any self-similar Lorentzian metric $g$.} to replace the term $\left(v+1\right)^2\Omega^2\slashed{\nabla}^A\slashed{\nabla}^B\left(\slashed{\nabla}\hat{\otimes}n\right)_{AB} $ on the right hand side of the modified~\eqref{kjsdjfoijoi3}:
	\begin{align}\label{om2omo2320}
	&\left(v+1\right)^2\Omega^2\slashed{\nabla}^A\slashed{\nabla}^B\left(\slashed{\nabla}\hat{\otimes}n\right)_{AB} = 
	\\ \nonumber &\qquad \Omega^2\left(v+1\right)^2\left(\slashed{\Delta}+2(v+1)^{-2}\right)\left(\mathcal{P}_{\ell \leq \ell_0}+\mathcal{P}_{\ell > \ell_0}\right)\slashed{\rm div}\mathfrak{n} + 2\Omega^2\left(v+1\right)^2\slashed{\nabla}^A\left(\left(K-(v+1)^{-2}\right)\mathfrak{n}_A\right).
	\end{align}
	\item Then we may further replace each instance of $\mathcal{P}_{\ell > \ell_0}\mathfrak{d}$  with $\mathcal{P}_{\ell > \ell_0}$ applied to~\eqref{jivjoiewoipo3} (and use the modified~\eqref{k2mo1mo3mo2p3p} to simplify). 
	\item We modify the term $ \Omega^2\left(v+1\right)^2\left(\slashed{\Delta}+2(v+1)^{-2}\right)\mathcal{P}_{\ell > \ell_0}\slashed{\rm div}\mathfrak{n}$  by using the modified~\eqref{knvnieo2o23} to substitute $\mathcal{P}_{\ell > \ell_0}\slashed{\rm div}\mathfrak{n}$.
\end{enumerate}
The resulting equation is of the following form:
\begin{align}\label{mo2omo1013}
&\mathcal{P}_{\ell > \ell_0}\Bigg((-v)\mathcal{L}_{\partial_v}^2\left(\left(v+1\right)^2\left(\mathfrak{k}-\left(v+1\right)^{-2}\right)\right)-\mathcal{L}_b\mathcal{L}_{\partial_v}\left((v+1)^2\left(\mathfrak{k}-\left(v+1\right)^{-2}\right)\right) 
\\ \nonumber &\qquad + \frac{2(-v)\Omega^2}{v+1}\mathcal{L}_{\partial_v}\left((v+1)^2\left(\mathfrak{k}-\left(v+1\right)^{-2}\right)\right)  + \Omega^2\left(\slashed{\Delta} +2(v+1)^{-2}\right)\mathcal{P}_{\ell \geq 1}\left((v+1)^2\left(\mathfrak{k}-\left(v+1\right)^{-2}\right)\right)\Bigg)
\\ \nonumber &\qquad  = \mathcal{P}_{\ell > \ell_0}\left(-\mathcal{E}_1+\mathcal{E}_2 + \mathcal{E}_3\right)
\end{align}
where $\mathcal{E}_1$ denotes the difference of the right hand of the modified~\eqref{kjsdjfoijoi3} and 
\[\Omega^2\left(v+1\right)^2\left(\slashed{\Delta}+2(v+1)^{-2}\right)\mathcal{P}_{\ell > \ell_0}\slashed{\rm div}\mathfrak{n},\]
$\mathcal{E}_2$ is $ \Omega^2\left(v+1\right)^2\left(\slashed{\Delta}+2(v+1)^{-2}\right)$ applied to the left hand side of the modified version of~\eqref{knvnieo2o23}, and $\mathcal{E}_3$  
is the sum of $\left((-v)\mathcal{L}_{\partial_v}-\mathcal{L}_b + (-v)\mathfrak{a}\right)\left[\left(v+1\right)^2\cdot \right]$ applied to the right hand side of~\eqref{jivjoiewoipo3} and 
\[(-v)\Omega^2\left(\Omega^{-2}\mathfrak{a} - \frac{2}{v+1}\right)\mathcal{L}_{\partial_v}\left((v+1)^2\left(\mathfrak{k}-\frac{1}{(v+1)^2}\right)\right).\]
\end{lemma}
\begin{proof}This is a straightforward if tedious calculation.
\end{proof}

In the next lemma we derive a useful equation for $ \left(v+1\right)^2\mathcal{P}_{\ell > \ell_0}\slashed{\rm curl}\mathfrak{n}$.
\begin{lemma}\label{k2ni2ni3}Suppose that $\left(\slashed{g},\mathfrak{a},\mathfrak{d},\mathfrak{e},\mathfrak{k},\mathfrak{n},\Phi\right)$ solves the modified $\left(\slashed{g},\mathfrak{n}\right)$ sub-system. We may derive an equation for $\mathcal{P}_{\ell > \ell_0}\slashed{\rm curl}\mathfrak{n}$, whose linearization is a model second order equation of type $III$, as follows:
\begin{enumerate}
\item We use the following general formula\footnote{See footnote~\ref{genform}.} to replace the term $\left(v+1\right)^2\Omega^2\slashed{\epsilon}^{CA}\slashed{\nabla}_C\slashed{\nabla}^B\left(\slashed{\nabla}\hat{\otimes}n\right)_{AB} $ on the right hand side of the modified~\eqref{jfijoijoo2}:
	\begin{align}\label{jji3io32oi}
	&\left(v+1\right)^2\Omega^2\slashed{\epsilon}^{CA}\slashed{\nabla}_C\slashed{\nabla}^B\left(\slashed{\nabla}\hat{\otimes}n\right)_{AB} = 
	\\ \nonumber &\qquad \Omega^2\left(v+1\right)^2\left(\slashed{\Delta}+2(v+1)^{-2}\right)\left(\mathcal{P}_{\ell \leq \ell_0} + \mathcal{P}_{\ell > \ell_0}\right)\slashed{\rm curl}\mathfrak{n} + 2\Omega^2\left(v+1\right)^2\slashed{\nabla}^A\left(\left(K-(v+1)^{-2}\right)\mathfrak{n}_A\right).
	\end{align}
\item We  apply $v\mathcal{L}_{\partial_v} +\mathcal{L}_b + v\mathfrak{a}$ to the modified~\eqref{akmso082hjoj2om3}, and use the modified~\eqref{jfijoijoo2} (with the substitution described above) to simplify the term involving $\mathfrak{e}$.
\end{enumerate}
We eventually obtain an equation of the form
\begin{align}\label{oo2ok3ok2}
&(-v)\mathcal{L}_{\partial_v}^2\nu + \mathcal{P}_{\ell > \ell_0}\mathcal{L}_b\mathcal{L}_{\partial_v}\nu +\mathcal{P}_{\ell > \ell_0}\frac{4(-v)\Omega^2}{v+1}\mathcal{L}_{\partial_v}\nu 
\\ \nonumber &\qquad +\mathcal{P}_{\ell > \ell_0}\left[\Omega^2\left(\slashed{\Delta}+2(v+1)^{-2}\right)+(-v)\left(8\Omega^4-2\Omega^2\right)(v+1)^{-2}\right]\nu = 
\mathcal{P}_{\ell > \ell_0}\left(\mathcal{E}_1 + \mathcal{E}_2 + \mathcal{E}_3\right),
\end{align}
where $\nu \doteq \left(v+1\right)^2\mathcal{P}_{\ell > \ell_0}\slashed{\rm curl}\mathfrak{n}$, $\mathcal{E}_1$  is $\mathcal{P}_{\ell > \ell_0}\left(v\mathcal{L}_{\partial_v} +\mathcal{L}_b + v\mathfrak{a}\right)$ applied to the right hand side of~\eqref{akmso082hjoj2om3}, $\mathcal{E}_2$ is the right hand side of~\eqref{jfijoijoo2} minus $\Omega^2\left(v+1\right)^2\left(\slashed{\Delta}+2(v+1)^{-2}\right)\mathcal{P}_{\ell > \ell_0}\slashed{\rm curl}\mathfrak{n}$, and $\mathcal{E}_3$ is a linear combination of terms of the form
\begin{align*}
 \left(v\mathfrak{a}- 2(-v)\Omega^2(v+1)^{-1}\right)\nu,\ \left(v\mathcal{L}_{\partial_v} +\mathcal{L}_b + v\mathfrak{a}\right)\mathcal{P}_{\ell \leq \ell_0}\mathfrak{e},\ (v+1)^{-1}\left(v\mathcal{L}_{\partial_v}\right)\left(\Omega^2\right) \nu,\ (v+1)^{-1}\mathcal{L}_b\left(\Omega^2\nu\right).
\end{align*}

Finally, we consider the case when we do \underline{not} yet know that $\left(\slashed{g},\mathfrak{a},\mathfrak{d},\mathfrak{e},\mathfrak{k},\mathfrak{n},\Phi\right)$ solves the modified $\left(\slashed{g},\mathfrak{n}\right)$ sub-system; but we do know that $\nu$ solves~\eqref{oo2ok3ok2} and that the modified~\eqref{jfijoijoo2} holds. Then, if we set $\mathcal{F}$ to be the difference of the left hand side and right hand side of the modified~\eqref{akmso082hjoj2om3}, we will have that $\mathcal{F}$ satisfies
\[\left(v\mathcal{L}_{\partial_v} +\mathcal{L}_b + v\Omega{\rm tr}\chi\right)  \mathcal{F} = 0.\]
In particular, if $\lim_{v\to -1}\left(v+1\right)^2\mathcal{F} = 0$, then $\mathcal{F}$ vanishes identically. 
\end{lemma}
\begin{proof}This is a straightforward if tedious calculation.
\end{proof}
\begin{remark}Since the equations~\eqref{mo2omo1013} and~\eqref{oo2ok3ok2} concern quantities which have been projected to be supported on $\ell > \ell_0$, we will treat these as model second order equations of type $III$. One consequence is that the detailed structure of the linear lower-order terms is not relevant to our later analysis. 
\end{remark}

In the next proposition we construct suitable solutions to the the modified $\left(\slashed{g},\mathfrak{n}\right)$ sub-system.
\begin{proposition}\label{3imomo1}Let $\slashed{g}_{AB}$ be a given family of Riemannian metrics along $\mathbb{S}^2_{-1,v}$ for $v \in (-1,0)$ such that~\eqref{2om2omo492849294124} holds. Then there exists a solution to the modified $\left(\slashed{g},\mathfrak{n}\right)$ sub-system such that moreoever
\begin{align}\label{2omomo1ji28}
& \left\vert\left\vert \mathfrak{A}\right\vert\right\vert_{\mathscr{Q}\left(N_1,-1/2+\check{\delta},50\check{p}\right)}+ \left\vert\left\vert \mathfrak{A}\right\vert\right\vert_{\mathscr{Q}\left(N_1-1,-1/2+\check{\delta},-1/2+\check{\delta}\right)} 
+\sum_{j=0}^1 \left\vert\left\vert \mathcal{L}^{1+j}_{\partial_v}\mathfrak{A}\right\vert\right\vert_{\mathscr{Q}\left(N_1-1-j,1/2+\check{\delta}+j,50\check{p}+1/2 +j \right)}
\\ \nonumber &\qquad +\sum_{j=0}^1 \left\vert\left\vert \left(1,\mathcal{L}_b\right)\mathcal{L}^{1+j}_{\partial_v}\mathfrak{A}\right\vert\right\vert_{\mathscr{Q}_{-1/2}^0\left(N_1-2-j,0,50\check{p}+j \right)}
 + \left\vert\left\vert \left(1,\mathcal{L}_b\right)\mathfrak{A}\right\vert\right\vert_{\mathscr{S}_{-1/2}^0\left(N_1-2,0,0\right)}
\\ \nonumber &\qquad +\sum_{j=0}^1\left\vert\left\vert \left(v\mathcal{L}_{\partial_v}\right)^j\mathcal{L}_{\partial_v}\mathfrak{A}\right\vert\right\vert_{\check{\mathscr{S}}_{-1/2}^0\left(N_2-1-j,0,10\check{p}+2\tilde{\kappa},0\right)}+\sum_{j=0}^2\left\vert\left\vert \mathcal{L}_{\partial_v}^j\mathfrak{A}\right\vert\right\vert_{\mathscr{S}_{-1}^{-1/2}\left(N_1-1-j,\check{\delta}+j,0\right)},
\\ \nonumber &\qquad +\left\vert\left\vert \left(v+1\right)^2\left(\mathfrak{k}-(v+1)^{-2}\right) \right\vert\right\vert_{L,III\left(1/2-\check{\delta},\check{p},0\right),N_1-2}
 + \left\vert\left\vert \left(v+1\right)\mathfrak{n}\right\vert\right\vert_{L,III\left(1/2-\check{\delta},100 N_1 \check{p},0\right),N_1-1}
\\ \nonumber &\qquad +\sum_{j=0}^1\left\vert\left\vert \mathcal{L}_{\partial_v}^j\left(\mathfrak{d},\mathfrak{e}\right)\right\vert\right\vert_{\mathscr{Q}\left(N_1-3-j,-1/2+\check{\delta}+j,1/2+j\right)} + \left\vert\left\vert\left(1, v\mathcal{L}_{\partial_v},\mathcal{L}_b\right)\left(\mathfrak{d},\mathfrak{e}\right)\right\vert\right\vert_{\mathscr{Q}_{-1/2}^0\left(N_1-4,0,\kappa+j\right)}
\\ \nonumber &\qquad +\sum_{j=0}^1\left\vert\left\vert \mathcal{L}_{\partial_v}^j\left(\mathfrak{d},\mathfrak{e}\right)\right\vert\right\vert_{\check{\mathscr{S}}_{-1/2}^0\left(N_2-3-j,0,\check{p} +2\kappa,0\right)} + \sum_{j=0}^1\left\vert\left\vert \mathcal{L}_{\partial_v}^j\left(\mathfrak{d},\mathfrak{e}\right)\right\vert\right\vert_{\mathscr{S}_{-1}^{-1/2}\left(N_1-4-j,\check{\delta}+j,0\right)}
\\ \nonumber &\qquad +\sum_{j=0}^1\left\vert\left\vert \mathcal{L}_{\partial_v}^j\Phi\right\vert\right\vert_{\mathscr{Q}\left(N_1-1-j,-1/2+\check{\delta}+j,1/2+j\right)} + \left\vert\left\vert \left(1,v\mathcal{L}_{\partial_v},\mathcal{L}_b\right)\Phi\right\vert\right\vert_{\mathscr{Q}_{-1/2}^0\left(N_1-2,0,\kappa\right)}
\\ \nonumber &\qquad +\sum_{j=0}^1\left\vert\left\vert \mathcal{L}_{\partial_v}^j\Phi\right\vert\right\vert_{\check{\mathscr{S}}_{-1/2}^0\left(N_2-1-j,0,3\check{p} +2\kappa,0\right)}+\sum_{j=0}^1\left\vert\left\vert \mathcal{L}_{\partial_v}^j\Phi\right\vert\right\vert_{\mathscr{S}_{-1}^{-1/2}\left(N_1-2-j,\check{\delta}+j,0\right)}
\\ \nonumber & \lesssim \mathcal{D},
\end{align}
where $\mathfrak{A} \doteq \Omega^{-2}\mathfrak{a} - 2(v+1)^{-1}$.
\end{proposition}
\begin{proof}
We will inductively define a sequence $\left\{\left(\mathfrak{a}^{(i)},\mathfrak{d}^{(i)},\mathfrak{e}^{(i)},\mathfrak{k}^{(i)},\nu^{(i)},\mathfrak{n}^{(i)},\Phi^{(i)}\right)\right\}_{i=0}^{\infty}$ where all quantities are as in the definition of the modified $\left(\slashed{g},\mathfrak{n}\right)$ sub-system. For $i = 0$ we set $\mathfrak{a}^{(0)} = 2\Omega^2(v+1)^{-1}$ and all other quantities to $0$.

For $i \geq 1$, we define the sequence as follows (the order reflects the order in which we solve for the unknowns):
\begin{enumerate}
		\item\label{2kn339u2} We define $\mathfrak{a}^{(i)}$ by setting $\mathfrak{A}^{(i)} \doteq \Omega^{-2}\mathfrak{a}^{(i)}-2(v+1)^{-1}$ and solving the transport equation
	\begin{align}\label{jdiji2mo1}
&\mathcal{L}_{\partial_v}\mathfrak{A}^{(i)} + 2(v+1)^{-1}\Omega^2\mathfrak{A}^{(i)} = 
 -\Omega^{-2}\left|\Phi^{(i-1)}\right|^2+2\left(1-\Omega^2\right)(v+1)^{-2} - \frac{1}{2}\Omega^2\left(\mathfrak{A}^{(i)}\right)^2,
\end{align}
with the boundary condition that
\[\mathfrak{A}^{(i)} = O(1)\text{ as }v\to -1.\]
\item We solve for $\mathcal{P}_{\ell > \ell_0}\mathfrak{k}^{(i)}$ by using the equation~\eqref{mo2omo1013} with $\mathfrak{k}$ replaced by $\mathfrak{k}^{(i)}$ and applying $\mathcal{P}_{\ell > \ell_0}$:
\begin{align}\label{k2m2omo3o2}
&(-v)\mathcal{L}_{\partial_v}^2\mathcal{P}_{\ell > \ell_0}\left(\left(v+1\right)^2\left(\mathfrak{k}^{(i)}-\left(v+1\right)^{-2}\right)\right)-\mathcal{P}_{\ell > \ell_0}\mathcal{L}_b\mathcal{L}_{\partial_v}\mathcal{P}_{\ell > \ell_0}\left((v+1)^2\left(\mathfrak{k}^{(i)}-\left(v+1\right)^{-2}\right)\right) 
\\ \nonumber &\qquad + \frac{2(-v)\mathcal{P}_{\ell > \ell_0}\Omega^2}{v+1}\mathcal{L}_{\partial_v}\mathcal{P}_{\ell > \ell_0}\left((v+1)^2\left(\mathfrak{k}^{(i)}-\left(v+1\right)^{-2}\right)\right) 
\\ \nonumber &\qquad  + \mathcal{P}_{\ell > \ell_0}\Omega^2\left(\slashed{\Delta} +2(v+1)^{-2}\right)\mathcal{P}_{\ell > \ell_0}\left((v+1)^2\left(\mathfrak{k}^{(i)}-\left(v+1\right)^{-2}\right)\right)
\\ \nonumber &\qquad  = \mathcal{P}_{\ell > \ell_0}\left(-\mathcal{E}_1+\mathcal{E}_2+\mathcal{E}_3\right)+\mathcal{P}_{\ell > \ell_0}\mathcal{L}_b\mathcal{L}_{\partial_v}\mathcal{P}_{\ell \leq \ell_0}\left((v+1)^2\left(\mathfrak{k}-\left(v+1\right)^{-2}\right)\right)
\\ \nonumber &\qquad -\frac{2(-v)\mathcal{P}_{\ell > \ell_0}\Omega^2}{v+1}\mathcal{L}_{\partial_v}\mathcal{P}_{\ell \leq \ell_0}\left((v+1)^2\left(\mathfrak{k}-\left(v+1\right)^{-2}\right)\right) 
\\ \nonumber &\qquad -\mathcal{P}_{\ell > \ell_0}\Omega^2\left(\slashed{\Delta} +2(v+1)^{-2}\right)\mathcal{P}_{\ell \leq \ell_0}\left((v+1)^2\left(\mathfrak{k}-\left(v+1\right)^{-2}\right)\right)
\end{align}
with all other quantities computed with respect to the $i-1$th iterate except $\mathfrak{a}$ which is replaced with $\mathfrak{a}^{(i)}$. This is a model second order equation of type $III$ and we shall use Proposition~\ref{fo3p39iwu88u} to obtain the existence of $\mathcal{P}_{\ell > \ell_0}\mathfrak{k}^{(i)}$ with the boundary condition
	\[\lim_{v\to 0}\mathcal{P}_{\ell > \ell_0}\mathfrak{k}^{(i)} = \mathcal{P}_{\ell > \ell_0}\mathfrak{o}.\]
	\item For $\nu^{(i)}$ we follow the analogous procedure as we did for $\mathfrak{k}^{(i)}$ except that we use the equation~\eqref{oo2ok3ok2} in place of the equation~\eqref{mo2omo1013}, and we use the boundary condition:
	\[\lim_{v\to 0}\mathcal{P}_{\ell > \ell_0}\nu^{(i)} = \mathfrak{w}.\]

\item We solve for $\mathcal{P}_{\ell \leq \ell_0}\mathfrak{d}^{(i)}$ by applying $\mathcal{P}_{\ell \leq \ell_0}$ to the modified version of~\eqref{kjsdjfoijoi3} and treating the resulting equation as a transport equation for $\mathcal{P}_{\ell \leq \ell_0}\mathfrak{d}^{(i)}$ and using the $i-1$ iterate for the other quantities. For $\mathcal{P}_{\ell > \ell_0}\mathfrak{d}^{(i)}$ we use $\mathcal{P}_{\ell > \ell_0}$ to the modified~\eqref{jivjoiewoipo3} with the $i$th iterate for $\mathcal{P}_{\ell > \ell_0}\mathfrak{k}$ and $\mathfrak{a}$ and the $i-1$th iterate for the other quantities.

\item We solve for $\mathcal{P}_{\ell \leq \ell_0}\mathfrak{k}^{(i)}$ by applying $\mathcal{P}_{\ell \leq \ell_0}$ to~\eqref{jivjoiewoipo3}  with the $i$th iterate for $\mathcal{P}_{\ell > \ell_0}\mathfrak{k}$, $\mathfrak{d}$, and $\mathfrak{a}$ and the $i-1$th iterate for the other quantities. We use the boundary condition that
\[\left(\left(v+1\right)^2\mathfrak{k}^{(j)}-1\right) \to 0\text{ as }v\to -1.\]

	\item We solve for $\mathcal{P}_{\ell > \ell_0}\slashed{\rm div}\mathfrak{n}^{(i)}$ by using the modified equation~\eqref{knvnieo2o23} with $\mathfrak{n}$ replaced by $\mathfrak{n}^{(i)}$, $\mathfrak{k}$ replaced by $\mathfrak{k}^{(i)}$, and $\mathfrak{a}$ replaced by $\mathfrak{a}^{(i)}$, and $\Phi$ replaced with $\Phi^{(i-1)}$.  We then solve for $\mathfrak{n}^{(i)}$ from this and the fact that $\left(v+1\right)^2\mathcal{P}_{\ell > \ell_0}\slashed{\rm curl}\mathfrak{n}^{(i)} = \nu^{(i)}$ and that $\mathcal{P}_{\ell \leq \ell_0}\mathfrak{n}^{(i)} = \pi$. 
	\item For $\mathfrak{e}^{(i)}$  we use~\eqref{jfijoijoo2} with $\mathfrak{n}$ replaced by $\mathfrak{n}^{(i)}$ and all other quantities computed with respect to the $i-1$th iterate. We use the boundary condition that $\mathfrak{e}^{(i)}$ is bounded as $v\to -1$. We will see also that this implies that $\mathfrak{e}^{(i)}$ satisfies the modified~\eqref{akmso082hjoj2om3} with $\mathfrak{n}$ replaced by $\mathfrak{n}^{(i)}$ and all other quantities computed with respect to the $i-1$th iterate.
	\item For $\Phi^{(i)}$ we set
	\begin{equation}\label{o32jo3i4}
	\Phi^{(i)} = \slashed{\mathcal{D}}_2^{-1}\Pi_{{\rm ker}\left({}^*\slashed{\mathcal{D}}_2\right)^{\perp}}\slashed{\mathcal{D}}_1^{-1}\Pi_{{\rm ker}\left({}^*\slashed{\mathcal{D}}_1\right)^{\perp}}\left(-\left(v+1\right)^{-2}\mathfrak{d}^{(i)},\left(v+1\right)^{-2}\mathfrak{e}^{(i)}\right).
	\end{equation}
	
\end{enumerate}

We will now prove by induction on $i$ that this sequence is well-defined and moreover satisfies the following uniform bound:
\begin{equation}\label{2knj23r}
\mathscr{O}\left(i\right) \leq  C_{\rm boot}\mathcal{D},
\end{equation}
for a suitable bootstrap constant $C_{\rm boot} \lesssim 1$ and where $\mathscr{O}\left(i\right)$ is defined by the left hand side of~\eqref{2omomo1ji28} except that we replace each quantity with the corresponding $i$th iterate. In the rest of this proof, the implied constants are assumed to not depend on $C_{\rm boot}$. 

The base case $i = 0$ is immediate, so we take $j \geq 1$, assume that 
\[\left\{\left(\mathfrak{a}^{(i)},\mathfrak{d}^{(i)},\mathfrak{e}^{(i)},\mathfrak{k}^{(i)},\mathfrak{n}^{(i)},\Phi^{(i)}\right)\right\}_{i=0}^{j-1}\]
is defined, and assume that~\eqref{2knj23r} holds for all $0 \leq i \leq j-1$. We now proceed sequentially down the hierarchy of the equations we presented above. 

The equation~\eqref{jdiji2mo1} for $\mathfrak{A}^{(j)}$ is a standard transport equation (after conjugation by $(v+1)^2$). In particular, we may conjugate by $(v+1)^2$, inductively commute with $\mathcal{L}_{\mathcal{Z}^{(\alpha)}}$ for $\left|\alpha\right| \leq N_1-1$, multiply by $\left(v+1\right)^{-2+2\check{\delta}}(-v)^{1/2+2\check{\delta}}\left(\mathfrak{A}^{(j)}\right)^{(\alpha)}$ and integrate by parts to obtain\footnote{One may in fact appeal directly to Lemma~\ref{l2l2degtranstrans} in the region $v \in [-1,-1/2]$. In particular, Lemma~\ref{l2l2degtranstrans} justifies that there is no contribution from the boundary term at $v = -1$ in the integration by parts.} that 
\[\left\vert\left\vert \mathfrak{A}\right\vert\right\vert_{\mathscr{Q}\left(N_1-1,-1/2+\check{\delta},-1/2+\check{\delta}\right)}  \lesssim \mathcal{D}.\]
It is also straightforward to then directly use the equation to obtain that
\[\sum_{j=0}^1 \left\vert\left\vert \mathcal{L}^{1+j}_{\partial_v}\mathfrak{A}\right\vert\right\vert_{\mathscr{Q}\left(N_1-1-j,1/2+\check{\delta}+j,50\check{p}+1/2 +j \right)} +\sum_{j=0}^1 \left\vert\left\vert \left(1,\mathcal{L}_b\right)\mathcal{L}^{1+j}_{\partial_v}\mathfrak{A}\right\vert\right\vert_{\mathscr{Q}\left(N_1-2-j,1/2+\check{\delta}+j,50\check{p}+j \right)} \lesssim \mathcal{D}.\]
However, this still leaves us one angular derivative below the estimate we desire for $\mathfrak{A}$. The term which prevents us from naively commuting with one extra angular derivative is the term proportional to $\Phi^{(j-1)}$ since we only control $N_1-1$ angular derivatives of $\Phi^{(j-1)}$. 

The key to handling this difficulty is the realization that due to the equations which $\mathfrak{d}^{(j-1)}$ and $\mathfrak{e}^{(j-1)}$ satisfy, after an application of a suitable elliptic operator, we can write $\Phi^{(j-1)}$, to leading order, as a total $\mathcal{L}_{\partial_v}$ derivative. We now explain the details. Let $\alpha$ be a multi-index with $\left|\alpha\right| = N_1-4$ and commute~\eqref{jdiji2mo1} with $\left(v+1\right)^4\slashed{\Delta}^2\mathcal{L}_{\mathcal{Z}^{(\alpha)}}$ and derive the following:
\begin{equation}\label{ij3io23joi23}
\left(\mathcal{L}_{\partial_v}+2\Omega^2(v+1)^{-1}\right)\left(\left(v+1\right)^4\slashed{\Delta}^2\left(\mathfrak{A}^{(j)}\right)^{(\alpha)}\right) = -2\Omega^2\left( \left(v+1\right)^4\slashed{\Delta}^2\left(\Phi^{(j-1)}\right)^{(\alpha)}\right)\cdot\Phi^{(j-1)} + \mathscr{R}_1,
\end{equation}
where $\mathscr{R}_1$ is lower order and, in particular, satisfies 
\begin{equation}\label{om2o2o3}
\left\vert\left\vert \mathscr{R}_1\right\vert\right\vert_{\mathscr{Q}\left(N_1,1/2+\check{\delta},1+50\check{p}\right)} \lesssim \mathcal{D}.
\end{equation}
In what follows, we will use $\mathscr{R}_1$ to stand for any term which satisfies~\eqref{om2o2o3}. We next observe that by Lemma~\ref{jioio2o2lkj3lijoi392},~\eqref{o32jo3i4}, the modified~\eqref{jivjoiewoipo3}, and the modified~\eqref{akmso082hjoj2om3}, we have (dropping the index $j-1$):
\begin{align}\label{2mo3mo2}
&-2\Omega^2\left( \left(v+1\right)^4\slashed{\Delta}^2\Phi^{(\alpha)}\right)\cdot\Phi  
\\ \nonumber &\qquad =-2\Omega^2\left(v+1\right)^4\left(2{}^*\slashed{\mathcal{D}}_2{}^*\slashed{\mathcal{D}}_1\slashed{\mathcal{D}}_1\slashed{\mathcal{D}}_2\Phi^{(\alpha)} -4{}^*\slashed{\mathcal{D}}_2 K \slashed{\mathcal{D}}_2\Phi^{(\alpha)} 
-4K{}^*\slashed{\mathcal{D}}_2\slashed{\mathcal{D}}_2 \Phi^{(\alpha)} + 2\slashed{\Delta} K \Phi^{(\alpha)}\right)\cdot \Phi 
\\ \nonumber &\qquad =-4\Omega^2\left(v+1\right)^4\left({}^*\slashed{\mathcal{D}}_2{}^*\slashed{\mathcal{D}}_1\slashed{\mathcal{D}}_1\slashed{\mathcal{D}}_2\Phi^{(\alpha)}\right)\cdot \Phi + \mathscr{R}_1
\\ \nonumber &\qquad = -4\Omega^2\left(v+1\right)^2\left({}^*\slashed{\mathcal{D}}_2{}^*\slashed{\mathcal{D}}_1\left(-\mathfrak{d}^{(\alpha)},\mathfrak{e}^{(\alpha)}\right)\right)\cdot \Phi + \mathscr{R}_1
\\ \nonumber &\qquad =\mathcal{L}_{\partial_v}\underbrace{\left(-4\Omega^2\left(v+1\right)^2\left({}^*\slashed{\mathcal{D}}_2{}^*\slashed{\mathcal{D}}_1\left(\left(-\left(v+1\right)^2\mathfrak{k}+1\right)^{(\alpha)},\left(v+1\right)^2\mathcal{P}_{\ell \geq 1}\slashed{\rm curl}\mathfrak{n}^{(\alpha)}\right)\right)\cdot \Phi\right)}_{\doteq \mathscr{R}_2} + \mathscr{R}_1
\end{align}
In particular we can re-write~\eqref{ij3io23joi23} as
\begin{equation}\label{ij3io23joi23}
\left(\mathcal{L}_{\partial_v}+2\Omega^2(v+1)^{-1}\right)\left(\left(v+1\right)^4\slashed{\Delta}^2\left(\mathfrak{A}^{(j)}\right)^{(\alpha)} - \mathscr{R}_2\right) = 2\Omega^2\left(v+1\right)^{-1}\mathscr{R}_2 + \mathscr{R}_1.
\end{equation}
Our induction hypothesis now allow us to control $\mathscr{R}_2$. We have, in particular from~\eqref{ij3io23joi23}, after contracting with $\left(v+1\right)^{2\check{\delta}}(-v)^{1+50\check{p}}\left(\left(v+1\right)^4\slashed{\Delta}^2\left(\mathfrak{A}^{(j)}\right)^{(\alpha)} - \mathscr{R}_2\right)$ and integrating by parts that
\[\left\vert\left\vert \left(v+1\right)^4\slashed{\Delta}^2\left(\mathfrak{A}^{(j)}\right)^{(\alpha)} - \mathscr{R}_2\right\vert\right\vert_{\mathscr{Q}\left(0,-1/2+\check{\delta},50\check{p}\right) }\lesssim  \left\vert\left\vert 2\Omega^2\left(v+1\right)^{-1}\mathscr{R}_2 + \mathscr{R}_1\right\vert\right\vert_{\mathscr{Q}\left(0,1/2+\check{\delta},1+50\check{p}\right)} \Rightarrow\]
\[ \left\vert\left\vert \mathfrak{A}^{(j)}\right\vert\right\vert_{\mathscr{Q}\left(N_1,-1/2+\check{\delta},50\check{p}\right)} \lesssim \mathcal{D}.\]
The remaining estimates for $\mathfrak{A}^{(j)}$ are established by using analogous of the above estimates and also the fundamental theorem of calculus.

We now come to $\mathcal{P}_{\ell > \ell_0}\mathfrak{k}^{(j)}$, we observe that~\eqref{k2m2omo3o2} is a model second order equation of type $III$, and we may thus apply Theorem~\ref{fo3p39iwu88u} with $J_1 = N_1 - 2$.  One then needs to check  that the right hand of~\eqref{k2m2omo3o2} may be estimated in $\left\vert\left\vert \cdot\right\vert\right\vert_{R,III\left(1/2-\check{\delta},\check{p},0\right),N_1-3}$. Checking term by term and and analyzing each with the inequalities from Section~\ref{iio98987923}, we see that the only term which requires a special analysis are ones for $v \geq -1/2$ and which are given schematically by
\[\left(\mathfrak{q},\mathfrak{r}\right)\cdot \slashed{\nabla}^2\Phi.\]
(Such nonlinear terms are produced in various places on the right hand side of the modified~\eqref{kjsdjfoijoi3}.) More specifically, this term cannot be estimated directly because if after commutation with $\mathcal{L}_{Z^{(\alpha)}}$ for $\left|\alpha\right| = N_1-3$, all angular derivatives fall on $\Phi$, then we have to multiply $\Phi$ with a $(-v)^{1/2}$ before it is in $L^2\left(dv\mathring{\rm dVol}\right)$ while the norm $\left\vert\left\vert \cdot\right\vert\right\vert_{R,III\left(1/2-\check{\delta},\check{p},0\right),N_1-3}$ will only multiply by $(-v)^{1/2-\check{\delta}}$. The resolution, as usual, is to re-write the term as a total $\mathcal{L}_{\partial_v}$ derivative, and then write the term as $\mathcal{L}_{\partial_v}H_2$ for a suitable $H_2$ which will then be controlled when we take the norm $\left\vert\left\vert \cdot\right\vert\right\vert_{R,III\left(1/2-\check{\delta},\check{p},0\right),N_1-3}$ (see Definition~\ref{2kmo2o3}). More concretely, we may mimic the calculation~\eqref{2mo3mo2} and write 
\begin{align}\label{4i4io}
&\left(\mathfrak{q},\mathfrak{r}\right)\cdot \slashed{\nabla}^2\Phi =
\left(\mathfrak{q},\mathfrak{r}\right)\cdot \slashed{\nabla}^2\left((v+1)^2\slashed{\Delta}\right)^{-2} \left((v+1)^2\slashed{\Delta}\right)^2\Phi
\\ \nonumber &=\mathcal{L}_{\partial_v} \left(\left(\mathfrak{q},\mathfrak{r}\right)\cdot \slashed{\nabla}^2\left((v+1)^2\slashed{\Delta}\right)^{-2}\left(v+1\right)^2\left({}^*\slashed{\mathcal{D}}_2{}^*\slashed{\mathcal{D}}_1\left(\left(-\left(v+1\right)^2\mathfrak{k}+1\right),\left(v+1\right)^2\mathcal{P}_{\ell \geq 1}\slashed{\rm curl}\mathfrak{n}\right)\right) \right)
\\ \nonumber &\qquad + \mathscr{R}_3,
\end{align}
for a suitable lower order term $\mathscr{R}_3$. It is then straightforward to control these terms in the norm $\left\vert\left\vert \cdot\right\vert\right\vert_{R,III\left(1/2-\check{\delta},\check{p},0\right),N_1-3}$. (See Lemma~\ref{2im3om2o} for estimates associated to $\slashed{\Delta}^{-1}$ and $\left[\slashed{\Delta}^{-1},\mathcal{L}_{\partial_v}\right]$.) We thus obtain
\[\left\vert\left\vert\left(v+1\right)^2 \mathcal{P}_{\ell > \ell_0}\left(\mathfrak{k}^{(j)}-2(v+1)^{-1}\right)\right\vert\right\vert_{L,III\left(1/2-\check{\delta},\check{p},0\right),N_1-2}^2  \lesssim \mathcal{D}^2. \]
The estimates for $\nu^{(j)}$ are analogous to those for $\mathcal{P}_{\ell > \ell_0}\mathfrak{k}^{(j)}$ with the only difference being that we adjust the weights in the $\sup_v$ estimates near $v = 0$ because of the presence of nonlinear terms proportional to $\slashed{\nabla}\log\Omega \cdot \mathcal{L}_{\partial_v}b$ on the right hand side. We end up with 
\[\left\vert\left\vert \nu^{(j)}\right\vert\right\vert^2_{L,III\left(1/2-\check{\delta},100 N_1\check{p},0\right),N_1-2}  \lesssim \mathcal{D}^2.\]
Elliptic theory along $\mathbb{S}^2$ then immediately yields the existence of and desired estimate for $\mathfrak{n}^{(j)}$:
\begin{equation}\label{jji3}
\left\vert\left\vert \left(v+1\right)\mathfrak{n}^{(j)}\right\vert\right\vert_{L,III\left(1/2-\check{\delta},100 N_1 \check{p},0\right),N_1-1}^2 \lesssim  \mathcal{D}^2.
\end{equation}
It is similarly straightforward to obtain the desired estimates for $\mathfrak{d}^{(j)}$ and $\mathfrak{e}^{(j)}$ using the modified versions of $\mathcal{P}_{\ell > \ell_0}$ applied to~\eqref{jivjoiewoipo3} and the equations~\eqref{jfijoijoo2}, and~\eqref{akmso082hjoj2om3}.  One then obtains estimates for $\mathcal{P}_{\ell \leq \ell_0}\mathfrak{k}^{(j)}$ by using $\mathcal{P}_{\ell \leq \ell_0}$ applied to~\eqref{jivjoiewoipo3}. Finally, the estimates for $\Phi^{(j)}$ are immediate consequences of elliptic estimates and the established estimates for $\mathfrak{d}^{(j)}$ and $\mathfrak{e}^{(j)}$.

As usual, having established a uniform bound on the iterates, it is then straightforward to carry out a compactness argument and extract a suitable limit which solves the equation.

\end{proof}

In the next proposition we construct a solution to the modified $\left(\slashed{g},\mathfrak{n}\right)$-system.
\begin{proposition}\label{solvetheprojected} There exists a solution to the modified $\left(\slashed{g},\mathfrak{n}\right)$ system such that
\begin{align}\label{2km24i2j}
&\left\vert\left\vert \slashed{g} \right\vert\right\vert_{\mathscr{A}_2\left(\kappa,b,\Omega\right)}^2 + \left\vert\left\vert \slashed{g}\right\vert\right\vert_{\mathscr{B}_2\left(\kappa,b,\Omega\right)}^2 
+\left\vert\left\vert \left(v+1\right)\mathfrak{n}\right\vert\right\vert_{L,III\left(1/2-\check{\delta},100N_1 \check{p},0\right),N_1-1}^2
\lesssim \mathcal{D}^2,
\end{align}
where $\mathcal{D}$ is as in Proposition~\ref{3imomo1}. 

We also have that $\slashed{g}$ extends continuously to $v = 0$ and
\begin{equation}\label{32oi32oi2}
\left\vert\left\vert \slashed{g}|_{v=0}-\slashed{\mathring{g}}\right\vert\right\vert_{\mathring{H}^{N_1-1}} \lesssim \mathcal{D}.
\end{equation}

If we are willing to apply $\mathcal{P}_{\ell > \ell_0}$ to $\slashed{g}$, then we may assume that dependence on $\pi$ is nonlinear:
\begin{align}\label{2kn3kn2k4}
&\left\vert\left\vert \mathcal{P}_{\ell > \ell_0}\slashed{g} \right\vert\right\vert_{\mathscr{A}_2\left(\kappa,b,\Omega\right)} + \left\vert\left\vert \mathcal{P}_{\ell > \ell_0}\slashed{g}\right\vert\right\vert_{\mathscr{B}_2\left(\kappa,b,\Omega\right)} 
\lesssim 
\\ \nonumber &\qquad \left\vert\left\vert \log\Omega_{\rm boun}\right\vert\right\vert_{\mathscr{A}_0\left(\kappa,b\right)}+  \left\vert\left\vert b\right\vert\right\vert_{\mathscr{A}^-_1\left(b,\kappa\right)} + \left\vert\left\vert \log\Omega_{\rm sing}\right\vert\right\vert_{\mathscr{B}_{00}\left(\kappa\right)}+\left\vert\left\vert \log\Omega_{\rm boun}\right\vert\right\vert_{\mathscr{B}_{01}\left(\kappa,b\right)} +\left\vert\left\vert b\right\vert\right\vert_{\mathscr{B}^-_1\left(\kappa\right)}
\\ \nonumber &\qquad+\boxed{\left\vert\left\vert \pi\right\vert\right\vert^2_{\mathscr{S}\left(0,\check{\delta},0\right)} +\left\vert\left\vert \pi\right\vert\right\vert^2_{\mathscr{Q}_{-1}^{-1/2}\left(0,-1/2+\check{\delta},0\right)}}  + \left\vert\left\vert \left(\mathfrak{o},\mathcal{L}_{b|_{v=0}}\mathfrak{o}\right)\right\vert\right\vert_{\mathring{H}^{N_1-3}\left(\mathbb{S}^2\right)} + \left\vert\left\vert \left(\mathfrak{w},\mathcal{L}_{b|_{v=0}}\mathfrak{w}\right)\right\vert\right\vert_{\mathring{H}^{N_1-3}\left(\mathbb{S}^2\right)},
\end{align}
where we have boxed the key term where there is a difference from the estimate without $\mathcal{P}_{\ell > \ell_0}$.

Finally, for any $\left|\alpha\right| \leq N_1-1$,
\begin{equation}\label{3io3i3iji}
\Phi^{(\alpha)} = \mathcal{L}_{\partial_v}\mathfrak{V}_1 + \mathfrak{V}_2,
\end{equation}
where
\begin{equation}\label{32om0902noo203}
\left\vert\left\vert \mathcal{L}_{\partial_v}\mathfrak{V}_1\right\vert\right\vert_{\mathscr{Q}_{-1/2}^0\left(0,0,1/2\right)} + \left\vert\left\vert \mathfrak{V}_1\right\vert\right\vert_{\mathscr{Q}_{-1/2}^0\left(1,0,-\kappa\right)} + \left\vert\left\vert \mathfrak{V}_2\right\vert\right\vert_{\mathscr{Q}_{-1/2}^0\left(0,0,-\kappa\right)} + \left\vert\left\vert \lim_{v\to 0}\mathfrak{V}_1\right\vert\right\vert_{L^2\left(\mathbb{S}^2\right)} \lesssim \mathcal{D}.
\end{equation}

\end{proposition}
\begin{proof}

As usual, we will solve this system via an iteration process. We will inductively define a sequence $\{\left(\slashed{g}^{(j)},\mathfrak{a}^{(j)},\Phi^{(j)}\right)\}_{j=0}^{\infty}$ of Riemannian metrics $\slashed{g}^{(j)}$, functions $\mathfrak{a}^{(j)}$, and symmetric $(0,2)$-tensors $\Phi^{(j)}_{AB}$ as follows: We set $\slashed{g}^{(0)} \doteq (v+1)^2\mathring{\slashed{g}}$, $\mathfrak{a}^{(0)} = \frac{2\Omega^2}{v+1}$, and $\Phi^{(0)} = 0$. For $j \geq 1$, we define $\mathfrak{a}^{(j)}$ and $\Phi^{(j)}$ by applying Proposition~\ref{3imomo1} with $\slashed{g} = \slashed{g}^{(j-1)}$. We then define $\slashed{g}^{(j)}$ by 
\begin{align}\label{i3ijj}
\mathcal{L}_{\partial_v}\slashed{g}^{(j)}_{AB} = \mathfrak{a}^{(j)}\slashed{g}^{(j)}_{AB} + 2\Phi^{(j)}_{AB},
\end{align}
with the boundary condition that, in the coordinate frame $\left(v+1\right)^{-2}\slashed{g}_{AB} - \mathring{\slashed{g}}_{AB} \to 0$ as $v\to -1$. 

We will now prove by induction that 
\begin{equation}\label{2m3om4}
\left\vert\left\vert \slashed{g}^{(i)} \right\vert\right\vert_{\mathscr{A}^{-}_2\left(b,\kappa\right)}^2 + \left\vert\left\vert \slashed{g}^{(i)}\right\vert\right\vert_{\mathscr{B}^{-}_2\left(\kappa\right)}^2  +\left\vert\left\vert \slashed{g}^{(i)}|_{v=0}-\slashed{\mathring{g}}\right\vert\right\vert^2_{\mathring{H}^{N_1-1}} \leq  C_{\rm boot}\mathcal{D}^2,
\end{equation}
for a suitable bootstrap constant $C_{\rm boot}$. The case $i = 0$ clearly holds, so we consider $j \geq 1$ and assume that~\eqref{2m3om4} holds for all $0 \leq i \leq j-1$. The induction hypothesis allows us to apply Proposition~\ref{3imomo1} and thus $\mathfrak{a}^{(j)}$ and $\Phi^{(j)}$ are well-defined (and satisfy the estimates of Proposition~\ref{3imomo1} and Lemma~\ref{1ioj3i4io}). We thus turn to $\slashed{g}^{(j)}$. Linearizing~\eqref{i3ijj} around $\left(v+1\right)^2\mathring{\slashed{g}}$ leads to  
\begin{align}\label{2ojm3ojm2}
&\mathcal{L}_{\partial_v}\left(\left(v+1\right)^{-2}\slashed{g}^{(j)} - \mathring{\slashed{g}}\right) =  \Omega^2\left(\Omega^{-2}\mathfrak{a}^{(j)}-2(v+1)^{-1}\right)\mathring{\slashed{g}} 
 \\ \nonumber &\qquad +\Omega^2\left(\Omega^{-2}\mathfrak{a}^{(j)}-2(v+1)^{-1}\right)\left(\left(v+1\right)^{-2}\slashed{g}^{(j)}-\mathring{\slashed{g}}\right)
 \\ \nonumber &\qquad  + \frac{2(\Omega^2-1)}{v+1}\mathring{\slashed{g}}+ \frac{2(\Omega^2-1)}{v+1}\left(\left(v+1\right)^{-2}\slashed{g}^{(j)}-\mathring{\slashed{g}}\right) +2\left(v+1\right)^{-2}\Phi^{(j)}.
\end{align}
This is a standard transport equation, and we thus immediately obtain the existence of $\slashed{g}^{(j)}$. Commuting inductively with $\mathcal{L}_{Z^{(\alpha)}}$  for $\left|\alpha\right| \leq N_1-1$, multiplying by $(-v)^{2\check{\delta}}(v+1)^{-2+2\check{\delta}}\mathring{\slashed{g}}^{AC}\mathring{\slashed{g}}^{BD}\left(\left(v+1\right)^{-2}\slashed{g}^{(j)} - \mathring{\slashed{g}}\right)_{CD}$, and integrating by parts leads to 
\[\left\vert\left\vert \left(\slashed{g} - \left(v+1\right)^2\mathring{\slashed{g}}\right)\right\vert\right\vert^2_{\mathscr{Q}\left(N_1-1,-3/2+\check{\delta},-1/2+\check{\delta}\right)} \lesssim \mathcal{D}^2.\]
It is also straightforward to integrate~\eqref{2ojm3ojm2} and obtain
\[\left\vert\left\vert \left(\slashed{g}  - (v+1)^2\mathring{\slashed{g}}\right)\right\vert\right\vert^2_{\mathscr{S}_{-1/2}^0\left(N_2,0,0\right)} +\left\vert\left\vert\left(\slashed{g} - (v+1)^2\mathring{\slashed{g}}\right)\right\vert\right\vert^2_{\mathscr{S}_{-1}^{-1/2}\left(N_1-1,-1+\check{\delta},0\right)} \lesssim \mathcal{D}^2.\]
The necessary estimates for $\mathcal{L}_{\partial_v}$-derivatives of $\slashed{g}^{(j)}$ are all straightforward consequence of the equation~\eqref{2ojm3ojm2} and the estimates we have already have for $\mathfrak{a}^{(j)}$ and $\Phi^{(j)}$. 

We still need an estimate in the $\mathscr{Q}$-norm for $\left(\slashed{g}^{(j)}\right)^{(\alpha)}$ when $\left|\alpha\right| \leq N_1$. Here we have a familiar problem: we cannot use directly transport estimates because we do not control $N_1$ angular derivatives of $\Phi$. We also will have the familiar resolution, that is, after the application of a suitable elliptic operator, we have $\Phi^{(\alpha)}$ will become a total $\mathcal{L}_{\partial_v}$ derivative plus a lower order term. This argument will be reminiscent of the one we have already seen in the proof of Proposition~\ref{3imomo1}. Repeating the argument that lead to~\eqref{2mo3mo2} yields
\begin{align}\label{kni2ni33202}
\left(v+1\right)^4\slashed{\Delta}^2\Phi^{(j)} = 2\left(v+1\right)^4{}^*\slashed{\mathcal{D}}_2{}^*\slashed{\mathcal{D}}_1\left(-(v+1)^{-2}\mathfrak{d}^{(j)},(v+1)^{-2}\mathfrak{e}^{(j)}\right)+\mathscr{T},
\end{align}
where $\left(\mathfrak{d}^{(j)},\mathfrak{e}^{(j)}\right)$ are the solutions produced by our earlier invocation of Proposition~\ref{3imomo1}, where here, and in what follows, the implicit $\slashed{g}$'s  are computed with respect to $\slashed{g}^{(j-1)}$, and where
\begin{equation}\label{2o2omo}
\left\vert\left\vert\mathscr{T} \right\vert\right\vert_{\mathscr{Q}\left(N_1-4,-1/2+\check{\delta},1-2\kappa\right)} \lesssim \mathcal{D}.
\end{equation}
In what follows, we denote by $\mathscr{T}$ any term which satisfies~\eqref{2o2omo}. In view of~\eqref{jivjoiewoipo3} and~\eqref{akmso082hjoj2om3} we obtain that
\begin{align}\label{kni2ni33202}
\left(v+1\right)^4\slashed{\Delta}^2\Phi^{(j-1)} = \mathcal{L}_{\partial_v}\left(2{}^*\slashed{\mathcal{D}}_2{}^*\slashed{\mathcal{D}}_1\left(-\left(v+1\right)^2\mathfrak{k}^{(j)},(v+1)^2\nu^{(j)}\right)\right)+\mathscr{T},
\end{align}
where $\left(\mathfrak{k}^{(j)},\mathfrak{\nu}^{(j)}\right)$ are the solutions produced by our previous application of Proposition~\ref{3imomo1}. Finally, we have 
\begin{align}\label{2knk2nk1}
\underbrace{\left(v+1\right)^4\left(\slashed{g}^{AB}\mathring{\nabla}_A\mathring{\nabla}_B\right)\left(\slashed{g}^{CD}\mathring{\nabla}_C\mathring{\nabla}_D\right)}_{\doteq \mathcal{Q}}\Phi^{(j-1)} = \mathcal{L}_{\partial_v}\left(2{}^*\slashed{\mathcal{D}}_2{}^*\slashed{\mathcal{D}}_1\left(-\left(v+1\right)^2\mathfrak{k}^{(j)},(v+1)^2\nu^{(j)}\right)\right)+\mathscr{T}.
\end{align}
We note that the operator $\mathcal{Q}$ is a $4$th order elliptic operator on each $\mathbb{S}^2_{-1,v}$, and moreover is uniformly elliptic  for $v \in [-1,0)$.  We further note that the difference of $\mathcal{Q}$ and $\left(v+1\right)^4\slashed{\Delta}^2$ involves at most four angular derivatives of $\left(v+1\right)^{-2}\slashed{g}-\mathring{\slashed{g}}$. 

With~\eqref{2knk2nk1} established, we now commute~\eqref{2ojm3ojm2} with $\mathcal{Q}$, note that the commutator of $\mathcal{Q}$ and $\mathcal{L}_{\partial_v}$ involves at most two angular derivatives of $\mathcal{L}_{\partial_v}\left((v+1)^{-2}\slashed{g}-\mathring{\slashed{g}}\right)$, and move the total $\mathcal{L}_{\partial_v}$ derivative to the left hand side of our equation. At the end we will have obtained
\begin{equation}\label{3i2ojjio3}
\mathcal{L}_{\partial_v}\underbrace{\left(\mathcal{Q}\left(\left(v+1\right)^{-2}\slashed{g}^{(j)}-\mathring{\slashed{g}}\right) -2{}^*\slashed{\mathcal{D}}_2{}^*\slashed{\mathcal{D}}_1\left(-\left(v+1\right)^2\mathfrak{k}^{(j)},(v+1)^2\nu^{(j)}\right)\right)}_{\doteq \mathcal{Z}} = \left(v+1\right)^{-2}\mathscr{T}.
\end{equation}
It is straightforward to commute this with $\mathcal{L}_{Z^{(\alpha)}}$ for $\left|\alpha\right| = N_1-4$, and separately also $\left|\alpha\right| = N_2 -4$, and apply transport estimates to obtain that 
\begin{equation*}
\left\vert\left\vert \left(v+1\right)^2\mathcal{Z} \right\vert\right\vert_{\mathscr{Q}\left(N_1-4,-3/2+\check{\delta},-2\kappa\right)}  \lesssim \mathcal{D}.
\end{equation*}
Finally, elliptic estimates with $\mathcal{Q}$ combined with the estimates for $\mathfrak{k}^{(j)}$ and $\mathfrak{n}^{(j)}$ from Proposition~\ref{3imomo1} then provide the final needed estimates in order to establish~\eqref{2m3om4}. Then, in order to estimate $\left\vert\left\vert \slashed{g}^{(i)}|_{v=0}-\mathring{g}\right\vert\right\vert^2_{\mathring{H}^{N_1-1}}$, we may also use~\eqref{3i2ojjio3}. Namely, we simply now commute~\eqref{3i2ojjio3} with $\mathcal{L}_{Z^{(\alpha)}}$ for $\left|\alpha\right| \leq N_1-5$ and apply the fundamental theorem of calculus.

With the uniform bounds established for $\slashed{g}^{(j)}$ and for the solutions produced by Proposition~\ref{3imomo1}, it is straightforward to extract a suitably convergent sub-sequence whose limits solve the modified $\left(\slashed{g},\mathfrak{n}\right)$-system. We will have
\begin{equation}\label{2kl3j2lkj2}
\left\vert\left\vert \slashed{g} \right\vert\right\vert_{\mathscr{A}^{-}_2\left(b,\kappa\right)}^2 + \left\vert\left\vert \slashed{g}\right\vert\right\vert_{\mathscr{B}^{-}_2\left(\kappa\right)}^2  +\left\vert\left\vert \slashed{g}|_{v=0}-\slashed{\mathring{g}}\right\vert\right\vert^2_{\mathring{H}^{N_1-1}} \lesssim \mathcal{D}^2,
\end{equation}
We now need to upgrade~\eqref{2kl3j2lkj2} to the estimate~\eqref{2km24i2j}. This requires us to establish that $\Omega^{-1}{\rm tr}\chi$ and $\mathcal{L}_bK$ satisfy certain improved estimates. For ${\rm tr}\chi$ it suffices to simply observe that, by the definition~\eqref{2k3mr4i2i4} of $\Phi$, that $\slashed{g}^{AB}\Phi_{AB} = 0$ and hence that ${\rm tr}\chi = \mathfrak{a}$. For $\mathcal{L}_bK$, we will need to relate $K$, the Gauss curvature of $\slashed{g}$ with the artificial variable $\mathfrak{k}$. To obtain the necessary comparison, we observe that from~\eqref{3l3omo29},~\eqref{jivjoiewoipo3}, and the definition~\eqref{2k3mr4i2i4} of $\Phi$, we obtain that 
\begin{align}\label{2i3mom2o}
&\mathcal{L}_{\partial_v}\left(\mathfrak{k}-K\right) + \left(\Omega{\rm tr}\chi\left(\mathfrak{k}-K\right)\right) 
\\ \nonumber &\qquad =\slashed{\rm div}\left(\Pi_{{\rm ker}\left({}^*\slashed{\mathcal{D}}_2\right)}\slashed{\mathcal{D}}_1^{-1}\Pi_{{\rm ker}\left({}^*\slashed{\mathcal{D}}_1\right)^{\perp}}\left(-\left(v+1\right)^{-2}\mathfrak{d},\left(v+1\right)^{-2}\mathfrak{e}\right)\right)+\Pi_{{\rm ker}\left({}^*\slashed{\mathcal{D}}_1\right)}\left(-\left(v+1\right)^{-2}\mathfrak{d},(v+1)^{-2}\mathfrak{e}\right).
\end{align}
Since we have that
\[\lim_{v\to -1}\left(v+1\right)^2\left(\mathfrak{k}-K\right) = 0,\]
in view of the estimates from Lemma~\ref{ijmom2om392}, it is straightforward to use this transport equation to obtain the needed estimates for $K$ from the estimates for $\mathfrak{k}$ provided by Proposition~\ref{3imomo1}.

The estimate~\eqref{2kn3kn2k4} follows by observing that $\pi$ is supported on spherical harmonics with $\ell \leq \ell_0$, and so, if we simply re-run all of the estimates in this proof for $\mathcal{P}_{\ell > \ell_0}$ applied to the various quantities, then $\pi$ cannot linearly couple to any of the terms being estimated. Then~\eqref{2kn3kn2k4} follows in a straightforward fashion.

Finally, the bound~\eqref{32om0902noo203} is established using the same techniques which lead to~\eqref{4i4io}. We omit the details.

 \end{proof}

In this next lemma, we will analyze more closely the solutions produced by Proposition~\ref{3imomo1} in the scenario where we additionally know that $\mathfrak{q} =\slashed{\rm div}b$ and $\mathfrak{r} =\slashed{\nabla}\hat{\otimes}b$. We will find, in particular, that the projections which we have used to define $\Phi$ may be dropped.
\begin{lemma}\label{1ioj3i4io}We consider a solution to the modified $\left(\slashed{g},\mathfrak{n}\right)$-system produced by Proposition~\ref{solvetheprojected}. Suppose further that we have $\mathfrak{q} =\slashed{\rm div}b$ and $\mathfrak{r} =\slashed{\nabla}\hat{\otimes}b$. Then $\left(\slashed{g},\mathfrak{n}\right)$ in fact solve the $\left(\slashed{g},\mathfrak{n}\right)$-system. 

More concretely,
\begin{align}\label{2om3ok3o}
&\Pi_{{\rm ker}\left({}^*\slashed{\mathcal{D}}_2\right)}\slashed{\mathcal{D}}_1^{-1}\Pi_{{\rm ker}\left({}^*\slashed{\mathcal{D}}_1\right)^{\perp}}\left(-\mathfrak{d},\mathfrak{e}\right) = 0,
\end{align}
\begin{equation}\label{kjn2ike4i3}
\Pi_{{\rm ker}\left({}^*\slashed{\mathcal{D}}_1\right)}\left(-\mathfrak{d},\mathfrak{e}\right) = 0.
\end{equation}
As a consequence, we have 
\begin{equation}\label{2i3in3ni}
\Phi = \slashed{\mathcal{D}}_2^{-1}\slashed{\mathcal{D}}_1^{-1}\left(-\left(v+1\right)^{-2}\mathfrak{d},\left(v+1\right)^{-2}\mathfrak{e}\right),
\end{equation}
and
\begin{align}\label{w3omre2om}
& v\Omega\nabla_4\Phi_{AB}+\mathcal{L}_b\Phi_{AB} +v(\Omega{\rm tr}\chi)\Phi_{AB} -\left(\slashed{\nabla}\hat{\otimes}b\right)^C_{\ \ (A}\Phi_{B)C}-\frac{1}{2}\slashed{\rm div}b \Phi_{AB}= 
\\ \nonumber &\qquad \Omega^2\left(\left(\slashed{\nabla}\hat\otimes \mathfrak{n}\right)_{AB} + \left(\mathfrak{n}\hat\otimes \mathfrak{n}\right)_{AB}\right) - \frac{1}{4}\left(\Omega{\rm tr}\chi\right)\left(\slashed{\nabla}\hat{\otimes}b\right)_{AB}.
 \end{align}
 
 We then have the following estimates for $\Phi$:
 \begin{align}\label{23ik4ij}
 \left\vert\left\vert \Omega^{-2}\Theta\right\vert\right\vert_{L,III\left(1/2-\check{\delta},100 N_1 \check{p},0\right),N_1-2}+ \left\vert\left\vert \Omega^{-2}\Theta\right\vert\right\vert_{\mathscr{S}_{-1/2}^0\left(N_2-2,0,0\right)}  \lesssim \mathcal{D},
 \end{align}
 where 
 \begin{align*}
& \Theta_{AB} \doteq
\\ \nonumber & \left(v+1\right)\left(v\Omega\nabla_4\Phi_{AB}+\mathcal{L}_b\Phi_{AB} +v(\Omega{\rm tr}\chi)\Phi_{AB} -\left(\slashed{\nabla}\hat{\otimes}b\right)^C_{\ \ (A}\Phi_{B)C}-\frac{1}{2}\slashed{\rm div}b\Phi_{AB}\right),
 \end{align*}
 and $\mathcal{D}$ is as in Proposition~\ref{3imomo1}.

Finally, for any solution to the $\left(\slashed{g},\mathfrak{n}\right)$-system we have
\[\mathfrak{k} = K,\]
where $K$ is the Gaussian curvature of $\slashed{g}$.
\end{lemma}
\begin{proof} We define a symmetric $(0,2)$-tensor $\tilde{\Phi}_{AB}$ by  solving~\eqref{w3omre2om} (see Remark~\ref{2km2om3o}) with $\Phi$ replaced by $\tilde{\Phi}$ and the boundary condition that $\tilde{\Phi}$ is bounded as $v\to -1$. If we trace the equation, we find, after a short calculation, that
\[\left(v\nabla_v+\mathcal{L}_b+v\left(\Omega{\rm tr}\chi\right) + \frac{1}{2}\slashed{\rm div}b\right)\left(\slashed{g}^{AB}\tilde{\Phi}_{AB}\right) = 0.\]
In particular, given the boundary condition for $\tilde{\Phi}$, we immediately conclude that $\tilde{\Phi}_{AB}$ is a trace-free symmetric tensor. We then define
\[ \left(-\left(v+1\right)^{-2}\tilde{\mathfrak{d}},\left(v+1\right)^{-2}\tilde{\mathfrak{e}}\right) \doteq \slashed{\mathcal{D}}_1\slashed{\mathcal{D}}_2\tilde{\Phi}.\]
In particular, if we define a map $\mathscr{P}$ from pairs of scalar functions to symmetric trace-free tensors by
\[\left(z,x\right) \mapsto_{\mathscr{P}}  \slashed{\mathcal{D}}_2^{-1}\Pi_{{\rm ker}\left({}^*\slashed{\mathcal{D}}_2\right)^{\perp}}\slashed{\mathcal{D}}_1^{-1}\Pi_{{\rm ker}\left({}^*\slashed{\mathcal{D}}_1\right)^{\perp}}\left(-\left(v+1\right)^{-2}z,(v+1)^{-2}x\right),\]
then we have both
\[\tilde{\Phi} = \mathscr{P}\left(\tilde{\mathfrak{d}},\tilde{\mathfrak{e}}\right),\qquad \Phi = \mathscr{P}\left(\mathfrak{d},\mathfrak{e}\right).\]
By applying $\slashed{\mathcal{D}}_1\slashed{\mathcal{D}}_2$ to the equation defining $\tilde{\Phi}$ and taking the difference with the equations of $\mathfrak{d}$ and $\mathfrak{e}$, we may derive
\begin{align}\label{ijmo2}
& v\Omega\nabla_4\mathfrak{R} +\mathcal{L}_b\mathfrak{R} +v\mathfrak{a}\mathfrak{R}= 
\left[v\Omega\nabla_4+\mathcal{L}_b+v\mathfrak{a},\slashed{\mathcal{D}}_1\right](v+1)\slashed{\mathcal{D}}_2\mathscr{P}\left[\mathfrak{R}\right]
\\ \nonumber &\qquad +(v+1)\slashed{\mathcal{D}}_1\Bigg(\left[v\Omega\nabla_4 + \mathcal{L}_b + v\mathfrak{a},(v+1)\slashed{\mathcal{D}}_2\right]\mathscr{P}\left[\mathfrak{R}\right]+
 (v+1)\slashed{\mathcal{D}}_2\left[\left(\slashed{\nabla}\hat{\otimes}b\right)^C_{\ \ (\cdot }\mathscr{P}\left[\mathfrak{R}\right]_{\cdot)C} + \frac{1}{2}\slashed{\rm div}b\mathscr{P}\left[\mathfrak{R}\right]\right]\Bigg),
\end{align}
where $\mathfrak{R} \doteq \left(\mathfrak{d},\mathfrak{e}\right) - \left(\tilde{\mathfrak{d}},\tilde{\mathfrak{e}}\right)$. Integrating from $v = -1$ easily yields that $\mathfrak{R}$ must then vanish identically. We thus obtain~\eqref{2om3ok3o}-\eqref{w3omre2om}.

The bound~\eqref{23ik4ij} is immediate from the already established estimates for $\mathfrak{n}$ in Proposition~\ref{3imomo1}. To see that $K = \mathfrak{k}$, we observe that from~\eqref{3l3omo29} and~\eqref{jivjoiewoipo3} we obtain that 
\[\mathcal{L}_{\partial_v}\left(\mathfrak{k}-K\right) + \left(\Omega{\rm tr}\chi\left(\mathfrak{k}-K\right)\right) = 0.\]
Since we have that
\[\lim_{v\to -1}\left(v+1\right)^2\left(\mathfrak{k}-K\right) = 0,\]
after integrating from $v = -1$ we obtain immediately that $\mathfrak{k} = K$. From the estimates for $\mathfrak{k}$ established in Proposition~\ref{3imomo1} we immediately obtain the desired estimate for $K$ and this concludes the proof.

\end{proof}

Finally, we observe the consequences for the Ricci tensor of solving the $\left(\slashed{g},\mathfrak{n}\right)$-system.
\begin{lemma}\label{Riccifromglsashsn}Let $\left(\mathcal{M},g\right)$ be a Lorentzian manifold with a self-similar foliation. Suppose that the corresponding $\left(\Omega,b,\slashed{g}\right)$ solve the $\left(\slashed{g},\mathfrak{n}\right)$-system along $\{v = -1\}$ for some $\mathbb{S}^2_{-1,v}$ $1$-form $\mathfrak{n}$. Then we have that 
\[{\rm Ric}_{44} = 0,\]
\[(\slashed{\nabla}\hat{\otimes}\left(\mathfrak{n}-\eta\right))_{AB}+\left(\mathfrak{n}\hat{\otimes}\mathfrak{n}\right)_{AB} - \left(\eta\hat{\otimes}\eta\right)_{AB} - \widehat{\rm Ric}_{AB} = 0,\]
\begin{align*}
&\Omega\nabla_4\left((v+1)^2\mathcal{P}_{\ell > \ell_0}\slashed{\rm curl}\left(\mathfrak{n}-\eta\right)\right)+\mathcal{P}_{\ell > \ell_0}\frac{2\Omega^2}{v+1}\left(v+1\right)^2\mathcal{P}_{\ell > \ell_0}\slashed{\rm curl}\left(\mathfrak{n}-\eta\right)-(v+1)^2\mathcal{P}_{\ell \geq 1}\slashed{\rm curl}\left(\Omega{\rm Ric}_{\cdot 4}\right) = 
\\ \nonumber &\qquad  \mathcal{P}_{\ell \geq 1}\Big( \left(v+1\right)^2\left[\Omega\nabla_4,(v+1)\slashed{\epsilon}^{CA}\slashed{\nabla}_C\right]\left(\mathfrak{n}_A-\eta_A\right)  -\left(v+1\right)^2\slashed{\epsilon}^{CA}\slashed{\nabla}_C\left(\Omega\hat{\chi}_{AB}\left(\mathfrak{n}^B-\eta^B\right)\right)
\\ \nonumber &\qquad -\frac{3}{2}\left(v+1\right)^2\slashed{\rm curl}\left(\Omega{\rm tr}\chi\right)\left(\mathfrak{n}_A -\eta_A\right)-\frac{1}{2}\left(\Omega{\rm tr}\chi - 2(v+1)^{-1}\right)\left(v+1\right)^2\slashed{\rm curl}\left(\mathfrak{n}-\eta\right)\\ \nonumber &\qquad - \mathfrak{a}\left(v+1\right)^2\mathcal{P}_{\ell \leq \ell_0}\slashed{\rm curl}\left(\mathfrak{n}-\eta\right) -\left(\Omega^{-2}\mathfrak{a}-2(v+1)^{-1}\right)\left(v+1\right)^2\mathcal{P}_{\ell > \ell_0}\slashed{\rm curl}\left(\mathfrak{n}-\eta\right)\Big),
\end{align*}
\end{lemma}
\begin{proof}This is immediate from the derivation of the $\left(\slashed{g},\mathfrak{n}\right)$-system and the fact, shown in Lemma~\ref{1ioj3i4io}, that $\mathfrak{k} = K$. 
\end{proof}

\section{Solving for $\mathring{\Pi}_{\rm div}b$}\label{k2m3mo492}
In this section we will discuss the types of equations which will eventually be used to solve for the divergence part of the vector $b$. Throughout this section we will let $\Omega_{\rm sing}(v) : (-1,0) \to (0,\infty)$ be a given spherically symmetric function, $\Omega_{\rm boun} : (-1,0) \times \mathbb{S}^2 \to (0,\infty)$ be a function, $\Omega \doteq \Omega_{\rm boun}\Omega_{\rm sing}$, and $\slashed{g}_{AB}$ be a positive definite symmetric $\mathbb{S}^2_{-1,v}$ $(0,2)$-tensor for $v\in (-1,0)$. We will assume that, for a suitable constant $\kappa$ satisfying $\left|\kappa\right| \lesssim \epsilon$ and some vector field $\tilde{b}$ we have 
\begin{align}\label{2knk2nk4n2k}
&\mathscr{C}_1 \doteq \left\vert\left\vert \log\Omega_{\rm boun}\right\vert\right\vert_{\mathscr{A}_0\left(\kappa,\tilde{b}\right)}+ \left\vert\left\vert \slashed{g}\right\vert\right\vert_{\mathscr{A}_2\left(\kappa,\tilde{b},\Omega\right)} 
+\left\vert\left\vert \log\Omega_{\rm boun}\right\vert\right\vert_{\mathscr{B}_{01}\left(\kappa,\tilde{b}\right)} \\ \nonumber &\qquad+ \left\vert\left\vert \log\Omega_{\rm sing}\right\vert\right\vert_{\mathscr{B}_{00}\left(\kappa\right)} + \left\vert\left\vert \slashed{g}\right\vert\right\vert_{\mathscr{B}_2\left(\kappa,\tilde{b},\Omega\right)}  \lesssim \epsilon.
\end{align}

We will further assume that we have an already specific scalar $\mathfrak{Y}$ which satisfies $\mathcal{P}_{\ell = 0}\mathfrak{Y} = 0$ and so that for the eventual $b$ we produce we will have that
\[\mathfrak{Y} = \mathring{\rm curl}b.\]
We also introduce a function $\mathfrak{y}$ so that $\mathcal{P}_{\ell = 0}\mathfrak{y} = 0$ and 
\[\mathfrak{Y} = \mathring{\Delta}\mathfrak{y}.\]
In particular, $\mathring{\Pi}_{\rm curl}b = -\mathring{\slashed{\epsilon}}^{AB}\mathring{\nabla}_B\mathfrak{y}$. We will assume that 
\begin{align}\label{oioioio2iou3}
& \mathscr{C}_2 \doteq \left\vert\left\vert \mathring{\Pi}_{\rm curl}b\right\vert\right\vert_{\mathscr{Q}\left(N_1,-3/2+\check{\delta},-\kappa\right)} + \left\vert\left\vert \mathring{\Pi}_{\rm curl}b\right\vert\right\vert_{\mathscr{Q}_{-1/2}^0\left(N_1-2,0,-\sqrt{\check{p}}\right)}
\\ \nonumber &\qquad  + \sum_{j=0}^1\left\vert\left\vert \mathcal{L}_{\partial_v}^{1+j} \mathring{\Pi}_{\rm curl}b\right\vert\right\vert_{\mathscr{Q}\left(N_1-1-j,-1/2+\check{\delta}+j,\kappa +j \right)}   +\left\vert\left\vert \mathring{\Pi}_{\rm curl}b\right\vert\right\vert_{\mathscr{B}_1\left(\kappa\right)}  \lesssim \epsilon.
\end{align}
We emphasize that none of the results in this section depend on the implied constants in~\eqref{2knk2nk4n2k},~\eqref{oioioio2iou3}, or in the inequality for $\kappa$ (though by our conventions for $\epsilon$, we may assume that $\epsilon$ is sufficiently small depending on the implied constants).

\subsection{Overview}\label{2o2o4mo292904nfj3k34i2}
As we have discussed in Section~\ref{joijoij2oi34} we will require in this section that $b$ satisfies an equation of the form: 
\begin{align}\label{kj2lkjlkjl32}
&\mathcal{P}_{\ell \geq 1}\slashed{\nabla}_A\left(-\frac{1}{4}\Omega^{-2}\mathcal{L}_{\partial_v}b^A\right) -\frac{1}{4} \mathcal{P}_{\ell \geq 1}\left(\Omega^{-1}{\rm tr}\chi \slashed{\rm div}b\right) + \mathcal{P}_{\ell \geq 1}\left( \left(v+1\right)^{-2}\mathring{\Delta}\slashed{\rm div}b\right) = H,
\\ \nonumber  \qquad \mathring{\rm curl}b &= \mathfrak{Y},
\end{align}
for a suitable right hand side $H$. Keeping in mind Lemma~\ref{othercommutelemma}, if we assume that~\eqref{kj2lkjlkjl32} holds, we may derive from~\eqref{kj2lkjlkjl32} the following equation for $\mathcal{P}_{\ell \geq 1}\slashed{\rm div}b$:
\begin{align}\label{ioiojoij2oiji3o}
&-\frac{1}{4}\mathcal{L}_{\partial_v}\mathcal{P}_{\ell \geq 1}\slashed{\rm div}b  -\frac{1}{4}\mathcal{P}_{\ell \geq 1}\left(\Omega{\rm tr}\chi \mathcal{P}_{\ell \geq 1}\slashed{\rm div}b\right) + \mathcal{P}_{\ell \geq 1}\left(\Omega^2\left(v+1\right)^{-2}\mathring{\Delta}\mathcal{P}_{\ell \geq 1}\slashed{\rm div}b\right) = 
\\ \nonumber &\qquad \mathscr{E}\left[b,\Omega^2,\Omega {\rm tr}\chi \right] + \mathcal{P}_{\ell \geq 1}\left(\Omega^2H\right),
\end{align}
where $\mathscr{E}$ is generated by various commutators, most importantly, $\left[\mathcal{P}_{\ell \geq 1}\slashed{\nabla}_A,-\frac{1}{4}\Omega^{-2}\mathcal{L}_{\partial_v}\right]b^A$. \underline{If} the term $\mathscr{E}$ can be treated perturbatively, then we could expect to use Proposition~\ref{2km2om1o} to solve for $\mathcal{P}_{\ell \geq 1}\slashed{\rm div}b$ from~\eqref{ioiojoij2oiji3o}.  This does indeed turn out to be a reasonable strategy when $|v| \gtrsim 1$; however, because $\slashed{\nabla}\log\Omega$ may blow-up as $v\to 0$, the term in $\mathscr{E}$ which is proportional to $\Omega^2\slashed{\nabla}_A\left(\Omega^{-2}\right)\mathcal{L}_{\partial_v}b^A$ would cause a problem with this strategy when $v$ is small. 

The resolution to this difficulty, at least as far as a priori estimates go, is as follows. Instead of only relying on the equation~\eqref{ioiojoij2oiji3o}, we should also keep in the mind the original equation~\eqref{kj2lkjlkjl32}. Here we can use the elliptic estimates from Lemma~\ref{om3om2} which have been exactly designed to apply in this scenario. They provide an a priori estimate for  $\mathcal{L}_{\partial_v}b$ in terms of $\mathring{\nabla}\slashed{\rm div}b$ (and $H$ and $\mathfrak{Y}$). This additional estimate would allow us to close our arguments.

Despite the above strategy for establishing estimates for $b$, there is an additional complication. Namely, we must also establish the \emph{existence} of the solution $b$ via a suitable iteration argument. In the course of iterating it is natural to define an iterate $\mathcal{P}_{\ell \geq 1}\slashed{\rm div}b^{(j)}$ by solving~\eqref{ioiojoij2oiji3o} with $\slashed{\rm div}b^{(j)}$ replacing $\slashed{\rm div}b$ on the left hand side, and using the $j-1$st iterates on the right hand side. However, this breaking of the nonlinear structure then does not allow us to access directly the original equation~\eqref{kj2lkjlkjl32}. We remedy this by the introduction of a new vector field $P^A$, which is defined for $v \geq -1/2$, by taking the equations~\eqref{kj2lkjlkjl32} and replacing $\mathcal{L}_{\partial_v}b^A$ with $P^A$ and $\mathring{\rm curl}b$ with $\mathring{\rm curl}P$. We then replace the $\mathcal{L}_{\partial_v}b^A$ in the problematic nonlinear term in~\eqref{ioiojoij2oiji3o} with $P^A$. Now there is no problem with running an iteration argument and, having found a solution to this new system, it turns out to be straightforward to show that we must in fact have $P^A = \mathcal{L}_{\partial_v}b^A$.  Hence solutions to the new system yield solutions to the original system.

\subsection{Main Result}
Now we are ready for the main result of this section.

\begin{proposition}\label{kljljl12}Let $H_5$ and $H_6$ be functions defined on $(-1,0)\times \mathbb{S}^2$ which lie in the closure of the norms determined by the right hand sides of~\eqref{3om4om2o4o3}  and so that the corresponding norms are sufficiently small. Then there exists a solution $b$ to the system
\begin{align}\label{2jk32ij23uhi2iu}
\mathcal{P}_{\ell \geq 1}\slashed{\nabla}_A\left(-\frac{1}{4}\Omega^{-2}\mathcal{L}_{\partial_v}b^A\right) -\frac{1}{4} \mathcal{P}_{\ell \geq 1}\left(\Omega^{-1}{\rm tr}\chi \slashed{\rm div}b\right) &+ \mathcal{P}_{\ell \geq 1}\left( \left(v+1\right)^{-2}\mathring{\Delta}\slashed{\rm div}b\right) 
\\ \nonumber &= \mathcal{P}_{\ell \geq 1}\left(H_5 + H_6\right),
\\ \nonumber  \qquad \mathring{\rm curl}b &= \mathfrak{Y},
\end{align}
so that we have 
\begin{align}\label{3om4om2o4o3}
& \left\vert\left\vert \mathring{\Pi}_{\rm div}b\right\vert\right\vert_{\mathscr{Q}\left(N_1,-3/2+\check{\delta},-\kappa\right)} + \sum_{j=0}^1\left\vert\left\vert \mathcal{L}_{\partial_v}^{1+j} \mathring{\Pi}_{\rm div}b\right\vert\right\vert_{\mathscr{Q}\left(N_1-1-j,-1/2+\check{\delta}+j,\kappa +j \right)} 
 + \left\vert\left\vert \mathring{\Pi}_{\rm div}b\right\vert\right\vert_{\mathscr{B}_1\left(\kappa\right)}
 \\ \nonumber &\qquad +\left\vert\left\vert \slashed{\rm div}b \right\vert\right\vert_{\mathscr{Q}\left(N_1-1,-1/2+\check{\delta},-1/2+\check{\delta}\right)} \lesssim \mathscr{C}_1\mathscr{C}_2 + \left\vert\left\vert \left((-v)^{-2\kappa}H_5,H_6\right)\right\vert\right\vert_{\mathscr{P}\mathscr{R}\left(\Omega,\kappa,50\check{p},50\check{p}\right)} \doteq \mathscr{D}.
\end{align}
\end{proposition}
\begin{proof} As usual, we will obtain our solution by an iteration process. We define a sequence $\left\{\Theta^{(i)}\right\}_{i=0}^{\infty}$, $\left\{\left(P^A\right)^{(i)}\right\}_{i=0}^{\infty}$,  and $\left\{\left(b^A\right)^{(i)}\right\}_{i=0}^{\infty}$ of scalar functions and $\mathbb{S}^2_{-1,v}$ vector fields for $v \in (-1,0)$ as follows. Set $P^{(0)} = \Theta^{(0)}  = b^{(0)} = 0$. For $i \geq 1$, we require that the following equations hold:
\begin{equation}\label{3in4in2i}
\mathcal{P}_{\ell \geq 1}\slashed{\rm div}b^{(i)} = \Theta^{(i)},\qquad \mathring{\rm curl}b^{(i)} = \mathfrak{Y},
\end{equation}
\begin{align}\label{2mom3omo5}
&\mathcal{P}_{\ell \geq 1}\slashed{\nabla}_A\left(-\frac{1}{4}\Omega^{-2}\left(P^A\right)^{(i)}\right) = \xi\Bigg[\frac{1}{4}\mathcal{P}_{\ell \geq 1}\left(\Omega^{-1}{\rm tr}\chi \slashed{\rm div}b^{(i)}\right) \\ \nonumber &\qquad + \mathcal{P}_{\ell \geq 1}\left(H_6 + H_5\right) - \mathcal{P}_{\ell \geq 1}\left(\left(v+1\right)^{-2}\mathring{\Delta}\slashed{\rm div}b^{(i)}\right)\Bigg],
\end{align}
\[\mathring{\rm curl}\left(P^{(i)}\right) =  \xi \mathfrak{Y},\]
\begin{align}\label{2o4ji3ij3ijr4o23}
&-\frac{1}{4}\mathcal{L}_{\partial_v}\Theta^{(i)}-\frac{1}{4}\left(2(v+1)^{-1}\Theta^{(i)}\right)+ \mathcal{P}_{\ell \geq 1}\left(\Omega^2\mathcal{P}_{\ell \geq 1}\left(v+1\right)^{-2}\mathring{\Delta}\Theta^{(i)}\right)  =\frac{1}{4}\mathcal{P}_{\ell \geq 1}\left(\Omega{\rm tr}\chi \mathcal{P}_{\ell = 0}\slashed{\rm div}b^{(i-1)}\right)
\\ \nonumber &\qquad +\mathcal{P}_{\ell \geq 1}\left( \left[\mathcal{P}_{\ell \geq 1},\Omega^2\right]\mathcal{P}_{\ell = 0}\left(\left(\frac{1}{4}\slashed{\nabla}_A\left(\Omega^{-2}\left( (P^A)^{(i-1)} + (1-\xi)\left(\mathcal{L}_{\partial_v}b^A\right)^{(i-1)}\right)\right)-\frac{1}{4}\left(\Omega^{-1}{\rm tr}\chi \slashed{\rm div}b^{(i-1)}\right)\right) \right)\right)
\\ \nonumber &\qquad  -\mathcal{P}_{\ell \geq 1}\left(\frac{1}{2}\slashed{\nabla}_A\log\Omega \left(\left(P^A\right)^{(i-1)}+\left(1-\xi\right)\mathcal{L}_{\partial_v}\left(b^A\right)^{(i-1)}\right)\right)-\frac{1}{4}\mathcal{P}_{\ell \geq 1}\mathcal{L}_{b^{(i-1)}}\left(\Omega{\rm tr}\chi\right)
\\ \nonumber &\qquad +\mathcal{P}_{\ell \geq 1}\left(\Omega^2\left(H_6 + H_5\right)\right)+\frac{1}{4}\mathcal{P}_{\ell \geq 1}\left(\left(\Omega{\rm tr}\chi -2(v+1)^{-1}\right)\Theta^{(i-1)}\right),
\end{align}
\begin{equation}\label{2pk32943}
\mathcal{P}_{\ell = 0}\Theta^{(i)} = 0.
\end{equation}
We note that the final equation~\eqref{2o4ji3ij3ijr4o23} is derived by multiplying through the equation with $\Omega^2$, then applying $\mathcal{P}_{\ell \geq 1}$, using~\eqref{3in4in2i}, and also the commutation formulas from Lemma~\ref{othercommutelemma}.\footnote{It is also useful to keep in mind that if $A$ and $B$ are two functions on $\mathbb{S}^2$ with $\mathcal{P}_{\ell = 0}B = 0$, then a short calculation shows that the equation 
\[\mathcal{P}_{\ell \geq 1}\left(\Omega^{-2}A\right) = B,\]
implies (and is, in fact, equivalent to)
\[\mathcal{P}_{\ell \geq 1}A + \left[\Omega^2,\mathcal{P}_{\ell \geq 1}\right]\mathcal{P}_{\ell = 0}\left(\Omega^{-2}A\right) = \mathcal{P}_{\ell \geq 1}\left(\Omega^2B\right).\]}

We will now prove by induction on $i$ that the sequences $\{\Theta^{(i)},P^{(i)},b^{(i)}\}_{i=0}^{\infty}$ are well-defined and moreover satisfy the following estimates:
\begin{align}\label{2om4om2o4}
& \sum_{j=0}^1\left[\left\vert\left\vert \mathcal{L}_{\partial_v}^j\mathring{\Pi}_{\rm div}P^{(i)}\right\vert\right\vert^2_{\mathscr{Q}\left(N_1-2-j,0,0,-\sqrt{p}+j\right)} + \left\vert\left\vert \mathcal{L}_{\partial_v}^j\mathring{\Pi}_{\rm div}P^{(i)}\right\vert\right\vert^2_{\check{\mathscr{S}}\left(N_2-1-j,0,50\check{p}\left(1+j\right)+j+2\kappa,50\check{p}\right)}\right]
\\ \nonumber &\qquad+ \left\vert\left\vert \Theta^{(i)}\right\vert\right\vert^2_{\mathscr{P}\mathscr{L}\left(\Omega,\kappa,100\check{p},50\check{p}\right)}  \leq C_{\rm boot}\mathscr{D}.
\end{align}
for a suitable bootstrap constant $C_{\rm boot}$. All implied constants which follow in this proof should be understood to be independent of $C_{\rm boot}$. 

We start by observing that $b^{(i)}$ itself is not present on the left hand side of~\eqref{2om4om2o4}. However, we may recover $b^{(i)}$ from $\Theta^{(i)}$ via elliptic theory on $\mathbb{S}^2$, commutation with $\mathcal{L}_{\partial_v}$, the fact that  the right hand side of formula~\eqref{2ijoij1oi2} does not involve $\Omega\hat{\chi}$, that $\left\vert\left\vert \slashed{g}\right\vert\right\vert_{\mathscr{A}_2\left(\kappa,\tilde{b},\Omega\right)}$ provides improved estimates for $\Omega{\rm tr}\chi$ at the highest level of angular derivatives relative to $\Omega\chi$, and the relation~\eqref{3in4in2i}. In particular, if $\Theta^{(i)}$ exists and satisfies~\eqref{2om4om2o4}, it is straightforward to obtain that there exists $b^{(i)}$ which satisfies~\eqref{3in4in2i} and moreover satisfies 
\begin{align}\label{2o3om2o3}
&\left\vert\left\vert b^{(i)}\right\vert\right\vert_{\mathscr{Q}\left(N_1,-3/2+\check{\delta},-\kappa\right)} + \sum_{j=0}^1\left\vert\left\vert \mathcal{L}_{\partial_v}^{1+j} b^{(i)}\right\vert\right\vert_{\mathscr{Q}\left(N_1-1-j,-1/2+\check{\delta}+j,\kappa +j \right)} 
+\left\vert\left\vert \left(1-\xi\right) b^{(i)}\right\vert\right\vert_{\mathscr{B}_1\left(\kappa\right)} \\ \nonumber &\qquad  + + \sum_{j=0}^1\left\vert\left\vert \left(v\mathcal{L}_{\partial_v}\right)^j\mathcal{L}_{\partial_v}\mathring{\nabla}b^{(i)}\right\vert\right\vert_{\check{\mathscr{S}}_{-1/2}^0\left(N_2-2-j,0,100\check{p}+50\check{p}j+2\kappa,50\check{p}\right)}\lesssim 
\mathscr{C}_2+ \left\vert\left\vert \Theta^{(i)}\right\vert\right\vert_{\mathscr{P}\mathscr{L}\left(\Omega,\kappa,100\check{p},50\check{p}\right)}.
\end{align}
We note two important facts about~\eqref{2o3om2o3}: We have not written $\left\vert\left\vert b^{(i)}\right\vert\right\vert_{\mathscr{B}_1\left(\kappa\right)}$ on the left hand side because we cannot obtain from~\eqref{3in4in2i} and available estimates for $\Theta^{(i)}$ the correct $L^{\infty}_vL^2\left(\mathbb{S}^2\right)$ estimate  $\mathcal{L}_{\partial_v}b$ if there is not an angular derivative applied to $\mathcal{L}_{\partial_v}b$. (We will remedy this defect at the end of the proof.) The second point is that there no $\mathscr{C}_1$ multiplying the term $\mathscr{C}_2$ on the right hand side of~\eqref{2o3om2o3}.

We now start the induction argument.  The base case $i =0$ is immediate so we assume that $\{\Theta^{(i)},P^{(i)},b^{(i)}\}_{i=0}^{j-1}$ have been shown to exist and that~\eqref{2om4om2o4} holds for $0 \leq i \leq j-1$. The estimates for $\Theta^{(i)}$ are obtained by applying Proposition~\ref{2km2om1o} to~\eqref{2o4ji3ij3ijr4o23}.  When we apply Proposition~\ref{2km2om1o} we treat all the terms on the right hand side of~\eqref{2o4ji3ij3ijr4o23} except the one involving $H_6$ as an ``$H_1$'' term. The desired estimate for $\Theta^{(i)}$ then follows from a straightforward analysis of each nonlinear term. Similarly, the desired estimates for $P^{(j)}$ are immediate consequences of Lemma~\ref{om3om2} and a straightforward analysis of the nonlinear terms (keeping in mind Remark~\ref{3o2ojio42}).

We may then run a compactness argument to find a limit along a subsequence $  \left(P^{(i)},\Theta^{(i)},b^{(i)}\right)\to \left(P,\Theta,b\right)$ which solves~\eqref{3in4in2i}-\eqref{2pk32943} with each $\left(P^{(i)},\Theta^{(i)},b^{(i)}\right)$ and $\left(P^{(i-1)},\Theta^{(i-1)},b^{(i-1)}\right)$ replaced by $\left(P,\Theta,b\right)$. We will now argue that we can extract from this system a solution to the original equation~\eqref{2jk32ij23uhi2iu}. First of all, we may multiply~\eqref{2mom3omo5} with $\Omega^2$ and derive the following equation:
\begin{align}\label{1n2knkn4k12}
&\mathcal{P}_{\ell \geq 1}\left(\Omega^2\slashed{\nabla}_A\left(-\frac{1}{4}\Omega^{-2}P^A\right)\right) + \left[\Omega^2,\mathcal{P}_{\ell \geq 1}\right]\mathcal{P}_{\ell = 0}\left(\slashed{\nabla}_A\left(-\frac{1}{4}\Omega^{-2}\left(P^A\right)\right)\right) =
\\ \nonumber &\qquad  \xi\Bigg[\frac{1}{4}\mathcal{P}_{\ell \geq 1}\left(\Omega{\rm tr}\chi \slashed{\rm div}b\right)  -\mathcal{P}_{\ell \geq 1}\left(\Omega^2\left(v+1\right)^{-2}\mathring{\Delta}\slashed{\rm div}b\right)\Bigg]
 + \frac{1}{4}\xi \left[\Omega^2,\mathcal{P}_{\ell \geq 1}\right]\mathcal{P}_{\ell = 0}\left(\Omega^{-1}{\rm tr}\chi \slashed{\rm div}b\right)
\\ \nonumber &\qquad + \xi\mathcal{P}_{\ell \geq 1}\left(\Omega^2H_6 + \Omega^2H_5\right).
\end{align}
Next, we observe that the following identity holds:
\begin{equation}\label{2om3mo3}
\mathcal{P}_{\ell \geq 1}\left(\frac{1}{2}\left(\slashed{\nabla}_A\log\Omega\right) P^A\right) = \mathcal{P}_{\ell \geq 1}\left(-\frac{1}{4}\Omega^2\slashed{\nabla}_A\left(\Omega^{-2}P^A\right) + \frac{1}{4}\slashed{\rm div}P\right).
\end{equation}
Now we may combine~\eqref{2om3mo3},~\eqref{1n2knkn4k12},~\eqref{2o4ji3ij3ijr4o23}, and Lemma~\ref{othercommutelemma} to obtain
\begin{align}\label{2m2omoo3}
&-\frac{1}{4}\mathcal{P}_{\ell \geq 1}\left(\slashed{\nabla}_A\mathcal{L}_{\partial_v}b^A-\slashed{\rm div}P\right)-\left(1-\xi\right)\frac{1}{4}\mathcal{P}_{\ell \geq 1}\left(\Omega{\rm tr}\chi\slashed{\rm div}b\right)+ \left(1-\xi\right)\mathcal{P}_{\ell \geq 1}\left(\Omega^2\mathcal{P}_{\ell \geq 1}\left(v+1\right)^{-2}\mathring{\Delta}\slashed{\rm div}b\right)  =
\\ \nonumber &\qquad +\mathcal{P}_{\ell \geq 1}\left( \left[\mathcal{P}_{\ell \geq 1},\Omega^2\right]\mathcal{P}_{\ell = 0}\left(\left(\left(\frac{1}{4}\slashed{\nabla}_A\left(\Omega^{-2} (1-\xi)\left(\mathcal{L}_{\partial_v}b^A\right)\right)-\left(1-\xi\right)\frac{1}{4}\left(\Omega^{-1}{\rm tr}\chi \slashed{\rm div}b\right)\right) \right)\right)\right)
\\ \nonumber &\qquad  -\mathcal{P}_{\ell \geq 1}\left(\frac{1}{2}\left(\slashed{\nabla}_A\log\Omega\right)\left(\left(1-\xi\right)\mathcal{L}_{\partial_v}\left(b^A\right)\right)\right)+\left(1-\xi\right)\mathcal{P}_{\ell \geq 1}\left(\Omega^2\left(H_6 + H_5\right)\right).
\end{align}
In particular, when $v \geq -1/4$ we have that $1-\xi = 0$ and we immediately obtain by elliptic estimates that $P = \xi \mathcal{L}_{\partial_v}b$. Now let $\mathscr{E}$ denote the difference between the left hand side and right hand side of~\eqref{2jk32ij23uhi2iu}. We may factor out $\Omega^2$ from~\eqref{2m2omoo3} (which is the reverse of the procedure by which~\eqref{2o4ji3ij3ijr4o23} was obtained) to achieve
\begin{align}\label{32o3o4in4}
-\frac{1}{4}\mathcal{P}_{\ell \geq 1}\left(\xi\slashed{\nabla}_A\mathcal{L}_{\partial_v}b^A-\slashed{\rm div}P\right)+\left(1-\xi\right)\left(\Omega^2\mathscr{E} + \left[\mathcal{P}_{\ell \geq 1},\Omega^2\right]\mathscr{E} \right) = 0.
\end{align}
On the other hand, it is immediate from~\eqref{2mom3omo5} that   
\begin{align}\label{2oj32oj3o3}
\mathcal{P}_{\ell \geq 1}\slashed{\nabla}_A\left(-\frac{1}{4}\Omega^{-2}\left( \xi\mathcal{L}_{\partial_v}b^A-P^A\right)\right) = \xi \mathscr{E},\qquad \mathring{\Pi}_{\rm curl}P = \mathring{\Pi}_{\rm curl}\mathcal{L}_{\partial_v}b.
\end{align}

Keeping in mind that we already know that $\xi\mathcal{L}_{\partial_v}b^A-P^A$ vanishes for $v \geq -1/4$, it is now straightforward to use~\eqref{32o3o4in4} and~\eqref{2oj32oj3o3} to conclude that $\xi\mathcal{L}_{\partial_v}b^A-P^A = \mathscr{E} = 0$ everywhere.

Lastly, it is now also straightforward to obtain~\eqref{3om4om2o4o3} from~\eqref{2om4om2o4} and the fact that $P^A = \xi\mathcal{L}_{\partial_v}b^A$.
\end{proof}

The actual equation we use to solve for $\mathring{\rm div}b$ will include terms on the right hand side will couple linearly to the lapse. It will be convenient to already include these terms already in this section's analysis:
\begin{proposition}\label{ij2ojo2432}
Let $H_4$, $H_5$ and $H_6$ be functions defined on $(-1,0)\times \mathbb{S}^2$ which lie in the closure of smooth functions under the norm on the right hand sides of~\eqref{3om4om2o4o32} and have a sufficiently small norm. Then there exists a solution $b$ to the system
\begin{align}\label{3lkl2jllk32lk}
&\mathcal{P}_{\ell \geq 1}\slashed{\nabla}_A\left(-\frac{1}{4}\Omega^{-2}\mathcal{L}_{\partial_v}b^A\right) -\frac{1}{4} \mathcal{P}_{\ell \geq 1}\left(\Omega^{-1}{\rm tr}\chi \slashed{\rm div}b\right) + \mathcal{P}_{\ell \geq 1}\left( \left(v+1\right)^{-2}\mathring{\Delta}\slashed{\rm div}b\right) 
\\ \nonumber &= \mathcal{P}_{\ell \geq 1}\Bigg(\left(H_5 -\slashed{\Delta}\log\Omega_{\rm boun}-2Y\right)  + H_6 + \tilde{H}_6\Bigg),
\\ \nonumber  \qquad \mathring{\rm curl}b &= \mathfrak{Y},
\end{align}
where $Y$ and $\tilde{H}_6$ are defined as in Lemma~\ref{2ij23ji3ijo32}. Then we have 
\begin{align}\label{3om4om2o4o32}
& \left\vert\left\vert \mathring{\Pi}_{\rm div}b\right\vert\right\vert_{\mathscr{Q}\left(N_1,-2,-\kappa\right)} + \sum_{j=0}^1\left\vert\left\vert \mathcal{L}_{\partial_v}^{1+j} \mathring{\Pi}_{\rm div}b\right\vert\right\vert_{\mathscr{Q}\left(N_1-1-j,-1+j,\kappa +j \right)} 
\\ \nonumber &\qquad + \left\vert\left\vert \mathring{\Pi}_{\rm div}b\right\vert\right\vert_{\mathscr{B}_1\left(\kappa\right)} +\left\vert\left\vert \slashed{\rm div}b \right\vert\right\vert_{\mathscr{Q}\left(N_1-1,-1,-1/2+\check{\delta}\right)} \lesssim \mathscr{D}
\\ \nonumber &\qquad \qquad  + \left\vert\left\vert H_4\right\vert\right\vert_{\mathscr{Q}\left(N_1-2,1/2+5\check{\delta},-\kappa\right)}
+ \left\vert\left\vert H_4\right\vert\right\vert_{\mathscr{S}_{-1}^{-1/2}\left(N_1-3,1+5\check{\delta},0\right)}+ \left\vert\left\vert H_4\right\vert\right\vert_{\mathscr{S}_{-1/2}^0\left(N_2-2,0,0\right)}.
\\ \nonumber &\qquad \qquad +\left\vert\left\vert \log\Omega_{\rm boun}\right\vert\right\vert_{\mathscr{A}\left(\kappa,b\right)} + \left\vert\left\vert \log\Omega_{\rm boun}\right\vert\right\vert_{\mathscr{B}_{01}\left(\kappa,b\right)} + \left\vert\left\vert \slashed{g} \right\vert\right\vert^2_{\mathscr{A}_2\left(\kappa,b,\Omega\right)} + \left\vert\left\vert \slashed{g}\right\vert\right\vert^2_{\mathscr{B}_2\left(\kappa,b,\Omega\right)} +  \left\vert\left\vert \log\Omega_{\rm sing}\right\vert\right\vert_{\mathscr{B}_{00}\left(\kappa\right)}^2,
\end{align}
where $\mathscr{D}$ is as in~\eqref{3om4om2o4o3}.
\end{proposition}
\begin{proof}This follows by an iteration argument and successively applying Proposition~\ref{kljljl12} and Lemma~\ref{2ij23ji3ijo32}. 
\end{proof}
\section{Solving for $\mathring{\Pi}_{\rm curl}b$ and the $\left(\mathfrak{o},\mathfrak{w}\right)$-Boundary System}\label{2oj3o}
In this section we will provide the ingredients by which we will eventually solve for $\mathring{\Pi}_{\rm curl}b$ and for the variables $\left(\mathfrak{o},\mathfrak{w}\right)$ which were used to determine boundary conditions at $\{v = 0\}$ in Section~\ref{ij3jr9j3}.

Throughout this section we will let $\Omega_{\rm sing}(v) : (-1,0) \to (0,\infty)$ be a given spherically symmetric function and $\pi_A$ be an $\mathbb{S}^2_{-1,v}$ vector field for $v\in (-1,0)$ satisfying $\left(1-\mathcal{P}_{\ell \leq \ell_0}\right)\pi = 0$. We will assume that, for a suitable constant $\kappa$ satisfying $\left|\kappa\right| \lesssim \epsilon$, we have
\begin{align}\label{wji3ij32ijo}
& \left\vert\left\vert \log\Omega_{\rm sing}\right\vert\right\vert_{\mathscr{B}_{00}\left(\kappa\right)}  + \left\vert\left\vert \pi\right\vert\right\vert_{\mathscr{S}\left(0,0,0\right)} \lesssim \epsilon.
\end{align}

We will also assume that we have two functions $T_{\rm low}: \mathbb{S}^2 \to \mathbb{R}$ and $T_{\rm high}: \mathbb{S}^2 \to \mathbb{R}$ which satisfy  $\left(1-\mathcal{P}_{1 \leq \ell \leq \ell_0}\right)T_{\rm low} = 0$, $\left(1-\mathcal{P}_{\ell > \ell_0}\right)T_{\rm high} = 0$, and 
\begin{equation}\label{3o1p9u390u1}
\sum_{j=0}^{N_1-3}\int_{\mathbb{S}^2}\left|\mathring{\nabla}^j\left(T_{\rm low},T_{\rm high}\right)\right|_{\mathring{\slashed{g}}}^2\mathring{\rm dVol} \lesssim 1.
\end{equation}
We emphasize that none of the results in this section depend on the implied constants in~\eqref{wji3ij32ijo},~\eqref{3o1p9u390u1}, or in the inequality for $\kappa$ (though by our conventions for $\epsilon$, we may assume that $\epsilon$ is sufficiently small depending on the implied constants).

\subsection{Overview of Solving for $\mathring{\Pi}_{\rm curl}b$}\label{3ij2oj2}
We start with an overview of the general strategy for solving for $\mathring{\Pi}_{\rm curl}b$. Our starting point is the equation~\eqref{thewavezetastartsojqj}. Dropping the Ricci curvature terms, restricting to $\{u=-1\}$, and denoting various nonlinear terms by $H$, we have 
\begin{align}\label{2kn32inioi2o3}
&(-v)\nabla_{\partial_v}\zeta_A - \frac{1}{2}\mathcal{L}_b\zeta_A + \left(1+3(v+1)^{-1}(-v)\Omega^2\right)\zeta_A  -\frac{1}{4}\slashed{\rm div}\left(\slashed{\nabla}\hat{\otimes}b\right)_A   +\frac{1}{4}\slashed{\nabla}_A\slashed{\rm div}b= H.
\end{align}
Keeping in mind that 
\begin{equation}\label{jijiojiojoi3joi23r221}
\zeta_A = -\frac{1}{4}\Omega^{-2}\slashed{g}_{AB}\mathcal{L}_{\partial_v}b^B,
\end{equation}
we may consider the linearization of the left hand side as a second order equation for $b$ (at least after multiplying through by $\Omega^2$). Unfortunately, we cannot use this equation to solve for all of $b$ for two reasons: First of all, the operator $-\frac{1}{4}\slashed{\rm div}\slashed{\nabla}\hat{\otimes} + \frac{1}{4}\slashed{\nabla}_A\slashed{\rm div}$ is not an elliptic differential operator along $\mathbb{S}^2$, and hence we cannot expect to use the model second order equation theory. Second of all, there are nonlinear terms in $H$ which involve second derivatives of the metric. More specifically, these terms are of the form $\mathcal{L}_b\slashed{\nabla}\log\Omega$. Thus, even if we had a left hand side which was amenable to our model second order equation theory, we would still have a derivative loss problem due to the nonlinear terms. 

If we apply $\slashed{\rm curl}$  to both sides of~\eqref{2kn32inioi2o3}, then, at the cost of only having an equation for  $\slashed{\rm curl}b$, it at first appears that we  have cured the two above problems. (Of course, it is not a problem to only control $\slashed{\rm curl}b$ as we have already a separate scheme for solving for $\slashed{\rm div}b$, see Section~\ref{k2m3mo492}.) Indeed, the application of $\slashed{\rm curl}$ will annihilate $\slashed{\nabla}_A\slashed{\rm div}b$ and the principle symbol of $\slashed{\rm curl}\slashed{\rm div}\slashed{\nabla}\hat{\otimes}$ is the same as the principle symbol of $\slashed{\Delta}\slashed{\rm curl}$ (see~\eqref{2m3momo2}), and $\slashed{\rm curl}\mathcal{L}_b\slashed{\nabla}\log\Omega$ may be controlled by an expression still only involving two derivatives of the metric (since $\slashed{\rm curl}\slashed{\nabla}\log\Omega = 0$). However, there are still three significant complications which we will now explain:
\begin{enumerate}
	\item The first issue concerns a still present nonlinear derivative loss, even after the application of $\slashed{\rm curl}$. Namely, if we inspect the full formula~\eqref{2m3momo2}, then we see that an application of $\slashed{\rm curl}$ will produce a nonlinear term proportional to $\left(\slashed{\nabla}K\right)\wedge b$.\footnote{We note that these type of nonlinear terms are only produced because $b$ is a tensor. In particular, when we studied the lapse $\Omega$ in Section~\ref{lapthesection}, this particular type of nonlinear difficulty was absent.} This involves three derivatives of $\slashed{g}$ and hence will produce a fatal derivative loss if we try to naively estimate it. The resolution is as follows:
	\begin{enumerate}
	\item This nonlinear term is not a problem for the ``bounded frequency'' part of $\slashed{\rm curl}b$, that is, $\mathcal{P}_{1 \leq \ell \leq \ell_0}\slashed{\rm curl}b$. Thus, we will solve for $\mathcal{P}_{1\leq \ell \leq \ell_0}\slashed{\rm curl}b$ by applying $\mathcal{P}_{1\leq \ell \leq \ell_0}\slashed{\rm curl}$ to~\eqref{2kn32inioi2o3} and treating the resulting equation as a model second order equation of type $II$ for the $\ell \geq 2$ part and as an ordinary differential equation for $\ell = 1$. 
	\item In order to solve for $\mathcal{P}_{\ell > \ell_0}\slashed{\rm curl}b$, we use a modification of the equation~\eqref{2kn32inioi2o3}. Namely, we first commute~\eqref{2kn32inioi2o3} with $\slashed{\nabla}\hat{\otimes}$ and use Lemma~\ref{emfkeo3} to simply the resulting expression (with the $\widehat{\rm Ric}$ term set to $0$). The use of Lemma~\ref{emfkeo3} allows us to avoid the previously described nonlinear derivative loss except for a term proportional to $\mathcal{L}_b\slashed{\nabla}\hat{\otimes}\slashed{\nabla}\log\Omega$. Then we apply $\left(v+1\right)^2\mathcal{P}_{\ell > \ell_0}\slashed{\rm curl}\slashed{\rm div}$, which annihilates $\mathcal{L}_b\slashed{\nabla}\hat{\otimes}\slashed{\nabla}\log\Omega$ to leading order. Keeping in mind the formula (derived from~\eqref{2m3momo2})
	\[\slashed{\rm curl}\slashed{\rm div}\slashed{\nabla}\hat{\otimes}\left(\slashed{\rm div}\left(\slashed{\nabla}\hat{\otimes}b\right)\right) = \left(\slashed{\Delta} + 2K\right)\slashed{\rm curl}\slashed{\rm div}\left(\slashed{\nabla}\hat{\otimes}b\right) + 2 \left(\slashed{\nabla}K\right)\wedge \slashed{\rm div}\left(\slashed{\nabla}\hat{\otimes}b\right),\]
	we see that we may consider the resulting equation as model second order equation of type $III$ for $\left(v+1\right)^2\mathcal{P}_{\ell > \ell_0}\slashed{\rm curl}\slashed{\rm div}\slashed{\nabla}\hat{\otimes}b$. Finally, we observe that given that we assume in this section that $\mathring{\rm div}b$ is already known, we will be able to recover $b$ from $\mathcal{P}_{\ell = 1}\slashed{\rm curl}b$, $\mathcal{P}_{2 \leq \ell \leq \ell_0}\slashed{\rm curl}b$ and $\mathcal{P}_{\ell > \ell_0}\left(v+1\right)^2\slashed{\rm curl}\slashed{\rm div}\slashed{\nabla}\hat{\otimes}b$ using elliptic theory along $\mathbb{S}^2$.
	\end{enumerate}
	\item Our commutation process described above produces also a nonlinear obstacle which is now of a ``low-frequency'' nature. Namely, in view of the formula~\eqref{jijiojiojoi3joi23r221}, when we apply $\slashed{\nabla}\otimes$ or $\slashed{\rm curl}$ to $\zeta$ and expand in order to see explicitly a term proportional to $\mathcal{L}_{\partial_v}\slashed{\rm curl}b$ or $\mathcal{L}_{\partial_v}\slashed{\nabla}\hat{\otimes}b$ we will produce a nonlinear term which involves a contraction of $\slashed{\nabla}\log\Omega$ and $\mathcal{L}_{\partial_v}b$. When $|v| \gtrsim 1$, the nonlinear analysis of this term is straightforward; however, when $|v| \ll 1$, then due to the singular behavior of the lapse as $v\to 0$, we cannot hope to treat this term in a purely perturbative fashion. An analogous difficulty occurred in Section~\ref{k2m3mo492} whenever we desired to expand the expression $\slashed{\rm div}\left(\Omega^{-2}\mathcal{L}_{\partial_v}b\right)$ (see the discussion specifically in Section~\ref{2o2o4mo292904nfj3k34i2}), and the resolution of the difficulty there will have similarities to the resolution in this section. For definiteness, let us consider the case of $\mathcal{P}_{2 \leq \ell \leq \ell_0}\slashed{\rm curl}b$. As in Section~\ref{2o2o4mo292904nfj3k34i2} the key idea is to replace the $\mathcal{L}_{\partial_v}b$ in the potentially problematic nonlinear term with an artificial variable $P$ and, for $P$, use the original equation to directly obtain an estimate for $\mathcal{P}_{2\leq \ell \leq \ell_0}\slashed{\rm curl}\left(\Omega^{-2}P\right)$ and then use the elliptic/transport estimates of Lemma~\ref{32m2omo4} to recover estimates directly for $P$. Thus when the nonlinear term containing $P$ is encountered later, $P$ has ``already been estimated'' and we will be able to use both the specific nonlinear structure of the terms containing $P$ and the interpolation trick of Lemma~\ref{3m2omo4}  to close our estimates.
	\item\label{3i3i2o2} The final main nonlinear difficulty involves yet another term produced when we apply Lemma~\ref{emfkeo3}. Namely, there will appear terms proportional to $\mathcal{L}_b\left(\Omega\hat{\chi}\right)$. At first it appears natural to treat this as a nonlinear term and consider it to be schematically of the form $\epsilon \cdot \slashed{\nabla}\hat{\chi}$.\footnote{Moreover, if we want to use an estimate for $\mathcal{L}_{\partial_v}\slashed{\nabla}\hat{\otimes}b$ to obtain an estimate for $\slashed{\nabla}\hat{\otimes}\mathcal{L}_{\partial_v}b$, then the corresponding commutator will also produce terms which, to leading order, are proportional to $\mathcal{L}_b\left(\Omega\hat{\chi}\right)$.} The problem with this is that at the highest level of derivatives our bootstrap assumptions only yield that $v^{1/2}\left(\Omega\hat{\chi}\right)^{(\alpha)}$ for $|\alpha| = N_1-1$ lies in $L^2_v$ for $|v| \ll 1$. This would in turn prevent us from establishing the estimates (at the highest level of derivatives) that we desire for $\mathcal{P}_{\ell > \ell_0}\slashed{\rm curl}\slashed{\rm div}\slashed{\nabla}\hat{\otimes}b$. The resolution is as follows: When we first obtain the existence of $b$, we weaken our estimates at the top order. (More specifically, we will control $\slashed{\rm curl}b$ in the norm $\left\vert\left\vert \slashed{\rm curl}b\right\vert\right\vert_{R,III,N_1-1}$ instead of $\left\vert\left\vert \slashed{\rm curl}b\right\vert\right\vert_{R,III',N_1-1}$, see Definition~\ref{i32ij3o990oinkijijo} and also the difference between~\eqref{2pk300jj0} and~\eqref{2pk300jj0123}.) Fortunately, in Section~\ref{ij3jr9j3} we have arranged for our procedure which solves for $\slashed{g}$, and hence $\hat{\chi}$, to only require this weaker control of $b$ (see the specific norms of $b$ used in~\eqref{3o2omoo2010injnn}). This then allows us to close an iteration argument which ends up solving for both $b$ and $\slashed{g}$ at the same time. Finally, at the end we can use that our estimates for $\slashed{g}$ from Section~\ref{ij3jr9j3} imply that $\mathcal{L}_b\left(\Omega\hat{\chi}\right)$ satisfies a stronger estimate than a generic angular derivative of $\Omega\hat{\chi}$ does. Using this \emph{linear} estimate for $\mathcal{L}_b\left(\Omega\hat{\chi}\right)$ we can then revisit the equations for $b$ and finally establish the stronger estimate we would like.

\end{enumerate}
\subsection{Overview of Solving for $\left(\mathfrak{o},\mathfrak{w}\right)$}
We now turn to the boundary variables $\mathfrak{o}$ and $\mathfrak{w}$. We start by recalling the role that $\mathfrak{o}$ and $\mathfrak{w}$ play in the context of solving the $\left(\slashed{g},\mathfrak{n}\right)$-system. We solve for $\mathcal{P}_{\ell > \ell_0}\slashed{\rm curl}\slashed{\rm div}\hat{\chi}$ indirectly by deriving a second order equation for the curl of artificial variable $\mathcal{P}_{\ell > \ell_0}\mathfrak{n}$ which after the entire iteration argument is closed will be shown to equal $\eta$. The function $\mathfrak{w}$ is used for the boundary condition $\mathcal{P}_{\ell > \ell_0}\slashed{\rm curl}\mathfrak{n}|_{v=0}$. We solve for $\mathcal{P}_{\ell > \ell_0}\slashed{\rm div}\slashed{\rm div}\hat{\chi}$ indirectly by deriving a second order equation satisfied by the Gauss curvature $\mathcal{P}_{\ell > \ell_0}K$ of the metric $\slashed{g}$. The function $\mathfrak{o}$ is used for the boundary condition $\mathcal{P}_{\ell > \ell_0}K|_{v=0}$.

The value of the function $\mathfrak{w}$ must be consistent with our expectation that $\{v = 0\}$ boundary values of $\slashed{\rm curl}\mathfrak{n}$ is equal to $\slashed{\rm curl}\eta$ in the final solution. It would thus appear natural, in the context of iteration, to update the value of $\slashed{\rm curl}\mathfrak{n}$ to simply be the $v\to 0$ limit of the current iterate corresponding to $\slashed{\rm curl}\eta$ (recall that when we solve the $\left(\slashed{g},\mathfrak{n}\right)$ system we expect linear coupling with the shift $b$ and lapse $\Omega$). However, it turns out to be difficult to directly implement such a scheme without encountering a problematic derivative loss.\footnote{More specifically, the equations which are convenient to use when we solve for $\slashed{\rm div}b$ in the bulk of the spacetime do not allow, on their own, for us to obtain sharp estimates from the point of view of regularity for $\slashed{\rm div}\eta$ at $\{v = 0\}$. If we had access to all of the equations of the double-null gauge in our iteration scheme this would not be a problem.} Instead, we introduce an artificial unknown $\mathfrak{j}$ to stand-in for $\eta|_{v=0}$, and derive an alternative set of equations for $\mathfrak{j}$. We derive the desired equation by taking Lemma~\ref{2mo2o3o2} and setting $\{v = 0\}$. We would like to use this equation to solve for an artificial variable $\mathfrak{j}$, and then use $\mathfrak{j}$ to suitably determine the boundary value of $\mathcal{P}_{\ell > \ell_0}\slashed{\rm curl}\mathfrak{n}$ which in turn determined $\mathfrak{w}$. This leads to
\begin{align}\label{1kj2ono}
&- \mathcal{L}_b\mathfrak{j}_A +\mathfrak{j}_A\left(2-\slashed{\rm div}b\right)=  \frac{1}{2}\slashed{\nabla}^B\left(\slashed{\nabla}\hat{\otimes}b\right)_{AB} - \frac{1}{2}\slashed{\nabla}_A\slashed{\rm div}b,
\end{align}
where we have dropped the term involving $\slashed{\nabla}\left(\Omega\underline{\omega}\right)$, because for our solutions we will have that $\Omega\underline{\omega}|_{v=0}$ is a spherically symmetric function. We now note two potential issues with~\eqref{1kj2ono}. Namely, the right hand side of~\eqref{1kj2ono} involves two derivatives of $b$ and also (via Christoffel symbols) two derivatives of $\slashed{g}$. In the context of our bootstrap argument, we can expect to control $b|_{v=0}$ in $\mathring{H}^{N_1-1}$ and $\slashed{g}|_{v=0}$ in $\mathring{H}^{N_1-1}$. This suggests that the best estimate we could hope for $\mathfrak{j}$ is to control $\left(\mathfrak{j},\mathcal{L}_b\mathfrak{j}\right) \in \mathring{H}^{N_1-3}$ and hence $\left(\mathfrak{w},\mathcal{L}_b\mathfrak{w}\right) \in \mathring{H}^{N_1-4}$. Unfortunately, see~\eqref{knefiomomo2}, this is a derivative lower than what we need to close our arguments. To overcome this potential derivative loss we need to exploit the specific structure on the right hand side of~\eqref{1kj2ono}. We start by separately apply $\mathcal{P}_{\ell \geq 1}\slashed{\rm div}$ and $\mathcal{P}_{\ell \geq 1}\slashed{\rm curl}$ to~\eqref{1kj2ono}. We obtain, using~\eqref{3o3oioi4}, that   
  \begin{align}\label{1i2niji4}
&\mathcal{P}_{\ell \geq 1}\left( -\mathcal{L}_b\mathcal{P}_{\ell \geq 1}\slashed{\rm div}\mathfrak{j}+\left(2-\slashed{\rm div}b\right)\mathcal{P}_{\ell \geq 1}\slashed{\rm div}\mathfrak{j} +\left[\mathcal{P}_{\ell \geq 1}\slashed{\rm div}, -\mathcal{L}_b+\left(2-\slashed{\rm div}b\right)\right]\mathfrak{j}\right) = 
\\ \nonumber &\qquad \mathcal{P}_{\ell \geq 1}\left(\left(\mathcal{P}_{\ell \leq \ell_0}K + \mathfrak{o}\right)\slashed{\rm div}b  + \mathcal{L}_b\left(\mathcal{P}_{\ell \leq \ell_0}K + \mathfrak{o}\right)\right),
 \end{align}
 \begin{align}\label{1kj2n4o1o3}
 &\mathcal{P}_{\ell \geq 1}\left( -\mathcal{L}_b\mathcal{P}_{\ell \geq 1}\slashed{\rm curl}\mathfrak{j}+\left(2-\slashed{\rm div}b\right)\mathcal{P}_{\ell \geq 1}\slashed{\rm curl}\mathfrak{j} +\left[\mathcal{P}_{\ell \geq 1}\slashed{\rm curl}, -\mathcal{L}_b+\left(2-\slashed{\rm div}b\right)\right]\mathfrak{j}\right) = 
 \\ \nonumber &\qquad  \mathcal{P}_{1 \leq \ell \leq \ell_0}\left(\slashed{\nabla}\left(\mathcal{P}_{\ell < \ell_0}K + \mathfrak{o}\right)\wedge b\right)+\frac{1}{2}\epsilon\left(\slashed{\Delta}+2K\right)T_{\rm low}+ \frac{1}{2}\epsilon T_{\rm high}
+\frac{1}{2}\left[\mathcal{P}_{1 \leq \ell \leq \ell_0},\slashed{\Delta}+2K\right]\slashed{\rm curl}b.
 \end{align}
 Here we also used the definitions of $T_{\rm low}$ and $T_{\rm high}$ as well the fact that $\mathfrak{o}|_{v=0}$ is intended to be equal to $\mathcal{P}_{\ell > \ell_0}K$. Since $T_{\rm high}$ is simply part of our initial data, in~\eqref{1kj2n4o1o3} we no longer need to worry about a derivative loss impacting our iteration scheme. In~\eqref{1i2niji4} we see that the terms involving the largest number of derivatives applied to $b$ have cancelled! At the highest level of derivatives, on the right hand side we are left with the nonlinear term $\mathcal{L}_b\mathfrak{o}$. This term, unfortunately, is still worrisome from a regularity point of view. We now pause our discussion of this equation to describe the equation we will use to solve for $\mathfrak{o}$ on $\{v = 0\}$. 
 
 We take~\eqref{2momoo3}, restrict to $v = 0$, apply $\mathcal{P}_{\ell > \ell_0}$, substitute $\eta$ with $\mathfrak{j}$, and drop the Ricci curvature terms to obtain:
 \begin{align}\label{2n3knk2}
\mathcal{P}_{\ell > \ell_0}\slashed{\rm div}\mathfrak{j} -\mathfrak{o} &= \mathcal{P}_{\ell > \ell_0}\Bigg(\frac{1}{2}\left(-1+\slashed{\rm div}b  -4\Omega\underline{\omega}\right) \left(\Omega^{-1}{\rm tr}\chi-\frac{2}{v+1}\right) + \frac{1}{2}\mathcal{L}_b\left(\Omega^{-1}{\rm tr}\chi\right)
\\ \nonumber &\qquad 
 - \left|\mathfrak{j}\right|^2
 +\frac{1}{v+1}\left(\slashed{\rm div}b - 4\Omega\underline{\omega}\right)\Bigg).
   \end{align}
  We thus see that at the highest level of derivatives, $\mathfrak{o}$ is given by a linear combination of $\mathcal{P}_{\ell > \ell_0}\slashed{\rm div}\mathfrak{j}$ and $\mathcal{L}_b\left(\Omega^{-1}{\rm tr}\chi\right)$. More significantly, we may effectively combine~\eqref{2n3knk2} and~\eqref{1i2niji4} by applying  $\left(2-\mathcal{P}_{\ell > \ell_0}\slashed{\rm div}b\right) - \mathcal{P}_{\ell > \ell_0}\mathcal{L}_b$ to~\eqref{2n3knk2} and using~\eqref{1i2niji4} to simplify. We obtain
   \begin{align}\label{kl2ljk1lkj21jkl}
& \left(2-2\mathcal{P}_{\ell > \ell_0}\left(\mathcal{L}_b+\slashed{\rm div}b\right)\right)\mathfrak{o}  =
\left(2-\mathcal{P}_{\ell > \ell_0}\left(\mathcal{L}_b+\slashed{\rm div}b\right)\right)\mathcal{P}_{\ell > \ell_0}\mathscr{N} 
\\ \nonumber &\qquad + \mathcal{P}_{\ell > \ell_0}\left(\mathcal{L}_b\mathcal{P}_{\ell \leq \ell_0}\slashed{\rm div}\mathfrak{j} + \left(\slashed{\rm div}b -2\right)\mathcal{P}_{\ell \leq \ell_0}\slashed{\rm div}\mathfrak{j}\right)
		\\ \nonumber &\qquad + \mathcal{P}_{\ell > \ell_0}\left(\left[\mathcal{P}_{\ell \geq 1}\slashed{\rm div}, -\mathcal{L}_b+\left(2-\slashed{\rm div}b\right)\right]\mathfrak{j}\right)		
		+\mathcal{P}_{\ell > \ell_0}\left(\mathcal{P}_{\ell \leq \ell_0}K\slashed{\rm div}b  + \mathcal{L}_b\mathcal{P}_{\ell \leq \ell_0}K \right). \end{align}
where
\begin{equation}
\mathscr{N} = \frac{1}{2}(\Omega^{-1}{\rm tr}\chi-2(v+1)^{-1})\left(1+\left(4\Omega\underline{\omega}-\slashed{\rm div}b\right)\right)-\frac{1}{2}\mathcal{L}_b\left[\Omega^{-1}{\rm tr}{\rm tr}\chi\right]
	 - \left|\mathfrak{j}\right|^2.
\end{equation}
Now, to highest order, on the right hand side of~\eqref{kl2ljk1lkj21jkl} we see $\mathcal{L}_b^2\left(\Omega^{-1}{\rm tr}\chi\right)$ which we can expect to control in $\mathring{H}^{N_1-3}$ (here it is crucial that one of the derivatives is $\mathcal{L}_b$!). In turn this suggests that we can expect that $\mathfrak{o}$ and $\mathcal{L}_b\mathfrak{o}$ are in $\mathring{H}^{N_1-3}$, which is exactly what we need for our arguments.  Returning to~\eqref{1i2niji4} we then can expect that $\slashed{\rm div}\mathfrak{j}$ and $\mathcal{L}_b\slashed{\rm div}\mathfrak{j}$ lies in $\mathring{H}^{N_1-3}$. This is again sufficient for our argument (in fact it is a better estimate than what we need). 

The conclusion from this heuristic analysis is that it is reasonable for us to use the equations~\eqref{1i2niji4},~\eqref{1kj2n4o1o3}, and~\eqref{kl2ljk1lkj21jkl} to solve for $\mathfrak{j}$ and $\mathfrak{o}$ along $\{v = 0\}$, and then set $\mathfrak{w} \doteq \mathcal{P}_{\ell > \ell_0}\slashed{\rm curl}\mathfrak{j}$.

\subsection{Iterating}\label{22oiiojioj32}
In our first iteration, we will fix $\mathfrak{q}$, $\mathfrak{r}$, and $\mathring{\Pi}_{\rm div}b$, and show that we may then solve for $b$, $\slashed{g}$, and the lapse $\Omega$. 

\begin{proposition}\label{3ij2oij3ij2} Let $H_1$, $H_2$, $\mathfrak{Z}$, $\mathfrak{q}$  be functions on $(-1,0)\times\mathbb{S}^2$ so that $\left(1-\mathcal{P}_{\ell = 0}\right)\mathfrak{Z} = 0$, $\pi$, $H_7$, and $H_8$  be $\mathbb{S}^2_{-1,v}$ $1$-forms so that $\left(1-\mathcal{P}_{\ell \leq \ell_0}\right)\pi = 0$, and $H_9$ and $\mathfrak{r}$ be $\mathbb{S}^2_{-1,v}$  $(0,2)$-tensors for $v \in (-1,0)$   so that  $\left(1-\mathcal{P}_{\ell \geq 1}\right)\left(H_1,H_2\right) = 0$,  $H_2$ is supported for $v \geq -1/2$, so that
\begin{align*}
\left\vert\left\vert \left(\mathfrak{q},\mathfrak{r}\right)\right\vert\right\vert_{\mathscr{Q}\left(N_1-1,-1/2+\check{\delta},-1/2+\check{\delta}\right)} + \left\vert\left\vert \left(\mathcal{L}_{\partial_v}\mathfrak{q},\mathcal{L}_{\partial_v}\mathfrak{r}\right)\right\vert\right\vert_{\mathscr{Q}\left(N_1-2,1/2+\check{\delta},\kappa\right)} + 
\left\vert\left\vert \left(\mathfrak{q},\mathfrak{r}\right)\right\vert\right\vert_{\mathscr{S}\left(N_2-1,\check{\delta},0\right)} \lesssim \epsilon,
\end{align*}
\[\left\vert\left\vert \pi\right\vert\right\vert_{\mathscr{S}\left(0,\check{\delta},0\right)} +\left\vert\left\vert \pi\right\vert\right\vert_{\mathscr{Q}_{-1}^{-1/2}\left(0,-1/2+\check{\delta},0\right)} \lesssim \epsilon,\]
and so that  the following 
\begin{align*}
 &\mathcal{D} \doteq \left\vert\left\vert \left(H_1,H_2,0\right)\right\vert\right\vert_{R,I,N_1-2} + \epsilon \left\vert\left\vert \left(T_{\rm low},T_{\rm high}\right)\right\vert\right\vert_{\mathring{H}^{N_1-3}} +
 \left\vert\left\vert \left((-v)^{-2\kappa}\mathring{\rm curl}H_7,0,0\right)\right\vert\right\vert_{R,II,N_1-1}  \\ \nonumber &\qquad +\left\vert\left\vert \left((-v)^{-2\kappa}\mathring{\Pi}_{\rm curl}H_8,0,0\right)\right\vert\right\vert_{R,III'\left(1/2-\check{\delta},50\check{p},0\right),N_1}+\sum_{j=0}^1\left\vert\left\vert \left(v\mathcal{L}_{\partial_v}\right)^j\mathring{\Pi}_{\rm curl}H_7\right\vert\right\vert_{\check{\mathscr{S}}_{-1/2}^0\left(0,0,50\check{p}\left(1+j\right),0\right)}
 \\ \nonumber &\qquad +\left\vert\left\vert \left((-v)^{-2\kappa}H_9,0,0\right)\right\vert\right\vert_{R,III'\left(1/2-\check{\delta},50\check{p},100\check{p}\right),N_1-1}+\left\vert\left\vert \log\Omega_{\rm sing}\right\vert\right\vert_{\mathscr{B}_{00}\left(\kappa\right)}
 \\ \nonumber &\qquad + \sum_{\left|\alpha\right| \leq 1}\sum_{j=0}^1\left\vert\left\vert \left(v\mathcal{L}_{\partial_v}\right)^j\mathcal{L}_{\mathcal{Z}^{(\alpha)}}\mathring{\Delta}^{-1}H_9\right\vert\right\vert_{\check{\mathscr{S}}_{-1/2}^0\left(N_2-3-j,0,50\check{p}\left(1+j\right),50\check{p}\right)},
 \end{align*}
 \begin{align*}
&\mathcal{F}  \doteq 
 \left\vert\left\vert \left((-v)^{-2\kappa}\mathring{\rm div}H_7,0,0\right)\right\vert\right\vert_{R,II,N_1-1}  \\ \nonumber &\qquad +\left\vert\left\vert \left((-v)^{-2\kappa}\mathring{\Pi}_{\rm div}H_8,0,0\right)\right\vert\right\vert_{R,III'\left(1/2-\check{\delta},50\check{p},0\right),N_1}+\sum_{j=0}^1\left\vert\left\vert \left(v\mathcal{L}_{\partial_v}\right)^j\mathring{\Pi}_{\rm div}H_7\right\vert\right\vert_{\check{\mathscr{S}}_{-1/2}^0\left(0,0,50\check{p}\left(1+j\right),0\right)},
 \end{align*}
 \begin{align*}
&\mathcal{G} \doteq \left\vert\left\vert\mathring{\nabla}\mathfrak{z}\right\vert\right\vert_{\mathscr{Q}\left(N_1,-3/2+\check{\delta},-\kappa\right)} + \sum_{j=0}^1\left\vert\left\vert \mathcal{L}_{\partial_v}^{1+j} \mathring{\nabla}\mathfrak{z}\right\vert\right\vert_{\mathscr{Q}\left(N_1-1-j,-1/2+\check{\delta}+j,\kappa +j \right)} + \left\vert\left\vert \mathring{\nabla}\mathfrak{z}\right\vert\right\vert_{\mathscr{B}_1\left(\kappa\right)}, 
\end{align*}
\[\mathfrak{z} \doteq \mathring{\Delta}^{-1} \mathfrak{Z},\]
are finite and moreover satisfy that $\mathcal{D},\mathcal{F} \lesssim \epsilon$.

Then there exist a function $\Omega_{\rm boun}: (-1,0)\times \mathbb{S}^2 \to (0,\infty)$, an $\mathbb{S}^2_{-1,v}$-vector field $b^A$, an $\mathbb{S}^2_{-1,v}$-symmetric $(0,2)$ tensor $\slashed{g}$, and an $\mathbb{S}^2_{-1,v}$  $1$-form $\mathfrak{n}_A$  so that the following hold:
\begin{enumerate}
\item We set $\Omega \doteq \Omega_{\rm sing}\Omega_{\rm boun}$.
\item The following equations hold:
\begin{align}\label{w3ifio32oi}
&\mathcal{P}_{1 \leq \ell \leq \ell_0}\slashed{\rm curl}\left[(-v)\nabla_{\partial_v}\zeta_A - \frac{1}{2}\mathcal{L}_b\zeta_A + \left(1+3(v+1)^{-1}(-v)\Omega^2\right)\zeta_A  -\frac{1}{4}\slashed{\rm div}\left(\slashed{\nabla}\hat{\otimes}b\right)_A \right] = \\ \nonumber &\qquad \mathcal{P}_{1 \leq \ell \leq \ell_0}\slashed{\rm curl}\left[H_7-\frac{1}{2}\mathcal{L}_b\slashed{\nabla}\log\Omega\right],
\end{align}
\begin{align}\label{oi32ioj32iojioj23}
&-\frac{1}{4}\mathcal{P}_{\ell > \ell_0}(v+1)^2\slashed{\rm curl}\slashed{\rm div}\Bigg[\left((-v)\nabla_v - \mathcal{L}_b\right)\left(\Omega^{-2}\nabla_v\left(\slashed{\nabla}\hat{\otimes}b\right)\right) + \left(1 + \frac{4(-v)\Omega^2}{v+1}\right)\Omega^{-2}\nabla_v\left(\slashed{\nabla}\hat{\otimes}b\right) 
\\ \nonumber &\qquad +\left(1 + 4\left(v+1\right)^{-1}(-v)\Omega^2\right)\left(-2\Omega^{-2}\mathcal{L}_b\left(\Omega\hat{\chi}\right) +8\slashed{\nabla}\log\Omega\hat{\otimes}\zeta -4 \Omega^{-2}\mathcal{E}\left[\slashed{g},b\right]\right)
\\ \nonumber &\qquad + \slashed{\nabla}\hat{\otimes}\left[\slashed{\rm div}\left(\slashed{\nabla}\hat{\otimes}b\right) - \slashed{\nabla}\slashed{\rm div}b\right]+\left((-v)\nabla_v-\mathcal{L}_b\right)\left(8\slashed{\nabla}\log\Omega\hat{\otimes}\zeta - 4\Omega^{-2}\mathcal{E}\left[\slashed{g},b\right]\right)\Bigg] 
\\ \nonumber &\qquad =\mathcal{P}_{\ell > \ell_0}(v+1)^2\slashed{\rm curl}\slashed{\rm div}\left(\slashed{\nabla}\hat{\otimes} \left[H_8-\frac{1}{2}\mathcal{L}_b\slashed{\nabla}\log\Omega\right] -2\Omega^{-2}\mathcal{L}_b\left(\Omega^2 \slashed{\nabla}\hat{\otimes}\slashed{\nabla}\log\Omega\right)+H_9 \right),
\end{align}
where
\begin{align}\label{2io34iojijo}
\mathcal{E}\left[\slashed{g},b\right]_{AB} \doteq -\frac{1}{2}\slashed{\rm div}b \left(\Omega\hat{\chi}\right)_{AB} - \frac{1}{2}\left(\Omega\hat{\chi}\right)^C_{\ \ (A}\left(\slashed{\nabla}\hat{\otimes}b\right)_{B)C},
\end{align}

\item We have $\mathring{\rm div}b = \mathfrak{Z}$.
\item We have the boundary conditions
\[\mathcal{P}_{1 \leq \ell \leq \ell_0}\slashed{\rm curl}b|_{v=0} = \epsilon T_{\rm low},\qquad \mathcal{P}_{\ell > \ell_0}\slashed{\rm curl}\slashed{\rm div}\slashed{\nabla}\hat{\otimes}b|_{v=0} = \epsilon T_{\rm high}.\]
\item We have that $\mathfrak{o}$, $\mathfrak{w}$, and $\mathfrak{j}$ solve the equations~\eqref{1i2niji4},~\eqref{1kj2n4o1o3},  and ~\eqref{kl2ljk1lkj21jkl}, and also $\mathfrak{w} \doteq \mathcal{P}_{\ell > \ell_0}\slashed{\rm curl}\mathfrak{j}$.
\item  We have that $\left(\slashed{g},\mathfrak{n}\right)$ are the solutions produced by Proposition~\ref{solvetheprojected} where the quantities $\left(\Omega,b,\mathfrak{o},\mathfrak{w},\pi,\mathfrak{q},\mathfrak{r}\right)$ are all defined from the corresponding quantities of this section.
\item We have that $\Omega_{\rm boun}$ is the solution produced by Proposition~\ref{23ini2999jn2j23i3j} with right hand sides $H_1$ and $H_2$. 
\end{enumerate}

We then have the following estimates:
\begin{align}\label{3o3pkp4}
&\left\vert\left\vert \log\Omega_{\rm boun}\right\vert\right\vert_{\mathscr{A}\left(\kappa,b\right)} + \left\vert\left\vert \log\Omega_{\rm boun}\right\vert\right\vert_{\mathscr{B}_{01}\left(\kappa,b\right)}  +\left\vert\left\vert \mathring{\Pi}_{\rm curl}b\right\vert\right\vert_{\mathscr{A}^-_1\left(\kappa\right)}+\left\vert\left\vert \mathring{\Pi}_{\rm curl}b\right\vert\right\vert_{\mathscr{B}^-_1\left(\kappa\right)}
\\ \nonumber &\qquad + \left\vert\left\vert \left(\mathfrak{o},\mathcal{L}_{b|_{v=0}}\mathfrak{o}\right)\right\vert\right\vert_{\mathring{H}^{N_1-3}\left(\mathbb{S}^2\right)} + \left\vert\left\vert \left(\mathfrak{j},\mathcal{L}_{b|_{v=0}}\mathfrak{j}\right)\right\vert\right\vert_{\mathring{H}^{N_1-2}\left(\mathbb{S}^2\right)}+ \left\vert\left\vert \left(\mathfrak{w},\mathcal{L}_{b|_{v=0}}\mathfrak{w}\right)\right\vert\right\vert_{\mathring{H}^{N_1-3}\left(\mathbb{S}^2\right)}
 \lesssim
 \\ \nonumber &\qquad \qquad  \mathcal{D}+ \mathcal{F}^2 + \mathcal{G}^2 +\left\vert\left\vert \pi\right\vert\right\vert^2_{\mathscr{S}\left(0,\check{\delta},0\right)} +\left\vert\left\vert \pi\right\vert\right\vert^2_{\mathscr{Q}_{-1}^{-1/2}\left(0,-1/2+\check{\delta},0\right)} 
\end{align}
\begin{align}\label{3ij32ijo23io}
&\left\vert\left\vert \slashed{g} \right\vert\right\vert_{\mathscr{A}_2\left(\kappa,b,\Omega\right)} + \left\vert\left\vert \slashed{g}\right\vert\right\vert_{\mathscr{B}_2\left(\kappa,b,\Omega\right)}
+\left\vert\left\vert \left(v+1\right)\mathfrak{n}\right\vert\right\vert_{L,III\left(1/2-\check{\delta},100\check{p} N_1,0\right),N_1-1}+\left\vert\left\vert \slashed{g}|_{v=0}-\mathring{\slashed{g}}\right\vert\right\vert_{\mathring{H}^{N_1-1}} \lesssim 
\\ \nonumber &\qquad \mathcal{D} + \mathcal{F}^2+\mathcal{G} + \left\vert\left\vert \pi\right\vert\right\vert_{\mathscr{S}\left(0,\check{\delta},0\right)} +\left\vert\left\vert \pi\right\vert\right\vert_{\mathscr{Q}_{-1}^{-1/2}\left(0,-1/2+\check{\delta},0\right)} ,
\end{align}
\begin{align}\label{2k3lnl2n3}
&\left\vert\left\vert \mathcal{P}_{\ell > \ell_0}\slashed{g} \right\vert\right\vert_{\mathscr{A}_2\left(\kappa,b,\Omega\right)} \lesssim  \mathcal{D} + \mathcal{F}^2+\mathcal{G} + \left\vert\left\vert \pi\right\vert\right\vert^2_{\mathscr{S}\left(0,\check{\delta},0\right)} +\left\vert\left\vert \pi\right\vert\right\vert^2_{\mathscr{Q}_{-1}^{-1/2}\left(0,-1/2+\check{\delta},0\right)} ,
\end{align}
\begin{align}\label{3j23moo4}
\left\vert\left\vert \mathring{\Pi}_{\rm curl}b\right\vert\right\vert_{\mathscr{B}_1\left(\kappa\right)}\lesssim \mathcal{D}+ \mathcal{F}^2 + \mathcal{G}^2 + \left\vert\left\vert \pi\right\vert\right\vert^2_{\mathscr{S}\left(0,\check{\delta},0\right)} +\left\vert\left\vert \pi\right\vert\right\vert^2_{\mathscr{Q}_{-1}^{-1/2}\left(0,-1/2+\check{\delta},0\right)} ,
\end{align}
\begin{align}\label{io23io32io32oi}
&\sum_{j=0}^1\left[\left\vert\left\vert \slashed{\rm curl}\slashed{\rm div}\slashed{\nabla}\hat{\otimes}\mathcal{L}_{\partial_v}^jb \right\vert\right\vert_{\mathscr{Q}_{-1/2}^0\left(N_1-3-j,0,-1/2+\check{\delta}+j\right)} + \left\vert\left\vert \slashed{\rm curl}\slashed{\rm div}\slashed{\nabla}\hat{\otimes}\mathcal{L}^{1+j}_{\partial_v}b\right\vert\right\vert_{\mathscr{Q}_{-1/2}^0\left(N_1-4-j,0,\kappa+j\right)}\right]
\\ \nonumber &\qquad \lesssim \mathcal{D}+ \mathcal{F}^2 + \mathcal{G} +  \left\vert\left\vert \pi\right\vert\right\vert^2_{\mathscr{S}\left(0,\check{\delta},0\right)} +\left\vert\left\vert \pi\right\vert\right\vert^2_{\mathscr{Q}_{-1}^{-1/2}\left(0,-1/2+\check{\delta},0\right)} .
\end{align}
\begin{proof}As usual, we will use an iteration argument. We also introduce three artificial scalar unknowns $\mathfrak{t}$, $\mathfrak{p}$, and $\mathfrak{W}$, as well an artificial vector field unknown $P^A$ and an artificial $(0,2)$-tensor $Q_{AB}$. (We will see later that $\mathfrak{t}$ corresponds to $\mathcal{P}_{\ell = 1}\slashed{\rm curl}b$, $\mathfrak{p}$ corresponds to $\mathcal{P}_{2 \leq \ell \leq \ell_0}\slashed{\rm curl}b$, $\mathfrak{W}$ corresponds to $\mathcal{P}_{\ell > \ell_0}\slashed{\rm curl}\slashed{\rm div}\slashed{\nabla}\hat{\otimes}b$, $P$ corresponds to $\mathcal{L}_{\partial_v}b$, and $Q$ corresponds to $\nabla_v\left(\slashed{\nabla}\hat{\otimes}b\right)$.) 

We will show inductively that the following sequence of iterates 
\[\{\Omega^{(i)},\Omega_{\rm boun}^{(i)},\mathfrak{t}^{(i)},\mathfrak{p}^{(i)},\mathfrak{W}^{(i)},P^{(i)},Q^{(i)},b^{(i)},\slashed{g}^{(i)},\mathfrak{n}^{(i)},\mathfrak{j}^{(i)},\mathfrak{w}^{(i)},\mathfrak{o}^{(j)}\}_{i=0}^{\infty}\] 
is well defined. For $i = 0$ we set
\[\mathfrak{t}^{(0)} = \mathfrak{p}^{(0)} = \mathfrak{W}^{(0)} = 0,\qquad \Omega^{(0)} = \Omega_{\rm sing},\qquad \Omega^{(0)}_{\rm boun} = 1,\qquad \mathfrak{o}^{(0)} = \mathfrak{w}^{(0)} = 0,\]
\[P^{(0)} = b^{(0)} =\mathfrak{n}^{(0)} = 0,\qquad Q^{(0)} = 0,\qquad \slashed{g}^{(0)} = \left(v+1\right)^2\mathring{\slashed{g}},\qquad \mathfrak{j}^{(0)} = 0.\]
For $i > 0$, we require that the following equations hold:
\begin{enumerate}
	\item For $\mathfrak{W}^{(i)}$ we solve 
	\begin{align}\label{2o2oo14829}
	&\left((-v)\mathcal{L}_{\partial_v}-\mathcal{P}_{\ell > \ell_0}\mathcal{L}_b\right)\mathcal{L}_{\partial_v}\mathfrak{W}^{(i)} 
	\\ \nonumber &\qquad + \left(1+\frac{4(-v)\mathcal{P}_{\ell > \ell_0}\Omega^2}{v+1}\right)\mathcal{L}_{\partial_v}\mathfrak{W}^{(i)} + \mathcal{P}_{\ell > \ell_0}\Omega^2\left(\slashed{\Delta}+2\left(v+1\right)^{-2}\right)\mathfrak{W}^{(i)} = 
	\\ \nonumber &\qquad -\mathcal{P}_{\ell > \ell_0}\Omega^2\left[\mathcal{P}_{\ell > \ell_0}(v+1)^2\slashed{\rm curl}\slashed{\rm div},\left((-v)\nabla_v - \mathcal{L}_b\right)\Omega^{-2}\right]Q
	\\ \nonumber &\qquad -\mathcal{P}_{\ell  > \ell_0}\Omega^2\left(\left((-v)\nabla_v - \mathcal{L}_b\right)\Omega^{-2}+ \left(1 + \frac{4(-v)}{v+1}\right)\right)\left[\mathcal{P}_{\ell > \ell_0}(v+1)^2\slashed{\rm curl}\slashed{\rm div},\nabla_v\right]\slashed{\nabla}\hat{\otimes}b
	\\ \nonumber &\qquad +2\mathcal{P}_{\ell > \ell_0}\left(\left((-v)\mathcal{L}_{\partial_v}-\mathcal{L}_b\right)\log\Omega\right)\mathcal{L}_{\partial_v}\mathfrak{W} + 2\mathcal{P}_{\ell > \ell_0}\Omega^2\mathcal{P}_{\ell > \ell_0}\left(v+1\right)^2\slashed{\nabla}K\wedge \slashed{\nabla}\slashed{\rm div}b
	\\ \nonumber &\qquad +\left(v+1\right)^2\mathcal{P}_{\ell > \ell_0}\Omega^2\mathcal{P}_{\ell > \ell_0}\left(2\left(K-(v+1)^{-2}\right)\slashed{\rm curl}\slashed{\rm div}\slashed{\nabla}\hat{\otimes}b + 2\slashed{\nabla}K\wedge \slashed{\rm div}\left(\slashed{\nabla}\hat{\otimes}b\right)\right)
	\\ \nonumber &\qquad -\mathcal{P}_{\ell > \ell_0}\Omega^2\mathcal{P}_{\ell > \ell_0}\left(v+1\right)^2\slashed{\rm curl}\slashed{\rm div}\left(\left(\Omega^{-2} + 4\left(v+1\right)^{-1}(-v)\right)\left(-2\mathcal{L}_b\left(\Omega\hat{\chi}\right) -2\slashed{\nabla}\log\Omega\hat{\otimes}P -4\mathcal{E}\right)\right)
	\\ \nonumber &\qquad + \left(v+1\right)^2\mathcal{P}_{\ell > \ell_0}\Omega^2\left[\mathcal{P}_{\ell > \ell_0}, \left(\slashed{\Delta} + 2(v+1)^{-2}\right)\right]\slashed{\rm curl}\slashed{\rm div}\slashed{\nabla}\hat{\otimes}b
	\\ \nonumber &\qquad -\mathcal{P}_{\ell > \ell_0}\Omega^2\mathcal{P}_{\ell > \ell_0}\left(v+1\right)^2\slashed{\rm curl}\slashed{\rm div}\left(\left((-v)\nabla_v-\mathcal{L}_b\right)\left(8\Omega^{-2}\slashed{\nabla}\log\Omega\hat{\otimes}P - 4\Omega^{-2}\mathcal{E}\right) \right)
	\\ \nonumber &\qquad +\mathcal{P}_{\ell > \ell_0}\Omega^2\mathcal{P}_{\ell > \ell_0}(v+1)^2\slashed{\rm curl}\slashed{\rm div}\slashed{\nabla}\left(\hat{\otimes}\left[H_8-\frac{1}{2}\mathcal{L}_b\slashed{\nabla}\log\Omega\right]+H_9\right),
	\end{align}
	where other than the already indicated $\mathfrak{W}^{(i)}$, we always use the $i-1$st iterate to define the various quantities. We moreover pose the boundary condition that
	\begin{equation}\label{23ji3joi}
	\mathfrak{W}^{(i)}|_{v=0} = \epsilon T_{\rm high}.
	\end{equation}
	\item For $\mathfrak{p}^{(i)}$ we solve
	\begin{align}\label{3oij3jio4tio39i}
	&\left((-v)\mathcal{L}_{\partial_v} - \frac{1}{2}\mathcal{P}_{2\leq \ell \leq \ell_0}\mathcal{L}_b\right)\mathcal{L}_{\partial_v}\mathfrak{p}^{(i)} 
	\\ \nonumber &\qquad + \left(1 + \frac{4(-v)\mathcal{P}_{2\leq \ell \leq \ell_0} \Omega^2}{v+1}\right)\mathcal{L}_{\partial_v}\mathfrak{p}^{(i)} + \mathcal{P}_{2 \leq \ell \leq \ell_0} \Omega^2\left(\slashed{\Delta} + 2\left(v+1\right)^{-2}\right)\mathfrak{p}^{(i)} = 
	\\ \nonumber &\qquad -\mathcal{P}_{2 \leq \ell \leq \ell_0}\Omega^2\left(\left[\mathcal{P}_{2\leq \ell \leq \ell_0}\slashed{\rm curl},\left((-v)\nabla_v - \frac{1}{2}\mathcal{L}_b\right)\Omega^{-2}+ \left(1 + \frac{3(-v)}{v+1}\right)\right] - \frac{(-v)}{v+1}\right)P
	\\ \nonumber &\qquad -\mathcal{P}_{2 \leq \ell \leq \ell_0}\Omega^2\left(\left((-v)\nabla_v - \frac{1}{2}\mathcal{L}_b\right)\Omega^{-2}+ \left(1 + \frac{4(-v)}{v+1}\right)\right)\left[\mathcal{P}_{2 \leq \ell \leq \ell_0}\slashed{\rm curl},\mathcal{L}_{\partial_v}\right]b	
	\\ \nonumber &\qquad +2\mathcal{P}_{2 \leq \ell \leq \ell_0}\left(\left((-v)\mathcal{L}_{\partial_v}-\frac{1}{2}\mathcal{L}_b\right)\log\Omega\right)\mathcal{L}_{\partial_v}\mathfrak{p} 
	\\ \nonumber &\qquad +\mathcal{P}_{2 \leq \ell \leq \ell_0}\Omega^2\mathcal{P}_{2 \leq \ell \leq \ell_0}\left(2\left(K-(v+1)^{-2}\right)\slashed{\rm curl}b + 2\slashed{\nabla}K\wedge b\right)
	\\ \nonumber &\qquad + \mathcal{P}_{2 \leq \ell \leq  \ell_0}\Omega^2\left[\mathcal{P}_{2 \leq \ell \leq \ell_0}, \left(\slashed{\Delta} + 2(v+1)^{-2}\right)\right]\slashed{\rm curl}b
	\\ \nonumber &\qquad +\mathcal{P}_{2 \leq \ell \leq \ell_0}\Omega^2 \mathcal{P}_{2\leq \ell \leq \ell_0}\slashed{\rm curl}\left[H_7-\frac{1}{2}\mathcal{L}_b\slashed{\nabla}\log\Omega\right],
	\end{align}
	where, except for the already indicated $\mathfrak{p}^{(i)}$,  we always use the $i-1$st iterate to define the various quantities. We moreover pose the boundary condition that
	\begin{equation}\label{2joijoij23}
	\mathfrak{p}^{(i)}|_{v=0} = \epsilon T_{\rm low}.
	\end{equation}
		\item For $\mathfrak{t}^{(i)}$ we have
		\begin{align}\label{2o2o293fm3o2}
		&\left((-v)\nabla_v +\left(1+4(v+1)^{-1}(-v)\mathcal{P}_{\ell = 1}\Omega^2\right)\right)\mathcal{L}_{\partial_v}\mathfrak{t}^{(i)} = 
		\\ \nonumber &\qquad \mathcal{P}_{\ell = 1}\Omega^2\left(\left[\left((-v)\nabla_v+3(v+1)^{-1}(-v)\Omega^2\right)\Omega^{-2},\mathcal{P}_{\ell = 1}\slashed{\rm curl}\right] -(-v)(v+1)^{-1}\right)P^{(i-1)}
		\\ \nonumber &\qquad -\left[\mathcal{P}_{\ell = 1}\Omega^2,(-v)\nabla_v\right]\Omega^{-2}\slashed{\rm curl}P^{(i-1)}+\mathcal{P}_{\ell = 1}\Omega^2\mathcal{P}_{\ell = 1}\slashed{\rm curl}\Bigg( \frac{1}{2}\mathcal{L}_b\left(\Omega^{-2}P^{(i-1)}\right) -\slashed{\rm div}\left(\slashed{\nabla}\hat{\otimes}b\right)_A+ H_7\Bigg),
		\end{align}
		where we impose the boundary condition that $\mathcal{L}_{\partial_v}\mathfrak{t}^{(i)}$  vanishes at $v = -1$ and that $\mathfrak{t}^{(i)}|_{v = 0} = \mathcal{P}_{\ell = 1}T_{\rm low}$. Other than the already indicated $\mathfrak{t}^{(i)}$, we use the $i-1$st iterate to define all of the relevant quantities.
	\item To determine $b^{(i)}$ we use elliptic theory along $\mathbb{S}^2$ and require that the following hold:
	\begin{equation}\label{2om3om92}
	\left(v+1\right)^2\mathcal{P}_{\ell > \ell_0}\slashed{\rm curl}\slashed{\rm div}\slashed{\nabla}\hat{\otimes} b^{(i)} = \mathfrak{W}^{(i)},\ \mathcal{P}_{2 \leq \ell \leq \ell_0}\slashed{\rm curl}b^{(i)} = \mathfrak{p}^{(i)},\ \mathcal{P}_{\ell = 1}\slashed{\rm curl}b^{(i)} = \mathfrak{t}^{(i)},\ \mathring{\Pi}_{\rm div}b^{(i)} = \mathring{\nabla}\mathfrak{z}.
	\end{equation}
	Here we use $\slashed{g}^{(i-1)}$ to define these various operators on $\mathbb{S}^2$.
	\item For $P^{(i)}$, we set $P^{(i)} = \mathscr{P}^{(i)} + \left(1-\xi\right)\mathcal{L}_{\partial_v}b$ for an unknown $\mathscr{P}^{(i)}$. For $\mathscr{P}^{(i)}$ we require that the following equations hold:
	\begin{align}\label{kn1kln21lk}
&\mathcal{P}_{1 \leq \ell \leq \ell_0}\slashed{\rm curl}\Bigg[(-v)\nabla_{\partial_v}\left(-\frac{1}{4}\Omega^{-2}\mathscr{P}^{(i)}\right)_A - \frac{1}{2}\mathcal{L}_b\left(-\frac{1}{4}\Omega^{-2}\mathscr{P}^{(i)}\right)_A - (-v)\xi'\left(-\frac{1}{4}\Omega^{-2}\mathcal{L}_{\partial_v}b\right)_A
\\ \nonumber &\qquad + \left(-\frac{1}{4}\Omega^{-2}\mathscr{P}^{(i)}\right)_A+3(v+1)^{-1}(-v)\Omega^2\left(-\frac{1}{4}\Omega^{-2}\mathcal{L}_{\partial_v}b\right)_A
  -\frac{1}{4}\xi \slashed{\rm div}\left(\slashed{\nabla}\hat{\otimes}b\right)_A \Bigg] = 
  \\ \nonumber &\qquad \xi \mathcal{P}_{1\leq \ell \leq \ell_0}\slashed{\rm curl} \left[H_7-\frac{1}{2}\mathcal{L}_b\slashed{\nabla}\log\Omega\right],
\end{align}
\begin{align}\label{ui2iu21iu12io}
&\mathcal{P}_{\ell > \ell_0}(v+1)^2\slashed{\rm curl}\slashed{\rm div}\Bigg[\left((-v)\nabla_v - \mathcal{L}_{b^{(i-1)}}\right)\slashed{\nabla}\hat{\otimes}\left(\Omega^{-2}\mathscr{P}^{(i)}\right) +\slashed{\nabla}\hat{\otimes}\left(\Omega^{-2}\mathscr{P}^{(i)}\right) 
\\ \nonumber &\qquad - (-v)\xi'\slashed{\nabla}\hat{\otimes}\left(\Omega^{-2}\mathcal{L}_{\partial_v}b\right) 
+\frac{4\xi(-v)\Omega^2}{v+1}\slashed{\nabla}\hat{\otimes}\left(\Omega^{-2}\mathcal{L}_{\partial_v}b\right)
\\ \nonumber &\qquad  + \xi\slashed{\nabla}\hat{\otimes}\left[\slashed{\rm div}\left(\slashed{\nabla}\hat{\otimes}b\right) - \xi\slashed{\nabla}\slashed{\rm div}b\right]+2\left((-v)\nabla_v-\mathcal{L}_{b^{(i-1)}}\right)\left(\xi\mathcal{L}_b\left(\Omega\hat{\chi}\right)\right) -2\xi \mathcal{E}_2\left[\slashed{g},b,\mathfrak{n}\right]\Bigg] 
\\ \nonumber &\qquad = \xi\mathcal{P}_{\ell > \ell_0}(v+1)^2\slashed{\rm curl}\slashed{\rm div}\left[\slashed{\nabla}\hat{\otimes} \left[H_8-\frac{1}{2}\mathcal{L}_b\slashed{\nabla}\log\Omega\right]+H_9\right]
\end{align}
\begin{equation}\label{2oj3oij2oo}
\mathring{\Pi}_{\rm div}\mathscr{P}^{(i)} = \xi\mathring{\nabla}\mathcal{L}_{\partial_v}\mathfrak{z}.
\end{equation}
(We note that Lemma~\ref{32m2omo4} has been specifically designed to solve this system.) Unless otherwise indicated, we use in the two above equations, $b^{(i)}$, $\slashed{g}^{(i-1)}$, and $\Omega^{(i-1)}$. 
\item For $Q^{(i)}$, we set
\begin{equation}\label{32m23o2}
Q^{(i)}_{AB} \doteq \slashed{\nabla}\hat{\otimes}P_{AB} + 2\mathcal{L}_b\left(\Omega\hat{\chi}\right)_{AB} - 2\slashed{\rm div}b\left(\Omega\hat{\chi}\right)_{AB} - 2\left(\Omega\hat{\chi}\right)^C_{\ \ (A}\left(\slashed{\nabla}\hat{\otimes}b\right)_{B)C},
\end{equation}
where on the right hand side we use $P^{(i)}$, $b^{(i)}$, and $\slashed{g}^{(i-1)}$. 
	\item We obtain $\Omega_{\rm boun}^{(i)}$ by applying Proposition~\ref{23ini2999jn2j23i3j} with $\slashed{g} = \slashed{g}^{(i-1)}$, $b = b^{(i)}$, and the right hand side $H_1$ and $H_2$.  Then we set $\Omega^{(i)} = \Omega_{\rm sing}\Omega^{(i)}_{\rm boun}$. 
	\item We solve for $\mathfrak{o}^{(i)}$ and $\mathfrak{j}^{(i)}$ by solving
	  \begin{align}\label{ijio2joij32}
&\mathcal{P}_{\ell \geq 1}\left( -\mathcal{L}_b\mathcal{P}_{\ell \geq 1}\slashed{\rm div}\mathfrak{j}^{(i)}+\left(2-\slashed{\rm div}b\right)\mathcal{P}_{\ell \geq 1}\slashed{\rm div}\mathfrak{j}^{(i)} +\left[\mathcal{P}_{\ell \geq 1}\slashed{\rm div}, -\mathcal{L}_b+\left(2-\slashed{\rm div}b\right)\right]\mathfrak{j}\right) = 
\\ \nonumber &\qquad \mathcal{P}_{\ell \geq 1}\left(\left(\mathcal{P}_{\ell \leq \ell_0}K + \mathfrak{o}\right)\slashed{\rm div}b  + \mathcal{L}_b\left(\mathcal{P}_{\ell \leq \ell_0}K + \mathfrak{o}\right)\right),
 \end{align}
 \begin{align}\label{12ij1iojoi12}
 &\mathcal{P}_{\ell \geq 1}\left( -\mathcal{L}_b\mathcal{P}_{\ell \geq 1}\slashed{\rm curl}\mathfrak{j}^{(i)}+\left(2-\slashed{\rm div}b\right)\mathcal{P}_{\ell \geq 1}\slashed{\rm curl}\mathfrak{j}^{(i)} +\left[\mathcal{P}_{\ell \geq 1}\slashed{\rm curl}, -\mathcal{L}_b+\left(2-\slashed{\rm div}b\right)\right]\mathfrak{j}\right) = 
 \\ \nonumber &\qquad \frac{1}{2}\mathcal{P}_{\ell = 1}\left(\left(\slashed{\Delta}+2K\right)\slashed{\rm curl}b\right) + \mathcal{P}_{1 \leq \ell \leq \ell_0}\left(\slashed{\nabla}\left(\mathcal{P}_{\ell < \ell_0}K + \mathfrak{o}\right)\wedge b\right)+\frac{1}{2}\epsilon\left(\slashed{\Delta}+2K\right)T_{\rm low}+ \frac{1}{2}\epsilon T_{\rm high}
 \\ \nonumber &\qquad +\frac{1}{2}\left[\mathcal{P}_{2 \leq \ell \leq \ell_0},\slashed{\Delta}+2K\right]\slashed{\rm curl}b, \end{align}
   \begin{align}\label{1ipo939209}
& \left(2-2\mathcal{P}_{\ell > \ell_0}\left(\mathcal{L}_b+\slashed{\rm div}b\right)\right)\mathfrak{o}^{(i)}  =
\left(2-\mathcal{P}_{\ell > \ell_0}\left(\mathcal{L}_b+\slashed{\rm div}b\right)\right)\mathcal{P}_{\ell > \ell_0}\mathscr{N} 
		\\ \nonumber &\qquad + \mathcal{P}_{\ell > \ell_0}\left(\left[2-\mathcal{P}_{\ell > \ell_0}\left(\mathcal{L}_b+\slashed{\rm div}b\right),\mathcal{P}_{\ell > \ell_0}\slashed{\rm div}\right] \mathfrak{j}\right)		
		+\mathcal{P}_{\ell > \ell_0}\left(\mathcal{P}_{\ell \leq \ell_0}K\slashed{\rm div}b  + \mathcal{L}_b\mathcal{P}_{\ell \leq \ell_0}K \right), \end{align}
where
\begin{equation}
\mathscr{N} \doteq \frac{1}{2}\mathfrak{A}^{(i)}|_{v=0}\left(1+\left(4\Omega\underline{\omega}-\slashed{\rm div}b\right)\right)-\frac{1}{2}\mathcal{L}_{b^{(i-1)}}\mathfrak{A}^{(i)}
	 - \left|\mathfrak{j}\right|^2,
\end{equation}
we define $\mathfrak{A}^{(i)}$ by solving 
\begin{align}\label{2lkkjl4k2}
&\mathcal{L}_{\partial_v}\mathfrak{A}^{(i)} + 2(v+1)^{-1}\Omega^2\mathfrak{A}^{(i)} = 
 -\Omega^{-2}\left|\left(\Omega\hat{\chi}\right)\right|^2+2\left(1-\left(\Omega^{(i)}\right)^2\right)(v+1)^{-2} - \frac{1}{2}\Omega^2\left(\mathfrak{A}^{(i)}\right)^2,
\end{align}
with the boundary condition that $\mathfrak{A}^{(i)}$ is bounded at $v= -1$ and, unless otherwise indicated, we use $\slashed{g}^{(i-1)}$, $b^{(i)}$, $\mathfrak{j}^{(i-1)}$, $\mathfrak{o}^{(i-1)}$, and $\Omega^{(i-1)}$. We then set
\[\mathfrak{w}^{(i)} \doteq \mathcal{P}_{\ell > \ell_0}\slashed{\rm curl}\mathfrak{j}^{(i)}.\]
We note that once that the iterates have converged, it will follow immediately that $\mathfrak{A} = \left(\Omega^{-1}{\rm tr}\chi - 2(v+1)^{-1}\right)$. It is convenient, however, to use $\mathfrak{A}^{(i)}$ when we run our iteration argument so that $\mathfrak{A}^{(i)}$ linearly couples to $\Omega^{(i)}$.
	\item We obtain $\left(\slashed{g}^{(i)},\mathfrak{n}^{(i)}\right)$ by applying Proposition~\ref{solvetheprojected} where we replace the use of $b$ with $b^{(i)}$, the use of $\Omega$ with $\Omega^{(i)}$, and the use of $\left(\mathfrak{w},\mathfrak{o}\right)$ with $\left(\mathfrak{w}^{(i)},\mathfrak{o}^{(i)}\right)$. 
\end{enumerate}

We also prove by induction that the following bounds hold for a suitable bootstrap constant $C_{\rm boot}$:
\begin{align}\label{l1kjkjljk12kl3}
&\left\vert\left\vert \log\Omega^{(i)}_{\rm boun}\right\vert\right\vert_{\mathscr{A}\left(\kappa,b^{(i)}\right)} + \left\vert\left\vert \log\Omega^{(i)}_{\rm boun}\right\vert\right\vert_{\mathscr{B}_{01}\left(\kappa,b^{(i)}\right)} +\left\vert\left\vert \mathfrak{W}^{(i)}\right\vert\right\vert_{L(b^{(i-1)}),III\left(1/2-\check{\delta},50\check{p},100\check{p}\right),N_1-3}
\\ \nonumber &\qquad + \left\vert\left\vert \mathfrak{p}^{(i)}\right\vert\right\vert_{L,II,N_1-1} + \sum_{j=0}^1\left\vert\left\vert \mathcal{L}^{1+j}_{\partial_v}\mathfrak{t}^{(i)}\right\vert\right\vert_{\check{\mathscr{S}}_{-1}^0\left(0,1+\check{\delta}+j,100\check{p}+j,0\right)} + \left\vert\left\vert \mathfrak{t}^{(i)}\right\vert\right\vert_{\mathscr{S}_{-1}^0\left(0,\check{\delta},0\right)} 
\\ \nonumber &\qquad + \sum_{j=0}^2 \left\vert\left\vert \mathcal{L}_{\partial_v}^j\mathfrak{t}^{(i)}\right\vert\right\vert_{\mathscr{Q}_{-1}^{-1/2}\left(0,-1/2+\check{\delta} + j,0\right)}
 +\left\vert\left\vert \mathring{\Pi}_{\rm curl}b^{(i)}\right\vert\right\vert_{\mathscr{A}^-_1\left(b^{(i-1)},\kappa\right)}+\left\vert\left\vert \mathring{\Pi}_{\rm curl}b^{(i)}\right\vert\right\vert_{\mathscr{B}^-_1\left(\kappa\right)} 
\\ \nonumber &\qquad + \sum_{j=0}^1\left\vert\left\vert \left(v\mathcal{L}_{\partial_v}\right)^j\mathring{\Pi}_{\rm curl}\mathscr{P}^{(i)}\right\vert\right\vert_{\check{\mathscr{S}}\left(N_2-1-j,0,50\check{p}\left(1+j\right)+2\kappa,50\check{p}\right)} 
\\ \nonumber &\qquad + \left\vert\left\vert \mathring{\Pi}_{\rm curl}\mathscr{P}^{(i)}\right\vert\right\vert_{W_{-1/2}^0\left(N_1-1,0,1/2\right)} + \left\vert\left\vert \left(1,v\mathcal{L}_{\partial_v},\mathcal{L}_{b^{(i)}}\right)\mathring{\Pi}_{\rm curl}\mathscr{P}^{(i)}\right\vert\right\vert_{W_{-1/2}^0\left(N_1-2,0,2\kappa\right)}
\\ \nonumber &\qquad + \left\vert\left\vert \left(\mathfrak{o}^{(i)},\mathcal{L}_{b^{(i)}|_{v=0}}\mathfrak{o}^{(i)}\right)\right\vert\right\vert_{\mathring{H}^{N_1-3}\left(\mathbb{S}^2\right)} + \left\vert\left\vert \left(\mathfrak{j}^{(i)},\mathcal{L}_{b^{(i)}|_{v=0}}\mathfrak{j}^{(i)}\right)\right\vert\right\vert_{\mathring{H}^{N_1-2}\left(\mathbb{S}^2\right)}+ \left\vert\left\vert \left(\mathfrak{w}^{(i)},\mathcal{L}_{b^{(i)}|_{v=0}}\mathfrak{w}^{(i)}\right)\right\vert\right\vert_{\mathring{H}^{N_1-3}\left(\mathbb{S}^2\right)}
\\ \nonumber &\qquad  \leq C_{\rm boot}\left(\mathcal{D} +\mathcal{F}^2+\mathcal{G}^2+ \left\vert\left\vert \pi\right\vert\right\vert^2_{\mathscr{S}\left(0,\check{\delta},0\right)} +\left\vert\left\vert \pi\right\vert\right\vert^2_{\mathscr{Q}_{-1}^{-1/2}\left(0,-1/2+\check{\delta},0\right)}\right),
\end{align}
\begin{align}\label{32oi32jio34}
&\left\vert\left\vert \slashed{g}^{(i)} \right\vert\right\vert_{\mathscr{A}_2\left(\kappa,b^{(i)},\Omega\right)} + \left\vert\left\vert \slashed{g}^{(i)}\right\vert\right\vert_{\mathscr{B}_2\left(\kappa,b^{(i)},\Omega\right)}
\\ \nonumber &\qquad +\left\vert\left\vert \left(v+1\right)\mathfrak{n}^{(i)}\right\vert\right\vert_{L,III\left(1/2-\check{\delta},100\check{p}N_1,0\right),N_1-1}+\left\vert\left\vert \slashed{g}^{(i)}|_{v=0}-\mathring{\slashed{g}}\right\vert\right\vert_{\mathring{H}^{N_1-1}} 
\\ \nonumber &\qquad \leq C_{\rm boot}\left(\mathcal{D} + \mathcal{F}^2+\mathcal{G}+ \left\vert\left\vert \pi\right\vert\right\vert_{\mathscr{S}\left(0,\check{\delta},0\right)} +\left\vert\left\vert \pi\right\vert\right\vert_{\mathscr{Q}_{-1}^{-1/2}\left(0,-1/2+\check{\delta},0\right)}\right),
\end{align}
\begin{align}\label{idjo2ij3oij2o}
&\left\vert\left\vert \mathcal{P}_{\ell > \ell_0}\slashed{g}^{(i)} \right\vert\right\vert_{\mathscr{A}_2\left(\kappa,b^{(i)},\Omega\right)} \leq C_{\rm boot}\left(\mathcal{D} + \mathcal{F}^2+\mathcal{G}+\left\vert\left\vert \pi\right\vert\right\vert^2_{\mathscr{S}\left(0,\check{\delta},0\right)} +\left\vert\left\vert \pi\right\vert\right\vert^2_{\mathscr{Q}_{-1}^{-1/2}\left(0,-1/2+\check{\delta},0\right)}\right).
\end{align}
In the rest of this proof, the implied constants should be understood to be independent of $C_{\rm boot}$.

We now proceed to the induction argument. The base case is immediate, so we assume that $j > 0$ and that~\eqref{l1kjkjljk12kl3} and~\eqref{32oi32jio34} holds for  all $0 \leq i \leq j-1$. Our first step is to note the following immediate consequence of elliptic estimates which holds for each $0 \leq i \leq j$ if the right hand is finite:
\begin{align}\label{2o3o2o4}
&\left\vert\left\vert b^{(i)}\right\vert\right\vert_{\mathscr{A}^-_1\left(b^{(i-1)},\kappa\right)}+\left\vert\left\vert b^{(i)}\right\vert\right\vert_{\mathscr{B}^-_1\left(\kappa\right)} + \sum_{j=0}^1\left\vert\left\vert \left(v\mathcal{L}_{\partial_v}\right)^jP^{(i)}\right\vert\right\vert_{\check{\mathscr{S}}\left(N_2-1-j,0,50\check{p}\left(1+j\right),0\right)} 
\\ \nonumber &\qquad + \left\vert\left\vert \mathscr{P}^{(i)}\right\vert\right\vert_{W_{-1/2}^0\left(N_1-1,0,1/2\right)} + \left\vert\left\vert \left(1,v\mathcal{L}_{\partial_v},\mathcal{L}_{b^{(i)}}\right)\mathscr{P}^{(i)}\right\vert\right\vert_{W_{-1/2}^0\left(N_1-2,0,2\kappa\right)}
\\ \nonumber &\qquad + \sum_{j=0}^1\left\vert\left\vert \left(v\mathcal{L}_{\partial_v}\right)^jQ^{(i)}\right\vert\right\vert_{\check{\mathscr{S}}\left(N_2-2-j,0,50\check{p}\left(2+j\right),50\check{p}\right)} 
\\ \nonumber &\qquad + \left\vert\left\vert Q^{(i)}\right\vert\right\vert_{W_{-1/2}^0\left(N_1-2,0,1/2\right)} + \left\vert\left\vert \left(1,v\mathcal{L}_{\partial_v},\mathcal{L}_{b^{(i)}}\right)Q^{(i)}\right\vert\right\vert_{W_{-1/2}^0\left(N_1-3,0,2\kappa\right)} \lesssim \mathscr{A}_1\left[i\right] +\left(\mathscr{A}_2\left[i\right]\right)^2+ \mathcal{G},
\end{align}
where $\mathscr{A}_1\left[i\right]$ denotes the left hand sides of~\eqref{l1kjkjljk12kl3} and $\mathscr{A}_2\left[i\right]$ denotes the left hand side of~\eqref{32oi32jio34}. It will be convenient to define 
\[\mathscr{A}\left[i\right] \doteq \mathscr{A}_1\left[i\right] + \mathscr{A}_2\left[i\right].\]

We solve~\eqref{2o2oo14829} for $\mathfrak{W}^{(j)}$ by applying Theorem~\ref{fo3p39iwu88u} for an equation of type $III$ with $\left(\tilde{p},\tilde{s},\tilde{w}\right) = \left(1/2-\check{\delta},50\check{p},150\check{p}\right)$. Using, in particular, that the highest order term on the right hand side containing $\mathcal{P}$ is a total $(-v)\mathcal{L}_{\partial_v} - \mathcal{L}_b$ derivative, after a term by term inspection of the nonlinear terms on the right hand side of~\eqref{2o2oo14829}, we end up with 
\[\left\vert\left\vert \mathfrak{W}^{(j)}\right\vert\right\vert_{L(b^{(j-1)}),III\left(1/2-\check{\delta},50\check{p},150\check{p}\right),N_1-3} \lesssim \mathcal{D} + \mathcal{F}^2+\mathcal{G}^2+\left\vert\left\vert \pi\right\vert\right\vert^2_{\mathscr{S}\left(0,\check{\delta},0\right)} +\left\vert\left\vert \pi\right\vert\right\vert^2_{\mathscr{Q}_{-1}^{-1/2}\left(0,-1/2+\check{\delta},0\right)}.\]
We emphasize here, as already mentioned in the introduction, we use the $III$ norm (as opposed to the $III'$ norm) due to the nonlinear term $\mathcal{L}_b\left(\Omega\hat{\chi}\right)$ on the right hand side.

Similarly, to solve for $\mathfrak{p}^{(j)}$ we apply Theorem~\ref{fo3p39iwu88u} to solve~\eqref{3oij3jio4tio39i} as an equation of type $II$. We obtain 
\[\left\vert\left\vert \mathfrak{p}^{(j)}\right\vert\right\vert_{L,II,N_1-1} \lesssim \mathcal{D} + \mathcal{F}^2+\mathcal{G}^2+\left\vert\left\vert \pi\right\vert\right\vert^2_{\mathscr{S}\left(0,\check{\delta},0\right)} +\left\vert\left\vert \pi\right\vert\right\vert^2_{\mathscr{Q}_{-1}^{-1/2}\left(0,-1/2+\check{\delta},0\right)}.\]

For $\mathfrak{t}^{(j)}$, we treat~\eqref{2o2o293fm3o2} first as a degenerate transport equation for $\mathfrak{t}^{(j)}$. From Lemma~\ref{linftofkwp3} and a straightforward nonlinear analysis, we obtain

\begin{align}\label{3oij32jio}
& \sum_{j=0}^1\left\vert\left\vert \mathcal{L}^{1+j}_{\partial_v}\mathfrak{t}^{(j)}\right\vert\right\vert_{\check{\mathscr{S}}_{-1}^0\left(0,1+\check{\delta}+j,100\check{p}+j,0\right)} +  \sum_{j=1}^2 \left\vert\left\vert \mathcal{L}_{\partial_v}^j\mathfrak{t}^{(j)}\right\vert\right\vert_{\mathscr{Q}_{-1}^{-1/2}\left(0,-1/2+\check{\delta} + j,0\right)}
\\ \nonumber &\qquad \lesssim \mathcal{D} + \mathcal{F}^2+\mathcal{G}^2+\left\vert\left\vert \pi\right\vert\right\vert^2_{\mathscr{S}\left(0,\check{\delta},0\right)} +\left\vert\left\vert \pi\right\vert\right\vert^2_{\mathscr{Q}_{-1}^{-1/2}\left(0,-1/2+\check{\delta},0\right)}.
\end{align}
We then apply the fundamental theorem of calculus to obtain the desired estimate for $\mathfrak{t}^{(j)}$:
\begin{equation}\label{3232ijo23}
 \left\vert\left\vert \mathfrak{t}^{(j)}\right\vert\right\vert_{\mathscr{Q}_{-1}^{-1/2}\left(0,-1/2+\check{\delta} ,0\right)}+\left\vert\left\vert \mathfrak{t}^{(j)}\right\vert\right\vert_{\mathscr{S}_{-1}^0\left(0,-1,0\right)} \lesssim \mathcal{D} + \mathcal{F}^2 + \mathcal{G}^2+\left\vert\left\vert \pi\right\vert\right\vert^2_{\mathscr{S}\left(0,\check{\delta},0\right)} +\left\vert\left\vert \pi\right\vert\right\vert^2_{\mathscr{Q}_{-1}^{-1/2}\left(0,-1/2+\check{\delta},0\right)}.
\end{equation}
From~\eqref{2om3om92} and elliptic theory along $\mathbb{S}^2$, we easily obtain 
\begin{align*}
&\left\vert\left\vert b^{(j)}\right\vert\right\vert_{\mathscr{A}^-_1\left(b^{(j-1)},\kappa\right)}+\left\vert\left\vert b^{(j)}\right\vert\right\vert_{\mathscr{B}^-_1\left(\kappa\right)} \lesssim \left\vert\left\vert \mathfrak{W}^{(j)}\right\vert\right\vert_{L(b^{(j-1)}),III\left(0,50\check{p},100\check{p}\right),N_1-3}  +\left\vert\left\vert \mathfrak{p}^{(j)}\right\vert\right\vert_{L,II,N_1-1} 
\\ \nonumber &+ \left\vert\left\vert \mathcal{L}_{\partial_v}\mathfrak{t}^{(j)}\right\vert\right\vert_{\check{\mathscr{S}}_{-1}^0\left(0,0,100\check{p},0\right)}+ \left\vert\left\vert \mathfrak{t}^{(j)}\right\vert\right\vert_{\mathscr{S}_{-1}^0\left(0,-1,0\right)}
+\left\vert\left\vert \mathring{\Pi}_{\rm div}b^{(j)}\right\vert\right\vert_{\mathscr{A}^-_1\left(b^{(j-1)},\kappa\right)}+\left\vert\left\vert \mathring{\Pi}_{\rm div}b^{(j)}\right\vert\right\vert_{\mathscr{B}^-_1\left(\kappa\right)},
\end{align*}
\begin{align*}
&\left\vert\left\vert \mathring{\Pi}_{\rm curl}b^{(j)}\right\vert\right\vert_{\mathscr{A}^-_1\left(b^{(j-1)},\kappa\right)}+\left\vert\left\vert \mathring{\Pi}_{\rm curl}b^{(j)}\right\vert\right\vert_{\mathscr{B}^-_1\left(\kappa\right)} \lesssim \left\vert\left\vert \mathfrak{W}^{(j)}\right\vert\right\vert_{L(b^{(j-1)}),III\left(0,50\check{p},100\check{p}\right),N_1-3}  +\left\vert\left\vert \mathfrak{p}^{(j)}\right\vert\right\vert_{L,II,N_1-1} 
\\ \nonumber &+ \left\vert\left\vert \mathcal{L}_{\partial_v}\mathfrak{t}^{(j)}\right\vert\right\vert_{\check{\mathscr{S}}_{-1}^0\left(0,0,100\check{p},0\right)}+ \left\vert\left\vert \mathfrak{t}^{(j)}\right\vert\right\vert_{\mathscr{S}_{-1}^0\left(0,-1,0\right)}
+\mathscr{A}\left[j-1\right]\left(\left\vert\left\vert \mathring{\Pi}_{\rm div}b^{(j)}\right\vert\right\vert_{\mathscr{A}^-_1\left(b^{(j-1)},\kappa\right)}+\left\vert\left\vert \mathring{\Pi}_{\rm div}b^{(j)}\right\vert\right\vert_{\mathscr{B}^-_1\left(\kappa\right)}\right).
\end{align*}

We turn now to $\mathscr{P}^{(j)}$. In order to estimate $\mathscr{P}^{(j)}$ and $\slashed{\rm curl}\mathscr{P}^{(j)}$ we apply~\eqref{jil2ij2ijij}-\eqref{jk32ji23iu32oi} from Lemma~\ref{32m2omo4} to~\eqref{kn1kln21lk}, \eqref{ui2iu21iu12io}, and~\eqref{2oj3oij2oo} with $q = 50\check{p}+2\kappa$ and further use elliptic estimates. We obtain
\begin{align}\label{32i32io23o}
&\sum_{j=0}^1\left\vert\left\vert \left(v\mathcal{L}_{\partial_v}\right)^j\mathring{\Pi}_{\rm curl}\mathscr{P}^{(j)}\right\vert\right\vert_{\check{\mathscr{S}}\left(N_2-1-j,0,50\check{p}\left(1+j\right),50\check{p}\right)} 
\\ \nonumber &\qquad  \lesssim \mathcal{D} + \mathcal{F}^2 + \mathcal{G}^2 + \left\vert\left\vert \pi\right\vert\right\vert^2_{\mathscr{S}\left(0,\check{\delta},0\right)} +\left\vert\left\vert \pi\right\vert\right\vert^2_{\mathscr{Q}_{-1}^{-1/2}\left(0,-1/2+\check{\delta},0\right)}. 
\end{align}

However, we cannot directly apply~\eqref{ijo2o1o1o1049} from Lemma~\ref{32m2omo4} to obtain the desired $\mathscr{Q}$-norm estimate for $\slashed{\rm curl}\mathscr{P}^{(j)}$ due to the term 
\[\xi\mathcal{P}_{\ell > \ell_0}\left(v+1\right)^2\slashed{\rm curl}\slashed{\rm div}\slashed{\nabla}\hat{\otimes}\slashed{\rm div}\left(\slashed{\nabla}\hat{\otimes}b^{(j)}\right).\]
The problem is that we can only estimate this term if we apply at most $N_1-5$ derivatives, while in order to apply~\eqref{ijo2o1o1o1049} from Lemma~\ref{32m2omo4} and recover the necessary estimates for $P$, we would need to apply $N_1-4$ derivatives.  The resolution is to rewrite this potentially problematic term as a sum of a total $(-v)\mathcal{L}_{\partial_v} - \mathcal{L}_b$ derivative and another term which can be angularly differentiated an additional time. In order to do this we first use~\eqref{2m3momo2} to write 
\begin{align}
&\xi\mathcal{P}_{\ell > \ell_0}\left(v+1\right)^2\slashed{\rm curl}\slashed{\rm div}\slashed{\nabla}\hat{\otimes}\slashed{\rm div}\left(\slashed{\nabla}\hat{\otimes}b^{(j)}\right) =\xi\mathcal{P}_{\ell > \ell_0}\left(\slashed{\Delta} + 2(v+1)^{-2}\right)\mathfrak{W}^{(j)} + 
\\ \nonumber &\qquad \xi\left(v+1\right)^2\mathcal{P}_{\ell > \ell_0}\left(\slashed{\Delta} + 2K\right)\mathcal{P}_{\ell \leq \ell_0}\slashed{\rm curl}\slashed{\rm div}\slashed{\nabla}\hat{\otimes}b^{(j)} + 2\left(v+1\right)^2\mathcal{P}_{\ell > \ell_0}\left(\left(\slashed{\nabla}K\right)\wedge \left(\slashed{\rm div}\left(\slashed{\nabla}\hat{\otimes}b^{(j)}\right)\right)\right)
\\ \nonumber &\qquad + 2\xi\mathcal{P}_{\ell > \ell_0}\left(\left(K-(v+1)^{-2}\right)\mathfrak{W}^{(j)}\right).
\end{align}
Focusing on the highest order term $\xi\mathcal{P}_{\ell > \ell_0}\left(\slashed{\Delta} + 2(v+1)^{-2}\right)\mathfrak{W}^{(j)}$ we then use~\eqref{2o2oo14829} to obtain that 
\begin{align}\label{ij3omo2}
&\xi\mathcal{P}_{\ell > \ell_0}\left(\slashed{\Delta} + 2(v+1)^{-2}\right)\mathfrak{W}^{(j)} =
\\ \nonumber &\qquad -\xi\left((-v)\mathcal{L}_{\partial_v}-\mathcal{P}_{\ell > \ell_0}\mathcal{L}_{b^{(j-1)}}\right)\left(\Omega^{-2}\mathcal{L}_{\partial_v}\mathfrak{W}^{(j)}\right) - \xi \left[\left((-v)\mathcal{L}_{\partial_v}-\mathcal{P}_{\ell > \ell_0}\mathcal{L}_{b^{(j-1)}}\right),\Omega^{-2}\right]\mathcal{L}_{\partial_v}\mathfrak{W}^{(j)}
\\ \nonumber &\qquad -\xi\Omega^{-2}\left(1+\frac{4(-v)\mathcal{P}_{\ell > \ell_0}\Omega^2}{v+1}\right)\mathcal{L}_{\partial_v}\mathfrak{W}^{(i)} + \xi \Omega^{-2}\left[\mathcal{P}_{\ell > \ell_0},\Omega^2\right]\left(\slashed{\Delta} + 2(v+1)^{-2}\right)\mathfrak{W}^{(j)} + \Omega^{-2}\xi\mathscr{H}(j),
\end{align}
where $\mathscr{H}(j)$ denotes the right hand side of~\eqref{2o2oo14829}. Then we will exploit that in~\eqref{ijo2o1o1o1049} we may write the terms on the right hand side as a total $(-v)\mathcal{L}_v - \mathcal{L}_b$ derivative and then gain a derivative in the corresponding estimate. We use this not only for the term $\xi\left((-v)\mathcal{L}_{\partial_v}-\mathcal{P}_{\ell > \ell_0}\mathcal{L}_{b^{(j-1)}}\right)\left(\Omega^{-2}\mathcal{L}_{\partial_v}\mathfrak{W}^{(j)}\right) $ on the right hand side of~\eqref{ij3omo2} but also for the term
\[-\mathcal{P}_{\ell > \ell_0}\Omega^2\mathcal{P}_{\ell > \ell_0}\left(v+1\right)^2\slashed{\rm curl}\slashed{\rm div}\left(\left((-v)\nabla_v-\mathcal{L}_b\right)\left(8\slashed{\nabla}\log\Omega\hat{\otimes}P - 4\Omega^{-2}\mathcal{E}\right) \right)\]
which is contained in $\mathscr{H}(j)$. Finally, we are able to apply~\eqref{ijo2o1o1o1049}, elliptic estimates, and the previously established estimates for $b^{(j)}$ to obtain the desired estimate
\begin{align*}
&\left\vert\left\vert \mathring{\Pi}_{\rm curl}\mathscr{P}^{(j)}\right\vert\right\vert_{W_{-1/2}^0\left(N_1-1,0,1/2\right)} + \left\vert\left\vert \left(1,v\mathcal{L}_{\partial_v},\mathcal{L}_{b^{(j)}}\right)\mathring{\Pi}_{\rm curl}\mathscr{P}^{(j)}\right\vert\right\vert_{W_{-1/2}^0\left(N_1-2,0,2\kappa\right)}
\\ \nonumber &\qquad  \lesssim \mathcal{D} + \mathcal{F}^2 + \mathcal{G}^2+\left\vert\left\vert \pi\right\vert\right\vert^2_{\mathscr{S}\left(0,\check{\delta},0\right)} +\left\vert\left\vert \pi\right\vert\right\vert^2_{\mathscr{Q}_{-1}^{-1/2}\left(0,-1/2+\check{\delta},0\right)}.
\end{align*}

The desired estimates for $\Omega_{\rm boun}^{(j)}$ are immediate from Proposition~\ref{23ini2999jn2j23i3j}:
\begin{equation}\label{3i3iojio49}
 \left\vert\left\vert \log\Omega^{(j)}_{\rm boun}\right\vert\right\vert_{\mathscr{A}\left(\kappa,b^{(i)}\right)} + \left\vert\left\vert \log\Omega^{(j)}_{\rm boun}\right\vert\right\vert_{\mathscr{B}_{01}\left(\kappa,b^{(i)}\right)}  \lesssim \mathcal{D}.
 \end{equation}

We now come to $\mathfrak{o}^{(j)}$ and $\mathfrak{j}^{(j)}$. Before we come directly to these, we first quickly collect some estimates at $\{v = 0\}$ for $b^{(j)}$, $\slashed{g}^{(j-1)}$ and $\mathfrak{A}^{(j)}$. From our induction hypothesis we have that 
\begin{align}\label{3opj2po2}
&\left\vert\left\vert \left(\slashed{g}^{(j-1)}-\mathring{\slashed{g}}\right)|_{v=0}\right\vert\right\vert_{\mathring{H}^{N_1-1}}  \leq C_{\rm boot}\left(\mathcal{D} + \mathcal{F}^2+\mathcal{G}+\left\vert\left\vert \pi\right\vert\right\vert_{\mathscr{S}\left(0,\check{\delta},0\right)} +\left\vert\left\vert \pi\right\vert\right\vert_{\mathscr{Q}_{-1}^{-1/2}\left(0,-1/2+\check{\delta},0\right)}\right).
\end{align}
As an immediate consequence of the fundamental theorem of calculus  we have that $b^{(j)}$ and $\mathfrak{A}^{(j)}$ extend continuously to $\{v = 0\}$, and moreover satisfy
\begin{align}\label{32oi32oi}
&\left\vert\left\vert \left(1,\mathcal{L}_{b^{(j-1)}}\right)\mathfrak{A}^{(j)}|_{v=0} \right\vert\right\vert_{\mathring{H}^{N_1-2}} \lesssim
\\ \nonumber &\qquad \left(\left\vert\left\vert \slashed{g}^{(j-1)} \right\vert\right\vert^2_{\mathscr{A}_2\left(\kappa,b^{(j-1)},\Omega\right)} + \left\vert\left\vert \slashed{g}^{(j-1)}\right\vert\right\vert^2_{\mathscr{B}_2\left(\kappa,b^{(j-1)},\Omega\right)} \right)+
\\ \nonumber &\qquad \qquad  \left\vert\left\vert \log\Omega^{(j)}_{\rm boun}\right\vert\right\vert_{\mathscr{A}\left(\kappa,b^{(i)}\right)} + \left\vert\left\vert \log\Omega^{(j)}_{\rm boun}\right\vert\right\vert_{\mathscr{B}_{01}\left(\kappa,b^{(i)}\right)} + \left\vert\left\vert \log\Omega_{\rm sing}\right\vert\right\vert_{\mathscr{B}_{00}\left(\kappa\right)}
\\ \nonumber &\qquad \lesssim \mathcal{D} + \mathcal{F}^2 + \mathcal{G}^2+\left\vert\left\vert \pi\right\vert\right\vert^2_{\mathscr{S}\left(0,\check{\delta},0\right)} +\left\vert\left\vert \pi\right\vert\right\vert^2_{\mathscr{Q}_{-1}^{-1/2}\left(0,-1/2+\check{\delta},0\right)},
\end{align}
\begin{align}\label{2om3o2}
&\left\vert\left\vert b^{(j)}|_{v=0}\right\vert\right\vert_{\mathring{H}^{N_1-2}}  \lesssim
\left\vert\left\vert b^{(j)}\right\vert\right\vert_{\mathscr{A}^-_1\left(b^{(j-1)},\kappa\right)}+\left\vert\left\vert b^{(j)}\right\vert\right\vert_{\mathscr{B}^-_1\left(\kappa\right)}
 \\ \nonumber &\qquad  \lesssim C_{\rm boot2}\left(\mathcal{D} + \mathcal{F}^2+\mathcal{G} + \left\vert\left\vert \pi\right\vert\right\vert^2_{\mathscr{S}\left(0,\check{\delta},0\right)} +\left\vert\left\vert \pi\right\vert\right\vert^2_{\mathscr{Q}_{-1}^{-1/2}\left(0,-1/2+\check{\delta},0\right)}\right).
\end{align}
We will need, however, to estimate one more derivative of $b$ than what is obtained from~\eqref{2om3o2}. Note that the problem with obtaining an estimate for $b^{(j)}|_{v=0}$ in $\mathring{H}^{N_1-1}$ from the fundamental theorem of calculus argument is that finiteness of the  the $\mathscr{A}^-_1$ norm does not yield that $\mathcal{L}_{\partial_v}b$ is integrable after an application of $N_1-1$ derivatives. However, we do not have this problem if we estimate $\mathring{\Pi}_{\rm div}b$ and thus 
\begin{equation}\label{2io3ijooij2}
\left\vert\left\vert \mathring{\Pi}_{\rm div}b^{(j)}|_{v=0}\right\vert\right\vert_{\mathring{H}^{N_1-1}}  \lesssim \mathcal{G}.
\end{equation}
Combining~\eqref{2io3ijooij2} with the restriction of the system~\eqref{2om3om92} to $\{v = 0\}$, the boundary conditions~\eqref{23ji3joi} and~\eqref{2joijoij23}, elliptic estimates, and~\eqref{3oij32jio} and~\eqref{3232ijo23} leads to
\begin{equation}\label{23ijo3oij}
\left\vert\left\vert b^{(j)}|_{v=0}\right\vert\right\vert_{\mathring{H}^{N_1-1}} \lesssim \mathcal{D}+\mathcal{F}^2 + \mathcal{G}+\left\vert\left\vert \pi\right\vert\right\vert^2_{\mathscr{S}\left(0,\check{\delta},0\right)} +\left\vert\left\vert \pi\right\vert\right\vert^2_{\mathscr{Q}_{-1}^{-1/2}\left(0,-1/2+\check{\delta},0\right)}.
\end{equation}
Finally, we also note that it follows from the boundary conditions used for $\Omega_{\rm boun}$ and ~\eqref{wji3ij32ijo} that $\left(\Omega\underline{\omega}\right)^{(j-1)}$ extends to $\{v = 0\}$ as a spherically symemtric function which satisfies 
\begin{equation}\label{3lj3jio34}
\left\vert\left\vert \left(\Omega\underline{\omega}\right)^{(j-1)}|_{v=0}\right\vert\right\vert_{L^{\infty}} \lesssim \mathcal{D}.
\end{equation}

Keeping~\eqref{3opj2po2},~\eqref{32oi32oi},~\eqref{23ijo3oij} and~\eqref{3lj3jio34} in mind, we apply Proposition~\ref{somestuimdie} to~\eqref{1ipo939209} and obtain that  
\begin{equation}\label{ioj23oij32ioj}
\left\vert\left\vert \left(1,\mathcal{L}_{b^{(j)}}\right)\mathfrak{o}^{(j)}\right\vert\right\vert_{\mathring{H}^{N_1-3}} \lesssim \mathcal{D} + \mathcal{F}^2 + \mathcal{G}^2+\left\vert\left\vert \pi\right\vert\right\vert^2_{\mathscr{S}\left(0,\check{\delta},0\right)} +\left\vert\left\vert \pi\right\vert\right\vert^2_{\mathscr{Q}_{-1}^{-1/2}\left(0,-1/2+\check{\delta},0\right)}.
\end{equation}
From~\eqref{ijio2joij32} and~\eqref{12ij1iojoi12}, an application of Proposition~\ref{somestuimdie}, and elliptic theory along $\mathbb{S}^2$, we then obtain
\begin{align}\label{3lj3jio4}
&\left\vert\left\vert \left(1,\mathcal{L}_{b^{(j)}}\right)\left(\mathcal{P}_{\ell \geq 1}\slashed{\rm div}\mathfrak{j}^{(j)},\mathcal{P}_{\ell \geq 1}\slashed{\rm curl}\mathfrak{j}^{(j)}\right)\right\vert\right\vert_{\mathring{H}^{N_1-3}} 
\\ \nonumber &\qquad \lesssim \mathcal{D} + \mathcal{F}^2+\mathcal{G}^2 +\left\vert\left\vert \pi\right\vert\right\vert^2_{\mathscr{S}\left(0,\check{\delta},0\right)} +\left\vert\left\vert \pi\right\vert\right\vert^2_{\mathscr{Q}_{-1}^{-1/2}\left(0,-1/2+\check{\delta},0\right)}\Rightarrow
\\ \nonumber  \left\vert\left\vert \left(1,\mathcal{L}_{b^{(j)}}\right)\mathfrak{j}^{(j)}\right\vert\right\vert_{\mathring{H}^{N_1-2}} &\lesssim \mathcal{D} + \mathcal{F}^2 +\mathcal{G}^2+\left\vert\left\vert \pi\right\vert\right\vert^2_{\mathscr{S}\left(0,\check{\delta},0\right)} +\left\vert\left\vert \pi\right\vert\right\vert^2_{\mathscr{Q}_{-1}^{-1/2}\left(0,-1/2+\check{\delta},0\right)}\Rightarrow 
\\ \nonumber  \left\vert\left\vert \left(1,\mathcal{L}_{b^{(j)}}\right)\mathfrak{w}^{(j)}\right\vert\right\vert_{\mathring{H}^{N_1-3}} &\lesssim \mathcal{D} + \mathcal{F}^2+\mathcal{G}^2+\left\vert\left\vert \pi\right\vert\right\vert^2_{\mathscr{S}\left(0,\check{\delta},0\right)} +\left\vert\left\vert \pi\right\vert\right\vert^2_{\mathscr{Q}_{-1}^{-1/2}\left(0,-1/2+\check{\delta},0\right)}. 
\end{align}

Finally, we may obtain the existence of  $\left(\slashed{g}^{(j)},\mathfrak{n}^{(j)}\right)$ with a suitable application Proposition~\ref{solvetheprojected}. It is an immediate consequence that 
\begin{align}
&\left\vert\left\vert \slashed{g}^{(j)} \right\vert\right\vert_{\mathscr{A}_2\left(\kappa,b^{(j)},\Omega\right)} + \left\vert\left\vert \slashed{g}^{(j)}\right\vert\right\vert_{\mathscr{B}_2\left(\kappa,b^{(j)},\Omega\right)}
\\ \nonumber &\qquad +\left\vert\left\vert \left(v+1\right)\mathfrak{n}^{(j)}\right\vert\right\vert_{L,III\left(0,100\check{p}N_1,0\right),N_1-1}+\left\vert\left\vert \slashed{g}^{(j)}|_{v=0}-\mathring{\slashed{g}}\right\vert\right\vert_{\mathring{H}^{N_1-1}}
\\ \nonumber&\qquad  \lesssim \mathcal{D} + \mathcal{F}^2+\mathcal{G} +\left\vert\left\vert \pi\right\vert\right\vert_{\mathscr{S}\left(0,\check{\delta},0\right)} +\left\vert\left\vert \pi\right\vert\right\vert_{\mathscr{Q}_{-1}^{-1/2}\left(0,-1/2+\check{\delta},0\right)},
\end{align}
\begin{align}
&\left\vert\left\vert \mathcal{P}_{\ell > \ell_0}\slashed{g}^{(j)} \right\vert\right\vert_{\mathscr{A}_2\left(\kappa,b^{(j)},\Omega\right)} \lesssim \mathcal{D} + \mathcal{F}^2+\mathcal{G} +\left\vert\left\vert \pi\right\vert\right\vert^2_{\mathscr{S}\left(0,\check{\delta},0\right)} +\left\vert\left\vert \pi\right\vert\right\vert^2_{\mathscr{Q}_{-1}^{-1/2}\left(0,-1/2+\check{\delta},0\right)}.
\end{align}

This concludes the induction argument, and we see that the bounds~\eqref{l1kjkjljk12kl3} and~\eqref{32oi32jio34} hold for all $i$. As usual, it is then straightforward to run a compactness argument and find corresponding solutions to~\eqref{2o2oo14829}-\eqref{2lkkjl4k2} with all of the $i$'s removed. To see that $\mathfrak{A} = \Omega^{-2}\left(\Omega{\rm tr}\chi\right)$ it suffices to observe that they will both solve the same transport equation~\eqref{2lkkjl4k2} and that this transport equation is easily seen to have a unique solution which is bounded as $v\to -1$. In order to see that we have solved the desired equations, it remains to show that $P = \mathcal{L}_{\partial_v}b$.  However, this may be done by adapting the argument at the end of the proof of Proposition~\ref{ij2ojo2432}. We omit the details. 

By taking the $i \to \infty$ limits of the bounds~\eqref{l1kjkjljk12kl3} and~\eqref{32oi32jio34}, we see that~\eqref{3o3pkp4} and~\eqref{3ij32ijo23io} hold. We now discuss how to establish~\eqref{3j23moo4} and~\eqref{io23io32io32oi}. First of all, the estimate for $\left\vert\left\vert \mathring{\Pi}_{\rm curl}b\right\vert\right\vert_{\mathscr{B}_1\left(\kappa\right)}$ follows immediately from the the fact that $P = \mathcal{L}_{\partial_v}b$ and the $i \to \infty$ limit of the estimate~\eqref{32i32io23o}. Thus we have~\eqref{3j23moo4}. Turning to~\eqref{io23io32io32oi}, we may obtain that
\begin{align}\label{oij2oij4}
&\sum_{j=0}^1\left[\left\vert\left\vert \mathcal{L}_{\partial_v}^j\slashed{\rm curl}\slashed{\rm div}\slashed{\nabla}\hat{\otimes}b \right\vert\right\vert_{\mathscr{Q}_{-1/2}^0\left(N_1-3-j,0,-1/2+\check{\delta}+j\right)} + \left\vert\left\vert \mathcal{L}^{1+j}_{\partial_v}\slashed{\rm curl}\slashed{\rm div}\slashed{\nabla}\hat{\otimes}b\right\vert\right\vert_{\mathscr{Q}_{-1/2}^0\left(N_1-4-j,0,\kappa+j\right)}\right] \lesssim
\\ \nonumber &\qquad  \mathcal{D}+ \mathcal{F}^2 + \mathcal{G}+\left\vert\left\vert \pi\right\vert\right\vert^2_{\mathscr{S}\left(0,\check{\delta},0\right)} +\left\vert\left\vert \pi\right\vert\right\vert^2_{\mathscr{Q}_{-1}^{-1/2}\left(0,-1/2+\check{\delta},0\right)},
\end{align}
by revisiting the ($i \to \infty$ limit of) the equation for $\mathfrak{W}^{(i)}$ and instead of estimating $\mathfrak{W}$ in the $III$ norm, using instead the corresponding $III'$ norm (see Theorem~\ref{fo3p39iwu88u}). Indeed, the only reason we did not do this in the iteration scheme was, see the discussion of difficulty \#\ref{3i3i2o2} in Section~\ref{3ij2oj2}, due to the presence of (angular derivatives of) the term $\mathcal{L}_b\left(\Omega\hat{\chi}\right)$ and our desire to treat this term as \emph{nonlinear}. However, given that our iteration scheme has now already closed, we will be satisfied with a \emph{linear} estimate. In particular, as a consequence of~\eqref{3ij32ijo23io} and~\eqref{2k3lnl2n3} we have
\begin{equation}\label{3i3i3j2}
\left\vert\left\vert \mathcal{L}_b\left(\Omega\hat{\chi}\right)\right\vert\right\vert_{\mathscr{Q}_{-1/2}^0\left(N_1-2,0,\kappa\right)} \lesssim \mathcal{D}+ \mathcal{F}^2 + \mathcal{G}+\left\vert\left\vert \pi\right\vert\right\vert^2_{\mathscr{S}\left(0,\check{\delta},0\right)} +\left\vert\left\vert \pi\right\vert\right\vert^2_{\mathscr{Q}_{-1}^{-1/2}\left(0,-1/2+\check{\delta},0\right)}.
\end{equation}
Thus we obtain~\eqref{oij2oij4}. To then obtain~\eqref{io23io32io32oi} it suffices to commute the $\mathcal{L}_{\partial_v}$ through the angular derivatives. In principle this could produce a term with three angular derivatives applied to $\Omega\chi$ which would, in principle, pose a problem to such a naive commutation scheme. However, due to the formula~\eqref{3k2ki2i3i}, to highest order, we see two angular derivatives applied to $\mathcal{L}_b\left(\Omega\hat{\chi}\right)$. Thus, we may again appeal to~\eqref{3i3i3j2} and we establish~\eqref{io23io32io32oi}.
\end{proof}

\end{proposition}

Now we run a second iteration argument where we will now use Proposition~\ref{ij2ojo2432} to solve for $\mathring{\Pi}_{\rm div}b$.  We will obtain finally our strongest estimates for $b$.
\begin{proposition}\label{3ij2oio23} Let $H_1$, $H_2$, $H_4$, $H_5$, $H_6$, and $\mathfrak{q}$  be functions on $(-1,0)\times\mathbb{S}^2$, $\pi$, $H_7$, and $H_8$  be $\mathbb{S}^2_{-1,v}$ $1$-forms and $H_9$ and $\mathfrak{r}$ be $\mathbb{S}^2_{-1,v}$ $(0,2)$-tensor on $(-1,0)\times \mathbb{S}^2$   so that  $\left(1-\mathcal{P}_{\ell \geq 1}\right)\left(H_1,H_2\right) = 0$,  $H_2$ is supported for $v \geq -1/2$, so that
\begin{align}\label{32l342lj42oij34}
\left\vert\left\vert \left(\mathfrak{q},\mathfrak{r}\right)\right\vert\right\vert_{\mathscr{Q}\left(N_1-1,-1/2+\check{\delta},-1/2+\check{\delta}\right)} + \left\vert\left\vert \left(\mathcal{L}_{\partial_v}\mathfrak{q},\mathcal{L}_{\partial_v}\mathfrak{r}\right)\right\vert\right\vert_{\mathscr{Q}\left(N_1-2,1/2+\check{\delta},\kappa\right)} + 
\left\vert\left\vert \left(\mathfrak{q},\mathfrak{r}\right)\right\vert\right\vert_{\mathscr{S}\left(N_2-1,\check{\delta},0\right)} \lesssim \epsilon,
\end{align}
\[\left\vert\left\vert \pi\right\vert\right\vert_{\mathscr{S}\left(0,\check{\delta},0\right)} +\left\vert\left\vert \pi\right\vert\right\vert_{\mathscr{Q}_{-1}^{-1/2}\left(0,-1/2+\check{\delta},0\right)} \lesssim \epsilon,\]

and so that  the following expressions
\begin{align*}
 &\mathcal{D} \doteq \left\vert\left\vert \left(H_1,H_2,0\right)\right\vert\right\vert_{R,I,N_1-2} + \epsilon \left\vert\left\vert \left(T_{\rm low},T_{\rm high}\right)\right\vert\right\vert_{\mathring{H}^{N_1-3}} +
 \left\vert\left\vert \left((-v)^{-2\kappa}\mathring{\rm curl}H_7,0,0\right)\right\vert\right\vert_{R,II,N_1-1}  \\ \nonumber &\qquad +\left\vert\left\vert \left((-v)^{-2\kappa}\mathring{\Pi}_{\rm curl}H_8,0,0\right)\right\vert\right\vert_{R,III'\left(1/2-\check{\delta},50\check{p},0\right),N_1}+\sum_{j=0}^1\left\vert\left\vert \left(v\mathcal{L}_{\partial_v}\right)^j\mathring{\Pi}_{\rm curl}H_7\right\vert\right\vert_{\check{\mathscr{S}}_{-1/2}^0\left(0,0,50\check{p}\left(1+j\right),0\right)}
 \\ \nonumber &\qquad +\left\vert\left\vert \log\Omega_{\rm sing}\right\vert\right\vert_{\mathscr{B}_{00}\left(\kappa\right)}+\left\vert\left\vert \left((-v)^{-2\kappa}H_5,H_6\right)\right\vert\right\vert_{\mathscr{P}\mathscr{R}\left(\Omega,\kappa,50\check{p},50\check{p}\right)}
 \\ \nonumber &\qquad +   \left\vert\left\vert H_4\right\vert\right\vert_{\mathscr{Q}\left(N_1-2,1/2+5\check{\delta},-\kappa\right)}
\\ \nonumber &\qquad + \left\vert\left\vert H_4\right\vert\right\vert_{\mathscr{S}_{-1}^{-1/2}\left(N_1-3,1+5\check{\delta},0\right)}+ \left\vert\left\vert H_4\right\vert\right\vert_{\mathscr{S}_{-1/2}^0\left(N_2-2,0,0\right)}  +\left\vert\left\vert \left((-v)^{-2\kappa}H_9,0,0\right)\right\vert\right\vert_{R,III'\left(1/2-\check{\delta},50\check{p},100\check{p}\right),N_1-3}
 \\ \nonumber &\qquad + \sum_{\left|\alpha\right| \leq 1}\sum_{j=0}^1\left\vert\left\vert \left(v\mathcal{L}_{\partial_v}\right)^j\mathcal{L}_{\mathcal{Z}^{(\alpha)}}\mathring{\Delta}^{-1}H_9\right\vert\right\vert_{\check{\mathscr{S}}_{-1/2}^0\left(N_2-1-j,0,50\check{p}\left(1+j\right),50\check{p}\right)},
 \end{align*}
 \begin{align*}
&\mathcal{F} \doteq 
 \left\vert\left\vert \left((-v)^{-2\kappa}\mathring{\rm div}H_7,0,0\right)\right\vert\right\vert_{R,II,N_1-1}+\sum_{j=0}^1\left\vert\left\vert \left(v\mathcal{L}_{\partial_v}\right)^j\mathring{\Pi}_{\rm div}H_7\right\vert\right\vert_{\check{\mathscr{S}}_{-1/2}^0\left(0,0,50\check{p}\left(1+j\right),0\right)}  \\ \nonumber &\qquad +\left\vert\left\vert \left((-v)^{-2\kappa}\mathring{\Pi}_{\rm div}H_8,0,0\right)\right\vert\right\vert_{R,III'\left(1/2-\check{\delta},50\check{p},0\right),N_1}+\sum_{j=0}^1\left\vert\left\vert \left(v\mathcal{L}_{\partial_v}\right)^j\mathring{\Pi}_{\rm div}H_7\right\vert\right\vert_{\check{\mathscr{S}}_{-1/2}^0\left(0,0,50\check{p}\left(1+j\right),0\right)},
 \end{align*}
are defined and satisfy $\mathcal{D},\mathcal{F} \lesssim \epsilon$.

Then there exist a function $\Omega_{\rm boun}: (-1,0)\times \mathbb{S}^2 \to (0,\infty)$, an $\mathbb{S}^2_{-1,v}$-vector field $b^A$, an $\mathbb{S}^2_{-1,v}$-symmetric $(0,2)$ tensor $\slashed{g}$, and an $\mathbb{S}^2_{-1,v}$ $1$-form $\mathfrak{n}_A$  such that all of the equations of Proposition~\ref{3ij2oij3ij2} hold except that instead of requiring $\mathring{\rm div}b = \mathfrak{Z}$, we require that $b$ solves equation~\eqref{3lkl2jllk32lk} from Proposition~\eqref{ij2ojo2432}.

We then have the following estimates:
\begin{align}\label{1kj21kj}
&\left\vert\left\vert \log\Omega_{\rm boun}\right\vert\right\vert_{\mathscr{A}\left(\kappa,b\right)} + \left\vert\left\vert \log\Omega_{\rm boun}\right\vert\right\vert_{\mathscr{B}_{01}\left(\kappa,b\right)}  +\left\vert\left\vert b\right\vert\right\vert_{\mathscr{A}_1\left(\kappa,\slashed{g}\right)}+\left\vert\left\vert b\right\vert\right\vert_{\mathscr{B}_1\left(\kappa\right)}+\left\vert\left\vert \mathcal{P}_{\ell > \ell_0}\slashed{g} \right\vert\right\vert_{\mathscr{A}_2\left(\kappa,b,\Omega\right)}
\\ \nonumber &\qquad + \left\vert\left\vert \left(\mathfrak{o},\mathcal{L}_{b|_{v=0}}\mathfrak{o}\right)\right\vert\right\vert_{\mathring{H}^{N_1-3}\left(\mathbb{S}^2\right)} +\left\vert\left\vert \left(\mathfrak{j},\mathcal{L}_{b|_{v=0}}\mathfrak{j}\right)\right\vert\right\vert_{\mathring{H}^{N_1-2}\left(\mathbb{S}^2\right)}+ \left\vert\left\vert \left(\mathfrak{w},\mathcal{L}_{b|_{v=0}}\mathfrak{w}\right)\right\vert\right\vert_{\mathring{H}^{N_1-3}\left(\mathbb{S}^2\right)}
\\ \nonumber &\qquad  \lesssim \mathcal{D}+ \mathcal{F}^2+\left\vert\left\vert \pi\right\vert\right\vert^2_{\mathscr{S}\left(0,\check{\delta},0\right)} +\left\vert\left\vert \pi\right\vert\right\vert^2_{\mathscr{Q}_{-1}^{-1/2}\left(0,-1/2+\check{\delta},0\right)},
\end{align}
\begin{align}\label{dkn2lkn3lkn2l}
&\left\vert\left\vert \slashed{g} \right\vert\right\vert_{\mathscr{A}_2\left(\kappa,b,\Omega\right)} + \left\vert\left\vert \slashed{g}\right\vert\right\vert_{\mathscr{B}_2\left(\kappa,b,\Omega\right)}
+\left\vert\left\vert \left(v+1\right)\mathfrak{n}\right\vert\right\vert_{L,III\left(1/2-\check{\delta},100\check{p}N_1,0\right),N_1-1}
 \\ \nonumber &\qquad \lesssim \mathcal{D}+ \mathcal{F}^2+\left\vert\left\vert \pi\right\vert\right\vert_{\mathscr{S}\left(0,\check{\delta},0\right)} +\left\vert\left\vert \pi\right\vert\right\vert_{\mathscr{Q}_{-1}^{-1/2}\left(0,-1/2+\check{\delta},0\right)}.
\end{align}

As usual, the implied constants in the estimate~\eqref{1kj21kj} and~\eqref{dkn2lkn3lkn2l} are independent of the implied constants in the various ``$\lesssim \epsilon$'' stated earlier in the proposition.
\end{proposition}
\begin{proof}As usual we employ an iteration argument: We define a sequence of iterates 
\[\{\Omega_{\rm boun}^{(i,k)},\slashed{g}^{(i,k)},\tilde{\mathfrak{n}}^{(i,k)},\mathfrak{j}^{(i,k)},\mathfrak{w}^{(i,k)},\mathfrak{o}^{(j,k)},b^{(i,k)}_{\rm curl},b^{(i)}_{\rm div}\}_{i \in \mathbb{Z}_{\geq 0},k \in \{0,1\}},\]
by setting, for $i = 0$:
\[\Omega_{\rm boun}^{(0,k)} = 1,\qquad \slashed{g}^{(0,k)}  = (v+1)^2\mathring{\slashed{g}},\qquad \mathfrak{n}^{(0,k)} = 0,\qquad \mathfrak{j}^{(0,k)} = 0,\qquad \mathfrak{w}^{(0,k)} = 0,\] 
\[\mathfrak{o}^{(0,k)} = 0,\qquad b^{(0,k)}_{\rm curl} = 0,\qquad b^{(0)}_{\rm div} = 0,\]
while for $i > 0$ we define the iterates via a three step process: 
\begin{enumerate}
	\item We define $\{\Omega_{\rm boun}^{(i,0)},\slashed{g}^{(i,0)},\mathfrak{n}^{(i,0)},\mathfrak{j}^{(i,0)},\mathfrak{w}^{(i,0)},\mathfrak{o}^{(j,0)},b^{(i,0)}_{\rm curl}\}$ by applying Proposition~\ref{3ij2oij3ij2} with $\mathfrak{Z} = \mathring{\rm div}b^{(i-1)}_{\rm div}$. 
	\item  We define $b_{\rm div}^{(i)}$ by applying Proposition~\ref{ij2ojo2432} with $\Omega = \Omega_{\rm sing}\Omega_{\rm boun}^{(i,0)}$, $\slashed{g} = \slashed{g}^{(i,0)}$, and $\mathfrak{Y} = \mathring{\rm curl}b^{(i,0)}_{\rm curl}$.
	\item Then we define $\{\Omega_{\rm boun}^{(i,1)},\slashed{g}^{(i,1)},\mathfrak{n}^{(i,1)},\mathfrak{j}^{(i,1)},\mathfrak{w}^{(i,1)},\mathfrak{o}^{(j,1)},b^{(i,1)}_{\rm curl}\}$ by again applying Proposition~\ref{3ij2oij3ij2}, but this time with $\mathfrak{Z} = \mathring{\rm div}b^{(i)}_{\rm div}$.
\end{enumerate}
Before we state the inductive bound we shall prove, it is convenient to define
\begin{align}\label{3lk3ji4}
&\mathscr{A}^{(i,k)} \doteq \left\vert\left\vert \log\Omega^{(i,k)}_{\rm boun}\right\vert\right\vert_{\mathscr{A}\left(\kappa,b_{\rm curl}^{(i,k)}\right)} + \left\vert\left\vert \log\Omega^{(i,k)}_{\rm boun}\right\vert\right\vert_{\mathscr{B}_{01}\left(\kappa,b_{\rm curl}^{(i,k)}\right)} + \left\vert\left\vert \left(\mathfrak{o}^{(i,k)},\mathcal{L}_{b^{(i,k)}_{\rm curl}|_{v=0}}\mathfrak{o}^{(i,k)}\right)\right\vert\right\vert_{\mathring{H}^{N_1-3}\left(\mathbb{S}^2\right)}
\\ \nonumber &+\left\vert\left\vert \left(\mathfrak{j}^{(i,k)},\mathcal{L}_{b^{(i,k)}_{\rm curl}|_{v=0}}\mathfrak{j}^{(i,k)}\right)\right\vert\right\vert_{\mathring{H}^{N_1-2}\left(\mathbb{S}^2\right)}+ \left\vert\left\vert \left(\mathfrak{w}^{(i,k)},\mathcal{L}_{b^{(i,k)}_{\rm curl}|_{v=0}}\mathfrak{w}^{(i,k)}\right)\right\vert\right\vert_{\mathring{H}^{N_1-3}\left(\mathbb{S}^2\right)}+\left\vert\left\vert \mathcal{P}_{\ell > \ell_0}\slashed{g}^{(i,k)} \right\vert\right\vert_{\mathscr{A}_2\left(\kappa,b^{(i,k)}_{\rm curl},\Omega^{(i,k)}\right)},
\end{align}
\begin{align}\label{3ijo3jio23}
&\mathscr{B}^{(i,k)} \doteq \left\vert\left\vert b^{(i,k)}_{\rm curl}\right\vert\right\vert_{\mathscr{Q}\left(N_1,-3/2+\check{\delta},-\kappa\right)} + \sum_{j=0}^1\left\vert\left\vert \mathcal{L}_{\partial_v}^{1+j} b^{(i,k)}_{\rm curl}\right\vert\right\vert_{\mathscr{Q}\left(N_1-1-j,-1/2+\check{\delta}+j,\kappa +j \right)} + \left\vert\left\vert b^{(i,k)}_{\rm curl}\right\vert\right\vert_{\mathscr{B}_1\left(\kappa\right)},
\end{align}
\begin{align}\label{3oi3io32}
&\mathscr{C}^{(i)} \doteq \left\vert\left\vert b^{(i)}_{\rm div}\right\vert\right\vert_{\mathscr{Q}\left(N_1,-3/2+\check{\delta},-\kappa\right)} + \sum_{j=0}^1\left\vert\left\vert \mathcal{L}_{\partial_v}^{1+j} b^{(i)}_{\rm div}\right\vert\right\vert_{\mathscr{Q}\left(N_1-1-j,-1/2+\check{\delta}+j,\kappa +j \right)} + \left\vert\left\vert b^{(i)}_{\rm div}\right\vert\right\vert_{\mathscr{B}_1\left(\kappa\right)}.
\end{align}
\begin{align}\label{2o3om2o203}
&\mathscr{D}^{(i,k)} \doteq \left\vert\left\vert \slashed{g}^{(i,k)} \right\vert\right\vert_{\mathscr{A}_2\left(\kappa,b^{(i,k)}_{\rm curl},\Omega^{(i,k)}\right)} + \left\vert\left\vert \slashed{g}^{(i,k)}\right\vert\right\vert_{\mathscr{B}_2\left(\kappa,b^{(i,k)}_{\rm curl},\Omega^{(i,k)}\right)}+\left\vert\left\vert \left(v+1\right)\mathfrak{n}^{(i,k)}\right\vert\right\vert_{L,III\left(1/2-\check{\delta},100\check{p}N_1,0\right),N_1-1}
\end{align}

We then claim that the following bound holds for each $i \geq 0$:
\begin{align}\label{2kjh3kjh2}
&\mathscr{A}^{(i,0)} + \mathscr{A}^{(i,1)} + \mathscr{B}^{(i,1)} + \mathscr{C}^{(i)}  \leq C_{\rm boot2}\left( \mathcal{D}+ \mathcal{F}^2+\left\vert\left\vert \pi\right\vert\right\vert^2_{\mathscr{S}\left(0,\check{\delta},0\right)} +\left\vert\left\vert \pi\right\vert\right\vert^2_{\mathscr{Q}_{-1}^{-1/2}\left(0,-1/2+\check{\delta},0\right)}\right),
\end{align}
\begin{align}\label{2k3nl2nlk4n}
&\mathscr{D}^{(i,0)} + \mathscr{D}^{(i,1)}  \leq C_{\rm boot2}\left( \mathcal{D}+ \mathcal{F}^2+\left\vert\left\vert \pi\right\vert\right\vert_{\mathscr{S}\left(0,\check{\delta},0\right)} +\left\vert\left\vert \pi\right\vert\right\vert_{\mathscr{Q}_{-1}^{-1/2}\left(0,-1/2+\check{\delta},0\right)}\right),
\end{align}
\begin{align}\label{32oij32ijo32}
&\mathscr{B}^{(i,0)} \leq C_{\rm boot3}C_{\rm boot2}\left(\mathcal{D} +\mathcal{F}^2+\left\vert\left\vert \pi\right\vert\right\vert^2_{\mathscr{S}\left(0,\check{\delta},0\right)} +\left\vert\left\vert \pi\right\vert\right\vert^2_{\mathscr{Q}_{-1}^{-1/2}\left(0,-1/2+\check{\delta},0\right)}\right).
\end{align}
for  suitable bootstrap constants $C_{\rm boot2}$ and $C_{\rm boot3}$. 

We now proceed to the induction argument. The base case is immediate, so we assume that $j > 0$ and that~\eqref{2kjh3kjh2} and~\eqref{32oij32ijo32} hold for  all $0 \leq i \leq j-1$. 

Applying Proposition~\ref{3ij2oij3ij2} and elliptic theory along $\mathbb{S}^2$,\footnote{We use the elliptic estimates, for example, to go from an estimate on $\slashed{\rm curl}\slashed{\rm div}\slashed{\nabla}\hat{\otimes}\mathcal{L}_{\partial_v}b_{\rm curl}^{(j)}$ and $\mathring{\Pi}_{\rm div}\mathcal{L}_{\partial_v}b^{(j)}_{\rm curl}$ to an estimate for $\mathcal{L}_{\partial_v}b^{(j)}_{\rm curl}$.} we obtain the existence of 
\[\{\Omega_{\rm boun}^{(i,0)},\slashed{g}^{(i,0)},\mathfrak{n}^{(i,0)},\mathfrak{j}^{(i,0)},\mathfrak{w}^{(i,0)},\mathfrak{o}^{(j,0)},b^{(i,0)}_{\rm curl}\},\]
so that
\begin{align}\label{2kj3lkj2lkj34}
\mathscr{A}^{(i,0)}  &\lesssim  \mathcal{D}+ \mathcal{F}^2 +\left(\mathscr{C}^{(i-1)}\right)^2+\left\vert\left\vert \pi\right\vert\right\vert^2_{\mathscr{S}\left(0,\check{\delta},0\right)} +\left\vert\left\vert \pi\right\vert\right\vert^2_{\mathscr{Q}_{-1}^{-1/2}\left(0,-1/2+\check{\delta},0\right)}
\\ \nonumber &\lesssim \mathcal{D} + \mathcal{F}^2+\left\vert\left\vert \pi\right\vert\right\vert^2_{\mathscr{S}\left(0,\check{\delta},0\right)} +\left\vert\left\vert \pi\right\vert\right\vert^2_{\mathscr{Q}_{-1}^{-1/2}\left(0,-1/2+\check{\delta},0\right)},
\end{align}
\begin{align}\label{k23ljlj23}
\mathscr{D}^{(i,0)}  &\lesssim  \mathcal{D}+ \mathcal{F}^2 +\left(\mathscr{C}^{(i-1)}\right)^2+\left\vert\left\vert \pi\right\vert\right\vert_{\mathscr{S}\left(0,\check{\delta},0\right)} +\left\vert\left\vert \pi\right\vert\right\vert_{\mathscr{Q}_{-1}^{-1/2}\left(0,-1/2+\check{\delta},0\right)}
\\ \nonumber  &\lesssim \mathcal{D} + \mathcal{F}^2+\left\vert\left\vert \pi\right\vert\right\vert_{\mathscr{S}\left(0,\check{\delta},0\right)} +\left\vert\left\vert \pi\right\vert\right\vert_{\mathscr{Q}_{-1}^{-1/2}\left(0,-1/2+\check{\delta},0\right)},
\end{align}
\begin{align}\label{3i3oi3}
\mathscr{B}^{(i,0)} &\lesssim \mathcal{D} + \mathcal{F}^2 + \mathscr{C}^{(i-1)}+  \left\vert\left\vert \pi\right\vert\right\vert^2_{\mathscr{S}\left(0,\check{\delta},0\right)} +\left\vert\left\vert \pi\right\vert\right\vert^2_{\mathscr{Q}_{-1}^{-1/2}\left(0,-1/2+\check{\delta},0\right)} 
\\ \nonumber &\lesssim C_{\rm boot2}\left(\mathcal{D}+\mathcal{F}^2+\left\vert\left\vert \pi\right\vert\right\vert^2_{\mathscr{S}\left(0,\check{\delta},0\right)} +\left\vert\left\vert \pi\right\vert\right\vert^2_{\mathscr{Q}_{-1}^{-1/2}\left(0,-1/2+\check{\delta},0\right)}\right). 
\end{align}

We then may apply Proposition~\ref{ij2ojo2432}  and use~\eqref{2kj3lkj2lkj34} and~\eqref{32oij32ijo32} to obtain the existence of $b_{\rm div}^{(j)}$ which satisfies
\begin{align}\label{i2k2mii3}
&\mathscr{C}^{(i)} \lesssim \mathcal{D} +\mathscr{A}^{(i,0)}+ \left(\mathcal{D} + \mathcal{F}^2\right)\mathscr{B}^{(i,0)} \lesssim
\mathcal{D} + \mathcal{F}^2+\left\vert\left\vert \pi\right\vert\right\vert^2_{\mathscr{S}\left(0,\check{\delta},0\right)} +\left\vert\left\vert \pi\right\vert\right\vert^2_{\mathscr{Q}_{-1}^{-1/2}\left(0,-1/2+\check{\delta},0\right)}.
\end{align}

With~\eqref{i2k2mii3}, we apply Proposition~\ref{3ij2oij3ij2} and elliptic estimates along $\mathbb{S}^2$ one more time and obtain the existence of
\[\{\Omega_{\rm boun}^{(i,1)},\slashed{g}^{(i,1)},\mathfrak{n}^{(i,1)},\mathfrak{j}^{(i,1)},\mathfrak{w}^{(i,1)},\mathfrak{o}^{(j,1)},b^{(i,1)}_{\rm curl}\},\]
which satisfy
\begin{align}\label{3l3ji4}
\mathscr{A}^{(i,1)}  &\lesssim  \mathcal{D}+ \mathcal{F}^2 +\left(\mathscr{C}^{(i)}\right)^2+\left\vert\left\vert \pi\right\vert\right\vert^2_{\mathscr{S}\left(0,\check{\delta},0\right)} +\left\vert\left\vert \pi\right\vert\right\vert^2_{\mathscr{Q}_{-1}^{-1/2}\left(0,-1/2+\check{\delta},0\right)}
\\ \nonumber &\lesssim \mathcal{D} + \mathcal{F}^2+\left\vert\left\vert \pi\right\vert\right\vert^2_{\mathscr{S}\left(0,\check{\delta},0\right)} +\left\vert\left\vert \pi\right\vert\right\vert^2_{\mathscr{Q}_{-1}^{-1/2}\left(0,-1/2+\check{\delta},0\right)},
\end{align}
\begin{align}\label{2kknkn22}
\mathscr{D}^{(i,1)}  &\lesssim  \mathcal{D}+ \mathcal{F}^2 +\left(\mathscr{C}^{(i)}\right)^2+\left\vert\left\vert \pi\right\vert\right\vert_{\mathscr{S}\left(0,\check{\delta},0\right)} +\left\vert\left\vert \pi\right\vert\right\vert_{\mathscr{Q}_{-1}^{-1/2}\left(0,-1/2+\check{\delta},0\right)}
\\ \nonumber &\lesssim \mathcal{D} + \mathcal{F}^2+\left\vert\left\vert \pi\right\vert\right\vert_{\mathscr{S}\left(0,\check{\delta},0\right)} +\left\vert\left\vert \pi\right\vert\right\vert_{\mathscr{Q}_{-1}^{-1/2}\left(0,-1/2+\check{\delta},0\right)},
\end{align}
\begin{align}\label{kljjijoijo1ijo}
\mathscr{B}^{(i,1)} &\lesssim \mathcal{D} + \mathcal{F}^2 + \mathscr{C}^{(i)} +\left\vert\left\vert \pi\right\vert\right\vert^2_{\mathscr{S}\left(0,\check{\delta},0\right)} +\left\vert\left\vert \pi\right\vert\right\vert^2_{\mathscr{Q}_{-1}^{-1/2}\left(0,-1/2+\check{\delta},0\right)}
\\ \nonumber  &\lesssim \mathcal{D}+\mathcal{F}^2+\left\vert\left\vert \pi\right\vert\right\vert^2_{\mathscr{S}\left(0,\check{\delta},0\right)} +\left\vert\left\vert \pi\right\vert\right\vert^2_{\mathscr{Q}_{-1}^{-1/2}\left(0,-1/2+\check{\delta},0\right)}. 
\end{align}
Thus, for suitable choices of $C_{\rm boot2}$ and $C_{\rm boot3}$ it is clear that we have closed the induction argument.

As usual, we may easily arrange to take the limit $i \to \infty$ along a suitable sequence and the resulting limiting quantities are the desired solutions. (We note, in particular that one immediately obtains that $b_{\rm curl} = b_{\rm div}$.) However, in order to obtain~\eqref{1kj21kj} it still remains to establish the estimate
\[\left\vert\left\vert \slashed{\rm div}b,\slashed{\nabla}\hat{\otimes}b\right\vert\right\vert_{\mathscr{Q}_{-1/2}^0\left(N_1-1,0,-1/2+\check{\delta}\right)}  \lesssim \mathcal{D} + \mathcal{F}^2+\left\vert\left\vert \pi\right\vert\right\vert^2_{\mathscr{S}\left(0,\check{\delta},0\right)} +\left\vert\left\vert \pi\right\vert\right\vert^2_{\mathscr{Q}_{-1}^{-1/2}\left(0,-1/2+\check{\delta},0\right)},\]
for these limiting quantities. For $\slashed{\rm div}b$ this already follows immediately from Proposition~\ref{ij2ojo2432}. However, from Proposition~\ref{3ij2oij3ij2}, we only have that 
\begin{equation}\label{3oi4oi3}
\left\vert\left\vert \slashed{\rm curl}\slashed{\rm div}\slashed{\nabla}\hat{\otimes}b\right\vert\right\vert_{\mathscr{Q}_{-1/2}^0\left(N_1-3,0,-1/2+\check{\delta}\right)}  \lesssim \mathcal{D} + \mathcal{F}^2+\left\vert\left\vert \pi\right\vert\right\vert^2_{\mathscr{S}\left(0,\check{\delta},0\right)} +\left\vert\left\vert \pi\right\vert\right\vert^2_{\mathscr{Q}_{-1}^{-1/2}\left(0,-1/2+\check{\delta},0\right)}.
\end{equation}
The key observation is that in view of the formula~\eqref{3o3oioi4} and the fact that we have already established
\[\left\vert\left\vert \mathcal{P}_{\ell > \ell_0}\slashed{g} \right\vert\right\vert_{\mathscr{A}_2\left(\kappa,b,\Omega\right)}  \lesssim \mathcal{D} + \mathcal{F}^2+\left\vert\left\vert \pi\right\vert\right\vert^2_{\mathscr{S}\left(0,\check{\delta},0\right)} +\left\vert\left\vert \pi\right\vert\right\vert^2_{\mathscr{Q}_{-1}^{-1/2}\left(0,-1/2+\check{\delta},0\right)},\]
we have that 
\begin{align}\label{3k2oo4}
&\left\vert\left\vert \slashed{\rm div}\slashed{\rm div}\slashed{\nabla}\hat{\otimes}b\right\vert\right\vert_{\mathscr{Q}_{-1/2}^0\left(N_1-3,0,-1/2+\check{\delta}\right)} 
\\ \nonumber &\qquad  \lesssim \left\vert\left\vert \slashed{\rm div}b\right\vert\right\vert_{\mathscr{Q}_{-1/2}^0\left(N_1-1,0,-1/2+\check{\delta}\right)} + \left\vert\left\vert \mathcal{L}_bK \right\vert\right\vert_{\mathscr{Q}_{-1/2}^0\left(N_1-3,0,-1/2+\check{\delta}\right)}+\mathcal{D} + \mathcal{F}^2
\\ \nonumber &\qquad \qquad +\left\vert\left\vert \pi\right\vert\right\vert^2_{\mathscr{S}\left(0,\check{\delta},0\right)} +\left\vert\left\vert \pi\right\vert\right\vert^2_{\mathscr{Q}_{-1}^{-1/2}\left(0,-1/2+\check{\delta},0\right)}
\\ \nonumber &\qquad \lesssim \mathcal{D} + \mathcal{F}^2+\left\vert\left\vert \pi\right\vert\right\vert^2_{\mathscr{S}\left(0,\check{\delta},0\right)} +\left\vert\left\vert \pi\right\vert\right\vert^2_{\mathscr{Q}_{-1}^{-1/2}\left(0,-1/2+\check{\delta},0\right)}.
\end{align}
(As we have seen in many other places, our estimate is rescued by a potentially dangerous term being a $\mathcal{L}_b$ derivative to leading order.)

Now, by yet another elliptic estimate, we have
\begin{align*}
&\left\vert\left\vert \slashed{\nabla}\hat{\otimes}b\right\vert\right\vert_{\mathscr{Q}_{-1/2}^0\left(N_1-1,0,-1/2+\check{\delta}\right)} \\ \nonumber &\qquad \lesssim \left\vert\left\vert \slashed{\rm curl}\slashed{\rm div}\slashed{\nabla}\hat{\otimes}b\right\vert\right\vert_{\mathscr{Q}_{-1/2}^0\left(N_1-3,0,-1/2+\check{\delta}\right)}+ \left\vert\left\vert \slashed{\rm div}\slashed{\rm div}\slashed{\nabla}\hat{\otimes}b\right\vert\right\vert_{\mathscr{Q}_{-1/2}^0\left(N_1-3,0,-1/2+\check{\delta}\right)}+\mathcal{D} + \mathcal{F}^2+\\ \nonumber &\qquad \qquad\left\vert\left\vert \pi\right\vert\right\vert^2_{\mathscr{S}\left(0,\check{\delta},0\right)} +\left\vert\left\vert \pi\right\vert\right\vert^2_{\mathscr{Q}_{-1}^{-1/2}\left(0,-1/2+\check{\delta},0\right)}
\\ \nonumber &\qquad \lesssim \mathcal{D} + \mathcal{F}^2+\left\vert\left\vert \pi\right\vert\right\vert^2_{\mathscr{S}\left(0,\check{\delta},0\right)} +\left\vert\left\vert \pi\right\vert\right\vert^2_{\mathscr{Q}_{-1}^{-1/2}\left(0,-1/2+\check{\delta},0\right)}.
\end{align*}
This concludes the proof.
\end{proof}

Finally, we come to our last iteration argument where we will argue that we can take $\mathfrak{q} = \slashed{\rm div}b$ and $\mathfrak{r} = \slashed{\nabla}\hat{\otimes}b$.
\begin{lemma}Let $H_1$, $H_2$, $H_4$, $H_5$, $H_6$,  $\pi$, $H_7$, $H_8$, and $H_9$ be as in Proposition~\ref{3ij2oio23}. Then we can find a solution $\Omega_{\rm boun}$, $\slashed{g}$, $b$, and $\mathfrak{n}_A$ which satisfy the same equations and estimates as in Proposition~\ref{3ij2oio23} except that we now also have
\begin{equation}\label{23m4rion2}
\mathfrak{q} = \slashed{\rm div}b,\qquad \mathfrak{r} =\slashed{\nabla}\hat{\otimes}b.
\end{equation}
\end{lemma}
\begin{proof}The key observation is that if we consider the shift vector $b$ produced by an application of Proposition~\ref{3ij2oio23}, then if we define $\mathfrak{q}$ and $\mathfrak{r}$ by~\eqref{23m4rion2}, then $\mathfrak{q}$ and $\mathfrak{r}$ will satisfy the required condition~\eqref{32l342lj42oij34}. 

This observation then leads to the following iteration. We define $\{\Omega_{\rm boun}^{(i)},\slashed{g}^{(i)},b^{(i)},\mathfrak{n}^{(i)},\mathfrak{q}^{(i)},\mathfrak{r}^{(i)}\}_{i=0}^{\infty}$ by setting 
\[\Omega_{\rm boun}^{(0)} = 0,\ \slashed{g}^{(0)} = (v+1)^2\mathring{\slashed{g}},\ b^{(0)} = 0, \mathfrak{n}^{(0)} = 0,\]
then setting
\[\mathring{q}^{(i)} \doteq \slashed{\rm div}^{(i)}b^{(i)},\qquad \mathfrak{r}^{(i)} \doteq \left(\slashed{\nabla}\hat{\otimes}\right)^{(i)}b^{(i)},\]
and finally defining $\{\Omega_{\rm boun}^{(i)},\slashed{g}^{(i)},b^{(i)},\mathfrak{n}^{(i)}\}_{i=1}^{\infty}$ by an application on Proposition~\ref{3ij2oio23} with $\mathfrak{q} = \mathfrak{q}^{(i-1)}$, $\mathfrak{r} = \mathfrak{r}^{(i-1)}$. One easily extracts a suitable convergent subsequence, and the limit provides the desired solution.
\end{proof}

\section{Adding in the Nonlinear Terms for $\Omega_{\rm boun}$ and $b$}\label{2i3oij2o3}
In this section we carry out some further steps in our iteration process. Throughout this section we will let $\Omega_{\rm sing}(v) : (-1,0) \to (0,\infty)$ be a given spherically symmetric function and $\pi_A$ be an $\mathbb{S}^2_{-1,v}$ $1$-form for $v\in (-1,0)$ satisfying $\left(1-\mathcal{P}_{\ell \leq \ell_0}\right)\pi = 0$. We will assume that for a suitable constant $\kappa$ satisfying $\left|\kappa\right| \lesssim \epsilon$, we have
\begin{align}\label{jk32kj32k}
& \left\vert\left\vert \log\Omega_{\rm sing}\right\vert\right\vert_{\mathscr{B}_{00}\left(\kappa\right)}  + \left\vert\left\vert \pi\right\vert\right\vert_{\mathscr{S}\left(0,0,0\right)}  \lesssim \epsilon.
\end{align}
We emphasize that none of the results in this section depend on the implied constants in~\eqref{jk32kj32k} or in the inequality for $\kappa$ (though by our conventions for $\epsilon$, we may assume that $\epsilon$ is sufficiently small depending on the implied constants).

We start by revisiting Proposition~\ref{3ij2oio23} and replacing $H_7$, $H_8$, and $H_9$ with suitable nonlinear expressions. In this next lemma we record the relevant nonlinear estimate.
\begin{lemma}\label{3io3oioi23}Let $\left(b,\slashed{g},\Omega_{\rm boun}\right)$ be given and assumed to satisfy 
\begin{align}\label{3o23o2}
&\left\vert\left\vert \log\Omega_{\rm boun}\right\vert\right\vert_{\mathscr{A}\left(\kappa,b\right)} + \left\vert\left\vert \log\Omega_{\rm boun}\right\vert\right\vert_{\mathscr{B}_{01}\left(\kappa,b\right)} 
\\ \nonumber&\qquad +\left\vert\left\vert \slashed{g} \right\vert\right\vert_{\mathscr{A}_2\left(\kappa,b,\Omega\right)} + \left\vert\left\vert \slashed{g}\right\vert\right\vert_{\mathscr{B}_2\left(\kappa,b,\Omega\right)}
 +\left\vert\left\vert b\right\vert\right\vert_{\mathscr{A}_1\left(\kappa,\slashed{g}\right)}+\left\vert\left\vert b\right\vert\right\vert_{\mathscr{B}_1\left(\kappa\right)}
  \lesssim \epsilon,
\end{align}
Then we define $H_7\left[\Omega,b,\slashed{g}\right] $, $H_8\left[\Omega,b,\slashed{g}\right]$, and $H_9\left[\Omega,b,\slashed{g}\right]$ by
\begin{align}\label{i2omo4}
&H_7\left[\Omega,b,\slashed{g}\right] \doteq  H_8\left[\Omega,b,\slashed{g}\right] \doteq
\\ \nonumber & -\frac{1}{4}\zeta \cdot \left(\slashed{\nabla}\hat{\otimes}b\right) + \frac{1}{2}\slashed{\nabla}\log\Omega\left(-2+\frac{1}{2}\slashed{\rm div}b\right)-\zeta\left(\frac{1}{2}(-v)\Omega\hat{\chi} +\frac{3}{2}(-v)\Omega^2\left(\Omega^{-1}{\rm tr}\chi - 2(v+1)^{-1}\right)-\frac{1}{2}\Omega\hat{\underline{\chi}} - \frac{3}{4}\slashed{\rm div}b\right),
\end{align}
\begin{align}\label{2ij3iojjio3}
&2H_9\left[\slashed{g},b,\mathfrak{n},\Omega\right]_{AB} \doteq 
\\ \nonumber &\qquad\left(1+v\right)^{-1}\left(  \left[\frac{v}{u}\Omega\nabla_4-\frac{1}{2}\mathcal{L}_b,\Omega^{-2}\left(1+v\right)\mathcal{L}_b\right]\left(\Omega\hat{\chi}\right) + 2\frac{v}{u}\left[\left(1+v\right)\slashed{\nabla}\hat{\otimes},\Omega\nabla_4\right]\zeta + \left[\mathcal{L}_b,\left(1+v\right)\slashed{\nabla}\hat{\otimes}\right]\zeta\right)
\\ \nonumber &\qquad +2\Omega^{-2}\left(1-\frac{v}{u}\right)\mathcal{L}_b\log\Omega \left(\frac{v}{u}\Omega\nabla_4 - \frac{1}{2}\mathcal{L}_b\right)\left(\Omega\hat{\chi}\right)
\\ \nonumber &\qquad  -\mathcal{L}_b\Bigg(\left(\slashed{\nabla}\hat{\otimes}b\right)^C_{\ \ (A}\left(\Omega^{-1}\hat{\chi}\right)_{B)C} 
 +\frac{1}{2}{\rm div}b\left(\Omega^{-1}\hat{\chi}\right)_{AB}+\frac{v}{u}\Omega{\rm tr}\chi\left(\Omega^{-1} \hat{\chi}\right)_{AB} 
 \\ \nonumber &\qquad +\left(\left(\slashed{\nabla}\hat\otimes \slashed{\nabla}\log\Omega\right)_{AB} + \left(\mathfrak{n}\hat\otimes \mathfrak{n}\right)_{AB} - \frac{1}{2}\Omega^{-1}{\rm tr}\chi\left(\slashed{\nabla}\hat{\otimes}b\right)_{AB}\right)\Bigg)
 \\ \nonumber &\qquad -\frac{1}{2} 3\left(v+1\right)^{-2}(-v)\left(\slashed{\nabla}\hat{\otimes}\left(\Omega^{-2}\right) \right)\zeta
 \\ \nonumber &\qquad -2(-v)\left(\Omega^2-1\right)(v+1)^{-1}\left(\Omega^{-2}\nabla_v\left(\slashed{\nabla}\hat{\otimes}b\right)-2\Omega^{-2}\mathcal{L}_b\left(\Omega\hat{\chi}\right) +8\slashed{\nabla}\log\Omega\hat{\otimes}\zeta -4 \Omega^{-2}\mathcal{E}\right),
\end{align}
where $\mathcal{E}$ is defined by~\eqref{2io34iojijo}.

Let $\mathscr{Z}$ denote the sum of the left hand hand of~\eqref{1kj21kj} and~\eqref{dkn2lkn3lkn2l}. Then we have 

\begin{align*}
& \left\vert\left\vert \left((-v)^{-2\kappa}\mathring{\rm curl}H_7,0,0\right)\right\vert\right\vert_{R,II,N_1-1}  +\left\vert\left\vert \left((-v)^{-2\kappa}\mathring{\Pi}_{\rm curl}H_8,0,0\right)\right\vert\right\vert_{R,III'\left(1/2-\check{\delta},50\check{p},0\right),N_1}
\\ \nonumber &\qquad +\sum_{j=0}^1\left\vert\left\vert \left(v\mathcal{L}_{\partial_v}\right)^j\mathring{\Pi}_{\rm curl}H_7\right\vert\right\vert_{\check{\mathscr{S}}_{-1/2}^0\left(0,0,50\check{p}\left(1+j\right),0\right)}
 +\left\vert\left\vert \left((-v)^{-2\kappa}H_9,0,0\right)\right\vert\right\vert_{R,III'\left(1/2-\check{\delta},50\check{p},100\check{p}\right),N_1-3}
 \\ \nonumber &\qquad + \sum_{\left|\alpha\right| \leq 1}\sum_{j=0}^1\left\vert\left\vert \left(v\mathcal{L}_{\partial_v}\right)^j\mathcal{L}_{\mathcal{Z}^{(\alpha)}}\mathring{\Delta}^{-1}H_9\right\vert\right\vert_{\check{\mathscr{S}}_{-1/2}^0\left(N_2-1-j,0,50\check{p}\left(1+j\right),50\check{p}\right)} \lesssim \mathscr{Z}^2,
 \end{align*}
 \begin{align*}
& \left\vert\left\vert \left((-v)^{-2\kappa}\mathring{\rm div}H_7,0,0\right)\right\vert\right\vert_{R,II,N_1-1} +\sum_{j=0}^1\left\vert\left\vert \left(v\mathcal{L}_{\partial_v}\right)^j\mathring{\Pi}_{\rm div}H_7\right\vert\right\vert_{\check{\mathscr{S}}_{-1/2}^0\left(0,0,50\check{p}\left(1+j\right),0\right)} \\ \nonumber &\qquad +\left\vert\left\vert \left((-v)^{-2\kappa}\mathring{\Pi}_{\rm div}H_8,0,0\right)\right\vert\right\vert_{R,III'\left(1/2-\check{\delta},50\check{p},0\right),N_1}+\sum_{j=0}^1\left\vert\left\vert \left(v\mathcal{L}_{\partial_v}\right)^j\mathring{\Pi}_{\rm div}H_7\right\vert\right\vert_{\check{\mathscr{S}}_{-1/2}^0\left(0,0,50\check{p}\left(1+j\right),0\right)} \lesssim \mathscr{Z}.
 \end{align*}
The implied constants in these estimates are independent of the implied constant in~\eqref{3o23o2}.
\end{lemma}
\begin{proof}These are all straightforward if tedious applications of the inequalities from Section~\ref{inequalitysectionsection}.
\end{proof}

With the estimates from Lemma~\ref{3io3oioi23} in hand, it is now straightforward to revisit Proposition~\ref{3ij2oio23} and input the relevant expressions for $H_7$, $H_8$, and $H_9$. Furthermore, we will then be able to carry out a more refined analysis and find that $\eta$ has an improved $\sup_v$ bound than what is obtained by simply treating $\eta$ as a sum of $\zeta$ and $\slashed{\nabla}\log\Omega$. More specifically, we will establish an estimate which will eventually show that $\eta$ and a suitable number of angular derivatives thereof will remain bounded in $L^2\left(\mathbb{S}^2\right)$ uniformly as $v\to 0$. 
\begin{proposition}\label{ioj32iojjio} Let $H_1$, $H_2$, $H_4$, $H_5$, $H_6$  be functions in the closure of smooth functions under the norm on the right hand side of~\eqref{k2mkmk4k} and~\eqref{32ij32ijo} and have a sufficiently small norm. Then there exist a function $\Omega_{\rm boun}: (-1,0)\times \mathbb{S}^2 \to (0,\infty)$, an $\mathbb{S}^2_{-1,v}$-vector field $b^A$, an $\mathbb{S}^2_{-1,v}$-symmetric $(0,2)$ tensor $\slashed{g}$, and an $\mathbb{S}^2_{-1,v}$ $1$-form $\mathfrak{n}_A$  such that all of the equations of Proposition~\ref{3ij2oio23} hold except that we replace $H_7$, $H_8$, and $H_9$ with the corresponding expressions from Lemma~\ref{3io3oioi23}. Define $\mathscr{V}$ by
\begin{align*}
&\mathscr{V} \doteq  \left\vert\left\vert \left(H_1,H_2,0\right)\right\vert\right\vert_{R,I,N_1-2} + \epsilon \left\vert\left\vert \left(T_{\rm low},T_{\rm high}\right)\right\vert\right\vert_{\mathring{H}^{N_1-3}} 
 +\left\vert\left\vert \log\Omega_{\rm sing}\right\vert\right\vert_{\mathscr{B}_{00}\left(\kappa\right)}+\left\vert\left\vert \left((-v)^{-2\kappa}H_5,H_6\right)\right\vert\right\vert_{\mathscr{P}\mathscr{R}\left(\Omega,\kappa,50\check{p},50\check{p}\right)}
 \\ \nonumber &\qquad  + \left\vert\left\vert H_4\right\vert\right\vert_{\mathscr{Q}\left(N_1-2,1/2+5\check{\delta},-\kappa\right)}
+ \left\vert\left\vert H_4\right\vert\right\vert_{\mathscr{S}_{-1}^{-1/2}\left(N_1-3,1+5\check{\delta},0\right)}+ \left\vert\left\vert H_4\right\vert\right\vert_{\mathscr{S}_{-1/2}^0\left(N_2-2,0,0\right)} \
\end{align*}

Assume that $\mathscr{V} \lesssim \epsilon$. We then have 
\begin{align}\label{k2mkmk4k}
&\left\vert\left\vert \log\Omega_{\rm boun}\right\vert\right\vert_{\mathscr{A}\left(\kappa,b\right)} + \left\vert\left\vert \log\Omega_{\rm boun}\right\vert\right\vert_{\mathscr{B}_{01}\left(\kappa,b\right)} 
\\ \nonumber &\qquad  +\left\vert\left\vert b\right\vert\right\vert_{\mathscr{A}_1\left(\kappa,\slashed{g}\right)}+\left\vert\left\vert b\right\vert\right\vert_{\mathscr{B}_1\left(\kappa\right)}
+\left\vert\left\vert \left(v+1\right)\mathfrak{n}\right\vert\right\vert_{L,III\left(1/2-\check{\delta},100\check{p}N_1,0\right),N_1-1}
+ \left\vert\left\vert \left(\mathfrak{o},\mathcal{L}_{b|_{v=0}}\mathfrak{o}\right)\right\vert\right\vert_{\mathring{H}^{N_1-3}\left(\mathbb{S}^2\right)}
\\ \nonumber &\qquad  +\left\vert\left\vert \left(\mathfrak{j},\mathcal{L}_{b|_{v=0}}\mathfrak{j}\right)\right\vert\right\vert_{\mathring{H}^{N_1-2}\left(\mathbb{S}^2\right)}+ \left\vert\left\vert \left(\mathfrak{w},\mathcal{L}_{b|_{v=0}}\mathfrak{w}\right)\right\vert\right\vert_{\mathring{H}^{N_1-3}\left(\mathbb{S}^2\right)}
 \lesssim \mathscr{V} + \left\vert\left\vert \pi\right\vert\right\vert^2_{\mathscr{S}\left(0,\check{\delta},0\right)} +\left\vert\left\vert \pi\right\vert\right\vert^2_{\mathscr{Q}_{-1}^{-1/2}\left(0,-1/2+\check{\delta},0\right)},
\end{align}
\begin{align}\label{kj1lkjlk}
& \left\vert\left\vert \slashed{g} \right\vert\right\vert_{\mathscr{A}_2\left(\kappa,b,\Omega\right)} + \left\vert\left\vert \slashed{g}\right\vert\right\vert_{\mathscr{B}_2\left(\kappa,b,\Omega\right)}
+\left\vert\left\vert \left(v+1\right)\mathfrak{n}\right\vert\right\vert_{L,III\left(1/2-\check{\delta},100\check{p}N_1,0\right),N_1-1} 
\\ \nonumber &\qquad \lesssim \mathscr{V} + \left\vert\left\vert \pi\right\vert\right\vert_{\mathscr{S}\left(0,\check{\delta},0\right)} +\left\vert\left\vert \pi\right\vert\right\vert_{\mathscr{Q}_{-1}^{-1/2}\left(0,-1/2+\check{\delta},0\right)}.
\end{align}

We furthermore have the following improved bound for  $\eta = -\frac{1}{4}\mathcal{L}_{\partial_v}b + \slashed{\nabla}\log\Omega$:
\begin{align}\label{32ij32ijo}
&\sum_{j=0}^1\left\vert\left\vert \left(v\mathcal{L}_{\partial_v}\right)^j\eta\right\vert\right\vert_{\mathscr{S}_{-1/2}^0\left(N_2-1-j,0,0\right)} \\ \nonumber &\qquad \lesssim \mathscr{V} + \sum_{j=0}^1\left\vert\left\vert \left(v\mathcal{L}_{\partial_v}\right)^j\left(H_5,H_6\right)\right\vert\right\vert_{\mathscr{S}_{-1/2}^0\left(N_2-2-j,0,0\right)} + \left\vert\left\vert \pi\right\vert\right\vert^2_{\mathscr{S}\left(0,\check{\delta},0\right)} +\left\vert\left\vert \pi\right\vert\right\vert^2_{\mathscr{Q}_{-1}^{-1/2}\left(0,-1/2+\check{\delta},0\right)}.
\end{align}
The bounds in these inequalities are independent of the implied constant in the assumption $\mathscr{V} \lesssim \epsilon$. 
\end{proposition}
\begin{proof}The existence of our solution and the bounds~\eqref{k2mkmk4k} and~\eqref{kj1lkjlk} are an immediate consequence of an iteration argument and the estimates from Lemma~\ref{3io3oioi23}. We thus turn to establishing~\eqref{32ij32ijo}. 

We will separately estimate $\mathcal{P}_{\ell \geq 1}\slashed{\rm div}\eta$ and $\mathcal{P}_{\ell \geq 1}\slashed{\rm curl}\eta$. The argument for the $\slashed{\rm div}\eta$ is more straightforward. We simply re-write the equation~\eqref{3lkl2jllk32lk} as  
\begin{align}\label{iio3j2ijo3ioj}
&\mathcal{P}_{\ell \geq 1}\slashed{\rm div}\eta -\frac{1}{4} \mathcal{P}_{\ell \geq 1}\left(\Omega^{-1}{\rm tr}\chi \slashed{\rm div}b\right) + \mathcal{P}_{\ell \geq 1}\left( \left(v+1\right)^{-2}\mathring{\Delta}\slashed{\rm div}b\right) 
 = \mathcal{P}_{\ell \geq 1}\Bigg(\left(H_5-2Y\right) + H_6 + \tilde{H}_6\Bigg).
\end{align}
Using also Lemma~\ref{2ij23ji3ijo32} and the inequality~\eqref{k2mkmk4k}, we thus obtain
\begin{align}\label{23oi4io4}
&\sum_{j=0}^1\left\vert\left\vert \left(v\mathcal{L}_{\partial_v}\right)^j\mathcal{P}_{\ell \geq 1}\slashed{\rm div}\eta \right\vert\right\vert_{\mathscr{S}_{-1/2}^0\left(N_2-2-j,0,0\right)}
\\ \nonumber &\qquad  \lesssim \mathscr{V} + \sum_{j=0}^1\left\vert\left\vert \left(v\mathcal{L}_{\partial_v}\right)^j\left(H_5,H_6\right)\right\vert\right\vert_{\mathscr{S}_{-1/2}^0\left(N_2-2-j,0,0\right)} +\left\vert\left\vert \pi\right\vert\right\vert^2_{\mathscr{S}\left(0,\check{\delta},0\right)} +\left\vert\left\vert \pi\right\vert\right\vert^2_{\mathscr{Q}_{-1}^{-1/2}\left(0,-1/2+\check{\delta},0\right)}.
\end{align}

We now turn to $\slashed{\rm curl}\eta$. We may form a Lorentzian metric $g$ from our $\left(\Omega,b,\slashed{g}\right)$. It is a consequence of the specific forms $H_7$, $H_8$, and $H_9$ and the formulas from Lemmas~\ref{3kdo2} and~\ref{emfkeo3} that the Ricci tensor of this metric will satisfy
\begin{equation}\label{32io32io2o2}
\mathcal{P}_{\ell > \ell_0}\slashed{\rm curl}\slashed{\rm div}\left[\slashed{\nabla}\hat{\otimes}\mathscr{W} +\mathcal{L}_b\widehat{\rm Ric}\right] = 0,\qquad \mathcal{P}_{1 \leq \ell \leq \ell_0}\slashed{\rm curl}\mathscr{W} = 0,
\end{equation}
where $\mathscr{W}_A \doteq v\Omega{\rm Ric}_{A4}-\Omega{\rm Ric}_{3A}$ and $\widehat{\rm Ric}_{AB}$ is the trace-free part of the symmetric $(0,2)$-$\mathbb{S}^2_{-1,v}$ tensor induced by the Ricci tensor. We further have from Lemma~\ref{Riccifromglsashsn} that 
\begin{equation}\label{2io12ijo12io}
(\slashed{\nabla}\hat{\otimes}\left(\mathfrak{n}-\eta\right))_{AB}+\left(\mathfrak{n}\hat{\otimes}\mathfrak{n}\right)_{AB} - \left(\eta\hat{\otimes}\eta\right)_{AB} - \widehat{\rm Ric}_{AB} = 0.
\end{equation}
On the other hand, in view of Lemma~\ref{2mo2o3o2},~\eqref{4eta}, and~\eqref{tcod1}, we also have that
\begin{align}\label{32ioj3ijo32}
&2(-v)\nabla_4\eta - \mathcal{L}_b\eta -\eta\left(-2+\slashed{\rm div}b +v\Omega{\rm tr}\chi+(-v)\Omega\chi\right) = \\ \nonumber &\qquad 4\slashed{\nabla}\left(\Omega\underline{\omega}\right) + \slashed{\rm div}\left(\frac{1}{2}\slashed{\nabla}\hat{\otimes}b +v\Omega\hat{\chi}\right) -\frac{1}{2}\slashed{\nabla}\left(\slashed{\rm div}b+v\Omega{\rm tr}\chi\right) +v\Omega\chi\left(\eta-\underline{\eta}\right) + (-v)\Omega\slashed{\rm div}\hat{\chi} 
\\ \nonumber &\qquad +v\frac{1}{2}\Omega\slashed{\nabla}{\rm tr}\chi +v\frac{1}{2}\Omega {\rm tr}\chi \zeta -v \Omega \zeta \hat{\chi} + \mathscr{W} \doteq \mathscr{J} + \mathscr{W}.
\end{align}
Next, we observe that as a consequence of~\eqref{k2mkmk4k}, we have that
\begin{equation}\label{3ioj3io32}
\left\vert\left\vert \mathscr{J}\right\vert\right\vert_{\mathscr{S}_{-1/2}^0\left(N_2-1,0,0\right)} \lesssim \mathscr{V}+ \left\vert\left\vert \pi\right\vert\right\vert^2_{\mathscr{S}\left(0,\check{\delta},0\right)} +\left\vert\left\vert \pi\right\vert\right\vert^2_{\mathscr{Q}_{-1}^{-1/2}\left(0,-1/2+\check{\delta},0\right)}.
\end{equation}
Combining the information in~\eqref{32io32io2o2}-\eqref{3ioj3io32}, we may commute~\eqref{32ioj3ijo32} with $\mathcal{P}_{1\leq \ell \leq \ell_0}\slashed{\rm curl}$ and $\mathcal{P}_{\ell > \ell_0}\slashed{\rm curl}\slashed{\rm div}\slashed{\nabla}\hat{\otimes}$  and obtain that  
\begin{equation}\label{32ijo32jio}
2(-v)\mathcal{L}_{\partial_v}\mathcal{P}_{1\leq \ell \leq \ell_0}\slashed{\rm curl}\eta -\mathcal{P}_{1\leq \ell \leq \ell_0}\mathcal{L}_b\mathcal{P}_{1\leq \ell \leq \ell_0}\slashed{\rm curl}\eta +2\mathcal{P}_{1\leq \ell \leq \ell_0}\slashed{\rm curl}\eta = \mathcal{P}_{1\leq \ell \leq \ell_0}\mathscr{G}_1,
\end{equation}
\begin{equation}\label{32joi3ijo4}
2(-v)\mathcal{L}_{\partial_v}\mathcal{P}_{\ell > \ell_0}\slashed{\rm curl}\slashed{\rm div}\slashed{\nabla}\hat{\otimes}\eta -2\mathcal{P}_{\ell > \ell_0}\mathcal{L}_b\mathcal{P}_{\ell > \ell_0}\slashed{\rm curl}\slashed{\rm div}\slashed{\nabla}\hat{\otimes}\eta +2\mathcal{P}_{\ell > \ell_0}\slashed{\rm curl}\slashed{\rm div}\slashed{\nabla}\hat{\otimes}\eta = \mathscr{G}_2,
\end{equation}
where, in view of elliptic estimates,~\eqref{23oi4io4},~\eqref{3ioj3io32}, and~\eqref{k2mkmk4k}, we have that  $\mathscr{G}_1$ and $\mathscr{G}_2$ satisfy, for each $v \in (-1/2,0)$:
\begin{align}
&\left\vert\left\vert \mathscr{G}_1\right\vert\right\vert_{L^2\left(\mathbb{S}^2_{-1,v}\right)} + \left\vert\left\vert \mathscr{G}_2\right\vert\right\vert_{\mathring{H}^{N_2-4}\left(\mathbb{S}^2_{-1,v}\right)}  \lesssim
\\ \nonumber &\qquad  \mathscr{V} + \left\vert\left\vert \pi\right\vert\right\vert^2_{\mathscr{S}\left(0,\check{\delta},0\right)} +\left\vert\left\vert \pi\right\vert\right\vert^2_{\mathscr{Q}_{-1}^{-1/2}\left(0,-1/2+\check{\delta},0\right)}+ \left\vert\left\vert \left(H_5,H_6\right)\right\vert\right\vert_{\mathscr{S}_{-1/2}^0\left(N_2-2,0,0\right)} 
\\ \nonumber &\qquad + \left\vert\left\vert \mathcal{P}_{1 \leq \ell \leq \ell_0}\slashed{\rm curl}\eta \right\vert\right\vert_{L^2\left(\mathbb{S}^2_{-1,v}\right)}^2 + \left\vert\left\vert \mathcal{P}_{\ell > \ell_0}\slashed{\rm curl}\slashed{\rm div}\slashed{\nabla}\hat{\otimes} \eta \right\vert\right\vert_{\mathring{H}^{N_2-4}\left(\mathbb{S}^2_{-1,v}\right)}^2.
\end{align}
In particular, after conjugating~\eqref{32ijo32jio} and~\eqref{32joi3ijo4}  with a cut-off which vanishes near $v = -1$ and is identically $1$ near $v = 0$, we see that~\eqref{32ij32ijo} now follows from the transport equation estimate~\eqref{21jnkjnkjn} of Lemma~\ref{linftofkwp3} and elliptic estimates.

\end{proof}

Now we will prove a version of Proposition~\ref{ioj32iojjio} where we will now replace $H_1$, $H_2$, $H_4$, $H_5$, and $H_6$ with suitable expressions involving $\Omega$, $b$, and $\slashed{g}$. The following lemma will be the analogue of Lemma~\ref{3io3oioi23} for this step.
\begin{lemma}\label{3oi32io}Let $\left(b,\slashed{g},\Omega_{\rm boun}\right)$ be given and assumed to satisfy 
\begin{align}\label{1kl2lkj124}
&\left\vert\left\vert \log\Omega_{\rm boun}\right\vert\right\vert_{\mathscr{A}\left(\kappa,b\right)} + \left\vert\left\vert \log\Omega_{\rm boun}\right\vert\right\vert_{\mathscr{B}_{01}\left(\kappa,b\right)} 
\\ \nonumber&\qquad +\left\vert\left\vert \slashed{g} \right\vert\right\vert_{\mathscr{A}_2\left(\kappa,b,\Omega\right)} + \left\vert\left\vert \slashed{g}\right\vert\right\vert_{\mathscr{B}_2\left(\kappa,b,\Omega\right)}
 +\left\vert\left\vert b\right\vert\right\vert_{\mathscr{A}_1\left(\kappa,\slashed{g}\right)}+\left\vert\left\vert b\right\vert\right\vert_{\mathscr{B}_1\left(\kappa\right)}
  \lesssim \epsilon.
\end{align}
Then we define
\begin{align}\label{3oij32io3iot}
&H_1\left[\Omega,b,\slashed{g}\right] = \left(1+v\right)^2\mathcal{P}_{\ell \geq 1}\Bigg(\Omega^2\Bigg[H_4\left[\Omega,b,\slashed{g}\right]\Bigg]\Bigg)
 \\ \nonumber &\qquad + \left(v+1\right)^2\mathcal{P}_{\ell \geq 1}\left(\Omega^2\left(\left((-v)\Omega{\rm tr}\chi - 2(-v)(v+1)^{-1}\right)+\frac{3}{2}\left[\mathcal{P}_{\ell \geq 1},\slashed{\rm div}b\right]\right)Y \right),
\end{align}
\begin{align}\label{iw4ieiegra}
&H_2\left[\Omega,b,\slashed{g}\right] \doteq \left(v+1\right)^2\mathcal{P}_{\ell \geq 1}\left(\left(4\Omega {\rm tr}\chi - 2(v+1)^{-1}\right)\Omega\underline{\omega} + 4\Omega^2\underline{\eta}\cdot\eta -\frac{1}{2}\left(\Omega\hat{\chi}\right)\cdot\left(\slashed{\nabla}\hat{\otimes}b\right)\right),
\end{align}
\begin{align}\label{3ijo3jio23io2}
&H_4\left[\Omega,b,\slashed{g}\right] =
\\ \nonumber &\qquad v\left(\mathcal{L}_{\partial_v}b\right)^A\slashed{\nabla}_A\log\Omega\left(\Omega^{-1}{\rm tr}\chi\right) -2\mathcal{L}_b\left(\left(\Omega\underline{\omega}\right)\left(\Omega^{-1}{\rm tr}\chi\right)\right) 
\\ \nonumber &\qquad \qquad + \mathcal{L}_b\log\Omega\left((-v)\mathcal{L}_{\partial_v}\left(\Omega^{-1}{\rm tr}\chi\right) +v\left(\Omega^{-1}{\rm tr}\chi\right)^2\right)
\\ \nonumber &\qquad \qquad  -2\left(\Omega\underline{\omega}\right)\left(\Omega^{-1}{\rm tr}\chi\right)\slashed{\rm div}b -\eta^A\slashed{\nabla}_A\slashed{\rm div}b + \underline{\eta}^A\slashed{\nabla}_A
\left((-v)\Omega{\rm tr}\chi\right) - 2\slashed{\nabla}^A\left(\Omega\hat{\underline{\chi}}\right)_{AB}\eta^B+2v\Omega{\rm tr}\chi\left( \eta\cdot\underline{\eta}\right)
\\ \nonumber &\qquad \qquad -\left(\slashed{\nabla}\hat{\otimes}\eta\right)\cdot\left(\slashed{\nabla}\hat{\otimes}b\right) + 2v\left(\Omega\hat{\chi}\right)\cdot\left(\slashed{\nabla}\hat{\otimes}\slashed{\nabla}\log\Omega\right) -\frac{1}{2}\left(\eta\hat{\otimes}\eta\right)\cdot\left(\slashed{\nabla}\hat{\otimes}b\right)+\frac{1}{2}\left(\underline{\eta}\hat{\otimes}\underline{\eta}\right)\cdot\left(v\Omega\hat{\chi}\right)
\\ \nonumber &\qquad  + \frac{1}{2}\Omega^{-2}\Theta \cdot\left(\slashed{\nabla}\hat{\otimes}b\right)
  -\frac{1}{2}\mathcal{L}_b\left(\Omega^{-1}{\rm tr}\chi\right)\slashed{\rm div}b+\frac{1}{4}\left(\Omega^{-1}{\rm tr}\chi\right)\left(\slashed{\rm div}b\right)^2 
 \\ \nonumber &\qquad - \Omega^2v\left(\Omega^{-1}\hat{\chi}\right)\cdot\left(\slashed{\nabla}\hat{\otimes}b\right)\left(-\frac{1}{2}\Omega^{-1}{\rm tr}\chi\right) + \frac{1}{8}\left(\Omega^{-1}{\rm tr}\chi\right)\left|\slashed{\nabla}\hat{\otimes}b\right|^2 - \frac{1}{2}\Omega^2\left(\Omega^{-1}{\rm tr}\chi\right)(-v)\mathcal{L}_b\left(\Omega^{-1}{\rm tr}\chi\right)
 \\ \nonumber &\qquad +\left(\left(v\mathcal{L}_{\partial_v} + \mathcal{L}_b\right)+\left(v\Omega{\rm tr}\chi + \slashed{\rm div}b\right)\right)\left(\left|\eta\right|^2-\frac{1}{2}\mathcal{L}_b\left(\Omega^{-1}{\rm tr}\chi\right)\right),
\end{align}
\begin{align}
H_5 \doteq \frac{1}{2}\mathcal{L}_b\left(\Omega^{-1}{\rm tr}\chi\right),
\end{align}
where $\Theta$ is defined as in Lemma~\ref{1ioj3i4io}.

Finally, we define $H_6$ to be the solution to the following transport equation
\begin{align}\label{32oij32jio32io49101i3k2}
&\left(v\mathcal{L}_{\partial_v}+\mathcal{P}_{\ell \geq 1}\mathcal{L}_b\right)H_6 + \left(-1+2v(v+1)^{-1}+v\mathcal{P}_{\ell \geq 1}\Omega{\rm tr}\chi\right)H_6 =
\\ \nonumber &\qquad -\mathcal{P}_{\ell \geq 1}\left[\left(v+1\right)^{-2}\mathring{\Delta},\mathcal{L}_b\right]\slashed{\rm div}b -\mathcal{P}_{\ell \geq 1}\left[\left(v+1\right)^{-2}\mathring{\Delta},v\Omega{\rm tr}\chi\right]\slashed{\rm div}b
\\ \nonumber &\qquad +\mathcal{P}_{\ell \geq 1}\left((v+1)^{-2}\mathring{\Delta}\left(-\frac{1}{2}\left(\slashed{\rm div}b\right)^2 -4\left(\Omega\underline{\omega}\right)\slashed{\rm div}b + \Omega(-v)\hat{\chi}\cdot\slashed{\nabla}\hat{\otimes}b - \frac{1}{4}\left|\slashed{\nabla}\hat{\otimes}b\right|^2 + \Omega^2(-v)\mathcal{L}_b\left(\Omega^{-1}{\rm tr}\chi\right)\right)\right),
\end{align}
with the boundary condition that 
\[\left(v+1\right)^2H_6 \to 0\text{ as }v\to -1.\]

We then have 
\begin{align}\label{32ji32oij2}
& \left\vert\left\vert \left(H_1,H_2,0\right)\right\vert\right\vert_{R,I,N_1-2} +\left\vert\left\vert \left((-v)^{-2\kappa}H_5,H_6\right)\right\vert\right\vert_{\mathscr{P}\mathscr{R}\left(\Omega,\kappa,50\check{p},50\check{p}\right)} + \left\vert\left\vert H_4\right\vert\right\vert_{\mathscr{Q}\left(N_1-2,1/2+5\check{\delta},-\kappa\right)}
\\ \nonumber &\qquad + \left\vert\left\vert H_4\right\vert\right\vert_{\mathscr{S}_{-1}^{-1/2}\left(N_1-3,1+5\check{\delta},0\right)}+ \left\vert\left\vert H_4\right\vert\right\vert_{\mathscr{S}_{-1/2}^0\left(N_2-2,0,0\right)} 
 +\sum_{j=0}^1\left\vert\left\vert \left(v\mathcal{L}_{\partial_v}\right)^j\left(H_5,H_6\right)\right\vert\right\vert_{\mathscr{S}_{-1/2}^0\left(N_2-2-j,0,0\right)} 
 \\ \nonumber &\qquad  \lesssim
 \mathscr{P}^2 + \sum_{j=0}^1\left\vert\left\vert \left(v\mathcal{L}_{\partial_v}\right)^j\eta\right\vert\right\vert^2_{\mathscr{S}_{-1/2}^0\left(N_2-1-j,0,0\right)},
 \end{align}
 where $\mathscr{P}$ is the left hand side of~\eqref{k2mkmk4k}. As usual, the constants in these estimates do not depend on the implicit constant in~\eqref{1kl2lkj124}.
\end{lemma}
\begin{proof}These are all straightforward if tedious applications of the inequalities from Section~\ref{inequalitysectionsection} and transport equation estimates. 
\end{proof}

Lastly, we can now plug in these nonlinear expressions into our iteration scheme.
\begin{proposition}\label{kjkjkj3kjkj}
There exist a function $\Omega_{\rm boun}: (-1,0)\times \mathbb{S}^2 \to (0,\infty)$, an $\mathbb{S}^2_{-1,v}$-vector field $b^A$, an $\mathbb{S}^2_{-1,v}$-symmetric $(0,2)$ tensor $\slashed{g}$, and an $\mathbb{S}^2_{-1,v}$ $1$-form $\mathfrak{n}_A$  such that all of the equations of Proposition~\ref{3ij2oio23} hold except that we replace $H_1$, $H_2$, $H_4$, $H_5$, and $H_6$ with the corresponding expressions from Lemma~\ref{3oi32io} and $H_7$, $H_8$, and $H_9$ with the corresponding expressions from Lemma~\ref{3io3oioi23}. We then have 
\begin{align}\label{k1nk1nl1}
&\left\vert\left\vert \log\Omega_{\rm boun}\right\vert\right\vert_{\mathscr{A}\left(\kappa,b\right)} + \left\vert\left\vert \log\Omega_{\rm boun}\right\vert\right\vert_{\mathscr{B}_{01}\left(\kappa,b\right)}  +\left\vert\left\vert b\right\vert\right\vert_{\mathscr{A}_1\left(\kappa,\slashed{g}\right)}+\left\vert\left\vert b\right\vert\right\vert_{\mathscr{B}_1\left(\kappa\right)}+ \sum_{j=0}^1\left\vert\left\vert \left(v\mathcal{L}_{\partial_v}\right)^j\eta\right\vert\right\vert^2_{\mathscr{S}_{-1/2}^0\left(N_2-1-j,0,0\right)}
\\ \nonumber &\qquad + \left\vert\left\vert \left(\mathfrak{o},\mathcal{L}_{b|_{v=0}}\mathfrak{o}\right)\right\vert\right\vert_{\mathring{H}^{N_1-3}\left(\mathbb{S}^2\right)} +\left\vert\left\vert \left(\mathfrak{j},\mathcal{L}_{b|_{v=0}}\mathfrak{j}\right)\right\vert\right\vert_{\mathring{H}^{N_1-2}\left(\mathbb{S}^2\right)}+ \left\vert\left\vert \left(\mathfrak{w},\mathcal{L}_{b|_{v=0}}\mathfrak{w}\right)\right\vert\right\vert_{\mathring{H}^{N_1-3}\left(\mathbb{S}^2\right)}
 \lesssim
 \\ \nonumber &\qquad   \epsilon \left\vert\left\vert \left(T_{\rm low},T_{\rm high}\right)\right\vert\right\vert_{\mathring{H}^{N_1-3}} 
 +\left\vert\left\vert \log\Omega_{\rm sing}\right\vert\right\vert_{\mathscr{B}_{00}\left(\kappa\right)}+\left\vert\left\vert \pi\right\vert\right\vert^2_{\mathscr{S}\left(0,\check{\delta},0\right)} +\left\vert\left\vert \pi\right\vert\right\vert^2_{\mathscr{Q}_{-1}^{-1/2}\left(0,-1/2+\check{\delta},0\right)},
\end{align}
\begin{align}\label{1joijoijio1o9}
&\left\vert\left\vert \slashed{g} \right\vert\right\vert_{\mathscr{A}_2\left(\kappa,b,\Omega\right)} + \left\vert\left\vert \slashed{g}\right\vert\right\vert_{\mathscr{B}_2\left(\kappa,b,\Omega\right)}
+\left\vert\left\vert \left(v+1\right)\mathfrak{n}\right\vert\right\vert_{L,III\left(0,100\check{p}N_1,0\right),N_1-1} \lesssim
 \\ \nonumber &\qquad   \epsilon \left\vert\left\vert \left(T_{\rm low},T_{\rm high}\right)\right\vert\right\vert_{\mathring{H}^{N_1-3}} 
 +\left\vert\left\vert \log\Omega_{\rm sing}\right\vert\right\vert_{\mathscr{B}_{00}\left(\kappa\right)}+\left\vert\left\vert \pi\right\vert\right\vert_{\mathscr{S}\left(0,\check{\delta},0\right)} +\left\vert\left\vert \pi\right\vert\right\vert_{\mathscr{Q}_{-1}^{-1/2}\left(0,-1/2+\check{\delta},0\right)}.
\end{align}

\end{proposition}
\begin{proof}This is an immediate consequence of an iteration argument and the nonlinear estimates we have established.
\end{proof}
\section{Solving for $\Omega_{\rm sing}$ and $\pi$ and Finishing the Iteration}\label{foim2io34o5u92hj5991}
In this section we will finally complete our main iteration argument and solve for $\Omega_{\rm sing}$ and $\kappa$. We start with some nonlinear estimates.
\begin{lemma}\label{21p019993jijm2omo}Let $\left(b,\slashed{g},\Omega\right)$ and $\tilde{\kappa} \in \mathbb{R}$ be given and assumed to satisfy 
\begin{align}\label{ij2kljlkj2lkj3}
&\left\vert\left\vert \log\Omega_{\rm boun}\right\vert\right\vert_{\mathscr{A}\left(\tilde{\kappa},b\right)} + \left\vert\left\vert \log\Omega_{\rm boun}\right\vert\right\vert_{\mathscr{B}_{01}\left(\tilde{\kappa},b\right)} + \left\vert\left\vert \log\Omega_{\rm sing}\right\vert\right\vert_{\mathscr{B}_{01}\left(\tilde{\kappa},b\right)} + \left|\tilde{\kappa}\right|
\\ \nonumber&\qquad +\left\vert\left\vert \slashed{g} \right\vert\right\vert_{\mathscr{A}_2\left(\tilde{\kappa},b,\Omega\right)} + \left\vert\left\vert \slashed{g}\right\vert\right\vert_{\mathscr{B}_2\left(\tilde{\kappa},b,\Omega\right)}
 +\left\vert\left\vert b\right\vert\right\vert_{\mathscr{A}_1\left(\tilde{\kappa},\slashed{g}\right)}+\left\vert\left\vert b\right\vert\right\vert_{\mathscr{B}_1\left(\tilde{\kappa}\right)}
  \lesssim \epsilon.
\end{align}
Then we define
\begin{align}\label{oijoijjio32o}
H_3\left[\Omega,b,\slashed{g}\right] \doteq \mathcal{P}_{\ell = 0}\left(\Omega^2\left(2Y-2\Omega^{-1}{\rm tr}\chi \mathcal{P}_{\ell \geq 1}\left(\Omega\underline{\omega}\right)-8\eta\cdot\underline{\eta} + \left(\Omega^{-1}\hat{\chi}\right)\cdot\left(\slashed{\nabla}\hat{\otimes}b\right)\right)\right)
\end{align}
We have
\begin{equation}\label{3iojo3ih3iuhij}
\sum_{j=0}^1\left\vert\left\vert \mathcal{L}_{\partial_v}^jH_3\right\vert\right\vert_{\mathscr{S}_{-1}^{-1/2}\left(0,1+\check{\delta}+j,0\right)}+ \sum_{j=0}^1\left\vert\left\vert \left(v\mathcal{L}_{\partial_v}\right)^jH_3\right\vert\right\vert_{\check{\mathscr{S}}_{-1/2}^0\left(0,0,200\check{p}\left(1+j\right)\right)} \lesssim \mathscr{Z}^2,
\end{equation}
where $\mathscr{Z}$ is the left hand side of~\eqref{ij2kljlkj2lkj3}. The implied constant in~\eqref{3iojo3ih3iuhij} is independent of the implied constant in~\eqref{ij2kljlkj2lkj3}. 

\end{lemma}
\begin{proof}This is a straightforward consequence of Lemma~\ref{2ij23ji3ijo32} and the assumption~\eqref{ij2kljlkj2lkj3}. 
\end{proof}

Finally, we can now complete the main iteration argument.
\begin{proposition}\label{k3jkl11}
There exist a function $\Omega = \Omega_{\rm sing}\Omega_{\rm boun}: (-1,0)\times \mathbb{S}^2 \to (0,\infty)$, an $\mathbb{S}^2_{-1,v}$-vector field $b^A$, an $\mathbb{S}^2_{-1,v}$-symmetric $(0,2)$ tensor $\slashed{g}$, and an $\mathbb{S}^2_{-1,v}$ $1$-form $\mathfrak{n}_A$  such that all of the equations of Proposition~\ref{3ij2oio23} hold except that we replace $H_1$, $H_2$, $H_4$, $H_5$, and $H_6$ with the corresponding expressions from Lemma~\ref{3oi32io}, $H_7$, $H_8$, and $H_9$ with the corresponding expressions from Lemma~\ref{3io3oioi23}, and $H_3$ with the expression from Lemma~\ref{21p019993jijm2omo}, and so that we set
\[\pi \doteq \mathcal{P}_{\ell \leq \ell_0}\eta.\]
We then have 
\begin{align}\label{k1nk1nl1}
&\left\vert\left\vert \log\Omega_{\rm boun}\right\vert\right\vert_{\mathscr{A}\left(\kappa,b\right)} + \left\vert\left\vert \log\Omega_{\rm boun}\right\vert\right\vert_{\mathscr{B}_{01}\left(\kappa,b\right)} +\left\vert\left\vert \log\Omega_{\rm sing}\right\vert\right\vert_{\mathscr{B}_{00}\left(\kappa\right)} +\left\vert\left\vert b\right\vert\right\vert_{\mathscr{A}_1\left(\kappa,\slashed{g}\right)}+\left\vert\left\vert b\right\vert\right\vert_{\mathscr{B}_1\left(\kappa\right)}
\\ \nonumber &\qquad +\left\vert\left\vert \slashed{g} \right\vert\right\vert_{\mathscr{A}_2\left(\kappa,b,\Omega\right)} + \left\vert\left\vert \slashed{g}\right\vert\right\vert_{\mathscr{B}_2\left(\kappa,b,\Omega\right)}
+\left\vert\left\vert \left(v+1\right)\mathfrak{n}\right\vert\right\vert_{L,III\left(1/2-\check{\delta},100\check{p}N_1,0\right),N_1-1}
\\ \nonumber &\qquad + \left\vert\left\vert \left(\mathfrak{o},\mathcal{L}_{b|_{v=0}}\mathfrak{o}\right)\right\vert\right\vert_{\mathring{H}^{N_1-3}\left(\mathbb{S}^2\right)} +\left\vert\left\vert \left(\mathfrak{j},\mathcal{L}_{b|_{v=0}}\mathfrak{j}\right)\right\vert\right\vert_{\mathring{H}^{N_1-2}\left(\mathbb{S}^2\right)}+ \left\vert\left\vert \left(\mathfrak{w},\mathcal{L}_{b|_{v=0}}\mathfrak{w}\right)\right\vert\right\vert_{\mathring{H}^{N_1-3}\left(\mathbb{S}^2\right)}
 \lesssim
 \\ \nonumber &\qquad   \epsilon \left\vert\left\vert \left(T_{\rm low},T_{\rm high}\right)\right\vert\right\vert_{\mathring{H}^{N_1-3}}.
\end{align}

\end{proposition}
\begin{proof}This follows immediately from Proposition~\ref{iojoij2}, Proposition~\ref{kjkjkj3kjkj}, an iteration argument, and Lemma~\ref{21p019993jijm2omo}.
\end{proof}

Now we can provide the proof of Theorem~\ref{thisiswhatishholdingatthenehdne}:
\begin{proof}We let $\left(\Omega,b,\slashed{g}\right)$ be those that are produced by Proposition~\ref{k3jkl11}. Then items~\ref{anitemitem2} and~\ref{anitemitem3} from the statement of Theorem~\ref{thisiswhatishholdingatthenehdne} are immediate consequences of the equations which we have used to solve for $\left(\Omega,b,\slashed{g}\right)$. 

Next we turn to the estimates from item~\ref{anitemitem1} of the statement of Theorem~\ref{thisiswhatishholdingatthenehdne}. This all already follows from Proposition~\ref{k3jkl11} except for the estimates for $\mathcal{L}_{\partial_v}^2\eta$ in the last term on the left hand side of~\eqref{ijoi9919njnkko2o201nbhjnbghj} and the estimate~\eqref{2om3om1oijtionhoin1}. One may establish the improved estimate for $\eta$ by revisiting the proofs of Proposition~\ref{ioj32iojjio} and Lemma~\ref{2ij23ji3ijo32} and applying Lemma~\ref{2om2omo4} along with commutation with $\mathcal{L}_{\partial_v}$ to estimate $\eta$ and $Y$. We omit the details. For estimate~\eqref{2om3om1oijtionhoin1} we start by noting that as a consequence of Propositions~\ref{23ini2999jn2j23i3j},~\ref{iojoij2}, Lemma~\ref{3oi32io}, Lemma~\ref{21p019993jijm2omo}, and  Proposition~\ref{kjkjkj3kjkj} that 
\begin{equation}\label{3oijoij23}
\left\vert\left\vert \log\Omega_{\rm boun}\right\vert\right\vert_{\mathscr{A}\left(\kappa,b\right)} + \left\vert\left\vert \log\Omega_{\rm boun}\right\vert\right\vert_{\mathscr{B}_{01}\left(\kappa,b\right)} +\left\vert\left\vert \log\Omega_{\rm sing}\right\vert\right\vert_{\mathscr{B}_{00}\left(\kappa\right)} \lesssim \epsilon^2\left\vert\left\vert \left(T_{\rm low},T_{\rm high}\right)\right\vert\right\vert_{\mathring{H}^{N_1-3}}^2.
\end{equation}
Next, we observe that in the equation satisfied by $\Omega^{-1}{\rm tr}\chi - 2(v+1)^{-1}$ (see~\eqref{jdiji2mo1}) the only term which is linearly coupled to $\Omega^{-1}{\rm tr}\chi-2(v+1)^{-1}$ is the one proportional to $1-\Omega^2$.  However,~\eqref{3oijoij23} shows that the lapse has a quadratic dependence on the data $\left(T_{\rm low},T_{\rm high}\right)$ and thus the desired estimate~\eqref{2om3om1oijtionhoin1} follows easily.

Finally, we come to item~\ref{anitemitem4} of the statement of Theorem~\ref{thisiswhatishholdingatthenehdne}. This follows by re-running all of the estimates behind the proof of Proposition~\ref{k3jkl11} after commuting each equation with $\mathcal{L}_X$. 
\end{proof} 
\section{Constraint Propagation}\label{propagatetheconstraintsforever}

In this section we will  show that given a solution to the set of iterated equations provided by Theorem~\ref{thisiswhatishholdingatthenehdne}, the corresponding metric $g$ determined by $\left(\Omega,b,\slashed{g}\right)$ actually solves the Einstein equations (see Proposition~\ref{constraintsaredone}). 

\emph{Unless indicated otherwise, throughout this section we will assume that $\left(\Omega,b,\slashed{g}\right)$ are given by Theorem~\ref{thisiswhatishholdingatthenehdne}, and we then let $g$ denote the corresponding Lorentzian metric.}

\subsection{Preliminary Estimates and Definitions of Norms}
In the next lemma upgrade the estimates of Theorem~\ref{thisiswhatishholdingatthenehdne} to involve additional $\mathcal{L}_{\partial_v}$ derivatives.
\begin{lemma}\label{3ioj2oijio45482}For any positive integer $N$ satisfying $1 \ll N \ll N_2$, we have that 
\begin{align}\label{2moc30vkd}
&\sum_{j+\left|\alpha\right| \leq N}\sup_{\left(v,\theta^A\right)\in (-1,0)\times \mathbb{S}^2}\Bigg[\left(v+1\right)^{\check{\delta}+j}(-v)^{\check{p}+2\kappa+j}\left|(\Omega\nabla_4)^j\left(\Omega\hat{\chi}\right)^{(\alpha)}\right|
\\ \nonumber &\qquad+ (v+1)^{\check{\delta}}\left|
\left(\Omega^{-1}{\rm tr}\chi - \frac{2}{v+1}\right)^{(\alpha)}\right|+ (v+1)^{\check{\delta}+1}\left|
\left((-v)^{10\check{p}+2\kappa}\Omega\nabla_4\right)\left(\Omega^{-1}{\rm tr}\chi - \frac{2}{v+1}\right)^{(\alpha)}\right|
\\ \nonumber &\qquad + (v+1)^{\check{\delta}+1+j}\left|
\left(v\Omega\nabla_4\right)^j\left((-v)^{10\check{p}+2\kappa}\Omega\nabla_4\right)\left(\Omega^{-1}{\rm tr}\chi - \frac{2}{v+1}\right)^{(\alpha)}\right|
\\ \nonumber &\qquad + \left(v+1\right)^{-1+\check{\delta}}\left|\left(\slashed{g}-\left(v+1\right)^2\mathring{\slashed{g}}\right)^{(\alpha)}\right|  
\\ \nonumber &\qquad+ \left(v+1\right)^{-1+\check{\delta}}(-v)^{\left|\alpha\right|\epsilon^{1/2}}\left| \left(\left(\mathcal{L}_b,1\right)\left(\log\Omega - \kappa\log(-v)\right)\right)^{(\alpha)}\right|
  +\left(v+1\right)^{\check{\delta}}(-v)^{\left|\alpha\right|\epsilon^{1/2}}\left|\left((-v)\Omega\nabla_4\right)\left(\log\Omega\right)^{(\alpha)}\right| 
 \\ \nonumber &\qquad  +(v+1)^{\check{\delta}}\left|\left(\Omega\underline{\omega}\right)^{(\alpha)}\right|  +
(v+1)^{\check{\delta}+1}\left|\left((-v)^{500\check{p}\left(\left|\alpha\right|+1\right)}\Omega\nabla_4\right) \left(\Omega\underline{\omega}\right)^{(\alpha)}\right|  
\\ \nonumber &\qquad +(v+1)^{\check{\delta}+1+j}\left|\left(v\Omega\nabla_4\right)^j\left((-v)^{500\check{p}\left(\left|\alpha\right|+1+j\right)}\Omega\nabla_4\right) \left(\Omega\underline{\omega}\right)^{(\alpha)}\right|  
\\ \nonumber&\qquad + \sum_{k=0}^1\left(v+1\right)^{\check{\delta}+j}\left|\left(v\Omega\nabla_4\right)^j\left((v+1)(-v)^{500\check{p}\left(\left|\alpha\right|+1+j\right)}\Omega\nabla_4\right)^k\left(\eta^{(\alpha)},\mathfrak{n}^{(\alpha)}\right)\right|
  \\ \nonumber &\qquad
+ \left(v+1\right)^{\check{\delta}+j}(-v)^{\left(1+\left|\alpha\right|+j\right)100\check{p}}\left|\left((-v)\Omega\nabla_4\right)^j \zeta^{(\alpha)}\right| 
+ \left(v+1\right)^{-1+\check{\delta}}\left|b^{(\alpha)}\right|\Bigg]
\\ \nonumber &\qquad \lesssim \epsilon\left\vert\left\vert \left(T_{\rm low},T_{\rm high}\right)\right\vert\right\vert_{\mathring{H}^{N_1-3}}.
\end{align}

We also have the following improved estimates near $v = -1$ for certain combinations of metric components:
\begin{align}\label{3iojo2iji4oj}
&\sup_{v \in (-1,-1/2)}\sum_{j=0}^2\sum_{j+\left|\alpha\right| \leq N}\left(v+1\right)^{10\check{\delta}+j}\left[\left|\mathcal{L}^j_{\partial_v}\left(\slashed{\rm div}b\right)^{(\alpha)}\right|+\left|\mathcal{L}^j_{\partial_v}\left(Y+2\slashed{\Delta}\log\Omega\right)^{(\alpha)}\right|\right] \\ \nonumber &\qquad \lesssim  \epsilon\left\vert\left\vert \left(T_{\rm low},T_{\rm high}\right)\right\vert\right\vert_{\mathring{H}^{N_1-3}},
\end{align}
where we recall that $Y$ is defined by 
\[Y \doteq \Omega^{-1}\nabla_4\left(\Omega\underline{\omega}\right)+\Omega^{-1}{\rm tr}\chi \left(\Omega\underline{\omega}\right)+4\eta\cdot\underline{\eta} - \frac{1}{2}\left(\Omega^{-1}\hat{\chi}\right)\cdot\slashed{\nabla}\hat{\otimes}b.\]
\end{lemma}
\begin{proof}We start with the proof of~\eqref{2moc30vkd}. If we restrict the total number of $\mathcal{L}_{\partial_v}$ derivatives to be at most two applied to $\left(\Omega,b,\slashed{g}\right)$, then this statement would already follow directly from Theorem~\ref{thisiswhatishholdingatthenehdne}. In order to obtain estimates for higher number of $\mathcal{L}_{\partial_v}$ derivatives, we use the fact that the quantities $\Omega^{-1}{\rm tr}\chi$, $\hat{\chi}$, $\Omega\underline{\omega}$, $\mathcal{P}_{1 \leq \ell \leq \ell_0}\slashed{\rm curl}\zeta$, and $\mathcal{P}_{\ell > \ell_0}\slashed{\rm curl}\slashed{\rm div}\slashed{\nabla}\hat{\otimes}\zeta$  all satisfy equations which directly allow one to trade additional $v\Omega\nabla_4$ derivatives for angular derivatives. For $\slashed{\rm div}\zeta$ we may use Lemma~\ref{k4oij2oijo3} and the equation~\eqref{klj3lk2jkl3j2}. The result then follows from a straightforward induction argument in the total number of $\mathcal{L}_{\partial_v}$ derivatives. 

The improved estimate~\eqref{3iojo2iji4oj} for $Y$ with the sum in $j$ restricted to $j \in \{0,1\}$ already follows from Lemma~\ref{2ij23ji3ijo32} and Lemma~\ref{3oi32io}. The full estimate~\eqref{3iojo2iji4oj} for $Y$ follows by simply repeating the method of proof with additional commutation by $\mathcal{L}_{\partial_v}$. The improved estimate for $\slashed{\rm div}b$, then follows by revisiting the equation~\eqref{2jk32ij23uhi2iu},  and using the improved estimate for $Y$, Lemma~\ref{k4oij2oijo3}, and the expressions for $H_5$ and $H_6$ from Lemma~\ref{3oi32io}.
\end{proof}

From the estimates of Lemma~\ref{3ioj2oijio45482} we may deduce various a prior bounds on the components of the Ricci tensor. We record these here:
\begin{lemma}\label{3joj2} We have, for a suitably large $N \in \mathbb{Z}_{> 0}$ and some $0 < \check{p} \ll 1$ and $\left|\kappa\right|\lesssim \epsilon$ that the Ricci tensor is $C^N$ for $\left(v,\theta^A\right) \in (-1,0)\times\mathbb{S}^2$ and moreover satisfies the following bounds:
\begin{align}\label{2kn2kn1k}
&\sum_{j=0}^1\sum_{k+j+\left|\alpha\right| \leq N}\sup_{\left(v,\theta^A\right)\in (-1,0)\times \mathbb{S}^2}\Bigg[\left(v+1\right)^{1+\check{\delta}+j+k}(-v)^{\left(j+k\right)500\check{p}+k}\left|\mathcal{L}^{j+k}_{\partial_v}\Phi\right| 
\\ \nonumber &\qquad +\left(v+1\right)^{20\check{\delta}+j+k}(-v)^{\left(j+k\right)500\check{p}+k}\left|\mathcal{L}^{j+k}_{\partial_v}\widetilde{\Phi}\right| 
\\ \nonumber &\qquad + \left(v+1\right)^{1+\check{\delta}+j+k}(-v)^{500\check{p}k+j+k}\left| \mathcal{L}^{j+k}_{\partial_v}\left(\Omega{\rm Ric}_{A4}\right)^{(\alpha)}\right|\Bigg]
 \lesssim \epsilon,
\end{align}
where $\Phi$ stands for any of $\Omega{\rm Ric}_{A3}$, $\widehat{\rm Ric}_{AB}$, or $R$ and $\widetilde{\Phi}$ stands for one of $\Omega^2{\rm Ric}_{33}$ or ${\rm Ric}_{34}$.
\end{lemma}
\begin{proof}These are immediate consequences of Theorem~\ref{thisiswhatishholdingatthenehdne} and the null-structure equations. Note in particular that the improved estimates for $\widetilde{\Phi}$ as $v\to -1$ are a consequence of~\eqref{3iojo2iji4oj}, Lemma~\ref{2m2om3o3923332}, and~\eqref{eqnyaydivb}. 
\end{proof}

The basic idea behind our approach to showing that ${\rm Ric} = 0$ is to define a certain norm $\left\vert\left\vert \left({\rm Ric},\mathfrak{n}-\eta\right)\right\vert\right\vert_{\mathscr{Z}}$ so that using a ``propagation of constraints'' argument, we can show that $\left\vert\left\vert \left({\rm Ric},\mathfrak{n}-\eta\right)\right\vert\right\vert_{\mathscr{Z}} \lesssim \epsilon\left\vert\left\vert \left({\rm Ric},\mathfrak{n}-\eta\right)\right\vert\right\vert_{\mathscr{Z}} $. This of course then implies that ${\rm Ric} = 0$ and that $\mathfrak{n} = \eta$. We now turn to the specifics. We start by defining a ``bulk'' norm for the various Ricci components (and for $\mathfrak{n}$). 
\begin{definition}We define
\begin{align*}
&\left\vert\left\vert {\rm Ric}_{33}\right\vert\right\vert_{\mathscr{Z}} \doteq \left\vert\left\vert \Omega^2{\rm Ric}_{33}\right\vert\right\vert_{\mathscr{Q}_{-1}^0\left(6,0,-\kappa\right)},
\end{align*}
\begin{align*}
&\left\vert\left\vert {\rm Ric}_{34}\right\vert\right\vert_{\mathscr{Z}} \doteq \left\vert\left\vert {\rm Ric}_{34}\right\vert\right\vert_{\mathscr{Q}_{-1}^0\left(4,0,-\kappa\right)} +\left\vert\left\vert {\rm Ric}_{34}\right\vert\right\vert_{\mathscr{Q}_{-1/2}^0\left(3,0,-1/2+\check{\delta}\right)} 
 +\left\vert\left\vert \mathcal{L}_{\partial_v}{\rm Ric}_{34}\right\vert\right\vert_{\mathscr{Q}_{-1}^0\left(3,1,1/2+\sqrt{\check{\delta}}\right)}
 \\ \nonumber &\qquad + \left\vert\left\vert \left((-v)\mathcal{L}_{\partial_v}-\mathcal{L}_b\right){\rm Ric}_{34}\right\vert\right\vert_{\mathscr{Q}_{-1}^0\left(4,1,-\kappa\right)} + \left\vert\left\vert \left((-v)\mathcal{L}_{\partial_v}-\mathcal{L}_b\right){\rm Ric}_{34}\right\vert\right\vert_{\mathscr{Q}_{-1/2}^0\left(3,0,-1/2+\check{\delta}\right)} +  \left\vert\left\vert \mathcal{L}_{\partial_v}{\rm Ric}_{34}\right\vert\right\vert_{\mathscr{Q}_{-1}^0\left(2,1,\kappa\right)}
 \\ \nonumber &\qquad + \left\vert\left\vert \mathcal{L}_{\partial_v} \left((-v)\mathcal{L}_{\partial_v}-\mathcal{L}_b\right){\rm Ric}_{34}\right\vert\right\vert_{\mathscr{Q}_{-1}^0\left(2,2,\kappa\right)},
\end{align*}
\begin{align*}
&\left\vert\left\vert {\rm Ric}_{3A}\right\vert\right\vert_{\mathscr{Z}} = \left\vert\left\vert \Omega{\rm Ric}_{3A}\right\vert\right\vert_{\mathscr{Q}_{-1}^0\left(4,1,-1/2+\sqrt{\check{\delta}}\right)}+\left\vert\left\vert \left(1,\mathcal{L}_b\right)\Omega{\rm Ric}_{3A}\right\vert\right\vert_{\mathscr{Q}_{-1}^0\left(3,1,-1/2+\check{\delta}\right)}  +  \left\vert\left\vert \mathcal{L}_{\partial_v}\left(\Omega{\rm Ric}_{3A}\right)\right\vert\right\vert_{\mathscr{Q}_{-1}^0\left(3,2,\kappa\right)},
\end{align*}
\begin{align*}
&\left\vert\left\vert {\rm Ric}_{4A}\right\vert\right\vert_{\mathscr{Z}} \doteq \left\vert\left\vert \Omega{\rm Ric}_{4A}\right\vert\right\vert_{\mathscr{Q}_{-1}^0\left(4,1,1/2+\sqrt{\check{\delta}}\right)} + \left\vert\left\vert \left(1,\mathcal{L}_b,v\left(v+1\right)\mathcal{L}_{\partial_v}\right)\Omega{\rm Ric}_{4A}\right\vert\right\vert_{\mathscr{Q}_{-1}^0\left(3,1,\kappa\right)},
\end{align*}
\begin{align*}
&\left\vert\left\vert {\rm R} + 2{\rm Ric}_{34} \right\vert\right\vert_{\mathscr{Z}} \doteq \left\vert\left\vert {\rm R} + 2{\rm Ric}_{34} \right\vert\right\vert_{\mathscr{Q}_{-1}^0\left(4,1,-\kappa\right)} + \left\vert\left\vert \left(1,\mathcal{L}_b\right)\left({\rm R} + 2{\rm Ric}_{34}\right) \right\vert\right\vert_{\mathscr{Q}_{-1}^0\left(3,1,-1/2+\check{\delta}\right)}
\\ \nonumber &\qquad +\left\vert\left\vert \mathcal{L}_{\partial_v}\left({\rm R} + 2{\rm Ric}_{34}\right) \right\vert\right\vert_{\mathscr{Q}_{-1}^0\left(3,2,1/2+\sqrt{\check{\delta}}\right)} + \left\vert\left\vert \left(1,\mathcal{L}_b,v\left(v+1\right)\mathcal{L}_{\partial_v}\right)\mathcal{L}_{\partial_v}\left({\rm R} + 2{\rm Ric}_{34}\right) \right\vert\right\vert_{\mathscr{Q}_{-1}^0\left(2,2,\kappa \right)},
\end{align*}
\begin{align*}
&\left\vert\left\vert \widehat{\rm Ric}_{AB} \right\vert\right\vert_{\mathscr{Z}} \doteq \left\vert\left\vert \widehat{\rm Ric}_{AB} \right\vert\right\vert_{\mathscr{Q}_{-1}^0\left(4,1,-\kappa\right)} + \left\vert\left\vert \left(1,\mathcal{L}_b\right)\widehat{\rm Ric}_{AB}  \right\vert\right\vert_{\mathscr{Q}_{-1}^0\left(3,1,-1/2+\check{\delta}\right)}
\\ \nonumber &\qquad +\left\vert\left\vert \mathcal{L}_{\partial_v}\widehat{\rm Ric}_{AB}  \right\vert\right\vert_{\mathscr{Q}_{-1}^0\left(3,2,1/2+\sqrt{\check{\delta}}\right)} + \left\vert\left\vert \left(1,\mathcal{L}_b,v\left(v+1\right)\mathcal{L}_{\partial_v}\right)\mathcal{L}_{\partial_v}\widehat{\rm Ric}_{AB}  \right\vert\right\vert_{\mathscr{Q}_{-1}^0\left(2,2,\kappa \right)},
\end{align*}
\begin{align*}
&\left\vert\left\vert \mathfrak{n}-\eta \right\vert\right\vert_{\mathscr{Z}} \doteq \left\vert\left\vert \mathfrak{n}-\eta \right\vert\right\vert_{\mathscr{Q}_{-1}^0\left(5,1,-\kappa\right)} + \left\vert\left\vert \left(1,\mathcal{L}_b\right)\left( \mathfrak{n}-\eta\right) \right\vert\right\vert_{\mathscr{Q}_{-1}^0\left(4,1,-1/2+\check{\delta}\right)}
\\ \nonumber &\qquad +\left\vert\left\vert \mathcal{L}_{\partial_v}\left( \mathfrak{n}-\eta\right)  \right\vert\right\vert_{\mathscr{Q}_{-1}^0\left(4,2,1/2+\sqrt{\check{\delta}}\right)} + \left\vert\left\vert \left(1,\mathcal{L}_b,v\left(v+1\right)\mathcal{L}_{\partial_v}\right)\mathcal{L}_{\partial_v}\left( \mathfrak{n}-\eta\right)  \right\vert\right\vert_{\mathscr{Q}_{-1}^0\left(3,2,\kappa \right)},
\end{align*}
\begin{align*}
\left\vert\left\vert {\rm Ric}\right\vert\right\vert_{\mathscr{Z}} &\doteq \left\vert\left\vert {\rm Ric}_{3A}\right\vert\right\vert_{\mathscr{Z}}+\left\vert\left\vert {\rm Ric}_{4A}\right\vert\right\vert_{\mathscr{Z}}+\left\vert\left\vert \widehat{\rm Ric}_{AB}\right\vert\right\vert_{\mathscr{Z}}+\left\vert\left\vert {\rm Ric}_{34}\right\vert\right\vert_{\mathscr{Z}}
+\left\vert\left\vert {\rm R}+2{\rm Ric}_{34}\right\vert\right\vert_{\mathscr{Z}}  + \left\vert\left\vert {\rm Ric}_{33}\right\vert\right\vert_{\mathscr{Z}}.
\end{align*}
\end{definition}
\begin{remark}\label{3ioj2oiu492}It is a consequence of Lemmas~\ref{3joj2} and~\ref{3ioj2oijio45482} that $\left\vert\left\vert {\rm Ric}\right\vert\right\vert_{\mathscr{Z}}$ and $\left\vert\left\vert \mathfrak{n}-\eta\right\vert\right\vert_{\mathscr{Z}}$ are finite and satisfy
\[\left\vert\left\vert {\rm Ric}\right\vert\right\vert_{\mathscr{Z}} +\left\vert\left\vert \mathfrak{n}-\eta\right\vert\right\vert_{\mathscr{Z}} \lesssim \epsilon.\]
\end{remark}

Next we define a norm which governs the ``boundary'' behavior of the Ricci components and $\nu$:
\begin{align}\label{k3j93jco2jn2oddskkw2222}
&\left\vert\left\vert {\rm Ric}\right\vert\right\vert_{\mathscr{M}} \doteq  \left\vert\left\vert \left(\mathcal{L}_b,1\right)\widehat{\rm Ric}_{AB}|_{v=0}\right\vert\right\vert_{\mathring{H}^3\left(\mathbb{S}^2\right)} +  \left\vert\left\vert \left(\mathcal{L}_b,1\right)\left(R+{\rm Ric}_{34}\right)|_{v=0}\right\vert\right\vert_{\mathring{H}^3\left(\mathbb{S}^2\right)} 
\\ \nonumber &\qquad + \left\vert\left\vert \left(1,\mathcal{L}_b\right){\rm Ric}_{34}|_{v=0}\right\vert\right\vert_{\mathring{H}^3\left(\mathbb{S}^2\right)} + \left\vert\left\vert \Omega{\rm Ric}_{3A}|_{v=0}\right\vert\right\vert_{\mathring{H}^4\left(\mathbb{S}^2\right)}.
\end{align}
\begin{remark}It follows immediately from Lemmas~\ref{3joj2} and~\ref{3ioj2oijio45482} that $\left\vert\left\vert {\rm Ric}\right\vert\right\vert_{\mathscr{M}}$ and $\left\vert\left\vert \left(1,\mathcal{L}_b\right)\left(\mathfrak{n}-\eta\right)|_{v=0}\right\vert\right\vert_{\mathring{H}^4\left(\mathbb{S}^2\right)}$ are finite and moreover satisfy
\[\left\vert\left\vert {\rm Ric}\right\vert\right\vert_{\mathscr{M}} +\left\vert\left\vert \left(1,\mathcal{L}_b\right)\left(\mathfrak{n}-\eta\right)|_{v=0}\right\vert\right\vert_{\mathring{H}^4\left(\mathbb{S}^2\right)} +\left\vert\left\vert \left(1,\mathcal{L}_b\right)\left(\mathfrak{j}-\eta\right)|_{v=0}\right\vert\right\vert_{\mathring{H}^4\left(\mathbb{S}^2\right)}\lesssim \epsilon.\]
\end{remark}

 \subsection{Useful Equations}
 
The fact that the Einstein tensor is always divergence free imposes certain differential relations between the Ricci curvature components. We list these in the next lemma.
\begin{lemma}\label{ljkj23121jn2in4}Every $3+1$ dimensional Lorentzian spacetime satisfies the following:
\begin{align}\label{3kcom1}
&-\frac{1}{2}\nabla_3{\rm Ric}_{44} +2\underline{\omega}{\rm Ric}_{44} + 2\eta^A{\rm Ric}_{4A} -\frac{1}{2}\nabla_4\left({\rm Ric}_{34}+ R\right) + \underline{\eta}^A{\rm Ric}_{A4}
\\ \nonumber &\qquad +\slashed{\nabla}^A{\rm Ric}_{A4} - \frac{1}{2}{\rm tr}\underline{\chi}{\rm Ric}_{44} - \frac{1}{2}{\rm tr}\chi\left({\rm Ric}_{34} + R\right) + \zeta^A{\rm Ric}_{A4}-\frac{1}{2}{\rm tr}\chi{\rm Ric}_{34} - \hat{\chi}^{AB}\widehat{{\rm Ric}}_{AB} = 0,
\end{align}

\begin{align}\label{3kcom2}
&-\frac{1}{2}\nabla_4{\rm Ric}_{33} +2\omega{\rm Ric}_{33} + 2\underline{\eta}^A{\rm Ric}_{3A} -\frac{1}{2}\nabla_3\left({\rm Ric}_{34}+ R\right) + \eta^A{\rm Ric}_{A3}
\\ \nonumber &\qquad +\slashed{\nabla}^A{\rm Ric}_{A3} - \frac{1}{2}{\rm tr}\chi{\rm Ric}_{33} - \frac{1}{2}{\rm tr}\underline{\chi}\left({\rm Ric}_{34} + R\right) - \zeta^A{\rm Ric}_{A3}-\frac{1}{2}{\rm tr}\underline{\chi}{\rm Ric}_{34} - \hat{\underline{\chi}}^{AB}\widehat{{\rm Ric}}_{AB} = 0,
\end{align}

\begin{align}\label{3kcom3}
&-\frac{1}{2}\nabla_3{\rm Ric}_{4A} + \underline{\omega}{\rm Ric}_{4A} + \eta^B{\rm Ric}_{BA} + \frac{1}{2}\eta_A{\rm Ric}_{34} -\frac{1}{2}\nabla_4{\rm Ric}_{3A} + \omega{\rm Ric}_{3A} + \underline{\eta}^B{\rm Ric}_{BA} + \frac{1}{2}\underline{\eta}_A{\rm Ric}_{34}
\\ \nonumber &\qquad +\slashed{\nabla}^B\widehat{{\rm Ric}}_{BA} + \frac{1}{2}\slashed{\nabla}_A{\rm Ric}_{34}-\frac{1}{2}{\rm tr}\underline{\chi}{\rm Ric}_{4A} - \frac{1}{2}{\rm tr}\chi{\rm Ric}_{3A} - \frac{1}{2}\underline{\chi}_A^{\ \ B}{\rm Ric}_{B4}-\frac{1}{2}\chi_A^{\ \ B}{\rm Ric}_{B3} = 0,
\end{align}
\begin{equation}\label{3kmom2}
{\rm Ric}_{AB} = \frac{1}{2}\left({\rm R} + {\rm Ric}_{34}\right)\slashed{g}_{AB} + \widehat{\rm Ric}_{AB}.
\end{equation}
\end{lemma}
\begin{proof}We have 
\[D^{\beta}\left({\rm Ric}_{\alpha\beta} - \frac{1}{2}g_{\alpha\beta}{\rm R}\right) = 0.\]
The formulas~\eqref{3kcom1}-\eqref{3kcom3} follow from patiently expanding out this identity for $\alpha \in \{3,4,A\}$. Finally,~\eqref{3kmom2} follows directly from the definition of scalar curvature.
\end{proof}

In the next lemma we will use Lemmas~\ref{thisiswhatishholdingatthenehdne} and~\ref{ljkj23121jn2in4} to derive various useful equations.
\begin{lemma}The following equations then hold for the Ricci tensor of $g$:
\begin{enumerate}
\item \begin{align}\label{2poj3om2}
&\mathcal{L}_{\partial_v}\left(R + {\rm Ric}_{34}\right) + \Omega {\rm tr}\chi \left(R + {\rm Ric}_{34}\right)
\\ \nonumber &\qquad  - 2\slashed{\nabla}^A\left(\Omega{\rm Ric}_{4A}\right) - 4\eta^A\left(\Omega{\rm Ric}\right)_{4A} + \Omega{\rm tr}\chi {\rm Ric}_{34} + 2\Omega \hat{\chi}^{AB}\widehat{\rm Ric}_{AB} = 0,
\end{align}
\item \begin{align}\label{2o3om24}
&\left((-v)\mathcal{L}_{\partial_v}-\mathcal{L}_b\right)\left({\rm R} + 3{\rm Ric}_{34}\right) + \left((-v)\Omega{\rm tr}\chi - \slashed{\rm div}b\right)\left({\rm R} + 3{\rm Ric}_{34}\right) - \Omega{\rm tr}\underline{\chi} {\rm Ric}_{34} 
\\ \nonumber &\qquad +2\slashed{\nabla}^A\left(\Omega{\rm Ric}_{3A}\right) -2 \left(\Omega\hat{\underline{\chi}}\right)^{AB}\widehat{\rm Ric}_{AB} = 0,
\end{align}
\item \begin{align}\label{oj2oj4oj2452}
&\Omega\nabla_4\left(v\Omega{\rm Ric}_{4A} + \Omega{\rm Ric}_{3A}\right) +\mathcal{L}_b\left(\Omega{\rm Ric}\right)_{4A}+ \left(-2-\frac{1}{2}\slashed{\rm div}b\right)\Omega{\rm Ric}_{4A}
\\ \nonumber &\qquad  + \frac{3}{2}\Omega{\rm tr}\chi \left(v\Omega{\rm Ric}_{4A}+\Omega{\rm Ric}_{3A}\right)
-2\Omega^2\slashed{\nabla}^B\widehat{\rm Ric}_{AB} - \Omega^2\slashed{\nabla}_A{\rm Ric}_{34} +\Omega\underline{\hat{\chi}}_A^{\ \ B}\left(\Omega{\rm Ric}\right)_{4B} + \Omega\hat{\chi}_A^{\ \ B}\left(\Omega{\rm Ric}\right)_{3B}
\\ \nonumber &\qquad  -4\Omega^2\left(\slashed{\nabla}^B\log\Omega\right){\rm Ric}_{BA} - \Omega^2\underline{\eta}_A{\rm Ric}_{34} = 0,
\end{align}
\item \begin{align}\label{2om3omo4}
(-v)\mathcal{L}_{\partial_v}^2X -\mathcal{L}_b\mathcal{L}_{\partial_v}X + (-v)\Omega{\rm tr}\chi \mathcal{L}_{\partial_v}X +\Omega^2\left(\slashed{\Delta}+2K\right)\mathcal{P}_{\ell > \ell_0}X  = \mathcal{L}_{\partial_v}\mathcal{E}_1 + \mathcal{E}_2 + \mathcal{E}_3+\mathcal{E}_4,
\end{align}
where $X \doteq \left(v+1\right)^2\left({\rm R} + 2{\rm Ric}_{34}\right)$, $\mathcal{E}_1$ is a linear combination of terms of the form
\[\left(v+1\right)^2\left((-v)\Omega{\rm tr}\chi-2(v+1)^{-1},\slashed{\rm div}b,v\Omega\hat{\chi},\Omega\hat{\underline{\chi}},\eta\right)\cdot\left(R+2{\rm Ric}_{34},{\rm Ric}_{34},\widehat{\rm Ric}_{AB},v\Omega{\rm Ric}_{A4}\right),\]
$\mathcal{E}_2$ is a linear combination of terms of the form 
\begin{equation}\label{2m3omo212rf4}
\left(v+1\right)^2\slashed{\nabla}\cdot\left(\left(\slashed{\rm div}b,\underline{\hat{\chi}}\right)\cdot \Omega{\rm Ric}_{A4}\right),
\end{equation}
\begin{equation}\label{2o2om41kio312}
\left(v+1\right)^2\Omega^2\slashed{\nabla}\cdot\left(\left(\Omega^{-1}{\rm tr}\chi - 2(v+1)^{-1}\Omega^{-2},\Omega^{-1}\hat{\chi},\eta,\underline{\eta}\right)\cdot\left(v\Omega{\rm Ric}_{A4},\Omega{\rm Ric}_{A3},\widehat{\rm Ric}_{AB},{\rm R} + 2{\rm Ric}_{34},{\rm Ric}_{34}\right)\right),
\end{equation}
 $\mathcal{E}_3$ is a linear combination of terms of the form
\[\Omega^2{\rm Ric}_{34}, \left((v+1),(-v)\right){\rm Ric}_{34}, (-v)(v+1)^{-1}\mathcal{L}_{\partial_v}\left(\left(v+1\right)^2{\rm Ric}_{34}\right),  \]
and 
\begin{align*}
&\mathcal{E}_4 = 2\left(\slashed{\Delta}+ 2K\right)\mathcal{P}_{\ell > \ell_0}\left(\left|\eta\right|^2-\left|\mathfrak{n}\right|^2\right) + 2\slashed{\nabla}^A\slashed{\nabla}^B\left(\mathfrak{n}\hat{\otimes}\mathfrak{n} - \eta\hat{\otimes}\eta\right)_{AB}+ 4\slashed{\nabla}K\cdot \left(\mathfrak{n}-\eta\right)
\\ \nonumber &\qquad \qquad +\left(\slashed{\Delta}+2K\right)\left(\mathcal{P}_{\ell \leq \ell_0}\slashed{\rm div}\left(\mathfrak{n}-\eta\right)\right).
\end{align*}
\item 
\begin{align}\label{2poj3ojp3}
&(-v)\mathcal{L}_{\partial_v}^2Y -\mathcal{P}_{\ell > \ell_0}\mathcal{L}_b\mathcal{L}_{\partial_v}Y + 4(-v)\mathcal{P}_{\ell > \ell_0}\Omega{\rm tr}\chi \mathcal{L}_{\partial_v}Y 
\\ \nonumber &\qquad + \mathcal{P}_{\ell > \ell_0}\left(\Omega^2\left(\slashed{\Delta}+2K\right)+2(-v)\Omega^2(v+1)^{-2}\right)Y = \mathcal{P}_{\ell > \ell_0}\mathcal{E}_6,
\end{align}
where 
\[Y \doteq \left(v+1\right)^3\mathcal{P}_{\ell > \ell_0}\slashed{\rm curl}\slashed{\rm div}\widehat{\rm Ric},\]
where $\mathcal{E}_5$ is given by 
\begin{align*}
&\mathcal{E}_5 = \left(v+1\right)^2\left(\slashed{\Delta}+2K\right)\mathcal{P}_{\ell > \ell_0}\mathfrak{H} - \left[\left(v+1\right)^2\left(\slashed{\Delta}+2K\right),\mathcal{L}_{\partial_v}\right]\left(v+1\right)^2\mathcal{P}_{\ell >\ell_0}\slashed{\rm curl}\left(\mathfrak{n}-\eta\right) 
\\ \nonumber &\qquad+ \mathcal{L}_{\partial_v}\left(\slashed{\nabla}^A\slashed{\nabla}^B\left(\mathfrak{n}\hat{\otimes}\mathfrak{n} - \eta\hat{\otimes}\eta\right)_{AB} + 2\slashed{\nabla}K\wedge \left(\mathfrak{n}-\eta\right)- \left(\slashed{\Delta}+2K\right)\mathcal{P}_{\ell \leq \ell_0}\slashed{\rm curl}\left(\mathfrak{n}-\eta\right) \right)
\\ \nonumber &\qquad +\mathcal{P}_{\ell > \ell_0}\frac{2\Omega^2}{v+1}(v+1)^2\mathcal{P}_{\ell > \ell_0}\left(\slashed{\nabla}^A\slashed{\nabla}^B\left(\mathfrak{n}\hat{\otimes}\mathfrak{n} - \eta\hat{\otimes}\eta\right)_{AB} + 2\slashed{\nabla}K\wedge \left(\mathfrak{n}-\eta\right)- \left(\slashed{\Delta}+2K\right)\mathcal{P}_{\ell \leq \ell_0}\slashed{\rm curl}\left(\mathfrak{n}-\eta\right) \right)
\\ \nonumber &\qquad -\left[\left(v+1\right)^2\left(\slashed{\Delta}+2K\right),\mathcal{P}_{\ell > \ell_0}\left(\frac{2\Omega^2}{v+1}(v+1)^2\mathcal{P}_{\ell > \ell_0}\cdot \right)\right]\slashed{\rm curl}\left(\mathfrak{n}-\eta\right)
\end{align*}
where $\mathfrak{H}$ is the right hand side of~\eqref{32ojpojp231}, and $\mathcal{E}_6$ is given by by a linear combination of the terms of the sort in~\eqref{2m3omo212rf4} and~\eqref{2o2om41kio312} with the $(v+1)^2$ replaced with a  $\left(v+1\right)^3\mathcal{P}_{\ell > \ell_0}$ and $\slashed{\nabla}$ replaced by $\slashed{\nabla}^3$ and also these additional terms
\[\left((-v)\mathcal{L}_{\partial_v},\Omega{\rm tr}\chi,1\right)\mathcal{E}_5,\ \left(\mathcal{L}_{\partial_v},\Omega{\rm tr}\chi\right)\mathcal{P}_{\ell > \ell_0}\left(v\slashed{\rm curl}\left(\Omega{\rm Ric}\right)_{4A}-\slashed{\rm curl}\left(\Omega{\rm Ric}\right)_{3A}\right),\ \mathcal{P}_{\ell > \ell_0}\left(\left(v+1\right)^3\mathcal{L}_{\partial_v}b \cdot \slashed{\nabla}^3\widehat{\rm Ric}\right).\]
\end{enumerate}
\end{lemma}
\begin{proof}The equation~\eqref{2poj3om2} follows from by multiplying~\eqref{3kcom1} through with $\Omega$ and using that ${\rm Ric}_{44} = 0$. 

We obtain~\eqref{2o3om24} by multiplying the equation~\eqref{3kcom2} with $\Omega$, using~\eqref{32oij32o3}, and then using self-similarity to convert all $\Omega\nabla_3$ derivatives into $\mathcal{L}_{\partial_v}$ and $\mathcal{L}_b$ derivatives plus lower order terms. 

We obtain~\eqref{oj2oj4oj2452} by multiplying the equation~\eqref{3kcom3} through by $\Omega^2$ and then using self-similarity to convert all $\Omega\nabla_3$ derivatives into $\mathcal{L}_{\partial_v}$ and $\mathcal{L}_b$ derivatives plus lower order terms.

Multiplying~\eqref{2poj3om2} with $\frac{1}{2}(-v)$ and~\eqref{2o3om24} with $\frac{1}{2}$ and adding the results together yields
\begin{align}\label{kjoino134}
& (-v)\mathcal{L}_{\partial_v}\left({\rm R} + 2{\rm Ric}_{34}\right) - \frac{1}{2}\mathcal{L}_b\left(R + 3{\rm Ric}_{34}\right) + (-v)\Omega{\rm tr}\chi \left({\rm R} + 2{\rm Ric}_{34}\right) -\frac{1}{2}\slashed{\rm div}b\left({\rm R} + 3{\rm Ric}_{34}\right) 
\\ \nonumber &\qquad  -2(-v)\eta^A\left(\Omega{\rm Ric}\right)_{A4}+\left((-v)\Omega{\rm tr}\chi + 1 - \frac{1}{2}\slashed{\rm div}b\right){\rm Ric}_{34} 
\\ \nonumber &\qquad +\slashed{\nabla}^A\left(v\Omega{\rm Ric}_{A4} + \Omega {\rm Ric}_{A3}\right) +\left((-v)\Omega\hat{\chi}-\Omega\hat{\underline{\chi}}\right)^{AB}\widehat{\rm Ric}_{AB} = 0. 
\end{align}
From~\eqref{ini3moo} and~\eqref{32ojojo1} we obtain 
\begin{align}\label{1im3inm4}
&\slashed{\nabla}^A\slashed{\nabla}^B\widehat{\rm Ric}_{AB} = \frac{1}{2}\left(\slashed{\Delta}+2K\right)\mathcal{P}_{\ell > \ell_0}\left(R+{\rm Ric}_{34}\right) + \frac{1}{2}\left(\slashed{\Delta}+2K\right)\left(\mathcal{P}_{\ell \leq \ell_0}\slashed{\rm div}\left(\mathfrak{n}-\eta\right)\right)
\\ \nonumber &\qquad+ \left(\slashed{\Delta} + 2K\right)\mathcal{P}_{\ell > \ell_0}\left(\left|\eta\right|^2-\left|\mathfrak{n}\right|^2\right) + \slashed{\nabla}^A\slashed{\nabla}^B\left(\mathfrak{n}\hat{\otimes}\mathfrak{n} - \eta\hat{\otimes}\eta\right)_{AB}
 + 2\slashed{\nabla}K\cdot \left(\mathfrak{n}-\eta\right).
\end{align}
Now we multiply~\eqref{kjoino134} through by $\left(v+1\right)^2$, apply $\mathcal{L}_{\partial_v}$, and then use~\eqref{oj2oj4oj2452},~\eqref{2poj3om2},~\eqref{kjoino134}, and~\eqref{1im3inm4} to simplify. After some computation we arrive at the formula~\eqref{2om3omo4}.

From~\eqref{ini3moo} and~\eqref{32ojojo1} we obtain 
\begin{align}\label{klj1lkjl124}
&\slashed{\epsilon}^{AC}\slashed{\nabla}_C\slashed{\nabla}^B\widehat{\rm Ric}_{AB} = \left(\slashed{\Delta}+2K\right)\mathcal{P}_{\ell > \ell_0}\slashed{\rm curl}\left(\mathfrak{n}-\eta\right) + \slashed{\nabla}^A\slashed{\nabla}^B\left(\mathfrak{n}\hat{\otimes}\mathfrak{n} - \eta\hat{\otimes}\eta\right)_{AB} + 2\slashed{\nabla}K\wedge \left(\mathfrak{n}-\eta\right)
\\ \nonumber &\qquad \qquad + \left(\slashed{\Delta}+2K\right)\mathcal{P}_{\ell < \ell_0}\slashed{\rm curl}\left(\mathfrak{n}-\eta\right).
\end{align}
Combining this with $\left(v+1\right)^2\left(\slashed{\Delta}+2K\right)$ applied~\eqref{32ojpojp231} leads to the equation
\begin{align}\label{2ioj4o}
&\mathcal{L}_{\partial_v}\left(\left(v+1\right)^3\mathcal{P}_{\ell > \ell_0}\slashed{\rm curl}\slashed{\rm div}\widehat{\rm Ric}\right) + \mathcal{P}_{\ell > \ell_0}\left(\frac{2\Omega^2}{v+1}\left(v+1\right)^3\mathcal{P}_{\ell > \ell_0}\slashed{\rm curl}\slashed{\rm div}\widehat{\rm Ric}\right)= 
\\ \nonumber &\qquad \left(v+1\right)^2\mathcal{P}_{\ell > \ell_0}\left(\slashed{\Delta}+2K\right)\mathcal{P}_{\ell > \ell_0}\slashed{\rm curl}\left(\Omega{\rm Ric}_{4\cdot}\right) + \mathcal{P}_{\ell > \ell_0}\mathcal{E}_5
\end{align}
Now we apply $\mathcal{L}_{\partial_v}(-v)$ to~\eqref{2ioj4o} and use~\eqref{oj2oj4oj2452} to simplify. After some computations one arrives at~\eqref{2poj3ojp3}.

\end{proof}

\subsection{Estimates for the $\ell = 0$ Projections}
In this section we establish estimates for the spherically symmetric parts of ${\rm Ric}_{33}$, ${\rm Ric}_{34}$, and ${\rm R}$.

We start with ${\rm Ric}_{34}$ and ${\rm R}$. 
\begin{lemma}\label{2omo102jo103}We have
\begin{align*}
&\left\vert\left\vert \mathcal{P}_{\ell = 0}{\rm Ric}_{34}\right\vert\right\vert_{\mathscr{Q}\left(0,0,-1/2+\check{\delta}\right)}^2 
+ \left\vert\left\vert \mathcal{L}_{\partial_v}\mathcal{P}_{\ell = 0}{\rm Ric}_{34}\right\vert\right\vert_{\mathscr{Q}\left(0,1,\kappa\right)}^2+\left\vert\left\vert \mathcal{P}_{\ell = 0}{\rm Ric}_{34}|_{v=0}\right\vert\right\vert_{L^2\left(\mathbb{S}^2\right)}^2
\\ \nonumber &\qquad  + \left\vert\left\vert \mathcal{P}_{\ell = 0}\left({\rm R}+2{\rm Ric}_{34}\right)\right\vert\right\vert_{\mathscr{Q}\left(0,0,-1/2+\check{\delta}\right)}^2
+ \left\vert\left\vert \mathcal{P}_{\ell = 0}\mathcal{L}_{\partial_v}\left({\rm R} + 2{\rm Ric}_{34}\right)\right\vert\right\vert_{\mathscr{Q}\left(0,1,\kappa\right)}^2  + \left\vert\left\vert \mathcal{P}_{\ell = 0}\left({\rm R} +2{\rm Ric}_{34}\right)|_{v=0}\right\vert\right\vert_{L^2\left(\mathbb{S}^2\right)}^2
\\ \nonumber&\qquad \qquad \lesssim \epsilon^2\left\vert\left\vert {\rm Ric}\right\vert\right\vert^2_{\mathscr{Z}} + \epsilon^2\left\vert\left\vert {\rm Ric}\right\vert\right\vert_{\mathscr{M}}^2.
\end{align*}
\end{lemma}
\begin{proof}From~\eqref{2poj3om2} and~\eqref{2o3om24} we find that
\begin{equation}\label{2o3momo2}
\mathcal{L}_{\partial_v}\mathcal{P}_{\ell = 0}\left({\rm R} + {\rm Ric}_{34}\right) + \frac{1}{v+1}\mathcal{P}_{\ell = 0}\left({\rm R} + {\rm Ric}_{34}\right) + \frac{1}{v+1}\mathcal{P}_{\ell = 0}\left({\rm R} + 3{\rm Ric}_{34}\right) = H_1,
\end{equation} 
\begin{equation}\label{2oj3iojn3woij2}
(-v)\mathcal{L}_{\partial_v}\mathcal{P}_{\ell = 0}\left({\rm R} + 3{\rm Ric}_{34}\right) + \left(\frac{2(-v)}{v+1} + \frac{1}{v+1}\right)\mathcal{P}_{\ell = 0}\left({\rm R} + 3{\rm Ric}_{34}\right) - \frac{1}{v+1}\mathcal{P}_{\ell = 0}\left({\rm R} + {\rm Ric}_{34}\right) = H_2,
\end{equation}
where $H_1$ and $H_2$ satisfy 
\[\left\vert\left\vert H_1\right\vert\right\vert^2_{\mathscr{Q}\left(0,1,\kappa\right)} + \left\vert\left\vert H_2\right\vert\right\vert^2_{\mathscr{Q}\left(0,1,-1/2+\check{\delta}\right)} + \left\vert\left\vert \mathcal{L}_{\partial_v}H_2\right\vert\right\vert^2_{\mathscr{Q}\left(0,2,\kappa\right)}\lesssim \epsilon^2\left\vert\left\vert {\rm Ric}\right\vert\right\vert_{\mathscr{Z}}^2,\qquad \left\vert\left\vert H_2|_{v=0}\right\vert\right\vert_{L^2\left(\mathbb{S}^2\right)}^2 \lesssim \epsilon^2\left\vert\left\vert {\rm Ric}\right\vert\right\vert_{\mathscr{M}}^2,\]
\[\sum_{j=0}^1\sup_{v \in [-1,-1/2]}\left(v+1\right)^{1+j+2\check{\delta}}\left[\left|\mathcal{L}_{\partial_v}^jH_1\right| + \left|\mathcal{L}_{\partial_v}^jH_2\right|\right] < \infty.\]
For the analysis near $v = -1$ it is further convenient to write~\eqref{2o3momo2} and~\eqref{2oj3iojn3woij2} as 
\begin{equation}\label{2o3momo2ijijui31}
\mathcal{L}_{\partial_v}\left(\left(v+1\right)\mathcal{P}_{\ell = 0}\left({\rm R} + {\rm Ric}_{34}\right)\right)  = \left(v+1\right)H_1-\mathcal{P}_{\ell = 0}\left({\rm R} + 3{\rm Ric}_{34}\right),
\end{equation} 
\begin{align}\label{2oj3iojn3woij2uiui1iuu}
&\mathcal{L}_{\partial_v}\left(\left(v+1\right)^2\mathcal{P}_{\ell = 0}\left({\rm R} + 3{\rm Ric}_{34}\right)\right) +  \left(v+1\right)^{-1}\left(\left(v+1\right)^2\mathcal{P}_{\ell = 0}\left({\rm R} + 3{\rm Ric}_{34}\right)\right)
\\ \nonumber &\qquad - \left(v+1\right)\mathcal{P}_{\ell = 0}\left({\rm R} + {\rm Ric}_{34}\right) 
 + \frac{(v+1)^2}{(-v)}\left(\mathcal{P}_{\ell = 0}\left({\rm R} + 3{\rm Ric}_{34}\right)-\mathcal{P}_{\ell = 0}\left({\rm R} + {\rm Ric}_{34}\right)\right) = (-v)^{-1}\left(v+1\right)^2H_2.
\end{align}

Now we apply $\mathcal{L}_{\partial_v}$ to~\eqref{2oj3iojn3woij2uiui1iuu}. We end up with 
\begin{align}\label{2om3om4o}
&\left(v+1\right)^{-1}\mathcal{L}_{\partial_v}\underbrace{\left((v+1)\mathcal{L}_{\partial_v}\left((v+1)^2\mathcal{P}_{\ell = 0}\left({\rm R} + 3{\rm Ric}_{34}\right)\right)\right)}_{\doteq X}\\ \nonumber &\qquad + \mathcal{L}_{\partial_v}\left(\frac{(v+1)^2}{(-v)}\mathcal{P}_{\ell = 0}\left(\left({\rm R} + 3{\rm Ric}_{34}\right) - \left({\rm R} +{\rm Ric}_{34}\right)\right)\right) = \left(v+1\right)H_1 + \mathcal{L}_{\partial_v}\left((v+1)^2(-v)^{-1}H_2\right).
\end{align}
From Lemma~\ref{3joj2}, we have that $|X| = O\left(v+1\right)^{1-\check{\delta}}$ as $v\to -1$.  We also have that the terms on the second line of~\eqref{2om3om4o} are  $O(\left(v+1\right)^{-2\check{\delta}})$ as $v\to -1$. We thus immediately see, by integrating~\eqref{2om3om4o}, that we in fact have $X = O\left((v+1)^{2-2\check{\delta}}\right)$ as $v\to -1$. With this improved boundary condition as $v = -1$, it is straightforward to contract~\eqref{2om3om4o} with $\left(v+1\right)^{-2}\left((v+1)\mathcal{L}_{\partial_v}\left((v+1)^2\mathcal{P}_{\ell = 0}\left({\rm R} + 3{\rm Ric}_{34}\right)\right)\right)$, integrate by parts, apply Hardy inequalities, and combine with~\eqref{2o3momo2} to establish, for some $c \ll 1$:  
\begin{align}\label{2om3om2o4324}
&\left\vert\left\vert \mathcal{P}_{\ell = 0}{\rm Ric}_{34}\right\vert\right\vert_{\mathscr{Q}_{-1}^{-1+c}\left(0,0,0\right)}^2 
+ \left\vert\left\vert \mathcal{L}_{\partial_v}\mathcal{P}_{\ell = 0}{\rm Ric}_{34}\right\vert\right\vert_{\mathscr{Q}_{-1}^{-1+c}\left(0,1,0\right)}^2
\\ \nonumber &\qquad  + \left\vert\left\vert \mathcal{P}_{\ell = 0}\left({\rm R}+2{\rm Ric}_{34}\right)\right\vert\right\vert_{\mathscr{Q}_{-1}^{-1+c}\left(0,0,0\right)}^2
+ \left\vert\left\vert \mathcal{P}_{\ell = 0}\mathcal{L}_{\partial_v}\left({\rm R} + 2{\rm Ric}_{34}\right)\right\vert\right\vert_{\mathscr{Q}_{-1}^{-1+c}\left(0,1,0\right)}^2 \lesssim \epsilon^2\left\vert\left\vert {\rm Ric}\right\vert\right\vert^2_{\mathscr{Z}}.
\end{align}
Furthermore, we can then use Gr\"{o}nwall's inequality and simply integrate the ODE's, to replace the $-1+c$ in~\eqref{2om3om2o4324} with $-d$ for a positive constant $d$ satisfying $|d| \ll 1$. 

The equation~\eqref{2om3om4o} is not useful for establishing uniform estimates up to $v = 0$. For this we return to the original pair of equations~\eqref{2o3momo2} and~\eqref{2oj3iojn3woij2}.  Integrating~\eqref{2o3momo2} and using the previously established estimates immediately yields
\begin{align}\label{2omo2444}
&\left\vert\left\vert \mathcal{P}_{\ell = 0}\left({\rm R} + {\rm Ric}_{34}\right)\right\vert\right\vert_{\mathscr{S}_{-1/2}^0\left(0,0,0\right)} + \left\vert\left\vert \mathcal{P}_{\ell = 0}\left({\rm R} + {\rm Ric}_{34}\right)|_{v=0}\right\vert\right\vert_{L^2\left(\mathbb{S}^2\right)} \lesssim 
\\ \nonumber &\qquad  \left\vert\left\vert \mathcal{P}_{\ell = 0}\left({\rm R} + 3{\rm Ric}_{34}\right)\right\vert\right\vert_{\mathscr{Q}_{-d}^0\left(0,0,1/2-\check{\delta}\right)} + \epsilon \left\vert\left\vert {\rm Ric}\right\vert\right\vert_{\mathscr{Z}}.
\end{align}
On the other hand, using Lemma~\ref{l2l2degtranstrans}, from~\eqref{2oj3iojn3woij2} we may establish that
 \begin{equation}\label{3om2omo3}
 \left\vert\left\vert \mathcal{P}_{\ell = 0}\left({\rm R} + 3{\rm Ric}_{34}\right)\right\vert\right\vert_{\mathscr{Q}_{-d}^0\left(0,0,-1/2+\check{\delta}\right)} \lesssim \epsilon\left\vert\left\vert {\rm Ric}\right\vert\right\vert_{\mathscr{Z}} + \left\vert\left\vert \mathcal{P}_{\ell = 0}\left({\rm R} + {\rm Ric}_{34}\right)\right\vert\right\vert_{\mathscr{Q}_{-d}^0\left(0,0,-1/2+\check{\delta}\right)}. 
 \end{equation}
 We may then use the smallness of $d$ to combine these and establish
 \begin{align}\label{i2omo2}
& \left\vert\left\vert \mathcal{P}_{\ell = 0}\left({\rm R} + {\rm Ric}_{34}\right)\right\vert\right\vert_{\mathscr{S}_{-1/2}^0\left(0,0,0\right)} + \left\vert\left\vert \mathcal{P}_{\ell = 0}\left({\rm R} + {\rm Ric}_{34}\right)|_{v=0}\right\vert\right\vert_{L^2\left(\mathbb{S}^2\right)}
\\ \nonumber &\qquad + \left\vert\left\vert \mathcal{P}_{\ell = 0}\left({\rm R} + 3{\rm Ric}_{34}\right)\right\vert\right\vert_{\mathscr{Q}_{-d}^0\left(0,0,-1/2+\check{\delta}\right)}  \lesssim \epsilon^2\left\vert\left\vert {\rm Ric}\right\vert\right\vert_{\mathscr{Z}}^2.
 \end{align}
 Revisiting the equation~\eqref{2o3momo2} then leads directly to the desired remaining estimate for $\mathcal{L}_{\partial_v}\mathcal{P}_{\ell = 0}\left({\rm R} + {\rm Ric}_{34}\right)$. Finally, the remaining estimates for $\mathcal{P}_{\ell = 0}\left({\rm R} + {\rm Ric}_{34}\right)$ are obtained by restricting~\eqref{2oj3iojn3woij2} to $\{v = 0\}$ (while keeping in mind that $(-v)\mathcal{L}_{\partial_v}\left({\rm R}+3{\rm Ric}_{34}\right)$ vanishes at $v = 0$), and then also differentiating~\eqref{2oj3iojn3woij2} with $\mathcal{L}_{\partial_v}$ and applying again Lemma~\ref{l2l2degtranstrans}.

\end{proof}

It is now straightforward to estimate $\mathcal{P}_{\ell = 0}\left(\Omega^2{\rm Ric}_{33}\right)$. 
\begin{lemma}We have
\begin{align*}
&\left\vert\left\vert \mathcal{L}_{\partial_v}\mathcal{P}_{\ell = 0}\left(\Omega^2{\rm Ric}_{33}\right)\right\vert\right\vert_{\mathscr{Q}\left(0,1,\kappa\right)}^2 + \left\vert\left\vert \mathcal{P}_{\ell = 0}\left(\Omega^2{\rm Ric}_{33}\right)\right\vert\right\vert_{\mathscr{Q}\left(0,0,-\kappa\right)}^2+\left\vert\left\vert \mathcal{P}_{\ell = 0}\left(\Omega^2{\rm Ric}_{33}\right)|_{v=0}\right\vert\right\vert_{L^2\left(\mathbb{S}^2\right)}^2 
\\ \nonumber &\qquad \lesssim \epsilon^2\left[\left\vert\left\vert {\rm Ric}\right\vert\right\vert_{\mathscr{Z}}^2+\left\vert\left\vert {\rm Ric}\right\vert\right\vert_{\mathscr{M}}^2\right].
\end{align*}
\end{lemma}
\begin{proof}In view of the already established estimates from Lemma~\ref{2omo102jo103}, this estimate follows easily by integrating the equation~\eqref{32oij32o3}, after applying $\mathcal{P}_{\ell = 0} \Omega^2$.   
\end{proof}

\subsection{Estimates for $\mathcal{P}_{\ell \geq 1}\left(\Omega^2{\rm Ric}_{33}\right)$ and $\mathcal{P}_{\ell \geq 1}{\rm Ric}_{34}$}\label{om1o1mo2}
In this section we will study the system of equations~\eqref{32oij32o3} and~\eqref{32oi32jio32123fuhn24} in order to establish estimates for $\mathcal{P}_{\ell \geq 1}\left(\Omega^2{\rm Ric}_{33}\right)$ and $\mathcal{P}_{\ell \geq 1}{\rm Ric}_{34}$.
\begin{lemma}We have
\[\left\vert\left\vert {\rm Ric}_{33}\right\vert\right\vert_{\mathscr{Z}} + \left\vert\left\vert {\rm Ric}_{34}\right\vert\right\vert_{\mathscr{Z}}+ \left\vert\left\vert {\rm Ric}_{34}\right\vert\right\vert_{\mathscr{M}} \lesssim \epsilon\left[\left\vert\left\vert {\rm Ric}\right\vert\right\vert_{\mathscr{Z}} + \left\vert\left\vert {\rm Ric}\right\vert\right\vert_{\mathscr{M}}\right].\]
\end{lemma}
\begin{proof}We start by observing that~\eqref{32oij32o3} and~\eqref{32oi32jio32123fuhn24} imply the following:
\begin{align}\label{o3omomo34920}
&\mathcal{L}_{\partial_v}\left(\mathcal{P}_{\ell \geq 1}\left(\Omega^2{\rm Ric}_{33}\right)\right) + \frac{2}{v+1}\mathcal{P}_{\ell \geq 1}\left(\Omega^2{\rm Ric}_{33}\right) 
\\ \nonumber&\qquad -2\Omega^2\left(\left((-v)\mathcal{L}_{\partial_v}-\mathcal{P}_{\ell \geq 1}\mathcal{L}_b\right)\mathcal{P}_{\ell \geq 1}{\rm Ric}_{34} + \frac{2(-v)}{v+1}\mathcal{P}_{\ell \geq 1}{\rm Ric}_{34}\right) = H_1,
\end{align}
\begin{align}\label{2o3momro1o49}
\left(\left((-v)\mathcal{L}_{\partial_v}-\mathcal{P}_{\ell \geq 1}\mathcal{L}_b\right) + \left(1 + \frac{4(-v)}{v+1}\right)\right)\mathcal{P}_{\ell \geq 1}{\rm Ric}_{34} - \left(v+1\right)^{-2}\mathring{\Delta}\mathcal{P}_{\ell \geq 1}\left(\Omega^2{\rm Ric}_{33}\right) = H_2,
\end{align}
where $H_1$ and $H_2$ satisfy the following:
\begin{align}\label{lm3om2om3}
&\left\vert\left\vert H_1\right\vert\right\vert_{\mathscr{Q}\left(4,1,-1/2+10\sqrt{\check{\delta}}\right)} + \left\vert\left\vert H_1\right\vert\right\vert_{\mathscr{Q}\left(3,1,-1/2+10\check{\delta}\right)}+ \left\vert\left\vert H_2\right\vert\right\vert_{\mathscr{Q}\left(4,0,-1/2+\check{\delta}\right)}+\left\vert\left\vert \mathcal{L}_{\partial_v}H_2\right\vert\right\vert_{\mathscr{Q}\left(2,1,-1/2+\check{\delta}\right)}
\\ \nonumber &\qquad \lesssim \epsilon\left\vert\left\vert {\rm Ric}\right\vert\right\vert_{\mathscr{Z}}.
\end{align}
Then, setting $X \doteq \left(v+1\right)^4\mathcal{P}_{\ell \geq 1}{\rm Ric}_{34}$, we may conjugate~\eqref{2o3momro1o49} with $\left(v+1\right)^4$ and then apply $\mathcal{L}_{\partial_v}$ and use~\eqref{o3omomo34920} to simplify. We obtain
\begin{align}\label{2pom3onm2}
\left((-v)\mathcal{L}_{\partial_v}-\mathcal{P}_{\ell \geq 1}\mathcal{L}_b\right)Z = H_3,\qquad Z \doteq \mathcal{L}_{\partial_v}X - \Omega^2\left(v+1\right)^{-2}\mathring{\Delta}X,
\end{align}
where $H_3$ satisfies 
\begin{equation}\label{2om3om1o4r591j4}
\left\vert\left\vert H_3\right\vert\right\vert_{\mathscr{Q}\left(2,-1,-1/2+10\sqrt{\check{\delta}}\right)} + \left\vert\left\vert H_3\right\vert\right\vert_{\mathscr{Q}\left(1,-1,-1/2+10\check{\delta}\right)}  \lesssim \epsilon \left\vert\left\vert {\rm Ric}\right\vert\right\vert_{\mathscr{Z}}.
\end{equation}
We now apply Lemma~\ref{2oj3omo1mo323} and immediately obtain the desired estimates ${\rm Ric}_{34}$. (The requirements~\eqref{omom2o4} and~\eqref{1mo2mo4o2} are easily verified.) 

To estimate ${\rm Ric}_{33}$, we re-write the equation~\eqref{o3omomo34920} as 
\begin{align}\label{2kll3kj2lj}
&\mathcal{L}_{\partial_v}\left(\mathcal{P}_{\ell \geq 1}\left(\Omega^2{\rm Ric}_{33}\right)\right) + \frac{2}{v+1}\mathcal{P}_{\ell \geq 1}\left(\Omega^2{\rm Ric}_{33}\right) -2\Omega^2\left(v+1\right)^{-2}\mathring{\Delta}\mathcal{P}_{\ell \geq 1}\left(\Omega^2{\rm Ric}_{33}\right)
\\ \nonumber&\qquad -2\Omega^2\left(-\left(1 + \frac{4(-v)}{v+1}\right)\mathcal{P}_{\ell \geq 1}{\rm Ric}_{34} + \frac{2(-v)}{v+1}\mathcal{P}_{\ell \geq 1}{\rm Ric}_{34}\right) = H_1+2\Omega^2H_2.
\end{align}
Then the desired estimate for ${\rm Ric}_{33}$ follows easily by energy estimates and the already established estimates for ${\rm Ric}_{34}$. 
 
\end{proof}

\subsection{Estimates for the Remaining $1 \leq \ell \leq \ell_0$ Projections}
In this section we will establish estimates for $\mathcal{P}_{1\leq \ell \leq \ell_0}\left({\rm R} + 2{\rm Ric}_{34},\widehat{\rm Ric},\Omega{\rm Ric}_{A4},\Omega{\rm Ric}_{A3}\right)$.

We start by applying $\mathcal{P}_{\ell \leq \ell_0}$ to~\eqref{2om3omo4} and using the resulting equation to estimate $
\mathcal{P}_{\ell \leq \ell_0}\left({\rm R} + {\rm Ric}_{34}\right)$.
\begin{lemma}We have
\begin{align*}
&\left\vert\left\vert \mathcal{P}_{1 \leq \ell \leq \ell_0}\left({\rm R} + 2{\rm Ric}_{34}\right)\right\vert\right\vert_{\mathscr{Q}\left(0,1,-1/2+\check{\delta}\right)} + \left\vert\left\vert \mathcal{L}_{\partial_v}\mathcal{P}_{1 \leq \ell \leq \ell_0}\left({\rm R} + 2{\rm Ric}_{34}\right)\right\vert\right\vert_{\mathscr{Q}\left(0,2,\kappa\right)}
\\ \nonumber &\qquad  + \left\vert\left\vert \mathcal{P}_{1 \leq \ell \leq \ell_0}\left({\rm R} + 2{\rm Ric}_{34}\right)|_{v=0}\right\vert\right\vert_{L^2\left(\mathbb{S}^2\right)} \lesssim  \epsilon\left[\left\vert\left\vert {\rm Ric}\right\vert\right\vert_{\mathscr{Z}}+\left\vert\left\vert {\rm Ric}\right\vert\right\vert_{\mathscr{M}}\right].
\end{align*}
\end{lemma}
\begin{proof}Applying $\mathcal{P}_{\ell \leq \ell_0}$ to~\eqref{2om3omo4} leads to an equation of the form
\begin{equation}
(-v)\mathcal{L}_{\partial_v}^2\mathcal{P}_{\ell \leq \ell_0}X + \frac{2(-v)}{v+1}\mathcal{L}_{\partial_v}\mathcal{P}_{\ell \leq \ell_0}X = \mathcal{E},
\end{equation}
where $\mathcal{E}$ satisfies
\[\left\vert\left\vert \mathcal{E}\right\vert\right\vert_{\mathscr{Q}\left(0,0,\kappa\right)} \lesssim \epsilon \left[\left\vert\left\vert {\rm Ric}\right\vert\right\vert_{\mathscr{Z}}+\left\vert\left\vert {\rm Ric}\right\vert\right\vert_{\mathscr{M}}\right].\]
Furthermore, we have that 
\[\mathcal{E} = O\left(\left(v+1\right)^{-2\check{\delta}}\right)\text{ as }v\to -1,\qquad \mathcal{P}_{\ell \leq \ell_0}X = O\left(\left(v+1\right)^{1-2\check{\delta}}\right)\text{ as }v\to -1,\]
\[ \mathcal{L}_{\partial_v}\mathcal{P}_{\ell \leq \ell_0}X = O\left(\left(v+1\right)^{-2\check{\delta}}\right)\text{ as }v\to -1\]
In particular, we can apply Lemma~\ref{l2l2degtranstrans} to obtain the following estimate for $\mathcal{L}_{\partial_v}\mathcal{P}_{\ell \leq \ell_0}X$:
\[\left\vert\left\vert \mathcal{L}_{\partial_v}\mathcal{P}_{\ell \leq \ell_0}X\right\vert\right\vert_{\mathscr{Q}\left(0,0,\kappa\right)} \lesssim \epsilon \left[\left\vert\left\vert {\rm Ric}\right\vert\right\vert_{\mathscr{Z}}+\left\vert\left\vert {\rm Ric}\right\vert\right\vert_{\mathscr{M}}\right].\]
The fundamental theorem of calculus then yields that
\[\left\vert\left\vert \mathcal{P}_{\ell \leq \ell_0}X\right\vert\right\vert_{\mathscr{Q}\left(0,-1,-1/2+\check{\delta}\right)} + \left\vert\left\vert \mathcal{P}_{\ell \leq \ell_0}X |_{v=0}\right\vert\right\vert_{L^2\left(\mathbb{S}^2\right)} \lesssim \epsilon \left[\left\vert\left\vert {\rm Ric}\right\vert\right\vert_{\mathscr{Z}}+\left\vert\left\vert {\rm Ric}\right\vert\right\vert_{\mathscr{M}}\right].\]
Converting these into estimates for ${\rm R} + 2{\rm Ric}_{34}$ finishes the proof.

\end{proof}

Next we turn to $\mathcal{P}_{\ell \leq \ell_0}\widehat{\rm Ric}$.
\begin{lemma}We have
\begin{align*}
&\left\vert\left\vert \mathcal{P}_{1 \leq \ell \leq \ell_0}\widehat{\rm Ric}\right\vert\right\vert_{\mathscr{Q}\left(0,1,-1/2+\check{\delta}\right)} + \left\vert\left\vert \mathcal{L}_{\partial_v}\mathcal{P}_{1 \leq \ell \leq \ell_0}\widehat{\rm Ric}\right\vert\right\vert_{\mathscr{Q}\left(0,2,\kappa\right)}
 + \left\vert\left\vert \mathcal{P}_{1 \leq \ell \leq \ell_0}\widehat{\rm Ric}|_{v=0}\right\vert\right\vert_{L^2\left(\mathbb{S}^2\right)} \\\ \nonumber&\qquad \lesssim  \epsilon\left[\left\vert\left\vert {\rm Ric}\right\vert\right\vert_{\mathscr{Z}}+\left\vert\left\vert {\rm Ric}\right\vert\right\vert_{\mathscr{M}}+\left\vert\left\vert \left(\mathfrak{n}-\eta\right)|_{v=0}\right\vert\right\vert_{L^2\left(\mathbb{S}^2\right)}\right].
\end{align*}
\end{lemma}
\begin{proof}This is an immediate consequence of~\eqref{ini3moo} and~\eqref{2om3pom2p}.
\end{proof}

Finally, we come to $\mathcal{P}_{\ell \leq \ell_0}\left(\Omega{\rm Ric}_{A4}\right)$ and $\mathcal{P}_{\ell \leq \ell_0}\left(\Omega{\rm Ric}_{3A}\right)$.
\begin{lemma}We have
\begin{align*}
&\left\vert\left\vert \mathcal{P}_{\ell \leq \ell_0}\Omega{\rm Ric}_{A4}\right\vert\right\vert_{\mathscr{Q}\left(0,1,\kappa\right)} + \left\vert\left\vert \mathcal{L}_{\partial_v}\mathcal{P}_{\ell \leq \ell_0}\Omega{\rm Ric}_{A4}\right\vert\right\vert_{\mathscr{Q}\left(0,2,1+\kappa\right)} + 
\\ \nonumber &\qquad \left\vert\left\vert \mathcal{P}_{\ell \leq \ell_0}\Omega{\rm Ric}_{A3}\right\vert\right\vert_{\mathscr{Q}\left(0,1,-1/2+\check{\delta}\right)} + \left\vert\left\vert \mathcal{L}_{\partial_v}\mathcal{P}_{\ell \leq \ell_0}\Omega{\rm Ric}_{A3}\right\vert\right\vert_{\mathscr{Q}\left(0,2,\kappa\right)} + \left\vert\left\vert \Omega{\rm Ric}_{3A}|_{v=0}\right\vert\right\vert_{L^2\left(\mathbb{S}^2\right)} 
\\ \nonumber&\qquad \lesssim  \epsilon\left[\left\vert\left\vert {\rm Ric}\right\vert\right\vert_{\mathscr{Z}}+\left\vert\left\vert {\rm Ric}\right\vert\right\vert_{\mathscr{M}}+\left\vert\left\vert \left(\mathfrak{n}-\eta\right)|_{v=0}\right\vert\right\vert_{L^2\left(\mathbb{S}^2\right)}\right].
\end{align*}
\end{lemma}
\begin{proof}From~\eqref{2poj3om2} and~\eqref{2o3om24} and our previous estimates we immediately obtain
 \begin{align*}
&\left\vert\left\vert \mathcal{P}_{\ell \leq \ell_0}\slashed{\nabla}^A\Omega{\rm Ric}_{A4}\right\vert\right\vert_{\mathscr{Q}\left(0,2,\kappa\right)} + \left\vert\left\vert \mathcal{L}_{\partial_v}\mathcal{P}_{\ell \leq \ell_0}\slashed{\nabla}^A\Omega{\rm Ric}_{A4}\right\vert\right\vert_{\mathscr{Q}\left(0,3,1+\kappa\right)} + 
\\ \nonumber &\qquad \left\vert\left\vert \mathcal{P}_{\ell \leq \ell_0}\slashed{\nabla}^A\Omega{\rm Ric}_{A3}\right\vert\right\vert_{\mathscr{Q}\left(0,1,-1/2+\check{\delta}\right)} + \left\vert\left\vert \mathcal{L}_{\partial_v}\mathcal{P}_{\ell \leq \ell_0}\slashed{\nabla}^A\Omega{\rm Ric}_{A3}\right\vert\right\vert_{\mathscr{Q}\left(0,3,\kappa\right)} + \left\vert\left\vert \slashed{\nabla}^A\Omega{\rm Ric}_{3A}|_{v=0}\right\vert\right\vert_{L^2\left(\mathbb{S}^2\right)} 
\\ \nonumber&\qquad \lesssim  \epsilon\left[\left\vert\left\vert {\rm Ric}\right\vert\right\vert_{\mathscr{Z}}+\left\vert\left\vert {\rm Ric}\right\vert\right\vert_{\mathscr{M}}+\left\vert\left\vert \left(\mathfrak{n}-\eta\right)|_{v=0}\right\vert\right\vert_{L^2\left(\mathbb{S}^2\right)}\right].
\end{align*}
In order to estimate the $\slashed{\rm curl}$ of ${\rm Ric}_{A4}$ we apply $\left(v+1\right)\mathcal{P}_{\ell \leq \ell_0}\slashed{\rm curl}$ to~\eqref{oj2oj4oj2452} and use~\eqref{m3mini43jn2o} to obtain an equation
\[(-v)\mathcal{L}_{\partial_v}\left((v+1)\mathcal{P}_{\ell \leq \ell_0}\slashed{\rm curl}{\rm Ric}_{\cdot 4}\right) + \frac{3(-v)}{v+1}\left((v+1)\mathcal{P}_{\ell \leq \ell_0}\slashed{\rm curl}{\rm Ric}_{\cdot 4}\right) = \mathcal{F},\]
where $\mathcal{F}$ satisfies
\[\left\vert\left\vert \mathcal{F}\right\vert\right\vert_{\mathscr{Q}\left(0, 1, \kappa\right)} \lesssim \epsilon\left[\left\vert\left\vert {\rm Ric}\right\vert\right\vert_{\mathscr{Z}}+\left\vert\left\vert {\rm Ric}\right\vert\right\vert_{\mathscr{M}}+\left\vert\left\vert \left(\mathfrak{n}-\eta\right)|_{v=0}\right\vert\right\vert_{L^2\left(\mathbb{S}^2\right)}\right],\qquad \mathcal{F} = O\left((v+1)^{-1-2\check{\delta}}\right)\text{ as }v\to -1.\]
Since have that ${\rm Ric}_{A4} = O\left((v+1)\right)^{-1-\check{\delta}}$ as $v\to -1$, we may apply Lemma~\ref{l2l2degtranstrans} to obtain that
\[\sum_{j=0}^1\left\vert\left\vert \mathcal{L}_{\partial_v}^j\mathcal{P}_{\ell \leq \ell_0}\slashed{\rm curl}\Omega{\rm Ric}_{\cdot 4}\right\vert\right\vert_{\mathscr{Q}\left(0,1+j,j+\kappa\right)} \lesssim \epsilon\left[\left\vert\left\vert {\rm Ric}\right\vert\right\vert_{\mathscr{Z}}+\left\vert\left\vert {\rm Ric}\right\vert\right\vert_{\mathscr{M}}+\left\vert\left\vert \left(\mathfrak{n}-\eta\right)|_{v=0}\right\vert\right\vert_{L^2\left(\mathbb{S}^2\right)}\right].\]
By way of~\eqref{m3mini43jn2o} we get the analogous estimates for $\mathcal{P}_{\ell \leq \ell_0}\slashed{\rm curl}\Omega{\rm Ric}_{\cdot 3}$. Combining all of these estimates and using elliptic estimates finishes the proof.
\end{proof}
\subsection{Estimates along the boundary $\{v = 0\}$}
In the next lemma, we will establish the desired boundary estimates for the Ricci curvature, $\mathfrak{n}$, and $\mathfrak{j}$.
\begin{lemma}We have
\[\left\vert\left\vert {\rm Ric}\right\vert\right\vert_{\mathscr{M}} +\left\vert\left\vert \left(1,\mathcal{L}_b\right)\left(\mathfrak{n}-\eta\right)|_{v=0}\right\vert\right\vert_{\mathring{H}^4\left(\mathbb{S}^2\right)} +\left\vert\left\vert \left(1,\mathcal{L}_b\right)\left(\mathfrak{n}-\eta\right)|_{v=0}\right\vert\right\vert_{\mathring{H}^4\left(\mathbb{S}^2\right)}\lesssim \epsilon\left\vert\left\vert {\rm Ric}\right\vert\right\vert_{\mathscr{Z}},\]
\end{lemma}
\begin{proof}

Restricting~\eqref{2o3om24} to $\{v = 0\}$ yields the following equation
\begin{align}\label{k2l3lkjl2k3j}
&-\mathcal{L}_b\left({\rm R} + 3{\rm Ric}_{34}\right) -\slashed{\rm div}b\left({\rm R} + 3{\rm Ric}_{34}\right) - \Omega{\rm tr}\underline{\chi} {\rm Ric}_{34} +2\slashed{\nabla}^A\left(\Omega{\rm Ric}_{3A}\right) -2 \left(\Omega\hat{\underline{\chi}}\right)^{AB}\widehat{\rm Ric}_{AB} = 0,
\end{align}

Combining this with the restrictions of~\eqref{m3mini43jn2o} and~\eqref{plqowune33342} to $\{v = 0\}$ and using the previously established estimates leads immediately to the estimate
\begin{equation}\label{kommo3m20fu93}
\left\vert\left\vert \Omega{\rm Ric}_{3A}|_{v=0}\right\vert\right\vert_{\mathring{H}^3\left(\mathbb{S}^2\right)} \lesssim \epsilon\left[\left\vert\left\vert {\rm Ric}\right\vert\right\vert_{\mathscr{M}}+ \left\vert\left\vert {\rm Ric}\right\vert\right\vert_{\mathscr{Z}}+ \left\vert\left\vert \left(1,\mathcal{L}_b\right)\left(\mathfrak{n}-\eta\right)|_{v=0}\right\vert\right\vert_{\mathring{H}^4\left(\mathbb{S}^2\right)}\right].
\end{equation}
Note, however, that we cannot use this particular argument to control four derivatives of $\Omega{\rm Ric}_{3A}$ because our norms do not allow us to treat $\slashed{\nabla}^3\mathcal{L}_b\left({\rm R}+3{\rm Ric}_{34}\right)$ as a nonlinear term.

 It then follows from~\eqref{2nk3knk2}, ~\eqref{2n3knk2}, Proposition~\ref{somestuimdie}, and elliptic estimates that
\begin{align}\label{mcom3o293}
&\left\vert\left\vert \left(1,\mathcal{L}_b\right)\left(\mathfrak{j}-\eta\right)|_{v=0} \right\vert\right\vert_{\mathring{H}^3\left(\mathbb{S}^2\right)}  \lesssim \left\vert\left\vert \Omega{\rm Ric}_{3A}|_{v=0}\right\vert\right\vert_{\mathring{H}^3\left(\mathbb{S}^2\right)}\\ \nonumber  &\lesssim \epsilon\left[\left\vert\left\vert {\rm Ric}\right\vert\right\vert_{\mathscr{M}}+ \left\vert\left\vert {\rm Ric}\right\vert\right\vert_{\mathscr{Z}}+ \left\vert\left\vert \left(1,\mathcal{L}_b\right)\left(\mathfrak{n}-\eta\right)|_{v=0}\right\vert\right\vert_{\mathring{H}^4\left(\mathbb{S}^2\right)}+\left\vert\left\vert \left(1,\mathcal{L}_b\right)\left(\mathfrak{j}-\eta\right)|_{v=0} \right\vert\right\vert_{\mathring{H}^4\left(\mathbb{S}^2\right)}\right].
\end{align}
Then, from~\eqref{2ojoijo2}, we obtain that 
\begin{align}\label{kiijm3mi3}
&\left\vert\left\vert \left(1,\mathcal{L}_b\right)\left({\rm R}+ {\rm Ric}_{34}\right)|_{v=0}\right\vert\right\vert_{\mathring{H}^2\left(\mathbb{S}^2\right)}  \lesssim \left\vert\left\vert \left(1,\mathcal{L}_b\right)\left(\mathfrak{j}-\eta\right)|_{v=0}\right\vert\right\vert_{\mathring{H}^3\left(\mathbb{S}^2\right)}
\\ \nonumber  &\lesssim  \epsilon\left[\left\vert\left\vert {\rm Ric}\right\vert\right\vert_{\mathscr{M}}+ \left\vert\left\vert {\rm Ric}\right\vert\right\vert_{\mathscr{Z}}+ \left\vert\left\vert \left(1,\mathcal{L}_b\right)\left(\mathfrak{n}-\eta\right)|_{v=0}\right\vert\right\vert_{\mathring{H}^4\left(\mathbb{S}^2\right)}+\left\vert\left\vert \left(1,\mathcal{L}_b\right)\left(\mathfrak{j}-\eta\right)|_{v=0} \right\vert\right\vert_{\mathring{H}^4\left(\mathbb{S}^2\right)}\right].
\end{align}
Finally, it is then straightforward to use~\eqref{ini3moo},~\eqref{2om3pom2p}, and~\eqref{32ojojo1} to then establish that  \begin{align}\label{slmfo3mo2p}
&\left\vert\left\vert \left(1,\mathcal{L}_b\right)\widehat{\rm Ric}|_{v=0} \right\vert\right\vert_{\mathring{H}^2\left(\mathbb{S}^2\right)} +\left\vert\left\vert \left(1,\mathcal{L}_b\right)\left(\mathfrak{n}-\eta\right)|_{v=0}\right\vert\right\vert_{\mathring{H}^3\left(\mathbb{S}^2\right)} 
\\ \nonumber &\qquad \lesssim \epsilon\left[\left\vert\left\vert {\rm Ric}\right\vert\right\vert_{\mathscr{M}}+ \left\vert\left\vert {\rm Ric}\right\vert\right\vert_{\mathscr{Z}}+ \left\vert\left\vert \left(1,\mathcal{L}_b\right)\left(\mathfrak{n}-\eta\right)|_{v=0}\right\vert\right\vert_{\mathring{H}^4\left(\mathbb{S}^2\right)}+\left\vert\left\vert \left(1,\mathcal{L}_b\right)\left(\mathfrak{j}-\eta\right)|_{v=0} \right\vert\right\vert_{\mathring{H}^4\left(\mathbb{S}^2\right)}\right].
\end{align}

For the top-order estimates we will have to argue differently. From~\eqref{2nk3knk2},~\eqref{2ojoijo2}, and~\eqref{k2l3lkjl2k3j} one derives along $\{v = 0\}$
\begin{equation}\label{mofem3omo4}
\mathcal{L}_b\mathcal{P}_{\ell > \ell_0}\left({\rm R} + {\rm Ric}_{34}\right) -\mathcal{P}_{\ell > \ell_0}\left({\rm R} + {\rm Ric}_{34}\right) = \mathcal{F},
\end{equation}
for $\mathcal{F}$ which satisfies
\[\left\vert\left\vert \mathcal{F}\right\vert\right\vert_{\mathring{H}^3\left(\mathbb{S}^2\right)} \lesssim \epsilon\left[\left\vert\left\vert {\rm Ric}\right\vert\right\vert_{\mathscr{M}}+ \left\vert\left\vert {\rm Ric}\right\vert\right\vert_{\mathscr{Z}}+ \left\vert\left\vert \left(1,\mathcal{L}_b\right)\left(\mathfrak{n}-\eta\right)|_{v=0}\right\vert\right\vert_{\mathring{H}^4\left(\mathbb{S}^2\right)}+\left\vert\left\vert \left(1,\mathcal{L}_b\right)\left(\mathfrak{j}-\eta\right)|_{v=0} \right\vert\right\vert_{\mathring{H}^4\left(\mathbb{S}^2\right)}\right].\]
In particular, we obtain from Proposition~\ref{somestuimdie} that 
\begin{align}\label{kiijm3mi323453}
&\left\vert\left\vert \left(1,\mathcal{L}_b\right)\left({\rm R} + {\rm Ric}_{34}\right)|_{v=0}\right\vert\right\vert_{\mathring{H}^3\left(\mathbb{S}^2\right)} 
\\ \nonumber &\lesssim \epsilon\left[\left\vert\left\vert {\rm Ric}\right\vert\right\vert_{\mathscr{M}}+ \left\vert\left\vert {\rm Ric}\right\vert\right\vert_{\mathscr{Z}}+ \left\vert\left\vert \left(1,\mathcal{L}_b\right)\left(\mathfrak{n}-\eta\right)|_{v=0}\right\vert\right\vert_{\mathring{H}^4\left(\mathbb{S}^2\right)}+\left\vert\left\vert \left(1,\mathcal{L}_b\right)\left(\mathfrak{j}-\eta\right)|_{v=0} \right\vert\right\vert_{\mathring{H}^4\left(\mathbb{S}^2\right)}\right].
\end{align}
Next we turn to $\widehat{\rm Ric}$. From~\eqref{ini3moo},~\eqref{2om3pom2p},~\eqref{32ojojo1},~\eqref{2ojoijo2}, and~\eqref{32ijo32ijo32ioj} we may derive the following pair of equations along $\{v= 0\}$:
\begin{equation}\label{j30ido3}
\slashed{\nabla}^A\slashed{\nabla}^B\widehat{\rm Ric}_{AB} = \frac{1}{2}\left(\slashed{\Delta}+2\right)\left({\rm R} + {\rm Ric}_{34}\right) + \mathcal{G}_1,
\end{equation}
\begin{equation}\label{ewijlewli}
\slashed{\epsilon}^{AB}\slashed{\nabla}_A\slashed{\nabla}^C\widehat{\rm Ric}_{BC} = \slashed{\Delta}\slashed{\rm curl}\left(j-\eta\right) + \mathcal{G}_2,
\end{equation}
where
\[\left\vert\left\vert \left(\mathcal{G}_1,\mathcal{G}_2\right)|_{v=0}\right\vert\right\vert_{\mathring{H}^2\left(\mathbb{S}^2\right)} \lesssim\epsilon\left[\left\vert\left\vert {\rm Ric}\right\vert\right\vert_{\mathscr{M}}+ \left\vert\left\vert {\rm Ric}\right\vert\right\vert_{\mathscr{Z}}+ \left\vert\left\vert \left(1,\mathcal{L}_b\right)\left(\mathfrak{n}-\eta\right)|_{v=0}\right\vert\right\vert_{\mathring{H}^4\left(\mathbb{S}^2\right)}+\left\vert\left\vert \left(1,\mathcal{L}_b\right)\left(\mathfrak{j}-\eta\right)|_{v=0} \right\vert\right\vert_{\mathring{H}^4\left(\mathbb{S}^2\right)}\right].\]
Furthermore, from~\eqref{plqowune33342} and~\eqref{2n3knk2} we have
\begin{equation}\label{ije4jo3od}
\left(2-\mathcal{L}_b\right)\slashed{\Delta}\slashed{\rm curl}\left(\mathfrak{j}-\eta\right)= \mathcal{L}_b\left(\slashed{\epsilon}^{AB}\slashed{\nabla}_A\slashed{\nabla}^C\widehat{\rm Ric}_{BC}\right) + \mathcal{G}_3,
\end{equation}
where
\[\left\vert\left\vert \mathcal{G}_3\right\vert\right\vert_{\mathring{H}^1\left(\mathbb{S}^2\right)} \lesssim  \epsilon\left[\left\vert\left\vert {\rm Ric}\right\vert\right\vert_{\mathscr{M}}+ \left\vert\left\vert {\rm Ric}\right\vert\right\vert_{\mathscr{Z}}+ \left\vert\left\vert \left(1,\mathcal{L}_b\right)\left(\mathfrak{n}-\eta\right)|_{v=0}\right\vert\right\vert_{\mathring{H}^4\left(\mathbb{S}^2\right)}+\left\vert\left\vert \left(1,\mathcal{L}_b\right)\left(\mathfrak{j}-\eta\right)|_{v=0} \right\vert\right\vert_{\mathring{H}^4\left(\mathbb{S}^2\right)}\right].\]
Then, we may use~\eqref{ije4jo3od} and~\eqref{ewijlewli} to derive
\begin{equation}\label{ij3ij3}
\mathcal{L}_b\left(\slashed{\Delta}\slashed{\rm curl}(\mathfrak{j}-\eta)\right) -2\slashed{\Delta}\slashed{\rm curl}\left(\mathfrak{j}-\eta\right) = \mathcal{G}_4,
\end{equation}
where $\mathcal{G}_4$ satisfies the same estimate $\mathcal{G}_3$. In particular, applying Proposition~\ref{somestuimdie} to~\eqref{ij3ij3} and then using~\eqref{j30ido3},~\eqref{ewijlewli}, and elliptic estimates leads to
\[\left\vert\left\vert \left(1,\mathcal{L}_b\right)\widehat{\rm Ric} \right\vert\right\vert_{\mathring{H}^3\left(\mathbb{S}^2\right)} \lesssim \epsilon\left[\left\vert\left\vert {\rm Ric}\right\vert\right\vert_{\mathscr{M}}+ \left\vert\left\vert {\rm Ric}\right\vert\right\vert_{\mathscr{Z}}+ \left\vert\left\vert \left(1,\mathcal{L}_b\right)\left(\mathfrak{n}-\eta\right)|_{v=0}\right\vert\right\vert_{\mathring{H}^4\left(\mathbb{S}^2\right)}+\left\vert\left\vert \left(1,\mathcal{L}_b\right)\left(\mathfrak{j}-\eta\right)|_{v=0} \right\vert\right\vert_{\mathring{H}^4\left(\mathbb{S}^2\right)}\right].\]

Since we have now the top order estimate for ${\rm R}  +{\rm Ric}_{34}$ and $\widehat{\rm Ric}$, we can revisit~\eqref{k2l3lkjl2k3j}, use the restrictions of~\eqref{m3mini43jn2o} and~\eqref{plqowune33342} to $\{v = 0\}$, and apply elliptic estimates to establish that
\[\left\vert\left\vert \Omega{\rm Ric}_{3A}\right\vert\right\vert_{\mathring{H}^4\left(\mathbb{S}^2\right)} \lesssim \epsilon\left[\left\vert\left\vert {\rm Ric}\right\vert\right\vert_{\mathscr{M}}+ \left\vert\left\vert {\rm Ric}\right\vert\right\vert_{\mathscr{Z}}+ \left\vert\left\vert \left(1,\mathcal{L}_b\right)\left(\mathfrak{n}-\eta\right)|_{v=0}\right\vert\right\vert_{\mathring{H}^4\left(\mathbb{S}^2\right)}+\left\vert\left\vert \left(1,\mathcal{L}_b\right)\left(\mathfrak{j}-\eta\right)|_{v=0} \right\vert\right\vert_{\mathring{H}^4\left(\mathbb{S}^2\right)}\right].\]

Finally, the remaining estimates for $\mathfrak{j}-\eta$ and $\eta-\mathfrak{n}$ are now immediate consequences of the already established estimates and the equations~\eqref{ini3moo},~\eqref{2om3pom2p},~\eqref{2nk3knk2}, and~\eqref{2n3knk2}.  This then yields
\begin{align*}
&\left\vert\left\vert {\rm Ric}\right\vert\right\vert_{\mathscr{M}} +\left\vert\left\vert \left(1,\mathcal{L}_b\right)\left(\mathfrak{n}-\eta\right)|_{v=0}\right\vert\right\vert_{\mathring{H}^4\left(\mathbb{S}^2\right)} +\left\vert\left\vert \left(1,\mathcal{L}_b\right)\left(\mathfrak{n}-\eta\right)|_{v=0}\right\vert\right\vert_{\mathring{H}^4\left(\mathbb{S}^2\right)}
\\ \nonumber &\qquad \lesssim \epsilon\left[\left\vert\left\vert {\rm Ric}\right\vert\right\vert_{\mathscr{M}}+ \left\vert\left\vert {\rm Ric}\right\vert\right\vert_{\mathscr{Z}}+ \left\vert\left\vert \left(1,\mathcal{L}_b\right)\left(\mathfrak{n}-\eta\right)|_{v=0}\right\vert\right\vert_{\mathring{H}^4\left(\mathbb{S}^2\right)}+\left\vert\left\vert \left(1,\mathcal{L}_b\right)\left(\mathfrak{j}-\eta\right)|_{v=0} \right\vert\right\vert_{\mathring{H}^4\left(\mathbb{S}^2\right)}\right].
\end{align*}
We can then absorb the boundary terms on the right hand side into the left had side. This completes the proof.

\end{proof}
\subsection{Finishing the Argument: Estimates for the $\ell > \ell_0$ Projections}

In this next result we finally show that ${\rm Ric} = 0$.
\begin{theorem}\label{constraintsaredone}We have
\[{\rm Ric} = 0.\]
\end{theorem}
\begin{proof}
It remains to estimate the various components of $\mathcal{P}_{\ell > \ell_0}{\rm Ric}$.

We will start with $\mathcal{P}_{\ell_0 < \ell}\left({\rm R} + {\rm Ric}_{34}\right)$. After applying $\mathcal{P}_{\ell > \ell_0}$ to~\eqref{2om3omo4} we find that $X$ satisfies a model second order equation of type $III$. Moreover, it is straightforward to see that the hypothesis of Lemma~\ref{asdf2nini3i9iun12} hold with $p = 0$. Thus, applying Lemma~\ref{asdf2nini3i9iun12} with  $p = 0$, using our previous estimates leads, and a straightforward nonlinear analysis of the terms on the right hand side leads to
\begin{equation}\label{jiwjie3o3ok}
\left\vert\left\vert {\rm R} + {\rm Ric}_{34}\right\vert\right\vert_{\mathscr{Z}} \lesssim  \epsilon\left\vert\left\vert {\rm Ric}\right\vert\right\vert_{\mathscr{Z}}.
\end{equation}

We may similarly treat the equation~\eqref{2poj3ojp3} as a model second order equation of type $III$ for $Y$. Applications of Lemma~\ref{asdf2nini3i9iun12} (with $p = 1$) thus provide control of $\slashed{\rm curl}\slashed{\rm div}\widehat{\rm Ric}$. On the other hand, from~\eqref{ini3moo} and~\eqref{32ojojo1} and the estimate~\eqref{jiwjie3o3ok}, we may obtain estimates for $\slashed{\rm div}\slashed{\rm div}\widehat{\rm Ric}$. All together (we omit the straightforward details) we end up with 
\begin{equation}\label{2kj3kjk2}
\left\vert\left\vert \widehat{\rm Ric} \right\vert\right\vert_{\mathscr{Z}} \lesssim  \epsilon\left\vert\left\vert {\rm Ric}\right\vert\right\vert_{\mathscr{Z}}.
\end{equation}

Since we now have successfully estimated ${\rm R}$, ${\rm Ric}_{34}$, and $\widehat{\rm Ric}$, we can obtain the desired estimates for ${\rm Ric}_{3A}$, ${\rm Ric}_{4A}$, and $\mathfrak{n}$ simply by revisiting the equations~\eqref{m3mini43jn2o}-\eqref{2om3pom2p} and also the equations~\eqref{2poj3om2} and~\eqref{2o3om24}. (We omit the straightforward details.)

We end up finally with the estimate
\[\left\vert\left\vert {\rm Ric}\right\vert\right\vert_{\mathscr{Z}} \lesssim \epsilon \left\vert\left\vert {\rm Ric}\right\vert\right\vert_{\mathscr{Z}} \Rightarrow \left\vert\left\vert {\rm Ric}\right\vert\right\vert_{\mathscr{Z}} = 0\Rightarrow {\rm Ric} = 0.\]

\end{proof}

\section{Regular Coordinate Near the Axis}\label{axiscoordinates}
In this section, we will discuss regular coordinates which allow extensions of our solution to the axis $\{u = v\}$.  We will build our regular coordinates near $\{u = v\}$ in three steps. First, in Section~\ref{iooiio1194392} we improve our estimates for the different metric components as $v\to -1$. In particular, we remove the various $\check{\delta}$'s in certain of our previous estimates. Second we introduce ``by hand'' standard coordinates $(t,x,y,z)$ in which we obtain a Lipschitz extension of $g$ to the axis, which corresponds to $(x,y,z) = (0,0,0)$. Lastly, we will define a suitable set of self-similar wave coordinates, and in these wave coordinates we will be able to obtain smoothness for the metric $g$.

\subsubsection{Improving the Estimates Near the Axis}\label{iooiio1194392}

The following simple lemma concerning ODE's with regular singularities plays a crucial role.
\begin{lemma}\label{indicial}Suppose that
\begin{equation}\label{kodwok2399}
\frac{d^2f}{ds^2} + \frac{p}{s}\frac{df}{ds} + \frac{q}{s^2}f = F,
\end{equation}
for some fixed $p,q \in \mathbb{R}$ and all $s \in K \subset \mathbb{R}\setminus\{0\}$.  

The corresponding indicial equation is 
\[\alpha\left(\alpha-1\right) + \alpha p + q = 0.\]
If $\alpha_0$ and $\alpha_1$ are (any ordering of) the two roots of the indicial equation, then~\eqref{kodwok2399} is equivalent to 
\begin{equation}\label{kodwkow20okjj}
\frac{d}{ds}\left(s^{1+\alpha_0-\alpha_1}\frac{d}{ds}\left(s^{-\alpha_0}f\right)\right) = s^{1-\alpha_1}F,
\end{equation}
for all $s \in K$. 
\end{lemma}
\begin{proof}This is a straightforward computation.
\end{proof}

In the next proposition we will analyze the asymptotics as $v\to -1$ of $\left(\Omega,b,\slashed{g}\right)$. In particular, we remove a $\left(v+1\right)^{\check{\delta}}$ loss in our estimates for $\left(\Omega,b,\slashed{g}\right)$ near $v = -1$.
\begin{proposition}\label{momm3o30cm39jom3od123} Let $\left(\Omega,b,\slashed{g}\right)$ be produced by Theorem~\ref{thisiswhatishholdingatthenehdne}. For any positive integer $N$ satisfying $1 \ll N \ll N_2$, we have that 
\begin{align}
&\sum_{j=0}^1\sum_{\left|\alpha\right| \leq N}\sup_{v \in (-1,-1/2)}\Bigg[ \left(v+1\right)^{-1+j}\left|\mathcal{L}_{\partial_v}^j\left(\slashed{g}-\left(v+1\right)^2\mathring{\slashed{g}}\right)^{(\alpha)}\right|  
+ \left(v+1\right)^{-1+j}\left|\mathcal{L}_{\partial_v}^j\left(\log\Omega\right)^{(\alpha)}\right|
\\ \nonumber &\qquad + \left(v+1\right)^{-1+j}\left|\mathcal{L}_{\partial_v}^jb^{(\alpha)}\right|\Bigg]
\lesssim \epsilon\left\vert\left\vert \left(T_{\rm low},T_{\rm high}\right)\right\vert\right\vert_{\mathring{H}^{N_1-3}}.
\end{align}

\end{proposition}
\begin{proof}We start with an estimate for $\slashed{\rm div}b$. In view of Theorem~\ref{constraintsaredone}, we have that~\eqref{eqnyaydivb} holds with all of the Ricci curvature terms set to $0$.  In view of Lemma~\ref{l2l2degtranstrans} and Lemma~\ref{3ioj2oijio45482}, we immediately obtain that
\begin{equation}\label{3po2opjopoi4}
\sup_{v\in (-1,-1/2)}\sum_{j=0}^1\sum_{\left|\alpha\right| \leq N}\left(v+1\right)^{-1+2\check{\delta}+j}\left|\mathcal{L}_{\partial_v}^j\left(\slashed{\rm div}b\right)^{(\alpha)}\right|\lesssim \epsilon\left\vert\left\vert \left(T_{\rm low},T_{\rm high}\right)\right\vert\right\vert_{\mathring{H}^{N_1-3}}.
\end{equation} 

Next we turn to the lapse $\Omega$. It turns out to be useful to first establish an improved estimate for $\Omega\underline{\omega}$. In view of Theorem~\ref{constraintsaredone}, Lemma~\ref{3ioj2oijio45482}, Lemma~\ref{2km2omo34}, and Lemma~\ref{l2l2degtranstrans} we have that $X \doteq \mathcal{P}_{\ell \geq 1}\left(\left(v+1\right)^2\Omega\underline{\omega}\right)$ will satisfy an equation of the form
\begin{equation}\label{3ojo2ij4oijoi2}
\mathcal{L}_{\partial_v}^2X + \left(v+1\right)^{-2}\mathring{\Delta}X = \mathscr{E},
\end{equation}
where $\mathscr{E}$ satisfies 
\begin{equation}\label{ioj3ioj94891901}
\sum_{\left|\alpha\right| \leq \tilde{N}}\sup_{v \in (-1,-1/2)}\left|\mathscr{E}^{(\alpha)}\right|\left(v+1\right)^{-1+2\check{\delta}} \lesssim \epsilon\left\vert\left\vert \left(T_{\rm low},T_{\rm high}\right)\right\vert\right\vert_{\mathring{H}^{N_1-3}}.
\end{equation}
We further have that
\begin{equation}\label{k3jklj2lkjkl23}
\sum_{\left|\alpha\right| \leq \tilde{N}}\sup_{v \in (-1,-1/2)}\left|X^{(\alpha)}\right|\left(v+1\right)^{-2+\check{\delta}} \lesssim \epsilon\left\vert\left\vert \left(T_{\rm low},T_{\rm high}\right)\right\vert\right\vert_{\mathring{H}^{N_1-3}}.
\end{equation}
Here $\tilde{N}$ is any positive integer which satisfies $1 \ll \tilde{N} \ll N_2$. (In particular, we can assume that $N \ll \tilde{N}$.)

Now we let $X_{\ell}$ and $\mathscr{E}_{\ell}$ denote the coefficient of the projections of $X$ and $\mathscr{E}$ to a spherical harmonic of eigenvalue $\ell\left(\ell+1\right)$ for $\ell \in \mathbb{Z}_{\geq 1}$. We then obtain the  ordinary differential equation
\begin{equation}\label{komo3o32349r}
\frac{d^2X_{\ell}}{dv^2} - \frac{\ell(\ell+1)}{(v+1)^2}X_{\ell} = E_{\ell},
\end{equation}
The indicial equation corresponding to the left hand side of~\eqref{komo3o32349r} is
\[\alpha^2 - \alpha - \ell(\ell+1) = 0 \Leftrightarrow \alpha = \frac{1\pm \sqrt{1+4\ell(\ell+1)}}{2} \Leftrightarrow \alpha \in \left\{\ell+1,-\ell\right\}.\]
In particular, integrating twice the formula from Lemma~\ref{indicial} with $\alpha_0 = -\ell$ and $\alpha_1 = \ell+1$, we obtain that there exist two constants $D_1(\ell)$ and $D_2(\ell)$ so that
\begin{align}\label{fok4ko4o3o320}
&X_{\ell}\left(v\right) = 
\\ \nonumber &\qquad \left(v+1\right)^{-\ell}\int_{-1}^v\left[\left(x+1\right)^{2\ell}\left(\int_{-1/4}^x\left(1+y\right)^{-\ell}\mathscr{E}_{\ell}(y)\, dy\right)\, dx\right] + D_1(\ell)\left(v+1\right)^{-\ell} + D_2(\ell)\left(v+1\right)^{\ell+1}.
\end{align}
For $v \in [-1,-1/2]$, in view of~\eqref{ioj3ioj94891901}, it is straightforward to establish the following bound
\begin{align}\label{k3mox9202939}
&\left(v+1\right)^{-\ell}\int_{-1}^v\left[\left(x+1\right)^{2\ell}\left(\int_{-1/4}^x\left(1+y\right)^{-\ell}\mathscr{E}_{\ell}(y)\, dy\right)\, dx\right]  
\\ \nonumber &\qquad \lesssim \ell^{-2}\left(\left(v+1\right)^{3-2\check{\delta}} + \left(v+1\right)^{\ell+1}\right)\sup_{s \in (-1,-1/2)}\left(\left(v+1\right)^{-1+2\check{\delta}}\mathscr{E}_{\ell}\right).
\end{align}
In particular,~\eqref{fok4ko4o3o320},~\eqref{k3mox9202939}, and~\eqref{k3jklj2lkjkl23} imply that $D_1(\ell)$ vanishes. Next, we may simply set $v = -1/2$ in~\eqref{fok4ko4o3o320} and use the estimate~\eqref{k3mox9202939} to obtain the bound:
\begin{equation}\label{3oijoijoi23}
\left|D_2(\ell)\right|2^{-\ell} \lesssim \ell^{-2} \sup_{s \in (-1,-1/2)}\left(\left(v+1\right)^{-1+2\check{\delta}}\mathscr{E}_{\ell}\right) + \left|X_{\ell}\left(1/2\right)\right|.
\end{equation}
In view of~\eqref{k3mox9202939} and~\eqref{3oijoijoi23} we then obtain from~\eqref{fok4ko4o3o320} that for $v \in (-1,-1/2)$
\begin{align}\label{oiiuiu3io2}
\ell^{-1}\left(v+1\right)\left|\frac{dX_{\ell}}{dv}\right|+\left|X_{\ell}\left(v\right)\right| &\lesssim \left(\ell^{-2}\left(v+1\right)^{3-2\check{\delta}} +\ell^{-2}2^{\ell}\left(v+1\right)^{\ell+1}\right)\sup_{s \in (-1,-1/2)}\left(\left(v+1\right)^{-1+2\check{\delta}}\mathscr{E}_{\ell}\right)
\\ \nonumber &\qquad + 2^{\ell}\left(v+1\right)^{\ell+1}\left|X_{\ell}\left(1/2\right)\right|. 
\end{align}
Keeping in mind that the fact that $v \in [-1,-1/2]$ implies that $\sup_{\ell \geq 1}2^{\ell}\left(v+1\right)^{\ell+1} \lesssim \left(v+1\right)^2$, we may use Sobolev inequalities on $\mathbb{S}^2$ and sum the estimate~\eqref{oiiuiu3io2} in $\ell$ (after multiplication by suitable powers of $\ell$) and use the estimates~\eqref{ioj3ioj94891901} and~\eqref{k3jklj2lkjkl23} to obtain (cf.~the end of the proof of Lemma~\ref{k4oij2oijo3}) that
\begin{equation}\label{oiioi3298198312kjhbnj}
\sum_{j=0}^1\sum_{\left|\alpha\right|\leq \check{N}}\sup_{v \in [-1,-1/2]}\left(v+1\right)^j\left|\mathcal{L}_{\partial_v}^j\mathcal{P}_{\ell \geq 1}\left(\Omega\underline{\omega}\right)\right| \lesssim \epsilon\left\vert\left\vert \left(T_{\rm low},T_{\rm high}\right)\right\vert\right\vert_{\mathring{H}^{N_1-3}},
\end{equation}
for any $\check{N}$ satisfying $1 \ll \check{N} \ll \tilde{N}$. Integrating from $v = -1$ then yields
\begin{equation}\label{2klj3ou2ioj4ioj2}
\sum_{\left|\alpha\right|\leq \check{N}}\sup_{v \in [-1,-1/2]}\left(v+1\right)^{-1}\left|\log\Omega_{\rm boun}\right| \lesssim \epsilon\left\vert\left\vert \left(T_{\rm low},T_{\rm high}\right)\right\vert\right\vert_{\mathring{H}^{N_1-3}}.
\end{equation}

For the analogous estimate of $\Omega_{\rm sing}$, we note that as a consequence of~\eqref{2o2o4ji2},~\eqref{oijoijjio32o}, and Lemma~\ref{3ioj2oijio45482}, we have that for $v \in (-1,-1/2)$
\begin{equation}\label{3oij2ioj3ioj2}
\left|\mathcal{L}^2_{\partial_v}\log\Omega_{\rm sing} + \frac{2}{v+1}\mathcal{L}_{\partial_v}\log\Omega_{\rm sing}\right| \lesssim \left(v+1\right)^{-10\check{\delta}}.
\end{equation}
Since Lemma~\ref{3ioj2oijio45482} implies that $\left|\log\Omega_{\rm sing}\right| \lesssim \left(v+1\right)^{1-\check{\delta}}$, it is clear that~\eqref{3oij2ioj3ioj2} implies that
\begin{equation}\label{2lkj3kl2j3}
\sum_{j=0}^1\sup_{v \in [-1,-1/2]}\left(v+1\right)^{-1+j}\left|\mathcal{L}_{\partial_v}^j\log\Omega_{\rm sing}\right| \lesssim \epsilon\left\vert\left\vert \left(T_{\rm low},T_{\rm high}\right)\right\vert\right\vert_{\mathring{H}^{N_1-3}}.
\end{equation} 
(We have in fact an even strong estimate, but we've only written what is necessary for the proof.) Since $\log\Omega = \log\Omega_{\rm sing} + \log\Omega_{\rm boun}$ where $\Omega_{\rm sing}$ is spherically symmetric, we thus obtain, after combining~\eqref{2klj3ou2ioj4ioj2},~\eqref{oiioi3298198312kjhbnj}, and~\eqref{2lkj3kl2j3} that   
\begin{equation}\label{2k3jl2kjl23}
\sum_{j=0}^1\sum_{\left|\alpha\right|\leq \check{N}}\sup_{v \in [-1,-1/2]}\left(v+1\right)^{-1+j}\left|\mathcal{L}_{\partial_v}^j\log\Omega\right| \lesssim \epsilon\left\vert\left\vert \left(T_{\rm low},T_{\rm high}\right)\right\vert\right\vert_{\mathring{H}^{N_1-3}}.
\end{equation}

Next we come to $\slashed{\rm curl}b$. We will argue in an analogous fashion to how we estimated $\mathcal{P}_{\ell \geq 1}\left(\Omega\underline{\omega}\right)$. As a consequence of Theorem~\ref{constraintsaredone}, the equation~\eqref{thewavezetastartsojqj} holds with all of the Ricci curvature terms set to $0$. We may then apply $\slashed{\rm curl}$ to the equation and in view of Lemma~\ref{l2l2degtranstrans}, Lemma~\ref{3ioj2oijio45482}, Lemma~\ref{othercommutelemma},~\eqref{acommut}, and~\eqref{2m3momo2}, we obtain that $Y = \slashed{\rm curl}b$ satisfies
\begin{equation}\label{3ij2oi48989892}
\mathcal{L}_{\partial_v}^2Y + \frac{4}{v+1}\mathcal{L}_{\partial_v}Y + \left(v+1\right)^{-2}\left(\mathring{\Delta}+2\right)Y = \mathscr{M},
\end{equation}
where $\mathscr{M}$ satisfies 
\begin{equation}\label{2k3j2joijio3jio2}
\sum_{\left|\alpha\right| \leq \tilde{N}}\sup_{v \in (-1,-1/2)}\left|\mathscr{M}^{(\alpha)}\right|\left(v+1\right)^{2\check{\delta}} \lesssim \epsilon\left\vert\left\vert \left(T_{\rm low},T_{\rm high}\right)\right\vert\right\vert_{\mathring{H}^{N_1-3}}.
\end{equation}
We further have that
\begin{equation}\label{k3jklj2lkjkl23}
\sum_{\left|\alpha\right| \leq \tilde{N}}\sup_{v \in (-1,-1/2)}\left|Y^{(\alpha)}\right|\left(v+1\right)^{\check{\delta}} \lesssim \epsilon\left\vert\left\vert \left(T_{\rm low},T_{\rm high}\right)\right\vert\right\vert_{\mathring{H}^{N_1-3}}.
\end{equation}
Here $\tilde{N}$ is any positive integer which satisfies $1 \ll \tilde{N} \ll N_2$. (In particular, we can assume that $N \ll \tilde{N}$.) The corresponding spherical harmonic projected equation is 
\begin{equation}\label{mkcm30wj123091231233}
\frac{d^2Y_{\ell}}{dv^2} + \frac{4}{v+1}\frac{dY_{\ell}}{dv} + \left(v+1\right)^{-2}\left(2-\ell(\ell+1) \right)Y_{\ell} = \mathscr{M}_{\ell},
\end{equation}
and the indicial equation corresponding to the left hand side of~\eqref{mkcm30wj123091231233} is
\[\alpha^2+3\alpha + \left(2-\ell(\ell+1)\right) = 0\Leftrightarrow \alpha = \frac{-3\pm \sqrt{9 + 4(\ell(\ell+1)-2)}}{2} \Leftrightarrow \alpha \in \{ \ell-1,-\ell-2\}.\]
Keeping in mind that $\ell \in \mathbb{Z}_{\geq 1}$, it is clear that we can argue just as we did for $\mathcal{P}_{\ell \geq 1}\left(\Omega\underline{\omega}\right)$ and eventually obtain that
\begin{equation}\label{2lk3jlk2jp21}
\sum_{j=0}^1\sum_{\left|\alpha\right|\leq \check{N}}\sup_{v \in [-1,-1/2]}\left(v+1\right)^j\left|\mathcal{L}_{\partial_v}^j\slashed{\rm curl}b\right| \lesssim \epsilon\left\vert\left\vert \left(T_{\rm low},T_{\rm high}\right)\right\vert\right\vert_{\mathring{H}^{N_1-3}}.
\end{equation}
From~\eqref{3po2opjopoi4},~\eqref{2lk3jlk2jp21}, and elliptic estimates, we then obtain
\begin{equation}\label{2klj3kljl23k}
\sum_{j=0}^1\sum_{\left|\alpha\right|\leq \check{N}}\sup_{v \in [-1,-1/2]}\left(v+1\right)^{-1+j}\left|\mathcal{L}_{\partial_v}^jb\right| \lesssim \epsilon\left\vert\left\vert \left(T_{\rm low},T_{\rm high}\right)\right\vert\right\vert_{\mathring{H}^{N_1-3}}.
\end{equation}

It remains to establish the improved estimates for $\slashed{g}$. As with the lapse, it is in fact more natural to estimate $\mathcal{L}_{\partial_v}\slashed{g}$. We start with ${\rm tr}\chi$. In view of Theorem~\ref{constraintsaredone}, Lemma~\ref{3ioj2oijio45482},~\eqref{2k3jl2kjl23}, and~\eqref{1opj3opjoii9459o} we have that
\begin{equation}\label{l3ij3jiioio3909035332}
\mathcal{L}_{\partial_v}\left(\Omega^{-1}{\rm tr}\chi - \frac{2}{v+1}\right) + \frac{2}{v+1}\left(\Omega^{-1}{\rm tr}\chi - \frac{2}{v+1}\right) = \mathscr{H},
\end{equation}
where $\mathscr{H}$ satisfies 
\begin{equation}\label{ij3oijioj3o2}
\sum_{\left|\alpha\right| \leq \check{N}}\sup_{v \in (-1,-1/2)}\left|\mathscr{H}^{(\alpha)}\right|\left(v+1\right) \lesssim \epsilon\left\vert\left\vert \left(T_{\rm low},T_{\rm high}\right)\right\vert\right\vert_{\mathring{H}^{N_1-3}}.
\end{equation}
We further have that
\begin{equation}\label{k3jklj2lkjkl23}
\sum_{\left|\alpha\right| \leq \tilde{N}}\sup_{v \in (-1,-1/2)}\left|\left(\Omega^{-1}{\rm tr}\chi - \frac{2}{v+1}\right)^{(\alpha)}\right|\left(v+1\right)^{\check{\delta}} \lesssim \epsilon\left\vert\left\vert \left(T_{\rm low},T_{\rm high}\right)\right\vert\right\vert_{\mathring{H}^{N_1-3}}.
\end{equation}
It then an immediate consequence that we in fact have
\begin{equation}\label{k3jklj2lkjkl23}
\sum_{\left|\alpha\right| \leq \check{N}}\sup_{v \in (-1,-1/2)}\left|\left(\Omega^{-1}{\rm tr}\chi - \frac{2}{v+1}\right)^{(\alpha)}\right| \lesssim \epsilon\left\vert\left\vert \left(T_{\rm low},T_{\rm high}\right)\right\vert\right\vert_{\mathring{H}^{N_1-3}}.
\end{equation}
Similarly, we obtain
\begin{equation}\label{3kj2oj3o2}
\sum_{\left|\alpha\right| \leq \check{N}}\sup_{v \in (-1,-1/2)}\left|\left(\hat{\chi}\right)^{(\alpha)}\right| \lesssim \epsilon\left\vert\left\vert \left(T_{\rm low},T_{\rm high}\right)\right\vert\right\vert_{\mathring{H}^{N_1-3}},
\end{equation}
by considering the equation~\eqref{kwdkodwok23dg} instead of~\eqref{1opj3opjoii9459o}.  Integrating these estimates then immediately yields the desired estimates for $\slashed{g}$ and completes the proof.

\end{proof}

\subsubsection{The $(t,x,y,z)$-coordinates}
Our first goal will be to introduce  a set of standard $\left(t,x,y,z\right)$ coordinates in terms of the double-null coordinates $\left(u,v,\theta^A\right)$ covering the region $\mathscr{U} \doteq \left\{\left(u,v,\theta^A\right) \in \left(-\infty,0\right)\times \left(-\infty,0\right) \times \mathbb{S}^2 : v > u\right\}$. We start by defining a $t$-coordinate and an $r$-coordinate in terms of the double-null $\left(u,v\right)$ coordinates by
\[t \doteq v+u,\qquad r \doteq v-u.\]
The region $\mathscr{U}$ now corresponds to $\left\{\left(t,r,\theta^A\right) \in (-\infty,0) \times (0,\infty) \times \mathbb{S}^2 :t < -r\right\}$. In the $\left(t,r,\theta^A\right)$ coordinates, the metric becomes
\begin{align}\label{2oj3o2jn4o5in2}
&g = \left(-\Omega^2+\frac{1}{4}\left|b\right|^2\right) dt\otimes dt + \left(\Omega^2 +\frac{1}{4}\left|b\right|^2\right)dr\otimes dr +\slashed{g}_{AB}d\theta^Ad\theta^B 
\\ \nonumber &\qquad -\frac{1}{2}b_A\left(d\theta^A\otimes dt + dt\otimes d\theta^A\right) + \frac{1}{2}b_A\left(d\theta^A\otimes dr + dr\otimes d\theta^A\right).
\end{align}
Let $0 < c \ll 1$ be a small constant. On our reference round metric $\left(\mathbb{S}^2,\mathring{\slashed{g}}\right)$ we can introduce standard spherical coordinates $\left(\varphi,\phi\right) \in (c,\pi-c) \times [0,2\pi)$\footnote{We are slightly abusing notation in a standard way by using the interval $[0,2\pi)$ and referring to this as a single coordinate. Of course, strictly speaking we should use two separate coordinate charts where one corresponds to, say, $\phi \in (0,2\pi)$, and another corresponds to, say, $\phi \in (-\frac{\pi}{2},\frac{3\pi}{2})$, but this point should not cause any confusion.}  so that in the domain of these coordinates, we have $\mathring{\slashed{g}} = d\varphi^2 + \sin^2\varphi d\phi^2$. If we then take out a suitable neighborhood of the north and south pole of each copy of $\mathbb{S}^2$, our spacetime will be covered by the region $\mathscr{P}_c \doteq \left\{(t,r,\varphi,\phi) \in (-\infty,0) \times (0,\infty)\times (c,\pi-c) \times [0,2\pi) : t < -r\right\}$. It will be useful to define an $\mathbb{S}^2_{u,v}$ $(0,2)$-tensor $e_{AB}$ by
\[e_{AB} \doteq \slashed{g}_{AB} - \left(v-u\right)^2\mathring{\slashed{g}}_{AB} = \slashed{g}_{AB} - r^2\mathring{\slashed{g}}_{AB}.\]
In the $\left(\varphi,\phi\right)$ coordinate chart, we have that $\left(\partial_{\varphi},\sin^{-1}\varphi \partial_{\phi}\right)$ forms an orthonormal frame for the round metric on $\mathbb{S}^2$. It thus follows from Proposition~\ref{momm3o30cm39jom3od123} that we have that, in this chart,
\begin{equation}\label{skjfnimo3}
\left|e\right|_{\mathring{\slashed{g}}} \lesssim \epsilon (-u)^{-1}\left(v-u\right)^3 = 2\epsilon \left(r-t\right)^{-1}r^3,\qquad \left|e_{\varphi\varphi}\right| \lesssim \epsilon \left(r-t\right)^{-1}r^3,
\end{equation}
\begin{equation}\label{2m3omo2}
\left|e_{\varphi\phi}\right| \lesssim \epsilon \sin\varphi \left(r-t\right)^{-1}r^3,\qquad \left|e_{\phi\phi}\right| \lesssim \epsilon \sin^2\varphi \left(r-t\right)^{-1}r^3
\end{equation}
Similarly, we  observe that the following bounds hold for the shift $1$-form $b_A$ in this chart:
\begin{equation}\label{2k3n3k4}
\left|b\right|_{\mathring{\slashed{g}}} \lesssim \epsilon \left(1-\frac{v}{u}\right)^2 =\epsilon  4\left(r-t\right)^{-2}r^2,\qquad \left|b_{\varphi}\right| \lesssim \epsilon \left(r-t\right)^{-2}r^2,\qquad \left|b_{\phi}\right| \lesssim \epsilon \sin\varphi \left(r-t\right)^{-2}r^2.
\end{equation}
We also have the following bounds for the lapse $\Omega$ in this chart:
\begin{equation}\label{2kn3kk243in2i2}
\left|\Omega-1\right| \lesssim \epsilon \left(1-\frac{v}{u}\right) = 2\epsilon (r-t)^{-1}r.
\end{equation}

In terms of the $\left(t,r,\varphi,\phi\right)$ coordinates, we now define $\left(t,x,y,z\right)$ coordinates via the usual formulas:
\begin{equation}\label{2oi3oin2oi4}
x \doteq r\cos\phi \sin\varphi,\qquad y \doteq r\sin\phi \sin\varphi,\qquad z\doteq r\cos\varphi.
\end{equation}
We note that in the region $\mathscr{P}_c$ we will have the bound 
\begin{equation}\label{2n3om2oo}
|z| \leq c'\left(|x| + |y|\right)
\end{equation}
for a suitable postive constant $c' > 0$.

In view of~\eqref{skjfnimo3},~\eqref{2m3omo2},~\eqref{2k3n3k4},~\eqref{2kn3kk243in2i2},~\eqref{2n3om2oo}, and the usual form of the Minkowski metric in spherical coordinates, it is straightforward to see that in the $(t,x,y,z)$ coordinates, the metric $g$ takes the following form
\begin{align}\label{2km3imo2}
&g = \left(-1+A_{tt}\right) dt\otimes dt  +\left(1+A_{xx}\right)dx\otimes dx + \left(1+A_{yy}\right)dy\otimes dy + \left(1+A_{zz}\right)dz\otimes dz
\\ \nonumber &\qquad + A_{tx}dt\otimes  dx + A_{xt} dx\otimes dt + A_{ty} dt\otimes dy + A_{yt} dy\otimes dt + A_{tz} dt\otimes dz + A_{zt} dz\otimes dt 
\\ \nonumber &\qquad + A_{xy} dx\otimes dy + A_{yz} dy\otimes dx + A_{xz} dx\otimes dz + A_{zx} dz\otimes dx + A_{yz} dy\otimes dz + A_{zy} dz\otimes dy,
\end{align}
where, for any $t_0 < 0$, in the region $\{t \leq t_0\} \cap \mathscr{P}_c$ the functions $A_{\cdot \cdot}$ all satisfy the bound
\begin{equation}\label{i2n3imo23o2j}
\left|A_{\cdot\cdot}\right| \lesssim_{t_0,c} \epsilon\left(|x| + |y| + |z|\right).
\end{equation}
We can extend the validity of~\eqref{2km3imo2} and~\eqref{i2n3imo23o2j} in the usual fashion. That is, we rotate our reference sphere by $90$ degrees to define an alternative set of coordinates $\left(\varphi',\phi'\right)$. Then we may consider the region $\mathscr{P}_c' \doteq  \left\{(t,r,\varphi',\phi') \in (-\infty,0) \times (0,\infty)\times (c,\pi'-c) \times [0,2\pi'): t < -r\right\}$. Note that $\mathscr{P}_c \cup \mathscr{P}_c'$ is now equal to the entire region $\mathscr{U}$. By a suitable modification of the formulas~\eqref{2oi3oin2oi4}, we may then define the $(t,x,y,z)$ coordinates in terms of $\left(t,r,\varphi',\phi'\right)$ such that they agree with the previous definition on the overlap of $\mathscr{P}_c$ and $\mathscr{P}_c'$. We then have that
\[\mathscr{P}_c \cup \mathscr{P}_c' = \left\{(t,x,y,z)  \in (-\infty,0) \times\mathbb{R}^3 : |x|+|y| + |z| > 0\text{ and } t < -\sqrt{x^2+y^2+z^2}\right\}.\]
Moreover, on $\{t \leq t_0\} \cap \left(\mathscr{P}_c \cup \mathscr{P}_c' \right) = \{t \leq t_0\} \cap \mathscr{U}$ we have that~\eqref{2km3imo2} and~\eqref{i2n3imo23o2j} hold. We then define
\[\mathscr{M} \doteq \left\{(t,x,y,z)  \in (-\infty,0) \times\mathbb{R}^3 :  t < -\sqrt{x^2+y^2+z^2}\right\},\]
and identity $\mathscr{U} \doteq \mathscr{P}_c \cup \mathscr{P}_c'$ with $\mathscr{M}\setminus \{(x,y,z) = (0,0,0)\}$. We now obtain that our metric extends to $\mathscr{M}$.
\begin{lemma}\label{23j2oimoi2oi293}The metric $\left(\mathscr{U},g\right)$ extends to the Lorentzian manifold $\left(\mathscr{M},g\right)$ as a Lipschitz metric. Moreover the self-similar vector field $K = u\partial_u + v\partial_v = t\partial_t + x\partial_x + y\partial_y + z\partial_z$ extends to a smooth vector field on $\mathscr{M}$ by setting $K|_{(x,y,z) = (0,0,0)} = t\partial_t$ and $\mathcal{L}_Kg = 2g$ holds on all of $\mathscr{M}$. Lastly, if we let $m = -dt^2+dx^2+dy^2+dz^2$ denote the Minkowski metric on $\mathscr{M}$, then the coefficients of $g-m$ satisfy
\[\left\vert\left\vert g-m\right\vert\right\vert_{W^{1,\infty}\left(\mathscr{M}\cap \{t \leq t_0\}\right)} \lesssim_{t_0} \epsilon,\]
for any  $t_0 < 0$.
\end{lemma}
\begin{proof}In the region $\mathscr{U}$, it easily follows from our already established estimates for $g$ that $g$ is Lipschitz continuous. From~\eqref{2km3imo2} we then see that a Lipschitz extension of $g$ to $\mathscr{M}$ is given by setting
\[g|_{(x,y,z) = (0,0,0)} = -dt^2 +dx^2+dy^2+dz^2.\]

The statements about the self-similar vector field $K$ are immediate. The estimates on $g-m$ follow then from the estimates for $g$ listed in Lemma~\ref{3ioj2oijio45482} and Proposition~\ref{momm3o30cm39jom3od123} and the formulas which link $\partial_x$, $\partial_y$, and $\partial_z$ derivatives with $\partial_r$, $\partial_{\vartheta}$, and $\partial_{\phi}$. (We omit the straightforward calculations.) 
\end{proof}

\begin{definition}\label{axisaxis}
We refer to the timelike curve $\left\{(x,y,z) = (0,0,0)\right\}$ in $\mathscr{M}$ as the ``axis'' and denote it by $\mathscr{A}$.
\end{definition}
\subsubsection{Self-Similar Wave Coordinates Near the Axis}
Let $r_0 \in (0,1)$ and define
\[\mathscr{M}_{r_0} \doteq \left\{(t,x,y,z) \in \mathscr{M} : \sqrt{x^2+y^2+z^2} < -tr_0 \right\}\]
This is a dilation invariant subset of $\mathscr{M}$ which still includes the axis $(x,y,z) = (0,0,0)$ but excises a neighborhood of $\{v = 0\}$. We note also that, by our convention for choosing $\epsilon$, we may assume that $K$ is timelike in $\mathscr{M}_{r_0}$. Lastly, it will be convenient to introduce the notation $\{x^{\alpha}\}_{\alpha = 0}^3$ to refer to the coordinate functions $(t,x,y,z)$.

In the next lemma we construct self-similar wave coordinates.
\begin{lemma}\label{23oij3ijo23ijo2393nck}Let $\left(\mathscr{M},g\right)$ be as in Lemma~\ref{23j2oimoi2oi293}. There exists functions $\{\xi^{\alpha}\}_{\alpha=0}^3$ on $\mathscr{M}_{r_0}$ so that for any $p \in [1,\infty)$ each $\xi^{\alpha} \in W^{2,p}\left(\mathscr{M}_{r_0}\cap \left\{x^0=-1\right\}\right)$, so that $\{\xi^{\alpha}\}$ form a wave coordinate system, and so that $K\left(\xi^{\alpha}\right) = \xi^{\alpha}$. When we define the $W^{2,p}$ space here we are identifying $\mathscr{M}_{r_0}\cap \left\{x^0=-1\right\}$ in the natural way with a ball of radius $r_0$ in $\mathbb{R}^3$. 

For each $\alpha$ we have the estimate
\begin{align}\label{23mlm2}
&\left\vert\left\vert \xi^{\alpha}-x^{\alpha}\right\vert\right\vert_{W^{2,p}\left(\mathscr{M}_{r_0}\cap \left\{x^0=-1\right\}\right)}  \lesssim_p \epsilon.
\end{align}
Moreover, the hypersurface $\{\xi^0 = -1\}$ is a $W^{2,p}$ graph over $\left\{x^0 = -1\right\}$  and every integral curve of $K$ intersects $\left\{\xi^0 = -1\right\}$ exactly once.
\end{lemma}
\begin{proof}Since the metric $g$ is Lipschitz continuous, it is differentiable almost everywhere, and its almost everywhere derivative is almost everywhere bounded by the corresponding Lipschitz constant at that point. The operator $\Box_g$ acting on scalar functions involves at most one derivative of the metric. Hence it make sense to talk about strong solutions $f$ to the equation $\Box_g f = 0$. 

Next, we note that $\mathscr{M}_{r_0}$ is foliated by the dilation invariant rays 
\[\gamma_{\left(x^1_0,x^2_0,x^3_0\right)} \doteq \left\{\left(x^0,x^1,x^2,x^3\right) : \left(\frac{x^1}{|x^0|},\frac{x^2}{|x^0|},\frac{x^3}{|x^0|}\right) = \left(x^1_0,x^2_0,x^3_0\right)\right\},\] where $\left(x^1_0,x^2_0,x^4_0\right) \in \mathbb{R}^3$ satisfies $\left(x^1_0\right)^2+\left(x^2_0\right)^2+\left(x_0^3\right)^2 \leq r_0^2$. Each such ray has a unique intersection with $\{x^0 = -1\} \cap \mathscr{M}_{r_0}$ at the point $\left(-1,x^1_0,x^2_0,x^3_0\right)$. These rays are, of course, integral curves of the vector $K$. In particular, for any function $f \in W^{2,p} \left(\mathscr{M}_{r_0}\cap \left\{x^0= -1\right\}\right)$ (which is thus $C^1$ on $\mathscr{M}_{r_0}$), we may uniquely extend $f$ to a function $f: \mathscr{M}_{r_0} \to \mathbb{R}$ by requiring $f$ to satisfy $Kf = f$. The resulting function will lie in $W^{2,p}_{\rm loc}\left(\mathscr{M}_{r_0}\right)$ where we define the $W^{2,p}_{\rm loc}$ by considering $\mathscr{M}_{r_0}$ as a subset of $\mathbb{R}^4$ in the natural way. 

Having defined a way to extend functions from $\mathscr{M}_{r_0}\cap \left\{x^0 = -1\right\}$ to $\mathscr{M}_{r_0}$, we may apply our extension notion to solving $\Box_gf = 0$. That is, suppose that $f \in W^{2,p}_{\rm loc}\left(\mathscr{M}_{r_0}\right) \cap W^{2,p}\left(\mathscr{M}_{r_0}\cap \left\{x^0= -1\right\}\right)$ satisfies the following: 
\[\Box_gf|_{\mathscr{M}_{r_0}\cap \left\{x^0=-1\right\}} = 0,\qquad Kf_{\mathscr{M}_{r_0}\cap \left\{x^0=-1\right\}} = f.\]
(Note that the equation $Kf = f$ allows us to trade $\partial_{x^0}$ derivatives of $f$ with $\partial_{x^{\alpha}}$ derivatives for $\alpha > 0$. Hence we automatically have that $\partial_{x^0}f \in W^{1,p} \left(\mathscr{M}_{r_0}\cap \left\{x^0 = -1\right\}\right)$ and $\partial_{x^0}^2f \in L^p\left(\mathscr{M}_{r_0}\cap \left\{x^0= -1\right\}\right)$.) Then we claim that we must have that 
\[\Box_gf = 0\text{ almost everywhere in }\mathscr{M}_{r_0}.\]
Indeed, this follows from the fact that, since $\mathcal{L}_Kg = 2g$ and $Kf = f$, we have the following transport equation for $\Box_gf$:
\[K\left(\Box_gf\right) -3\left(\Box_gf\right) = 0,\]
and $\Box_gf$ vanishes almost everywhere along $\left\{x^0= -1\right\}$. 

We now turn to constructing the functions $\xi^{\alpha}$. By the above discussion it will suffice to find $\xi^{\alpha}$ which solves
\begin{equation}\label{2lkn3oin2o}
\Box_g\xi^{\alpha}|_{\mathscr{M}_{r_0}\cap \left\{x^0 = -1\right\}} = 0,\qquad K\left(\xi^{\alpha}\right)_{\mathscr{M}_{r_0}\cap \left\{x^0=-1\right\}} = \xi^{\alpha}.
\end{equation}
The equation $K\left(\xi^{\alpha}\right) = \xi^{\alpha}$ implies a relation between the time derivative of $\xi^{\alpha}$ and spatial derivatives of $\xi^{\alpha}$. Namely,
\begin{equation}\label{23mom2o4o2}
\partial_{x^0}\xi^{\alpha} = -(x_0)^{-1}\left(\left(\sum_{\beta =1}^3x^{\beta}\partial_{\beta}\xi^{\alpha}\right)-\xi^{\alpha}\right). 
\end{equation}
Similarly, we may differentiate the relation $K\left(\xi^{\alpha}\right) = \xi^{\alpha}$ to obtain that $K\left(\partial_{x^{\beta}}\xi^{\alpha}\right) = 0$ for $\beta \in \{0,\cdots,3\}$. Thus,
\begin{equation}\label{23omp2mo5}
\partial_{x^0}\partial_{x^{\beta}}\xi^{\alpha} = -(x_0)^{-1}\sum_{\gamma = 1}^3 x^{\gamma}\partial_{x^{\gamma}}\partial_{x^{\beta}}\xi^{\alpha}.
 \end{equation}
By systematically using the relations from~\eqref{23omp2mo5} and~\eqref{23mom2o4o2} we can define a second order operator $\mathscr{L}_g$ which only involves derivatives with respect to $\{x^{\beta}\}_{\beta = 1}^3$ and so that~\eqref{2lkn3oin2o} holds if and only if the following holds:
\begin{equation}\label{2k3jhrk2hrk3}
\mathscr{L}_g\xi_t|_{\mathscr{M}_{r_0}\cap \left\{x^0=-1\right\}} = 0,\qquad K(\xi_t)_{\mathscr{M}_{r_0}\cap \left\{x^0=-1\right\}} = \xi_t.
\end{equation}
Moreover, because $K$ is timelike along $\mathscr{M}_{r_0}\cap \left\{x^0=-1\right\}$, we have that $\mathscr{L}_g$ is an elliptic operator along $\mathscr{M}_{r_0}\cap \left\{x^0=-1\right\}$. Let us denote by $\mathscr{L}_m$ the corresponding operator associated to the Minkowski metric $m$. A computation yields the following formula for $\mathscr{L}_m$:
\begin{equation}\label{23kmomo4}
\mathscr{L}_m = \left(1-x^2\right)\partial_x^2 + \left(1-y^2\right)\partial_y^2 + \left(1-z^2\right)\partial_z^2 -2xy\partial^2_{x,y} -2xz\partial_{x,z}^2 -2yz\partial^2_{y,z}.
\end{equation}

Now we want to define $\xi^{\alpha}|_{\mathscr{M}_{r_0}\cap \left\{x^0=-1\right\}}$ by solving the elliptic boundary problem:
\begin{equation}\label{2o3mo2mo44}
\mathscr{L}_g\xi^{\alpha}|_{\mathscr{M}_{r_0}\cap \left\{x^0=-1\right\}} = 0,\qquad \xi^{\alpha}|_{\mathscr{M}_{r_0}\cap \left\{x^0=-1\right\}\cap\{r = r_0\}} = x^{\alpha}|_{\mathscr{M}_{r_0}\cap \{t=-1\}\cap\{r = r_0\}}.
\end{equation}
Observe that in the case of $\mathscr{L}_m$ it follows easily from the maximum principle that $\xi^{\alpha}= x^{\alpha}$ is the unique solution to this problem. It follows from our estimates for the metric $g$, that the operator $\mathscr{L}_m - \mathscr{L}_g$ has coefficients which are bounded in $L^{\infty}$ by $\epsilon$ in $\mathscr{M}_{r_0}\cap \left\{x^0=-1\right\}$. Thus, by a standard perturbation argument and $L^p$ elliptic theory, there exists a unique solution $\xi^{\alpha}$ to~\eqref{2o3mo2mo44} such that for any $p \in [1,\infty)$, $\xi^{\alpha} \in W^{2,p}\left(\mathscr{M}_{r_0}\cap \left\{x^0=-1\right\}\right)$. Moreover, we have the estimate
\[\left\vert\left\vert \xi^{\alpha}-x^{\alpha}\right\vert\right\vert_{W^{2,p}\left(\mathscr{M}_{r_0}\cap \left\{x^0=-1\right\}\right)} \lesssim_p \epsilon.\]

We then extend the function $\xi^{\alpha}$ to all of $\mathscr{M}_{r_0}$ via the extension procedure described above. Since, for $p$ sufficiently large, the $W^{2,p}$ norm controls, via Sobolev inequalities, the $C^1$ norm, it is straightforward to use~\eqref{23mlm2} to see that the Jacobian of the coordinate transform $\left\{x^{\alpha}\right\} \mapsto \left\{\xi^{\alpha}\right\}$ is non-degenerate and hence that $\left\{\xi^{\alpha}\right\}$ indeed forms a set of wave coordinates. Lastly, the statements about the hypersurface $\left\{\xi^0 = -1\right\}$ follow in a straightforward way from the estimate~\eqref{23mlm2} and the fact that $K\xi^0 = \xi^0$. 

\end{proof}

In the next lemma, we show that the metric $g$ is smooth when expressed the  self-similar wave coordinates. 
\begin{lemma}The metric $g$ from Lemma~\ref{23j2oimoi2oi293} is smooth on $\mathscr{M}_{r_0}$ when given in terms of the self-similar wave coordinates from Lemma~\ref{23oij3ijo23ijo2393nck}. 
\end{lemma}
\begin{proof}Let $\hat{\mathscr{M}}_{r_0} \doteq \mathscr{M}_{r_0}\setminus \mathscr{A}$. In view of Lemma~\ref{3ioj2oijio45482} and elliptic estimates in $\hat{\mathscr{M}}_{r_0}$ it is clear that $g$ is at least $C^2$ when expressed in the self-similar wave coordinates in $\hat{\mathscr{M}}_{r_0}$. In particular, in view of Theorem~\ref{constraintsaredone}, in the region $\hat{\mathscr{M}}_{r_0}$, ${\rm Ric}\left(g\right) = 0$ holds in a classical sense and, in view of the well-known formula for the Ricci tensor in wave coordinates, we have that
\begin{equation}\label{232oi3}
\Box_g\left(g_{ij}\right) = \mathcal{N}_{ij}\left(\partial g,g\right), \qquad 0 \leq i,j \leq 3
\end{equation}
where $\Box_g$ denotes the scalar wave equation, $g_{ij} \doteq g\left(\partial_{\xi^i},\partial_{\xi^j}\right)$, and $\mathcal{N}_{ij}$ stands for a linear combination of suitable nonlinear contractions between $g$ and its derivatives in the self-similar wave coordinates $\{\xi^i\}$.

Next, in view of the fact that $K\left(\xi^i\right) = \xi^i$, we observe that the self-similar vector field $K$ takes the following form in the $\{\xi^i\}$ coordinates:
\begin{equation}\label{2m3m23o3}
K = \xi^i\partial_{\xi^i}.
\end{equation}
In particular, $K$ is a smooth vector field in the self-similar wave coordinates. 

Now we follow a similar strategy to the one we used in our construction of the wave coordinates, except that we will replace the hypersurface $\{t = 0\}$ with $\{\xi^0 = 0\}$. Note, in particular, that in view of Lemma~\ref{23oij3ijo23ijo2393nck}, for each $0 \leq i,j \leq 3$ and $p \in [1,\infty)$, we have
\begin{equation}\label{2io3oi2}
g_{ij} \in W^{1,p}\left(\mathscr{M}_{r_0}\cap \left\{\xi^0 =-1\right\}\right),\qquad \partial_{\xi^0}g_{ij} \in L^p\left(\mathscr{M}_{r_0}\cap \left\{\xi^0 = -1\right\}\right),\end{equation}
In a similar fashion to Lemma~\ref{23oij3ijo23ijo2393nck}, these Sobolev spaces are defined by using the wave coordinates $\left\{\xi^{\beta}\right\}_{\beta = 1}^3$ to identify $\mathscr{M}_{r_0} \cap \left\{\xi^0 =-1\right\}$ with the ball of radius $r_0$ in $\mathbb{R}^3$.

The relation $\mathcal{L}_Kg = 2g$ allows us to convert $\partial_{\xi^0}g_{ij}$ into a first order expression involving only the derivatives $\partial_{\xi^k}g_{ij}$ for $k > 0$. Thus, in a similar fashion to how we derived the elliptic operator $\mathscr{L}_g$ in the proof of Lemma~\ref{23oij3ijo23ijo2393nck}, we may eliminate all $\xi^0$ derivatives in the equation~\eqref{232oi3} and restrict the result to $\{\xi^0 = 0\} \cap \hat{\mathscr{M}}_{r_0}$ to obtain an equation of the following form for $g_{ij}$ along $\left\{\xi^0 = 0\right\}\cap \hat{\mathscr{M}}_{r_0}$:
\begin{equation}\label{2k3om2o}
\sum_{\alpha,\beta = 1}^3 \partial_{\xi^{\alpha}}\left(P^{\alpha\beta}\partial_{\xi^{\beta}}g_{ij}\right) = H_{ij},
\end{equation}
where matrix $P$ is a symmetric matrix which is a smooth function of $g_{ij}$ and $H_{ij}$ is a smooth function of $g_{ij}$ and $\partial_{\xi^{\beta}}g_{ij}$ where $\beta > 0$. In view of~\eqref{2io3oi2} and the fact that $K$ is timelike, we moreover have that~\eqref{2k3om2o} is uniformly elliptic, $P^{\alpha\beta}$ extends to $\mathscr{M}_{r_0}\cap \left\{\xi^0 = 0\right\}$ as a  $W^{1,p}$ function for any $p \in [1,\infty)$, and $H_{ij}$ extends to $\mathscr{M}_{r_0}\cap \left\{\xi^0 = 0\right\}$ as a $L^p$ function for any $p \in [1,\infty)$. 

We next claim that the equation~\eqref{2k3om2o} holds weakly on $\mathscr{M}_{r_0} \cap \left\{\xi^0 = -1\right\}$. This is equivalent to the statement that for every smooth compactly supported function $\varphi : \mathscr{M}_{r_0} \cap \left\{\xi^0 = -1\right\} \to \mathbb{R}$, we have
\begin{equation}\label{2io3oim1oi3}
\int_{\mathscr{M}_{r_0} \cap \left\{\xi^0 = -1\right\}}\left(\sum_{\alpha,\beta = 1}^3P^{\alpha\beta}\left(\partial_{\xi^{\beta}}g_{ij}\right)\left(\partial_{\xi^{\alpha}}\varphi\right) + H_{ij}\varphi\right) d\xi^1\, d\xi^2\, d\xi^3 = 0.
\end{equation}
Let $q\left(\xi^1,\xi^2,\xi^3\right) : \mathbb{R}^3 \to [0,1]$ be a smooth function so that $q$ is identically $1$ for $\sqrt{|\xi^1|^2 + |\xi^2|^2+|\xi^3|^2} \geq 2$ and identically $0$ for $\sqrt{|\xi^1|^2 + |\xi^2|^2 + |\xi^3|^2} \leq 1$. Then, letting $p$ be sufficiently large, an integration by parts, Sobolev inequalities, and H\"{o}lder's inequality yields the following: 
\begin{align}\label{2om3o2123}
&\left|\int_{\mathscr{M}_{r_0} \cap \left\{\xi^0 = -1\right\}}\left(\sum_{\alpha,\beta = 1}^3P^{\alpha\beta}\left(\partial_{\xi^{\beta}}g_{ij}\right)\left(\partial_{\xi^{\alpha}}\varphi\right) + H_{ij}\varphi\right) d\xi^1\, d\xi^2\, d\xi^3 \right| =
\\ \nonumber &\qquad  \left|\lim_{\tilde{\epsilon} \to 0}\int_{\mathscr{M}_{r_0} \cap \left\{\xi^0 = -1\right\}}q\left(\tilde{\epsilon}^{-1}\xi^1,\tilde{\epsilon}^{-1}\xi^2,\tilde{\epsilon}^{-1}\xi^3\right)\left(\sum_{\alpha,\beta = 1}^3P^{\alpha\beta}\left(\partial_{\xi^{\beta}}g_{ij}\right)\left(\partial_{\xi^{\alpha}}\varphi\right) + H_{ij}\varphi\right) d\xi^1\, d\xi^2\, d\xi^3 \right| \lesssim 
\\ \nonumber &\qquad \limsup_{\tilde{\epsilon} \to 0}\tilde{\epsilon}^{-1}\sum_{\beta = 1}^3\int_{\mathscr{M}_{r_0} \cap \left\{\xi^0 = -1\right\}\cap \left\{ \tilde{\epsilon} \leq \sqrt{|\xi^1|^2+|\xi^2|^2+|\xi^3|^2} \leq 2\tilde{\epsilon}\right\}}\left|\partial_{\xi^{\beta}}g_{ij}\right|\, d\xi^1\, d\xi^2\, d\xi^3 \lesssim 
\\ \nonumber &\qquad \left\vert\left\vert g_{ij}\right\vert\right\vert_{W^{1,p}\left(\mathscr{M}_{r_0}\cap \left\{\xi^0 =-1\right\}\right)}\limsup_{\tilde{\epsilon} \to 0}\tilde{\epsilon}^{\frac{2p-3}{p}} = 0.
\end{align}

Having established that $g_{ij}$ is a weak solution to~\eqref{2k3om2o} on $\mathscr{M}_{r_0} \cap \left\{\xi^0 = -1\right\}$, $L^p$-elliptic regularity yields that $g_{ij} \in W^{2,p}\left(\mathscr{M}_{r_0}\cap \left\{\xi^0 =-1\right\}\right)$ for every $p \in [1,\infty)$. This in turn implies that $P^{\alpha\beta} \in W^{2,p}\left(\mathscr{M}_{r_0}\cap \left\{\xi^0 =-1\right\}\right)$ and that $H_{ij} \in W^{1,p}\left(\mathscr{M}_{r_0}\cap \left\{\xi^0 =-1\right\}\right)$ for every $p \in [1,\infty)$. It is then clear by iteratively applying $L^p$-elliptic regularity and then using Sobolev inequalities that we will obtain that the restriction of each $g_{ij}$ to $\mathscr{M}_{r_0}\cap \left\{\xi^0 =-1\right\}$ is a smooth function. In view of $\mathcal{L}_Kg = 2g$, we immediately obtain that $g$ is a smooth metric in $\mathscr{M}_{r_0}$. 
\end{proof}

\section{Regular Coordinates Near $\{v = 0\}$ and Gluing to the Exterior Solution of~\cite{nakedone}}\label{toexterior}
In this section we discuss the procedure of gluing the interior solution constructed in this paper with the exterior solution constructed in the paper~\cite{nakedone}. In our previous work~\cite{nakedone} we have already understood how to carry a detailed analysis of the regularity of self-similar solutions along $\{v = 0\}$. However, the analysis of~\cite{nakedone} was carried out in a double-null gauge which was initialized so that $\slashed{g}|_{v=0} = e^{2\varphi}\mathring{\slashed{g}}$ for a suitable function $\varphi$ and under the assumption that, in these coordinates, $\mathring{\Pi}_{\rm curl}b$ takes the form of a specific ansatz. We thus start in Section~\ref{23imo2j1234tr2} by showing that after fine-tuning our seed data $T_{\rm high}$ and $T_{\rm low}$ and undertaking a suitable angular coordinate change, we can put ourselves in a similar situation to that of~\cite{nakedone}. Then in Section~\ref{12jn1oiodo009u2jimkqwss} we leverage the analysis from~\cite{nakedone} to define a new coordinate system near $\{v = 0\}$ where we obtain the desired regularity for the metric $g$. Finally, in Section~\ref{yayglue} we explicitly glue our interior solution to the exterior solution.
\subsubsection{Conformal Coordinates along $\{v = 0\}$}\label{23imo2j1234tr2}
We start by defining a procedure by which a choice of diffeomorphism along $\mathbb{S}^2$ induces a corresponding double-null coordinate change. 
\begin{definition}\label{itsinduced}Let us denote by $\left(\mathscr{U},g\right)$ and $\left(u,v,\theta^A\right)$ the spacetime and its associated double-null coordinate system produced by Theorem~\ref{thisiswhatishholdingatthenehdne}. (See~\eqref{2oj3omio2} for the definition of $\mathscr{U}$.) It is convenient to then let $\overline{\mathscr{U}} \doteq \mathscr{U} \cap \{v = 0\}$ which is made into a manifold with boundary in a natural way. As with $\mathscr{U}$, we may use the notation $\mathbb{S}^2_{u_0,v_0}$ to the refer to the copy of $\mathbb{S}^2$ in $\overline{\mathscr{U}}$ corresponding to $u = u_0$ and $v = v_0$.

Let $X^A \in \mathring{H}^{N_1}\left(\mathbb{S}^2\right)$ be a vector field along $\mathbb{S}^2$ and $\mathcal{F}_X$ denote the corresponding diffeomorphism generated by the time-$1$ flow of $X$. We then define a coordinate system $\left\{\tilde{\theta}^A\right\}$ on $\mathbb{S}^2_{-1,0}$ by setting, for $p \in \mathbb{S}^2_{-1,0}$: 
\[\tilde{\theta}^A\left(p\right) \doteq \theta^A\left(\mathcal{F}_X^{-1}\left(p\right)\right),\]
where the coordinate charts associated to the $\left\{\tilde{\theta}^A\right\}$ are the image under $\mathcal{F}_X^{-1}$ of the coordinate charts associated to $\left\{\theta^A\right\}$. We then extend $\left\{\tilde{\theta}^A\right\}$ to the hypersurface $\{v = 0\} = \left(-\infty,0\right) \times \mathbb{S}^2$, by requiring that $\mathcal{L}_{\partial_u}\tilde{\theta}^A = 0$ holds along $(-\infty,0)\times U$ for suitable coordinate charts $U$ in the $\mathbb{S}^2_{u,0}$.  Finally, we then extend $\left\{\tilde{\theta}^A\right\}$ to the entire manifold $\overline{\mathscr{U}}$ by requiring that $\mathcal{L}_{\partial_v}\tilde{\theta}^A = 0$ holds when restricted to suitable coordinate charts in the $\mathbb{S}^2_{u,v}$. 

We then say that the resulting double-null coordinate coordinate system $\left(u,v,\tilde{\theta}^A\right)$ is induced by the vector field $X^A$. 
\end{definition}
\begin{remark}Observe that the coordinate system $\left(u,v,\tilde{\theta}^A\right)$ preserves the double-null form~\eqref{doublenullisg}. Moreover, because we have that $\left[\partial_v,\partial_u\right] =0$ in the original coordinate system, we will have that $\partial_u\tilde{\theta}^A = \partial_v\tilde{\theta}^A = 0$. This implies that in the induced coordinate system $\left(u,v,\tilde{\theta}^A\right)$, the self-similar vector field $K$ maintains the familiar form
\[K = u\partial_u + v\partial_v,\]
and we retain all of the algebraic consequences of self-similarity.
\end{remark}

As we have mentioned in the introduction to the section, in oder to carry out our analysis near $\{v = 0\}$ it will be useful to work in coordinates where $\slashed{g}|_{v=0}$ is conformal to the round metric in those coordinates. The following definition connects the imposition of the seed data $T_{\rm low}$ and $T_{\rm high}$ with the form of $b$ after being pulled backed by a diffeomorphism which makes $\slashed{g}|_{v=0}$ conformal to a round metric.
\begin{definition}\label{2in3moi2o4}Let $z^A$ be a vector field on $\mathbb{S}^2$ which satisfies 
\begin{equation}\label{23m2omo2124tgf}
\mathring{\nabla}_Az^A = 0,\qquad \left\vert\left\vert z\right\vert\right\vert_{\mathring{H}^{N_1}\left(\mathbb{S}^2\right)} \lesssim \epsilon,
\end{equation}
and let us fix a choice of coordinates $\left(\vartheta,\phi\right)$ on $\mathbb{S}^2$ so that the round metric $\mathring{\slashed{g}}$ takes the form $d\vartheta^2 + \sin^2\vartheta d\phi^2$.

Then we say that a choice of seed data $T_{\rm low}$ and $T_{\rm high}$ (see Section~\ref{seeddatasection}) is conformally compatible with $z^A$ if there exists a vector field $X \in \mathring{H}^{N_1}\left(\mathbb{S}^2\right)$ satisfying $\left\vert\left\vert X\right\vert\right\vert_{\mathring{H}^{N_1}\left(\mathbb{S}^2\right)} \lesssim \epsilon$ such if we take the $\slashed{g}$ corresponding to the solution produced by Theorem~\ref{thisiswhatishholdingatthenehdne} we have that 
\begin{enumerate}
\item $\mathcal{F}_X^*\slashed{g} = e^{2\varphi}\mathring{\slashed{g}}$ for some function $\varphi$.
\item $\mathring{\rm curl}\mathcal{F}_X^*b = \mathring{\rm curl}z$. 
\end{enumerate}

\end{definition}

Our main goal in this section is to show that we can always find seed data which is conformally compatible with $z^A$. We will find the desired seed data and vector field $X$ by running a suitable iteration argument. The following definition will be useful for this.
\begin{definition}Let $y^A$ be a given vector field on $\mathbb{S}^2$ which satisfies~\eqref{23m2omo2124tgf} (with $z$ replaced by $y$), $\varphi_{\rm low}:\mathbb{S}^2 \to \mathbb{R}$ satisfy $\left(1-\mathcal{P}_{\ell \leq 1}\right)\varphi_{\rm low} = 0$, $\tilde{b}^A$ be a vector field on $\mathbb{S}^2$, and $\Omega^{-1}{\rm tr}\chi : \mathbb{S}^2\to\mathbb{R}$ be a function. Then we say that the $4$-tuple $\left(\eta_A,f,\varphi,\hat{c}\right)$ of a $1$-form $\eta$, two functions $f$ and $\varphi$, and a constant $\hat{c}$ satisfy the ``conformal coordinate boundary system'' if after setting
\begin{equation}\label{2pij3oijoijoi2}
b^A \doteq \mathring{\nabla}^Af + y^A,\qquad \slashed{g} \doteq e^{2\varphi}\mathring{\slashed{g}},\qquad \Omega\underline{\omega} \doteq \hat{c},
\end{equation}
we have
\begin{align}\label{1k2jkl1jk}
&\slashed{\rm div}b - \mathcal{L}_b\slashed{\rm div}b = \frac{1}{2}\left(\slashed{\rm div}b\right)^2 + \frac{1}{4}\left|\slashed{\nabla}\hat{\otimes}b\right|^2 + 8\left(\Omega\underline{\omega}\right)\left(-1+\frac{1}{2}\slashed{\rm div}b\right),
\end{align}
\begin{align}\label{2oj3pomp}
\eta_A\left(2-\slashed{\rm div}b\right) -\mathcal{L}_b\eta_A = \frac{1}{2}\slashed{\nabla}^B\left(\slashed{\nabla}\hat{\otimes}b\right)_{AB} -\frac{1}{2}\slashed{\nabla}_A\slashed{\rm div}b,
\end{align}
\begin{align}\label{2j4oij2oij4oij2}
&\mathcal{P}_{\ell \geq 2}\left(-2e^{2\varphi}K\left[\slashed{g}\right] + 2e^{2\varphi}\slashed{\rm div}\eta\right) = 
 \mathcal{P}_{\ell \geq 2}\Bigg(e^{2\varphi}\Bigg(\left(-1+\slashed{\rm div}b -4\Omega\underline{\omega}\right)\left(\Omega^{-1}{\rm tr}\chi - 2\right)
\\ \nonumber &\qquad \qquad + \mathcal{L}_{\tilde{b}}\left(\Omega^{-1}{\rm tr}\chi -2\right)-2\left|\eta\right|^2 + 2\left(\slashed{\rm div}b -4\Omega\underline{\omega}\right)\Bigg)\Bigg),
\end{align}
\[\mathcal{P}_{\ell \leq 1}\varphi = \varphi_{\rm low}.\]
(The reader should note the presence of $\tilde{b}$ in the equation~\eqref{2j4oij2oij4oij2}.)
\end{definition}

In the next lemma we show that the conformal coordinate boundary system has a unique solution under suitable smallness assumptions for $y$, $\Omega^{-1}{\rm tr}\chi$, and $\mathcal{P}_{\ell \leq 1}\varphi$.
\begin{lemma}\label{2io3ouhui2}Let $y^A$ be a given vector field on $\mathbb{S}^2$ which satisfies~\eqref{23m2omo2124tgf} (with $z$ replaced by $y$), $\varphi_{\rm low}:\mathbb{S}^2 \to \mathbb{R}$ satisfy $\left(1-\mathcal{P}_{\ell \leq 1}\right)\varphi_{\rm low} = 0$, $\tilde{b}^A$ be a vector field on $\mathbb{S}^2$, and $\Omega^{-1}{\rm tr}\chi : \mathbb{S}^2\to\mathbb{R}$ be a function. Additionally assume that
\begin{equation}\label{2o3mp2oo01010040212}
\left\vert\left\vert \varphi_{\rm low}\right\vert\right\vert_{L^2\left(\mathbb{S}^2\right)} + \left\vert\left\vert \left(1,\mathcal{L}_{\tilde{b}}\right)\left(\Omega^{-1}{\rm tr}\chi-2\right)\right\vert\right\vert_{\mathring{H}^{N_1-2}} \lesssim \epsilon. 
\end{equation}
Then there exists a unique solution $\left(\eta_A,f,\varphi,\hat{c}\right)$ to the conformal coordinate boundary system which satisfies the estimates:
\begin{align}\label{2o3momo1joi2901901}
&\left\vert\left\vert f\right\vert\right\vert_{\mathring{H}^{N_1+1}} + \left\vert\left\vert \left(1,\mathcal{L}_b\right)\mathcal{P}_{\ell \geq 2}\varphi\right\vert\right\vert_{\mathring{H}^{N_1-1}} + \left|\hat{c}\right| + \left\vert\left\vert \left(1,\mathcal{L}_b\right)\eta\right\vert\right\vert_{\mathring{H}_{N_1-2}} \lesssim
\\ \nonumber &\qquad \left\vert\left\vert y\right\vert\right\vert_{\mathring{H}^{N_1}}+ \left\vert\left\vert \varphi_{\rm low}\right\vert\right\vert^2_{L^2\left(\mathbb{S}^2\right)} + \left\vert\left\vert \left(1,\mathcal{L}_{\tilde{b}}\right)\left(\Omega^{-1}{\rm tr}\chi-2\right)\right\vert\right\vert_{\mathring{H}^{N_1-2}}.
\end{align}
Moreover if $\{\left(\eta_i,f_i,\varphi_i,\hat{c}_i\right)\}_{i=1}^2$ are two different solutions corresponding to $\left\{\left(y_i,\left(\varphi_{\rm low}\right)_i,\tilde{b}_i,\left(\Omega^{-1}{\rm tr}\chi\right)_i\right)\right\}_{i=1}^2$, then we have 
\begin{align}\label{2kj3lkj219i3}
&\left\vert\left\vert f_1-f_2\right\vert\right\vert_{\mathring{H}^{N_1}} + \left\vert\left\vert \left(1,\mathcal{L}_{b_1},\mathcal{L}_{b_2}\right)\mathcal{P}_{\ell \geq 2}\left(\varphi_1-\varphi_2\right)\right\vert\right\vert_{\mathring{H}^{N_1-2}} + \left|\hat{c}_1-\hat{c}_2\right| + \left\vert\left\vert \left(1,\mathcal{L}_{b_1},\mathcal{L}_{b_2}\right)\eta_1-\eta_2\right\vert\right\vert_{\mathring{H}_{N_1-3}} \lesssim
\\ \nonumber &\qquad \left\vert\left\vert y_1-y_2\right\vert\right\vert_{\mathring{H}^{N_1-1}}+ \left\vert\left\vert \left(\varphi_{\rm low}\right)_1 - \left(\varphi_{\rm low}\right)_2\right\vert\right\vert^2_{L^2\left(\mathbb{S}^2\right)} 
\\ \nonumber &\qquad + \left\vert\left\vert \left(\Omega^{-1}{\rm tr}\chi\right)_1 - \left(\Omega^{-1}{\rm tr}\chi\right)_2\right\vert\right\vert_{\mathring{H}^{N_1-3}}+ \left\vert\left\vert \mathcal{L}_{\tilde{b}_1}\left(\Omega^{-1}{\rm tr}\chi\right)_1 - \mathcal{L}_{\tilde{b}_2}\left(\Omega^{-1}{\rm tr}\chi\right)_2\right\vert\right\vert_{\mathring{H}^{N_1-3}}.
\end{align}
\end{lemma}
\begin{proof}It is useful to observe that
\[\left(\slashed{\nabla}\hat{\otimes}b\right)^{AB} = e^{-2\varphi}\left(\mathring{\nabla}\hat{\otimes}b\right)^{AB}.\]
In particular,~\eqref{1k2jkl1jk} becomes 
\begin{align}\label{2lkj32kj32}
&\slashed{\rm div}b - \mathcal{L}_b\slashed{\rm div}b = \frac{1}{2}\left(\slashed{\rm div}b\right)^2 + \frac{1}{4}\left|\mathring{\nabla}\hat{\otimes}b\right|_{\slashed{\mathring{g}}}^2 + 8\left(\Omega\underline{\omega}\right)\left(-1+\frac{1}{2}\slashed{\rm div}b\right).
\end{align}
Note that this has the affect that $\slashed{g}$ only appears in~\eqref{2lkj32kj32} through its role in defining the operator $\slashed{\rm div}$. 

It is also useful to recall that since $\slashed{g} = e^{2\varphi}\mathring{\slashed{g}}$, we have that
\[-\mathring{\Delta}\varphi - 2\varphi = e^{2\varphi}\left(K-1\right) + \left(e^{2\varphi}-1-2\varphi\right).\]

In order to establish the existence statement, we run an iteration argument. Let $\left\{A_i,B_i,\eta_i,f_i,\varphi_i,\hat{c}_i\right\}_{i=0}^{\infty}$ be a sequence of  $6$-tuples consisting of four functions $A_i$, $B_i$, $\varphi_i$, and  $f_i$ which satisfy $\mathcal{P}_{\ell = 0}\left(A_i,f_i\right) = 0$, a $1$-form $\eta_i$, and a constant $\hat{c}_i$ such that all quantities vanish when $i = 0$, and, for $i \geq 1$:
\begin{align}\label{2ij2in42ijm5}
&A_i - \mathcal{P}_{\ell \geq 1}\mathcal{L}_{b_{i-1}} A_i = 
\left[\mathcal{P}_{\ell \geq 1},\mathcal{L}_{b_{i-1}}\right]B_{i-1}+\mathcal{P}_{\ell \geq 1}\left(\frac{1}{2}B_{i-1}^2 + \frac{1}{4}\left|\mathring{\nabla}\hat{\otimes}b_{i-1}\right|^2_{\mathring{\slashed{g}}} + 4\hat{c}_{i-1}B_{i-1}\right),
\end{align}
\begin{equation}\label{23o4oij5oi2}
\mathring{\Delta} f_i = \mathcal{P}_{\ell \geq 1}\left(A_{i-1}-2\mathcal{L}_{b_{i-2}}\varphi_{i-1}\right),
\end{equation}
\begin{equation}\label{o3ij2oij4io1jo4}
\mathcal{P}_{\ell = 0}\left(B_{i-1}- \mathcal{L}_{b_{i-1}}B_{i-1}\right) = 8\hat{c}_i + \mathcal{P}_{\ell = 0}\left(\frac{1}{2}B_{i-1}^2 + \frac{1}{4}\left|\mathring{\nabla}\hat{\otimes}b_{i-1}\right|_{\slashed{\mathring{g}}}^2 + 4\hat{c}_{i-1}B_{i-1}\right),
\end{equation}
\[B_{i} \doteq A_{i} + \mathcal{P}_{\ell = 0}\slashed{\rm div}b_{i},\]
\[\eta_i\left(2-B_{i-1}\right) -\mathcal{L}_{b_{i-1}}\eta_i = \frac{1}{2}\slashed{\rm div}_{i-1}\left(e^{-2\varphi_{i-1}}\left(\mathring{\nabla}\hat{\otimes}b_{i-1}\right)\right) -\frac{1}{2}\slashed{\nabla}B_{i},\]
\begin{align}\label{2j4oij2oij4oij2123}
&\mathcal{P}_{\ell \geq 2}\left(\mathring{\Delta}\varphi_i + 2\varphi_i  +\left(e^{2\varphi_{i-1}}-1-2\varphi_{i-1}\right)+ e^{2\varphi_{i-1}}\slashed{\rm div}_{i-1}\eta_{i}\right) = 
\\ \nonumber &\qquad \frac{1}{2} \mathcal{P}_{\ell \geq 2}\Bigg(e^{2\varphi_{i-1}}\Bigg(\left(-1+B_{i-1} -4\hat{c}_{i-1}\right)\left(\Omega^{-1}{\rm tr}\chi - 2\right)
\\ \nonumber &\qquad \qquad + \mathcal{L}_{\tilde{b}}\left(\Omega^{-1}{\rm tr}\chi -2\right)-2\left|\eta_{i-1}\right|_{\slashed{g}_{i-1}}^2 + 2\left(B_i -4\hat{c}_{i}\right)\Bigg)\Bigg),
\end{align}
\[\mathcal{P}_{\ell \leq 1}\varphi_i = \varphi_{\rm low},\]
and where we set $b_{-1} = 0$ and, for $i \geq 0$ we set
\begin{equation}\label{i2j3iono2}
b^A_{i} \doteq \mathring{\nabla}^Af_{i} + z^A,\qquad \slashed{g}_i \doteq e^{2\varphi_i}\mathring{\slashed{g}}.
\end{equation}
It is then straightforward to show that these iterates converge (using elliptic estimates and Proposition~\ref{somestuimdie}) and lead to the desired solution after setting $A = \mathcal{P}_{\ell \geq 1}\slashed{\rm div}b$, $B = \slashed{\rm div}b$, and $\hat{c} = \Omega\underline{\omega}$. The uniqueness statement and the estimate~\eqref{2kj3lkj219i3} follows similarly by considering differences between the equations for the quantities associated to $i=1$ and $i=2$.

\end{proof}

By the uniformization theorem, after the application of a suitable diffeomorphism, every metric on $\mathbb{S}^2$ is conformally equivalent to a round metric in standard spherical coordinates. As is well known, there is non-uniqueness in the uniformization process which is generated by the $6$-dimensional conformal symmetry group of the round sphere. This next lemma may be thought of as providing a canonical way to parametrize the choice of a uniformization in the special setting which the metric $\slashed{g}$ is already assumed close to a fixed round metric $\mathring{\slashed{g}}$.  
\begin{lemma}\label{fixthespheregauge}Let $k$ be sufficiently large, $\mathring{\slashed{g}}$ be a round metric on $\mathbb{S}^2$, $\Theta_{AB}$ be a symmetric $(0,2)$-tensor on $\mathbb{S}^2$ which satisfies $\left\vert\left\vert \Theta - \mathring{\slashed{g}}\right\vert\right\vert_{\mathring{H}^k\left(\mathbb{S}^2\right)} \ll 1$, $\{y_i\}_{i=-1}^1 \in \mathbb{R}^3$ satisfy $\left|y_i\right| \ll 1$, and $\{w_i\}_{i=-1}^1 \in \mathbb{R}^3$ satisfy $\left|w_i\right| \ll 1$. Then there exists a vector field $X \in \mathring{H}^{k+1}\left(\mathbb{S}^2\right)$ satisfying 
\[\left\vert\left\vert X \right\vert\right\vert_{\mathring{H}^{k+1}\left(\mathbb{S}^2\right)} \lesssim \left\vert\left\vert \Theta - \mathring{\slashed{g}}\right\vert\right\vert_{\mathring{H}^k\left(\mathbb{S}^2\right)}+\sum_{i=-1}^1\left[\left|y_i\right| + \left|w_i\right|\right],\qquad \mathring{\rm tf}\left(\mathcal{F}_X^*\Theta\right) = 0,\]
\[ \mathcal{P}_{\ell = 1} \log\left[\mathring{\rm tr}\left(\mathcal{F}_X^* \Theta\right)\right] = \sum_{i=-1}^1y_iY^1_i,\qquad \int_{\mathbb{S}^2}\mathring{\slashed{g}}^{AB}X_A\left(\mathring{\slashed{\epsilon}}_B^{\ \ C}\nabla_CY^1_i\right)\mathring{\rm dVol} = w_i,\]
where $\mathcal{F}_X$ denotes the diffeomorphism generated by the time-$1$ flow of $X$, $\mathring{\rm tf}$ denotes the trace-free part relative to $\mathring{\slashed{g}}$, $\mathring{\rm tr}$ denotes the trace relative to $\mathring{\slashed{g}}$, and we recall that $Y^1_{-1}$, $Y^1_0$, $Y^1_1$ are the spherical harmonics corresponding to $\ell = 1$.  

In particular, there exists a function $\varphi$ so that 
\[\mathcal{F}_X^*\Theta_{AB} = e^{2\varphi}\mathring{\slashed{g}}_{AB},\qquad \mathcal{P}_{\ell = 1}\varphi =  \sum_{i=-1}^1y_iY^1_i.\]
The choice of $w \in \mathbb{R}^3$ corresponds to the freedom to apply a diffeomorphism generated by a (suitably small) element of $SO\left(3\right)$. 
\end{lemma}
\begin{proof}This may be deduced from a specialization of the proof of Theorem 3.6 from Lecture 3 of~\cite{viacnotes} to the case of small perturbations of the round metric on $\mathbb{S}^2$ (replacing the use of H\"{o}lder spaces there with Sobolev spaces) and then using the fact that all  trace-free symmetric $(0,2)$ tensors on the round sphere with a vanishing divergence must vanish identically.
\end{proof}

Now we are ready for the main result of the section.
\begin{proposition}\label{23moo49281}Let $z^A$ be a given vector field on $\mathbb{S}^2$ which satisfies~\eqref{23m2omo2124tgf}. Then there exists a choice of seed data $\left(T_{\rm low},T_{\rm high}\right)$ which is conformally compatible with $z^A$. Moreover, we have the estimate
\[\left\vert\left\vert \epsilon \left(T_{\rm low},T_{\rm high}\right)\right\vert\right\vert_{\mathring{H}^{N_1-3}} + \left\vert\left\vert X\right\vert\right\vert_{\mathring{H}^{N_1}} \lesssim \left\vert\left\vert z\right\vert\right\vert_{\mathring{H}^{N_1}},\]
where $X$ is the vector field satisfying the requirements from Definition~\ref{2in3moi2o4}.
\end{proposition}
\begin{proof}As usual, we run an iteration argument. We will define a sequence
\[\left\{\left(\varphi_{\rm low}\right)_i,\varphi_i,\tilde{b}_i,b_i,\eta_i,\left(\Omega\underline{\omega}\right)_i,\left(T_{\rm low}\right)_i,\left(T_{\rm high}\right)_i,X_i,\left(\Omega^{-1}{\rm tr}\chi\right)_i\right\}_{i=0}^{\infty}\]
of vector fields $\tilde{b}_i$, $b_i$, and $X_i$, $1$-forms $\eta_i$, functions $(T_{\rm low})_i$, $(T_{\rm high})_i$, $(\varphi_{\rm low})_i$, $\varphi_i$, and $\left(\Omega^{-1}{\rm tr}\chi\right)_i$, and  a constant $\left(\Omega\underline{\omega}\right)_i$. We will have that $\left(1-\mathcal{P}_{\ell \leq 1}\right)\varphi_i = \left(1-\mathcal{P}_{\ell \leq \ell_0}\right)T_{\rm low} = 0$. We initialize the values of the quantities by setting
\[\tilde{b}_0 = b_0 = X_0 = 0,\qquad \eta_0 = 0,\qquad (T_{\rm low})_0 = (T_{\rm high})_0 = (\varphi_{\rm low})_0 =  0,\]
\[\left(\Omega^{-1}{\rm tr}\chi\right)_0 = 2,\qquad \left(\Omega\underline{\omega}\right)_0 = A_0 = 0.\]
It is convenient to set
\begin{align*}
&\mathscr{C}\left[i\right] \doteq \left\vert\left\vert \tilde{b}_i\right\vert\right\vert_{\mathring{H}^{N_1-1}} + \left\vert\left\vert b_i\right\vert\right\vert_{\mathring{H}^{N_1}} + \left\vert\left\vert X_i\right\vert\right\vert_{\mathring{H}^{N_1}}
\\ \nonumber &\qquad + \left\vert\left\vert \left(1,\mathcal{L}_{b_i}\right)\eta_i\right\vert\right\vert_{\mathring{H}^{N_1-2}}+
 \left\vert\left\vert T_{\rm low}\right\vert\right\vert_{L^2} + \left\vert\left\vert T_{\rm high}\right\vert\right\vert_{\mathring{H}^{N_1-3}}+\left\vert\left\vert \left(\varphi_{\rm low}\right)_i\right\vert\right\vert_{L^2} + \left\vert\left\vert \left(1,\mathcal{L}_{b_i}\right)\mathcal{P}_{\ell \geq 2}\varphi_i\right\vert\right\vert_{\mathring{H}^{N_1-1}} + \left|\left(\Omega\underline{\omega}\right)_i\right|,
\end{align*}
\begin{align*}
\mathscr{D}\left[i\right] \doteq \left\vert\left\vert \left(1,\mathcal{L}_{\tilde{b}}\right)\left(\Omega^{-1}{\rm tr}\chi - 2\right)\right\vert\right\vert_{H^{N_1-2}} + \left|A_i\right|.
\end{align*}
We will now inductively construct the sequence and also establish that
\begin{equation}\label{2oijo4ij29582u2i11}
\mathscr{C}\left[i\right] \leq C_{\rm boot}\left\vert\left\vert z\right\vert\right\vert_{\mathring{H}^{N_1}},\qquad \mathscr{D}\left[i\right] \leq C_{\rm boot}\left\vert\left\vert z\right\vert\right\vert^2_{\mathring{H}^{N_1}},
\end{equation}
for a suitable bootstrap constant $C_{\rm boot}$. Now suppose that $j \geq 1$ and we have constructed our sequence and verified the estimates~\eqref{2oijo4ij29582u2i11} for $0 \leq i \leq j-1$.

\begin{enumerate}
	\item We apply Lemma~\ref{2io3ouhui2} with $y = z$, $\varphi_{\rm low} = \left(\varphi_{\rm low}\right)_{i-1}$, $\tilde{b} = \mathcal{F}_{X_{i-1}}^*\tilde{b}_{i-1}$, and $\Omega^{-1}{\rm tr}\chi = \mathcal{F}_{X_{i-1}}^*\left(\Omega^{-1}{\rm tr}\chi\right)_{i-1}$. We let the corresponding solutions define $\left(\eta_i, f_i,\left(\Omega\underline{\omega}\right)_i,\varphi_i\right)$. If we also set $b_i \doteq \mathring{\nabla}f + y$, then we will have
	\begin{equation}\label{2o3mpompo4522}
	\left\vert\left\vert b\right\vert\right\vert_{\mathring{H}^{N_1}}+\left\vert\left\vert f_i\right\vert\right\vert_{\mathring{H}^{N_1+1}} + \left\vert\left\vert \left(1,\mathcal{L}_{b_i}\right)\eta_i\right\vert\right\vert_{\mathring{H}^{N_1-2}} + \left\vert\left\vert \left(1,\mathcal{L}_{b_i}\right)\mathcal{P}_{\ell \geq 2}\varphi_i\right\vert\right\vert_{\mathring{H}^{N_1-1}}+\left|\left(\Omega\underline{\omega}\right)_i\right| \lesssim \left\vert\left\vert z\right\vert\right\vert_{\mathring{H}^{N_1}}.
	\end{equation}
	\item Next we define $\left(T_{\rm low}\right)_i$ and $\left(T_{\rm high}\right)_i$ by 
	\[\left(T_{\rm low}\right)_i \doteq \mathcal{P}_{1 \leq \ell \leq \ell_0}\left(\mathcal{F}_{X_{i-1}}^{-1}\right)^*\slashed{\rm curl}b_i,\qquad \left(T_{\rm high} \right)_i \doteq \mathcal{P}_{\ell > \ell_0}\left(\mathcal{F}_{X_{i-1}}^{-1}\right)^*\slashed{\rm curl}\slashed{\rm div}\slashed{\nabla}\hat{\otimes}b_i,\]
	where the angular operators are defined with respect to $\slashed{g}_i \doteq e^{2\varphi_i}\mathring{\slashed{g}}$. Noting that we have $\left(\slashed{\nabla}\hat{\otimes}b_i\right)^{AB} = e^{-2\varphi_i}\left(\mathring{\nabla}\hat{\otimes}b_i\right)^{AB}$, we see that, in view of~\eqref{2o3mpompo4522}, that  
	\[\left\vert\left\vert \mathcal{P}_{\ell = 1}\slashed{\rm curl}b\right\vert\right\vert_{L^2} + \left\vert\left\vert \slashed{\rm curl}\slashed{\rm div}\slashed{\nabla}\hat{\otimes}b_i\right\vert\right\vert_{\mathring{H}^{N_1-3}} \lesssim \left\vert\left\vert z\right\vert\right\vert_{\mathring{H}^{N_1}}.\]
	This then yields that
	\[\left\vert\left\vert \left(T_{\rm low}\right)_i \right\vert\right\vert_{L^2\left(\mathbb{S}^2\right)} + \left\vert\left\vert \left(T_{\rm high}\right)_i\right\vert\right\vert_{\mathring{H}^{N_1-3}} \lesssim \left\vert\left\vert z\right\vert\right\vert_{\mathring{H}^{N_1}}.\]
	\item Now we apply Theorem~\ref{thisiswhatishholdingatthenehdne} with the seed data $\left(\left(T_{\rm low}\right)_i,\left(T_{\rm high}\right)_i\right)$.  We call the outputted $\slashed{g}$ by $\tilde{\slashed{g}}_i$, the outputted shift by $\tilde{b}_i$, and the outputted $\Omega^{-1}{\rm tr}\chi$ by $\left(\Omega^{-1}{\rm tr}\chi\right)_i$.  Restricting all these quantities to $\{v = 0\}$ and keeping~\eqref{2om3om1oijtionhoin1} in mind, we will have
	\[\left\vert\left\vert \left(1,\mathcal{L}_{\tilde{b}_i}\right)\left(\left(\Omega^{-1}{\rm tr}\chi\right)_i-2\right)\right\vert\right\vert_{\mathring{H}^{N_1-2}} \lesssim \left\vert\left\vert z\right\vert\right\vert^2_{\mathring{H}^{N_1}}, \]
	\begin{equation}\label{3m2ojio4j2}
	\left\vert\left\vert \tilde{\slashed{g}}_i - \mathring{\slashed{g}}\right\vert\right\vert_{\mathring{H}^{N_1-1}} + \left\vert\left\vert \tilde{b}_i \right\vert\right\vert_{\mathring{H}^{N_1-1}} \lesssim \left\vert\left\vert z\right\vert\right\vert_{\mathring{H}^{N_1}}.
	\end{equation}
	\item Now we apply Lemma~\ref{fixthespheregauge} to a find a vector field $X_i$ so that 
	\begin{equation}\label{2oo9582}
	\left\vert\left\vert X_i\right\vert\right\vert_{\mathring{H}^{N_1}} \lesssim \left\vert\left\vert z\right\vert\right\vert_{\mathring{H}^{N_1}},
	\end{equation}
	and
	\[\mathcal{F}_{X_i}^*\tilde{\slashed{g}} = e^{2\tilde{\varphi}_i}\mathring{\slashed{g}}.\]
	It then follows from~\eqref{3m2ojio4j2} and~\eqref{2oo9582} that
	\[\left\vert\left\vert \tilde{\varphi}_i\right\vert\right\vert_{\mathring{H}^{N_1-1}} \lesssim \left\vert\left\vert z\right\vert\right\vert_{\mathring{H}^{N_1}}.\] 
	We then set $\left(\varphi_{\rm low}\right)_i \doteq \mathcal{P}_{\ell \leq 1}\tilde{\varphi}_i$.

\end{enumerate}

This completes the induction step. We may then extract a suitably convergent subsequence. We denote the corresponding limits by 
\[\left\{\varphi_{\rm low},\varphi,\tilde{b},b,q,\eta,\Omega\underline{\omega},T_{\rm low},T_{\rm high},X,\left(\Omega^{-1}{\rm tr}\chi\right)\right\}.\]
To conclude the proof it suffices to verify that $b = \mathcal{F}_X^*\tilde{b}$. Let $\tilde{\slashed{g}}$, $\widetilde{\Omega\underline{\omega}}$, and $\tilde{\eta}$ be the quantities associated to the application of Proposition~\ref{k3jkl11} to the seed data $T_{\rm low}$ and $T_{\rm high}$. Then we define $\tilde{\varphi}$ by $\mathcal{F}_X^*\tilde{\slashed{g}} = e^{2\tilde{\varphi}}\mathring{\slashed{g}}$. We note that as a consequence of the iteration argument, we have that $\mathcal{P}_{\ell \leq 1}\tilde{\varphi} = \varphi_{\rm low}$. Furthermore, in view of how we have set up the iteration argument, and also the restriction to $\{v = 0\}$ of~\eqref{eqnyaydivb},~\eqref{3pk2o294}, and~\eqref{2momoo3} we will have that $\left(\tilde{\varphi},\mathcal{F}_X^*\tilde{b},\mathcal{F}_X^*\tilde{\eta},\widetilde{\Omega\underline{\omega}}\right)$ and  $\left(\varphi,b,\eta,\Omega\underline{\omega}\right)$ both satisfy the conformal coordinate boundary system associated to $\left(\mathring{\Pi}_{\rm curl}\mathcal{F}_X^*\tilde{b},\varphi_{\rm low},\mathcal{F}_X^*\left(\Omega^{-1}{\rm tr}\chi\right),\mathcal{F}_X^*\tilde{b}\right)$ and $\left(y,\varphi_{\rm low},\mathcal{F}_X^*\left(\Omega^{-1}{\rm tr}\chi\right),\mathcal{F}_X^*\tilde{b}\right)$ respectively. It then follows from Lemma~\ref{2io3ouhui2} that
\begin{equation}\label{23omoo19}
\left\vert\left\vert \mathring{\Pi}_{\rm div}\left(\mathcal{F}_X^*\tilde{b} - b\right)\right\vert\right\vert_{\mathring{H}^{N_1-1}} + \left\vert\left\vert \mathcal{P}_{\ell \geq 2}\left(\tilde{\varphi}-\varphi\right)\right\vert\right\vert_{\mathring{H}^{N_1-2}} \lesssim \left\vert\left\vert \mathring{\Pi}_{\rm curl}b - \mathring{\Pi}_{\rm curl}\left(\mathcal{F}_X^*\tilde{b}\right)\right\vert\right\vert_{\mathring{H}^{N_1-1}}
\end{equation}

On the other hand, as a consequence of the iteration scheme, we have that
\begin{equation}\label{2o4iio482}
\mathcal{P}_{1 \leq \ell \leq \ell_0}\left(\widetilde{\slashed{\rm curl}}\tilde{b}\right) = \mathcal{P}_{1 \leq \ell \leq \ell_0}\left(\mathcal{F}_{X}^{-1}\right)^*\slashed{\rm curl}b,\qquad \mathcal{P}_{\ell > \ell_0}\left(\widetilde{\slashed{\rm curl}}\widetilde{\slashed{\rm div}}\widetilde{\slashed{\nabla}\hat{\otimes}}\tilde{b}\right)= \mathcal{P}_{\ell > \ell_0}\left(\mathcal{F}_{X}^{-1}\right)^*\slashed{\rm curl}\slashed{\rm div}\slashed{\nabla}\hat{\otimes}b.
\end{equation}
Keeping in mind that
\[\left(\slashed{\nabla}\hat{\otimes}b\right)^{AB} = e^{-2\varphi}\left(\mathring{\nabla}\hat{\otimes}b\right)^{AB},\qquad \mathcal{F}_X^*\left(\widetilde{\slashed{\nabla}\hat{\otimes}}b\right)^{AB} = e^{-2\tilde{\varphi}}\left(\mathring{\nabla}\hat{\otimes}\left(\mathcal{F}_X^*b\right)\right)^{AB}\]
the relations~\eqref{2o4iio482} and elliptic estimates yield 
\begin{equation}
\left\vert\left\vert \mathring{\Pi}_{\rm curl}\left(\mathcal{F}_X^*\tilde{b} -b\right)\right\vert\right\vert_{\mathring{H}^{N_1-1}} \lesssim \epsilon\left[\left\vert\left\vert \mathring{\Pi}_{\rm div}\left(\mathcal{F}_X^*\tilde{b} - b\right)\right\vert\right\vert_{\mathring{H}^{N_1-1}} + \left\vert\left\vert \mathcal{P}_{\ell \geq 2}\left(\tilde{\varphi}-\varphi\right)\right\vert\right\vert_{\mathring{H}^{N_1-2}}\right].
\end{equation}
We thus conclude that $\mathcal{F}_X^*\tilde{b} = b$ and that $\tilde{\varphi} = \varphi$.

\end{proof}

In this next proposition we observe that if we take $\mathcal{L}_{\partial_{\phi}}z$ to satisfy a suitable smallness condition, then this smallness will propagate through the solution.
\begin{proposition}\label{1098uhnjuygbn}Let $z^A$ be a given vector field on $\mathbb{S}^2$ which satisfies~\eqref{23m2omo2124tgf}. Further suppose that $\hat{\epsilon} > 0$ satisfies $\hat{\epsilon} \ll \epsilon$ and that
\[\left\vert\left\vert \mathcal{L}_{\partial_{\phi}}z\right\vert\right\vert_{\mathring{H}^{N_1-1}} \lesssim \hat{\epsilon}.\]
Then, if we let $T_{\rm low}$ and $T_{\rm high}$ be produced by Proposition~\ref{23moo49281}, we will have
\[\left\vert\left\vert \epsilon\left(\mathcal{L}_{\partial_{\phi}}T_{\rm low},\mathcal{L}_{\partial_{\phi}}T_{\rm high}\right)\right\vert\right\vert_{\mathring{H}^{N_1-4}} \lesssim \hat{\epsilon}.\]
\end{proposition}
\begin{proof}This follows by repeating the proof of Proposition~\ref{23moo49281} and also using~\eqref{3kj2oi4iojir5i2io2}.
\end{proof}

In this final definition of the section we define the class of $z$ which will allow us to exploit various results from~\cite{nakedone}.
\begin{definition}\label{2kn3i1i42}Let $0 < \gamma \ll 1$. We say that a vector field $z$ on $\mathbb{S}^2$ satisfying~\eqref{23m2omo2124tgf} is ``fine-tuned'' if it additionally satisfies that 
\[z^A = \epsilon\left(\left(\int_{\pi/2}^0\frac{a\left(\tilde{\theta}\right)}{\sin\left(\tilde{\theta}\right)}\, d\tilde{\theta}\right) + r\right)\partial_{\phi} + \tilde{z}^A,\]
where $|r| \lesssim 1$, $a(\theta)$ satisfies
\begin{enumerate}
	\item $a$ is a smooth function of $\theta$,
	\item $a$ is identically $1$ for $\theta \in [2\gamma,\pi-2\gamma]$,
	\item $a$ vanishes identically for $\theta \in [0,\gamma] \cup [\pi-\gamma,\pi]$,
	\item $\left|\partial_{\theta}^ka\right| \lesssim \gamma^{-k}$,
\end{enumerate} 
and $\tilde{z}^A$ satisfies
\[\mathring{\nabla}_A\tilde{z}^A = 0,\qquad \left\vert\left\vert \tilde{z}\right\vert\right\vert_{\mathring{H}^{M_1}}\lesssim \epsilon^{M_0},\]
where $M_1 \gg N_1$ and $M_0 \gg 1$. 

Finally, we say that $\left(\mathcal{M},g\right)$ is a fine-tuned self-similar solution if $\left(\mathcal{M},g\right)$ results from applying Proposition~\ref{23moo49281} with a fine-tuned vector field $z$. We then further assume that $\left(\mathcal{M},g\right)$ is then equipped with the double-null coordinate system induced from the vector field $X$ produced by Proposition~\ref{23moo49281} (in the sense of Definition~\ref{itsinduced}) .
\end{definition}
\begin{remark}The reader should compare with Definition 4.4 from~\cite{nakedone}. In particular such ``fine-tuned'' choices of $z$ are exactly those which allow for the detailed analysis in~\cite{nakedone}. 
\end{remark}
\subsubsection{Refined Asymptotics for $\Omega$ and $\Omega^{-1}\hat{\chi}$}
In this section we will further study the solution produced by Proposition~\ref{23moo49281} associated to a fine-tuned vector field $z^A$ in the sense of Definition~\ref{2kn3i1i42}. For such solutions we can appeal to various parts of the analysis from~\cite{nakedone}. We will exploit this now to establish various improved estimates for the lapse $\Omega$ and also for $\Omega^{-1}\hat{\chi}$.

In this first lemma, we will appeal to results from~\cite{nakedone} to give a refined description of the vector field $b$ along the cone $\{v = 0\}$. 
\begin{lemma}\label{2opj4p209905} \underline{Let us introduce the convention that all constants in this lemma are independent of $\gamma$.} 

Let $\left(\mathcal{M},g\right)$ be a fine-tuned self-similar solution (see Definition~\ref{2kn3i1i42}). Then we have that
\begin{equation}\label{2ij3oij09995}
b|_{\mathbb{S}^2_{-1,0}} = \check{b} + \epsilon^2 h(\theta)\partial_{\theta} + e,
\end{equation}
where 
\begin{equation}\label{3ijoi2j3}
h(\theta) \doteq \frac{1}{2\sin\theta}\left[\left(\int_0^{\theta}a^2(x)\sin(x)\, dx\right)-\left(\int_0^{\pi}a^2(x)\sin(x)\, dx\right)\left(\frac{1-\cos(\theta)}{2}\right)\right].
\end{equation}
and we moreover have that 
\begin{enumerate}
	\item We have
	\[\check{b} = \epsilon\left(\left(\int_{\pi/2}^{\theta}\frac{a\left(\tilde{\theta}\right)}{\sin\left(\tilde{\theta}\right)}\, d\tilde{\theta}\right) + r\right)\partial_{\phi},\]
	where $a$ and $r$ are as in Definition~\ref{2kn3i1i42}.
	\item 
	\begin{enumerate}
	\item	The function $h(\theta)$ has three simple zeros at $0$, $y_0$, and $\pi$ where $y_0$ satisfies $\left|y_0-\pi/2\right|\lesssim \gamma$.
	\item There exists a constant $0 < c \ll 1$ so that $\left|h(\theta)\right| \geq \frac{\gamma^2}{100}$ for $\theta \in [\gamma^2/4,y_0-c] \cup [y_0 + c,\pi-\gamma^2/4]$.
	\item For $\theta \in [0,\gamma]$ we have  $h = -\frac{1}{4}\theta\left(1+O\left(\gamma^2\right)\right) + O\left(\theta^2\right)$. For $\theta \in [\pi - \gamma,\pi]$ we have $h = \frac{\pi-\theta}{4}\left(1+O\left(\gamma^2\right)\right) + O\left((\pi-\theta)^2\right)$. 
	\item For every $\theta \in [0,\pi]$ and $k \in \mathbb{Z}_{\geq 0}$ we have $\left|\frac{d^k}{d\theta^k}\left(\sin\theta h\right)\right| \lesssim_k \gamma^{2-k}$. For every $\theta \in [3\gamma,\pi-3\gamma]$ and $k \in \mathbb{Z}_{\geq 0}$ we have $\left|\frac{d^k}{d\theta^k}\left(\sin\theta h\right)\right| \lesssim_k \gamma^2$.
	\end{enumerate}
	\item We have $\left\vert\left\vert e\right\vert\right\vert_{\mathring{H}^3} \lesssim \epsilon^{5/2}$ and $\left\vert\left\vert \mathcal{L}_{\phi}e\right\vert\right\vert_{\mathring{H}^{N_0}} \lesssim \epsilon^{M_1}$.  In particular, by interpolating with bound $\left\vert\left\vert b\right\vert\right\vert_{\mathring{H}^{N_1}} \lesssim \epsilon \gamma^{N_1}$, for any $\tilde{N}$ satisfying $1 \ll \tilde{N} \ll N_2$ there exists a constant $m > 2$ such that $\left\vert\left\vert e\right\vert\right\vert_{\mathring{H}^{\tilde{N}}} \lesssim \epsilon^m$.
\end{enumerate}

Moreover, we have that
\begin{equation}\label{2oij4oijio592}
\left(\Omega\underline{\omega}\right)|_{v=0} \sim \epsilon^2(-u)^{-1}.
\end{equation}

We also have the following formula for $h(\theta)$ when $\theta \in [0,\gamma]$:
\begin{equation}\label{3oij4iojoij2io3}
h(\theta) = -\frac{1}{4}\left(\int_0^{\pi}a^2(x)\sin(x)\, dx\right)\frac{1-\cos(\theta)}{\sin(\theta)}.
\end{equation}
\end{lemma}
\begin{proof}Recall that the analysis from Sections 4.2 and 4.3 of~\cite{nakedone} starts with a suitably given ``$\left(\epsilon,\gamma,\delta,N_0,M_0,M_1\right)$-regular'' $4$-tuples $\left(\slashed{g},b,\kappa,\Omega\right)$ of a metric, vector field, constant, and function on $\mathbb{S}^2$ (see Definition 4.5 of~\cite{nakedone}). It is an immediate consequence of Propositions~\ref{23moo49281} and~\ref{1098uhnjuygbn} that we obtain we obtain a corresponding  $\left(\epsilon,\gamma,\delta,N_0,M_0,M_1\right)$-regular $4$-tuple if we take $\left(e^{2\varphi}\mathring{\slashed{g}},b|_{\mathbb{S}^2_{-1,0}},2\Omega\underline{\omega}|_{v=0},1\right)$ as our tuple. The desired statements for the vector field $b$ and $\Omega\underline{\omega}$ then follow from Lemma 4.8, Lemma 4.10, and Lemma 4.11 of~\cite{nakedone}.
\end{proof}

In the next lemma, we use the extra information afforded by Lemma~\ref{2opj4p209905} to establish a more refined estimate (compared with the estimates from Theorem~\ref{thisiswhatishholdingatthenehdne}) for the lapse $\Omega$ as $v\to 0$.
\begin{lemma}\label{32oi3ijo32ijo} Let us introduce the convention that, unless said otherwise, all constants in this lemma are independent of $\gamma$.

Let $\left(\mathcal{M},g\right)$ be a fine-tuned self-similar solution (see Definition~\ref{2kn3i1i42}) and let $\tilde{N}$ satisfy $1 \ll \tilde{N} \ll N_2$. We may write the lapse $\Omega$ as $\log\Omega = -\kappa\log(-v) + \log\Omega_1$ so that there then exists constants $d > 0$ and $C\left(\epsilon,\gamma\right) > 0$ so that for every $v \in (-1/2,0)$:
\begin{equation}\label{i4ijo43ijo3ijo3ijo}
\sum_{j + \left|\alpha\right| \leq \tilde{N}}\left\vert\left\vert \left(v\mathcal{L}_{\partial_v}\right)^j\log\Omega^{(\alpha)}_1\right\vert\right\vert_{L^2\left(\mathbb{S}^2_{-1,v}\right)} \leq C\left(\epsilon,\gamma\right)(-v)^{-d\epsilon^2 \gamma},\qquad \kappa \sim \epsilon^2,
\end{equation}
\begin{equation}\label{3io3jiiooi89898}
\sup_{v \in (-1/2,0)}\left\vert\left\vert \log\Omega_1\right\vert\right\vert_{L^{\infty}\left(\mathbb{S}^2_{-1,v}\right)} \lesssim 1.
\end{equation}
We emphasize that $d$ is independent of both $\gamma$ and $\epsilon$.
\end{lemma} 
\begin{proof}

Along $\{v = -1\}$, we have
\begin{equation}\label{32oij3iojoi4}
\left((-v)\mathcal{L}_{\partial_v}-\mathcal{L}_b\right)\log\Omega  = 2\left(\Omega\underline{\omega}\right),\qquad \log\Omega|_{v=-1} = 0.
\end{equation}
Moreover, $\Omega\underline{\omega}|_{v=0}$ is a constant and, in view of Theorem~\ref{thisiswhatishholdingatthenehdne}  and Lemma~\ref{2opj4p209905} we have that
\begin{equation}\label{23m4oimo2}
\sup_{v \in (-1/2,0)}(-v)^{-3/4}\sum_{j+\left|\alpha\right| \leq \tilde{N}_0}\left\vert\left\vert\left(v\mathcal{L}_{\partial_v}\right)^j\mathcal{L}_{Z^{(\alpha)}} \left(\left(\Omega\underline{\omega} - \left(\Omega\underline{\omega}\right)|_{v=0}\right), \left(\mathcal{L}_b-\mathcal{L}_{b|_{v=0}}\right)\log\Omega\right) \right\vert\right\vert_{L^2} \lesssim 1, 
\end{equation}
for any $\tilde{N}_0$ satisyfing $1 \ll \tilde{N}_0 \ll N_2$. Let $\log\Omega_0$ be defined by solving 
\begin{equation}\label{23kjlk2j}
\left((-v)\mathcal{L}_{\partial_v}-\mathcal{L}_b\right)\log\Omega_0  = 2\left(\Omega\underline{\omega}\right)|_{v=0},\qquad \log\Omega_0|_{v=-1} = 0.
\end{equation}
By the uniqueness of solutions to transport equations, we have in fact that $\log\Omega_0$ is spherically symmetric and thus that
\begin{equation}\label{32iuou23}
(-v)\mathcal{L}_{\partial_v}\log\Omega_0  = 2\left(\Omega\underline{\omega}\right)|_{v=0},\qquad \log\Omega_0|_{v=-1} = 0.
\end{equation}
Setting $\kappa = \frac{1}{2}\left(\Omega\underline{\omega}\right)|_{v=0}$ we thus obtain that $\log\Omega_0 = -\kappa\log\left(-v\right)$ and $\kappa \sim \epsilon^2$ where the estimate for $\kappa$ follows from Lemma~\ref{2opj4p209905}. 

Setting $\log\Omega_1 \doteq \log\Omega - \log\Omega_0$, we then have that
\begin{equation}\label{3oij2ij5}
\left((-v)\mathcal{L}_{\partial_v}-\mathcal{L}_{b|_{v=0}}\right)\log\Omega_1 = 2\left(\left(\Omega\underline{\omega}\right) - \left(\Omega\underline{\omega}\right)|_{v=0}\right) + \left(\mathcal{L}_{b|_{v=0}}-\mathcal{L}_b\right)\log\Omega \doteq H_0,\qquad \log\Omega_1|_{v=-1} = 0.
\end{equation}
For notational convenience, let us set $b_0 \doteq b|_{v=0}$. For every $\hat{\epsilon} > 0$ we define $\log\Omega_{2,\hat{\epsilon}}$ by solving 
\begin{equation}\label{23lkj}
\left((-v)\mathcal{L}_{\partial_v}-\mathcal{L}_{b_0}\right)\log\Omega_{2,\hat{\epsilon}} =  H_0,\qquad \log\Omega_{2,\hat{\epsilon}}|_{v=-\hat{\epsilon}} = 0.
\end{equation}
Switching to the variable $s \doteq -\log\left(-v)\right)$ and commuting with $\mathcal{L}^{(\alpha)}$ for $\left|\alpha\right| \leq \tilde{N}$ leads to the equation
\begin{equation}\label{23koi4oiut4io}
\left(\mathcal{L}_{\partial_s}-\mathcal{L}_{b_0}-\frac{s}{10}\right)\left(e^{\frac{s}{10}}\log\Omega^{(\alpha)}_{2,\hat{\epsilon}}\right)=  e^{\frac{s}{10}}H^{(\alpha)}_0 + e^{\frac{s}{10}}W(\alpha),\qquad \log\Omega_2|_{v=-\hat{\epsilon}} = 0,
\end{equation}
where $\left|W(\alpha)\right| \lesssim_{\alpha} \epsilon \sum_{\left|\beta\right| \leq \left|\alpha\right|}\left|\log\Omega_{2,\hat{\epsilon}}^{(\beta)}\right|$. Now we may contract~\eqref{23koi4oiut4io} with $\left(e^{\frac{s}{10}}\log\Omega^{(\alpha)}_{2,\hat{\epsilon}}\right)$, integrate with respect to $ds\mathring{\rm dVol}$ from $s = -\log\left(\hat{\epsilon}\right)$ to $s = \tilde{s}$, integrate by parts, and sum over $\left|\alpha\right| \leq \tilde{N}$. We obtain, for every $\tilde{s} \in [-\log\left(-1/2\right),-\log\left(\hat{\epsilon}\right)]$, the estimate  
\begin{align}\label{23poj4oj2}
\sup_{s \in [\tilde{s},-\log\left(\hat{\epsilon}\right)]}\left[e^{\frac{s}{5}}\left\vert\left\vert \log\Omega_{2,\hat{\epsilon}}\right\vert\right\vert^2_{\mathring{H}^{\tilde{N}}}\right] \lesssim \int_{\tilde{s}}^{-\log\left(\hat{\epsilon}\right)} e^{\frac{s}{5}}\left\vert\left\vert H_0\right\vert\right\vert^2_{\mathring{H}^{\tilde{N}}}\, ds.
\end{align}
In view of~\eqref{23m4oimo2} we obtain from the estimate~\eqref{23poj4oj2} that
\begin{align*}
\sup_{v \in (-1/2,\hat{\epsilon})} (-v)^{-3/4}\left\vert\left\vert \log\Omega_{2,\hat{\epsilon}}\right\vert\right\vert_{\mathring{H}^{\tilde{N}}} \lesssim 1.
\end{align*}
It is then straightforward to use the equation~\eqref{23lkj} to obtain that 
\begin{align}\label{32oij4ijo43i}
\sup_{v \in (-1/2,\hat{\epsilon})} (-v)^{-3/4}\sum_{j+\left|\alpha\right| \leq \tilde{N}}\left\vert\left\vert \left(v\mathcal{L}_{\partial_v}\right)^j\log\Omega^{(\alpha)}_{2,\hat{\epsilon}}\right\vert\right\vert_{L^2} \lesssim 1.
\end{align}

By consider the equation for the differences, it is straightforward to then see that we may define $\log\Omega_2 \doteq \lim_{\hat{\epsilon} \to 0}\log\Omega_{2,\hat{\epsilon}}$ which will solve 
\begin{equation}\label{2lk3jlj}
\left((-v)\mathcal{L}_{\partial_v}-\mathcal{L}_{b_0}\right)\log\Omega_2 =  H_0,
\end{equation}
and satisfy the estimate~\eqref{32oij4ijo43i} with $\hat{\epsilon} = 0$. 

We now set $\log\Omega_3 \doteq \log\Omega_1 - \log\Omega_2$. This will satisfy
\begin{equation}\label{3oj4oij5ijo43}
\left((-v)\mathcal{L}_{\partial_v}-\mathcal{L}_{b_0}\right)\log\Omega_3 = 0,\qquad \left\vert\left\vert \log\Omega_3|_{v=-1/2}\right\vert\right\vert_{\tilde{H}^N} \lesssim 1.
\end{equation}
The estimate~\eqref{3io3jiiooi89898} with $\Omega_3$ replacing $\Omega_1$ is an immediate consequence on integrating along the integral curves of~\eqref{3oj4oij5ijo43}. Thus, it is clear that in order to finish the proof, it would suffice to establish that for any $ v \in (-1/2,0)$:
\begin{equation}\label{32oij4oij2oij3}
\sum_{\left|\alpha\right| \leq \tilde{N}}\left\vert\left\vert \log\Omega^{(\alpha)}_3\right\vert\right\vert_{L^{\infty}\left(\mathbb{S}^2_{-1,v}\right)} \leq C\left(\epsilon,\gamma\right)(-v)^{-d\epsilon^2 \gamma}.
\end{equation}
Establishing this estimate will be the most difficult part of the proof.

We first note that~\eqref{32oij4oij2oij3} clearly holds if we were to set $\tilde{N} = 0$ because $\log\Omega_3$ is transported along the integral curves of $(-v)\mathcal{L}_{\partial_v}-\mathcal{L}_{b_0}$. Thus we proceed by induction and assume, for some $k \in \mathbb{Z}_{\geq 0}$ which satisfies $k \ll N_2$, that there exists constants $C\left(k,\epsilon,\gamma\right)$ and $d\left(k\right)$ such that for all $v \in (-1/2,0)$:
\begin{equation}\label{23kljkjl}
\sum_{\left|\alpha\right| \leq k}\left\vert\left\vert \log\Omega^{(\alpha)}_3\right\vert\right\vert_{L^{\infty}\left(\mathbb{S}^2_{-1,v}\right)} \leq C\left(k,\epsilon,\gamma\right)(-v)^{-d\left(k\right)\epsilon^2 \gamma}.
\end{equation}
We will now show that~\eqref{23kljkjl} holds with $k$ replaced by $k+1$. 

For any $j \in [0,k+1]$, we may commute~\eqref{3oj4oij5ijo43} with $\mathcal{L}_{\phi}^j\mathcal{L}_{b_0}^{k+1-j}$ and obtain an equation of the form 
\begin{equation}\label{2klj3l2}
\left((-v)\mathcal{L}_{\partial_v}-\mathcal{L}_{b_0}\right)\left(\mathcal{L}_{\phi}^j\mathcal{L}_{b_0}^{k+1-j}\log\Omega_3\right) = \mathcal{E}_j,
\end{equation}
where, in view of the bound on $\mathcal{L}_{\partial_{\phi}}e$ from Lemma~\ref{2opj4p209905}, we have that  $\mathcal{E}_j$ satisfies the pointwise bound
\begin{equation}\label{3j2oijio4}
\left|\mathcal{E}_j\right| \lesssim \epsilon^{M_1}\sum_{\left|\alpha\right| \leq k+1}\left|\log\Omega_3^{(\alpha)}\right|.
\end{equation}
Next, in view of the estimates for $h(\theta)$ from Lemma~\ref{2opj4p209905}, we observe that as long we restrict to $\theta \in [\gamma/2,y_0-c] \cup [y_0+c,\pi-\gamma/2]$, then estimates for linear combinations of $\{\mathcal{L}_{\phi}^j\mathcal{L}_{b_0}^{k+1-j}\log\Omega_3\}_{j=0}^{k+1}$ suffice to control $\left(\gamma^2\epsilon\right)^{(k+1)\tilde{N}}\sum_{\left|\alpha\right| \leq k+1}\left|\log\Omega_3^{(\alpha)}\right|$. Since $M_1 \gg \tilde{N}$, this type of estimate is consistent with eventually using Gr\"{o}nwall's inequality on~\eqref{2klj3l2} to establish the desired estimate, as long as by some other means we have suitable control for $\theta \not\in [\gamma/2,y_0-c] \cup [y_0+c,\pi-\gamma/2]$.

We first turn to the region $\theta \in [y_0-c,y_0+c]$. In this region, from Lemma~\ref{2opj4p209905}, we have that $\sum_{j \leq k+1}\left|h^{(k)}(\theta)\right| \lesssim \gamma^2$. Letting $p(\theta)$ be a bump function which is identically $1$ for $\theta \in[y_0-c,y_0+c]$ and vanishes identically for $\theta \not\in [y_0-2c,y_0+2c]$, we have, for each $j \in [1,k+1]$:
\begin{equation}\label{2o3ijoij5io2}
\left((-v)\mathcal{L}_{\partial_v}-\mathcal{L}_{b_0}\right)\left(p\mathcal{L}_{\partial_{\theta}}^j\mathcal{L}_{\partial_{\phi}}^{k+1-j}\log\Omega_3\right) = \mathcal{H}_j,
\end{equation}
where $\mathcal{H}_j$ satisfies the estimate 
\begin{align}\label{23ojoij5ioj2ioj4}
&\left|\mathcal{H}_j\right| \lesssim 1_{\theta \in [\pi-2c,\pi+2c]}\left(\epsilon^2\gamma^2 \sum_{\left|\alpha\right| \leq k+1}\left|\log\Omega_3^{(\alpha)}\right| + \epsilon \sum_{\left|\alpha\right| \leq j-1,\beta \leq k+1-j }\left|\mathcal{L}^{1+\beta}_{\partial_{\phi}}\log\Omega_3^{(\alpha)}\right|\right)
\\ \nonumber &\qquad \qquad + \gamma^{-\tilde{N}}\left(1_{\theta 
\in [\pi-2c,\pi-c]}+1_{\theta \in [\pi+c,\pi+2c]}\right)\sum_{\left|\alpha\right| \leq k+1}\left|\log\Omega_3^{(\alpha)}\right|.
\end{align}

Next we come to the region $\theta \in [0,\gamma]$. Since the the $\left(\theta,\phi\right)$ coordinates degenerate here, it is useful to introduce a set $\left(\tilde{x},\tilde{y}\right)$ of regular coordinates by the usual formulas:
\[\tilde{x} \doteq \theta \cos\phi,\qquad \tilde{y} \doteq \theta \sin\phi.\]
In view of Lemma~\ref{2opj4p209905}, when $\theta \in [0,\gamma]$, we then have
\begin{equation}\label{2oi3joij}
b_0 = -\frac{\epsilon^2}{4}\left(A_0+f\left(\tilde{x},\tilde{y}\right)\right)\left(\tilde{x}\partial_{\tilde{x}} + \tilde{y}\partial_{\tilde{y}}\right) + \epsilon  A_1\left(\tilde{x}\partial_{\tilde{y}}-\tilde{y}\partial_{\tilde{x}}\right) +e,
\end{equation}
where $A_0$ and $A_1$ are constants satisfying
\[A_0 \sim \frac{1}{4},\qquad \left|A_1\right| \lesssim \left|\log\left(\gamma\right)\right|,\]
and we have that $f$ is a smooth function which satisfies 
\[\left|\partial_{\tilde{x}}^i\partial_{\tilde{y}}^jf\left(\tilde{x},\tilde{y}\right)\right| \lesssim_{i,j} \left[\left|\tilde{x}\right|^{{\rm max}\left(2-j,0\right)}+\left|\tilde{y}\right|^{{\rm max}\left(2-j,0\right)}\right],\qquad \forall (i,j) \in \mathbb{Z}_{\geq 0}\times\mathbb{Z}_{\geq 0}.\]
Now, for every $j \in [0,k+1]$ we may commute the equation~\eqref{3oj4oij5ijo43} with $\tilde{p}(\theta)\mathcal{L}^{k+1-j}_{\partial_{\tilde{x}}}\mathcal{L}^j_{\partial_{\tilde{y}}}$, where $\tilde{p}(\theta)$ is identically $1$ for $\theta \in [0,\gamma/2]$ and vanishes for $\theta \in [\gamma,\pi]$. We obtain 
\begin{align}\label{32oj4oijio2}
&\left((-v)\mathcal{L}_{\partial_v}-\mathcal{L}_{b_0}\right)\left(\tilde{p} \mathcal{L}^{k+1-j}_{\partial_{\tilde{x}}}\mathcal{L}^j_{\partial_{\tilde{y}}}\log\Omega_3\right) + \frac{\epsilon^2A_0}{4}\left(k+1\right)\left(\tilde{p} \mathcal{L}^{k+1-j}_{\partial_{\tilde{x}}}\mathcal{L}^j_{\partial_{\tilde{y}}}\log\Omega_3\right) 
\\ \nonumber &\qquad +\epsilon A_1 \left(k+1-j\right)\left(\tilde{p}\mathcal{L}_{\partial_{\tilde{x}}}^{k-j}\mathcal{L}_{\partial_{\tilde{y}}}^{j+1}\log\Omega_3\right) - \epsilon A_1 j \left(\tilde{p}\mathcal{L}_{\partial_{\tilde{x}}}^{k+2-j}\mathcal{L}_{\partial_{\tilde{y}}}^{j-1}\log\Omega_3\right) = \mathcal{W}_j,
\end{align}
where $\mathcal{W}_j$ satisfies 
\begin{align}\label{32ioj4oi2o2}
\left|\mathcal{W}_j\right| &\lesssim \epsilon^2  1_{\theta \in [0,\gamma/2]} \left(\gamma^2 \sum_{\left|\alpha\right| \leq k+1}\left|\log\Omega_3^{(\alpha)}\right| +  \sum_{\left|\alpha\right| \leq k}\left|\log\Omega_3^{(\alpha)}\right|\right) + 
\epsilon \gamma^{-1} 1_{\theta \in [\gamma/2,\gamma]}\sum_{\left|\alpha\right| \leq k+1}\left|\log\Omega_3^{(\alpha)}\right|.
\end{align}
The terms proportional to $\epsilon A_1$ in~\eqref{32oj4oijio2} (which are generated by the commutator with $\mathcal{L}_{\partial_{\phi}}$) may appear troublesome. However, it will turn out that we can effectively cancel these terms.  We start by defining a quantity:
\begin{equation}\label{oij32ioj4ijo5op2}
\mathfrak{h}[k+1] \doteq \sum_{j=0}^{k+1} \frac{(k+1)!}{(k+1-j)!j!}\left(\tilde{p}\mathcal{L}_{\partial_{\tilde{x}}}^{k+1-j}\mathcal{L}_{\partial_{\tilde{y}}}^j\log\Omega_3\right)^2.
\end{equation}
We then claim that the following holds
\begin{align}\label{32oij4iojio2o42}
\frac{1}{2}\left((-v)\mathcal{L}_{\partial_v}-\mathcal{L}_{b_0}\right)\mathfrak{h}\left[k+1\right] + \frac{\epsilon^2A_0}{4}\left(k+1\right)\mathfrak{h}\left[k+1\right] = \mathcal{Q}_j,
\end{align}
where $\mathcal{Q}_j$ satisfies 
\begin{align}\label{l3jio32io23jio32ij}
\left|\mathcal{Q}_j\right| &\lesssim \epsilon^2  1_{\theta \in [0,\gamma/2]} \left(\gamma^2 \sum_{\left|\alpha\right| \leq k+1}\left|\log\Omega_3^{(\alpha)}\right|^2 +  \sum_{\left|\alpha\right| \leq k}\left|\log\Omega_3^{(\alpha)}\right|^2\right)  
+\epsilon \gamma^{-1} 1_{\theta \in [\gamma/2,\gamma]}\sum_{\left|\alpha\right| \leq k+1}\left|\log\Omega_3^{(\alpha)}\right|^2.
\end{align}
To derive this, we multiply~\eqref{32oj4oijio2} by $\frac{(k+1)!}{(k+1-j)!j!}\left(\tilde{p}\mathcal{L}_{\partial_{\tilde{x}}}^{k+1-j}\mathcal{L}_{\partial_{\tilde{y}}}^j\log\Omega_3\right)$ and sum over $j$. One may then check that all of the terms proportional to $\epsilon A_1$ then cancel. One may, of course, derive an analogous equation in the region $\theta \in [\pi-\gamma,\pi]$. We denote the analogue of $\mathfrak{h}\left[k+1\right]$ in this region by $\underline{\mathfrak{h}}\left[k+1\right]$. 

We are now ready for the main equation and estimate. Let us define
\begin{align}\label{3ijoi2joij44}
&\mathfrak{p}\left[k+1\right] \doteq \sum_{j=0}^{k+1}\left|\mathcal{L}_{\phi}^j\mathcal{L}_{b_0}^{k+1-j}\log\Omega_3\right|^2 
\\ \nonumber &\qquad + \sum_{j=1}^{k+1}\left(\epsilon^{10\tilde{N} j}\left|p\mathcal{L}_{\partial_{\theta}}^j\mathcal{L}_{\partial_{\phi}}^{k+1-j}\log\Omega_3\right|^2\right) + \epsilon^{10\tilde{N}}\left(\mathfrak{h}\left[k+1\right] + \underline{\mathfrak{h}}\left[k+1\right]\right).
\end{align}
By contracting~\eqref{2klj3l2} with $\mathcal{L}_{\phi}^j\mathcal{L}_{b_0}^{k+1-j}\log\Omega_3$,~\eqref{2o3ijoij5io2} with $\epsilon^{10\tilde{N} j}\left(p\mathcal{L}_{\partial_{\theta}}^j\mathcal{L}_{\partial_{\phi}}^{k+1-j}\log\Omega_3\right)$, summing over $j$, then adding to $\epsilon^{10\tilde{N}}$ times the identity~\eqref{32oij4iojio2o42} (and the analogue for $\underline{\mathfrak{h}}\left[k+1\right]$), using the induction hypothesis to control up to $k$ derivatives of $\log\Omega_3$,  using the observation that linear combinations of $\mathcal{L}^j_{\partial_{\phi}}\mathcal{L}^{k+1-j}_{b_0}$ control all $k+1$ derivatives for $\theta \in [\gamma/2,\pi-c] \cup [\pi+c,\pi-\gamma/2]$, and finally using that $M_1 \gg \tilde{N}$, we obtain that 
\begin{equation}\label{i3ojio2jioj5o2}
\left((-v)\mathcal{L}_{\partial_v}-\mathcal{L}_{b_0}\right)\mathfrak{p}\left[k+1\right] \lesssim \gamma \epsilon^2 \mathfrak{p}\left[k+1\right] + \left(C\left(k,\epsilon,\gamma\right)(-v)^{- d\left(k\right) \gamma \epsilon^2}\right)^2.
\end{equation}
Integrating along the integral curves of $\left(\mathcal{L}_{\partial_s}-\mathcal{L}_{b_0}\right)$ for $s \doteq -\log\left(-v\right)$  and then using Gr\"{o}nwall's inequality leads to the estimate 
\[\mathfrak{p}\left[k+1\right] \leq \tilde{C}\left(\epsilon,\gamma,k+1\right) (-v)^{-\tilde{d}\left(k+1\right) \gamma \epsilon^2},\]
for some constants $\tilde{C}$ and $\tilde{d}$. Since $\mathfrak{p}$ in fact controls all $k+1$ derivatives of $\log\Omega_3$, we are in fact done.
\end{proof}

In the next lemma, we establish an estimate for $\Omega^{-1}\hat{\chi}$ which is more refined than that provided by Theorem~\ref{thisiswhatishholdingatthenehdne}.
\begin{lemma}\label{23oij3jio23ij} Let us introduce the convention that, unless said otherwise, all constants in this lemma are independent of $\gamma$.

Let $\left(\mathcal{M},g\right)$ be a fine-tuned self-similar solution (see Definition~\ref{2kn3i1i42}) and let $\tilde{N}$ satisfy $1 \ll \tilde{N} \ll N_2$. Then we have that $\Omega^{-1}\hat{\chi}|_{u=-1}$ extends to $\{v = 0\}$ as $H^{\tilde{N}}\left(\mathbb{S}^2\right)$ tensor. Furthermore, we have that 
\begin{align}\label{32oij43ijo5ioj}
&\left\vert\left\vert \Omega^{-1}\hat{\chi}|_{v=0}\right\vert\right\vert_{H^{\tilde{N}}\left(\mathbb{S}^2\right)}
\\ \nonumber &\qquad +\sup_{v \in (-1/2,0)}(-v)^{-\frac{\kappa}{10}}\sum_{j + \left|\alpha\right| \leq \tilde{N}}\left\vert\left\vert \left(v\mathcal{L}_{\partial_v}\right)^j\left(\Omega^{-1}\hat{\chi} - \Omega^{-1}\hat{\chi}|_{v=0}\right)^{(\alpha)}\right\vert\right\vert_{L^2\left(\mathbb{S}^2_{-1,v}\right)}  \lesssim C\left(\epsilon,\gamma\right),
\end{align}
where $\kappa \sim \epsilon^2$. In particular the tensor $\Omega^{-1}\hat{\chi}$ extends to $\{v=0\}$ as a H\"{o}lder continuous tensor. 
\end{lemma}
\begin{proof}In view of Lemma~\ref{03kdo3k5}, after multiplying through by $\Omega^{-2}$, we obtain the following equation for $\Omega^{-1}\hat{\chi}$ along $\{u = -1\}$:
\begin{align}\label{kl2j3k4jl20}
& v\left(\Omega\nabla_4\right)\left(\Omega^{-1}\hat{\chi}\right)_{AB} +\mathscr{L}\left(\Omega^{-1}\hat{\chi}\right)_{AB}+4\left(\Omega\underline{\omega}\right)\left(\Omega^{-1}\hat{\chi}\right)_{AB} -\frac{v}{u}\Omega^{-1}{\rm tr}\chi\left(\Omega\hat{\chi}\right)_{AB} = 
\\ \nonumber &\qquad \left(\slashed{\nabla}\hat\otimes \eta\right)_{AB} + \left(\eta\hat\otimes \eta\right)_{AB} - \frac{1}{4}\left(\Omega^{-1}{\rm tr}\chi\right)\left(\slashed{\nabla}\hat{\otimes}b\right),
 \end{align}
 where
 \[\mathscr{L}f_{AB} \doteq \mathcal{L}_bf_{AB}- \left(\slashed{\nabla}\hat{\otimes}b\right)^C_{\ \ (A}f_{B)C} -\frac{1}{2}\slashed{\rm div}bf_{AB}.\]
 In particular, we may derive the following equation for $\Omega^{-1}\hat{\chi}$ along $\{u = -1\}$:
 \begin{align}\label{kl2j3k4jl20}
& v\mathcal{L}_{\partial_v}\left(\Omega^{-1}\hat{\chi}\right)_{AB} +\mathscr{L}|_{v=0}\left(\Omega^{-1}\hat{\chi}\right)_{AB}+4\left(\Omega\underline{\omega}\right)|_{v=0}\left(\Omega^{-1}\hat{\chi}\right)_{AB}= H_{AB},
 \end{align}
 where in view of our already established estimates for $g$, the tensor $H_{AB}$ has a continuous limit $H|_{v=0}$ and satisfies the estimates:
 \begin{align}\label{2oj4oij2ioj4ioj2oi2}
 \left\vert\left\vert H|_{v=0}\right\vert\right\vert_{\mathring{H}^{\tilde{N}}\left(\mathbb{S}^2_{-1,0}\right)} +\sum_{j \leq \tilde{N}} \sup_{v \in (-1/2,0)}(-v)^{-3/4}\left\vert\left\vert \left(v\mathcal{L}_{\partial_v}\right)^j\left(H - H|_{v=0}\right)\right\vert\right\vert_{\mathring{H}^{\tilde{N}}\left(\mathbb{S}^2_{-1,v}\right)} \lesssim \epsilon.
 \end{align}
 As in the proof of Lemma~\ref{2opj4p209905} we  note that it is an immediate consequence of Propositions~\ref{23moo49281} and~\ref{1098uhnjuygbn} that we obtain we obtain a corresponding  $\left(\epsilon,\gamma,\delta,N_0,M_0,M_1\right)$-regular $4$-tuple in the sense of~\cite{nakedone} if we take $\left(e^{2\varphi}\mathring{\slashed{g}},b|_{\mathbb{S}^2_{-1,0}},2\Omega\underline{\omega}|_{v=0},1\right)$ as our tuple. In particular, we may  apply Proposition 4.18 from~\cite{nakedone} (after a change of variables $v\mapsto -v$) and use the equation~\eqref{kl2j3k4jl20} directly to obtain control $v\mathcal{L}_{\partial_v}$ derivatives to obtain the estimate~\eqref{32oij43ijo5ioj}. 
\end{proof}

\subsubsection{Lapse-Renormalized Coordinates Near $\{v = 0\}$}\label{12jn1oiodo009u2jimkqwss}
In this section we will define a new set of coordinates near $\{v = 0\}$ where we will have improved regularity properties of our spacetime. We will further show that for suitable seed data, we may exploit results from our previous work~\cite{nakedone} to obtain the full desired regularity. 

\begin{definition}\label{lapserenormaldef}Let $\left(\mathcal{M},g\right)$ be a solution to the Einstein vacuum equations which is self-similar and defined in the region $\mathscr{U}$. We further assume that the conclusions of Lemma~\ref{3ioj2oijio45482} hold for this spacetime for a suitably large integer $N$. We then define ``lapse-renormalized coordinates'' $\left(\hat{v},u,\theta^A\right)$ by setting
\[\hat{v}\left(v,u,\theta^A\right) \doteq -\int_v^0\Omega^2\left(\tilde{v},u,\theta^A\right)\, d\tilde{v}.\]
\end{definition}

In the next lemma, we compute the form of the metric in the lapse-renormalized coordinates.
\begin{lemma}Let $\left(\mathcal{M},g\right)$ be a solution to the Einstein vacuum equations which is self-similar and defined in the region $\mathscr{U}$. In the lapse-renormalized coordinates, the metric takes the form
\begin{align}\label{2o3omo29495}
&g = -2\left(du\otimes d\hat{v} + d\hat{v}\otimes du\right) + \slashed{g}_{AB}d\theta^A\otimes d\theta^B 
\\ \nonumber &\qquad + \left(-4\int_v^0\left(\Omega^2\partial_{\theta^A}\log\Omega\right)d\tilde{v}-b_A\right)\left(du\otimes d\theta^A + d\theta^A\otimes du\right)
\\ \nonumber &\qquad +\left(-8\int_v^0\left(\Omega^2\partial_u\log\Omega\right)d\tilde{v} + \left|b\right|^2\right)du\otimes du.
\end{align}
\end{lemma}
\begin{proof}A computation yields that
\[dv = \Omega^{-2}d\hat{v} +2\Omega^{-2}\left(\int_v^0\left(\Omega^2\partial_u\log\Omega\right)\, d\tilde{v}\right)du +2\Omega^{-2}\left(\int_v^0\left(\Omega^2\partial_{\theta^A}\log\Omega\right)\, d\tilde{v}\right)d\theta^A.\]
Plugging this into the metric form~\eqref{doublenullisg} leads to~\eqref{2o3omo29495}.
\end{proof}

Next we compute the new form of the self-similar vector field.
\begin{lemma}In the $\left(\hat{v},u,\theta^A\right)$ coordinates, we have that 
\[K = u\partial_u + \hat{v}\partial_{\hat{v}}.\]
\end{lemma}
\begin{proof}A computation yields
\[\partial_v = \Omega^2\partial_{\hat{v}},\  \partial_u = \partial_{\hat{u}} - \left(\int_v^0\partial_u\left(\Omega^2\right)\, d\tilde{v}\right) \partial_{\hat{v}},\]
where $\partial_{\hat{u}}$ denotes the $u$-derivative in the $\left(\hat{v},u,\theta^A\right)$ coordinates. Thus, starting from $K = u\partial_u + v\partial_v$, we have that in the $\left(\hat{v},u,\theta^A\right)$ coordinates, 
\begin{align*}
K &= u\partial_{\hat{u}} + \left(v\Omega^2 - u\left(\int_v^0\partial_u\left(\Omega^2\right)\right)\, d\tilde{v}\right)\partial_{\hat{v}}
\\ \nonumber &=u\partial_{\hat{u}} + \left(v\Omega^2 + \left(\int_v^0\tilde{v}\partial_{\tilde{v}}\left(\Omega^2\right)\, d\tilde{v}\right)\right)\partial_{\hat{v}}
\\ \nonumber &= u\partial_{\hat{u}} + \left(-\left(\int_v^0\Omega^2\, d\tilde{v}\right)\right)\partial_{\hat{v}}
\\ \nonumber &= u\partial_{\hat{u}} + \hat{v}\partial_{\hat{v}}
\end{align*}
\end{proof}

The key result of this section is the following.
\begin{proposition}\label{3ijo3iojio9}Let $z^A$ be a vector field on $\mathbb{S}^2$ which is fine-tuned in the sense of Definition~\ref{2kn3i1i42}. Let $\left(\mathcal{M},g\right)$ be the solution to the Einstein vacuum equations produced by Proposition~\ref{23moo49281} which is conformally compatible with $z$. Letting $X$ denote the vector field from Proposition~\ref{23moo49281}, we then equip $\left(\mathcal{M},g\right)$ with the induced coordinates from $X$ (in the sense of Definition~\ref{itsinduced}). Finally, we then introduce lapse-renormalized coordinates (in the sense of Definition~\ref{lapserenormaldef}).

We then claim that the there exists $s > 0$ so that the resulting spacetime extends to $\{v = 0\}$ as a $C^{1,s}_vC^N\left(\mathbb{S}^2\right)C^N_u$ manifold with boundary for any positive integer $N$ satisfying $1 \ll N \ll N_2$. 
\end{proposition}
\begin{proof}We will let $\tilde{N}$ stand for a possibly changing positive integer which at the end of the proof will satisfy $1 \ll \tilde{N} \ll N_2$. We observe that the estimates of Lemma~\ref{3ioj2oijio45482} continue to hold in double-null coordinates induced by the vector field $X$. In order to ease the notation, we will in the rest of this proof use $\left(\Omega,b,\slashed{g}\right)$ and the other standard double-null terminology to refer to the components of the metric in the double-null coordinates induced by $X$. In particular, $\slashed{g}|_{(u,v) = (-1,0)} = e^{2\varphi}\mathring{\slashed{g}}$ for a suitable function $\varphi$. 

We introduce the convention that when we take derivatives with respect to an angular coordinate $\theta^A$ in the lapse-renormalized coordinates, we will write $\mathcal{L}_{\partial_{\hat{\theta}^A}}$ and we will reserve $\mathcal{L}_{\partial_{\theta^A}}$ to stand for a derivative with respect to $\theta^A$ taken in the double-null coordinates induced by the vector field $X$. Similarly, we will use $\mathcal{L}_{\partial_{\hat{u}}}$ to denote a derivative with respect to $u$ in the lapse-renormalized coordinates and we will reserve $\mathcal{L}_{\partial_u}$ for the derivative with respect to $u$ in the double-null coordinates induced by the vector field $X$. We then have the following formulas for when the various derivatives act on scalars:
\begin{equation}\label{32ioj3oij}
\partial_{\hat{v}} = \Omega^{-2}\partial_v,\ \  \partial_{\hat{\theta}_A} = \partial_{\theta^A} + 2\Omega^{-2}\left(\int_v^0\Omega^2\partial_{\theta^A}\log\Omega\, d\tilde{v} \right)\partial_v,\ \  \partial_{\hat{u}} = \partial_u + 2\Omega^{-2}\left(\int_v^0\Omega^2\partial_u\log\Omega\, d\tilde{v}\right) \partial_v.
\end{equation}
It will be useful to observe that one consequence of Lemma~\ref{32oi3ijo32ijo} and Sobolev inequalities is that for any $1 \ll \tilde{N} \ll N_2$ and after restricting to $\{u = -1\}$:
\begin{equation}\label{3ioj3ioj}
\sum_{j+\left|\alpha\right| \leq \tilde{N}}\left|\left(v\mathcal{L}_{\partial_v}\right)^j\mathcal{L}_{Z^{(\alpha)}}\left(\Omega^{-2}\left(\int_v^0\Omega^2\partial_{\theta^A}\log\Omega\, d\tilde{v} \right),\Omega^{-2}\left(\int_v^0\Omega^2\partial_u\log\Omega\, d\tilde{v} \right)\right)\right| \lesssim_{\epsilon,\gamma} (-v)^{1-d\gamma \epsilon^2},  
\end{equation} 
for a constant $d$ which is independent of $\gamma$ and $\epsilon$. By self-similarity, one immediately obtains that the same estimate holds if $u$ is allowed to range over any compact subset of $(0,\infty)$.

Now we turn to the expression~\eqref{2o3omo29495} and examine the various components of the metric. We start with $\slashed{g}_{AB}$.  In view of Lemma~\ref{23oij3jio23ij}, the estimate~\eqref{3ioj3ioj},  the formulas~\eqref{32ioj3oij}, and Lemma~\ref{3ioj2oijio45482} we see that $\slashed{g}_{AB}$ lies in  $C^{1,s}_vC^N\left(\mathbb{S}^2\right)C^N_u$  for any positive integer $N$ satisfying $1 \ll N \ll N_2$ and a suitable $s > 0$. (We may take, in fact, $s \gtrsim \epsilon^2$.)

We next come to metric component $\left(-4\int_v^0\left(\Omega^2\partial_{\theta^A}\log\Omega\right)d\tilde{v}-b_A\right)$ from~\eqref{2o3omo29495}. The key computation is that 
\begin{align}\label{3ioj23ioj23io}
\mathcal{L}_{\partial_{\hat{v}}}\left(-4\int_v^0\left(\Omega^2\partial_{\theta^A}\log\Omega\right)d\tilde{v}-b_A\right) &= 4\left(\partial_{\theta^A}\log\Omega - \frac{1}{4}\Omega^{-2}\slashed{g}_{AB}\mathcal{L}_{\partial_v}b^B - \frac{1}{4}\Omega^{-2}b^B\mathcal{L}_{\partial_v}\slashed{g}_{AB}\right) 
\\ \nonumber &= 4\eta_A - \frac{1}{4}\Omega^{-2}b^B\mathcal{L}_{\partial_v}\slashed{g}_{AB}.
\end{align} 
In view of~\eqref{3ioj23ioj23io} and Lemma~\ref{3ioj2oijio45482} we thus obtain that $-4\int_v^0\left(\Omega^2\partial_{\theta^A}\log\Omega\right)d\tilde{v}-b_A$ lies in  $C^{1,s}_vC^N\left(\mathbb{S}^2\right)C^N_u$  for any positive integer $N$ satisfying $1 \ll N \ll N_2$ and a suitable $s > 0$.

Lastly, we come to the metric component $-8\int_v^0\left(\Omega^2\partial_u\log\Omega\right)d\tilde{v} + \left|b\right|^2$ from~\eqref{2o3omo29495}. For this term, the key computation is that 
\begin{align}\label{3oij32io3oij9}
&\mathcal{L}_{\partial_{\hat{v}}}\left(-8\int_v^0\left(\Omega^2\partial_u\log\Omega\right)d\tilde{v} + \left|b\right|^2\right)  \\ \nonumber &\qquad =8\left(\mathcal{L}_{\partial_u}\log\Omega+\mathcal{L}_b\log\Omega\right) - 8\left(\mathcal{L}_b\log\Omega -\frac{1}{4}\Omega^{-2}\left(\mathcal{L}_{\partial_v}b^A\right)b_A\right) + \left(\mathcal{L}_{\partial_{\hat{v}}}\slashed{g}_{AB}\right)b^Ab^B
\\ \nonumber &\qquad =-16\left(\Omega\underline{\omega}\right) - 8\eta^Ab_A + \left(\mathcal{L}_{\partial_{\hat{v}}}\slashed{g}_{AB}\right)b^Ab^B.
\end{align}
In view of~\eqref{3oij32io3oij9} and Lemma~\ref{3ioj2oijio45482} we thus obtain that $-8\int_v^0\left(\Omega^2\partial_u\log\Omega\right)d\tilde{v} + \left|b\right|^2$ lies in  $C^{1,s}_vC^N\left(\mathbb{S}^2\right)C^N_u$  for any positive integer $N$ satisfying $1 \ll N \ll N_2$ and a suitable $s > 0$.

\end{proof}
\subsubsection{Gluing}\label{yayglue}
In this section we discuss the gluing of our interior solution to a suitable exterior solution from~\cite{nakedone}.
\begin{theorem}Let $\left(\mathcal{M}_{\rm int},g\right)$ be the solution in lapse-renormalized coordinates produced by Proposition~\ref{3ijo3iojio9} where we restrict the $u$ coordinate to $u \in (0,-1]$. Let $\left(\mathcal{M}_{\rm ext},g\right)$ be the solution produced by Theorem 1 from~\cite{nakedone} in the self-similar coordinates $\left(u,v,\theta^A\right)$ which is associated to the $4$-tuple $\left(e^{2\varphi}\mathring{\slashed{g}},b|_{\mathbb{S}^2_{-1,0}},2\Omega\underline{\omega}|_{v=0},1\right)$. We may introduce lapse-renormalized coordinates for $\left(\mathcal{M}_{\rm ext},g\right)$ via the formula from Definition~\ref{lapserenormaldef}. The manifold $\mathcal{M}_{\rm ext}$ then corresponds to the region $\{\left(\hat{v},u,\theta^A\right) \in [0,\infty)\times [-1,0) \times \mathbb{S}^2\}$. Finally, we may glue $\mathcal{M}_{\rm int}$ to $\mathcal{M}_{\rm ext}$ by identifying the boundary hypersurface $\{\hat{v} = 0\} \cap \{u \in (0,-1]\}$. Then we have that
\begin{enumerate}
	\item The resulting spacetime is a $C^{1,s}_vC^N\left(\mathbb{S}^2\right)C^N_u$ manifold  for any positive integer $N$ satisfying $1 \ll N \ll N_2$ and some $s > 0$.
	\item The resulting spacetime solves the Einstein vacuum equations ${\rm Ric}\left(g\right) = 0$.
\end{enumerate}
\end{theorem}
\begin{proof}We start with the regularity statement. By repeating \emph{mutatis mutandis} the proof of Proposition~\ref{3ijo3iojio9}, we have that the conclusions of Lemma~\ref{3ioj2oijio45482} and Proposition~\ref{3ijo3iojio9} also hold for $\left(\mathcal{M}_{\rm ext},g\right)$.\footnote{In fact, one has stronger estimates than provided by Lemma~\ref{3ioj2oijio45482} since, in particular, $\underline{\eta}$ is also well-behaved as $v\to 0^+$. However, these improvements will not be needed in this proof.} Thus, for the regularity statements for the glued spacetime, it suffices to establish that each component and $\mathcal{L}_{\partial_{\hat{v}}}$-derivative and angular derivatives thereof of the metric in the formula~\eqref{2o3omo29495} has the same value as $\hat{v}$ tends to $0$ from the left and from the right. In view of~\eqref{3ioj23ioj23io} and~\eqref{3oij32io3oij9}, it in fact suffices to show that the following double-null quantities (and their angular derivatives) continuously extend to $\{\hat{v} = 0\}$ and have the same values from the left and right
\[\slashed{g}, \qquad \Omega^{-1}\hat{\chi}, \qquad \Omega^{-1}{\rm tr}\chi, \qquad b, \qquad \eta,\qquad \Omega\underline{\omega}.\]
It is immediate from the construction that the two limits are the same for $\slashed{g}$, $b$ and $\Omega\underline{\omega}$. To see that $\eta$ is the same, we simply observe that either of the limiting values of $\eta|_{\hat{v} = 0}$ must satisfy the equation~\eqref{3pk2o294} with ${\rm Ric}_{3A}$ set to $0$ and $v = 0$. By Proposition~\ref{somestuimdie} this equation has a unique solution and hence we have agreement for the limits. Similarly,~\eqref{2momoo3} implies that the limit from the left and right of $\Omega^{-1}{\rm tr}\chi$ agree. Finally, the limits of $\Omega^{-1}\hat{\chi}$ agree in view of the equation~\eqref{kwdkodwok23dg} and Proposition 4.17 from~\cite{nakedone}. 

Next we establish that the spacetime solves the Einstein vacuum equations ${\rm Ric}(g) = 0$. Away from $\{\hat{v} = 0\}$ this is immediate, so we focus on regions which include $\{\hat{v} = 0\}$.  In view of~\eqref{32ioj3oij}, we have that
\[\Omega^{-1}e_4 = \Omega^{-2}\partial_v = \partial_{\hat{v}},\]
is a smooth vector field on the glued spacetime. Moreover,
\[\Omega e_3 \doteq \mathcal{L}_{\partial_u} + \mathcal{L}_b,\qquad e_A\]
both uniquely extend to $\{\hat{v} = 0\}$ as H\"{o}lder continuous vector fields on the glued spacetime, where $e_A$ refers to $\partial_{\theta^A}$ in the $\left(u,v,\theta^A\right)$ double-null coordinate system. We will use $\hat{e}_A$ to refer to a vector field $\partial_{\theta^A}$ in the lapse-renormalized coordinates and $\mathcal{L}_b$ is taken with respect to the $\left(u,v,\theta^A\right)$ coordinate system. We have that $\hat{e}_A$ is a smooth vector field on the glued spacetime. We will similarly define $\hat{e}_u$ to be $\partial_u$ with respect to the lapse-renormalized coordinates. Again this is a smooth vector field on the glued spacetime.

In view of the fact that ${\rm Ric} = 0$ holds classically away from $\{\hat{v} = 0\}$ we will be able to repeatedly use the following principles:
\begin{enumerate}
	\item If ${\rm Ric}\left(X,Y\right)$ is given by an expression where each term extends continuously to $\{\hat{v} = 0\}$, then it follows that ${\rm Ric}\left(X,Y\right) = 0$ holds classically everywhere.
	\item If ${\rm Ric}\left(X,Y\right) = \Omega^{-1}e_4\left(f\right) + h$, where $f$ extends continuously to $\hat{v} = 0$ and $h$ is locally integrable near $\hat{v} = 0$, then ${\rm Ric}\left(X,Y\right) = 0$ holds in a weak sense. Furthermore, the distribution $\Omega^{-1}e_4\left(f\right)$ is in fact an $L^1_{\rm loc}$ function. As a particular consequence, $p{\rm Ric}\left(X,Y\right) = 0$ as an $L^1_{\rm loc}$ function for any bounded function $p$. 
	\item If ${\rm Ric}\left(X,Y\right) =  \Omega^{-1}e_4\left(f\right) + h$, where $f$ extends continuously to $\hat{v} = 0$ and $h$ is continuous up to $\hat{v} = 0$, then $\Omega^{-1}e_4\left(f\right)$ is in  fact continuous up to $\hat{v} = 0$, and the equation ${\rm Ric}\left(X,Y\right) = 0$ holds everywhere classically.
\end{enumerate}

We now go through the various components of the Ricci tensor one by one:
\begin{enumerate}
\item  From~\eqref{4trchi} we have that 
	\begin{align*}
	{\rm Ric}\left(\Omega^{-1}e_4,\Omega^{-1}e_4\right) = -\left(\Omega^{-1}e_4\right)\left(\Omega^{-1}{\rm tr}\chi\right) - \frac{1}{2}\left(\Omega^{-1}{\rm tr}\chi\right)^2-\left|\Omega^{-1}\hat{\chi}\right|^2.
	\end{align*}
	Since $-\frac{1}{2}\left(\Omega^{-1}{\rm tr}\chi\right)^2-\left|\Omega^{-1}\hat{\chi}\right|^2$ and also $\Omega^{-1}{\rm tr}\chi$ are continuous up to $\{\hat{v} = 0\}$, it is a consequence of this equation holding away from $\{\hat{v} = 0\}$ that we in fact have that $\left(\Omega^{-1}e_4\right)\left(\Omega^{-1}{\rm tr}\chi\right)$ extends continuously to $\{\hat{v} = 0\}$. In particular we see that ${\rm Ric}\left(\Omega^{-1}e_4,\Omega^{-1}e_4\right) = 0$ holds classically everywhere.
	\item From~\eqref{4eta} and~\eqref{tcod1} we have that 
\begin{align}\label{3iojioj2}
&{\rm Ric}\left(\Omega^{-1}e_4,e_A\right) = -\left(\Omega^{-1}e_4\right)\eta  -\left(\Omega^{-1}\chi\right)\cdot\left(\eta-\underline{\eta}\right) + \Omega^{-1}\slashed{\nabla}^B\hat{\chi}_{AB}
\\ \nonumber &\qquad -\frac{1}{2}\Omega^{-1}\slashed{\nabla}_A{\rm tr}\chi-  \frac{1}{2}\Omega^{-1}{\rm tr}\chi \zeta_A +\Omega^{-1} \zeta^B\hat{\chi}_{AB}.
\end{align} 
Since $\eta$ is continuous at $\{\hat{v} = 0\}$ and 
\[-\left(\Omega^{-1}\chi\right)\cdot\left(\eta-\underline{\eta}\right) + \Omega^{-1}\slashed{\nabla}^B\hat{\chi}_{AB}
 -\frac{1}{2}\Omega^{-1}\slashed{\nabla}_A{\rm tr}\chi- \frac{1}{2}\Omega^{-1}{\rm tr}\chi \zeta_A +\Omega^{-1} \zeta^B\hat{\chi}_{AB}\]
 is locally integrable near $\{v = 0\}$, we see that ${\rm Ric}\left(\Omega^{-1}e_4,e_A\right) = 0$ everywhere as an $L^1_{\rm loc}$ function. In view of~\eqref{32ioj3oij}, we note that we have $\hat{e}_A = e_A + O\left(|v|^{1-O\left(\epsilon\right)}\right)\left(\Omega^{-1}e_4\right)$. Given that ${\rm Ric}\left(\Omega^{-1}e_4,\Omega^{-1}e_4\right) = 0$, we thus have that ${\rm Ric}\left(\Omega^{-1}e_4,\hat{e}_A\right) = 0$ holds everywhere as an $L^1_{\rm loc}$ function.

 \item From~\eqref{4uomega} and~\eqref{3trchi} we have that
 \begin{align}\label{3oij3ijo23i4}
& \frac{1}{4}{\rm Ric}\left(\Omega e_3,\Omega^{-1}e_4\right)  = \left(\Omega^{-1}e_4\right)\left(\Omega\underline{\omega}\right) - \frac{1}{2}\left|\eta\right|^2 + \eta\cdot\underline{\eta}
\\ \nonumber &\qquad \frac{1}{4}\left(\left(\Omega e_3\right)\left(\Omega^{-1}{\rm tr}\chi\right) + \frac{1}{2}\left(\Omega^{-1}{\rm tr}\chi\right)\left(\Omega{\rm tr}\underline{\chi}\right)-2\left(\Omega\underline{\omega}\right)\left(\Omega^{-1}{\rm tr}\chi\right) - 2\slashed{\rm div}\eta + 2\left|\eta\right|^2 + \left(\Omega^{-1}\hat{\chi}\right)\cdot\left(\Omega\hat{\underline{\chi}}\right)\right).
 \end{align}
 Since $\Omega\underline{\omega}$ is continuous up to $\{\hat{v} = 0\}$ and all the other terms on the right hand side of~\eqref{3oij3ijo23i4} are integrable, we conclude that ${\rm Ric}\left(\Omega e_3,\Omega^{-1}e_4\right) = 0$ holds everywhere as an $L^1_{\rm loc}$ function. Since we have that $\Omega e_3 = \hat{e}_u + b^A\hat{e}_A + O\left(v^{1-O\left(\epsilon\right)}\right)\left(\Omega^{-1}e_4\right)$, we may combine with the above to see that ${\rm Ric}\left(\hat{e}_u,\Omega^{-1}e_4\right) = 0$ holds weakly everywhere as an $L^1_{\rm loc}$ function.
 \item From~\eqref{3trchi} and~\eqref{slashr} we have that
	\begin{align*}
	{\rm R} + {\rm Ric}\left(\Omega e_3,\Omega^{-1}e_4\right) &= \left(\Omega e_3\right)\left(\Omega^{-1}{\rm tr}\chi\right) + \frac{1}{4}\left(\Omega^{-1}{\rm tr}\chi\right)\left(\Omega{\rm tr}\underline{\chi}\right)-4\left(\Omega\underline{\omega}\right)\left(\Omega^{-1}{\rm tr}\chi\right) - 2\slashed{\rm div}\eta 
	\\ \nonumber &\qquad - 2\left|\eta\right|^2+ K.
	\end{align*} 
	In view of Lemma~\ref{thefirstrelations}, the established regularity of our spacetime, and the above above analysis of the Ricci tensor, we thus see that $R = 0$ holds everywhere as an $L^1_{\rm loc}$ function.
\item From~\eqref{3hatchi} we have that
	\begin{align*}
	&\widehat{\rm Ric}\left(e_A,e_B\right) = \left(\Omega e_3\right)\left(\Omega^{-1}\hat{\chi}\right)_{AB} +\frac{1}{2}\left(\Omega {\rm tr}\underline{\chi}\right)\left(\Omega^{-1}\hat{\chi}\right)_{AB}
	\\ \nonumber &\qquad -2\left(\Omega\underline{\omega}\right)\left(\Omega^{-1}\hat{\chi}\right)_{AB} - \left(\slashed{\nabla}\hat\otimes \eta\right)_{AB} - \left(\eta\hat\otimes \eta\right)_{AB} + \frac{1}{2}\left(\Omega^{-1}{\rm tr}\chi\right)\left(\Omega \hat{\underline{\chi}}\right)_{AB}.
	\end{align*}
	Here $\widehat{\rm Ric}\left(e_A,e_B\right)$ stands for the trace-free part of the Ricci tensor considered as a symmetric $2$-tensor on $\mathbb{S}^2_{u,v}$. In view of Lemma~\ref{thefirstrelations} and the established regularity of our spacetime every term on the right hand is continuous up to $\hat{v} = 0$. Thus $\widehat{\rm Ric}\left(e_A,e_B\right) = 0$ holds in a classical sense everywhere. Since we have $\hat{e}_A = e_A + O\left(|v|^{1-O\left(\epsilon\right)}\right)\left(\Omega^{-1}e_4\right)$ we also conclude that $\widehat{\rm Ric}\left(\hat{e}_A,\hat{e}_B\right) = 0$ holds everywhere as an $L^1_{\rm loc}$ function. 
	\item From~\eqref{3pk2o294} we have  
	\begin{align*}
&{\rm Ric}\left(\Omega e_3,e_A\right) = -\frac{v}{u}\mathcal{L}_{\partial_v}\eta_A + \mathcal{L}_b\eta_A +\eta_A\left(\Omega{\rm tr}\underline{\chi}\right)+ 4\slashed{\nabla}_A\left(\Omega\underline{\omega}\right) +  \slashed{\nabla}^B\left(\Omega\hat{\underline{\chi}}\right)_{AB} - \frac{1}{2}\slashed{\nabla}_A\left(\Omega{\rm tr}\underline{\chi}\right).
\end{align*}
In view of Lemma~\ref{thefirstrelations} and the established regularity of our spacetime, we see that every term on the right hand side is continuous up to $\hat{v}  = 0$. Hence ${\rm Ric}\left(\Omega e_3,e_A\right) = 0$ holds in a classical sense everywhere. Analogously to the arguments above, we also then have that ${\rm Ric}\left(e_{\hat{u}},\hat{e}_A\right) = 0$ everywhere as an $L^1_{\rm loc}$ function.
	\item From~\eqref{3truchi} we have that
	\[{\rm Ric}\left(\Omega e_3,\Omega e_3\right) = \left(\Omega e_3\right)\left(\Omega{\rm tr}\underline{\chi}\right) + \frac{1}{2}\left(\Omega{\rm tr}\underline{\chi}\right)^2 + 4\left(\Omega\underline{\omega}\right)\left(\Omega{\rm tr}\underline{\chi}\right) + \left|\Omega\underline{\hat{\chi}}\right|^2.\]
	In view of Lemma~\ref{thefirstrelations} and the established regularity of our spacetime all of the terms on the right hand side are continuous up to $\hat{v} = 0$. Hence the equation ${\rm Ric}\left(\Omega^{-1}e_3,\Omega^{-1}e_3\right) = 0$ holds in a classical sense everywhere. Just as above, we then also conclude that ${\rm Ric}\left(\hat{e}_u,\hat{e}_u\right) = 0$ everywhere as an $L^1_{\rm loc}$ function.

\end{enumerate}

Combining all of these assertions, we have shown, in particular, that ${\rm Ric}\left(X,Y\right) = 0$ holds everywhere as an $L^1_{\rm loc}$ function whenever $X,Y \in \left\{\Omega^{-1}e_4,\hat{e}_u,\hat{e}_A\right\}$. This concludes the proof.

\end{proof}

\appendix

\section{Controlling $\Pi_{{\rm ker}\left({}^*\slashed{\mathcal{D}}_2\right)}$ and $\Pi_{{\rm ker}\left({}^*\slashed{\mathcal{D}}_1\right)}$: Proof of Lemma~\ref{ijmom2om392}}\label{ioj2ojo2}
In this appendix we provide the proof of Lemma~\ref{ijmom2om392}. The argument is standard but we provide a proof for completeness.
\begin{proof}We first verify that ${\rm ker}\left({}^*\slashed{\mathcal{D}}_2\right)$ is $6$-dimensional. By the uniformization theorem, there exists a function $\varphi$ and a diffeomorphism $\Phi$ so that $\slashed{g} = e^{2\varphi}\Phi^*\left(\mathring{\slashed{g}}\right)$. In particular, if $X_A \in {\rm ker}\left({}^*\slashed{\mathcal{D}}_2\right)$, then $\slashed{g}^{AB}X_B$ is a conformal Killing vector field of $\Phi^*\left(\mathring{\slashed{g}}\right)$. This clearly yields an isomorphism between ${\rm ker}\left({}^*\slashed{\mathcal{D}}_2\right)$ and the conformal Killing fields of $\Phi^*\left(\mathring{\slashed{g}}\right)$. Since the latter group is $6$-dimensional, we conclude that ${\rm ker}\left({}^*\slashed{\mathcal{D}}_2\right)$ is also six dimensional. 

It follows from the formulas~\eqref{d2ids} that for any metric of positive Gauss curvature ${\rm ker}\left(\slashed{\mathcal{D}}_2\right) = \{0\}$. Since the assumption~\eqref{km2omo} of course implies that $\slashed{g}$ has positive curvature, we obtain from~\eqref{d2ids} that
\[{\rm ker}\left({}^*\slashed{\mathcal{D}}_2\right) = {\rm ker}\left(\slashed{\mathcal{D}}_2{}^*\slashed{\mathcal{D}}_2\right) = {\rm ker}\left(\slashed{\Delta}+K\right).\]
This is an elliptic operator and~\eqref{i3omromo3} follows now immediately from elliptic estimates.

Let us set $\slashed{L} \doteq -\slashed{\Delta}-K$ and $\mathring{L} \doteq -\mathring{\Delta}-1$. For the round sphere, the spectrum of $\mathring{L}$ is easily computed via the use of spherical harmonics (see Section~\ref{sechodge1form}). In particular, we see that the first six eigenvalues $\mathring{\lambda}_1$, $\cdots$, $\mathring{\lambda}_6$ of $\mathring{L}$ are all equal to $0$, and also that the seventh eigenvalue satisfies $\mathring{\lambda}_7 = 5$. It follows from the ``min-max principle'' and~\eqref{km2omo} that the seventh eigenvalue $\slashed{\lambda}_7$ of $\slashed{L}$ must satisfy $\slashed{\lambda}_7 -5 \gtrsim -\epsilon$. Since we have already established that the kernel of ${}^*\slashed{\mathcal{D}}_2$, and hence  $\slashed{L}$, is $6$-dimensional, we see that the first six eigenvalues of $\slashed{L}$ must all be $0$. 

We may, of course, extend the operator $\slashed{L}$ to the complexification of the space of $1$-forms. Let $\Gamma$ denote a circle of radius $1$ about the origin in the complex plane. Then, for any $1$-form $\vartheta$, we have the following consequence of Cauchy's integral formula, the eigenfunction expansion for $\vartheta$, the fact that $\slashed{\lambda}_i = 0$ for $i = 1,\cdots,6$, that fact that $\slashed{\lambda}_7 -5 \gtrsim -\epsilon$:
\begin{equation}\label{okm3omo32}
\Pi_{{\rm ker}\left(\slashed{L}\right)}\vartheta = \frac{-1}{2\pi i}\oint_{\Gamma}\left(\slashed{L}-\zeta\right)^{-1}\vartheta d\zeta.
\end{equation}
From this one may also obtain that 
\begin{equation}\label{okm3omo321234292}
\left[\partial_{\tau},\Pi_{{\rm ker}\left(\slashed{L}\right)}\right]\vartheta = \frac{-1}{2\pi i}\oint_{\Gamma}\left(\slashed{L}-\zeta\right)^{-1}\left(\left[\slashed{L},\partial_{\tau}\right]\left(\slashed{L}-\zeta\right)^{-1}\vartheta\right) d\zeta,
\end{equation}
\begin{equation}\label{k2om2om2r}
\left(\Pi_{{\rm ker}\left(\slashed{L}\right)}-\Pi_{{\rm ker}\left(\tilde{\slashed{L}}\right)}\right)\vartheta = \frac{-1}{2\pi i}\oint_{\Gamma}\left[\left(\slashed{L}-\zeta\right)^{-1}-\left(\tilde{\slashed{L}}-\zeta\right)^{-1}\right]\vartheta d\zeta,
\end{equation}
where $\tilde{\slashed{L}}$ is associated to the metric $\tilde{\slashed{g}}$. The estimates~\eqref{kjfewijoewfiowefio} and~\eqref{mfo3mo2p1} follow easily from these formulas and elliptic estimates.

Finally, the estimates~\eqref{khuhiu3iu2}-\eqref{ljfjii2p3o} are easily established from the simple formula
\[\Pi_{{\rm ker}\left({}^*\slashed{\mathcal{D}}_1\right)}\left(f,h\right) = \frac{1}{{\rm Vol}(\slashed{g})^{1/2}}\int_{\mathbb{S}^2}\left(f,h\right)\slashed{\rm dVol}.\]
\end{proof}

\end{document}